%% file: CCFW_2026_AER.tex
\documentclass[11pt]{article}
\usepackage{amsfonts,amssymb,amsmath,amsthm,dsfont,graphicx,enumitem,bm,subcaption,mathrsfs,xcolor}
\usepackage{hyperref}
\hypersetup{pdfborder = {0 0 0},colorlinks=true,linkcolor=blue,urlcolor=blue,citecolor=blue}
\input{arxiv-crossrefs.tex}
\usepackage[onehalfspacing]{setspace}
\usepackage{natbib}
\usepackage[top=1in,bottom=1in,left=1in,right=1in]{geometry}

\graphicspath{{figures/}}
\usepackage{algorithm}
\usepackage{algpseudocode}
\usepackage{comment}

\newtheorem{assumption}{Assumption}
\newtheorem{theorem}{Theorem}[section]

\newtheorem{corollary}{Corollary}[theorem]

\theoremstyle{definition}
\newtheorem{remark}[theorem]{Remark}

\numberwithin{equation}{section}

\DeclareMathOperator*{\len}{diam}

\renewcommand{\P}{\mathbb{P}}

\newcommand{\E}{\mathbb{E}}
\newcommand{\Var}{\mathbb{V}\mathrm{ar}}

\newcommand{\I}{\mathds{1}}
\newcommand{\IF}{\mathds{1}}

\begin{document}

\title{\vspace{-0.0in}Beta-Sorted Portfolios\thanks{We would like to thank Nina Boyarchenko, Mehmet Caner, Fernando Duarte, David Lucca, Olivier Scaillet, and participants at various seminars, workshops, and conferences for helpful comments and discussions. We also thank the editor and three anonymous referees for helpful comments and suggestions that substantially improved the paper.  The views expressed in this paper are those of the authors and do not necessarily represent those of the Federal Reserve Bank of New York or the Federal Reserve System.  Dennis Kongoli provided excellent research assistance. Cattaneo gratefully acknowledges financial support from the National Science Foundation (SES-1947662, DMS-2210561, and SES-2241575), and from the John Simon Guggenheim Memorial Foundation through a 2026 Guggenheim Fellowship.
Feng gratefully acknowledges financial support from the National Natural Science Foundation of China (NSFC) through grants 72522001 and 72203122.
Wang gratefully acknowledges financial support from the ESRC (Grant Reference: ES/T01573X/1).}
\bigskip }
\author{Matias D. Cattaneo\thanks{Department of Economics, Princeton University.} \and
	    Richard K. Crump\thanks{Capital Markets, Federal Reserve Bank of New York.} \and
        Yingjie Feng\thanks{School of Economics and Management, Tsinghua University.} \and
	    Weining Wang\thanks{School of Economics, University of Bristol.}}
\maketitle

\begin{abstract}
    Beta-sorted portfolios---portfolios comprised of assets with similar covariation with selected risk factors---are a popular tool in empirical finance to analyze models of (conditional) expected returns. Despite their widespread use, little is known of their econometric properties in contrast to comparable procedures such as two-pass regressions. We formally investigate the properties of beta-sorted portfolio returns by casting the procedure as a two-step nonparametric estimator with a nonparametric first step and a beta-adaptive portfolio construction. Our framework rationalizes the well-known estimation algorithm with precise economic and statistical assumptions on the general data-generating process. We provide conditions which ensure valid estimation and inference allowing for a range of hypotheses of interest in financial applications. We show that the rate of convergence of the estimator changes depending on the value of beta. We demonstrate that valid inference depends critically on the object of interest and discuss drawbacks of the widely used Fama-MacBeth variance estimator.  To address these limitations, we propose a new variance estimator.  We demonstrate the usefulness of our theoretical results in two empirical applications, including one in which we introduce a novel risk factor that captures the business credit cycle and show that it predicts both the cross-sectional and time-series behavior of U.S. stock returns.
\end{abstract}


\thispagestyle{empty}
\clearpage

\doublespacing
\setcounter{page}{1}
\pagestyle{plain}

\pagestyle{plain}

\section{Introduction}\label{sec:Introduction}

Deconstructing expected returns into asset-specific factor loadings and corresponding prices of risk for interpretable factors is an evergreen pursuit in the empirical finance literature. When factors are observable, there are two workhorse approaches that continue to enjoy widespread use. The first approach, Fama-MacBeth two-pass regressions, has been extensively studied in the financial econometrics literature.\footnote{See, for example, \cite{jagannathan1998asymptotic}, \cite{shanken2007estimating}, \cite{kleibergen2009tests}, \cite{AngLiuSchwarz2010}, \cite{gospodinov2014misspecification}, \cite{adrian2015regression}, \cite{gagliardini2016ecmta}, \cite{RRZ2020}, \cite{giglio2021asset}  and many others. For a recent survey, see \cite{gagliardini2020estimation}.}  The second approach, which we refer to as beta-sorted portfolios, has received scant attention in the econometrics literature despite its empirical popularity.\footnote{The empirical literature using beta-sorted portfolios is extensive. For a textbook treatment, see \cite{BEM2016}, and for a few recent papers see, for example, \cite{boons2020time}, \cite{ChenHanPan2021}, \cite{Eisdorferetal2021}, \cite{GoldbergNozawa2021}, \cite{FanLondonoXiao2022}, \cite{CDHW2023}, and
\cite{chen2024investor}.}

The implementation of beta-sorted portfolios entails the following two-step procedure, which incorporates a beta-adaptive portfolio construction (see, e.g., \citealt{BEM2016}).  In a first step, time-varying risk factor exposures (betas) are estimated through (backward-looking) weighted time-series regressions of asset returns on the observed factors. The most popular implementation uses rolling window regressions, often with a choice of a five-year window. In a second step, the estimated factor exposures, based on data up to the previous period, are ordered and used to group assets into portfolios. These portfolios then represent assets with a similar exposure to the risk factors, and the magnitude of the return differential for differently exposed assets is used to assess the compensation for bearing this common risk. Most frequently this is achieved by differencing the portfolio returns from the two most extreme portfolios.  Finally, an average over time of these return differentials is taken to infer whether the risk is priced unconditionally---whether the long/short portfolio earns systematic (and significant) excess returns. Notwithstanding the simple and intuitive nature of the methodology, little is known of the formal properties of this estimator and its associated inference procedures.

We provide a comprehensive framework to study the economic and statistical properties of beta-sorted portfolios.  We first translate the two-step estimation algorithm with beta-adaptive portfolio construction into a corresponding econometric model. We show that the model has key features which are important to consider for valid interpretation of the empirical results. Notably, we do not impose no-arbitrage conditions. Instead, the restrictions implied by no arbitrage become testable hypotheses.  Furthermore, our framework  precisely clarifies how the dynamics of the risk factors relate to the functional form of conditional expected returns.  Within this framework, we introduce general sampling assumptions allowing for time-varying factor loadings, serially-correlated factors, and conditional heteroskedasticity across time and assets.  We then study the asymptotic properties of the beta-sorted portfolio estimator and associated inference procedures in settings with large cross-sectional, $n$, and time-series, $T$, sample sizes (i.e., $n,T\to \infty$).

Our analysis yields new methodological and theoretical results for this popular procedure. At any fixed date, the beta-sorted portfolio estimator is centered on the sum of two terms: conditional expected returns and a linear combination of the common risk-factor innovations. By a law of large numbers, the time average of the second term converges to zero, so the estimator targets the sample average of conditional expected returns (SACER), which we argue is the natural object of interest. We establish uniform consistency for this time-averaged estimator under transparent high-level conditions. The key requirements are that the first-step beta estimates become uniformly accurate, the portfolios become sufficiently homogeneous in beta values, and each portfolio contains enough assets. Importantly, this consistency result does not impose a particular dynamic model for the betas and can accommodate commonly used parametric and nonparametric specifications (e.g., \citealt{adrian2015regression}, \citealt{gagliardini2016ecmta}, \citealt{GuKellyXiu2020}).

These findings have direct implications for how beta-sorted portfolios should be constructed and reported.  To estimate conditional expected returns for a fixed value of $\beta$, a researcher must track the portfolio containing that value in each period.  Mechanically averaging portfolios with the same rank over time, as is commonly done, can combine different beta ranges as the cross-sectional distribution of these risk-factor exposures generally changes.  The first-step window size, $H$, and the number of portfolios, $J$ in the single-factor case, should both adapt to the available time-series and cross-sectional sample sizes rather than be fixed at conventional choices such as five years and ten portfolios.  We develop data-driven choices for $H$ and $J$ and provide practical guidance for their use in estimation and inference.

Our distribution theory uncovers a sharp difference in the estimator's behavior at nonzero and at zero risk-factor exposures.  For any fixed nonzero evaluation point, the estimator retains a common risk-factor component that is unaffected by the number of assets in each cross-section, so that the time-averaged estimator converges at only a $\sqrt{T}$ rate.  When evaluated at the origin, this common component disappears and idiosyncratic variation---including the additional uncertainty from estimating the risk-factor exposures---becomes the dominant source of uncertainty.  Under our stated conditions, the estimator can attain the faster $\sqrt{nT/J^d}$ rate.  This special role of zero follows from the model’s factor structure and extends beyond the beta-sorted portfolio estimator.

We propose a variance estimator that separately estimates the common risk-factor and idiosyncratic components and accounts for the dependence created by overlapping first-step windows.  We show that it provides valid, although potentially conservative, inference.  To address the changing behavior near zero, we develop a new inference procedure that remains valid across different values of $\beta$, including at and near the origin. The procedure uses a data-driven rule to identify the region near zero and a conservative variance adjustment to ensure that the associated confidence intervals attain at least their nominal coverage asymptotically.  We also show that the ubiquitous Fama-MacBeth (FM) variance estimator \citep{FamaMacBeth1973} is generally conservative for inference on the SACER at nonzero evaluation points but may not be valid at the origin, where it can either overestimate or underestimate uncertainty.  

Our framework accommodates multiple risk factors and clarifies the trade-offs among common ways of constructing portfolios. Independent sorts form $J$ groups based on each of the $d$ estimated risk-factor exposures and then intersect them, producing $J^d$ portfolios that may contain too few assets as $d$ increases.  Estimating several betas but sorting on only one avoids this issue.  However, this approach changes the object being estimated. We characterize this alternative estimand and explain how our framework can be adapted to conduct inference on it. Conditional sorts first group assets by one estimated risk-factor exposure and then divide each group further using additional exposures. This approach can also help avoid empty portfolios, but does not eliminate the dimensionality problem.  In this case, recovering the SACER at a fixed beta requires tracking the exposure ranges represented by each portfolio. Averaging returns or spreads by sequential rank across groups, as is commonly done in practice, generally combines different exposure regions, complicating interpretation.

Certain linear combinations of beta-sorted portfolio returns eliminate the common risk-factor component and, when constructed appropriately, can attain the same faster rate. A butterfly portfolio is one such ``factor-cancelling contrast'' and measures a discrete second difference in the SACER, thereby capturing curvature in the relationship between expected returns and risk-factor exposures. Separately, we develop a test of the affine restriction implied by no arbitrage for raw returns and the corresponding linear restriction for excess returns.

From a theoretical perspective, beta-sorted portfolios present a number of technical challenges since the estimator relies on two nested nonparametric estimation steps together with a portfolio construction based on a first-step nonparametric generated regressor. More precisely, risk-factor exposures estimated nonparametrically in the first step enter directly into the nonsmooth partitioning scheme which complicates the analysis.\footnote{For analysis of partitioning-based nonparametric estimators see \cite{Cattaneo-Farrell-Feng_2020_AoS} and references therein. Partitioning-based estimators with random basis functions have been recently studied in \cite{cattaneo2020characteristic} and \cite{Cattaneo-Crump-Farrell-Feng_2024_AER}, but in those papers the conditioning variables are observed, while here the conditioning variable is generated using a preliminary time-series (nonparametric) regression, and therefore prior results are not applicable to the settings considered herein.} To our knowledge, we are the first to prove validity of such an approach.

This paper is most related to the large literature studying asset pricing models with observable factors.\footnote{See, for example, \cite{goyal2012empirical}, \cite{nagel2013empirical}, \cite{GospodinovRobotti2013}, or \cite{gagliardini2020estimation} for surveys. A related literature endeavors to jointly estimate factor loadings and \emph{latent} risk factors. See, for example, \cite{ConnorLinton2007}, \cite{connor2012efficient},  \cite{kelly2019characteristics}, and \cite{BCLT2024}.} Given our focus on conditional asset pricing models with large panels in both the cross-sectional and time-series dimensions, this paper is most closely related to \cite{gagliardini2016ecmta} \citep[see also][]{gagliardini2020estimation}. The authors introduce a general framework and econometric methodology for inference in large-dimensional conditional factor models under no-arbitrage restrictions. They model risk-factor exposures as parametric functions of observable variables and establish conditions for consistent estimation and inference on the prices of risk.  Although the statistical model under study shares important similarities with the setup of \cite{gagliardini2016ecmta}, there are substantial differences, and the models explored previously in the literature do not nest our setup. For example, our framework for the classical beta-sorted portfolio estimator accommodates data-generating processes that do not necessarily exclude arbitrage opportunities and allow risk-factor exposures to vary smoothly over time.  Furthermore, unlike the existing literature (e.g., \citealt{adrian2015regression}, \citealt{gagliardini2016ecmta}), we establish valid estimation and inference without specifying the functional form of the conditional mean of the risk factors.

Our paper is also related to the financial econometrics literature on nonparametric estimation and inference. In particular, the two steps of the beta-sorted portfolio algorithm align individually with \cite{ang2012testing}, who study kernel regression estimators of time-varying alphas and betas, and \cite{cattaneo2020characteristic}, who study portfolio sorting estimators given observed individual characteristic variables. However, the individual results from each of these papers cannot be applied in our multi-step setting.  Furthermore, the linkage between the two steps, through the role of the generated (nonparametrically estimated) regressor in the second-stage nonparametric partitioning estimator, represents a substantial technical challenge and has not been studied before.

We demonstrate the practical usefulness of our estimation and inference results in two empirical applications.  First, we introduce a new risk factor---a measure of the business credit cycle---based on the Federal Reserve's Senior Loan Officer Opinion Survey and the ISM Manufacturing Index.  We show that this factor is predictive of both the cross-sectional and time-series behavior of U.S. stock returns.  Second, we revisit the capital share factor from \cite{LLM2019}.  Using our new inference procedures, we find no evidence that the capital share risk factor is priced in the cross-section of U.S. stock returns.

We collect additional information and technical details in two Supplemental Appendices. The Supplemental Appendix: Theory (SA-Theory) develops the technical foundations for our results, including primitive conditions, additional theoretical results,
and proofs. The Supplemental Appendix: Applied (SA-Applied) provides implementation details, data-driven guidance for tuning-parameter selection, simulation evidence, and the second empirical application. Unless otherwise stated, asymptotic limits are taken as $n,T\to\infty$, with $H$ and $J$ varying as specified below. We use $O_\P(1)$ and $o_\P(1)$ for sequences that are bounded in probability and converge to zero in probability, respectively. For positive random sequences $a_n$ and $b_n$, $a_n\asymp_\P b_n$ means that both $a_n/b_n$ and $b_n/a_n$ are $O_\P(1)$.  For positive deterministic sequences $a_n\asymp b_n$ means both $a_n/b_n$ and $b_n/a_n$ are $O(1)$. Finally, for a set $S\subseteq\mathbb R^d$, define its diameter as $\len(S)=\sup\{\|s-s'\|:s,s'\in S\}$. 

\section{Model setup}\label{section2}

We introduce a general econometric model of asset returns and show how it aligns with the two steps that comprise the beta-sorted portfolio algorithm. We explain the rationale behind the form of the estimator and discuss relevant properties of the model.

\subsection{Modeling returns}

Let $R_{it}$ denote the return of asset $i$ at time $t$, and $f_t$ a vector of observable risk factors with $f_t \in \mathbb{R}^d$.  We assume that asset returns are generated by the linear stochastic coefficient model,
\begin{align}
        R_{it}  = \alpha_{it} + \beta_{it}^{\top} f_t + \varepsilon_{it}, 
        \qquad \E\left[ \left. \varepsilon_{it} \right| \mathcal{F}_{t-1}, f_t \right]=0,
        \quad i= 1, \cdots, n_t, \quad t = 1, \cdots, T, \label{mainmodel}
\end{align}
where $\alpha_{it}$ and $\beta_{it}$ are random coefficients, $ \varepsilon_{it}$ is an idiosyncratic error term, and $\mathcal{F}_{t}$ is the information set up to time $t$.\footnote{For an alternative example of a random coefficient model tailored to a financial application, see \cite{BarrasGagliardiniScaillet2022}.}  The sigma field $\mathcal{F}_{t}$ will, in general, depend on $T$ and $\{n_s: s<t\}$ but we suppress this dependence for simplicity of notation. We assume that $\alpha_{it}$ and $\beta_{it}$ are both $\mathcal{F}_{t-1}$-measurable so we have that $\E[ \left. R_{it} \right| \mathcal{F}_{t-1},f_t] = \alpha_{it} + \beta_{it}^\top f_t$.    Finally, notice that equation \eqref{mainmodel} accommodates an unbalanced panel.  Although each $n_t$ may be different, we assume that they all grow at the same rate which ensures that each cross-section contributes to the asymptotic properties of the estimator (i.e.,  $n_t \asymp n $ for $t = 1,\cdots, T$).

To obtain the structural form of our model, we assume that there exists a non-random function $\mu_0(\cdot)$ such that
\begin{equation}\label{csmodel}
    \E(R_{it}|\mathcal{F}_{t-1}) = \mu_0(\beta_{it}; \mathcal{G}_{t-1}) = \mu_{t}(\beta_{it}),
\end{equation}  
where $\mathcal{G}_{t}$ is the sub-sigma field of $\mathcal{F}_{t}$ generated only by $i$-invariant random variables such as time-series like $f_t$.  Equation \eqref{csmodel} restricts conditional expected returns to be a function of $\beta_{it}$ only, but it allows the functional form to be random and to vary with past realizations of economy-wide random variables.\footnote{This allows considerable flexibility in conditional expected returns, including time variation in the importance of economy-wide variables and in the strength of return predictability \citep[e.g.,][]{FST2023}.} This restriction requires assets with the same risk-factor exposures (as measured by $\beta_{it}$) to have the same conditional expected return in a given period.  For the discussion in this subsection, we take $R_{it}$ to denote excess returns for simplicity of presentation. If $R_{it}$ instead denotes raw returns, $\mu_t(0)$ may be interpreted as the zero-beta rate at time $t$.

Since the economy's state variables will reside in $\mathcal{G}_{t}$, a natural restriction is to impose that
\begin{equation}\label{condexp}
\E[f_t|\mathcal{F}_{t-1}]=\E[f_t|\mathcal{G}_{t-1}], \qquad t=1,2,\dots,T.
\end{equation}
This economic restriction also ensures that the systematic component of $R_{it}$ is determined only by $\beta_{it}$ and economy-wide variables.  Using equation \eqref{condexp} along with equations \eqref{mainmodel} and \eqref{csmodel}, we obtain,
\begin{equation}\label{model}
    \alpha_{it} = \alpha_t(\beta_{it}), \qquad 
      \alpha_t(\beta) = \mu_{t}(\beta) - \beta^{\top}\E(f_t|\mathcal{G}_{t-1}),
\end{equation}
which clarifies the restriction on $\alpha_{it}$ implied by equation \eqref{csmodel}.  Equation \eqref{model} is the nonparametric analog to a reduced-rank restriction in parametric models as it imposes cross-equation restrictions between the linear coefficients in equation \eqref{mainmodel}.  In fact, equation \eqref{model} nests as a special case the assumption of no arbitrage which directly implies a parametric reduced-rank regression model \citep[e.g.,][]{adrian2015regression}.  As shown in \cite{gagliardini2016ecmta}, in the absence of arbitrage opportunities, there exists a $\lambda_t$ such that $\mu_{t}(\beta) = \beta^\top\lambda_t$ for all $\beta$ which implies $\alpha_{t}(\beta) =  \beta^{\top} \left( \lambda_t - \E(f_t|\mathcal{G}_{t-1}) \right)$.

Finally, combining equations \eqref{mainmodel} and \eqref{model}, we arrive at the structural form
\begin{equation}\label{structmodel}
    R_{it} = \mu _{t}(\beta _{it}) + \beta_{it}^{\top}(f_{t}-\E[f_{t}\vert \mathcal{G}_{t-1}]) + \varepsilon_{it}. 
\end{equation}
Equation \eqref{structmodel} may be compared to the standard beta pricing model \citep[e.g.,][Chapter 12]{Cochrane2005} and generalizations thereof \citep[e.g.,][]{cochrane1996jpe,adrian2015regression,gagliardini2016ecmta}. The most noteworthy difference is the presence of the (possibly) nonlinear, time-varying function $\mu _{t}\left( \beta _{it}\right)$ in equation \eqref{structmodel}.  When $\mu_t(\cdot)$ is nonlinear, the model departs from the no-arbitrage pricing restriction discussed above. Such departures can be motivated by structural models featuring violations of the law of one price.  For example, equation \eqref{structmodel} could arise in the margin-constraints model of \cite{garleanu2011margin} under the assumption that the security's margin is a nonlinear function of its (past) beta.

\subsection{Relation to the two-step estimator}

To understand the rationale behind the beta-sorted portfolios algorithm, it is instructive to (temporarily) make the simplifying assumption that $\beta_{it}$ is observable.  Let us  define systematic realized returns as
\begin{equation}
\label{sysrealrets}
  {M}_t(\beta) = \mu_{t}(\beta) + \beta^{\top}(f_{t}-\E[f_{t}\vert \mathcal{G}_{t-1}]),
\end{equation}
which implies that ${M}_t(\beta_{it}) = \alpha_{it} + \beta_{it}^{\top}f_t$ and 
\begin{align}
    R_{it}  = M_t(\beta_{it}) +  \varepsilon_{it}. \label{eq:Mt}
\end{align}
This equation makes clear that, for a fixed time period, we can only consistently (nonparametrically as $n_t \to \infty$) estimate the unknown function $M_t(\cdot)$ rather than the direct object of interest $\mu_t(\cdot)$.  This comes about because both terms in $M_t(\beta)$ are an unknown transformation of $\beta$ as they involve $\mu_t(\cdot)$ and $\E[f_t|\mathcal{G}_{t-1}]$ neither of which are observed.  However, its sample average may be written as,
\begin{equation}\label{M_decomp}
    \frac{1}{T} \sum_{t=1}^{T} {M}_t(\beta) = \frac{1}{T} \sum_{t=1}^{T} \mu_t(\beta) + \frac{1}{T} \sum_{t=1}^T \beta^\top(f_t - \E[f_t|\mathcal{G}_{t-1}]).
\end{equation}
Importantly, even though we do not observe $\E[f_t|\mathcal{G}_{t-1}]$, the second term has summands, $\beta^\top(f_t - \E[f_t|\mathcal{G}_{t-1}])$, which are a martingale difference sequence with respect to $\mathcal{G}_{t}$. Under mild conditions a law of large numbers will hold, and this sample average will converge to zero in probability so that
\begin{equation}\label{M_decomp2}
    \frac{1}{T} \sum_{t=1}^{T} {M}_t(\beta) = \frac{1}{T} \sum_{t=1}^{T} \mu_t(\beta) + o_\P(1).
\end{equation}
This demonstrates how $T^{-1} \sum_{t=1}^T \mu_t(\beta)$, the sample average of conditional expected returns, can be consistently estimable even if the individual $\mu_t(\beta)$ are not.  A notable exception occurs when $\beta=0$ as the linear term in $M_t(\beta)$ is eliminated (see Remark \ref{rem:beta0} for further discussion). 

In practice, we must estimate $M_t(\cdot)$ from the data (which requires estimates of the $\beta_{it}$) to obtain a feasible estimator of $T^{-1} \sum_{t=1}^T \mu_t(\beta)$.  In this setting the choice of estimator is informed by additional considerations beyond just statistical properties.  Specifically, empirical practice in finance has revolved around two considerations.  First, the estimator should have the interpretation of a feasible investment strategy, so that estimates of expected returns and their statistical significance can be directly assessed in economic terms.
Second, the estimator should have a ``real-time'' formulation, i.e., when an additional time period of data is added, prior estimates are unaffected.

Taking these two considerations in mind translates to the following two-step procedure which is ubiquitous in the empirical finance literature:

\paragraph{Step One:} Using only information up to time $t-1$, construct $\widehat{\beta}_{it}$ for $i = 1, \ldots, n_t$.  The classical implementation uses rolling window OLS regressions,
\begin{equation}\label{eq:betahat_roll}
\begin{pmatrix}
\widehat{\alpha}_{it}  \\[-.9 ex] \widehat{\beta}_{it}
\end{pmatrix}
= \underset{a\in\mathbb R,\;b\in\mathbb R^d}{\arg\min}
\sum_{s=1}^{H}\bigl(R_{i(t-s)}-a-b^\top f_{t-s}\bigr)^2,
\end{equation}
where $H$ is the window size.  This approach can be formally justified under the assumption that $\alpha_{it}$ and $\beta_{it}$ are smoothly varying (random) coefficients which we discuss further in Section \ref{section4}.

\paragraph{Step Two:} Based on step one, group assets together with similar values of $\widehat{\beta}_{it}$ and form portfolios.  The classical implementation uses marginal quantiles from the cross-sectional distribution of $\widehat{\beta}_{it}$ to separate the assets.  Next, using the individual returns for time $t$, calculate the return for each portfolio.  Finally, average the portfolio returns over time to assess the unconditional risk properties of the investment strategy.
\medskip

Portfolios have been established as the fundamental investment object since at least \cite{Markowitz1952,Markowitz1959} and \cite{Tobin1958} and so are the natural choice to link returns, $R_{it}$, to risk exposures, $\beta_{it}$.  However, since the $\beta_{it}$ are unobserved they must first be estimated in a preliminary step.  Moreover, to ensure the portfolio returns could have been earned in real time (without look ahead bias) each  $\beta_{it}$ is estimated using only information up to time $t-1$.

To formally write the beta-sorted portfolio estimator, let $\widehat{\Delta}_t = \{\widehat{P}_{jt}: 1\leq j\leq \bar{J}_t\}$ be a collection of nonoverlapping cells covering the support $\mathcal B$ and all estimated betas, based on $\{\widehat{\beta}_{it}: i=1,\ldots,n_t\}$. The $\widehat{P}_{jt}$ are portfolio designations; specifically, each asset $i$ is allocated to a portfolio, $\widehat{P}_{jt}$, depending on its value for $\widehat{\beta}_{it}$. For example, in the classical implementation, portfolio designation is governed by the marginal quantiles of the (cross-sectional) distribution of $\widehat{\beta}_{it}$, i.e., when $d=1$,
\begin{align}
\widehat{P}_{1t} &= \Big[ \min_{1 \leq i \leq n_t} \widehat{\beta}_{it}, \;\widehat{F}_{\widehat{\beta},t}^{-1}\big(1/\bar{J}_{t}\big)  \Big), \nonumber \\[0.5 ex]
\widehat{P}_{jt} &= \Big[\widehat{F}_{\widehat{\beta},t}^{-1}\big((j-1)/\bar{J}_{t}\big),\;
\widehat{F}_{\widehat{\beta},t}^{-1}\big(j/\bar{J}_{t}\big)\Big), \qquad\qquad {j=2,\ldots \bar{J}_{t}-1}, \nonumber \\[0.5 ex]
\label{eq:classical_partition}
\widehat{P}_{\bar{J}_{t} t} &= \Big[\widehat{F}_{\widehat{\beta},t}^{-1}\big((\bar{J}_{t}-1)/\bar{J}_{t}\big), \;
\max_{1 \leq i \leq n_t} \widehat{\beta}_{it} \Big], 
\end{align}
where $\widehat{F}_{\widehat{\beta},t}^{-1}(\cdot)$ is the empirical quantile function.  In practice, the first and last cells are extended outward, as needed, so that the partition covers the full support of the betas.  When $d>1$, the classical implementation uses the tensor (Cartesian) product of the partition in equation \eqref{eq:classical_partition} across the $d$ dimensions.  It is important to emphasize that the portfolio that any specific $\beta$ resides in may change over time.  For example, when $d=1$, it may be that assets with values of $\beta$ near $1/2$ fall in the sixth portfolio at times and the fifth portfolio at other times and so on.

To formalize the beta-sorted portfolio estimator we need to keep track of which portfolio any specific value of $\beta$ resides in. We do this via
\begin{equation}
\widehat{p}_t(\beta)=\Big(\IF(\beta\in\widehat{P}_{1t}), \cdots, \IF(\beta\in\widehat{P}_{\bar{J}_tt})\Big)^\top.
\end{equation}
The vector, $\widehat{p}_t(\beta)$, is comprised of a single element equal to one and all other elements equal to zero.  For example, if the value of $\widehat{\beta}_{it}$ resides in the first portfolio, $\widehat{P}_{1t}$, then $\widehat{p}_t(\widehat{\beta}_{it}) = (1,0,\ldots,0)^\top$, and the return for the first portfolio at time $t$ is
\begin{equation}\label{eq:Mhat1}
\frac{1}{\widehat{n}_{1t}} \sum_{i=1}^{n_t} \IF\big\{\widehat{\beta}_{it} \in \widehat{P}_{1t}\big\} R_{it}, \qquad \widehat{n}_{1t} =\sum_{i=1}^{n_t} \IF\big\{\widehat{\beta}_{it} \in \widehat{P}_{1t}\big\}. 
\end{equation}

More generally, the beta-sorted portfolio  estimator at time $t$ is
\begin{equation}\label{eq:Mhat}
 \widehat{M}_t(\beta)=\widehat{p}_{t}(\beta)^\top \Bigg(\sum_{i=1}^{n_t} \widehat{p}_t(\widehat{\beta}_{it}) \widehat{p}_t\big(\widehat{\beta}_{it})^\top\Bigg)^{-1} \sum_{i=1}^{n_t} \widehat{p}_t(\widehat{\beta}_{it}) R_{it}
\end{equation}
and the corresponding sample average estimator is
\begin{equation}\label{eq:muhat}
\widehat{\mu}(\beta)=\frac{1}{T-H} \sum_{t=H+1}^T\widehat{M}_t(\beta).
\end{equation}
The estimator is comprised of $T-H$ individual estimates, $\widehat{M}_t(\beta)$, motivated by equation \eqref{M_decomp}.  Notice that  $\widehat{M}_t(\beta)$ can be interpreted as the fitted value from a cross-sectional OLS regression of $R_{it}$ on $\widehat{p}_t(\widehat{\beta}_{it})$ and evaluated at $\widehat{p}_t(\beta)$.  Furthermore, it is straightforward to show that if $\beta \in \widehat{P}_{jt}$, then $\widehat{M}_t(\beta)$ may be equivalently written as,
\begin{equation}\label{eq:Mhat_alt}
\widehat{M}_t(\beta) = \frac{1}{\widehat{n}_{jt}} \sum_{i=1}^{n_t} \IF\big\{\widehat{\beta}_{it} \in \widehat{P}_{jt}\big\} R_{it}, \qquad \widehat{n}_{jt} =\sum_{i=1}^{n_t} \IF\big\{\widehat{\beta}_{it} \in \widehat{P}_{jt}\big\}. 
\end{equation}
The sample average in the definition of $\widehat{\mu}(\beta)$ begins at time $H+1$ to reflect the need for an initial sample period (of size $H$) to enable estimation of $\beta_{i(H+1)}$.\footnote{We use $H$ as the length of the initial sample for convenience of notation.  A longer burn-in period could also be used.}

We will provide conditions under which $\widehat{\mu}(\beta)$, given by equation \eqref{eq:muhat}, is uniformly consistent for the sample average of conditional expected returns (SACER) defined as
\begin{equation}\label{eq:SACER}
\bar{\mu}_T(\beta;H) = \frac{1}{T-H} \sum_{t=H+1}^T \mu_t(\beta).
\end{equation}
The estimator, $\widehat{\mu}(\beta)$, thus provides an estimate of the SACER for a generic asset with risk exposures given by the evaluation point, $\beta$.

\begin{remark}[Comparison to the typical implementation approach] There is one subtle but important difference between equation \eqref{eq:muhat} and the typical implementation of beta-sorted portfolios.\footnote{We distinguish between ``classical'' beta-sorted portfolios and the ``typical'' implementation.  By the former we mean the use of rolling window regressions to estimate $\beta_{it}$ and marginal quantiles of the distribution of $\widehat{\beta}_{it}$ to form portfolios.  By the latter we mean choosing a fixed number of portfolios (often ten) and reporting only the average returns of those portfolios.} In existing practice, $ \bar{J}_t = \bar{J}$ for all $t$, and the reported $\widehat{\mu}(\cdot)$ takes on only $\bar{J}$ values as the specific evaluation point $\beta$ is not considered.  For example, the first element of the reported $\widehat{\mu}(\cdot)$ is the average return across time associated with the first portfolio, $\widehat{P}_{1t}$.  However, if the cross-sectional distribution of the $\beta_{it}$ is changing over time then there is no guarantee that a chosen portfolio (e.g., the fifth portfolio) always represents assets with sufficiently similar betas.\footnote{In fact, there may be no value of $\beta$ that resides in the same ordered portfolio across all time periods.}  Therefore, the typical implementation does not generally estimate the SACER at a fixed value of $\beta$; rather, it averages returns over beta regions that may change over time. This distinction is particularly important when we are interested in expected returns at intermediate values of $\beta$ or in tests of monotonicity or other shape restrictions.

\end{remark}

\begin{remark}[Evaluation at $\beta=0$]\label{rem:beta0}
As discussed above, the case where $\beta=0$ is distinctive in this model.  Observe that 
\begin{equation}
\mathbb{E}[R_{it} | \mathcal{G}_{t-1}, \beta_{it}= 0 ] = M_t(0) = \mu_t(0),
\end{equation}
so we no longer require the time averaging described by equation \eqref{M_decomp} for $\mu_t(0)$ to be consistently estimable.  The disappearance of the common risk-factor component at zero is a property of the data-generating process and is not specific to the choice of nonparametric estimator.  This distinction will be made throughout the presentation of the theoretical results in the subsequent sections.  However, it is important to emphasize that primary interest is generally in values of $\beta$ which are of larger magnitude.
\end{remark}

\section{Consistency of beta-sorted portfolios}\label{section3}

We first prove consistency of the beta-sorted portfolio estimator under minimal conditions.  It is important to emphasize the challenging structure of the estimator as the (generally nonparametrically) generated regressors, $\widehat{\beta}_{it}$,  enter the second stage through non-smooth indicator functions, $\widehat{p}_t(\widehat{\beta}_{it})$.  Further, the first step and the second step share the same underlying data used to estimate $\beta_{it}$ and managing this dependence presents a number of  technical challenges.

We first provide conditions on the DGP.

\begin{assumption}[DGP]\label{ass1:DGP}
The data satisfy equations \eqref{mainmodel}--\eqref{condexp} and $(i)$ $\beta_{it}$ has the compact support $\mathcal{B}$ for each $i=1,\ldots,n_t$ and $t = 1,\ldots,T$;  $(ii)$ 
         there exists a finite constant $C>0$ such that $|\mu_t(\beta)-\mu_t(\beta')|\leq C\|\beta-\beta'\|$ for all $\beta,\beta'\in\mathcal{B}$ and $H+1 \leq t \leq T$; $(iii)$ 
        $\varepsilon_{it}$ is independent over $i$ for each $t$ conditional on $\mathcal{F}_{t-1}$; $(iv)$
         $\max_{1\leq t\leq T}\max_{1\leq i \leq n_t}\E[|\varepsilon_{it}|^q|\mathcal{F}_{t-1}]$ and $\max_{1\leq t\leq T}\E[\|\tilde{f}_t\|^q|\mathcal{F}_{t-1}]$ are bounded for some $q>2$ where $\tilde{f}_t = f_{t}-\E[f_{t}\vert \mathcal{G}_{t-1}]$.
\end{assumption}

Assumption \ref{ass1:DGP}$(i)$--$(ii)$ are standard assumptions in the nonparametric literature which rule out discontinuities and other pathologies that would invalidate standard nonparametric estimation approaches.  Assumption \ref{ass1:DGP}$(iii)$ restricts the cross-sectional dependence properties of $\varepsilon_{it}$.  This type of sampling assumption was introduced in \cite{Andrews2005} and has been utilized in the financial econometrics literature by \cite{gagliardini2016ecmta} and \cite{cattaneo2020characteristic}.  The assumption is quite general and allows for higher-order dependence in the distribution of $\varepsilon_{it}$; for example, it accommodates conditional heteroskedasticity both cross sectionally and over time.  In Remark \ref{rem:csdependence} of Section \ref{section4}, we discuss weakening this assumption.  Finally, Assumption \ref{ass1:DGP}$(iv)$ comprises standard moment conditions on the factors and idiosyncratic errors.  

Since the beta-sorted portfolio estimator consists of two steps, we must impose some conditions on the first-step estimator.  However, we only impose minimal assumptions on the estimators of $\beta_{it}$.

\begin{assumption}[First step estimator]\label{ass2:firststep}
        $\widehat{\beta}_{it}$ is $\mathcal{F}_{t-1}$-measurable,
        $H\leq cT$ for a constant $c\in (0,1)$, and there exists some sequence of positive constants $\mathsf{R}_{\tt FS}$ such that 
	\[
        \max_{H + 1 \leq t\leq T} \max_{1\leq i\leq n_t}\|\widehat{\beta}_{it}-\beta_{it}\| = O_\P\big( \mathsf{R}_{\tt FS} \big).
        \]
\end{assumption}

Assumption \ref{ass2:firststep} allows for an arbitrary collection of first-step estimators. It requires each $\widehat{\beta}_{it}$ to only use information available before $R_{it}$ is realized and defines their uniform convergence rate, $\mathsf{R}_{\tt FS}$. We will impose that $\mathsf{R}_{\tt FS} = o(1)$ which is a natural (and generally necessary) assumption when working with generated regressors. For the rolling window OLS estimator in equation \eqref{eq:betahat_roll}, Lemma~\ref*{theory-lem:first-step-rate} of SA-Theory provides primitive conditions under which
\begin{equation}
\mathsf{R}_{\tt FS}
=
\sqrt{\frac{\log(nT)}{H}}
+\frac{(nT)^{1/q}\log(nT)}{H}
+\mathsf{B}_{\tt FS},
\end{equation}
where the first two terms capture the variability of the estimator and $\mathsf{B}_{\tt FS}$ captures the bias from approximating the time-varying coefficients by a constant function in the window.  Thus, we require $H$ to diverge with $T$ in such a way that ensures that all of these terms vanish asymptotically.

Importantly, we only require that uniformly consistent estimators of the $\beta_{it}$ are available but they are not required to be of the form given by equation \eqref{eq:betahat_roll}.  Assumption \ref{ass2:firststep} covers, for example, cases where a parametric or nonparametric form for the $\beta_{it}$ is specified in terms of additional variables \citep[e.g.][]{connor2012efficient,gagliardini2016ecmta,GuKellyXiu2020} so long as suitable estimators are available.

We now introduce the following assumption which ensures that the probability of an empty portfolio  vanishes as the sample size grows.

\begin{assumption}[Partition]\label{ass3:partition}
Let $\widehat{n}_{\min} = \min_{H+1\leq t\leq T}\min_{1\leq j\leq \bar{J}_t}\widehat{n}_{jt}$. There exists some positive sequence $n_{\min}>0$ such that $\widehat{n}_{\min}\geq n_{\min}$ with probability approaching one.
\end{assumption}

Our first main result is the following:

\begin{theorem}\label{thm:consistency}
Suppose Assumptions \ref{ass1:DGP}--\ref{ass3:partition} hold and the following conditions are satisfied:\vspace{-1.5ex}
\begin{enumerate}[label=\emph{(}\alph*\hspace{0.04em}\emph{)}]  
\itemsep 0pt

\item $\mathsf{R}_{\tt FS} = o(1)$,

\item $\widehat{\mathsf{R}}_{\tt D}  = o_\P(1) $ where $\widehat{\mathsf{R}}_{\tt D} = \max_{H+1\leq t\leq T}\max_{1\leq j\leq \bar{J}_t} \len(\widehat{P}_{jt})$,

\item $\frac{\log (nT)}{n_{\min}} =o(1) $,

\item $\limsup_{n,T \to \infty}\frac{(nT)^{2/q}\log (nT)}{n_{\min}} \leq C$ for some constant $C$. 
\end{enumerate}
Then, 
\[
\sup_{\beta \in \mathcal{B}}\, \Big| \widehat{\mu}(\beta) - \bar{\mu}_T(\beta;H)  \Big| =o_\P(1).
\]
\end{theorem}
\medskip

Theorem \ref{thm:consistency} shows that the beta-sorted portfolio estimator is uniformly consistent under general conditions that follow naturally from the structure of the estimator:  Condition (\emph{a}\hspace{0.04em}) requires that the first-step estimators of the $\beta_{it}$ approximate their targets sufficiently well. Combining this requirement with condition (\emph{b}\hspace{0.04em}) ensures that assets within a portfolio have increasingly similar true betas.
Condition ($\emph{c}\hspace{0.04em}$) ensures that all portfolios are comprised of a sufficient number of assets.  Together, conditions (\emph{c}\hspace{0.04em}) and (\emph{d}\hspace{0.04em}) ensure that the contribution of the idiosyncratic errors vanishes asymptotically.

If the partition is ``quasi-uniform'' (so that no portfolio covers an arbitrarily large or small portion of the support) and uses a number of intervals of order $J$ along each of the $d$ coordinates, and each portfolio contains roughly the same number of assets, then it can be shown that
\begin{equation}
\max_{H+1\leq t\leq T}\max_{1\leq j\leq \bar{J}_t}\len(\widehat{P}_{jt}) = O_\P(J^{-1}), \qquad n_{\min} \asymp \frac{n}{J^d},
\end{equation}
where $\bar{J}_t\asymp J^d$ uniformly in $t$.  In this case, conditions (\emph{b}\hspace{0.04em})--(\emph{d}\hspace{0.04em}) require that $J\to\infty$, $J^d\log(nT)/n \to 0$, and $J^d(nT)^{2/q}\log(nT)/n = O(1)$.  Consequently, $J$ performs a similar balancing act as $H$; specifically, $J$ must grow fast enough to eliminate the approximation bias from grouping assets with different betas together, but slowly enough to ensure that each portfolio contains a sufficient number of assets.  For the marginal-quantile portfolio construction introduced in equation \eqref{eq:classical_partition}, Lemma~\ref*{theory-lem:partition-basis-moments} of SA-Theory provides primitive conditions under which the resulting partition is quasi-uniform over appropriately defined interior portfolios.

Apart from assuming a compact support and the availability of uniformly consistent first-step estimators, consistency of $\widehat{\mu}(\cdot)$ requires no direct restrictions on the time-series or cross-sectional properties of $\beta_{it}$.  For example, the $\beta_{it}$ can be random, can be correlated in the cross-section and the time-series, and could depend on additional variables in a (possibly) nonlinear fashion.  Furthermore, the factors, $f_t$, are only required to satisfy the (conditional) moment restrictions given by Assumption \ref{ass1:DGP}$(iv)$, but otherwise have no  restrictions on their serial dependence properties.

\begin{remark}[Technical details on Theorem \ref{thm:consistency}]
Under the conditions of Theorem \ref{thm:consistency}, Theorem~\ref*{theory-thm:general-consistency} of SA-Theory establishes the stronger result that, for any arbitrarily slowly diverging sequence $\ell_n$ (e.g., $\ell_n=\log\log(nT)$), there exists a fixed, finite constant $C>0$ such that, with probability approaching one, simultaneously for all $\beta\in\mathcal B$,
\begin{equation}
\big|\widehat{\mu}(\beta)-\bar{\mu}_T(\beta;H)\big|
\leq C
\ell_n\left[
\mathsf{R}_{\tt FS}
+\widehat{\mathsf{R}}_{\tt D}
+\sqrt{\frac{\log(nT)}{n_{\min}}}
+\frac{\|\beta\|}{\sqrt{T}}
\right].
\end{equation}
Unlike the uniform consistency result given in Theorem \ref{thm:consistency}, this bound identifies the components that can  determine the accuracy of the estimator.   It is noteworthy that the result does not require a specific partitioning rule (i.e., portfolio construction) since only the maximum diameter and the minimum number of assets in any one portfolio enter the bound.  The final term arises from the unforecastable component of the risk factors and is proportional to the norm of the evaluation point $\beta$.  For a fixed nonzero $\beta$, this term is unaffected by the cross-sectional sample size, whereas it vanishes at $\beta=0$.  The disappearance of this term at $\beta=0$ is the source of the differential convergence rates of the estimator that we establish in Section~\ref{section4}.
\end{remark}

\subsection{Object of interest: SACER vs. PACER}

In Theorem \ref{thm:consistency} we showed that $\widehat{\mu}(\beta)$ is a consistent estimator of the SACER.  Following the arguments in Section \ref{section2}, the SACER is the natural candidate for our object of interest.  However, it is important to point out that the SACER 
is a sample average of (generally) random quantities and, hence, random itself.  The randomness arises from the presence of $\mu_t(\cdot)$ which depends on the $i$-invariant sigma field $\mathcal{G}_{t-1}$. We can contrast SACER to its population counterpart, the population average of conditional expected returns (PACER).  To do so we need the following assumption:

\renewcommand{\theassumption}{\Alph{assumption}} \setcounter{assumption}{15}

\begin{assumption}[PACER]\label{b0:PACER}
There exists a non-random function ${\mu}(\cdot)$ such that
$
\sup_{\beta\in\mathcal{B}}
\big|\bar{\mu}_T(\beta;H)-{\mu}(\beta)\big|
=o_\P(1).
$
\end{assumption}

\renewcommand{\theassumption}{\arabic{assumption}} \setcounter{assumption}{3}

Assumption \ref{b0:PACER} imposes a uniform convergence condition on the SACER and defines the PACER as its non-random (probability) limit. In the special case where $\mu_t(\cdot)=\mu(\cdot)\,\, \forall t$, the SACER and PACER are equal.

The SACER and the PACER represent two distinct objects of interest.  To make this distinction concrete, suppose that there exists a finite-dimensional state vector $\mathsf{S}_{t-1}$ that is $\mathcal{G}_{t-1}$-measurable and $\mu_t(\beta)=\mu_0(\beta;\mathsf{S}_{t-1})$. Further, assume that $\{\mathsf{S}_t\}$ is stationary and ergodic and $\mu_0(\cdot;\cdot)$ satisfies the regularity conditions required for a uniform law of large numbers.\footnote{Equation \eqref{csmodel} allows for more generality; however, we form our discussion around a finite set of state variables governing the economy for simplicity of exposition and to cement intuition.}  Assumption \ref{b0:PACER} then holds with $\mu(\beta)=\E\!\left[\mu_0(\beta;\mathsf{S}_0)\right]$.  Thus, the PACER averages conditional expected returns over the stationary distribution of the state variables, whereas the SACER averages conditional expected returns over the particular realizations of those state variables observed during the sample period. Consequently, the SACER and PACER may differ meaningfully in any particular sample.

We recommend that empirical researchers focus on the SACER rather than the PACER for a few reasons. First, even if Assumption \ref{b0:PACER} fails to hold, the SACER is well-defined under our remaining assumptions.  The PACER may not exist if, for example, some state variables driving $\mu_t(\cdot)$ are not stationary.  Second, and more importantly, the SACER is a more intuitive and interpretable object of interest as we have a wealth of data summarizing what occurred over our sample.  For example, we observe the behavior of macroeconomic aggregates, changes in the legal and regulatory landscape, or the presence of unusual events (e.g., financial crises or natural disasters). Conversely, interpretation of the PACER will generally hinge on the stationary distribution of the economy-wide state variables and the mapping $\mu_0(\cdot;\cdot)$ from those states to conditional expected returns.  Characterizing the properties of this distribution would be challenging (e.g., assessing the probability of recessions or financial crises) and the appeal of the approach herein is that we can avoid making specific assumptions about the functional form of $\mu_t(\cdot)$.  Third, inference on the PACER requires additional assumptions beyond those needed for inference on the SACER.  In addition to the estimation error $\widehat{\mu}(\beta)-\bar{\mu}_T(\beta;H)$, inference on the PACER must characterize the time-series variation in $\bar{\mu}_T(\beta;H)-\mu(\beta)$.  Despite our preference for the SACER, Section \ref{sec:PACER} discusses this additional source of uncertainty and the resulting challenges for inference on the PACER.

\section{Inference for beta sorted portfolios}\label{section4}

In the previous section we established consistency under minimal conditions, requiring only that the first-step estimators are well behaved and the portfolio-formation rule, using these estimators, results in portfolios with sufficiently similar $\beta$ values and sufficiently many assets.  To move the theory beyond consistency, however, we will now specialize to the family of first-step estimators and portfolio-formation rules that are in widespread practice.

\subsection{Classical beta-sorted portfolio estimator}
\label{sec:classical}

The classical beta-sorted portfolio estimator is comprised of $(i)$ rolling window OLS regressions to estimate $\beta_{it}$ as given by equation \eqref{eq:betahat_roll} and $(ii)$ forming portfolios based on the marginal quantiles of the cross-sectional distribution of $\widehat{\beta}_{it}$ as given by equation \eqref{eq:classical_partition} in the special case of $d=1$.  Taking these mainstays of empirical practice as our starting point, we will obtain an asymptotic linearization of $\widehat{\mu}(\cdot)$ that underpins our results on asymptotic normality and feasible inference.

The first ingredient of beta sorts is an estimate of $\beta_{it}$.  It will be convenient to introduce the following generalization of the conventional rolling window estimator. Let $1\leq s\leq H$ and define
\[
r_{\mathfrak p}(u)=(1,u,\ldots,u^{\mathfrak p})^\top,\qquad X_s(t)=
r_{\mathfrak p}\left(s/H\right)\otimes (1,f_{t-s}^{\top})^\top
\]
where $\otimes$ denotes the Kronecker product and  $\mathfrak p$ is a fixed integer.  The corresponding estimator of $\beta_{it}$ is formed from the following local polynomial regression estimator,
\begin{equation}\label{eq:betahat}
\widehat{\beta}_{it}=S\,\widehat{\theta}_{it},\qquad
\widehat{\theta}_{it}
= \Bigg(\sum_{s=1}^{H} X_s(t)X_s(t)^\top\Bigg)^{-1} \sum_{s=1}^{H} X_s(t) R_{i(t-s)},
\end{equation}
where $S$ selects elements $2,\ldots,d+1$ of $\widehat{\theta}_{it}$ which correspond to the estimator of $\beta_{it}$.  This is still a rolling-window least squares estimator except with higher-order polynomial terms.  In SA-Theory we also allow for kernel weights in equation \eqref{eq:betahat} but we suppress this generalization since we focus on equal weights to match empirical practice.  When $\mathfrak p=0$, we recover the rolling-window estimator in equation \eqref{eq:betahat_roll}.  For the SACER estimation and inference procedures discussed in this section, we use $\mathfrak p=0$ and $\mathfrak p=1$. The general formulation also accommodates the local quadratic regressions, $\mathfrak p=2$, used in the empirical affine specification test.

We require that $H\to\infty$ and $H/T \to 0$ to ensure that the number of observations in each window grows but the size of the window itself, relative to the full sample, shrinks.  The polynomial order, $\mathfrak p$, controls the flexibility of the fit within the window. A more flexible fit can reduce the (first-step) approximation bias. Indeed, we will use the first-step estimator with $\mathfrak{p}=1$ to conduct valid pointwise inference at fixed nonzero evaluation points based on a data-driven choice of $H$ (see Section~\ref*{applied-sec:tuning-pointwise} of SA-Applied for details).\footnote{Section~\ref*{theory-sec:classical-first-step} of SA-Theory also provides additional results, including primitive smoothness conditions and uniform convergence rates for the first-step estimator with arbitrary $\mathfrak p$.}

Just as the first step of the estimation procedure can be generalized to allow for more flexible fits (i.e., $\mathfrak{p}>0$), so too can the second step.  For example, rather than use a constant fit within the portfolio we can use the least-squares fit from a regression of $R_{it}$ on a constant and $\widehat{\beta}_{it}$ using \emph{only} the assets placed in that portfolio.  To distinguish the order of this second-step fit from the first-step polynomial order, let $\mathfrak q$ denote the polynomial order within each portfolio and write the corresponding estimators as $\widehat M_t(\beta;\mathfrak q)$ and $\widehat\mu(\beta;\mathfrak q)=(T-H)^{-1}\sum_{t=H+1}^T\widehat M_t(\beta;\mathfrak q)$.  The case $\mathfrak q=0$ is exactly the classical portfolio-mean estimator whereas  $\mathfrak q=1$ corresponds to running linear regressions within the portfolio.  This more flexible fit can reduce the magnitude of the second-step approximation bias.  In particular, we will need to use $\widehat\mu(\beta;1)$ to conduct inference when $\beta=0$.  Further, even when $\beta\neq0$, using $\mathfrak q=1$ can reduce the magnitude of the approximation bias. It also eliminates a higher-order remainder term exactly.\footnote{Specifically, $\mathfrak q\geq1$ implies that $\mathscr R_3(\beta)=0$; see Lemma~\ref*{theory-lem:general-linearization} of SA-Theory.} Throughout, we use $(\mathfrak p,\mathfrak q)=(0,0)$ for point estimation and $(\mathfrak p,\mathfrak q)=(1,1)$ for inference on the SACER.

\subsection{Asymptotic normality} 
\label{sec:asymptotic_normality}

In this section, we establish asymptotic normality of the estimator under the additional conditions collected in Assumption~\ref*{theory-ass:dgp} of SA-Theory. Most are standard regularity conditions, so we discuss only the more noteworthy restrictions here.  First, we impose more structure on the idiosyncratic errors by writing $\varepsilon_{it} = \varsigma_{it} \eta_{it}$, requiring the primitive innovations, $\eta_{it}$, to be independent across assets and time while allowing for  cross-sectional and time-series dependence in the scale factors, $\varsigma_{it}$ (Assumption \ref*{theory-ass:dgp}$(ii)$).  Second, we impose conditions on $\alpha_{it}$ and $\beta_{it}$ to ensure a well-behaved time-varying coefficient model and assume that the histories of these coefficients through date $t$ are conditionally independent across assets, while allowing their conditional distributions to differ across assets (Assumption~\ref*{theory-ass:dgp}$(iii)$). This permits systematic differences in factor exposures across groups such as industries. Third, we impose a near-epoch dependence assumption on the risk factors and require the time average of the conditional variance to converge to a deterministic positive-definite matrix (Assumption \ref*{theory-ass:dgp}$(iv)$).

The starting point for the distributional analysis is the stochastic linearization.  For fixed $\mathfrak{q} \in \{0,1\}$, we have that
\begin{equation}
    \label{eq:stochlin}
\widehat{\mu}(\beta; \mathfrak{q})-\bar{\mu}_T(\beta;H)
=
\bar{\mathscr S}_f(\beta)
+\bar{\mathscr S}_\varepsilon(\beta; \mathfrak{q})
+\mathscr R(\beta; \mathfrak{q}),
\end{equation}
where $\bar{\mathscr S}_f(\beta)$ is the risk factor component,
\begin{equation}
    \label{eq:factor_component}
\bar{\mathscr S}_f(\beta)
=
\frac{1}{T-H}\sum_{t=H+1}^T
\beta^\top\big(f_t-\E[f_t\mid\mathcal G_{t-1}]\big),
\end{equation}
$\bar{\mathscr S}_\varepsilon(\beta; \mathfrak{q})$ is the idiosyncratic component, and $\mathscr R(\beta; \mathfrak{q})$ collects all remaining approximation errors.  The idiosyncratic component captures two distinct sources of estimation error. The first source is the direct contribution of the idiosyncratic shocks, $\varepsilon_{it}$, to the second-step portfolio fit. This term also arises in standard nonparametric regression problems.  The second source is the indirect contribution of past idiosyncratic shocks through the estimated betas used in the second step. Because the rolling estimation windows overlap, the same shock can affect beta estimates in more than one period.  Formal expressions are provided in Section~\ref*{theory-sec:classical-linearization} of SA-Theory.

We can use equation \eqref{eq:stochlin} to introduce some key properties of $\widehat{\mu}(\beta; \mathfrak{q})$.  First, under the rate conditions in Section~\ref*{theory-sec:pointwise-distribution} of SA-Theory, it can be shown that $\mathscr R(\beta; \mathfrak{q})$ is negligible relative to the other terms in the linearization.  Second, the remaining terms $\bar{\mathscr S}_f(\beta)$ and $\bar{\mathscr S}_\varepsilon(\beta; \mathfrak{q})$ have different rates of convergence.  Specifically, $\bar{\mathscr S}_f(0)=0$ and
\begin{equation}
    \bar{\mathscr S}_f(\beta) = O_\P\bigg(\frac{1}{\sqrt{T}}\bigg), \qquad \beta \neq 0,
\end{equation}
whereas
\begin{equation}
   \bar{\mathscr S}_\varepsilon(0; \mathfrak{q}) = O_\P\bigg(\sqrt{\frac{J^d}{nT}}\bigg),
\end{equation}
and, importantly, $\bar{\mathscr S}_\varepsilon(\beta; \mathfrak{q}) = o_\P(T^{-1/2})$ when $\beta \neq 0$.  Thus, $\bar{\mathscr S}_f(\beta)$ is the dominant first-order term when $\beta \neq 0$.  This implies that, at any fixed nonzero evaluation point, the estimator achieves only a $\sqrt{T}$ rate of convergence even though it uses approximately $nT$  observations.  However, the pointwise convergence rate changes discontinuously at $\beta=0$. At the origin, $\bar{\mathscr S}_f(0)=0$, so, to first order, only $\bar{\mathscr S}_\varepsilon(0;\mathfrak q)$ contributes, yielding a faster convergence rate.

The $\sqrt{T}$ convergence rate is a consequence of the common factor structure imposed in equation \eqref{mainmodel}.  At any fixed $\beta\neq0$, each period's return contains the innovation to the risk factor, $\beta^\top(f_t-\E[f_t\mid\mathcal G_{t-1}])$.  Increasing the cross-sectional sample size averages away the effect of idiosyncratic shocks but has no effect on this common component, whose influence can only be reduced by averaging over time.  It is this feature that explains why $\bar{\mathscr S}_f(\beta)$ dominates the other terms in the linearization.  This slower rate of convergence is the cost of remaining agnostic about the functional form of $\E[f_t\mid\mathcal G_{t-1}]$.  If the conditional mean of the risk factors were known, the factor innovation could be removed from each portfolio return and a faster convergence rate could then be achieved.  However, this would require a functional-form assumption on the factor dynamics and would lead to an entirely different estimator.  As our focus is on the classical beta-sorted portfolio estimator, we do not discuss such an approach further.

The following theorem formalizes the distinction between the pointwise asymptotic distributions at zero and at nonzero evaluation points.  Let $\mathcal B^\circ\subset\operatorname{int}(\mathcal B)$ denote an
interior evaluation region containing zero.

\begin{theorem}[Pointwise asymptotic normality]
    \label{thm:CLT}
Fix $\beta\in\mathcal B^\circ$ and suppose the conditions of Theorem~\ref*{theory-thm:pointwise-clt} of SA-Theory hold. If $\beta\neq0$, then, for each $\mathfrak q\in\{0,1\}$,
\[
\frac{\widehat{\mu}(\beta;\mathfrak q)-\bar{\mu}_T(\beta;H)}
{\sqrt{\Sigma_f(\beta)}}
\rightsquigarrow
\mathcal N(0,1),
\qquad
\Sigma_f(\beta)\asymp_\P T^{-1},
\]
where $\rightsquigarrow$ denotes convergence in distribution and
\[
\Sigma_f(\beta)
=
\frac{1}{(T-H)^2}\sum_{t=H+1}^T
\beta^\top\Var(f_t\mid\mathcal G_{t-1})\beta.
\]
If $\beta=0$, then
\[
\frac{\widehat{\mu}(0;1)-\bar{\mu}_T(0;H)}
{\sqrt{\Sigma_\varepsilon(0; 1)}}
\rightsquigarrow\mathcal N(0,1),
\qquad
\Sigma_\varepsilon(0; 1)\asymp_\P\frac{J^d}{nT},
\]
where $\Sigma_\varepsilon(0;\mathfrak q)$ is the conditional variance of $\bar{\mathscr S}_\varepsilon(0;\mathfrak q)$ and is defined in Section~\ref*{theory-sec:pointwise-distribution} of SA-Theory. 
\end{theorem}

Theorem \ref{thm:CLT} establishes pointwise asymptotic normality of the beta-sorted portfolio estimator centered at the SACER and formalizes the distinction between the convergence rates described above. Although the leading asymptotic variance at any fixed nonzero evaluation point, $\Sigma_f(\beta)$, does not depend on the order $\mathfrak q$ of the second-step fit, the choice of $\mathfrak q$ affects the second-step approximation bias, and $\mathfrak q=1$ is required to conduct inference at the origin. For all values of $\beta$, asymptotic normality also requires the approximation errors from estimating the $\beta_{it}$ in the first step and fitting $\mu_t(\cdot)$ in the second step to be negligible relative to the corresponding standard error.

\begin{remark}[Technical details on Theorem \ref{thm:CLT}]
\label{rem:technical_CLT}
Section~\ref*{theory-sec:general-linearization} of SA-Theory develops a stochastic linearization  based on generic first-step estimators, and Section~\ref*{theory-sec:classical-linearization} specializes it to the classical estimator outlined in Section~\ref{sec:classical}.  Further, SA-Theory gives conditions under which $\mathscr R(\beta; \mathfrak{q})$ is negligible relative to $T^{-1/2}$ when $\beta\neq0$ and $\sqrt{J^d/(nT)}$ when $\beta=0$. To establish our pointwise inference result at $\beta=0$, we modify the marginal-quantile partition in equation \eqref{eq:classical_partition} so that zero is a fixed breakpoint in every marginal partition. This ensures that zero occupies the same boundary position in its selected portfolio at every date, providing the sharper control of the estimated partition needed at the origin. At a fixed $\beta\neq0$, the leading common risk-factor component, $\bar{\mathscr S}_f(\beta)$, does not depend on this adjustment. The partition anchored at zero can nevertheless be used at all evaluation points to provide a common implementation. Importantly, the marginal quantile partition of equation \eqref{eq:classical_partition} remains valid for consistent estimation of the SACER.
\end{remark}

\begin{remark}[Cross-sectional dependence]\label{rem:csdependence}
Equation~\eqref{eq:stochlin} clarifies the scope for weakening the cross-sectional independence assumptions.  For a fixed $\beta\neq0$, the leading term is $\bar{\mathscr S}_f(\beta)$, whose behavior is solely governed by the evaluation point and the time-series properties of the risk factors. Consequently, sufficiently weak cross-sectional dependence that preserves the orders of
$\bar{\mathscr S}_\varepsilon(\beta; \mathfrak{q})$ and $\mathscr R(\beta; \mathfrak{q})$ would leave the first-order distribution unchanged.  This suggests that the result could accommodate, for example, additional weak cross-sectional dependence in the idiosyncratic errors of the type considered by \citet{ChamberlainRothschild1983}.
\end{remark}

\begin{remark}[Time series properties of conditional expected returns]
One way to satisfy the first-step bias requirement in Assumption~\ref*{theory-ass:first-step-bias} of SA-Theory is to require $\alpha_{it}$ and $\beta_{it}$ to vary sufficiently smoothly within each rolling estimation window.  However,  equation \eqref{model} shows that the relevant restriction cannot be interpreted independently of the functional form for conditional expected returns because $
\alpha_t(\beta)
=\mu_t(\beta)-\beta^\top\E[\left. f_t \right| \mathcal G_{t-1}]$. To illustrate, suppose that $\mathsf S_t$ is a vector of $\mathcal G_t$-measurable state variables which follows a stationary linear time-series process (with nondegenerate innovations) and $\E[f_t\mid\mathcal G_{t-1}]=\zeta+\Lambda\mathsf S_{t-1}$, with $\Lambda\neq0$ (as in, e.g., \citealt{adrian2015regression} or \citealt{gagliardini2016ecmta}).  Then, in order for $\alpha_{it}=\alpha_t(\beta_{it})$ to vary smoothly, $\mu_t(\cdot)$ \emph{must} vary over time in a way that offsets the nonsmooth variation induced by $\beta^\top\Lambda\mathsf S_{t-1}$.  In that case, the SACER and PACER are distinct objects and need not coincide in a given sample, which directly affects the interpretation of estimation and inference in empirical applications.
\end{remark}

\subsection{Feasible inference on the SACER}

We have established conditions which ensure asymptotic normality of the beta-sorted portfolio estimator when centered at the SACER.  In this section, we investigate the properties of two feasible inference procedures.  The first is the so-called Fama-MacBeth (FM) variance estimator which is ubiquitous in existing applications of beta-sorted portfolios.  The FM variance estimator can be  motivated by the classical sample variance estimator and constructed as
\begin{equation}
    \label{eq:FM}
\widehat\Sigma_{\mathtt{FM}}(\beta;\mathfrak q)
=
\frac{1}{(T-H)^2}\sum_{t=H+1}^T
\Big(\widehat M_t(\beta;\mathfrak q)-\widehat\mu(\beta;\mathfrak q)\Big)^2.
\end{equation}
Here, $\widehat{M}_t(\beta;\mathfrak q)$ for $t = H+1, \ldots, T$ serves as the sample ``observations''  and $\widehat{\mu}(\beta;\mathfrak q)$ serves as the sample ``mean.''  Consequently, apart from a degrees-of-freedom correction, the FM variance estimator also has an interpretation as a leave-$t$ out jackknife variance estimator. $\widehat\Sigma_{\mathtt{FM}}(\beta;0)$ is ubiquitous in the empirical finance literature. We will discuss the theoretical properties of this conventional FM variance estimator, and its generalization given by equation \eqref{eq:FM}, later in this section.

The second feasible inference procedure is based on the theoretical results we have presented. From equation \eqref{eq:stochlin} and Theorem \ref{thm:CLT}, sampling uncertainty has a common risk-factor component, $\Sigma_f(\beta)$, and an idiosyncratic component, $\Sigma_\varepsilon(\beta;1)$, which we will estimate separately.  At any fixed $\beta\neq0$, $\Sigma_f(\beta)$ is the leading variance component, which we estimate using the following plug-in estimator:
\begin{equation}
    \widehat{\Sigma}_f(\beta) = \frac{1}{(T-H)^2}\sum_{t=H+1}^T \Big[ \beta^\top \big(f_t - \widehat{f}_{t|t-1}\big) \Big]^2.
\end{equation}
Here, $\widehat{f}_{t|t-1}$ is an estimate of $\E[f_t|\mathcal{G}_{t-1}]$ using only information contained in $\mathcal{G}_{t-1}$.  For example, a parametric approach would lead to $\widehat{f}_{t|t-1} = h_{t-1}( {\widehat{\vartheta}})$ for some $h_{t-1}(\vartheta)$ with
$h_{t-1}( \widehat{\vartheta})$ a corresponding feasible estimate.  As a simple example, in the case of an AR(1) we would obtain $h_{t-1}( {\vartheta}) = \vartheta_0 + \vartheta_1 f_{t-1}$, and we would estimate the parameters, $(\vartheta_0,\vartheta_1)$, recursively.  Importantly, the estimator, $\widehat{f}_{t|t-1}$, need not be based on the correct specification.  If the conditional mean of the factors is misspecified, then Theorem \ref{thm:jn-variance} below shows that
\begin{equation}
    \widehat{\Sigma}_f(\beta) - \Sigma_f(\beta) = \mathscr B_{\mathtt F}(\beta) + o_\P(T^{-1}),
\end{equation}
where
\begin{equation}
\mathscr B_{\mathtt F}(\beta)
= \frac{1}{(T-H)^2}\sum_{t=H+1}^T
\Big[\beta^\top\big(\widehat{f}_{t|t-1} - \E[f_t|\mathcal{G}_{t-1}]\big)\Big]^2.
\end{equation}
Clearly, $\mathscr B_{\mathtt F}(\beta)$ is non-negative; thus, we can still conduct valid, albeit possibly conservative, inference based on $\widehat{\Sigma}_f(\beta)$.

We estimate the variance contribution from the  idiosyncratic component using a bin-restricted delete-$i$ jackknife.  Let $\widehat M_t^{(-i)}(\beta;\mathfrak q)$ be the period-$t$ portfolio estimator computed after deleting unit $i$ from the second-step cross-section while holding the portfolio formation rule represented by the partition $\widehat\Delta_t$ fixed, and define
\begin{equation}
\widehat d_{it}(\beta;\mathfrak q) =
\widehat M_t^{(-i)}(\beta;\mathfrak q)-\widehat M_t(\beta;\mathfrak q).
\end{equation}
Set $\widehat d_{it}(\beta;\mathfrak q)=0$ when unit $i$ is not observed at date $t$.  The jackknife variance estimator is
\begin{equation}
\label{eq:jackknife-variance-proposed}
\widehat\Sigma_\varepsilon(\beta;\mathfrak q)
=
\frac{1}{(T-H)^2}
\sum_i\sum_{t=H+1}^T\sum_{s=H+1}^T
\widehat d_{it}(\beta;\mathfrak q)\widehat d_{is}(\beta;\mathfrak q)
\I\{|t-s|\leq H\}.
\end{equation}
The truncation at lag $H$ captures the serial dependence induced by the overlapping first-step estimation windows.  Pairs farther than $H$ periods apart do not share primitive idiosyncratic shocks after the first-step correction is isolated. Importantly, calculating $\widehat\Sigma_\varepsilon(\beta;\mathfrak q)$ does not require any re-estimation steps in practice.  Using properties of leave-one-out OLS estimation, we can express $\widehat d_{it}(\beta;\mathfrak q)$ as a function of the fitted residuals from the second-step regression and the leverage of unit $i$ in that regression.  Section~\ref*{theory-sec:pointwise-variance} of SA-Theory gives equivalent computationally simple formulations of this  jackknife variance estimator.

Combining the two components gives our proposed variance estimator
\begin{equation}
\label{eq:combined-variance-proposed}
\widehat\Sigma(\beta)
=\widehat\Sigma_f(\beta)+\widehat\Sigma_\varepsilon(\beta;1).
\end{equation}

\begin{theorem}[Plug-in variance estimator]
\label{thm:jn-variance}
Fix $\beta\in\mathcal B^\circ$ and suppose the conditions of
Theorem~\ref*{theory-thm:variance-estimation} of SA-Theory hold.
If $\beta\neq0$, then
\[
\widehat\Sigma(\beta)-\Sigma_f(\beta)
-\Sigma_\varepsilon(\beta;1)-\mathscr B_{\mathtt F}(\beta)
=o_\P(T^{-1}).
\]
If $\beta=0$, then
\[
\widehat\Sigma(0)-\Sigma_\varepsilon(0;1)
=o_\P\left(\frac{J^d}{nT}\right).
\]
\end{theorem}

Theorem \ref{thm:jn-variance} shows that the newly-proposed variance estimator satisfies the required rate restrictions to ensure feasible inference.  When the conditional mean of the risk factors is correctly specified and estimated sufficiently accurately, the plug-in variance estimator is consistent for the first-order variance.  When the conditional mean is misspecified, the plug-in variance
estimator is conservative and overestimates the first-order
variance by $\mathscr B_{\mathtt F}(\beta)$.  The jackknife variance estimator is consistent for the idiosyncratic component of the first-order variance at $\beta=0$.  Thus, Theorem \ref{thm:jn-variance} establishes that $\widehat\Sigma(\beta)$ for $\beta\neq0$ and $\widehat\Sigma(0)$ at the origin can be used to conduct valid, possibly conservative, pointwise inference.

We now discuss the properties of the Fama-MacBeth variance estimator.

\begin{theorem}[Fama--MacBeth variance estimator]
\label{thm:fm-variance}
Fix $\beta\in\mathcal B^\circ$ and $\mathfrak q\in\{0,1\}$, and
suppose the conditions of
Theorem~\ref*{theory-thm:fama-macbeth-variance} of SA-Theory hold.
Define
\[
\Sigma_\mu(\beta)
=
\frac{1}{(T-H)^2}\sum_{t=H+1}^T
\big(\mu_t(\beta)-\bar\mu_T(\beta;H)\big)^2.
\]
If $\beta\neq0$, then
\[
\widehat\Sigma_{\mathtt{FM}}(\beta;\mathfrak q)
=
\Sigma_f(\beta)+\Sigma_\mu(\beta)+o_\P(T^{-1}).
\]
\end{theorem}

Theorem \ref{thm:fm-variance} shows that the FM variance estimator is generally conservative for inference on the SACER at a fixed nonzero $\beta$ as it includes the time-series variation in conditional expected returns, $\Sigma_\mu(\beta)$, which is not part of the first-order variance.  This additional term is nonnegative and can be strictly positive when $\mu_t(\beta)$ varies over time.  The intuition for this result can be gleaned from the following decomposition of the summands of the estimator:
\begin{equation}
\widehat M_t(\beta;\mathfrak q)-\widehat\mu(\beta;\mathfrak q)
=
[\widehat M_t(\beta;\mathfrak q)-\mu_t(\beta)]
+[\mu_t(\beta)-\bar\mu_T(\beta;H)]
-[\widehat\mu(\beta;\mathfrak q)-\bar\mu_T(\beta;H)].
\end{equation}
After squaring and averaging this decomposition, the first component generates the leading sampling variance, $\Sigma_f(\beta)$, while the second generates $\Sigma_\mu(\beta)$.  The final component and the relevant cross-products are asymptotically negligible.

The properties of the FM variance estimator at $\beta=0$ are very different.  As we saw in Section \ref{sec:asymptotic_normality}, the  uncertainty of the estimator depends importantly on the fact that the second-step procedure relies on first-step estimation.   The FM estimator uses the period-by-period dispersion of the portfolio estimates but does not account for the dependence induced by the overlapping windows in the first step.  As such, it need not estimate the appropriate variance at the origin. This is the case even when the no-arbitrage condition holds and the conditional expected return function is linear in $\beta$.  Depending on the DGP, $\widehat\Sigma_{\mathtt{FM}}(0;\mathfrak{q})$ may either overestimate or underestimate the true variance.

\begin{corollary}[Pointwise feasible inference]
\label{corr:feasinf}
Suppose the conditions of Theorems~\ref{thm:CLT},
\ref{thm:jn-variance}, and \ref{thm:fm-variance} hold. Let
$\mathsf{z}_{1-\alpha/2}$ be  the $1-\alpha/2$ quantile of a standard Gaussian random variable.  For any nonnegative variance estimator $\widehat V$, define
\[
\mathcal I_{1-\alpha}(\beta;\mathfrak q,\widehat V)
=
\left[
\widehat\mu(\beta;\mathfrak q)
-\mathsf{z}_{1-\alpha/2}\widehat V^{1/2},
\ \widehat\mu(\beta;\mathfrak q)
+\mathsf{z}_{1-\alpha/2}\widehat V^{1/2}
\right]
\]
and $
\widehat\Sigma(\beta)_+
=
\max\big\{\widehat\Sigma(\beta),0\big\}
$. Then:
\begin{enumerate}[label=(\roman*)]
\item For each fixed $\beta\neq0$ and $\mathfrak q\in\{0,1\}$,
\[
\liminf_{n,T\to\infty}
\P\left(
\bar\mu_T(\beta;H)
\in
\mathcal I_{1-\alpha}\big(\beta;\mathfrak q,
\widehat\Sigma(\beta)_+\big)
\right)
\geq1-\alpha.
\]
The coverage probability converges to $1-\alpha$ if
$T\mathscr B_{\mathtt F}(\beta)=o_\P(1)$.

\item For each fixed $\beta\neq0$ and $\mathfrak q\in\{0,1\}$,
\[
\liminf_{n,T\to\infty}
\P\left(
\bar\mu_T(\beta;H)
\in
\mathcal I_{1-\alpha}\big(\beta;\mathfrak q,
\widehat\Sigma_{\mathtt{FM}}(\beta;\mathfrak q)\big)
\right)
\geq1-\alpha,
\]
The coverage probability converges to $1-\alpha$ if
$T\Sigma_\mu(\beta)=o_\P(1)$.

\item For $\beta=0$,
\[
\P\left(
\bar\mu_T(0;H)
\in
\mathcal I_{1-\alpha}\big(0;1,\widehat\Sigma(0)_+\big)
\right)
\to1-\alpha.
\]
\end{enumerate}
\end{corollary}

Corollary \ref{corr:feasinf} shows that for $\beta \neq 0$ confidence intervals based on either the newly-proposed variance estimator or the FM variance estimator attain at least their nominal coverage probability.  The former variance estimator is  asymptotically exact when $T\mathscr B_{\mathtt F}(\beta)$ is asymptotically negligible, whereas the FM interval is asymptotically exact when the time-series variation in conditional expected returns is asymptotically negligible.  When $\beta=0$, the newly-proposed variance estimator is asymptotically exact.  The use of $\widehat\Sigma(\beta)_+$ in Corollary \ref{corr:feasinf} arises because the jackknife component of the variance estimator, $\widehat\Sigma_\varepsilon(\beta;\mathfrak q)$, can be negative in finite samples.\footnote{Alternatively, we could add weights to the jackknife variance estimator to ensure non-negativity as is done in the HAC literature \citep[e.g.,][]{NeweyWest1987}.  However, in empirically-relevant settings we do not observe negative values of $\widehat\Sigma_\varepsilon(\beta;\mathfrak q)$.}  This ensures that the standard error is well defined.  In the limit, this adjustment is unnecessary with probability approaching one.

\begin{remark}[Uniformity over $\beta$]
    \label{rem:uniform_beta}
Corollary \ref{corr:feasinf} establishes coverage at each fixed evaluation point, but it does not guarantee that coverage remains valid uniformly as $\beta$ approaches zero.  This distinction matters because of the differential behavior of the limiting distribution at the origin that we have discussed.  Section~\ref*{theory-sec:uniform-beta-inference} of SA-Theory develops a uniformly valid procedure that uses the newly-proposed variance estimator away from zero and a conservative variance adjustment near zero.  Let $\mathcal I^{\mathtt{UI}}_{1-\alpha}(\beta)$ denote the resulting pointwise confidence interval.  These confidence intervals attain at least their nominal coverage probability uniformly over $\beta\in\mathcal B^\circ$, i.e.,
\[
\liminf_{n,T\to\infty}
\inf_{\beta\in\mathcal B^\circ}
\P\left(
\bar\mu_T(\beta;H)
\in
\mathcal I^{\mathtt{UI}}_{1-\alpha}(\beta)
\right)
\geq1-\alpha.
\]
\end{remark}

\begin{remark}
\label{rem:lowbound}
Theorems \ref{thm:jn-variance} and \ref{thm:fm-variance} imply that we can construct an asymptotic lower bound on the time-series variation in conditional expected returns, as measured by $\Sigma_\mu(\beta)$, by taking the difference between $\widehat\Sigma_{\mathtt{FM}}(\beta;\mathfrak q)$ and $\widehat\Sigma(\beta)$.  Under the assumptions of these theorems, for each fixed $\beta\neq0$ and $\mathfrak q\in\{0,1\}$,
\begin{equation}
\label{eq:lowbound}
\widehat\Sigma_{\mathtt{FM}}(\beta; \mathfrak{q})-\widehat\Sigma(\beta)
=
\Sigma_\mu(\beta)-\mathscr B_{\mathtt F}(\beta)+o_\P(T^{-1}).
\end{equation}
Because $\mathscr B_{\mathtt F}(\beta)\geq0$, equation \eqref{eq:lowbound} shows that this difference provides an asymptotic lower bound on $\Sigma_\mu(\beta)$.  When $T\mathscr B_{\mathtt F}(\beta)=o_\P(1)$, the bound is asymptotically sharp.  Thus, the contrast between the two variance estimators provides information about the time-series variation in conditional expected returns.  In both our empirical applications, we find that this difference is relatively large, making the bound informative.
\end{remark}

\begin{remark}
    Theorem \ref{thm:jn-variance} and Corollary \ref{corr:feasinf} imply that valid, although potentially conservative, inference on the SACER does not require correct specification of $\E[f_t\mid\mathcal G_{t-1}]$.  A misspecified predictor increases the estimated variance through $\mathscr B_{\mathtt F}(\beta)$ but does not invalidate coverage.  This contrasts with, for example, \citet{adrian2015regression} and \citet{gagliardini2016ecmta}, which impose a first-order Markov structure on the factor dynamics.  In practice, specifying the correct functional form and the appropriate conditioning variables for the factor dynamics can be challenging.  This robustness is an advantage of the beta-sorted portfolio estimator.
\end{remark}

The choices of $H$ and $J$ govern the amount of smoothing in the two estimation steps. Longer first-step windows reduce the variability of the estimated betas but can increase approximation bias when risk-factor exposures vary over time, while forming more portfolios reduces within-portfolio approximation bias but leaves fewer assets in each portfolio. Part~\ref*{applied-part:tuning} of SA-Applied develops data-driven choices, based on mean squared error criteria, at fixed nonzero evaluation points, together with a near-origin adjustment for uniform-in-$\beta$ inference. It also explains why factor-cancelling inference requires dedicated undersmoothing. Part~\ref*{applied-part:numerical} reports Monte Carlo evidence on the finite-sample performance of the SACER estimators, the proposed and FM variance estimators, and the data-driven tuning rules.  The simulation results are consistent with our theoretical predictions. We find that confidence intervals based on our proposed variance
estimator perform well when the evaluation point is away from the origin.  Near zero, the uniform-in-$\beta$ adjustment substantially improves coverage relative to the pointwise procedure.

\subsubsection{Estimation and Inference on the PACER}
\label{sec:PACER}

Although we recommend focusing on the SACER, it is useful to compare the requirements for estimation and inference on the PACER.  For $\mathfrak q\in\{0,1\}$, decompose
\begin{equation}
    \label{eq:PACER_decomp}
    \widehat\mu(\beta;\mathfrak q)-\mu(\beta)
=
\big[\widehat\mu(\beta;\mathfrak q)-\bar\mu_T(\beta;H)\big]
+
\big[\bar\mu_T(\beta;H)-\mu(\beta)\big].
\end{equation}
The first term is $o_\P(1)$ under the conditions in Section~\ref{section3} and Theorem~\ref*{theory-thm:general-consistency} of SA-Theory.  At any fixed $\beta$, consistency for the PACER therefore requires only a law of large numbers for $\bar\mu_T(\beta;H)$.  Assumption \ref{b0:PACER} imposes this convergence uniformly over $\beta\in\mathcal B$.

Inference on the PACER, however, requires substantially more structure.  When the estimator is centered at the SACER, its leading component has a martingale-difference structure that underlies Theorem \ref{thm:CLT}.  Centering at the PACER introduces the additional term $\bar\mu_T(\beta;H)-\mu(\beta)$, which reflects the time-series variation in conditional expected returns and is not generally a martingale difference.  A PACER central limit theorem must, therefore,  characterize the asymptotic distribution of this time-series average, including its long-run variance, as well as its covariance with the SACER-centered estimation error (the first term in equation \eqref{eq:PACER_decomp}).  Assumption \ref{b0:PACER} alone is insufficient for such results.

Consequently, the current variance results do not by themselves provide feasible inference on the PACER.  At a fixed nonzero $\beta$, Theorem \ref{thm:fm-variance} shows that the FM variance estimator captures the period-by-period variation in conditional expected returns through $\Sigma_\mu(\beta)$.  It does not, in general, estimate the long-run variance of this process when conditional expected returns are serially dependent, nor does it account for their possible covariance with the SACER-centered estimation error.  Establishing a PACER central limit theorem and constructing a feasible variance estimator would therefore require substantially more structure on the time-series dynamics of the DGP.

\section{Extensions and further discussion}
\label{sec:extensions}

\subsection{Multiple factors and portfolio construction}
\label{sec:multiple-factors}

The beta-sorted portfolio estimator that we have studied throughout the paper is the one based on so-called independent sorts, i.e., forming portfolios based on tensor products of the marginal partitions of each factor exposure (using marginal quantiles as breakpoints) as described in earlier sections.  This is the multivariate analog of the estimator when $d=1$.  A drawback of this approach is that, as we showed, the number of portfolios grows exponentially in the number of factors (at the rate $J^d$), which can lead to sparsely populated portfolios and even empty portfolios.  In practice, researchers often use a small number of factors to avoid this problem; in particular, applications with $d>2$ are rare (one exception is \cite{ChangChristoffersenJacobs2013}, who used $d=4$).

An alternative approach that avoids multidimensional sorting is to form portfolios based on a single target risk-factor exposure, while controlling for the other factors in the first-step regression.\footnote{For recent examples, see \cite{BBT2017}, \cite{ChenHanPan2021}, \cite{FanLondonoXiao2022}, and
\cite{CDHW2023}.} However, such an approach targets a different object than the SACER and the distinction is analogous to the difference between the ``short'' and ``long'' regression when studying omitted variables in OLS regressions.  Suppose that the first risk factor is the target and is used for portfolio construction whereas the other $d-1$ factors serve only as control variables in the first-step regression.  Partition the evaluation point $\beta$ as $\beta=(\beta_1, \beta_{-1}^\top)^\top$. For simplicity of presentation, suppose the $\beta_{it}$ are conditionally \emph{i.i.d.} across assets rather than just conditionally independent. Let $\E^\beta_t[\,\cdot\mid\beta_{it,1}=b_1]$ denote the expectation
with respect to their common period-$t$ conditional distribution,
further conditioning on the first component being equal to $b_1$.\footnote{Formally, $\E^\beta_t[\,\cdot\mid\beta_{it,1}=b_1] = \E[\,\cdot\mid\mathcal D_{t-1},\beta_{it,1}=b_1]$ where $\mathcal D_{t-1}$ is the design sigma field which we suppress for simplicity.  The design sigma field and the related filtrations are defined in Section~\ref*{theory-sec:classical-filtrations} of SA-Theory.} The resultant estimator targets the following object
\begin{equation}
\bar{\mu}_{T\mid 1}(b_1;H)
=
\frac{1}{T-H}
\sum_{t=H+1}^{T}
\mu_{t \mid 1}(b_1),
\qquad
\mu_{t \mid 1}(b_1) =
\E^\beta_t\!\left[
\mu_t(\beta_{it})
\mid \beta_{it,1}=b_1
\right].
\end{equation}
Like the SACER, $\bar{\mu}_{T\mid 1}(b_1;H)$ is generally random because both conditional expected returns and the cross-sectional distribution of the remaining betas, $\beta_{it,-1}$, may vary over time. Sorting on a single target variable, therefore, averages over the remaining exposures rather than holding them fixed and does not generally recover the SACER at any specified value of $(b_1,b_{-1}^\top)^\top$.

We can use the theoretical structure introduced in earlier sections to describe the asymptotic distribution of this marginal sorting estimator which we label as $\widehat{\mu}_{\mid 1}(b_1; \mathfrak{q})$.  Define the average beta of assets with target risk factor $\beta_1 = b_1$ at time $t$ as
$
\vartheta_{t\mid1}(b_1)
=
\E^\beta_t\!\left[
\beta_{it}
\mid \beta_{it,1}=b_1
\right]$.
Under corresponding regularity and rate conditions, the leading expansion of the marginal beta-sorted portfolio estimator is
\begin{equation}
\widehat\mu_{\mid 1}(b_1;\mathfrak q)
-\bar\mu_{T \mid 1}(b_1;H)
=
\frac{1}{T-H}\sum_{t=H+1}^T
\vartheta_{t \mid 1}(b_1)^{\top}
\big(f_t-\E[f_t\mid\mathcal G_{t-1}]\big)
+\bar{\mathscr S}_{\mid 1}(b_1)
+\mathscr R_{\mid 1}(b_1),
\end{equation}
where $\bar{\mathscr S}_{\mid 1}(b_1)=O_\P\big(\sqrt{J/(nT)}\big)$ represents the contribution from the idiosyncratic component and $\mathscr R_{\mid 1}(b_1)$ collects all other remainder terms.  As before,
when the risk factor component is nondegenerate (and the remainder is negligible) the estimator possesses a $\sqrt{T}$ rate.  Importantly, setting $b_1=0$ does not generally eliminate the factor component because the average exposures to the remaining factors may be nonzero.  The stronger condition $\vartheta_{t\mid1}(0)=0$ for all $t$
eliminates the common-factor component at $b_1=0$.  Our theoretical framework could be extended to conduct inference
on $\bar{\mu}_{T\mid1}(b_1;H)$ using a variance estimator analogous
to $\widehat{\Sigma}(\beta)$.  However, this marginal target depends on the joint distribution of risk-factor exposures and can change even when the conditional expected return function remains unchanged.  Thus, care must be taken when translating empirical results based on marginal sorting to economic conclusions.

A more principled alternative for mitigating the sparse portfolio problem is to impose additive separability of conditional expected returns across the components of $\beta$.  Specifically, suppose that
\begin{equation}
    \label{eq:addsep}
\mu_t(\beta)
=
\mu_{0t}+\sum_{k=1}^d\mu_{kt}(\beta_k),
\qquad
\mu_{kt}(0)=0,
\quad k=1,\ldots,d.
\end{equation}
Additive separability rules out interactions among factor exposures because the effect of changing $\beta_k$ cannot depend on the remaining components of $\beta$. However, it still retains substantial flexibility as each $\mu_{kt}(\cdot)$ may be nonlinear and time varying.  Notice that the no-arbitrage restriction is nested as the special case where $\mu_{kt}(\beta_k)=\beta_k\lambda_{kt}$.

Under additive separability, a lower-dimensional estimator is available.  To provide intuition, consider the case where $\mathfrak{q}=0$.  Without additive separability, $\widehat{M}_t(\cdot;0)$ can be obtained, at each $t$, via a cross-sectional regression of $R_{it}$ on approximately $J^d$ indicator functions.  In contrast, additive separability of $\mu_t(\cdot)$ directly implies additive separability of the realized systematic return function $M_t(\cdot)$.  We can therefore rely on a cross-sectional regression of $R_{it}$ on approximately $dJ$ indicator functions instead.\footnote{A fully additively separable specification is not necessary.  Partially additive specifications can allow interactions within low-dimensional groups of risk factor exposures while retaining much of the reduction in dimensionality. In that case, the dimensionality cost is governed by the largest group of exposures allowed to interact rather than by the full dimension $d$. For example, we could have $\mu_t(\beta) = \mu_{0t}+\mu_{12t}(\beta_1,\beta_2)+\sum_{k=3}^d\mu_{kt}(\beta_k)$ so the dimension would be of the order $J^2 + (d-2)J$.}

Importantly, additive separability alone does not imply that the marginal sorting estimator discussed above recovers the component $\mu_{1t}(b_1)$.  Under the additive model in equation \eqref{eq:addsep},
\begin{equation}
\mu_{t\mid1}(b_1)
=
\mu_{0t}+\mu_{1t}(b_1)
+
\sum_{k=2}^d
\E_t^\beta\!\left[
\mu_{kt}(\beta_{it,k})
\mid \beta_{it,1}=b_1
\right].
\end{equation}
When the risk-factor exposures are dependent, each summand in the final term can  vary with $b_1$, which can alter both the level and the shape of $\mu_{t\mid1}(b_1)$.  In the special case in which each summand is (almost surely) constant in $b_1$,  marginal sorting recovers the shape and contrasts of $\mu_{1t}(b_1)$.  However, neither zero correlation nor zero rank correlation across the dimensions of risk-factor exposures is sufficient for this conclusion.  Joint additive estimation avoids this restriction by controlling flexibly for the other risk-factor exposure components.

A further useful feature of the additive estimator is that it retains a well-defined interpretation when additive separability is misspecified.  At each $t$, the conditional expected return target becomes the best mean-square approximation to $\mu_t(\cdot)$ within the space of additively separable functions, where mean-square error is computed using $\E_t^\beta[\cdot]$.  This interpretation follows because the factor component of $M_t(\cdot)$ is already linear in $\beta$ (and therefore already additively separable).  The time-averaged estimator can thus be understood as targeting the sample average of these additive approximations rather than the unrestricted SACER.

\begin{remark}[Conditional sorts]
\label{rem:conditional-sorts}    
Another common approach to portfolio construction is a conditional, or sequential, sort.\footnote{For examples of conditional sorts involving estimated risk-factor exposures, see 
\cite{HerskovicKellyLustigVanNieuwerburgh2016}, \cite{BollerslevLiTodorov2016}, and \cite{BBT2017}.}  For example, to study exposure $\beta_1$ while controlling for exposure $\beta_2$, researchers first sort assets into portfolios using the marginal quantiles of $\widehat{\beta}_{it,2}$ and then, within each of those portfolios, sort assets using \emph{conditional} quantiles of $\widehat{\beta}_{it,1}$.  The procedure can be repeated recursively with additional risk-factor exposures.  Because the breakpoints at each stage depend on portfolio membership at the preceding stages, the resulting partition is not a tensor product of marginal-quantile partitions and generally depends on the ordering of the procedure.  However, the complete collection of final portfolios generated by the sequential sorts partitions the observed cross-section.\footnote{Although the final portfolios partition the observed cross-section, covering the entire support $\mathcal B$ requires extending the outer conditional intervals within every parent portfolio to the corresponding boundaries of $\mathcal B$.}  Thus, provided that the resulting cells shrink appropriately and their boundaries within the support are recorded, an estimator evaluated at a fixed $\beta$ continues to target the SACER.  Importantly, this statement requires keeping track of the part of the support each portfolio represents.  For example, the fifth $\beta_1$ portfolio within a low-$\beta_2$ group can cover a very different range of $\beta_1$ from the fifth $\beta_1$ portfolio within a high-$\beta_2$ group.  If instead portfolio returns or high-minus-low spreads are averaged across parent portfolios based only on their sequential ranks, without explicitly keeping track of the $\beta$ ranges represented by each portfolio, as is common in empirical applications, the target becomes a weighted average of conditional expected returns or return differences,
respectively, and does not generally identify the SACER or a SACER contrast at a fixed beta.

Although conditional sorts can ensure nonempty portfolios when the cross-section is sufficiently large, they do not eliminate the curse of dimensionality.  With $J$ portfolios formed for each of $d$ exposures, there are still $J^d$ final portfolios containing approximately $n_t/J^d$ assets each.  Thus, conditional sorts alleviate empty portfolios caused by dependence among the risk-factor exposures, but not the sparsity caused by increasing the number of exposures.  Our high-level framework can accommodate conditional sorts, but providing primitive conditions under which our results apply is challenging because the recursively estimated breakpoints require, throughout the sequential sorting procedure, uniform control over both the $\beta$ ranges represented by the portfolios and the number of assets they contain.
\end{remark}

\subsection{Factor-cancelling portfolios and linear-pricing restrictions}
\label{sec:factor-cancelling-discussion}

The stochastic expansion in Section~\ref{sec:asymptotic_normality} shows that, for $\beta \neq 0$, the common risk-factor component is the leading source of uncertainty for the beta-sorted portfolio estimator.  Inspecting equation \eqref{eq:factor_component}, we observe that this term is linear in $\beta$; consequently, it can be eliminated for judiciously chosen linear contrasts.  Let $\mathsf{b}_1,\ldots,\mathsf{b}_{K_{\Psi}}$ be fixed exposure vectors, let $a_1,\ldots,a_{K_{\Psi}}$ be fixed weights, and define $\widehat{\Psi}(\mathfrak{q})
=\sum_{k=1}^{K_{\Psi}}a_k\widehat{\mu}(\mathsf{b}_k;\mathfrak{q})$ and its target $\bar{\Psi}_T
=\sum_{k=1}^{K_{\Psi}}a_k\bar{\mu}_T(\mathsf{b}_k;H)$.  If the weights and evaluation points satisfy
\begin{align}
\sum_{k=1}^{K_{\Psi}}a_k\mathsf{b}_k=0,
\label{eq:main-factor-cancelling}
\end{align}
then the leading common risk-factor component in the stochastic expansion cancels exactly.

A specific example of such a linear contrast is a ``butterfly'' trade where one goes long one unit of each of two assets, one with risk-factor exposure $\mathsf{b}_1$ and one with risk-factor exposure $\mathsf{b}_3$, and short two units of an asset with risk-factor exposure $\mathsf{b}_2$. Then, if $\mathsf{b}_1-\mathsf{b}_2 = \mathsf{b}_2 - \mathsf{b}_3$ and we take weights as $(1,-2,1)$, equation \eqref{eq:main-factor-cancelling} is satisfied.  The estimator for the sample average of the conditional expected returns of such a trade is then
\begin{align}
\widehat{\Psi}_{\mathtt{BF}}(\mathfrak{q})
&=\widehat{\mu}(\mathsf{b}_1;\mathfrak{q})
-2\widehat{\mu}(\mathsf{b}_2;\mathfrak{q})
+\widehat{\mu}(\mathsf{b}_3;\mathfrak{q}).
\end{align}
This contrast is a discrete second difference and therefore a measure of curvature in the sample-average conditional expected return function.

Consistency of $\widehat{\Psi}(\mathfrak{q})$ follows directly
under the conditions in Section~\ref{section3} and
Theorem~\ref*{theory-thm:general-consistency} of SA-Theory because the contrast is just a fixed finite linear
combination of consistent estimators.  In fact, consistency holds regardless of whether equation \eqref{eq:main-factor-cancelling} is satisfied.  Inference, on the other hand, does not follow directly from the pointwise results in Section~\ref{sec:asymptotic_normality}.  Because the factor-cancelling restriction eliminates the leading common risk-factor component, the distributional properties of $\widehat{\Psi}(\mathfrak{q}) - \bar{\Psi}_T$ in the limit are akin to those of $\widehat{\mu}(\beta) - \bar{\mu}_T(\beta;H)$ when $\beta=0$.\footnote{Ordinary high-minus-low portfolios do not generally satisfy condition~\eqref{eq:main-factor-cancelling} and therefore retain the common risk-factor component.}   The rate of convergence is now governed by the idiosyncratic component in equation \eqref{eq:stochlin}.  Under the conditions in Theorem~\ref*{theory-thm:factor-cancelling-inference} of SA-Theory, it is shown that the estimator enjoys the faster $\sqrt{nT/J^d}$ rate.  In order to derive these properties, however, we require a partition anchored at the prespecified evaluation points, as detailed in Section~\ref*{theory-sec:factor-cancelling} of SA-Theory.  This construction is the analogue, for factor-cancelling linear contrasts, of the partition anchored at zero (see Remark \ref{rem:technical_CLT}).

A natural hypothesis of interest is whether conditional expected returns satisfy the restrictions implied by the absence of arbitrage opportunities.  As we have discussed, no-arbitrage conditions restrict the form of $\mu_t(\beta)$ to be linear (for excess returns) or affine (for raw returns).  One approach would be to test whether the time-averaged conditional expected return of a butterfly trade equals zero. A broader approach tests the linear or affine specification jointly across a collection of values for the risk-factor exposures, which is the approach we propose.  

Here, we will focus on the affine case.  Consider the following null hypothesis,
\begin{align}
\mathsf{H}_{0}^{\mathtt{LIN}}:\,
\mu_t(b)
&=a_t+c_t^\top b,
\qquad
b\in\mathcal B,\quad t=1,\ldots,T,
\label{eq:main-linearity-null}
\end{align}
where the intercept and slopes may vary over time.  Under the null, realized returns satisfy
\begin{align}
R_{it}
=a_t+\varpi_t^\top\beta_{it}+\varepsilon_{it},
\end{align}
where $\varpi_t=c_t+f_t-\E[f_t\mid\mathcal G_{t-1}]$.  Thus, the realized cross-sectional slope is $\varpi_t$ rather than $c_t$ because it absorbs the contemporaneous risk-factor innovation.  If the $\beta_{it}$ were known, then $a_t$ and $\varpi_t$ could be estimated by a cross-sectional regression at each $t$.

The procedure evaluates equation~\eqref{eq:main-linearity-null} at a fixed collection of values for the risk-factor exposures: $\mathsf b_1,\ldots,\mathsf b_{K_{\mathtt{LIN}}}$.  At each date $t$, we construct two beta estimates using adjacent, nonoverlapping first-step windows.  The earlier estimate, $\widehat\beta_{it}$, uses observations dated $t-H,\ldots,t-1$, and the later estimate, $\widehat\beta_{i(t+H)}$, uses observations dated $t,\ldots,t+H-1$ and is paired with the outcome $R_{i(t+H)}$.

Because the later beta estimate contains first-step estimation error, an OLS regression of $R_{i(t+H)}$ on $\widehat\beta_{i(t+H)}$ would generally exhibit errors-in-variables bias. To address this bias in the linear fit used for the test, we estimate the relation by instrumental variables, using $\widehat\beta_{it}$ as an instrument for $\widehat\beta_{i(t+H)}$.  Since the two windows do not overlap, the instrument is predetermined with respect to the innovations entering the later beta estimate and asset return.  Relevance additionally requires that the earlier beta estimate retain sufficient cross-sectional information about the later beta.  This condition is natural in our setting because risk-factor exposures would be expected to evolve gradually over short horizons, preserving the bulk of the cross-sectional information across adjacent estimation windows.

Let $\widehat{a}_{t+H}$ and $\widehat\varpi_{t+H}$ denote the resulting IV coefficient estimates and define
\begin{align}
\widehat e_{i(t+H)}
&=
R_{i(t+H)}
-\widehat a_{t+H}
-\widehat\varpi_{t+H}^\top\widehat\beta_{i(t+H)}.
\label{eq:main-linearity-residual}
\end{align}
For an evaluation point $b$, let $\widehat w_{it}(b)$ denote the weight assigned to asset $i$ by the beta-sorted portfolio estimator using the beta estimates from the earlier window.  The test averages the residual returns as follows:
\begin{align}
\widehat\nu(b)
&=
\frac{1}{T-2H}
\sum_{t=H+1}^{T-H}
\sum_{i\in\mathcal I_t^{\mathtt{LIN}}}
\widehat w_{it}(b)\widehat e_{i(t+H)},
\label{eq:main-linearity-moment}
\end{align}
where $\mathcal I_t^{\mathtt{LIN}}$ contains the assets observed in both windows and whose return, $R_{i(t+H)}$, is non-missing.  The IV regression, portfolio partition, and normalization of portfolio weights all use this matched sample.  Under the null, these residual portfolio returns should be close to zero.

Stacking the moments over the fixed grid gives $\widehat{\boldsymbol\nu}_{\mathtt{LIN}}=(\widehat\nu(\mathsf b_1),\ldots,\widehat\nu(\mathsf b_{K_{\mathtt{LIN}}}))^\top$.  The Wald statistic is
$\mathsf W_{\mathtt{LIN}}
=
\widehat{\boldsymbol\nu}_{\mathtt{LIN}}^\top
\widehat\Sigma_{\mathtt{LIN}}^{-1}
\widehat{\boldsymbol\nu}_{\mathtt{LIN}},
$
where $\widehat\Sigma_{\mathtt{LIN}}$ accounts for dependence across the grid points and the serial dependence from overlapping observations.  Under the null, $\mathsf W_{\mathtt{LIN}}$ converges to a chi-squared distribution with $K_{\mathtt{LIN}}$ degrees of freedom, so large values provide evidence against the affine specification at one or more of the selected exposure points.  Because the grid is fixed and the moments are averaged over time, this is not an omnibus test over the entire support and nonlinear departures from the affine specification that offset across dates need not be detected.  Technical conditions and the variance estimator are provided in Section~\ref*{theory-sec:fixed-grid-linearity} of SA-Theory.

\section{Empirical application}
\label{empapp}

In this section, we introduce a novel risk factor and show that it predicts both the cross-sectional and time-series behavior of U.S. stock returns. In Section~\ref*{applied-sec:empirical-application-capital-share} of SA-Applied, we present a second empirical application that revisits the capital share factor of \citet{LLM2019}. Together, these applications illustrate the practical value of the new theoretical and inferential results developed earlier in the paper.  

Our risk factor is a new measure of the business credit cycle.  Standard measures of the business credit cycle use ratios of credit aggregates to measures of output.  Although theoretically appealing, these ratios can be difficult to interpret because their movements may reflect changes in the composition of the underlying aggregates rather than changes in credit conditions relative to economic activity.  Here we take a different approach.  We rely on the Federal Reserve's Senior Loan Officer Opinion Survey\footnote{For background information on the SLOOS and its relationship to lending and economic activity, see \cite{LMR2000} and \cite{LM2006}.} (SLOOS) to measure credit conditions  and the ISM Manufacturing Index to measure economic activity.  This construction has three distinct advantages.  First, as the SLOOS and ISM can both be expressed as diffusion indices, their behavior is less sensitive to changes in the structure of the economy.  Second, they are much more timely than credit aggregates and national accounts data which tend to be released with a substantial lag.  Third, they are not subject to revision outside of updated seasonal factors for the ISM index.  Thus, we have a timely factor which we can evaluate in real time with no look-ahead bias.

Our factor is constructed as
\begin{equation}
\mathrm{BCF}_t = \left( \frac{1}{2} \cdot \mathrm{SLOOS}_t + 50\right) + \mathrm{ISM}_t,
\label{eq:CCWt}
\end{equation}
where $\mathrm{SLOOS}_t$ is the net percentage of large domestic banks tightening standards for commercial and industrial loans to all firms and $\mathrm{ISM}_t$ is the ISM Manufacturing Index.  For both series we date each observation to the month in which it became publicly available.\footnote{The Senior Loan Officer Opinion Survey is currently conducted on a quarterly basis.  To construct a monthly series we keep the SLOOS value constant until a new value is available.  For the period from 1984m1 through 1990m1, the credit standards question was not included in the SLOOS, and we use the net willingness to make consumer installment loans by large domestic banks, with its sign reversed to match the tightening convention.  For the initial period from 1965m1 through 1967m9, we use the credit-standards series for all banks and all firms.}
Although both the SLOOS and the ISM are based on questions with a similar format, they are scaled differently and so the affine transformation of the SLOOS is implemented so that they are both on the same scale (between 0 and 100).  To understand why this is (the inverse of) a credit-to-output type measure, note that a fall in the SLOOS corresponds to easier lending standards and a fall in the ISM to weaker economic activity. Thus, a low value of the business-credit factor is intended to capture greater credit availability relative to economic activity. The reverse interpretation applies when the business-credit factor is high. Our factor series begins in January 1965 representing the first SLOOS observation.  

As a preliminary check for the validity of our factor, we assess its ability to predict future market returns.   Motivated by equation (\ref{mainmodel}), we ask whether the persistence of the factor translates into predictability of future equity returns.  We estimate predictive regressions of the form
\begin{equation}
r_{t,t+\ell}=a^{(\ell)}+b^{(\ell)} z_t+v^{(\ell)}_{t+\ell},
\end{equation}
where $r_{t,t+\ell}$ is the cumulative market excess return over the following $\ell$ months and $z_t$ denotes the predictor under consideration. We compare our risk factor with the benchmark predictors studied by \cite{GoyalWelch2008}.  In Table \ref{tab:insample} we present in-sample $R^2$ from predictive regressions for forecast horizons of 1, 3, 6, and 12 months ahead.  The first fourteen rows report results for the benchmark predictors.  The row labeled ``CGP'' reports results using only the SLOOS component of our factor, as in \cite{CGP2015}, while the row labeled ``BCF'' reports results for our business-credit factor.  The results are stark.  The in-sample $R^2$ from our new risk factor far outstrips that of the other predictors considered.  The out-of-sample results, reported in Table~\ref*{applied-tab:outsample} of SA-Applied, lead to the same conclusion, namely, that our risk factor outperforms each of the other predictors by a wide margin.

\begin{table}[tbp]
\caption{\textbf{In-sample Predictive Regressions} This table reports $R^2$ (in percent) from predictive regressions of cumulative market excess returns on an individual predictor from \cite{GoyalWelch2008} for horizons of 1, 3, 6, and 12 months ahead.  The row labeled ``CGP'' reports results for the SLOOS only portion of our risk factor as studied in \cite{CGP2015}.  The row labeled ``BCF'' reports results for our proposed risk factor.  The sample period is 1965m1--2019m12.}
\label{tab:insample}
\centering
\begin{tabular}{|l|cccc|} \hline
 & $\ell=1$  & $\ell=3$  & $\ell=6$  & $\ell=12$  \\  \hline \hline 
\input tables/predictive-is.tex
\hline 
\end{tabular}
\end{table}

We now investigate how the business-credit factor performs in explaining the cross-section of equity returns.  We use monthly raw returns from the Center for Research in Security Prices (CRSP) from January 1965 through December 2019.  We retain common shares listed on the New York Stock Exchange, American Stock Exchange, or Nasdaq and follow \citet{BEM2016} in incorporating delisting returns.  The first 120 months are reserved for initial first-step estimation, so the reported expected-return estimates cover January 1975 through December 2019.  We select the first-step window $H_{it}$ and second-step number of portfolios $J_t$ 
using the data-driven approach described in Section~\ref*{applied-sec:tuning-pointwise} of SA-Applied.  Within each window, we use all available nonmissing observations for an asset and require at least 36 usable prior observations to estimate $\beta_{it}$ and include the asset in that month's portfolio sort.  Further implementation details are provided in Section~\ref*{applied-sec:implementation-defaults} of SA-Applied.

\begin{figure}[h!]
\centering
\caption{\textbf{Inference on the SACER.} The black line reports the $(\mathfrak{p},\mathfrak{q})=(0,0)$ estimate of the SACER.  The grey region is the 95\% JN pointwise interval.  The red dashed lines are the 95\% JN uniform-in-beta intervals, and the green dotted lines are the 95\% modified FM pointwise intervals.  All three intervals use $(\mathfrak{p},\mathfrak{q})=(1,1)$.  Results are reported for monthly returns (in percent).  The sample period is 1965m1--2019m12 with a 120-month burn-in period.}
\label{fig:grandmean_ptwise}
\includegraphics[width=.92\textwidth]{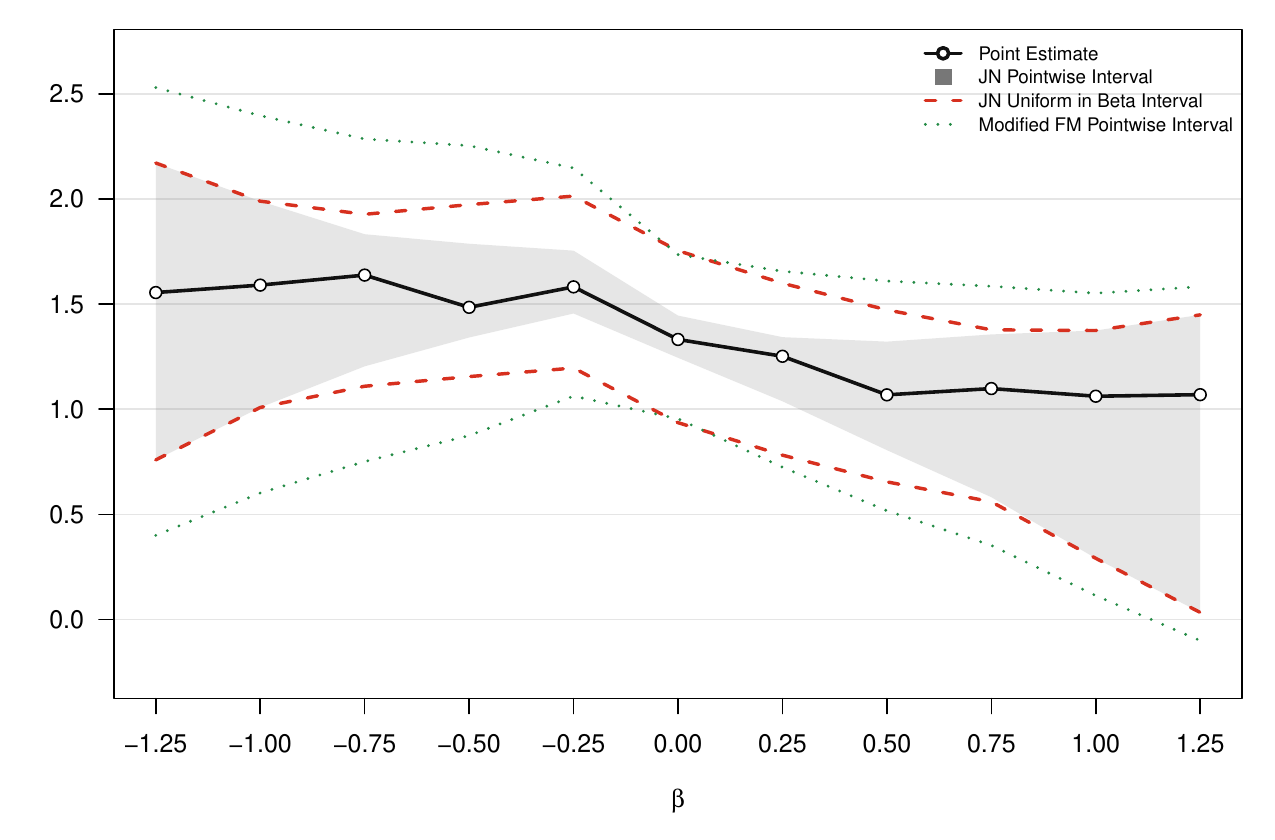}
\end{figure}

Figure~\ref{fig:grandmean_ptwise} reports the estimated SACER, $\widehat\mu(\beta)$, over a grid ranging from $-1.25$ to $1.25$ in increments of $0.25$.  The black line is the $(\mathfrak{p},\mathfrak{q})=(0,0)$ point estimate. The estimated SACER exhibits a clear negative relationship with beta. It is around 1.5 percent per month at negative beta values and around 1 percent at positive beta values. The overall pattern indicates that greater exposure to the business-credit factor is associated with lower expected returns.

We use our new theoretical results to conduct inference on the SACER.  In the figure, the grey region is the 95\% pointwise interval based on our full variance estimator, $\widehat\Sigma(\beta)$ (denoted JN), while the green dotted lines delimit the 95\% pointwise interval based on the modified FM variance estimator, $\widehat\Sigma_{\mathtt{FM}}(\beta;1)$.  Both pointwise intervals use $(\mathfrak{p},\mathfrak{q})=(1,1)$, and therefore need not be centered on the black line.  Using our data-driven choices of tuning parameters, these
intervals are asymptotically valid at fixed nonzero evaluation
points (see Lemma~\ref*{applied-lem:tuning-pointwise-validity}
of SA-Applied).  The red dashed lines report the 95\% JN uniform-in-beta interval discussed in Remark \ref{rem:uniform_beta}.  In this application, the estimated boundaries are approximately $\beta = \pm 0.79$, so the uniform adjustment applies to the seven reported grid points from $-0.75$ through $0.75$.  Consistent with the theory, Figure~\ref{fig:grandmean_ptwise} shows that the uniform-in-beta adjustment widens the interval near zero, where the risk-factor component of the variance becomes small, while leaving the JN pointwise interval unchanged away from zero.  The figure also shows that the modified FM pointwise intervals are wider than the JN pointwise intervals at every reported value of $\beta$.

We quantify the economic magnitude of the downward slope in Figure~\ref{fig:grandmean_ptwise} using a high-minus-low contrast.
Table~\ref{tab:credit-contrasts} provides results for the contrast between $\beta=-0.75$ and $0.75$.  We choose these two values as the contrast endpoints since they reside in the realized support in almost every month (both endpoints lie within the realized range in 94.6\% of the months).  The point estimate of this contrast, based on the $(\mathfrak{p},\mathfrak{q})=(0,0)$ estimator, is $-0.540$ percentage points per month corresponding to the difference between the two points on the black line in Figure~\ref{fig:grandmean_ptwise}.  Table~\ref{tab:credit-contrasts} presents the results from our new inference procedures.  The JN standard error is 0.304, giving a (two-sided) $p$-value of 0.070.  The modified FM variance estimator produces a slightly smaller standard error, giving a two-sided $p$-value of 0.054. Thus, both the JN and modified FM procedures reject a zero contrast at the 10\% level.  As a reference, using $\widehat\Sigma_f$ alone gives a smaller standard error and a $p$-value of 0.019. Although both endpoints are away from zero, so the idiosyncratic component is asymptotically of smaller order, we recommend the full JN variance estimator because it retains this source of uncertainty.  Finally, applying the affine specification test from Section~\ref{sec:factor-cancelling-discussion} to the seven central grid points yields a Wald statistic of $7.871$, with an associated chi-squared $p$-value of $0.344$.  Thus, we do not reject the null of an affine specification at conventional significance levels.

\begin{table}[h!]
\centering
\caption{\textbf{High Minus Low Contrast.} The table reports inference for $\widehat\mu(0.75)-\widehat\mu(-0.75)$ in monthly percentage points.  Each row reports the standard error, 90\% confidence interval, and two-sided normal-approximation $p$-value under the JN, modified FM, or $\widehat\Sigma_f$-only variance estimator.  The confidence intervals are centered at the $(\mathfrak{p},\mathfrak{q})=(1,1)$ estimate.}
\label{tab:credit-contrasts}
\small
\input{tables/CCFW_2026_AER--credit-main-contrasts}
\end{table}

\section{Conclusion}
\label{conclusion}

Beta-sorted portfolios are a commonly used empirical tool in asset pricing.  In a first step, time-varying factor exposures are estimated by rolling regressions of asset returns on observable risk factors to measure how returns co-vary with those factors. In a second step, assets with similar risk-factor exposures are grouped into portfolios, whose returns are compared across different exposure levels.  Yet the simple and intuitively appealing algorithm belies a more complicated statistical setting involving a two-step procedure in which both stages involve nonparametric estimation.

We provide a comprehensive econometric framework that rationalizes this commonly used estimator and links its properties to the first-step estimation window and the number of portfolios formed in the second step.  We establish uniform consistency and pointwise asymptotic normality for the time-averaged beta-sorted portfolio estimator, showing that its rate of convergence is $\sqrt{T}$ at fixed nonzero risk-factor exposures but a faster rate of $\sqrt{nT/J^d}$ at the origin.  We also show that certain weighted linear combinations can attain this faster rate of convergence.  We introduce a new variance estimator, characterize the properties of the ubiquitous Fama-MacBeth variance estimator, and provide inference procedures for hypotheses of interest.   Finally, we identify limitations of current practice and provide guidance on implementation and interpretation.

We apply our new theoretical results in two substantive empirical applications. First, we introduce a new risk factor designed to capture the business-credit cycle. We find a downward-sloping relationship between expected returns and exposure to this factor, with a statistically significant high-minus-low contrast using our proposed inference procedures. Second, we revisit the capital share risk factor proposed by \citet{LLM2019}. Using our new inference procedures, we do not find a statistically significant relation in the cross-section of U.S. equity returns.

\singlespacing
\section*{References} 
\begingroup
\renewcommand{\section}[2]{}	
\bibliography{CCFW_2026_AER--bib}
\bibliographystyle{jfe}
\endgroup

\end{document}


\pagenumbering{roman}

\title{\vspace{-0.0in}Beta-Sorted Portfolios \\Supplemental Appendix: Applied\bigskip }
\author{Matias D. Cattaneo\thanks{Department of Economics, Princeton University.} \and
        Richard K. Crump\thanks{Capital Markets, Federal Reserve Bank of New York.} \and
        Yingjie Feng\thanks{School of Economics and Management, Tsinghua University.} \and
        Weining Wang\thanks{School of Economics, University of Bristol.}}
\maketitle

\begin{abstract}
    This supplemental appendix provides practical guidance and numerical evidence for beta-sorted portfolio estimation and inference. Part~\ref{part:tuning} presents data-driven choices of the first-step bandwidth and the second-step number of portfolios and discusses their use in pointwise-in-$\beta$ and uniform-in-$\beta$ inference. Part~\ref{part:numerical} reports Monte Carlo evidence on confidence interval coverage and average length, additional empirical results for the business-credit factor, and an application to capital share risk. The companion Supplemental Appendix: Theory provides the theoretical foundations for these estimation and inference procedures.
\end{abstract}

\thispagestyle{empty}

\clearpage
\addtocontents{toc}{\protect\thispagestyle{empty}}
\tableofcontents
\thispagestyle{empty}

\newpage

\pagestyle{plain}

\clearpage
\pagenumbering{arabic}

\setcounter{part}{2}

\part{Tuning Parameter Selection}\label{part:tuning}

This part turns the rate restrictions in Part~\ref*{theory-part:classical}
into implementable tuning choices for the classical beta-sorted portfolio
estimator. Section~\ref{sec:tuning-pointwise} gives the
pointwise first-step plug-in rule for the bandwidth $h$ and the
integrated-MSE plug-in rule for the second-step number of portfolios $J$.
Section~\ref{sec:tuning-uniform} keeps those $h$ and
$J$ choices, and proposes a data-driven selector for the near-origin truncation parameter $\delta$.
Section~\ref{sec:tuning-factor-cancelling}
records the separate implication for factor-cancelling linear contrasts, whose
idiosyncratic normalization requires dedicated undersmoothing rather than the
default plug-in choices.

\section{Pointwise-in-\texorpdfstring{$\beta$}{beta}}\label{sec:tuning-pointwise}

This section summarizes a practical data-driven tuning rule for the classical beta-sorted
portfolio estimator and an associated robust bias-corrected confidence interval.
Throughout this section, braces on tuning parameters denote the relevant collection over time and units; for example, $\{J_t\}$ denotes $\{J_t:t=H+1,\ldots,T\}$.
\begin{itemize}
    \item We write the first-step estimator as $\widehat{\beta}_{it}(\mathfrak{p},h_{it})$, where $\mathfrak{p}$ is the local-polynomial order and $h_{it}$ is the bandwidth.
    \item We write the period-$t$ second-step fit as $\widehat{M}_t(\beta;\mathfrak{q},J_t)$, where $\mathfrak{q}$ is the portfolio polynomial order and $J_t$ is the number of portfolios per beta dimension at time $t$.
\end{itemize}
The full second-step estimator averages these period-specific fits:
\begin{align*}
    \widehat{\mu}(\beta;\mathfrak{q},\{J_t\})
    &=
    \frac{1}{T-H}\sum_{t=H+1}^{T}\widehat{M}_t(\beta;\mathfrak{q},J_t).
\end{align*}
Given choices of bandwidths $h_{it}$ and portfolio counts $J_t$,
the classical beta-sorted portfolio estimator is $\widehat{\mu}(\beta;0,\{J_t\})$ constructed using
$\widehat{\beta}_{it}(0,h_{it})$. The bias-corrected estimator used for inference is
$\widehat{\mu}(\beta;1,\{J_t\})$ constructed using $\widehat{\beta}_{it}(1,h_{it})$.

The bandwidths $h_{it}$ and portfolio counts $J_t$ are chosen once for the
classical point estimation objective and then reused for inference. Thus, the
empirical rule is simple: choose $h_{it}$ and $J_t$ as if the objective were the
classical estimator with $(\mathfrak{p},\mathfrak{q})=(0,0)$, but form confidence intervals with the
bias-corrected estimator with $(\mathfrak{p},\mathfrak{q})=(1,1)$. This is the usual robust
bias-correction logic: the lower-order fit determines the MSE-optimal tuning,
and the higher-order fit removes the leading smoothing bias at that same
tuning; see \citet{CCT2014} and \citet{Calonico-Cattaneo-Farrell_2018_JASA,Calonico-Cattaneo-Farrell_2022_Bernoulli}.

The plug-in logic is the same in both steps:
\begin{enumerate}
    \item run a pilot regression at the point estimation order and an auxiliary
    pilot regression at the higher order;
    \item estimate the variance constant from the point estimation pilot;
    \item estimate the leading bias constant from the difference between the two
    pilot fits; and
    \item plug the two constants into the MSE-optimal tuning formula.
\end{enumerate}

\subsection{First-Step Bandwidth Selection}\label{sec:tuning-first-step-bandwidth}

For first-step point estimation of order $\mathfrak{p}\geq0$, we use squared Euclidean loss for the vector $\widehat{\beta}_{it}(\mathfrak{p},h_{it})$. Accordingly, let $\mathcal{V}_{\mathtt{FS},it}(\mathfrak{p})$ denote the trace
of its leading conditional covariance matrix and let
$\mathcal{B}_{\mathtt{FS},it}^{2}(\mathfrak{p})$ denote the squared Euclidean
norm of its leading bias vector.
In the smooth-trend case considered in
Lemma~\ref*{theory-lem:first-step-bias}(i) with $s\geq \mathfrak{p}+1$, the point estimation
first-step (conditional) MSE has the expansion
\begin{align*}
    \MSE\bigl(\widehat{\beta}_{it}(\mathfrak{p},h_{it})\bigr)
    &=
    \mathcal{V}_{\mathtt{FS},it}(\mathfrak{p})(Th_{it})^{-1}
    +\mathcal{B}_{\mathtt{FS},it}^{2}(\mathfrak{p})h_{it}^{2(\mathfrak{p}+1)}
    +o_{\P}\big((Th_{it})^{-1}+h_{it}^{2(\mathfrak{p}+1)}\big).
\end{align*}
The MSE-optimal bandwidth is therefore
\begin{align*}
    h_{it}^{\star}(\mathfrak{p})
    &=
    \left(
    \frac{\mathcal{V}_{\mathtt{FS},it}(\mathfrak{p})}
    {2(\mathfrak{p}+1)\mathcal{B}_{\mathtt{FS},it}^{2}(\mathfrak{p})}
    \right)^{1/(2\mathfrak{p}+3)}
    T^{-1/(2\mathfrak{p}+3)},
\end{align*}
provided that $\mathcal{B}_{\mathtt{FS},it}^{2}(\mathfrak{p})>0$. Algorithm~\ref{alg:first-step-bandwidth} implements this rule for the classical $\beta$-estimation underlying the beta-sorted portfolio estimator.

\bigskip
\begin{algorithm}[H]
    \caption{First-step MSE-optimal bandwidth ($\mathfrak{p}=0$)}\label{alg:first-step-bandwidth}
    \SetAlgoLined
    \KwIn{Pilot bandwidth $h_{it,\mathrm{pilot}}$}
    \textbf{Pilot regressions at $h_{it,\mathrm{pilot}}$:}\;
    \quad order $\mathfrak{p}=0$ fit $\widehat{\beta}_{it}(0,h_{it,\mathrm{pilot}})$ of $R_{i,t-s}$ on $X_s(t)$ over $s\in[1,\lceil T h_{it,\mathrm{pilot}}\rceil]$\;
    \quad order $\mathfrak{p}=1$ fit $\widehat{\beta}_{it}(1,h_{it,\mathrm{pilot}})$ over the same window and with the same kernel\;
    \textbf{Variance estimate:} $\widehat{\mathcal{V}}_{\mathtt{FS},it}(0)\leftarrow$ trace of the sandwich covariance constant from the $\mathfrak{p}=0$ pilot\;
    \textbf{Squared-bias estimate:}
    $\widehat{\mathcal{B}}_{\mathtt{FS},it}^{2}(0)\leftarrow
    \left\|\bigl(\widehat{\beta}_{it}(1,h_{it,\mathrm{pilot}})
    -\widehat{\beta}_{it}(0,h_{it,\mathrm{pilot}})\bigr)    /h_{it,\mathrm{pilot}}\right\|^2$\;
    \textbf{Plug-in bandwidth:}
    \begin{align*}
        \widehat{h}_{it}^{\star}
        &\leftarrow
        \left(
        \frac{\widehat{\mathcal{V}}_{\mathtt{FS},it}(0)}
        {2\widehat{\mathcal{B}}_{\mathtt{FS},it}^{2}(0)}
        \right)^{1/3}
        T^{-1/3}
    \end{align*}\;
    \textbf{Estimates:} compute $\widehat{\beta}_{it}=\widehat{\beta}_{it}(0,\widehat h_{it}^{\star})$ and $\widehat{\beta}_{it}^{\mathrm{BC}}=\widehat{\beta}_{it}(1,\widehat h_{it}^{\star})$\;
    \KwOut{$\widehat{h}_{it}^{\star}$, $\widehat{\beta}_{it}$, and $\widehat{\beta}_{it}^{\mathrm{BC}}$.}
\end{algorithm}

\subsection{Second-Step Selection of the Number of Portfolios}\label{sec:tuning-second-step-portfolios}

For second-step point estimation using within-portfolio polynomials of order $\mathfrak{q}\geq0$, we use the same marginal portfolio count $J_t$ in every beta coordinate.  Thus $J_t$ is a single scalar selected for period $t$, and
the tensor-product partition contains at most $J_t^d$ cells. Let
$\mathcal{V}_{\mathtt{SS},t}(\beta;\mathfrak{q})$ denote the leading variance constant of
$\widehat{M}_t(\beta;\mathfrak{q},J_t)$, and let $\mathcal{B}_{\mathtt{SS},t}(\beta;\mathfrak{q})$ denote the
leading approximation-bias constant multiplying $J_t^{-(\mathfrak{q}+1)}$. Up to additive
terms that do not depend on the portfolio count, including the factor time-series component
present when $\beta\neq0$, we use the period-$t$ point estimation approximate MSE criterion
\begin{align*}
    \MSE_t\bigl(\widehat{M}_t(\beta;\mathfrak{q},J_t)\bigr)
    &=
    \mathcal{V}_{\mathtt{SS},t}(\beta;\mathfrak{q})\frac{J_t^d}{n_t}
    +\mathcal{B}_{\mathtt{SS},t}(\beta;\mathfrak{q})^2J_t^{-2(\mathfrak{q}+1)}
    +o_\P(J_t^d/n_t+J_t^{-2(\mathfrak{q}+1)}).
\end{align*}

To obtain one portfolio count for the whole beta domain, fix a nonnegative
weight function $w(\beta)$ on $\mathcal B^\circ$ satisfying
$0<\int_{\mathcal B^\circ}w(\beta)d\beta<\infty$, and define the integrated
variance and squared-bias constants
\begin{align*}
    \mathcal{V}_{\mathtt{SS},t}^{w}(\mathfrak{q})
    &:=
    \int_{\mathcal B^\circ}
    \mathcal{V}_{\mathtt{SS},t}(\beta;\mathfrak{q})
    w(\beta)d\beta,
    \\
    \mathcal{B}_{\mathtt{SS},t}^{2,w}(\mathfrak{q})
    &:=
    \int_{\mathcal B^\circ}
    \mathcal{B}_{\mathtt{SS},t}(\beta;\mathfrak{q})^2
    w(\beta)d\beta.
\end{align*}
The corresponding approximate integrated MSE is
\begin{align*}
    \operatorname{IMSE}_t(\mathfrak{q},J_t;w)
    &=
    \mathcal{V}_{\mathtt{SS},t}^{w}(\mathfrak{q})
    \frac{J_t^d}{n_t}
    +
    \mathcal{B}_{\mathtt{SS},t}^{2,w}(\mathfrak{q})
    J_t^{-2(\mathfrak{q}+1)}
    \\
    &\quad
    +o_\P(J_t^d/n_t+J_t^{-2(\mathfrak{q}+1)}).
\end{align*}
Thus the IMSE-optimal common number of portfolios per beta dimension is
\begin{align*}
    J_t^{\star}(\mathfrak{q};w)
    &=
    \left(
    \frac{2(\mathfrak{q}+1)
    \mathcal{B}_{\mathtt{SS},t}^{2,w}(\mathfrak{q})}
    {d\,\mathcal{V}_{\mathtt{SS},t}^{w}(\mathfrak{q})}
    \right)^{1/(d+2\mathfrak{q}+2)}
    n_t^{1/(d+2\mathfrak{q}+2)},
\end{align*}
provided that
$\mathcal{B}_{\mathtt{SS},t}^{2,w}(\mathfrak{q})>0$.
Algorithm~\ref{alg:second-step-portfolios} implements this rule for the classical beta-sorted portfolio estimator.

\bigskip
\begin{algorithm}[H]
    \caption{Second-step IMSE-optimal portfolios ($\mathfrak{q}=0$)}\label{alg:second-step-portfolios}
    \SetAlgoLined
    \KwIn{First-step point estimates $\{\widehat{\beta}_{it}\}$ from Algorithm~\ref{alg:first-step-bandwidth}; pilot portfolio counts $\{J_{t,\mathrm{pilot}}\}$ with common scale $J_{\mathrm{pilot}}$; grid $\mathcal G=\{\beta_\ell:\ell\in[L]\}\subset\mathcal B^\circ$; nonnegative grid weights $\{w_\ell:\ell\in[L]\}$}
    \textbf{Pilot portfolio regressions at $\{J_{t,\mathrm{pilot}}\}$:}\;
    \quad second-step order $\mathfrak{q}=0$ portfolio regression $\widehat{M}_t(\cdot;0,J_{t,\mathrm{pilot}})$\;
    \quad second-step order $\mathfrak{q}=1$ portfolio regression $\widehat{M}_t(\cdot;1,J_{t,\mathrm{pilot}})$\;
    \textbf{Pointwise variance estimates:} $\widehat{\mathcal{V}}_{\mathtt{SS},t}(\beta_\ell;0)\leftarrow$ period-$t$ sandwich variance constant from the $\mathfrak{q}=0$ pilot for each $\ell\in[L]$\;
    \textbf{Integrated variance estimate:}
    $\widehat{\mathcal{V}}_{\mathtt{SS},t}^{w}(0)
    \leftarrow
\sum_{\ell=1}^{L}w_\ell\widehat{\mathcal{V}}_{\mathtt{SS},t}(\beta_\ell;0)$\;
    \textbf{Integrated squared-bias estimate:}
    \begin{align*}
        \widehat{\mathcal{B}}_{\mathtt{SS},t}^{2,w}(0)
        &\leftarrow
        \sum_{\ell=1}^{L}w_\ell
        \left[
        \widehat{M}_t(\beta_\ell;1,J_{t,\mathrm{pilot}})
        -
        \widehat{M}_t(\beta_\ell;0,J_{t,\mathrm{pilot}})
        \right]^2
        J_{t,\mathrm{pilot}}^2
    \end{align*}\;
    \textbf{Plug-in number of portfolios:}
    \begin{align*}
        \widehat{J}_t^{\star}
        &\leftarrow
        \left(
        \frac{2\widehat{\mathcal{B}}_{\mathtt{SS},t}^{2,w}(0)}
        {d\,\widehat{\mathcal{V}}_{\mathtt{SS},t}^{w}(0)}
        \right)^{1/(d+2)}
        n_t^{1/(d+2)}
    \end{align*}\;
    \KwOut{The common marginal portfolio count $\widehat{J}_t^{\star}$.}
\end{algorithm}

Our default implementation takes $w_\ell=1$ for every grid point. Hence the
estimated squared-bias constant is the sum of the squared pointwise pilot
biases and the estimated variance constant is the sum of the pointwise pilot
variance constants. Normalizing the grid weights to sum to one would leave
$\widehat J_t^\star$ unchanged because the same normalization multiplies the
numerator and denominator.
On an irregular grid, the $w_\ell$ may instead combine the desired
$w(\beta_\ell)$ with quadrature weights; this changes the discrete
approximation to the integral but not the form of the selector.

\subsection{Estimation and Robust Bias-Corrected Inference}\label{sec:tuning-pointwise-rbc-inference}

The classical beta-sorted portfolio estimator constructed using the first- and second-step plug-in rules is
\begin{align*}
    \widehat{\mu}(\beta)
    &=
    \widehat{\mu}(\beta;0,\{\widehat J_t^{\star}\}),
\end{align*}
which is computed using the $\mathfrak{p}=0$ first-step estimates $\{\widehat{\beta}_{it}\}$ from Algorithm \ref{alg:first-step-bandwidth}.

A robust bias-corrected (RBC) confidence interval uses the $\mathfrak{p}=\mathfrak{q}=1$ fits with the selected tuning parameters $\{\widehat{h}_{it}^{\star}\}$ and $\{\widehat{J}_t^{\star}\}$, and the bias-corrected first-step point estimates $\{\widehat{\beta}^{\mathrm{BC}}_{it}\}$ from Algorithm~\ref{alg:first-step-bandwidth}. Specifically, a $100(1-\alpha)\%$ nominal level RBC confidence interval is
\begin{align*}
    \widehat{\mu}^{\mathrm{BC}}(\beta)
    \pm z_{1-\alpha/2}\sqrt{\widehat{\Sigma}^{\mathrm{BC}}(\beta)},
\end{align*}
where $\widehat{\mu}^{\mathrm{BC}}(\beta)=\widehat{\mu}(\beta;1,\{\widehat J_t^{\star}\})$ using $\{\widehat{\beta}^{\mathrm{BC}}_{it}\}$, $\{\widehat{h}_{it}^{\star}\}$, and $\{\widehat{J}_t^{\star}\}$, and $\widehat{\Sigma}^{\mathrm{BC}}(\beta)=\widehat{\Sigma}(\beta)$ as in Section~\ref*{theory-sec:pointwise-variance} for $\widehat{\mu}^{\mathrm{BC}}(\beta)$.
For each fixed $\beta\neq0$, under the pointwise Gaussian-approximation
conditions and the conditions of Theorem~\ref*{theory-thm:variance-estimation}, this
interval has exact pointwise coverage when
$T\mathscr{B}_{\mathtt{F}}(\beta)=o_\P(1)$; without this additional
factor-prediction rate it is conservative because
$\mathscr{B}_{\mathtt{F}}(\beta)\geq0$, and no anchoring condition is needed
because Lemma~\ref*{theory-lem:bin-restricted-jackknife} gives
$|\widehat{\Sigma}_{\varepsilon}(\beta)|
=O_\P(J^d/(nT))=o_\P(T^{-1})$.

\subsection{Validity of Tuning Parameter Selection}\label{sec:tuning-pointwise-validity}

The next lemma checks the polynomial-rate restrictions for the plug-in
scaling. The key distinction is the normalization. Pointwise inference at a
fixed nonzero $\beta$ is governed by the factor scale. Pointwise inference at
$\beta=0$ and factor-cancelling contrasts are governed by the idiosyncratic
scale, which magnifies the components of $\mathsf{R}_{\mathtt{LA}}$ that are
harmless at the factor scale.

\begin{lemma}[Validity of Pointwise-in-$\beta$ Tuning]\label{lem:tuning-pointwise-validity}
    Suppose $T=n^{\eta}$ for some $\eta>0$, and suppose that Assumptions~\ref*{theory-ass:dgp}, \ref*{theory-ass:first-step-ls}, \ref*{theory-ass:partition}, and \ref*{theory-ass:first-step-bias} hold and that $s_\mu\geq2$. {Suppose} the plug-in choices satisfy, uniformly over their indices,
    \begin{align*}
        \widehat h_{it}^{\star}\asymp h\asymp T^{-1/3},
        \qquad
        \widehat J_t^{\star}\asymp J\asymp n^{1/(d+2)},
    \end{align*}
    and take the moment exponent in Assumption~\ref*{theory-ass:dgp} sufficiently large. Then the following rate implications hold.
    \begin{enumerate}[label=(\roman*)]
        \item \emph{Pointwise inference for $\beta\neq0$.} The plug-in scaling verifies the factor-scale restrictions for fixed-nonzero-$\beta$ inference if
        \begin{align}
            0<\eta&<
            \frac{4}{d+2}.
            \label{eq:tuning-pointwise-nonzero-condition}
        \end{align}
        {In addition, the deterministic threshold in
        Assumption~\ref*{theory-ass:first-step-bias} must satisfy
        \begin{align*}
            \sqrt T\,\mathsf{B}_{\mathtt{FS}}
            &=o(1).
        \end{align*}}
        The upper bound is the factor-scale second-step smoothing-bias condition for the $\mathfrak{q}=1$ inference fit,
        \begin{align*}
            \sqrt T J^{-2}\to0,
        \end{align*}
        which is equivalent to $\eta<4/(d+2)$ under $J\asymp n^{1/(d+2)}$ and $T=n^\eta$.

        \item \emph{Pointwise inference for $\beta=0$ and factor-cancelling contrasts.} Pointwise inference at $\beta=0$ in Theorem~\ref*{theory-thm:pointwise-clt} and factor-cancelling contrast inference in Theorem~\ref*{theory-thm:factor-cancelling-inference} both require the linearization error to be negligible after normalization by $\sqrt{nT/J^d}$. Ignoring logarithmic factors, the third component of $\mathsf{R}_{\mathtt{LA}}$ in Lemma~\ref*{theory-lem:classical-linearization} requires
        \begin{align}
            \sqrt{\frac{nT}{J^d}}
            \sqrt{\frac{J^{2d}}
            {nHJ^{d-1}\max\{J,\sqrt H\}}}
            =
            \sqrt{\frac{TJ}{H\max\{J,\sqrt H\}}}
            &\to0.
            \label{eq:tuning-origin-linearization}
        \end{align}
        With $H\asymp T^{2/3}$ and $J\asymp n^{1/(d+2)}$, the right-hand side is equal to
        \begin{align*}
            \begin{cases}
                T^{1/6},&J\geq \sqrt H,\\
                J^{1/2},&J<\sqrt H,
            \end{cases}
        \end{align*}
        and therefore diverges. Thus the plug-in scaling cannot verify the natural origin-scale RBC rate restrictions. In addition, the second-step smoothing-bias condition alone would require
        \begin{align*}
            \frac{nT}{J^{d+4}}\to0
            \qquad\Longleftrightarrow\qquad
            \eta<\frac{2}{d+2},
        \end{align*}
        while the deterministic $H^{-1}$ component of
        $\mathsf{R}_{\mathtt{LA}}$ would require, up to logarithmic factors,
        \begin{align*}
            \eta>\frac{6}{d+2}.
        \end{align*}
        {The first-step bias component would separately
        require
        \begin{align*}
            \sqrt{\frac{nT}{J^d}}\,\mathsf{B}_{\mathtt{FS}}
            &=o(1).
        \end{align*}}
    \end{enumerate}
\end{lemma}

The stronger primitive condition $J\mathsf{R}_{\mathtt{FS}}=o(1)$ is not needed for this fixed-nonzero-$\beta$ factor-scale calculation; it is an idiosyncratic-scale sufficient condition for absorbing the basis-switching term into the local-support scale. Thus, the recommended choices of $h$ and $J$, {together with
$\sqrt T\,\mathsf{B}_{\mathtt{FS}}=o(1)$}, verify fixed-nonzero-$\beta$ pointwise inference within the growth window in \eqref{eq:tuning-pointwise-nonzero-condition}, but do not verify idiosyncratic-scale RBC inference at the origin or for factor-cancelling contrasts.

\subsection{Proof of Lemma~\ref{lem:tuning-pointwise-validity}}

Throughout the proof, exponent comparisons suppress logarithmic factors.
{The additional first-step bias restriction in part
\textnormal{(i)} gives
$\sqrt T\,\mathsf{B}_{\mathtt{FS}}=o(1)$.}
The first-step variance terms satisfy
\begin{align*}
    \sqrt{T}\frac{1}{Th}
    &\asymp
    T^{-1/6}\to0,
\end{align*}
and the remaining first-step terms involving the moment exponent are negligible by taking that exponent sufficiently large.
For the second-step portfolio fit with $\mathfrak{q}=1$ used for inference, the smoothing bias is of order
$J^{-2}$. Since $J\asymp n^{1/(d+2)}$ and
$T=n^\eta$,
\begin{align*}
    \sqrt{T}J^{-2}\to0
    \quad\Longleftrightarrow\quad
    n^{\eta/2-2/(d+2)}\to0
    \quad\Longleftrightarrow\quad
    \eta<\frac{4}{d+2},
\end{align*}
which gives the upper bound in \eqref{eq:tuning-pointwise-nonzero-condition}. The main portfolio variance restriction is
$J^d/(nh)=o(1)$. With the same substitutions,
\begin{align*}
    \frac{J^d}{nh}
    &\asymp
    n^{d/(d+2)-1+\eta/3},
\end{align*}
so this term is $o(1)$ whenever $\eta<6/(d+2)$, which is implied by \eqref{eq:tuning-pointwise-nonzero-condition}. This condition also implies $J^d/n=o(1)$, which is the effective nonempty-portfolio requirement in the consistency and variance-estimation arguments. The sharpened jackknife restriction for the $\mathfrak{q}=1$ inference fit requires $HJ^{-4}=o(1)$ up to logarithmic factors. Since
\begin{align*}
    HJ^{-4}
    &\asymp
    n^{2\eta/3-4/(d+2)},
\end{align*}
this condition follows from \eqref{eq:tuning-pointwise-nonzero-condition}. The remaining terms in \arxiveqref{theory-eq:jackknife-conditions} are smaller under the same condition and the first-step rates above.
The endpoint condition also holds, since
\begin{align*}
    \frac{J^2}{n}
    &\asymp
    n^{2/(d+2)-1}
    =o(1).
\end{align*}

The full linear-approximation remainder is negligible at the factor scale. The deterministic pieces satisfy
\begin{align*}
    {
    \sqrt T\left(\frac{1}{H}+\mathsf{B}_{\mathtt{FS}}\right)}
    &{\lesssim
    T^{-1/6}+\sqrt T\,\mathsf{B}_{\mathtt{FS}}
    =o(1)}.
\end{align*}
The empirical-process component in $\mathsf{R}_{\mathtt{LA}}$ satisfies
\begin{align*}
    \sqrt T
    \sqrt{\frac{J^{2d}}
    {nHJ^{d-1}\max\{J,\sqrt H\}}}
    &=
    \begin{cases}
        \sqrt{T J^d/(nH)},&J\geq\sqrt H,\\[0.4em]
        \sqrt{T J^{d+1}/(nH^{3/2})},&J<\sqrt H,
    \end{cases}
    =o(1),
\end{align*}
where the first case uses $\eta<6/(d+2)$ and the second case is $n^{-1/(2(d+2))}$. The Gram-perturbation term is also negligible. Indeed, Lemma~\ref*{theory-lem:partition-basis-moments} gives $J^d\mathsf{R}_{\mathtt{Q}}=o(1)$ and
\begin{align*}
    \sqrt{T}\,J^d\mathsf{R}_{\mathtt{Q}}
    \sqrt{\frac{J^d}{nH}}
    &\leq
    J^d\mathsf{R}_{\mathtt{Q}}
    \sqrt{\frac{J^d}{nh}}
    =o(1),
\end{align*}
where logarithmic factors are suppressed. The basis-switching term is also negligible at the factor scale without imposing the stronger idiosyncratic-scale condition $J^d\mathsf{R}_{\mathtt{P}}=o(1)$. Since the minimum operators in $\mathsf{R}_{\mathtt{P}}$ are bounded by one,
\begin{align*}
    \mathsf{R}_{\mathtt{P}}
    &\lesssim
    J^{-d}
    +\sqrt{\frac{J^{-d}\log(nT)}{n}}
    +\frac{\log(nT)}{n}.
\end{align*}
Therefore Lemma~\ref*{theory-lem:general-linearization} gives
\begin{align*}
    \sqrt{T}\,|\mathscr{R}_2(\beta)|
    &\lesssim_\P
    \sqrt{\frac{J^{2d}\mathsf{R}_{\mathtt{P}}}{n}}
    =
    o_\P(1),
\end{align*}
because $J\asymp n^{1/(d+2)}$ implies $J^d/n=o(1)$.
The same-cell nonconstant-basis component in $\mathsf{R}_{\mathtt{LA}}$ is negligible away from the origin. At the plug-in scaling, for fixed $\beta\neq0$,
\begin{align*}
    \sqrt{T}
    \sqrt{\frac{J^{d+2}}{nH^2}}
    &\asymp
    T^{-1/6}
    =o(1),
\end{align*}
because $J^{d+2}\asymp n$ and $H\asymp T^{2/3}$. The endpoint-interior term is smaller:
\begin{align*}
    \sqrt{T}\,
    \sqrt{\frac{J^{d+2}}{n^2H}}
    &\asymp
    \sqrt{\frac{TJ^{d+2}}{n^2H}}
    =
    \sqrt{\frac{T}{nH}}
    =o(1).
\end{align*}
Thus the infeasible Gaussian approximation holds for each fixed $\beta\neq0$ at the plug-in scaling.

At the origin, the relevant normalization is $\sqrt{nT/J^d}$. The third component of $\mathsf{R}_{\mathtt{LA}}$ gives the necessary condition in \eqref{eq:tuning-origin-linearization}. At the plug-in scaling,
\begin{align*}
    \sqrt{\frac{nT}{J^d}}
    \sqrt{\frac{J^{2d}}
    {nHJ^{d-1}\max\{J,\sqrt H\}}}
    &=
    \begin{cases}
        T^{1/6},&J\geq \sqrt H,\\
        J^{1/2},&J<\sqrt H,
    \end{cases}
\end{align*}
which diverges. Hence $\mathsf{R}_{\mathtt{CLT}}/\sqrt{\Sigma(0)}$ cannot vanish under the plug-in scaling. Even ignoring this obstruction, the smoothing-bias component in $\mathsf{R}_{\mathtt{CLT}}$ would require
\begin{align*}
    \sqrt{\frac{nT}{J^d}}J^{-2}\to0
    \quad\Longleftrightarrow\quad
    \frac{nT}{J^{d+4}}\to0
    \quad\Longleftrightarrow\quad
    \eta<\frac{2}{d+2}.
\end{align*}
The deterministic and smooth-interior components of $\mathsf{R}_{\mathtt{LA}}$ also give
\begin{align*}
    \sqrt{\frac{nT}{J^d}}
    \left(
    {\frac{1}{H}+\mathsf{B}_{\mathtt{FS}}}
    +\sqrt{\frac{J^{d+2}}{nH^2}}
    \right)
    &\gtrsim
    \frac{J\sqrt{T}}{H},
\end{align*}
up to logarithmic factors. Under the plug-in scaling,
\begin{align*}
    \frac{J\sqrt{T}}{H}
    &\asymp
    n^{1/(d+2)-\eta/6},
\end{align*}
so these components would require $\eta>6/(d+2)$, which is incompatible with the smoothing-bias requirement above.
{The first-step bias component would additionally
require
\begin{align*}
    \sqrt{\frac{nT}{J^d}}\,\mathsf{B}_{\mathtt{FS}}
    &=o(1).
\end{align*}}
The recommended tuning therefore does not verify natural origin-scale RBC
inference. The same obstruction applies to factor-cancelling contrasts. In
Theorem~\ref*{theory-thm:factor-cancelling-inference}, the factor term cancels, so the
condition \arxiveqref{theory-eq:factor-cancelling-rate} normalizes the residual at the idiosyncratic scale
$\sqrt{nT/J^d}$. By definition,
$\mathsf{R}_{\mathtt{\Psi}}=(\sum_{k=1}^{K_{\Psi}}|a_k|)
\mathsf{R}_{\mathtt{CLT}}$, and therefore it contains the same
$\mathsf{R}_{\mathtt{LA}}$ component.
Thus \arxiveqref{theory-eq:factor-cancelling-rate} imposes the same necessary condition
\eqref{eq:tuning-origin-linearization}, which diverges under the plug-in scaling.
\qed

\section{Uniform-in-\texorpdfstring{$\beta$}{beta}}\label{sec:tuning-uniform}

This section gives the tuning and implementation steps for the
uniform-in-$\beta$ confidence intervals of Theorem~\ref*{theory-thm:uniform-coverage}.
Throughout, ``uniform-in-$\beta$'' means that the coverage probability is
controlled uniformly over evaluation points
$\beta\in\mathcal{B}^\circ$, in the pointwise-uniform sense defined in
Section~\ref*{theory-sec:uniform-beta-inference}; it is not a simultaneous confidence band for
the full function $\mu(\cdot)$. The region $\mathcal{B}^\circ$ is the same
fixed interior region used in Theorem~\ref*{theory-thm:uniform-coverage}, and it
contains a fixed neighborhood of the origin.

We keep the first-step bandwidths and second-step portfolio counts
selected in Section~\ref{sec:tuning-pointwise}. The rate calculation
uses the same uniform orders as in
Lemma~\ref{lem:tuning-pointwise-validity}, namely
$\widehat h_{it}^{\star}\asymp T^{-1/3}$ and
$\widehat J_t^{\star}\asymp n_t^{1/(d+2)}$.
Algorithm~\ref{alg:second-step-portfolios} implements a grid approximation
to the period-specific weighted IMSE criterion.

We separate the final time-average burn-in from the selected
local first-step windows. Let $T_0$ denote a pre-specified implementation
burn-in, chosen as a deterministic cap for the selected first-step windows so
that, with probability approaching one,
\begin{align*}
    T_0
    &\geq
    \max_{1\leq s\leq T}\max_{i\in[n_s]}\widehat H_{is}^{\star},
    \qquad
    \widehat H_{is}^{\star}=\lceil T\widehat h_{is}^{\star}\rceil,
\end{align*}
and $T_0\asymp T^{2/3}$. The final time-series averages below use
$t=T_0+1,\ldots,T$, whereas each first-step local-polynomial fit still uses its own selected window
$\widehat H_{it}^{\star}$. The remaining tuning parameter is the near-origin truncation
parameter $\delta$, which determines where the usual feasible standard error
is replaced by the conservative variance floor from
Theorem~\ref*{theory-thm:uniform-coverage}.

\subsection{Near-Origin Truncation Parameter}\label{sec:tuning-uniform-delta}

Theorem~\ref*{theory-thm:uniform-coverage} uses a scalar threshold $\delta$ in
factor-variance units. Its role in the implementation is only to identify a
small neighborhood of the origin in which the usual factor-based standard
error is replaced by a conservative floor. We therefore use a deliberately
simple rule of thumb rather than attempting to estimate every remainder
constant.

For the degree-one second step used in the robust bias-corrected
implementation, we choose a threshold of the form
\begin{align}\label{eq:tuning-delta-rule}
    \widehat\delta
    &:=
    c_\delta\widehat s_R^2
    (\log T)^{d+1} (\log\log T)
    \max\left\{
    \widehat r_J,
    \widehat r_H,
    \widehat r_N
    \right\}.
\end{align}
Here $\widehat r_J$, $\widehat r_H$, and $\widehat r_N$ estimate the rate
components associated with the squared degree-one second-step approximation
bias, the squared local-window remainder, and the uniform first-step
stochastic effect, respectively; $\widehat s_R^2$ puts the threshold in return-
variance units.  Section~\ref{sec:implementation-defaults} gives the exact
finite-sample definitions of these quantities and the default choice of
$c_\delta$. The adjustment factor $(\log T)^{d+1}(\log\log T)$ ensures that the threshold rate restrictions required by Theorem~\ref*{theory-thm:uniform-coverage} are satisfied under our recommended tuning parameter choices, provided that $\mathsf B_{\mathtt{FS}}=O(H^{-1})$.

Under independent factors and homoskedastic first-step errors, converting
coefficient noise back to factor-variance units cancels the factor variance;
local portfolio averaging then gives the rate $J^d/(nH)$ underlying
$\widehat r_N$.

The rule is scale equivariant in the intended sense. Multiplying returns by
$a$ multiplies $\widehat s_R^2$ and $\widehat\delta$ by
$a^2$. A diagonal rescaling of a beta coordinate paired with the inverse
rescaling of its factor leaves the switching comparison unchanged. Coordinate
permutations leave the rate candidates unchanged. The rule is a finite-sample
calibration device, not a claim that the unknown
optimal threshold constant has been consistently estimated. Formal use of
Theorem~\ref*{theory-thm:uniform-coverage} still requires the resulting scalar
threshold to satisfy \arxiveqref{theory-eq:uniform-delta-conditions}.

\subsection{Estimation and Robust Bias-Corrected Inference}\label{sec:tuning-uniform-rbc-inference}

Given $\widehat h_{it}^{\star}$, the common portfolio-count scale
$\widehat J_t^{\star}$, and the threshold $\widehat\delta$, compute the
same robust bias-corrected estimator as in
Subsection~\ref{sec:tuning-pointwise-rbc-inference}, but with the common
uniform-in-$\beta$ portfolio-count sequence and the final averaging range
$t=T_0+1,\ldots,T$:
\begin{align*}
    \widehat{\mu}^{\mathrm{BC}}(\beta)
    &=
    \frac{1}{T-T_0}
    \sum_{t=T_0+1}^{T}
    \widehat{M}_t(\beta;1,\widehat J_t^{\star}),
\end{align*}
using the bias-corrected first-step estimates
$\{\widehat{\beta}_{it}^{\mathrm{BC}}\}$ and the selected bandwidths
$\{\widehat h_{it}^{\star}\}$. For this estimator, compute the feasible
factor component $\widehat\Sigma_f(\beta)$ and the idiosyncratic component
$\widehat\Sigma_{\varepsilon}^{\mathrm{BC}}(\beta)$ as in
Section~\ref*{theory-sec:pointwise-variance}, but over the same final averaging range
$t=T_0+1,\ldots,T$, with the idiosyncratic component using the residuals and
weights from the same robust bias-corrected fit. Set
\begin{align*}
    \widehat\Sigma^{\mathrm{BC}}(\beta)
    &=
    \widehat\Sigma_f(\beta)
    +
    \widehat\Sigma_{\varepsilon}^{\mathrm{BC}}(\beta).
\end{align*}

The conservative variance floor is
\begin{align*}
    \widehat{\Sigma}_{\widehat\delta}^{\mathrm{BC}}
    &=
    \sup_{\substack{b\in\mathcal{B}^\circ\\
    \widehat\Sigma_f(b)=\widehat\delta}}
    \max\{\widehat{\Sigma}^{\mathrm{BC}}(b),\widehat\delta\}
    =
    \widehat\delta+
    \sup_{\substack{b\in\mathcal{B}^\circ\\
    \widehat\Sigma_f(b)=\widehat\delta}}
    \max\{\widehat{\Sigma}_{\varepsilon}^{\mathrm{BC}}(b),0\}.
\end{align*}
If the level set is empty in a finite sample, set
$\widehat{\Sigma}_{\widehat\delta}^{\mathrm{BC}}=\widehat\delta$. For
numerical work, the boundary supremum is evaluated on the grid
$\mathcal{G}$, using interpolation or the nearest grid points to the level set
$\widehat\Sigma_f(b)=\widehat\delta$. This grid approximation is purely
computational; the theoretical statement above is the continuum version.

The variance used in the uniform-in-$\beta$ robust bias-corrected interval is
\begin{align*}
    \widehat{\Sigma}_{\mathtt{UI}}^{\mathrm{BC}}(\beta)
    &=
    \begin{cases}
        \widehat{\Sigma}^{\mathrm{BC}}(\beta)&
        \quad\text{if }\widehat\Sigma_f(\beta)>\widehat\delta\\
        \max\{\widehat{\Sigma}^{\mathrm{BC}}(\beta),
        \widehat{\Sigma}_{\widehat\delta}^{\mathrm{BC}}\}&
        \quad\text{if }\widehat\Sigma_f(\beta)\leq\widehat\delta
    \end{cases},
\end{align*}
The resulting $100(1-\alpha)\%$ interval is
\begin{align*}
    \widehat{\mu}^{\mathrm{BC}}(\beta)
    \pm
    z_{1-\alpha/2}
    \sqrt{\widehat{\Sigma}_{\mathtt{UI}}^{\mathrm{BC}}(\beta)}.
\end{align*}

\section{Factor-Cancelling Linear Contrast}\label{sec:tuning-factor-cancelling}

Factor-cancelling contrasts do not introduce a new threshold parameter. Their
distinctive feature is the normalization: after the factor component cancels,
Theorem~\ref*{theory-thm:factor-cancelling-inference} works at the idiosyncratic scale
$\sqrt{nT/J^d}$. Thus a factor-cancelling contrast is covered by the theory
only when the selected first-step bandwidths and second-step portfolio counts
verify
\begin{align*}
    \sqrt{\frac{nT}{J^d}}\mathsf{R}_{\mathtt{\Psi}}
    &\to0,
\end{align*}
together with the remaining conditions in Theorem~\ref*{theory-thm:factor-cancelling-inference}.
For the fixed finite contrast, the definition above yields
\begin{align*}
    \mathsf{R}_{\mathtt{\Psi}}
    &={}
    \left(\sum_{k=1}^{K_{\Psi}}|a_k|\right)
    \mathsf{R}_{\mathtt{CLT}},
\end{align*}
so this condition is equivalent to imposing the same normalized rate on
$\mathsf{R}_{\mathtt{CLT}}$.

The pointwise first-step and IMSE second-step plug-in choices from
Algorithm~\ref{alg:first-step-bandwidth} and
Algorithm~\ref{alg:second-step-portfolios} are therefore not a validated default for
factor-cancelling inference. Lemma~\ref{lem:tuning-pointwise-validity} shows that,
under those choices, the idiosyncratic-scale normalization magnifies the
linear-approximation remainder in $\mathsf{R}_{\mathtt{LA}}$, and the required
rate diverges. Consequently, a factor-cancelling confidence interval should use
the jackknife variance estimator in Theorem~\ref*{theory-thm:factor-cancelling-inference} only
with a dedicated undersmoothing choice of $h$ and $J$ that satisfies the
display above. Operationally, this means choosing $h$ and $J$ to make the
estimated non-leading remainder small relative to $\sqrt{J^d/(nT)}$, rather
than choosing $J$ to minimize the IMSE of the lower-order estimator.

\section{Default Implementation Choices}
\label{sec:implementation-defaults}

In this section, we describe the implementation conventions used in the two empirical applications and the Monte Carlo simulations. 

\paragraph{Tuning parameters} We follow the approach described in Section~\ref{sec:tuning-pointwise}.  The pilot choices are $H_{\mathrm{pilot}} = 10$ years (120 months or 40 quarters) and $J_{\mathrm{pilot}}=10$. Thus, $h_{\mathrm{pilot}}=120/T$ for monthly data and
$h_{\mathrm{pilot}}=40/T$ for quarterly data.  Here $h=H/T$, so the denominator is the full time-series sample size $T$, rather than the $T-H$ dates entering the second-step average.  We make a number of finite-sample regularizations to ensure that the
first-step and second-step estimators are well behaved. For the first step we restrict the selected window to $36\leq H_{it}\leq360$ for monthly data and $36\leq H_{it}\leq120$ for quarterly data.  For an unbalanced panel, we require at least 36 usable observations within the selected window to estimate a beta.  In addition, for inference, we use $\mathfrak{p}=1$ so long as $
\widehat{\operatorname{Var}}_{\mathrm{HC1}}
\bigl(\widehat\beta_{it}^{(1)}\bigr)
\leq
T^{1/10}
\widehat{\operatorname{Var}}_{\mathrm{HC1}}
\bigl(\widehat\beta_{it}^{(0)}\bigr)$,
and use $\mathfrak{p}=0$ otherwise.  Because $T\to\infty$, this regularization becomes asymptotically inactive under the conditions in SA-Theory. Similarly, if the $\mathfrak p=1$ design matrix is singular, we use the $\mathfrak p=0$ estimate instead.  We constrain the number of portfolios to $2\leq J_t\leq40$.  The lower bound ensures the data are split into separate portfolios whereas the upper bound helps prevent the formation of portfolios containing too few assets.  The grids used to select $J_t$ and to report estimates and inference results are chosen separately.  In each eligible post-burn-in period, the
implementation finds the largest origin-centered interval contained in the
cross-sectional range of the relevant estimated betas.  It then uses the
largest such symmetric interval that is contained in the realized range in
at least 90\% of the eligible periods, weighting periods equally.  The
$J_t$-selection grid consists of 25 equally spaced points on this interval
and is constructed from the point-estimation betas. The reporting and inference grid consists of 11 equally spaced points and is constructed from the first-step beta estimates used for inference.  Hence the reporting grid contains zero exactly but does not affect the choice of $J_t$.  For graphical clarity, the business-credit application uses a manually specified reporting grid from $-1.25$ to $1.25$ in increments of $0.25$, together with a separate grid of 25 equally spaced points on $[-1,1]$ for selecting $J_t$. For simplicity, we use the round endpoints $-1$ and $1$, which are close to those selected by the default support rule. This interval lies within the realized range of the point-estimation betas in $85.7\%$ of months, compared with the default target of $90\%$.

\paragraph{Variance estimation} 
For the common-factor component of the variance estimator, both
empirical applications forecast the risk factor using an AR(1)
with an intercept, estimated recursively by OLS using only prior
observations. Since the first-step window $H_{it}$ generally varies across assets and dates,
the implementation uses a pair-specific lag cutoff in the JN variance
estimator.  For observations on asset $i$ at dates $t$ and $s$, it retains the
jackknife cross-product whenever $|t-s|\leq \max\{H_{it},H_{is}\}$.  This rule accommodates the dependence induced by overlapping
first-step windows of different lengths without applying the largest selected window to every
asset-date pair.  If $H_{it}=H$ for all $i$ and $t$, it reduces to the common
lag-$H$ truncation in equation~\arxiveqref{main-eq:jackknife-variance-proposed} of
the main text.

\paragraph{Uniform-in-$\beta$ inference} For each post-burn-in date $t$, let $\mathcal I_t^\delta$ denote the set of assets included in the second-step estimator used for inference.  Let $n_t^\delta=|\mathcal I_t^\delta|$ and
$\mathcal I^\delta=\{(i,t):i\in\mathcal I_t^\delta\}$. Because returns can contain extreme observations, we use the square of the pooled median absolute deviation as a robust estimate of the return variance scale, rescaled to be consistent for the variance under Gaussianity:
\begin{equation*}
\widehat s_{R,\mathrm{MAD}}^2
=
\left\{
1.4826\mathop{\mathrm{median}}_{(i,t)\in\mathcal I^\delta}
\left|R_{it}-
\mathop{\mathrm{median}}_{(j,s)\in\mathcal I^\delta}R_{js}
\right|
\right\}^2.
\end{equation*}
We compute the three rate components as observation-level averages:
\begin{equation*}
\widehat r_J
=
\frac{1}{|\mathcal I^\delta|}
\sum_{(i,t)\in\mathcal I^\delta}J_t^{-4},
\qquad
\widehat r_H
=
\frac{1}{|\mathcal I^\delta|}
\sum_{(i,t)\in\mathcal I^\delta}H_{it}^{-2},
\qquad
\widehat r_N
=
\frac{1}{|\mathcal I^\delta|}
\sum_{(i,t)\in\mathcal I^\delta}
\frac{J_t^d}{n_t^\delta H_{it}}.
\end{equation*}
Finally, we choose $c_\delta = 1/160$ so that
\begin{equation*}
\widehat\delta
=
\frac{1}{160}\widehat s_{R,\mathrm{MAD}}^2
(\log T)^{d+1}(\log\log T)
\max\left\{\widehat r_J,\widehat r_H,\widehat r_N\right\}.
\end{equation*}
Each included observation receives equal weight, so periods with larger
effective cross sections receive proportionally more weight in these pooled
averages.  The calibration sample is determined before evaluating the
reporting grid; consequently, changing only the reporting grid does not alter
$\widehat\delta$.

\paragraph{Affine Specification Test}  For the business-credit application, we implement the affine specification test described in
Section~\ref*{main-sec:factor-cancelling-discussion} of the main text, as follows. We estimate both betas using local quadratic regressions, $\mathfrak{p}=2$, over two adjacent, nonoverlapping windows of $H=63$ months each.  We choose $H=63$ as it is the time-series median of the cross-sectional median series shown in Figure~\ref{fig:credit-tuning-diagnostics}.  We continue to require at least 36 nonmissing observations in each window. At each date, we regress returns on an intercept and the most-recent beta by using the earlier beta as the instrument. We then form equal-weighted portfolios of the resulting residuals by sorting on the earlier beta, with $\mathfrak{q}=0$.  We select $J_t$ on the matched sample using the MSE criterion with earlier local-constant beta estimates and bounds of 2 and 40 portfolios. The testing grid consists of the seven points from $-0.75$ to $0.75$ in increments of $0.25$. The covariance estimator retains cross-products of the same firm's weighted residual contributions for outcome dates at most 63 calendar months apart, accounting for overlap between recent estimation windows and outcome dates.  The test sample comprises 534 months, from July 1975
through December 2019. The resulting Wald statistic is $7.871$; comparison with a chi-squared distribution with seven degrees of freedom gives a $p$-value of $0.344$.

\part{Numerical Evidence}\label{part:numerical}

\section{Simulations}\label{sec:simulations}

\subsection{Data-Generating Process}

We consider two specifications for the risk factor and one volatility state. Let $k\in\{1,2\}$ index the risk factor specification, with
$\rho_1=0.5$ and $\rho_2=0.9$.
For each $k$, the risk factor follows the stationary Gaussian AR(1)
\begin{equation}
f_{kt}
=
\rho_k f_{k,t-1}+u_{kt},\qquad
u_{kt}
\overset{\mathrm{iid}}{\sim}N(0,\sigma_{u,k}^2).
\label{eq:simulation-factor-ar1}
\end{equation}
We fix the stationary factor standard deviation at
$\sigma_f=0.04$ in both specifications and therefore set
$
\sigma_{u,k}^2
=
(1-\rho_k^2)\sigma_f^2$.
For each specification, we draw the initial condition from $N(0,\sigma_f^2)$.

The volatility state follows
\begin{equation}
z_t
=
\rho_z z_{t-1}+\sqrt{1-\rho_z^2}\,v_t, \qquad 
v_t
\overset{\mathrm{iid}}{\sim}N(0,1),
\label{eq:simulation-volatility-state}
\end{equation}
where $\rho_z=0.975$. We initialize $z_t$ from its stationary $N(0,1)$
distribution. The $z_t$
process is independent of the risk-factor process.

The return equation under factor specification $k$ is
\begin{align}
R_{it}^{(k)}
&=
\mu_{\mathrm F}(\beta_{it})
+\beta_{it}f_{kt}
+\varsigma_{it}^{(k)}\eta_{it}.
\label{eq:simulation-return-dgp}
\end{align}
We specify the function $\mu_{\mathrm F}(\cdot)$, the paths of $\beta_{it}$, and the volatility scale $\varsigma_{it}^{(k)}$ below.

\paragraph{Functional form of $\mu_{\mathrm F}(\cdot)$}

For $\beta\in[-1,2]$, define the normalized exposure
$\widetilde\beta=(\beta+1)/3\in[0,1]$ and let
\begin{align*}
\mu_{\mathrm F}(\beta)
&=
-\frac{7}{200}+\frac{1}{20}
\left\{
6\widetilde\beta^5
-15\widetilde\beta^4
+10\widetilde\beta^3
\right\}.
\end{align*}
Thus, for all eight designs, the conditional expected return is
\begin{align}
\mathbb E\left[R_{it}^{(k)}\mid
\mathcal F_{t-1},\beta_{it}=\beta\right]
&=
\mu_{\mathrm F}(\beta)
+\beta\rho_k f_{k,t-1}.
\label{eq:simulation-mu-f}
\end{align}
Conditional expected returns therefore vary with the lagged value of the risk factor even though the structural expected-return curve $\mu_{\mathrm F}(\cdot)$ has no additional time-varying component.

\paragraph{Paths of $\beta_{it}$}

We consider two specifications for the beta paths. In the time-varying specification, we generate smooth paths using the low-frequency components of a persistent time-series process \citep[see][]{CG2022}:

\begin{enumerate}
\item Constant betas: $\beta_{it}=\beta_i$ for $t=1,\ldots,T$.

\item Time-varying betas:
\begin{align*}
\beta_{it}
&=
\beta_i+\sum_{j=1}^{K_\beta} a_{ij}\varphi_{j,t},
&
\varphi_{j,t}
&=
\sqrt{\frac{2}{T}}
\cos\left(\frac{t(j-\frac{1}{2})\pi}{T}\right),
\\
a_{ij}
&=
\sqrt{\lambda_j}\,\xi_{ij},
&
\lambda_j
&=
\frac{
v_\beta(1-\rho_\beta^2)
}{
1+\rho_\beta^2-2\rho_\beta\cos(\omega_j)
},
\qquad
\omega_j=\frac{(j-\frac{1}{2})\pi}{T}.
\end{align*}
Here $a_{ij}$ and $\xi_{ij}$ are scalar.  
\end{enumerate}
We set $\mathcal B=[-1,2]$ and draw
$\beta_i\overset{\mathrm{iid}}{\sim}\operatorname{Unif}[-1,2]$.
For the time-varying beta specification, we set $K_\beta=5$,
$\rho_\beta=0.95$, $v_\beta=0.04$, and
$\xi_{ij}\overset{\mathrm{iid}}{\sim}N(0,1)$. After drawing the low-frequency coefficients, for each unit we scale the
time-varying component by the smaller of one and $99.9\%$ of the largest
factor that keeps the entire resulting beta path in $\mathcal B$.

\paragraph{Functional form of $\protect\varsigma_{it}^{(k)}$ and $\eta_{it}$}

We assume that
$\eta_{it}\overset{\mathrm{iid}}{\sim}N(0,1)$ for
$i=1,\ldots,n$ and $t=1,\ldots,T$.  We
consider two volatility specifications:

\begin{enumerate}
\item Constant volatility:
$\left(\varsigma_{it}^{(k)}\right)^2=(1/50)^2$.

\item State-dependent volatility:
\begin{align*}
\left(\varsigma_{it}^{(k)}\right)^2
&=
\left(\frac{1}{50}\right)^2
\exp\left\{
\frac{1}{2}z_{t-1}-\frac{1}{8}
\right\}.
\end{align*}
\end{enumerate}
Because $z_t\sim N(0,1)$ in the stationary distribution,
$\mathbb E[(\varsigma_{it}^{(k)})^2]=(1/50)^2$ under both specifications.
\smallskip

Within each design, the risk-factor process, volatility-state process, beta paths, and idiosyncratic return shocks are mutually independent. The complete set of experiments crosses the two persistence choices, two types of beta paths, and two volatility specifications, for a total of eight data-generating processes (DGPs) summarized in Table \ref{tab:simulation-designs}.

\begin{table}[ht!]
\centering
\begin{tabular}{ccll}
\hline
Design & $\rho_k$ & Beta path & Idiosyncratic volatility \\
\hline
S1 & $0.5$ & Constant & Constant \\
S2 & $0.5$ & Constant & State-dependent \\
S3 & $0.5$ & Time-varying & Constant \\
S4 & $0.5$ & Time-varying & State-dependent \\
S5 & $0.9$ & Constant & Constant \\
S6 & $0.9$ & Constant & State-dependent \\
S7 & $0.9$ & Time-varying & Constant \\
S8 & $0.9$ & Time-varying & State-dependent \\
\hline
\end{tabular}
\caption{Simulation designs. All other calibration choices are common across
designs; the factor-innovation variance varies with $\rho_k$ as described above
to hold the unconditional risk-factor variance fixed.}
\label{tab:simulation-designs}
\end{table}

\subsection{Monte Carlo Implementation}

We generate $S=1,000$ Monte Carlo replications for each of the eight DGPs. Each replication is a 
panel dataset with $n=5{,}000$ and $T=800$. The first 240 months are excluded from second-step estimation and inference as a burn-in period. In each replication, we select the first-step windows $H_{it}$ and the second-step number of portfolios $J_t$
using the procedure described in Section~\ref{sec:tuning-pointwise}. Apart from the longer 240-month burn-in and the evaluation grid specified
below, we use the default implementation for monthly data described in
Section~\ref{sec:implementation-defaults}.

For the factor component of the proposed covariance estimator, the
implementation recursively fits via OLS the univariate AR(1)
\begin{align*}
f_{k\tau}
&=
a_{k,t}+\rho_{k,t}f_{k,\tau-1}+e_{k\tau},
&&
\tau<t,
\end{align*}
using only the history of the selected factor $f_k$, and uses the resulting
one-step residual at date $t$.  The implementation estimates the intercept
and persistence recursively and does not use the true value of $\rho_k$.

We evaluate the SACER at
\begin{align*}
\mathcal B_{\mathrm{eval}}
&=
\left\{
-1+\frac{\ell}{7}:\ell=1,\ldots,20
\right\}.
\end{align*}
For replication $s$ and evaluation point $\beta$, the corresponding inferential target is
\begin{align*}
\overline\mu_s^{(k)}(\beta)
&=
\frac{1}{T-240}
\sum_{t=241}^T
\left\{\mu_{\mathrm F}(\beta)+\beta\rho_k f^{(s)}_{k,t-1}\right\}.
\end{align*}

Since our focus is on inference, our primary estimators are the infeasible oracle with known $\beta_{it}$ and $\mathfrak q=1$, and the feasible estimator with $(\mathfrak p,\mathfrak q)=(1,1)$. For pointwise inference, we report the following procedures:
\begin{enumerate}
\item the proposed JN variance estimator;
\item the FM variance estimator formed from the
$(\mathfrak p,\mathfrak q)=(0,0)$ estimates; and
\item the FM variance estimator formed from the $(\mathfrak{p},\mathfrak{q})=(1,1)$ estimates.
\end{enumerate}
For the proposed JN variance estimator, we report the oracle and feasible estimators separately.  For every reported procedure, we record pointwise coverage and average confidence-interval length.

All reported confidence intervals have nominal coverage of $95\%$. For the feasible estimator paired with the proposed JN variance estimator, we also report pointwise confidence intervals that are uniformly valid over $\beta$, as developed in Section~\ref{sec:tuning-uniform}. We choose $\widehat\delta$ using the default implementation described in Section~\ref{sec:implementation-defaults}. We record coverage and average interval length at every evaluation point.

\subsection{Results}

Figure~\ref{fig:simulation-coverage} reports empirical coverage for the proposed pointwise and uniform-in-$\beta$ confidence intervals, together with the infeasible oracle procedure.  The oracle procedure---which treats the
$\beta_{it}$ as observed---performs well across all eight designs, with average pointwise coverage across the evaluation grid ranging from 93.9\% to 96.0\% (blue solid line).\footnote{Recall that, in our simulation design, $\widehat\Sigma_f$ uses the correctly specified dynamic model for $f_t$.} The feasible JN pointwise interval also performs well away from the origin but can undercover near zero, particularly when the factor is highly persistent and betas vary over time (black dashed line).  The uniform-in-$\beta$ procedure eliminates this sharp deterioration (orange solid line).  Its minimum coverage across the interior grid ranges from 93.2\% to 95.6\%, compared with 72.1\% to 90.5\% for pointwise inference.  This contrast illustrates the distinction between pointwise validity at each fixed evaluation point in Corollary~\ref*{main-corr:feasinf} and the uniform-in-$\beta$ coverage guarantee discussed in Remark~\ref*{main-rem:uniform_beta} of the main text.  The uniform-in-$\beta$ intervals are conservative at the origin, where empirical coverage is 100\% in every design.  Comparing the left and right columns shows that allowing for state-dependent idiosyncratic volatility has only a modest effect on coverage.  Figure~\ref{fig:simulation-interval-length} reports the corresponding average confidence interval lengths for all three procedures.  The feasible and oracle JN pointwise intervals have nearly identical average lengths throughout the evaluation grid.  The uniform-in-$\beta$ adjustment increases interval length in a neighborhood of the origin, where the pointwise procedure can undercover.  Outside that neighborhood, the uniform-in-$\beta$ and pointwise intervals coincide exactly.  Finally, note that $\widehat\Sigma_\varepsilon(\beta;1)$ was never negative at any evaluation point in the simulations across all eight designs.

Figure~\ref{fig:simulation-fm-coverage} reports pointwise coverage for the two Fama-MacBeth procedures we consider.  The procedure based on $(\mathfrak p,\mathfrak q)=(0,0)$ for both point and variance estimation corresponds to the conventional implementation in empirical work. In contrast, the modified procedure is based on $(\mathfrak p,\mathfrak q)=(1,1)$.  Both procedures exhibit a sharp deterioration at $\beta=0$.  Coverage at the origin ranges from 26.0\% to 39.2\% for the procedure based on $(\mathfrak p,\mathfrak q)=(0,0)$ and from 60.7\% to 85.7\% for the modified procedure based on $(\mathfrak p,\mathfrak q)=(1,1)$.  Thus, the modified procedure performs substantially better near the origin, but it remains below the nominal level.  This undercoverage is consistent with the discussion following Theorem~\ref*{main-thm:fm-variance} of the main text, which explains
why the FM variance estimator is not generally valid at the origin.  Figure~\ref{fig:simulation-fm-interval-length} reports the corresponding average confidence interval lengths.  The curves for the two Fama-MacBeth procedures are nearly indistinguishable in every design.  This similarity is consistent with Theorems~\ref*{main-thm:CLT} and~\ref*{main-thm:fm-variance} of the main text, which show that, at each fixed nonzero evaluation point, $\mathfrak q=0$ and $\mathfrak q=1$ have the same leading asymptotic variance and the same first-order Fama-MacBeth variance target.  Consequently, the improvement from the modified procedure near the origin reflects the use of the higher-order point estimator rather than wider confidence intervals.

Finally, Figure~\ref{fig:simulation-jn-uniform-fm-interval-length} directly compares the average lengths of the JN uniform-in-$\beta$ intervals and the modified Fama-MacBeth pointwise intervals.  Away from the origin, the Fama-MacBeth intervals are wider, especially when the factor is highly persistent. Their greater width is reflected in the overcoverage shown in Figure~\ref{fig:simulation-fm-coverage}. As shown in Theorem~\ref*{main-thm:fm-variance} of the main text, this difference arises because the FM variance estimator includes period-by-period variation in conditional expected returns that is not part of the estimation uncertainty around the SACER.  Near the origin, the Fama-MacBeth intervals can instead be shorter, but Figure~\ref{fig:simulation-fm-coverage} shows that this ostensible gain in precision is accompanied by substantial undercoverage.  Taken together, the results support the uniform-in-$\beta$ JN procedure as a reliable approach to inference across both zero and nonzero evaluation points.

\begin{figure}[p!]
\centering
\caption{\textbf{Coverage probability.} Each panel reports empirical coverage
at the 20 interior evaluation points for one of the eight designs in
Table~\ref{tab:simulation-designs}. The black dashed line is the feasible JN
pointwise interval, the orange solid line is the feasible JN uniform-in-$\beta$
interval using the default calibration of $\widehat\delta$, and the blue solid
line is the infeasible oracle JN pointwise interval. The horizontal grey dotted
line denotes the nominal coverage probability of 0.95. These results are based
on 1,000 Monte Carlo replications per design.}
\label{fig:simulation-coverage}
\simulationfigure{CCFW_2026_AER--simulation-coverage}
\end{figure}

\begin{figure}[p!]
\centering
\caption{\textbf{Pointwise coverage using Fama-MacBeth variance estimators.}
Each panel reports empirical coverage at the 20 interior evaluation points for
one of the eight designs in Table~\ref{tab:simulation-designs}. The blue solid
line uses the $(\mathfrak p,\mathfrak q)=(0,0)$ estimates and the corresponding
Fama-MacBeth variance estimator. The orange dashed line is the modified
Fama-MacBeth procedure based on $(\mathfrak p,\mathfrak q)=(1,1)$. The horizontal
grey dotted line denotes the nominal coverage probability of 0.95. These
results are based on 1,000 Monte Carlo replications per design.}
\label{fig:simulation-fm-coverage}
\simulationfigure{CCFW_2026_AER--simulation-fm-coverage}
\end{figure}

\begin{figure}[p!]
\centering
\caption{\textbf{Average confidence-interval length.} Each panel reports the
average length of the 95\% confidence interval at the 20 interior evaluation
points for one of the eight designs in Table~\ref{tab:simulation-designs}. The
black dashed line is the feasible JN pointwise interval, the orange solid line is
the feasible JN uniform-in-$\beta$ interval using the default calibration of
$\widehat\delta$, and the blue solid line is the infeasible oracle JN
pointwise interval. These results are based on 1,000 Monte Carlo replications
per design.}
\label{fig:simulation-interval-length}
\simulationfigure{CCFW_2026_AER--simulation-interval-length}
\end{figure}

\begin{figure}[p!]
\centering
\caption{\textbf{Average confidence-interval length using Fama-MacBeth
variance estimators.} Each panel reports the average length of the 95\%
confidence interval at the 20 interior evaluation points for one of the eight
designs in Table~\ref{tab:simulation-designs}. The blue solid line uses the
$(\mathfrak p,\mathfrak q)=(0,0)$ estimates and the corresponding
Fama-MacBeth variance estimator. The orange dashed line is the modified
Fama-MacBeth procedure based on $(\mathfrak p,\mathfrak q)=(1,1)$. These results
are based on 1,000 Monte Carlo replications per design.}
\label{fig:simulation-fm-interval-length}
\simulationfigure{CCFW_2026_AER--simulation-fm-interval-length}
\end{figure}

\begin{figure}[p!]
\centering
\caption{\textbf{Average confidence-interval length: JN uniform-in-$\beta$
and modified Fama-MacBeth.} Each panel reports the average length of the
95\% confidence interval at the 20 interior evaluation points for one of the
eight designs in Table~\ref{tab:simulation-designs}. The orange solid line is the
JN uniform-in-$\beta$ interval using the default calibration of
$\widehat\delta$. The blue dashed line is the modified Fama-MacBeth
pointwise interval. Both procedures use the same feasible estimates. Results are based on 1,000 Monte Carlo
replications per design.}
\label{fig:simulation-jn-uniform-fm-interval-length}
\simulationfigure{CCFW_2026_AER--simulation-jn-uniform-fm-interval-length}
\end{figure}

\clearpage
\section{Empirical Applications}\label{sec:empirical-application}

In this section, we provide additional results illustrating our new procedures.  First, we provide additional evidence and implementation details for the business-credit factor introduced in the main text.  Second, we revisit the capital share risk factor of \citet{LLM2019}, providing another application of our new estimation and inference procedures.

\subsection{Additional Results for Business Credit Cycle Factor}
\label{sec:empirical-application-credit-extra}

\subsubsection{Out-of-Sample Predictive Regressions}

As an additional check on the predictive performance of the business-credit factor, Table~\ref{tab:outsample} reports out-of-sample $R^2$ from expanding-window predictive regressions.  The out-of-sample $R^2$ compares the mean squared forecast error from each predictive regression with that from the recursively estimated historical mean forecast.  We use data through 1989 as the initial training sample and evaluate the forecasts from January 1990 through December 2019.  The business-credit factor produces positive out-of-sample $R^2$ at all four horizons and outperforms each of the benchmark predictors.

\begin{table}[h!]
\caption{\textbf{Out-of-Sample Predictive Regressions} This table reports out-of-sample $R^2$ (in percent) from expanding window predictive regressions of cumulative market excess returns on an individual predictor from \cite{GoyalWelch2008} for horizons of 1, 3, 6, and 12 months ahead.   The row labeled ``CGP'' reports results for the SLOOS only portion of our risk factor as studied in \cite{CGP2015}.   The row labeled ``BCF'' reports results for our proposed risk factor. Positive values have been bolded. The training period is 1965m1--1989m12 and the evaluation sample is 1990m1--2019m12.}
\label{tab:outsample}
\centering
\bigskip
\begin{tabular}{|l|cccc|} \hline
 & $\ell=1$  & $\ell=3$  & $\ell=6$  & $\ell=12$  \\  \hline \hline 
\input tables/predictive-oos.tex
\hline 
\end{tabular}
\end{table}

\subsubsection{Implementation and Tuning}

\begin{figure}[h!]
\centering
\caption{\textbf{Tuning parameters and effective sample size over time.} Panel A reports the monthly median of the selected first-step window $H_{it}$ across firms, with the shaded region denoting the interquartile range.  Panel B reports the selected number of portfolios, $J_t$.  Panel C reports the effective cross-sectional sample size, $n_t$, defined as the number of firms with a finite beta estimate that are assigned to a portfolio; the grey line reports the number of firms with an observed return before imposing the first-step estimation requirement.  The sample period is January 1975 through December 2019.}
\label{fig:credit-tuning-diagnostics}
\includegraphics[width=.82\textwidth]{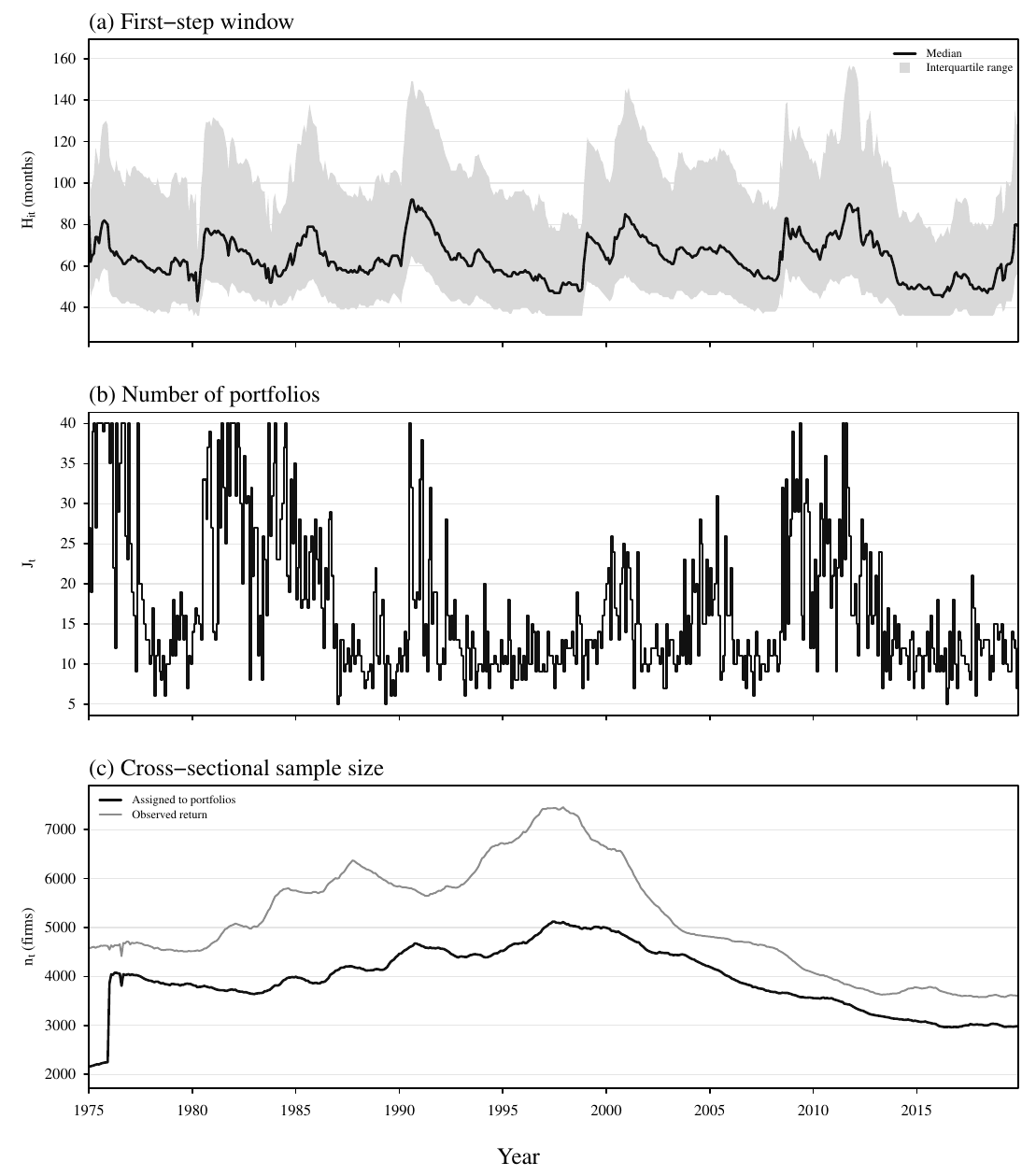}
\end{figure}

Here we provide further details on the implementation of our main empirical application. The values of $H_{it}$ and $J_t$ are obtained following the procedure described in Section~\ref{sec:tuning-pointwise}. The grid used to select $J_t$ is described in Section~\ref{sec:implementation-defaults}. Figure~\ref{fig:credit-tuning-diagnostics} summarizes the resulting data-driven tuning choices and the effective cross-sectional sample size over time.  The selected first-step windows vary substantially across firms and months: $H_{it}$ equals its lower bound for 15.5\% and its upper bound for 2.2\% of firm-month observations assigned to portfolios.  The selected number of portfolios ranges from 5 to 40 and reaches its upper bound in 4.8\% of the sample months.

The effective cross-sectional sample size ranges from 2,153 to 5,122 firms, with an average of 3,949 firms per month.  Even when the upper bound on $J_t$ is attained, the smallest realized portfolio contains 54 firms.  The sharp increase in the effective cross-sectional sample size in January 1976 reflects the beginning of CRSP's Nasdaq coverage in December 1972. Because many Nasdaq firms have their first usable return in January 1973, they first satisfy the requirement of 36 prior monthly observations in January 1976 and then enter the portfolio sorts.

Finally, the uniform-in-$\beta$ calibration described in
Section~\ref{sec:implementation-defaults} gives
$\widehat\delta=0.0151$.  With this threshold, the uniform variance floor applies at the seven
reporting grid points from $\beta=-0.75$ through $\beta=0.75$.

\subsection{Capital Share Risk Factor}
\label{sec:empirical-application-capital-share}

In this section, we revisit the capital share risk factor proposed by \citet{LLM2019} using our new estimation and inference procedures. \citet{LLM2019} argue that an empirical model in which growth in the capital share of aggregate income is the single source of macroeconomic risk explains a substantial fraction of the cross-sectional variation in expected returns across a range of equity characteristic portfolios and non-equity asset classes, with positive exposure associated with higher expected returns.  Their benchmark capital share measure is constructed from quarterly data on the labor share of the nonfarm business sector from the Bureau of Labor Statistics' Productivity and Costs release.\footnote{We obtain these data from the paper's replication package available at \url{https://onlinelibrary.wiley.com/doi/full/10.1111/jofi.12772}.}  Specifically, they define the capital share as $KS_t=1-LS_t$ and measure its growth over eight quarters. In our implementation, we multiply this log growth rate by 100 and construct the risk factor as $f_t=100[\log(KS_t)-\log(KS_{t-8})]$, where $LS_t$ denotes the labor share at time $t$.\footnote{\citet{GospodinovRobotti2021Reappraisal} reexamine these findings and argue that the original evidence of statistical significance may reflect the persistence induced by using overlapping long-horizon returns and capital-share growth. These authors recommend using nonoverlapping data or alternative empirical methods.  This critique motivates our alternative implementation: we retain the eight-quarter capital-share growth factor of \citet{LLM2019} but use nonoverlapping one-quarter returns and our beta-sorted portfolio procedure rather than their two-pass risk-premium procedure based on overlapping returns.}

\begin{figure}[h!]
\centering
\caption{\textbf{Tuning parameters and effective sample size over time for the capital share application.} Panel A reports the quarterly median of the selected first-step window $H_{it}$ across firms, with the shaded region denoting the interquartile range.  Panel B reports the selected number of portfolios, $J_t$.  Panel C reports the effective cross-sectional sample size, $n_t$, defined as the number of firms with a finite beta estimate that are assigned to a portfolio; the grey line reports the number of firms with an observed complete quarterly return before imposing the first-step estimation requirement.  The estimation period is 1975Q2 through 2013Q4.}
\label{fig:llm-tuning-diagnostics}
\includegraphics[width=.82\textwidth]{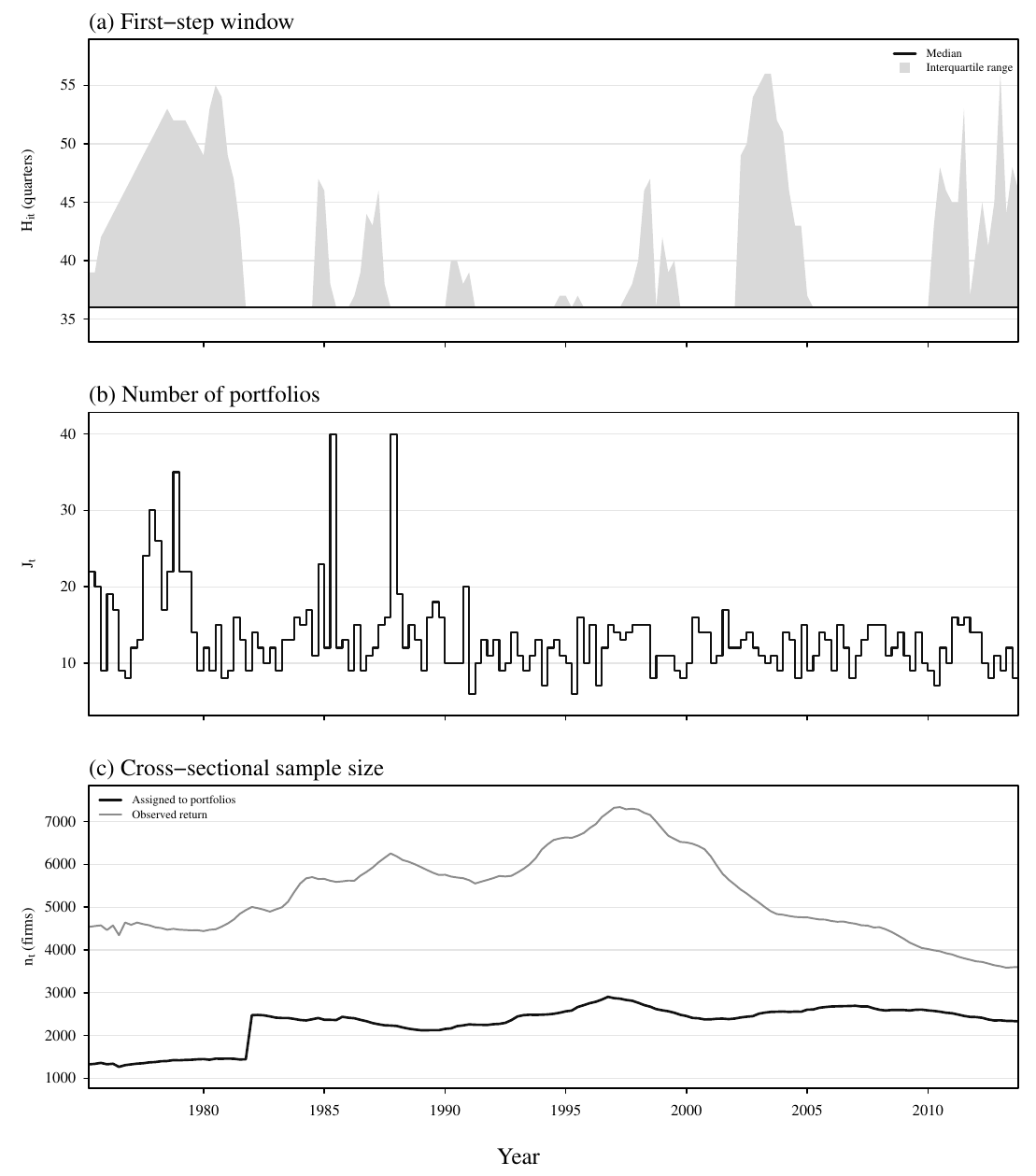}
\end{figure}

We use the same CRSP stock universe and treatment of delisting returns as in Section~\ref*{main-empapp} of the main text. We form quarterly raw returns by compounding the three monthly returns within each quarter and use the one-factor quarterly implementation choices described in Section~\ref{sec:implementation-defaults}.  The values of $H_{it}$ and $J_t$ are obtained following the procedure described in Section~\ref{sec:tuning-pointwise}.  Applying the default support rule to the point-estimation betas yields a $J_t$-selection grid of 25 equally spaced points with endpoints $-1.92$ and $1.92$, while the corresponding rule applied to the inferential betas yields a reporting and inference grid of 11 equally spaced points with endpoints $-2.37$ and $2.37$, where all endpoints are rounded. The full reporting interval lies within the realized cross-sectional range of the $\widehat\beta_{it}$ values used for inference in 90.3\% of the estimation quarters.  Figure~\ref{fig:llm-tuning-diagnostics} summarizes the resulting data-driven tuning choices and effective cross-sectional sample size over time.  The selected first-step window $H_{it}$ equals its lower bound for 73.3\% and its upper bound for 2.5\% of firm-quarter observations assigned to portfolios.  The selected number of portfolios ranges from 6 to 40 and reaches its upper bound in 1.3\% of the sample quarters.  The effective cross-sectional sample size ranges from 1,269 to 2,905 firms, with an average of 2,285 firms per quarter; the smallest realized portfolio contains 40 firms.  Finally, the uniform-in-$\beta$ calibration gives $\widehat\delta=0.0548$.  The
estimated risk-factor variance component reaches this threshold at approximately
$\beta=-1.27$ and $\beta=1.27$.  Accordingly, the uniform variance floor
applies at the five reporting grid points from $\beta=-0.95$ through
$\beta=0.95$.

\begin{figure}[h!]
\centering
\caption{\textbf{Inference on the SACER.} The black line reports the $(\mathfrak{p},\mathfrak{q})=(0,0)$ estimate of the SACER.  The grey region is the 95\% JN pointwise interval.  The red dashed lines are the 95\% JN uniform-in-beta intervals, and the green dotted lines are the 95\% modified FM pointwise intervals.  All three intervals use $(\mathfrak p,\mathfrak q)=(1,1)$.  Results are reported for quarterly raw returns (in percent), formed by compounding the three monthly returns within each quarter.  The sample period is 1965Q2--2013Q4 with a 40-quarter burn-in period, so the estimation period is 1975Q2--2013Q4.}
\label{fig:llm-grandmean-ptwise}
\includegraphics[width=.92\textwidth]{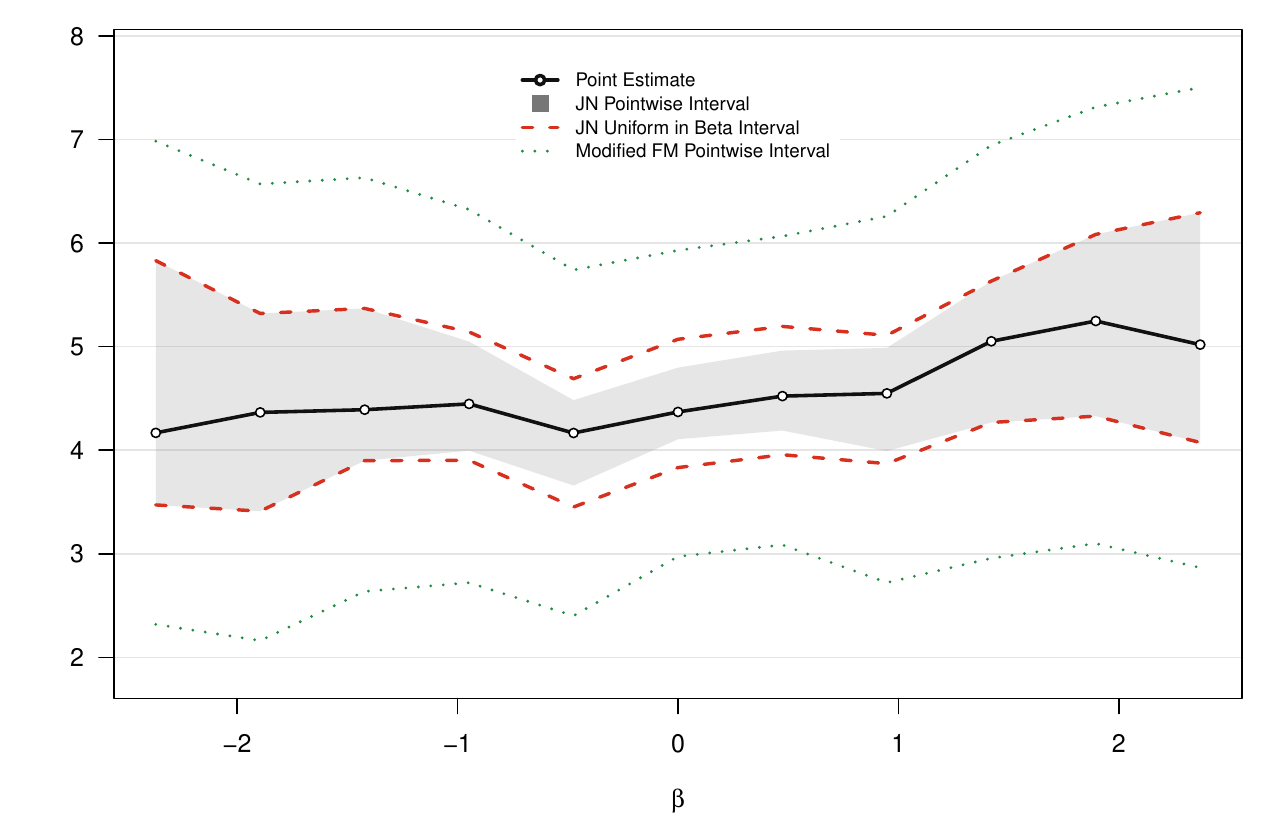}
\end{figure}

Figure~\ref{fig:llm-grandmean-ptwise} reports the estimated SACER and associated inference using the capital share risk factor.  The point estimate based on $(\mathfrak{p},\mathfrak{q})=(0,0)$ is given by the black line.  The estimated relation generally slopes upward; specifically, the estimated SACER increases from 4.17\% per quarter at $\beta=-2.37$ to 5.02\% at $\beta=2.37$.  This pattern is qualitatively consistent with the positive price of capital share risk documented by \citet{LLM2019}.  The corresponding high-minus-low contrast point estimate based on $(\mathfrak{p},\mathfrak{q})=(0,0)$ is 0.85 percentage points.  Table~\ref{tab:llm-contrasts} reports inference for the same contrast using the $(\mathfrak{p},\mathfrak{q})=(1,1)$ estimate; consequently, the confidence intervals need not be centered on the black-line estimate.  The JN standard error is 1.030, giving a two-sided $p$-value of 0.605, while the modified FM standard error is 0.634, giving a two-sided $p$-value of 0.401.  The $\widehat\Sigma_f$-only benchmark also fails to reject, so none of the three procedures rejects a zero contrast.

\begin{table}[h!]
\centering
\caption{\textbf{High Minus Low Contrast.} The table reports inference for $\widehat\mu(2.37)-\widehat\mu(-2.37)$ in quarterly percentage points.  Each row reports the standard error, 90\% confidence interval, and two-sided normal-approximation $p$-value under the JN, modified FM, or $\widehat\Sigma_f$-only variance estimator.  The confidence intervals are centered at the $(\mathfrak{p},\mathfrak{q})=(1,1)$ estimate.}
\label{tab:llm-contrasts}
\small
\input{tables/CCFW_2026_AER--llm-capital-share-contrasts}
\end{table}

\clearpage
\bibliography{CCFW_2026_AER--bib}
\bibliographystyle{jfe}


\pagenumbering{roman}

\title{\vspace{-0.0in}Beta-Sorted Portfolios \\Supplemental Appendix: Theory\bigskip }
\author{Matias D. Cattaneo\thanks{Department of Economics, Princeton University.} \and
        Richard K. Crump\thanks{Capital Markets, Federal Reserve Bank of New York.} \and
        Yingjie Feng\thanks{School of Economics and Management, Tsinghua University.} \and
        Weining Wang\thanks{School of Economics, University of Bristol.}}
\maketitle

\begin{abstract}
    This supplemental appendix gives the theoretical details for beta-sorted portfolio estimation and inference.
    Part~\ref{part:general} develops general high-level consistency and stochastic linearization results for beta-sorted estimators based on generated regressors and data-dependent portfolios.
    Part~\ref{part:classical} specializes these results to the classical beta-sorted portfolio construction based on rolling-window first-step regressions and marginal-quantile portfolios, including convergence rates, distributional approximations, feasible variance estimation, uniform-in-$\beta$ inference, and factor-cancelling contrasts. The companion Supplemental Appendix: Applied provides implementation
    guidance, simulation evidence, and empirical applications.
\end{abstract}

\thispagestyle{empty}

\clearpage
\addtocontents{toc}{\protect\thispagestyle{empty}}
\tableofcontents
\thispagestyle{empty}

\newpage

\pagestyle{plain}

\clearpage
\pagenumbering{arabic}

\part{General Beta-Sorted Portfolios}
\label{part:general}

This part of the supplemental appendix considers a general portfolio scheme and establishes two main results under high-level conditions for the beta-sorted portfolio estimator: (i) uniform consistency, and (ii) stochastic linearization. Part \ref{part:classical} specializes this setup to the classical beta-sorted portfolio estimator, where portfolios are constructed using first-step local rolling-window regressions and second-step quantile-spaced partitions. Part \ref{part:classical} also provides primitive conditions for the high-level conditions imposed here.

\section{Setup}\label{sec:general-setup}

We consider the following asset return model throughout:
\begin{align*}
    R_{it}
    &= \alpha_{it}+\beta_{it}^\top f_t+\varepsilon_{it},\qquad
    \E[\varepsilon_{it}|\mathcal{F}_{t-1},f_t]=0,\qquad
    i=1,\dots,n_t,\quad t=1,\dots,T,
\end{align*}
where $\alpha_{it}$ and $\beta_{it}$ are $\mathcal{F}_{t-1}$-measurable, $f_t\in\R^d$ is a column vector of observable factors, and $\beta_{it}$ takes values in a common compact support $\mathcal{B}\subset\R^d$ containing $0$ as an interior point.
For each finite sample, let $n$ be a cross-sectional index bound such that $[n_t]\subseteq[n]$ for all $t\leq T$, where for any positive integer $m$, $[m]:=\{1, \ldots, m\}$.
We define
\begin{align*}
    \mathcal{G}_t
    &=
    \sigma\!\left(
        \{f_s\}_{s=1}^{t},
        \{\aleph_{2s}\}_{s=1}^{t}
    \right),
    \\
    \mathcal{F}_t
    &=
    \sigma\!\left(
        \{f_s\}_{s=1}^{t},
        \{\{\varepsilon_{is}\}_{i=1}^{n_s}\}_{s=1}^{t},
        \{\aleph_{1i}\}_{i=1}^{n},
        \{\aleph_{2s}\}_{s=1}^{t},
        \{\{\aleph_{3is}\}_{i=1}^{n_s}\}_{s=1}^{t}
    \right).
\end{align*}
Here $\aleph_{1i}$, $\aleph_{2t}$, and $\aleph_{3it}$ are auxiliary random vectors that may enter the data-generating process.
The predictable coefficients $\alpha_{it}$ and $\beta_{it}$ are $\mathcal{F}_{t-1}$-measurable. Part~\ref{part:classical} gives a primitive innovation-free representation in terms of factor, scale, and auxiliary histories dated at most $t-1$.
Thus $\{\mathcal{F}_t\}$ and $\{\mathcal{G}_t\}$ are filtrations with $\mathcal{G}_t\subseteq\mathcal{F}_t$ and $\sigma(f_s:1\leq s\leq t)\subseteq\mathcal{G}_t$.
We interpret $\mathcal{F}_t$ as the full information set and $\mathcal{G}_t$ as the factor-state information set.
We assume that there is a fixed constant $c>0$ such that
$cn\leq n_t\leq n$ for all $t=1,\dots,T$, and asymptotic approximations are
taken as both $n$ and $T$ diverge to infinity.

We assume that there exists a nonrandom function $\mu_0(\cdot\,;\cdot)$ such that
\begin{align*}
    \E[R_{it}|\mathcal{F}_{t-1}]
    &= \mu_0(\beta_{it};\mathcal{G}_{t-1}) =: \mu_t(\beta_{it}),
\end{align*}
and the structural form is
\begin{align*}
    R_{it}
    &= \mu_t(\beta_{it})+\beta_{it}^\top(f_t-\E[f_t|\mathcal{F}_{t-1}])+\varepsilon_{it}.
\end{align*}
We impose $\E[f_t|\mathcal{F}_{t-1}]=\E[f_t|\mathcal{G}_{t-1}]$ and therefore write
\begin{align*}
    R_{it}
    &=\mu_t(\beta_{it})+\beta_{it}^\top\widetilde{f}_t+\varepsilon_{it},
    \qquad
    \widetilde{f}_t=f_t-\E[f_t|\mathcal{G}_{t-1}].
\end{align*}
It follows that
\begin{align*}
    \mu_t(\cdot)
    =\mu_0(\cdot;\mathcal{G}_{t-1}),\qquad
    \alpha_t(b):=\mu_t(b)
    -b^\top\E[f_t\mid\mathcal{G}_{t-1}],\quad b\in\mathcal B,
    \qquad
    \alpha_{it} = \alpha_t(\beta_{it}).
\end{align*}

Let $H=H_{nT}$ be an integer satisfying $1\leq H<T$. In this general part,
$H$ records the first-step initialization period, so second-step estimation
uses dates $t=H+1,\ldots,T$. Part~\ref{part:classical} specializes $H$ to the
rolling-window length and imposes its asymptotic rate conditions.

At a fixed date, a cross-sectional regression of returns on risk exposures recovers the realized systematic return function
\begin{align*}
    M_t(\beta)&=\mu_t(\beta)+\beta^\top\widetilde{f}_t,
\end{align*}
rather than $\mu_t(\beta)$ itself. The time average of the second term,
$(T-H)^{-1}\sum_{t=H+1}^T\beta^\top\widetilde{f}_t$, has martingale-difference summands with respect to $\{\mathcal{G}_t\}$ because $\E[\beta^\top\widetilde{f}_t|\mathcal{G}_{t-1}]=0$. Hence this average is asymptotically negligible for consistency under the conditions below, although it contributes to the first-order distribution when $\beta\neq 0$.

The goal is to estimate and conduct inference on the \textit{sample average of conditional expected returns (SACER)}:
\begin{align*}
    \bar{\mu}_T(\beta;H)
    &= \frac{1}{T-H}\sum_{t=H+1}^T\mu_t(\beta).
\end{align*}

\subsection{Estimator}

The beta-sorted portfolio procedure targets $\bar{\mu}_T(\cdot)$ by averaging over time cross-sectional piecewise polynomial regression fits constructed using estimated partitions (portfolios). More precisely, the estimator is constructed in two steps:
\begin{enumerate}[label=(\roman*)]
    \item obtain an estimator $\widehat{\beta}_{it}$ of $\beta_{it}$ for each $i=1,\dots,n_t$ and $t=H+1,\dots,T$; and
    \item for each $t=H+1,\ldots,T$, construct a collection of $\bar{J}_t$ disjoint cells
    $\widehat{\Delta}_t=\{\widehat{P}_{jt}: j\in[\bar{J}_t]\}$, and then define the following estimator of $\bar{\mu}_T(\beta;H)$:
    \begin{align*}
        \widehat{\mu}(\beta)
        =\frac{1}{T-H} \sum_{t=H+1}^T\widehat{M}_t(\beta),
        \qquad
        \widehat{M}_t(\beta)
        =\widehat{p}_{t}(\beta)^\top
        \left(
        \sum_{i=1}^{n_t}\widehat{p}_{t}(\widehat{\beta}_{it})
        \widehat{p}_{t}(\widehat{\beta}_{it})^\top
        \right)^{-1}
        \sum_{i=1}^{n_t}\widehat{p}_{t}(\widehat{\beta}_{it})R_{it},
    \end{align*}
    where $\widehat{p}_t(b)$ denotes a piecewise polynomial expansion based on the partition $\widehat{\Delta}_t$.
\end{enumerate}

Without loss of generality, we write this piecewise polynomial expansion in cell-normalized coordinates. For each cell $\widehat{P}_{jt}$, choose a center $\widehat{c}_{jt}$ and a nonsingular scaling/rotation matrix $\widehat{D}_{jt}$ such that $\widehat{D}_{jt}^{-1}(b-\widehat{c}_{jt})$ is uniformly bounded over $b\in\widehat{P}_{jt}$. Let $\widehat{\jmath}_t(b)$ denote the index of the cell containing $b$. For a nonnegative integer $\mathfrak{q}$, let $\check{p}_{\mathfrak{q}}\!\big(z\big)=(z^s: |s|\leq \mathfrak{q})^\top$ denote the column vector collecting, in a fixed order, all $d$-variate monomials in $z$ with total degree no greater than $\mathfrak{q}$, using the multi-index notation collected in Section~\ref{sec:general-notation}. For any cell $P$ equipped with a center $c_P$ and nonsingular scaling/rotation matrix $D_P$, define the localized degree-$\mathfrak{q}$ block basis by
\begin{align*}
    p_{\mathfrak{q}}(b,P)
    &=
    \IF(b\in P)\check{p}_{\mathfrak{q}}\!\big(D_P^{-1}(b-c_P)\big).
\end{align*}
Cell indicators are understood with the fixed boundary tie-breaking convention for the corresponding partition, so a point contributes to at most one block.
For any admissible partition $\Delta=\{P_j:j\in[\bar J]\}$ equipped with its cell normalizations, define the associated stacked basis by
\begin{align*}
    p(b;\Delta)
    &=
    \big(
    p_{\mathfrak{q}}(b,P_1)^\top,\ldots,
    p_{\mathfrak{q}}(b,P_{\bar J})^\top
    \big)^\top .
\end{align*}
We then take $\widehat p_t(b)=p(b;\widehat\Delta_t)$.
This stacked-block form emphasizes that $\widehat{p}_t(b)$ is the product of a cell-indicator and the polynomial basis used within the selected cell. When $\mathfrak{q}=0$, $\check{p}_{\mathfrak{q}}$ is the scalar $1$, so $p_{\mathfrak{q}}(b,P)=\IF(b\in P)$. If $K_{\mathfrak{q}}$ denotes the number of monomials, then $\widehat{p}_t(\beta)\in\mathbb{R}^{\bar{J}_tK_{\mathfrak{q}}}$.

At this level of generality, we allow for generic choices of the first-step estimator $\widehat{\beta}_{it}$ of $\beta_{it}$ and the partitioning scheme $\{\widehat{\Delta}_t\}_{t=H+1}^{T}$, while in the second step, for each $t=H+1,\ldots,T$, we allow for a piecewise polynomial regression of total degree $\mathfrak{q}$ to obtain $\widehat{M}_t(\cdot)$.

Although $\beta_{it}$ is assumed to have bounded support $\mathcal{B}$, the first-step estimator $\widehat{\beta}_{it}$ may lie outside $\mathcal{B}$. Thus, we fix $\delta_{\mathcal{B}}>0$ and introduce a compact convex set $\mathcal{B}^+$ satisfying
\begin{align*}
    \left\{b\in\mathbb{R}^d:\inf_{a\in\mathcal{B}}\|b-a\|\leq \delta_{\mathcal{B}}\right\}
    &\subseteq \mathcal{B}^+.
\end{align*}
When $\widehat{\beta}_{it}$ is uniformly consistent for $\beta_{it}$ in the sense that
\begin{align*}
    \max_{H+1\leq t\leq T}\max_{i\in[n_t]}\|\widehat{\beta}_{it}-\beta_{it}\|
    =o_\P(1),
\end{align*}
the estimated $\widehat{\beta}_{it}$ will belong to $\mathcal{B}^+$ for all $i\in[n_t]$ and $t=H+1,\ldots,T$, with probability approaching one.
In the theoretical analysis below, $\mu_t$ is understood as a bounded continuous extension from $\mathcal{B}$ to $\mathcal{B}^+$ whenever it is evaluated at $\widehat{\beta}_{it}$ outside $\mathcal{B}$.
Thus, $\mathcal{B}^+$ is only an extension device for evaluating functions at first-step estimates and along Taylor paths; it does not change the support $\mathcal{B}$.

{
For the interior results below, fix an evaluation region
$\mathcal B^\circ\subset\mathrm{int}(\mathcal B)$. For each $t$, define
\begin{align*}
    \mathcal J_t
    &=
    \{j\in[\bar J_t]:\widehat P_{jt}\cap\mathcal B^\circ\neq\emptyset\}.
\end{align*}
Thus, $\mathcal J_t$ collects precisely the estimated cells that may contain
evaluation points in $\mathcal B^\circ$.
}

\subsection{Notation}\label{sec:general-notation}

For a positive integer $m$, let $[m]=\{1,\ldots,m\}$. All vectors are column vectors unless explicitly transposed. For a generic vector $v\in\mathbb{R}^d$, $v_j$ denotes its $j$-th element and $\|v\|$ denotes its Euclidean norm. For a matrix $A\in\mathbb{R}^{m\times n}$, $\|A\|$ denotes its operator norm; for a square symmetric matrix $A$, $\lambda_{\min}(A)$ and $\lambda_{\max}(A)$ denote its minimum and maximum eigenvalues. For a scalar- or vector-valued random variable $Y$ and $r\geq1$, write $\|Y\|_r=(\E\|Y\|^r)^{1/r}$, where the inner norm is absolute value for scalars and Euclidean norm for vectors. For a set $S\subseteq\mathbb{R}^d$, define its diameter as $\diam(S)=\sup\{\|s-s'\|:s,s'\in S\}$. We write $\otimes$ for the Kronecker product.

For $\ell,u\in\mathbb{R}^d$ such that $\ell\leq u$ elementwise, define the rectangle
\begin{align*}
    [\ell,u]
    &=\{b=(b_1,\ldots,b_d)^\top: b_l\in[\ell_l,u_l],\ l\in[d]\},
\end{align*}
where $\ell_l$ and $u_l$ denote the $l$-th elements of $\ell$ and $u$. Thus, $\diam([\ell,u])=(\sum_{l\in[d]}(u_l-\ell_l)^2)^{1/2}$.

For a multi-index $m\in\mathbb{Z}_+^d$, let $|m|=\sum_{j=1}^d m_j$, and for $z\in\mathbb{R}^d$, write $z^m=z_1^{m_1}\cdots z_d^{m_d}$. For a nonnegative integer $\mathfrak{q}$, $(z^m:|m|\leq\mathfrak{q})^\top$ denotes the column vector collecting, in a fixed order, all $d$-variate monomials in $z$ with total degree no greater than $\mathfrak{q}$. For a multivariate function $g:\mathbb{R}^d\to\mathbb{R}$, $\nabla g=\nabla^1 g$ denotes its gradient, $\nabla^2 g$ its Hessian, and
\begin{align*}
    g^{(m)}(z)
    &=\frac{\partial^{|m|}}{\partial z_1^{m_1}\cdots\partial z_d^{m_d}}g(z).
\end{align*}
For a univariate function $g$, we write $g\in L^1(\mathbb{R})$ if $\int_{\mathbb{R}}|g(z)|\d{z}<\infty$. For a function $g:\mathcal{X}\to\mathbb{R}$ on a compact set $\mathcal{X}\subseteq\mathbb{R}^d$ and some $s>0$, we say $g$ is H\"{o}lder-$s$ on $\mathcal{X}$ if all partial derivatives $g^{(m)}$ with $|m|\leq\underline{s}$ exist and are continuous on $\mathcal{X}$, and all partial derivatives with $|m|=\underline{s}$ are H\"{o}lder continuous with exponent $s-\underline{s}$ uniformly on $\mathcal{X}$, where $\underline{s}=\max\{k\in\mathbb{Z}_+:k<s\}$.
Whenever a function enters the analysis through its Hessian, we additionally
require it to be twice continuously differentiable with a uniformly bounded
Hessian on its domain. This requirement is additional to H\"older-$s$
smoothness when $s=2$ under the preceding convention.

For sequences, $a_n\lesssim b_n$ denotes that $\limsup_n|a_n/b_n|$ is finite; $a_n=O_\P(b_n)$ or $a_n\lesssim_\P b_n$ means that $\lim_{\epsilon\rightarrow\infty}\limsup_n \P[|a_n/b_n|\geq\epsilon]=0$; $a_n=o(b_n)$ or $a_n\ll b_n$ denotes $a_n/b_n\rightarrow0$; and $a_n=o_\P(b_n)$ denotes that $a_n/b_n$ converges to zero in probability. We write $a_n\asymp b_n$ when $a_n\lesssim b_n$ and $b_n\lesssim a_n$.

\section{Convergence Rate}

For the remaining results in this part, define
\begin{align*}
    \widehat{n}_{jt}
    &=\sum_{i=1}^{n_t}\IF(\widehat{\beta}_{it}\in \widehat{P}_{jt}),
    \qquad
    \widehat{n}_{\max}=\max_{H+1\leq t\leq T}\max_{j\in[\bar{J}_t]}\widehat{n}_{jt},
    \qquad
    \widehat{n}_{\min}=\min_{H+1\leq t\leq T}\min_{j\in[\bar{J}_t]}\widehat{n}_{jt}.
\end{align*}
For each cell, define the estimated Gram block
\begin{align*}
    \widehat{Q}_{jt}
    &= \frac{1}{n_t} \sum_{i=1}^{n_t}
    \IF(\widehat{\beta}_{it}\in\widehat{P}_{jt})
    \check{p}_{\mathfrak{q}}\!\big(\widehat{D}_{jt}^{-1}(\widehat{\beta}_{it}-\widehat{c}_{jt})\big)
    \check{p}_{\mathfrak{q}}\!\big(\widehat{D}_{jt}^{-1}(\widehat{\beta}_{it}-\widehat{c}_{jt})\big)^\top .
\end{align*}
When $\mathfrak{q}=0$, $\check{p}_{\mathfrak{q}}$ is the scalar $1$, and $\widehat{Q}_{jt}=\widehat n_{jt}/n_t$ on every nonempty cell.

\subsection{Result}

\begin{theorem}[Convergence Rate]\label{thm:general-consistency}
    Consider the second-step estimator with polynomial degree $\mathfrak{q}\geq 0$. Suppose that the following conditions hold:
    \begin{enumerate}[label=(\roman*)]
        \item \textbf{DGP}.
        $\{\beta_{it}: i\in[n_t], t\in[T]\}$ take values in compact support $\mathcal{B}$;
        $\mu_t$ admits a bounded extension to the neighborhood $\mathcal{B}^+$ of $\mathcal{B}$ and this extension is uniformly H\"{o}lder-$s_\mu$ on $\mathcal{B}^+$ for some $s_\mu\geq1$;
        $\E[f_t|\mathcal{F}_{t-1}]=\E[f_t|\mathcal{G}_{t-1}]$,
        $\E[\varepsilon_{it}|\mathcal{F}_{t-1}, f_t]=0$,
        and $\varepsilon_{it}$ is independent over $i$ for each $t$ conditional on $\mathcal{F}_{t-1}$.
        For some $q>2$, $\max_{t\in[T]}\max_{i\in[n_t]}
            \E[|\varepsilon_{it}|^q|\mathcal{F}_{t-1}]\lesssim 1$ and $\max_{t\in[T]}
            \E[\|\widetilde{f}_t\|^q|\mathcal{F}_{t-1}]
            \lesssim 1$.

        \item \textbf{First-Step}.
        $\widehat{\beta}_{it}$ is $\mathcal{F}_{t-1}$-measurable, and
        $H\leq cT$ for a constant $c\in (0,1)$. With probability approaching one,
        $\widehat{\beta}_{it}\in\mathcal{B}^+$ for all $i\in[n_t]$ and $t=H+1,\ldots,T$.
        Define the realized first-step error
        \begin{align*}
            \widehat{\mathsf{R}}_{\mathtt{FS}}
            =\max_{H+1\leq t\leq T}\max_{i\in[n_t]}
            \|\widehat{\beta}_{it}-\beta_{it}\|.
        \end{align*}

        \item \textbf{Partition and Design}.
        The estimated partition $\widehat{\Delta}_t$ and its associated centers and scaling matrices are $\mathcal{F}_{t-1}$-measurable for each $t$.
        With probability approaching one, $\mathcal{B}\cup\{\widehat{\beta}_{it}:i\in[n_t]\}$ is contained in $\cup_{j\in[\bar{J}_t]}\widehat{P}_{jt}$ for each $t=H+1,\ldots,T$, $\widehat{n}_{\min}>0$, and, for all $t=H+1,\ldots,T$ and $j\in[\bar{J}_t]$,
        \begin{align*}
            1
            &\lesssim
            \frac{n_t}{\widehat n_{jt}}\lambda_{\min}(\widehat{Q}_{jt})
            \leq
            \frac{n_t}{\widehat n_{jt}}\lambda_{\max}(\widehat{Q}_{jt})
            \lesssim
            1.
        \end{align*}
        Define the realized partition diameter
        \begin{align*}
            \widehat{\mathsf{R}}_{\mathtt{D}}
            =\max_{H+1\leq t\leq T}\max_{j\in[\bar{J}_t]}
            \diam(\widehat{P}_{jt}).
        \end{align*}
    \end{enumerate}
    If $\frac{(nT)^{2/q}\log (nT)}{\widehat{n}_{\min}}\lesssim_\P 1$, then for any $\ell_n\rightarrow\infty$, with probability approaching one, the following bound holds for all $\beta\in\mathcal{B}$:
    \begin{align*}
        \big|\widehat{\mu}(\beta)-\bar{\mu}_T(\beta;H)\big|
        &\lesssim \ell_n\left[
            \widehat{\mathsf{R}}_{\mathtt{FS}}
            + \widehat{\mathsf{R}}_{\mathtt{D}}^{\min\{s_\mu,\mathfrak{q}+1\}}
            + \sqrt{\frac{\log (nT)}{\widehat{n}_{\min}}}
            + \frac{\|\beta\|}{\sqrt{T}}\right].
    \end{align*}
    All implicit constants in this theorem are independent of the evaluation point $\beta$ and of $n,T$, but may depend on the H\"{o}lder and design constants.
\end{theorem}

\subsection{Discussion}

\begin{enumerate}[label=(\roman*)]
    \item The sequence $\ell_n\to\infty$ is an arbitrary high-probability slack sequence. Equivalently, the same display can be read as the corresponding $\lesssim_\P$ rate, while retaining the explicit dependence of the factor term on the evaluation point $\beta$.

    \item When $\mathfrak{q}=0$, the within-cell polynomial basis $\check{p}_{\mathfrak{q}}$ is the scalar $1$ and $\widehat{Q}_{jt}=\widehat n_{jt}/n_t$ on every nonempty cell. Thus, the design condition in (iii) reduces to the nonempty-cell condition $\widehat{n}_{\min}>0$, and since $s_\mu\geq1$, the second-step approximation rate is governed by the cell-diameter term $\widehat{\mathsf{R}}_{\mathtt{D}}$.

    \item If the partition is quasi-uniform and each cell roughly contains the same number of observations, then $\widehat{\mathsf{R}}_{\mathtt{D}}\lesssim_\P J^{-1}$ and $n/J^d\lesssim_\P\widehat n_{\min}\leq\widehat n_{\max}\lesssim_\P n/J^d$. In that case, the theorem gives the rate
    \begin{align*}
        &\ell_n\left[
            \widehat{\mathsf{R}}_{\mathtt{FS}}
            + J^{-\min\{s_\mu,\mathfrak{q}+1\}}
            + \sqrt{\frac{J^d\log(nT)}{n}}
            + \frac{\|\beta\|}{\sqrt{T}}
        \right].
    \end{align*}

    \item With a boundary-specific modification of the assumptions and proof, the same argument can allow boundary cells that are large outside the support but have a small intersection with the feasible support $\mathcal{B}$. For example, if $\beta=\underline{\beta}$ is on the lower boundary and the first cell intersects $\mathcal{B}$ only on a shrinking neighborhood of $\underline{\beta}$, then any $\widehat{\beta}_{it}$ in that cell is still close to $\beta$ with high probability.
\end{enumerate}

\subsection{Proof of Theorem~\ref{thm:general-consistency}}

Work on the event where the first-step estimates belong to $\mathcal{B}^+$, the estimated partition covers $\mathcal{B}$ and the first-step estimates, and the design conditions in the theorem hold. This event has probability approaching one.

For each $\beta\in\mathcal{B}$, let $\widehat{\jmath}_t(\beta)$ denote the index of the cell containing $\beta$.
The fitted value can be written as a weighted average,
\begin{align*}
    \widehat{M}_t(\beta)
    &=\sum_{i=1}^{n_t}\widehat{w}_{it}(\beta)R_{it},
\end{align*}
where
\begin{align*}
    \widehat{w}_{it}(\beta)
    &=\frac{\IF(\widehat{\beta}_{it}\in\widehat{P}_{\widehat{\jmath}_t(\beta)t})}
    {n_t}
    \check{p}_{\mathfrak{q}}\!\big(\widehat{D}_{\widehat{\jmath}_t(\beta)t}^{-1}
    (\beta-\widehat{c}_{\widehat{\jmath}_t(\beta)t})\big)^\top
    \widehat{Q}_{\widehat{\jmath}_t(\beta)t}^{-1}
    \check{p}_{\mathfrak{q}}\!\big(\widehat{D}_{\widehat{\jmath}_t(\beta)t}^{-1}
    (\widehat{\beta}_{it}-\widehat{c}_{\widehat{\jmath}_t(\beta)t})\big).
\end{align*}
The boundedness of the normalized polynomial basis and the scaled eigenvalue condition imply that with probability approaching one, uniformly over $t$ and $\beta$,
\begin{align*}
    \sum_{i=1}^{n_t}|\widehat{w}_{it}(\beta)|
    &\lesssim1,\qquad
    \sum_{i=1}^{n_t}\widehat{w}_{it}(\beta)^2
    \lesssim \frac{1}{\widehat{n}_{\widehat{\jmath}_t(\beta)t}}.
\end{align*}
Applying the same algebra to the constant polynomial gives $\sum_{i=1}^{n_t}\widehat{w}_{it}(\beta)=1$. Moreover, the weights reproduce all polynomials of total degree no greater than $\mathfrak{q}$ on the selected cell:
\begin{align*}
    \sum_{i=1}^{n_t}\widehat{w}_{it}(\beta)\poly(\widehat{\beta}_{it})
    &=\poly(\beta)
\end{align*}
for every such polynomial $\poly$. This includes constant functions when $\mathfrak{q}=0$ and coordinate functions when $\mathfrak{q}\geq1$.

Using $R_{it}=\mu_t(\beta_{it})+\beta_{it}^\top\widetilde{f}_t+\varepsilon_{it}$,
\begin{align*}
    \widehat{\mu}(\beta)-\bar{\mu}_T(\beta;H)
    &=\frac{1}{T-H}\sum_{t=H+1}^T\sum_{i=1}^{n_t}\widehat{w}_{it}(\beta)\varepsilon_{it}\\
    &+\frac{1}{T-H}\sum_{t=H+1}^T\sum_{i=1}^{n_t}\widehat{w}_{it}(\beta)
    \big(\mu_t(\beta_{it})-\mu_t(\beta)\big)\\
    &+\frac{1}{T-H}\sum_{t=H+1}^T
    \left(\sum_{i=1}^{n_t}\widehat{w}_{it}(\beta)\beta_{it}\right)^\top\widetilde{f}_t .
\end{align*}

We first bound the approximation term. Since $s_\mu\geq1$, the uniform H\"{o}lder condition implies a uniform Lipschitz bound on $\mathcal{B}^+$. Therefore,
\begin{align*}
    \max_{H+1\leq t\leq T}\max_{i\in[n_t]}
    \big|\mu_t(\beta_{it})-\mu_t(\widehat{\beta}_{it})\big|
    &\lesssim \widehat{\mathsf{R}}_{\mathtt{FS}}.
\end{align*}
Together with the absolute-weight-sum bound, this controls the contribution from replacing $\beta_{it}$ by $\widehat{\beta}_{it}$. For the remaining within-cell term, local polynomial approximation gives, uniformly over $t$ and cells, a polynomial $\poly_{jt}$ of degree no greater than $\mathfrak{q}$ such that
\begin{align*}
    \sup_{b\in\widehat{P}_{jt}\cap\mathcal{B}^+}
    \big|\mu_t(b)-\poly_{jt}(b)\big|
    &\lesssim
    \diam(\widehat{P}_{jt})^{\min\{s_\mu,\mathfrak{q}+1\}}.
\end{align*}
Using polynomial reproduction,
\begin{align*}
    \left|\sum_{i=1}^{n_t}\widehat{w}_{it}(\beta)
    \big(\mu_t(\widehat{\beta}_{it})-\mu_t(\beta)\big)\right|
    &\lesssim
    \widehat{\mathsf{R}}_{\mathtt{D}}^{\min\{s_\mu,\mathfrak{q}+1\}}.
\end{align*}
Thus, uniformly over $\beta\in\mathcal{B}$,
\begin{align*}
    \left|
    \frac{1}{T-H}\sum_{t=H+1}^T\sum_{i=1}^{n_t}\widehat{w}_{it}(\beta)
    \big(\mu_t(\beta_{it})-\mu_t(\beta)\big)
    \right|
    &\lesssim
    \widehat{\mathsf{R}}_{\mathtt{FS}}
    +\widehat{\mathsf{R}}_{\mathtt{D}}^{\min\{s_\mu,\mathfrak{q}+1\}}.
\end{align*}

Next, consider the idiosyncratic-error term. For each component $k\in[K_{\mathfrak{q}}]$ of
$\check{p}_{\mathfrak{q}}\!\big(\widehat{D}_{jt}^{-1}(\widehat{\beta}_{it}-\widehat{c}_{jt})\big)$,
\begin{align*}
    &\frac{1}{\widehat{n}_{jt}}\sum_{i=1}^{n_t}
    \IF(\widehat{\beta}_{it}\in\widehat{P}_{jt})
    \left[\check{p}_{\mathfrak{q}}\!\big(\widehat{D}_{jt}^{-1}
    (\widehat{\beta}_{it}-\widehat{c}_{jt})\big)\right]_k\varepsilon_{it}
\end{align*}
is a conditionally mean-zero average, independent over $i$ given $\mathcal{F}_{t-1}$, with conditional variance bounded by a constant multiple of $1/\widehat{n}_{jt}$. On the event $\widehat{n}_{\min}>0$, the number of cell-component averages is of order $\sum_{t=H+1}^T\bar{J}_t\leq\sum_{t=H+1}^T n_t\lesssim nT$. Since the normalized regressors are bounded, a standard truncation and union-bound argument for conditionally independent arrays yields
\begin{align*}
    \max_{H+1\leq t\leq T}\max_{j\in[\bar{J}_t]}\max_{k\in[K_{\mathfrak{q}}]}
    \left|
    \frac{1}{\widehat{n}_{jt}}\sum_{i=1}^{n_t}
    \IF(\widehat{\beta}_{it}\in\widehat{P}_{jt})
    \left[\check{p}_{\mathfrak{q}}\!\big(\widehat{D}_{jt}^{-1}
    (\widehat{\beta}_{it}-\widehat{c}_{jt})\big)\right]_k\varepsilon_{it}
    \right|
    &\lesssim
    \ell_n\sqrt{\frac{\log(nT)}{\widehat{n}_{\min}}}
\end{align*}
with probability approaching one; the rate restriction involving $\widehat{n}_{\min}$ ensures that the truncation contribution is absorbed into the displayed bound. For any evaluation point $\beta$, the weighted error is a bounded linear combination of these component averages in the selected cell. The boundedness of the normalized polynomial basis and the scaled eigenvalue bound for $\widehat{Q}_{jt}$ then gives, uniformly over $\beta\in\mathcal{B}$,
\begin{align*}
    \left|
    \frac{1}{T-H}\sum_{t=H+1}^T\sum_{i=1}^{n_t}
    \widehat{w}_{it}(\beta)\varepsilon_{it}
    \right|
    &\lesssim
    \ell_n\sqrt{\frac{\log(nT)}{\widehat{n}_{\min}}}.
\end{align*}

Finally, decompose the factor term as
\begin{align*}
    &\frac{1}{T-H}\sum_{t=H+1}^T
    \left(\sum_{i=1}^{n_t}\widehat{w}_{it}(\beta)\beta_{it}\right)^\top\widetilde{f}_t\\
    &=\frac{1}{T-H}\sum_{t=H+1}^T\beta^\top\widetilde{f}_t\\
    &\qquad+\frac{1}{T-H}\sum_{t=H+1}^T
    \left(\sum_{i=1}^{n_t}\widehat{w}_{it}(\beta)(\beta_{it}-\widehat{\beta}_{it})\right)^\top\widetilde{f}_t\\
    &\qquad+\frac{1}{T-H}\sum_{t=H+1}^T
    \left(\sum_{i=1}^{n_t}\widehat{w}_{it}(\beta)\widehat{\beta}_{it}-\beta\right)^\top\widetilde{f}_t .
\end{align*}
By the martingale-difference property of $\widetilde{f}_t$ with respect to $\mathcal{G}_t$ and the conditional moment bound,
\begin{align*}
    \left\|
    \frac{1}{T-H}\sum_{t=H+1}^T\widetilde{f}_t
    \right\|
    &\lesssim_\P T^{-1/2},
\end{align*}
so the first term is $\lesssim_\P\|\beta\|T^{-1/2}$ uniformly over $\beta$. Also,
\begin{align*}
    \frac{1}{T-H}\sum_{t=H+1}^T\|\widetilde{f}_t\|
    &\lesssim_\P 1
\end{align*}
by Markov's inequality. The second term is $\lesssim_\P\widehat{\mathsf{R}}_{\mathtt{FS}}$ because the weights have uniformly bounded absolute sum. For the third term, polynomial reproduction gives
\begin{align*}
    \sum_{i=1}^{n_t}\widehat{w}_{it}(\beta)\widehat{\beta}_{it}
    &=\beta
\end{align*}
when $\mathfrak{q}\geq1$. When $\mathfrak{q}=0$, the same expression is the ordinary cell average of the $\widehat{\beta}_{it}$'s in the selected cell, and hence its distance from $\beta$ is bounded by the diameter of that cell. Therefore, uniformly over $\beta\in\mathcal{B}$,
\begin{align*}
    \left|
    \frac{1}{T-H}\sum_{t=H+1}^T
    \left(\sum_{i=1}^{n_t}\widehat{w}_{it}(\beta)\beta_{it}\right)^\top\widetilde{f}_t
    \right|
    &\lesssim_\P
    \frac{\|\beta\|}{\sqrt{T}}
    +\widehat{\mathsf{R}}_{\mathtt{FS}}
    +\widehat{\mathsf{R}}_{\mathtt{D}}^{\min\{s_\mu,\mathfrak{q}+1\}}.
\end{align*}

Combining the three bounds and absorbing the stochastic constants into the arbitrary high-probability slack sequence $\ell_n$ gives the claimed result, uniformly over $\beta\in\mathcal{B}$. \qed

\section{Stochastic Linearization}\label{sec:general-linearization}

This section derives a high-level stochastic linearization for second-step polynomial degree $\mathfrak{q}\geq0$. The result is stated in terms of generic first-step estimators and generic partitions, so it does not rely on the specific local-polynomial first step or marginal-quantile-spaced partition introduced later.
Let $J$ denote a generic partition-resolution sequence; the conditions below encode cell mass of order $J^{-d}$ and cell diameter of order $J^{-1}$.

Let $\{\mathcal{D}_t\}$ denote an auxiliary design filtration with $\mathcal{D}_{t-1}\subseteq\mathcal{F}_{t-1}$. In this high-level section, $\mathcal{D}_{t-1}$ records the information used to construct the conditionally nonrandom reference partition; Section~\ref{sec:classical-filtrations} gives its concrete definition for the classical beta-sorted estimator. To state the linearization, introduce {a reference partition},
\begin{align*}
    \Delta_t
    &=\{P_{jt}:j\in[\bar{J}_t]\}
\end{align*}
that is $\mathcal{D}_{t-1}$-measurable and has the same number of cells as the estimated partition $\widehat{\Delta}_t$ from Section~\ref{sec:general-setup}.
{
For each reference cell paired with an estimated cell intersecting
$\mathcal B^\circ$, choose a $\mathcal{D}_{t-1}$-measurable center $c_{jt}$
and nonsingular scaling/rotation matrix $D_{jt}$ so that the normalized
polynomial basis is uniformly bounded on the cell. Tail reference cells may
instead be assigned any $\mathcal D_{t-1}$-measurable nonsingular
normalization; no uniform boundedness or design condition is imposed on them
because they are never selected by an interior evaluation point.
}
Thus, $\mathcal{D}_{t-1}$ governs the reference partition and reference objects, whereas the realized estimated partition $\widehat{\Delta}_t$ may depend on first-step estimates and is only required below to be $\mathcal{F}_{t-1}$-measurable.
The reference cells and estimated cells are paired by their common index $j$.

Define the reference basis maps by
\begin{align*}
    p_t(b)
    &=p(b;\Delta_t),
    \qquad
    p_{jt}(b)
    =
    p_{\mathfrak{q}}(b,P_{jt}).
\end{align*}
{
Both basis maps are evaluated at the raw argument $b$; in particular, the
classical construction does not restrict the generated regressor to reside in
$\mathcal B$. The estimated marginal partition in
Section~\ref{sec:classical-partition} uses the coordinatewise sample extrema
as its outer endpoints and therefore contains every realized
$\widehat\beta_{it}$. Reference cells are paired with estimated cells by their
split-quantile ranks. Only pairs whose estimated cell intersects
$\mathcal B^\circ$ are normalized and used below; no reference design
condition is imposed on the cells outside $\mathcal{B}^\circ$.
}
Write $\widehat{p}_{jt}(b)$ for the $j$-th $K_{\mathfrak{q}}$-dimensional block of the estimated basis $\widehat{p}_t(b)$ defined in Section~\ref{sec:general-setup}. Thus $p_t(\widehat{\beta}_{it})$ denotes the reference basis evaluated at the first-step estimate, while $p_t(\beta_{it})$ denotes the reference basis evaluated at the true exposure. The reference Gram matrix below remains centered at the generated-regressor distribution, but the leading idiosyncratic term uses $p_t(\beta_{it})$ so that its coefficients are fixed by the innovation-free conditioning field in the classical verification.
The estimated and reference Gram matrices are
\begin{align*}
    \widehat{Q}_t
    =\frac{1}{n_t}\sum_{i=1}^{n_t}
    \widehat{p}_t(\widehat{\beta}_{it})\widehat{p}_t(\widehat{\beta}_{it})^\top
    \qquad\text{and}\qquad
    Q_t
    =\frac{1}{n_t}\sum_{i=1}^{n_t}
    \E\big[
    p_t(\widehat{\beta}_{it})p_t(\widehat{\beta}_{it})^\top
    \mid\mathcal{D}_{t-1}
    \big].
\end{align*}
Both matrices are block diagonal. Write $\widehat{Q}_{jt}$
and $Q_{jt}$ for their $j$-th
$K_{\mathfrak{q}}\times K_{\mathfrak{q}}$ diagonal blocks.
Thus $Q_t$ is the conditional reference Gram matrix for the generated regressor evaluated on the conditionally nonrandom reference partition. When $\mathfrak{q}=0$, $Q_{jt}=n_t^{-1}\sum_{i=1}^{n_t}\P(\widehat{\beta}_{it}\in P_{jt}\mid\mathcal{D}_{t-1})$.
{
In every expression involving $Q_t^{-1}$ or $\widehat Q_t^{-1}$ and an
evaluation point in $\mathcal B^\circ$, the inverse is blockwise shorthand for
the inverse of the selected block. Thus the algebra does not require
invertibility or a polynomial normalization for tail blocks outside
$\mathcal J_t$.
}

\subsection{Decomposition}

Using the block Gram matrices defined above, the second-step estimator admits the decomposition
\begin{align*}
    \widehat{\mu}(\beta)- \bar{\mu}_T(\beta;H)
    &= \frac{1}{T-H} \sum_{t=H+1}^T (\widehat{M}_t(\beta) - \mu_t(\beta))\\
    &= \mathscr{G}(\beta) + \mathscr{E}(\beta) - \mathscr{L}(\beta) + \mathscr{B}(\beta)\\
    &\qquad + \mathscr{R}_1(\beta)+\mathscr{R}_2(\beta)+\mathscr{R}_3(\beta)+\mathscr{R}_4(\beta) - \mathscr{R}_5(\beta),
\end{align*}
where the potentially leading terms are
\begin{align*}
    \mathscr{G}(\beta)
    &= \frac{1}{T-H}\sum_{t=H+1}^T\beta^\top\widetilde{f}_t,\\
    \mathscr{E}(\beta)
    &= \frac{1}{{T-H}}\sum_{t=H+1}^T\widehat{p}_t(\beta)^{\top}Q_{t}^{-1}
    \frac{1}{n_t}\sum_{i=1}^{n_t}p_t(\beta_{it}) \varepsilon_{it},\\
    \mathscr{L}(\beta)
    &= \frac{1}{T-H} \sum_{t=H+1}^T \widehat{p}_{t}(\beta)^\top\widehat{Q}_t^{-1}
    \frac{1}{n_t}\sum_{i=1}^{n_t} \widehat{p}_t(\widehat{\beta}_{it}) [\nabla\mu_t(\beta_{it})]^\top(\widehat{\beta}_{it}-\beta_{it}),\\
    \mathscr{B}(\beta)
    &= \frac{1}{T-H} \sum_{t=H+1}^T
    \widehat{p}_{t}(\beta)^\top\widehat{Q}_t^{-1}
    \frac{1}{n_t}\sum_{i=1}^{n_t}
    \widehat{p}_t(\widehat{\beta}_{it}) \big(\mu_t(\widehat{\beta}_{it}) - \mu_t(\beta) \big),
\end{align*}
and the higher-order remainder terms are
\begin{align*}
    \mathscr{R}_1(\beta)
    &= \frac{1}{T-H} \sum_{t=H+1}^T
    \widehat{p}_{t}(\beta)^\top
    \left(\widehat{Q}_t^{-1} - Q_{t}^{-1} \right)
    \frac{1}{n_t} \sum_{i=1}^{n_t} \widehat{p}_t(\widehat{\beta}_{it}) \varepsilon_{it},\\
    \mathscr{R}_2(\beta)
    &= \frac{1}{T-H} \sum_{t=H+1}^T
    \widehat{p}_{t}(\beta)^\top Q_{t}^{-1}
    \frac{1}{n_t} \sum_{i=1}^{n_t} \big(\widehat{p}_t(\widehat{\beta}_{it})-p_t(\beta_{it})\big) \varepsilon_{it},\\
    \mathscr{R}_3(\beta)
    &=\frac{1}{T-H}\sum_{t=H+1}^T
    \left[
    \widehat{p}_{t}(\beta)^\top \widehat{Q}_{t}^{-1}
    \left(\frac{1}{n_t}\sum_{i=1}^{n_t}\widehat{p}_t(\widehat{\beta}_{it})\widehat{\beta}_{it}^\top\right)
    -\beta^\top\right]\widetilde{f}_t,\\
    \mathscr{R}_4(\beta)
    &=\frac{1}{T-H}\sum_{t=H+1}^T
    \widehat{p}_{t}(\beta)^\top \widehat{Q}_{t}^{-1}
    \left(\frac{1}{n_t}\sum_{i=1}^{n_t}
    \widehat{p}_t(\widehat{\beta}_{it})(\beta_{it}-\widehat{\beta}_{it})^\top\right)\widetilde{f}_t,\\
    \mathscr{R}_5(\beta)
    &= \frac{1}{2(T-H)} \sum_{t=H+1}^T
    \widehat{p}_{t}(\beta)^\top\widehat{Q}_t^{-1}
    \frac{1}{n_t}\sum_{i=1}^{n_t}\widehat{p}_t(\widehat{\beta}_{it})
    \big(\widehat{\beta}_{it} - \beta_{it} \big)^\top
    \nabla^2\mu_t(\tilde{\beta}_{it})
    \big(\widehat{\beta}_{it} - \beta_{it} \big),
\end{align*}
with $\tilde{\beta}_{it}=a_{it}\widehat{\beta}_{it}+(1-a_{it})\beta_{it}$ for some $a_{it}\in[0,1]$.

The expanded set $\mathcal{B}^+$ from Section~\ref{sec:general-setup} handles first-step estimates and Taylor intermediate points that may lie outside the support $\mathcal{B}$. {Recall that $\mathcal B^\circ$ is the interior evaluation region and $\mathcal J_t$ is the set of estimated cells intersecting it.}
For each $j\in\mathcal{J}_t$, the reference block $Q_{jt}$ below is formed from the reference cell paired with the estimated cell $\widehat P_{jt}$; the partition-design condition requires these paired reference cells to satisfy the same interior design bounds.
Recall the previously defined quantity $\widehat{\mathsf{R}}_{\mathtt{FS}}$, and also define
\begin{align*}
    \widehat{\mathsf{R}}_{\mathtt{D}}^{\circ}
    &=\max_{H+1\leq t\leq T}\max_{j\in\mathcal{J}_t}\;
    \diam(\widehat{P}_{jt}),\\
    \widehat{\mathsf{R}}_{\mathtt{Q}}
    &=\max_{H+1\leq t\leq T}\max_{j\in\mathcal{J}_t}
    \|\widehat{Q}_{jt}-Q_{jt}\|,\\
    \widehat{\mathsf{R}}_{\mathtt{P}}
    &=\max_{H+1\leq t\leq T}\max_{j\in\mathcal{J}_t}
    \frac{1}{n_t}\sum_{i=1}^{n_t}
    \|\widehat{p}_{jt}(\widehat{\beta}_{it})-p_{jt}(\beta_{it})\|^2.
\end{align*}

\subsection{Result}

The following lemma establishes a general stochastic linearization for generic first-step estimators and partitioning schemes. Part \ref{part:classical} then verifies these high-level quantities for the classical beta-sorted portfolio estimator.

\begin{lemma}[Stochastic Linearization]\label{lem:general-linearization}
    Maintain the setup and notation of Section~\ref{sec:general-setup} and the definitions above. Suppose that conditions (i) and (ii) in Theorem~\ref{thm:general-consistency} hold, and that $\mu_t$ is uniformly H\"older-$s_\mu$ on $\mathcal{B}^+$ and twice continuously differentiable on $\mathcal{B}^+$ with uniformly bounded first and second derivatives. Assume further that the following conditions hold:
    \begin{enumerate}[label=(\roman*)]
    \item\label{cond:partition-design}
    (Partition and Design) the estimated partition $\widehat{\Delta}_t$ and its associated centers and scaling matrices are $\mathcal{F}_{t-1}$-measurable for each $t$,
    $\mathcal{B}^\circ\cup\{\widehat{\beta}_{it}:i\in[n_t]\}$ is contained in $\cup_{j\in[\bar{J}_t]}\widehat{P}_{jt}$ for each $t=H+1,\ldots,T$ w.p.a.1,
    $\widehat{\mathsf{R}}_{\mathtt{D}}^{\circ}\lesssim_\P J^{-1}$, and $J^{-d}
        \lesssim_\P \lambda_{\min}(Q_{jt}) \leq \lambda_{\max}(Q_{jt}) \lesssim_\P J^{-d}$ uniformly over $t=H+1,\ldots,T$ and $j\in\mathcal{J}_t$.
        Also, uniformly over the same indices,
        \begin{align*}
            \lambda_{\max}\!\left(
            \frac{1}{n_t}\sum_{i=1}^{n_t}
            p_{jt}(\beta_{it})p_{jt}(\beta_{it})^\top
            \right)
            &\lesssim_\P J^{-d}.
        \end{align*}

    \item\label{cond:gram-convergence}
    (Gram Convergence) $\widehat{\mathsf{R}}_{\mathtt{Q}} =o_\P(J^{-d})$.
    \end{enumerate}
    Then, for each fixed $\beta\in\mathcal{B}^\circ$,
    \begin{align*}
        |\mathscr{G}(\beta)|
        &\lesssim_\P \frac{\|\beta\|}{\sqrt{T}},&
        |\mathscr{E}(\beta)|
        &\lesssim_\P \sqrt{\frac{J^d}{nT}},\\
        |\mathscr{R}_1(\beta)|
        &\lesssim_\P \sqrt{\frac{J^{3d}(\widehat{\mathsf{R}}_{\mathtt{Q}})^2}{nT}},&
        |\mathscr{R}_2(\beta)|
        &\lesssim_\P \sqrt{\frac{J^{2d}\widehat{\mathsf{R}}_{\mathtt{P}}}{nT}},\\
        |\mathscr{R}_3(\beta)|
        &\lesssim_\P \IF(\mathfrak{q}=0)\frac{\widehat{\mathsf{R}}_{\mathtt{D}}^{\circ}}{\sqrt{T}},&
        |\mathscr{R}_4(\beta)|
        &\lesssim_\P \frac{\widehat{\mathsf{R}}_{\mathtt{FS}}}{\sqrt{T}}.
    \end{align*}
    Moreover,
    \begin{align*}
        \sup_{\beta\in\mathcal{B}^\circ}|\mathscr{L}(\beta)|
        &\lesssim_\P \widehat{\mathsf{R}}_{\mathtt{FS}},&
        \sup_{\beta\in\mathcal{B}^\circ}|\mathscr{R}_5(\beta)|
        &\lesssim_\P(\widehat{\mathsf{R}}_{\mathtt{FS}})^2,\\
        \sup_{\beta\in\mathcal{B}^\circ}|\mathscr{B}(\beta)|
        &\lesssim_\P (\widehat{\mathsf{R}}_{\mathtt{D}}^{\circ})^{\min\{s_\mu,\mathfrak{q}+1\}},&
        \sup_{\beta\in\mathcal{B}^\circ}|\mathscr{R}_3(\beta)|
        &\lesssim_\P \IF(\mathfrak{q}=0)\widehat{\mathsf{R}}_{\mathtt{D}}^{\circ}.
    \end{align*}
    The constants implicit in $\lesssim_\P$ do not depend on the evaluation point $\beta\in\mathcal{B}^\circ$.
\end{lemma}

\subsection{Discussion}

\begin{enumerate}[label=(\roman*)]
    \item The condition $\widehat{\mathsf{R}}_{\mathtt{Q}}=o_\P(J^{-d})$ is a relative Gram-convergence condition, because each reference block $Q_{jt}$ has eigenvalues of order $J^{-d}$. Together with the reference design condition, it implies $\widehat{Q}_{jt}\asymp_\P J^{-d}$ uniformly over $j\in\mathcal{J}_t$ and controls the inverse perturbation $\widehat{Q}_{jt}^{-1}-Q_{jt}^{-1}$ used in the bound for $\mathscr{R}_1(\beta)$. The realized size of $\widehat{\mathsf{R}}_{\mathtt{P}}$ controls the generated-regressor basis-switching term $\mathscr{R}_2(\beta)$.

    \item Since the variance bounds used in the proof do not depend on $\beta$, the pointwise stochastic bounds for $\mathscr{E}(\beta)$ and $\mathscr{R}_j(\beta)$ can be read uniformly in the following sense: if $G(\beta)$ denotes one of these terms and $r_n$ is the displayed rate, then for any deterministic sequence $\ell_n\to\infty$,
    \begin{align*}
        \limsup_{n,T\rightarrow\infty}\;
        \sup_{\beta\in\mathcal{B}^\circ}
        \P\Big(|G(\beta)|>\ell_n r_n\Big)
        &=0.
    \end{align*}

    \item The interior partition and design conditions may not hold for all bins. The index set $\mathcal{J}_t$ restricts attention to cells that can contain evaluation points in $\mathcal{B}^\circ$. Given the results in Section \ref{sec:classical-partition}, this can be guaranteed if we focus on interior points $\beta$ that are separated from the boundary of $\mathcal{B}$ by more than the cell width and the first-step smoothing scale.

    \item When $\mathfrak{q}=0$, the block basis is the scalar $1$. In that special case, $\widehat{\mathsf{R}}_{\mathtt{Q}}$ is the cell-average discrepancy and $\widehat{\mathsf{R}}_{\mathtt{P}}$ is the cell-switching discrepancy between the generated-regressor basis $\widehat p_t(\widehat\beta_{it})$ and the true-exposure reference basis $p_t(\beta_{it})$. The latter enters only through $\mathscr{R}_2(\beta)$.

    \item For the classical marginal-quantile partition, Part \ref{part:classical} verifies the high-level quantities appearing in this lemma under primitive conditions. In particular, Section \ref{sec:classical-partition} establishes the partition, Gram, and basis-switching bounds, while Section \ref{sec:classical-linearization} uses the structure of the rolling-window first step to control the generated-regressor terms. The rate substitutions are therefore postponed until the classical setup and its notation have been introduced.
\end{enumerate}

\subsection{Proof of Lemma~\ref{lem:general-linearization}}

Work on an event whose probability approaches one and on which the conditions in the lemma hold. By Weyl's inequality, for every $t=H+1,\ldots,T$ and $j\in\mathcal{J}_t$,
\begin{align*}
    J^{-d}
    &\lesssim \lambda_{\min}(\widehat{Q}_{jt})
    \leq \lambda_{\max}(\widehat{Q}_{jt})
    \lesssim J^{-d}.
\end{align*}
Because $\E[f_t|\mathcal{F}_{t-1}]=\E[f_t|\mathcal{G}_{t-1}]$ and $\mathcal{G}_{t-1}\subseteq\mathcal{F}_{t-1}$, $\widetilde{f}_t$ has mean zero conditional on $\mathcal{F}_{t-1}$.
Thus any $\mathcal{F}_{t-1}$-measurable coefficient multiplying $\widetilde{f}_t$ generates a martingale-difference summand with respect to $\{\mathcal{F}_t\}$.
The bound on $\mathscr{G}(\beta)$ is immediate from the martingale-difference property of $\widetilde{f}_t$ and the conditional moment bound.

For $\mathscr{E}(\beta)$, let $\widehat{\jmath}_t(\beta)$ denote the cell containing $\beta$ in $\widehat{\Delta}_t$. Since $\beta\in\mathcal{B}^\circ$, $\widehat{\jmath}_t(\beta)\in\mathcal{J}_t$, and the paired indexing identifies the corresponding reference block. By the martingale structure and $\E[\varepsilon_{it}^2|\mathcal{F}_{t-1}]\lesssim 1$, for each fixed $\beta\in\mathcal{B}^\circ$,
\begin{align*}
    &\frac{1}{(T-H)^2}\sum_{t=H+1}^T
    \Var\bigg[
    \widehat{p}_{t}(\beta)^\top Q_{t}^{-1}
    \frac{1}{n_t}\sum_{i=1}^{n_t} p_t(\beta_{it})\varepsilon_{it}
    \,\bigg|\,\mathcal{F}_{t-1}\bigg]\\
    &\lesssim
    \frac{1}{T^2}\sum_{t=H+1}^T
    \frac{1}{n_t}
    \widehat{p}_{\widehat{\jmath}_t(\beta)t}(\beta)^\top Q_{\widehat{\jmath}_t(\beta)t}^{-1}
    \left(
    \frac{1}{n_t}\sum_{i=1}^{n_t}
    p_{\widehat{\jmath}_t(\beta)t}(\beta_{it})
    p_{\widehat{\jmath}_t(\beta)t}(\beta_{it})^\top
    \right)
    Q_{\widehat{\jmath}_t(\beta)t}^{-1}\widehat{p}_{\widehat{\jmath}_t(\beta)t}(\beta)
    \lesssim \frac{1}{T^2} \frac{J^dT}{n} =\frac{J^d}{nT}.
\end{align*}
The bound on $\mathscr{E}(\beta)$ follows by the Lenglart--Rebolledo inequality.

For $\mathscr{L}(\beta)$ and $\mathscr{R}_5(\beta)$, bounded derivatives, bounded basis functions, and the eigenvalue bounds for $\widehat{Q}_{jt}$ give
\begin{align*}
    |\mathscr{R}_5(\beta)|
    &\leq
    \frac{1}{2(T-H)} \sum_{t=H+1}^T
    \left[
    \frac{1}{n_t}\sum_{i=1}^{n_t}
    \left|\widehat{p}_{t}(\beta)^\top \widehat{Q}_{t}^{-1}
    \widehat{p}_t(\widehat{\beta}_{it})\right|
    \|\nabla^2\mu_t(\tilde{\beta}_{it})\|
    \right]
    \widehat{\mathsf{R}}_{\mathtt{FS}}^2
    \lesssim \widehat{\mathsf{R}}_{\mathtt{FS}}^2,
\end{align*}
and
\begin{align*}
    |\mathscr{L}(\beta)|
    &\leq
    \frac{1}{T-H} \sum_{t=H+1}^T
    \left[
    \frac{1}{n_t}\sum_{i=1}^{n_t}
    \left|\widehat{p}_{t}(\beta)^\top\widehat{Q}_{t}^{-1}
    \widehat{p}_t(\widehat{\beta}_{it})\right|
    \|\nabla\mu_t(\beta_{it})\|
    \right]\widehat{\mathsf{R}}_{\mathtt{FS}}
    \lesssim \widehat{\mathsf{R}}_{\mathtt{FS}}.
\end{align*}
The same bounds hold uniformly over $\beta\in\mathcal{B}^\circ$.

Next, consider $\mathscr{B}(\beta)$. On each interior cell, let $\poly_{jt}$ denote a polynomial of degree at most $\mathfrak{q}$ satisfying the usual approximation bound
\begin{align*}
    \sup_{b\in\widehat{P}_{jt}\cap\mathcal{B}^+}|\mu_t(b)-\poly_{jt}(b)|
    &\lesssim \diam(\widehat{P}_{jt})^{\min\{s_\mu,\mathfrak{q}+1\}}.
\end{align*}
Because the second-step basis reproduces every polynomial of degree at most $\mathfrak{q}$ on the estimated cell containing $\beta$,
\begin{align*}
    \widehat{p}_{t}(\beta)^\top \widehat{Q}_{t}^{-1}
    \frac{1}{n_t}\sum_{i=1}^{n_t}\widehat{p}_t(\widehat{\beta}_{it})\poly_{\widehat{\jmath}_t(\beta)t}(\widehat{\beta}_{it})
    &=\poly_{\widehat{\jmath}_t(\beta)t}(\beta).
\end{align*}
The boundedness of the absolute sum of the local polynomial weights then gives
\begin{align*}
    \sup_{\beta\in\mathcal{B}^\circ}|\mathscr{B}(\beta)|
    &\lesssim(\widehat{\mathsf{R}}_{\mathtt{D}}^{\circ})^{\min\{s_\mu,\mathfrak{q}+1\}}.
\end{align*}

For $\mathscr{R}_3(\beta)$, if $\mathfrak{q}\geq1$, the global linear function $g(b)=b$ is reproduced exactly by the local polynomial basis, and therefore $\mathscr{R}_3(\beta)=0$. If $\mathfrak{q}=0$,
\begin{align*}
    \left\|
    \widehat{p}_{t}(\beta)^\top \widehat{Q}_{t}^{-1}
    \left(\frac{1}{n_t}\sum_{i=1}^{n_t}
    \widehat{p}_t(\widehat{\beta}_{it})\widehat{\beta}_{it}^\top\right)
    -\beta^\top
    \right\|
    &\leq \widehat{\mathsf{R}}_{\mathtt{D}}^{\circ},
\end{align*}
because the left-hand side is the distance between $\beta$ and the average of the estimated betas in the same cell as $\beta$. This gives
\begin{align*}
    |\mathscr{R}_3(\beta)|
    \lesssim_\P \IF(\mathfrak{q}=0)\frac{\widehat{\mathsf{R}}_{\mathtt{D}}^{\circ}}{\sqrt{T}}
    \qquad\text{and}\qquad
    \sup_{\beta\in\mathcal{B}^\circ}|\mathscr{R}_3(\beta)|
    \lesssim_\P \IF(\mathfrak{q}=0)\widehat{\mathsf{R}}_{\mathtt{D}}^{\circ}
\end{align*}
because the term is identically zero when $\mathfrak{q}\geq1$.

Finally, consider $\mathscr{R}_j(\beta)$, $j=1,2,4$. The block-diagonal structure gives
\begin{align*}
    \|\widehat{Q}_{jt}^{-1}-Q_{jt}^{-1}\|
    &\lesssim J^{2d}\widehat{\mathsf{R}}_{\mathtt{Q}}
\end{align*}
uniformly over $t=H+1,\ldots,T$ and $j\in\mathcal{J}_t$. Boundedness of the estimated basis and the eigenvalue bounds for $\widehat{Q}_{jt}$ control the conditional second moment of the selected cell average. Therefore,
\begin{align*}
    &\frac{1}{T}
    \max_{H+1\leq t\leq T}
    \Var\Big[
    \widehat{p}_{t}(\beta)^\top
    \left(\widehat{Q}_t^{-1}-Q_t^{-1}\right)
    \frac{1}{n_t}\sum_{i=1}^{n_t}\widehat{p}_t(\widehat{\beta}_{it})\varepsilon_{it}
    \,\Big|\,\mathcal{F}_{t-1}\Big]\\
    &\lesssim
    \frac{1}{nTJ^d}\big(J^{2d}\widehat{\mathsf{R}}_{\mathtt{Q}}\big)^2
    =
    \frac{J^{3d} \widehat{\mathsf{R}}_{\mathtt{Q}}^2}{nT}.
\end{align*}
For $\mathscr{R}_2(\beta)$,
\begin{align*}
    &\frac{1}{T}\max_{H+1\leq t\leq T}
    \Var\Big[
    \widehat{p}_{t}(\beta)^\top Q_t^{-1}
    \frac{1}{n_t}\sum_{i=1}^{n_t}
    \big(\widehat{p}_t(\widehat{\beta}_{it})-p_t(\beta_{it})\big)\varepsilon_{it}
    \,\Big|\,\mathcal{F}_{t-1}\Big]\\
    &\lesssim
    \frac{J^{2d}}{T}\max_{H+1\leq t\leq T}\max_{j\in\mathcal{J}_t}
    \frac{1}{n_t^2}\sum_{i=1}^{n_t}
    \|\widehat{p}_{jt}(\widehat{\beta}_{it})-p_{jt}(\beta_{it})\|^2
    \lesssim
    \frac{J^{2d}\widehat{\mathsf{R}}_{\mathtt{P}}}{nT}.
\end{align*}
For $\mathscr{R}_4(\beta)$,
\begin{align*}
    &\frac{1}{(T-H)^2}\sum_{t=H+1}^T
    \Var\bigg[
    \widehat{p}_{t}(\beta)^\top \widehat{Q}_{t}^{-1}
    \left(\frac{1}{n_t}\sum_{i=1}^{n_t}
    \widehat{p}_t(\widehat{\beta}_{it})(\beta_{it}-\widehat{\beta}_{it})^\top\right)
    \widetilde{f}_t
    \,\bigg|\,\mathcal{F}_{t-1}\bigg]
    \lesssim \frac{\widehat{\mathsf{R}}_{\mathtt{FS}}^2}{T}.
\end{align*}
The desired pointwise bounds follow again by the Lenglart--Rebolledo inequality. \qed

\part{Classical Beta-Sorted Portfolios}
\label{part:classical}

This part specializes the general setup in Part \ref{part:general} to the classical beta-sorted portfolio estimator, where the portfolios are constructed using local rolling-window least-squares estimates of $\beta_{it}$.

We keep the notation from Part \ref{part:general}, including the estimator $\widehat{\mu}(\beta)$, the estimated and reference partitions $\widehat{\Delta}_t$ and $\Delta_t$, and the basis maps $\widehat{p}_t(\cdot)$ and $p_t(\cdot)$. The specialization has two ingredients: the first-step betas are estimated by rolling-window least-squares regressions, and the second-step portfolios are formed by constructing tensor products based on marginal empirical quantiles of the estimated betas.
Throughout Part \ref{part:classical}, we take $\mathcal{B}=[\lb,\ub]$ to be a rectangle with $\lb=(\lb_1,\ldots,\lb_d)^\top$, $\ub=(\ub_1,\ldots,\ub_d)^\top$, and $0$ in its interior.

\section{Assumptions}
\label{sec:classical-assumptions}

All notation from Part \ref{part:general} remains in force.

We employ the following simple notion of time-series dependence. For $r\geq1$, a sequence of random variables $\{Y_t\}_{t\in\mathbb{Z}}$ with finite $r$-th moments is $L_r$-near-epoch dependent on a sequence $\{\xi_t\}_{t\in\mathbb{Z}}$ with approximation constants $a_m$ if, for $\mathcal{H}_{t,m}=\sigma(\xi_{t-m},\ldots,\xi_{t+m})$,
\begin{align*}
    \|Y_t-\E[Y_t|\mathcal{H}_{t,m}]\|_r
    &\leq a_m,
\end{align*}
where $a_m\rightarrow0$ as $m\rightarrow\infty$.

\subsection{Primitive Conditions}

The following primitive conditions specialize the high-level conditions from Part \ref{part:general} to the classical beta-sorted estimator. They govern the data-generating process, the predictability and cross-sectional distribution of the time-varying coefficients, and the dependence and moment conditions for the factors and idiosyncratic errors. The precise definitions of the filtrations and limit sigma-fields appearing in the assumption are collected in Section~\ref{sec:classical-filtrations}.

\begin{assumption}[DGP][DGP]\label{ass:dgp}\leavevmode
    The following conditions hold:
    \begin{enumerate}[label=(\roman*)]
        \item \textbf{Smoothness}.
        $\mu_t$ admits a bounded extension to the neighborhood $\mathcal{B}^+$ of $\mathcal{B}$, and this extension is uniformly H\"older-$s_\mu$ on $\mathcal{B}^+$ for some $s_\mu>0$.

        \item \textbf{Idiosyncratic Error}.
        $\varepsilon_{it}=\varsigma_{it}\eta_{it}$, where
        (a) $\eta_{it}$ is i.i.d. across $i$ and over $t$ with mean $0$ and variance $1$, is independent of $\mathcal{E}_\infty$, and
        has a Lebesgue density $g_\eta$ with weak derivative $g^{(1)}_\eta\in L^1(\mathbb{R})$; moreover, there exists a moment exponent $q>2(3+d)$ such that $\E[|\eta_{it}|^q]<\infty$;
        (b) $\varsigma_{it}$ is $\mathcal{E}_t$-measurable, bounded, and bounded away from zero for all $i\in[n_t]$ and $t\in[T]$.

        \item \textbf{Varying Coefficients}.
        \begin{enumerate}[label=(\alph*)]
            \item \textbf{Predictability and Boundedness}.
            The coefficients $\alpha_{it}$ and $\beta_{it}$ are
            $\mathcal E_{t-1}$-measurable, $\beta_{it}\in\mathcal B$, and
            $|\alpha_{it}|+\|\beta_{it}\|$ is bounded a.s.\ by a deterministic
            finite bound uniformly over $i$, $t$, and $T$.

            \item \textbf{Independence}.
            Conditional on $\mathcal{D}_{t-1}$, the unit-specific predictable histories
            \begin{align*}
                \big((\alpha_{i1},\beta_{i1}^\top),\ldots,
                (\alpha_{it},\beta_{it}^\top)\big)
            \end{align*}
            are independent (but not necessarily identically distributed) over $i\in[n_t]$, and $\beta_{it}$ has a density $f_{\beta_{it}}(b|\mathcal{D}_{t-1})$ conditional on $\mathcal{D}_{t-1}$, which is bounded, bounded away from zero, and Lipschitz continuous on $\mathcal{B}$.
        \end{enumerate}

        \item \textbf{Factors}.
        $\E[f_t|\mathcal{F}_{t-1}]=\E[f_t|\mathcal{G}_{t-1}]$,
        $\lambda_{\min}(\E[\bar{f}_t\bar{f}_t^\top])\asymp 1$ with $\bar{f}_t=(1,f_t^\top)^\top$, for all $t\in[T]$, and ${\max_{1\leq t\leq T}}\,\E[\|\widetilde{f}_t\|^4|\mathcal{F}_{t-1}]\lesssim_\P 1$.
        There exists a deterministic positive-definite matrix
        $\Omega_f\in\mathbb R^{d\times d}$ with
        $0<\lambda_{\min}(\Omega_f)\leq
        \lambda_{\max}(\Omega_f)<\infty$ such that
        \begin{align}
            \left\|
            \frac{1}{T-H}\sum_{t=H+1}^{T}
            \Var(f_t\mid\mathcal{G}_{t-1})
            -\Omega_f
            \right\|
            &=o_\P(1).
            \label{eq:factor-variance-stabilization}
        \end{align}
        For the moment exponent $q$ in part~\textnormal{(ii)}, there exists a fixed $\upsilon>0$ such that, for each coordinate $k\in[d]$,
        $\sup_{t\in\mathbb{Z}}\E[|f_{t,k}|^{q+\upsilon}]<\infty$ and the process $f_{t,k}$ is $L_{q+\upsilon}$-near-epoch dependent on a strictly stationary strongly mixing process $\{\xi_t\}$ with mixing coefficients $\alpha(m)$ and approximation constants $a_m\downarrow 0$.
        Assume there exist constants $C_a,C_\alpha>0$ and $\rho_a,\rho_\alpha\in(0,1)$ such that
        \begin{align*}
            a_m \le C_a \rho_a^{\,m},
            \qquad
            \alpha(m) \le C_\alpha \rho_\alpha^{\,m},
            \qquad m\ge 1.
        \end{align*}
    \end{enumerate}
\end{assumption}

\begin{remark}[Primitive verification of factor variance stabilization]
Condition~\eqref{eq:factor-variance-stabilization} holds, for example, if
$\{\widetilde f_t\}$ is strictly stationary, its coordinates are
$L_q$-near-epoch dependent on the same strongly mixing process with the
exponential rates in Assumption~\ref{ass:dgp}(iv), and
$\Omega_f=\E[\widetilde f_0\widetilde f_0^\top]$ is positive definite.
Indeed, because $q>4$, the NED law of large numbers applies to
$\widetilde f_t\widetilde f_t^\top$. Moreover,
$\widetilde f_t\widetilde f_t^\top-
\E[\widetilde f_t\widetilde f_t^\top\mid\mathcal G_{t-1}]$ is a
matrix-valued martingale difference whose average is
$O_\P((T-H)^{-1/2})$ by the fourth-moment bound and martingale
orthogonality. Hence the average conditional variance converges to
$\Omega_f$ whenever $T-H\to\infty$.
\end{remark}

{
\begin{assumption}[SEP][Interior separation]\label{ass:interior-separation}\leavevmode
    Let $J=J_{nT}\to\infty$ denote the common order of the coordinatewise
    portfolio counts used to construct the second-step partition. Writing
    $b=(b_1,\ldots,b_d)^\top$, the interior evaluation region
    $\mathcal B^\circ$ introduced in Section~\ref{sec:general-setup} satisfies
    \begin{align*}
        \inf_{b\in\mathcal B^\circ}
        \min_{l\in[d]}
        \min\{|b_l-\lb_l|,|\ub_l-b_l|\}
        &\gg
        J^{-1}+\frac{\log(nT)}{\sqrt H}.
    \end{align*}
\end{assumption}
}

\subsection{Filtrations}\label{sec:classical-filtrations}

The idiosyncratic-error condition in Assumption \ref{ass:dgp} separates predictable model information from the i.i.d. innovations. In Part~\ref{part:classical} we use four nested sigma-fields, ordered from common factor information to full information:
\begin{align*}
    \mathcal{G}_t
    &=
    \sigma\!\left(
        \{f_s\}_{s=1}^{t},
        \{\aleph_{2s}\}_{s=1}^{t}
    \right),
    \\
    \mathcal{D}_t
    &=
    \sigma\!\left(
        \{f_s\}_{s=1}^{t},
        \{\{\varsigma_{is}\}_{i=1}^{n_s}\}_{s=1}^{t},
        \{\aleph_{2s}\}_{s=1}^{t}
    \right),
    \\
    \mathcal{E}_t
    &=
    \sigma\!\left(
        \{f_s\}_{s=1}^{t},
        \{\{\varsigma_{is}\}_{i=1}^{n_s}\}_{s=1}^{t},
        \{\aleph_{1i}\}_{i=1}^{n},
        \{\aleph_{2s}\}_{s=1}^{t},
        \{\{\aleph_{3is}\}_{i=1}^{n_s}\}_{s=1}^{t}
    \right),
    \\
    \mathcal{F}_t
    &=
    \sigma\!\left(
        \{f_s\}_{s=1}^{t},
        \{\{\varsigma_{is}\}_{i=1}^{n_s}\}_{s=1}^{t},
        \{\{\eta_{is}\}_{i=1}^{n_s}\}_{s=1}^{t},
        \{\aleph_{1i}\}_{i=1}^{n},
        \{\aleph_{2s}\}_{s=1}^{t},
        \{\{\aleph_{3is}\}_{i=1}^{n_s}\}_{s=1}^{t}
    \right).
\end{align*}
Thus
\begin{align*}
    \mathcal{G}_t
    \subseteq
    \mathcal{D}_t
    \subseteq
    \mathcal{E}_t
    \subseteq
    \mathcal{F}_t.
\end{align*}
Here $\mathcal{G}_t$ is the common information used to form conditional factor means and variances. The design field $\mathcal{D}_t$ adds realized volatility scales and is the conditioning set for reference portfolios, conditional beta densities, and reference cell objects. It excludes the unit-specific auxiliary variables $\aleph_{1i}$ and $\aleph_{3is}$, so after conditioning on $\mathcal{D}_{t-1}$ the predictable unit histories in Assumption~\ref{ass:dgp}(iii)(b) remain independent across $i$. The environment field $\mathcal{E}_t$ adds those unit-specific auxiliary variables and therefore contains the realized predictable paths $\alpha_{is}$ and $\beta_{is}$, but still excludes the i.i.d. innovations. The full field $\mathcal{F}_t$ adds the realized innovations $\eta_{is}$; under the bounded-away-from-zero condition on $\varsigma_{it}$, this is equivalently the field obtained by augmenting $\mathcal{E}_t$ with the idiosyncratic errors $\varepsilon_{is}$.

The corresponding limit sigma-fields are
\begin{align*}
    \mathcal{G}_\infty
    = \sigma\!\left(\cup_{t\ge 1}\mathcal{G}_t\right),
    \qquad
    \mathcal{D}_\infty
    = \sigma\!\left(\cup_{t\ge 1}\mathcal{D}_t\right),
    \qquad \mathcal{E}_\infty
    = \sigma\!\left(\cup_{t\ge 1}\mathcal{E}_t\right),
    \qquad \mathcal{F}_\infty
    = \sigma\!\left(\cup_{t\ge 1}\mathcal{F}_t\right).
\end{align*}

\begin{remark}[Interpretation of the auxiliary variables]
\label{rem:classical-auxiliary-variables}
The variables $\aleph_{1i}$, $\aleph_{2t}$, and $\aleph_{3it}$ are bookkeeping devices for features of the data-generating process not otherwise specified. Their inclusion in the sigma-fields above does not require that they be observed by the econometrician or used to construct the estimator. The variable $\aleph_{1i}$ collects time-invariant unit-specific information, such as a fixed industry classification or a permanent firm type. The variable $\aleph_{2t}$ collects time-varying common information not already summarized by $f_t$, such as an aggregate volatility or liquidity state. Finally, $\aleph_{3it}$ collects time-varying unit-specific information, such as a predetermined exogenous firm characteristic (e.g., firm age) or an unobserved firm-specific state governing the evolution of its risk factor exposure. Thus, $\beta_{it}$ may depend on $\aleph_{1i}$ and on the histories of $\aleph_{2s}$, $\aleph_{3is}$, and $f_s$ through date $t-1$, with $\alpha_{it}$ then restricted to satisfy $\alpha_{it}=\alpha_t(\beta_{it})$.

All of the examples remain subject to Assumption~\ref{ass:dgp}. In particular, Assumption~\ref{ass:dgp}(iii)(b) separately requires the resulting predictable coefficient histories to be independent, although not necessarily identically distributed, across assets conditional on $\mathcal{D}_{t-1}$. For example, fixed industry membership may be included in $\aleph_{1i}$, whereas a time-varying state shared by firms in an industry should be included as a component of the common vector $\aleph_{2t}$, and hence in the conditioning information. Moreover, because the auxiliary variables belong to $\mathcal E_\infty$, Assumption~\ref{ass:dgp}(ii)(a) requires the primitive idiosyncratic innovations $\eta_{it}$ to be independent of them.
\end{remark}

\subsection{Discussion}\label{sec:classical-conditioning-discussion}

The filtration notation in this part should be read as the primitive
information structure for the classical beta-sorted estimator. Although the
same letters $\mathcal{F}_t$ and $\mathcal{G}_t$ are used in Part
\ref{part:general}, the high-level fields in Part
\ref{part:general} and the primitive fields in this
part are local to their respective arguments. Part
\ref{part:general} uses a compact two-field notation to
state abstract measurability and mean-zero conditions. This part refines that
structure by separating common factor information, design information,
innovation-free environment information, and full information.

This distinction matters for verifying the high-level idiosyncratic-error
arguments under Assumption~\ref{ass:dgp}. The
contemporaneous scale $\varsigma_{it}$ is $\mathcal{E}_t$-measurable and need
not be measurable with respect to the past full field $\mathcal{F}_{t-1}$ as
defined in this part. Thus, if the high-level Part
\ref{part:general} arguments were applied literally
under the primitive assumptions in this part, the conditioning field used for
period-$t$ idiosyncratic cross-sectional arrays would have to fix the current
innovation-free scales and design objects while still excluding the current
primitive innovations. Equivalently, one may introduce the intermediate field
\begin{align*}
    \mathcal{C}_t
    &=
    \mathcal{E}_t
    \vee
    \sigma\!\left(
        \{\{\eta_{is}\}_{i=1}^{n_s}\}_{s=1}^{t-1}
    \right),
\end{align*}
which contains the past full information and the current innovation-free
environment, but not the current shocks $\{\eta_{it}:i\in[n_t]\}$. Conditional
on $\mathcal{C}_t$, the variables
$\varepsilon_{it}=\varsigma_{it}\eta_{it}$ are independent over $i$ and have
conditional mean zero. The first-step estimates, partitions, and reference
weights are measurable with respect to subfields of
$\mathcal{C}_t$ in the arguments where they are held fixed, and
martingale or mean-zero statements relative to coarser past information follow
by iterated expectations. The specialized proofs below use this conditioning
logic, sometimes through proof-specific stabilized design fields, without
changing the rates or substantive conclusions.

\section{First-Step Beta Estimates}\label{sec:classical-first-step}

Our main inference results focus on a local-polynomial least-squares estimator of $\beta_{it}$, which is commonly used in practice.
When the panel is unbalanced, the notation $i\in[n_t]$ in the rolling-window formulas below refers to units observed throughout the window ending at $t$, so that every lagged observation $R_{i(t-s)}$ with $s\in[H]$ is available.
Specifically, let $\mathfrak{p}\geq0$, let $K(\cdot)$ be a kernel function supported on $[0,1]$, and define
\begin{align*}
    r_{\mathfrak{p}}(u)
    &=(1,u,\ldots,u^{\mathfrak{p}})^\top,\qquad
    X_s(t)=
    r_{\mathfrak{p}}\left(s/H\right)\otimes (1, f_{t-s}^{\top})^\top\in\R^{(d+1)(\mathfrak{p}+1)}.
\end{align*}
Let $S=(e_2^\top, e_3^\top, \ldots, e_{d+1}^\top)^\top$, where each $e_j\in\R^{(\mathfrak{p}+1)(d+1)}$ is the $j$-th standard basis vector. Define
\begin{align}\label{eq:first-step-estimator}
    \widehat{\beta}_{it}
    &=S\,\widehat{\theta}_{it},\qquad
    \widehat{\theta}_{it}
    =\argmin_{\theta\in\R^{(d+1)(\mathfrak{p}+1)}}
    \sum_{s=1}^{H}\Big(R_{i(t-s)}-X_s(t)^\top\theta\Big)^2K\!\left(\frac{s}{H}\right).
\end{align}
In this case, we can write
\begin{align*}
    \widehat{\beta}_{it}
    &=\frac{1}{H}\sum_{s=1}^H w_s(t)R_{i(t-s)},
    \qquad
    w_s(t)=S\widehat{A}(t)^{-1}X_s(t)K\!\left(\frac{s}{H}\right),
    \qquad
    \widehat{A}(t)
    =\frac{1}{H}\sum_{s=1}^HK\!\left(\frac{s}{H}\right)X_s(t)X_s(t)^\top.
\end{align*}

\begin{assumption}[FS][First step: least squares]
    \label{ass:first-step-ls}\leavevmode
    $\widehat{\beta}_{it}$ is the least-squares estimator defined in \eqref{eq:first-step-estimator}, where $H\rightarrow\infty$ and $h=H/T=o(1)$.
    The kernel $K(\cdot):[0,1]\to\mathbb{R}_+$ is bounded and Lipschitz continuous, and
    $\int_0^1 r_{\mathfrak{p}}(s)r_{\mathfrak{p}}(s)^\top K(s)\,\d{s}$ is positive definite.
\end{assumption}

\subsection{Decomposition}

The estimation error admits the decomposition
\begin{align}\label{eq:first-step-decomposition}
    \widehat{\beta}_{it}-\beta_{it}
    =\big(\breve{\beta}_{it}-\beta_{it}\big)
    +\frac{1}{H}\sum_{s=1}^Hw_s(t)\varepsilon_{i(t-s)},
\end{align}
where
\begin{align*}
    \breve{\beta}_{it}
    =\frac{1}{H}\sum_{s=1}^Hw_s(t)(\alpha_{i(t-s)}+\beta_{i(t-s)}^\top f_{t-s}).
\end{align*}
The term $\breve{\beta}_{it}-\beta_{it}$ is the first-step bias component, while the second term is the first-step variance component. To characterize the bias component, define
\begin{align*}
    \widehat{\mathsf{B}}_{\mathtt{FS}}
    &=\max_{H+1\le t\le T}\max_{i\in[n_t]}
    \|\breve{\beta}_{it}-\beta_{it}\|.
\end{align*}
{
\begin{assumption}[FSB][First step: bias]\label{ass:first-step-bias}\leavevmode
    There exists a deterministic sequence
    $\mathsf{B}_{\mathtt{FS}}>0$ such that
    $\mathsf{B}_{\mathtt{FS}}=o(H^{-1/2})$ and
    \begin{align*}
        H^{(d-1)/2}
        \max_{H+1\leq t\leq T}
        \P\!\left(
            \widehat{\mathsf{B}}_{\mathtt{FS}}
            >\mathsf{B}_{\mathtt{FS}}
            \,\middle|\,\mathcal D_{t-1}
        \right)
        &\lesssim_\P
        \mathsf{B}_{\mathtt{FS}}+H^{-1}.
    \end{align*}
\end{assumption}

The uniform rate is summarized by
\begin{align*}
    \mathsf{R}_{\mathtt{FS}}
    &=\sqrt{\frac{\log(nT)}{H}}
    +\frac{(nT)^{1/q}\log(nT)}{H}
    +\mathsf{B}_{\mathtt{FS}},\qquad
    h=\frac{H}{T}.
\end{align*}
Assumption~\ref{ass:first-step-bias} implies
\begin{align*}
    \widehat{\mathsf{B}}_{\mathtt{FS}}
    &\leq\mathsf{B}_{\mathtt{FS}}
    \qquad\text{w.p.a.1}.
\end{align*}
Consequently, $\mathsf{R}_{\mathtt{FS}}$ controls the uniform first-step error. If
$(nT)^{2/q}\log(nT)/H\to0$, the middle term is dominated by
$\sqrt{\log(nT)/H}$. Lemma~\ref{lem:first-step-bias} gives primitive
bounds for $\mathsf{B}_{\mathtt{FS}}$ under several smoothness models.

\begin{remark}[Stronger and weaker first-step bias conditions]
    A stronger sufficient condition for Assumption~\ref{ass:first-step-bias} is $\widehat{\mathsf{B}}_{\mathtt{FS}} \leq \mathsf{B}_{\mathtt{FS}}$ a.s.
    for all sufficiently large $n,T$.
    In addition, inspection of our proofs shows that the weaker conditions $\widehat{\mathsf{B}}_{\mathtt{FS}}\lesssim_\P\mathsf{B}_{\mathtt{FS}}$ and
    $H^{(d-1)/2}
        \max_{H+1\leq t\leq T}
        \max_{i\in[n_t]}
        \P\!\big(
            \|\breve\beta_{it}-\beta_{it}\|
            >\mathsf{B}_{\mathtt{FS}}
            \,|\,\mathcal D_{t-1}
        \big)
        \lesssim_\P \mathsf{B}_{\mathtt{FS}}+H^{-1}
    $ are sufficient. We impose the stronger condition in Assumption~\ref{ass:first-step-bias} for clarity of presentation.
\end{remark}

}

\subsection{Results}

The first result verifies the uniform concentration of the local-polynomial Gram matrix. This regularity condition is used below to control the inverse Gram matrix uniformly over the rolling windows.

\begin{lemma}[First-Step Gram Matrix]\label{lem:first-step-gram}
    Suppose that Assumptions~\ref{ass:dgp}(iv)
    and~\ref{ass:first-step-ls} hold.
    Then,
    \begin{align*}
        &\max_{H+1\le t\le T}
        \big\|\widehat{A}(t)-\E[\widehat{A}(t)]\big\|
        \;\lesssim_\P\;
        \frac{T^{2/q}}{H}+\sqrt{\frac{\log T}{H}}.
    \end{align*}
\end{lemma}

Together with Assumptions~\ref{ass:dgp}(iv)
and~\ref{ass:first-step-ls}, this implies that, if
$T^{2/q}/H+\sqrt{\log T/H}=o(1)$, then
$\sup_{H+1\le t\le T}\|\widehat{A}(t)^{-1}\|\lesssim_\P 1$.
The lower bound follows from the uniform positive definiteness of
$\E[\bar f_t\bar f_t^\top]$ and the positive definiteness of the kernel-weighted
polynomial Gram matrix in Assumption~\ref{ass:first-step-ls}.

The next result gives the uniform first-step estimation error rate $\mathsf{R}_{\mathtt{FS}}$ in the least-squares case, which provides the high-level bound
$\widehat{\mathsf{R}}_{\mathtt{FS}}\lesssim_\P\mathsf{R}_{\mathtt{FS}}$ for Theorem \ref{thm:general-consistency}.

\begin{lemma}[First-Step Convergence Rate]\label{lem:first-step-rate}
    Suppose Assumptions~
    \ref{ass:dgp}(ii) and (iv)
    and~\ref{ass:first-step-ls} and
    {\ref{ass:first-step-bias}} hold, and that
    $T^{2/q}/H+\sqrt{\log T/H}=o(1)$.
    Then,
    \begin{align*}
        \widehat{\mathsf{R}}_{\mathtt{FS}}
        =\max_{H+1\le t\le T}\max_{i\in[n_t]}
        \|\widehat{\beta}_{it}-\beta_{it}\|
        \lesssim_\P
        \sqrt{\frac{\log(nT)}{H}}
        +\frac{(nT)^{1/q}\log (nT)}{H}
        +{\mathsf{B}_{\mathtt{FS}}}
        = \mathsf{R}_{\mathtt{FS}}.
    \end{align*}
\end{lemma}

Assumption~\ref{ass:dgp} does not impose temporal smoothness on
the realized coefficient paths. The next result gives primitive bounds on the
deterministic threshold $\mathsf{B}_{\mathtt{FS}}$ under several smoothness
models.

{
\begin{lemma}[First-Step Bias Threshold]
    \label{lem:first-step-bias}
    \leavevmode\par\noindent
    Suppose Assumptions~\ref{ass:dgp}(iv)
    and~\ref{ass:first-step-ls} hold, and
    \begin{align*}
        \frac{T^{2/(q-d-1)}}{H}
        &=o(1).
    \end{align*}

    \begin{enumerate}[label=\textnormal{(\roman*)}]

    \item \textbf{Nonrandom $\beta_i(\cdot)$: H\"{o}lder-$s$.}
    Suppose that
    $\alpha_{it}=\alpha_i(t/T)$ and $\beta_{it}=\beta_i(t/T)$ for all
    $i$ and $t$, where $\alpha_i(\cdot)$ and $\beta_i(\cdot)$ are nonrandom,
    bounded, and H\"{o}lder-$s$ on $[0,1]$ uniformly in $i\in[n]$ for some
    $s>0$.
    If
    \begin{align*}
        \sqrt H\,h^{\min\{s,\mathfrak p+1\}}
        &\to0,
    \end{align*}
    then Assumption~\ref{ass:first-step-bias} holds with
    \begin{align*}
        \mathsf{B}_{\mathtt{FS}}
        &\lesssim h^{\min\{s,\mathfrak{p}+1\}}.
    \end{align*}

    \item \textbf{Stochastic $\beta_i(\cdot)$: Brownian Motion.}
    Suppose that
    $\beta_{it}=\beta_i^0+\sigma_i W_i(t/T)$ for all $i$ and $t$, where
    $\beta_i^0\in\mathbb{R}^d$ is nonrandom,
    $W_i(\cdot)$ is a $d$-dimensional standard Brownian motion on $[0,1]$,
    independent across $i$, and
    $0<\underline\sigma\le\sigma_i\le\overline\sigma<\infty$
    uniformly in $i$. Suppose also that the conditional intercept function
    satisfies
    \begin{align*}
        \alpha_t(b)=\alpha(t/T,b),\qquad b\in\mathcal B,
    \end{align*}
    where the possibly random map $\alpha(\cdot,\cdot)$ is bounded,
    H\"older-$s$ in its first argument uniformly over its second argument
    for some $s>1/2$, and Lipschitz in its second argument uniformly over
    its first argument, a.s.\ with deterministic uniform bounds.
    If
    \begin{align*}
        \sqrt{Hh\log(nT)}
        &\to0,
    \end{align*}
    then Assumption~\ref{ass:first-step-bias} holds with
    \begin{align*}
        \mathsf{B}_{\mathtt{FS}}
        &\lesssim h^{1/2}\bigl(\log(nT)\bigr)^{1/2}.
    \end{align*}

    \item \textbf{Stochastic $\beta_i(\cdot)$: Nonrandom Trend Plus Brownian Motion.}
    Suppose that the conditional intercept function satisfies
    \begin{align*}
        \alpha_t(b)=\alpha(t/T,b),\qquad b\in\mathcal B,
    \end{align*}
    where the possibly random map $\alpha(\cdot,\cdot)$ is bounded,
    H\"{o}lder-$s$ in its first argument uniformly over its second argument,
    and Lipschitz in its second argument uniformly over its first argument,
    a.s.\ with deterministic uniform bounds. Suppose also that
    \begin{align*}
        \beta_{it}
        = \bar\beta_i(t/T)+\sigma_i W_i(t/T),
    \end{align*}
    where $\bar\beta_i(\cdot)$ is nonrandom and H\"{o}lder-$s$
    as in case~(i) with $s\in(0,\mathfrak{p}+1]$,
    $W_i(\cdot)$ is a $d$-dimensional standard Brownian motion on $[0,1]$,
    independent across $i$, and
    the scale condition in case~(ii) holds.
    If
    \begin{align*}
        \sqrt H\,
        \max\left\{h^s,h^{1/2}\bigl(\log(nT)\bigr)^{1/2}\right\}
        &\to0,
    \end{align*}
    then Assumption~\ref{ass:first-step-bias} holds with
    \begin{align*}
        \mathsf{B}_{\mathtt{FS}}
        &\lesssim
        \max\left\{h^s,h^{1/2}\bigl(\log(nT)\bigr)^{1/2}\right\}.
    \end{align*}

    \item \textbf{Stochastic $\beta_i(\cdot)$: Integrated Brownian Motion.}
    Suppose that
    \begin{align*}
        \beta_{it}
        &=
        \beta_i^0
        +\sigma_i\int_0^{t/T}W_i(v)\,\d v,
    \end{align*}
    where $\beta_i^0\in\mathbb R^d$ is nonrandom,
    $W_i(\cdot)$ is a $d$-dimensional standard Brownian motion on $[0,1]$,
    independent across $i$, and the scale condition in case~(ii) holds.
    Suppose also that
    \begin{align*}
        \alpha_t(b)
        &=
        \alpha(t/T,b),\qquad b\in\mathcal B,
    \end{align*}
    where the possibly random map $\alpha(\cdot,\cdot)$ is twice continuously
    differentiable jointly in its arguments, a.s., and the function and all
    its derivatives through order two are uniformly bounded by deterministic
    constants. If $h\log(nT)=o(1)$ and
    \begin{align*}
        \sqrt H\,
        h^{\min\{\mathfrak p+1,\,3/2\}}
        \bigl(\log(nT)\bigr)^{1/2}
        &\to0,
    \end{align*}
    then Assumption~\ref{ass:first-step-bias} holds with
    \begin{align*}
        \mathsf{B}_{\mathtt{FS}}
        &\lesssim
        h^{\min\{\mathfrak p+1,\,3/2\}}
        \bigl(\log(nT)\bigr)^{1/2}.
    \end{align*}
    \end{enumerate}
\end{lemma}
}

\subsection{Discussion}

\begin{enumerate}[label=(\roman*)]
    \item \emph{First-step rate under smooth trends.}
    Lemma~\ref{lem:first-step-bias} translates the path restrictions
    stated directly in that lemma into primitive bounds on the deterministic
    threshold $\mathsf{B}_{\mathtt{FS}}$. If
    $(nT)^{2/q}\log(nT)/H\to 0$, then the term $(nT)^{1/q}\log(nT)/H$ in
    $\mathsf{R}_{\mathtt{FS}}$ is dominated by $\sqrt{\log(nT)/H}$, and
    Lemma~\ref{lem:first-step-rate} gives
    \begin{align*}
        \mathsf{R}_{\mathtt{FS}}
        &\lesssim_\P
        \sqrt{\frac{\log(nT)}{H}}
        +\mathsf{B}_{\mathtt{FS}}.
    \end{align*}
    In the nonrandom H\"older-$s$ case, writing $h=H/T$,
    Lemma~\ref{lem:first-step-bias}(i) gives
    $\mathsf{B}_{\mathtt{FS}}\lesssim
    h^{\min\{s,\mathfrak p+1\}}$.
    Thus the local-polynomial order reduces the first-step bias only up to the
    smoothness ceiling $s$.

    In the integrated-Brownian case~(iv), the analogous distinction is
    \begin{align*}
        \mathsf{B}_{\mathtt{FS}}
        &\lesssim
        \begin{cases}
            h\bigl(\log(nT)\bigr)^{1/2},
            &\mathfrak p=0,\\[0.3em]
            h^{3/2}\bigl(\log(nT)\bigr)^{1/2},
            &\mathfrak p\geq1.
        \end{cases}
    \end{align*}
    A degree-one fit absorbs the current random slope
    $\sigma_iW_i(t/T)$ of the integrated path. Higher polynomial orders do not
    further improve this component because that slope itself has Brownian
    local variation. In particular, the displayed primitive
    $\mathsf{B}_{\mathtt{FS}}$ bound is $o(H^{-1/2})$ whenever
    \begin{align*}
        \sqrt H\,
        h^{\min\{\mathfrak p+1,\,3/2\}}
        \bigl(\log(nT)\bigr)^{1/2}
        &\to0.
    \end{align*}
    {
    The restriction $T^{2/(q-d-1)}/H=o(1)$ in
    Lemma~\ref{lem:first-step-bias} is stronger than the concentration rate
    $T^{2/q}/H=o(1)$ used for the first-step Gram matrix because the lemma
    verifies the conditional-tail requirement in
    Assumption~\ref{ass:first-step-bias} through the stronger
    $H^{(d+1)/2}$-weighted global-tail bound. It is implied by
    $T^{(3+d)/q}/H=o(1)$, the rate used in the partition results, because
    $q>2(3+d)$ implies $2/(q-d-1)<(3+d)/q$.
    }

    \item \emph{Conditional intercepts in the Brownian cases.}
    The conditions in cases~(ii)--(iii) are imposed on the conditional
    intercept function $\alpha_t(b)=\mu_t(b)-b^\top\E[f_t\mid\mathcal
    G_{t-1}]$, rather than on a separate nonrandom path for $\alpha_{it}$. Because
    $\alpha_{it}=\alpha_t(\beta_{it})$, temporal smoothness of
    $\alpha_t(b)$ at fixed $b$ controls variation in the conditional
    intercept function, while Lipschitz continuity in $b$ transfers the
    local modulus of $\beta_{it}$ to $\alpha_{it}$.
    In case~(iv), the stronger joint twice-continuous differentiability
    condition permits a first-order expansion of
    $\tau\mapsto\alpha(\tau,\beta_i(\tau))$. The local polynomial can therefore
    absorb both its time derivative and the contribution of the current random
    beta slope, leaving a remainder of integrated-Brownian order.

    \item \emph{Compact support in the Brownian cases.}
    Cases~(ii)--(iii) use the Brownian component only through its local modulus of
    continuity. The proof requires
    \begin{align*}
        \max_{i\in[n]}\sup_{|u-v|\le h}\bigl\|W_i(u)-W_i(v)\bigr\|
        \;\lesssim_\P\; h^{1/2}\bigl(\log(nT)\bigr)^{1/2}
    \end{align*}
    and does not otherwise use the global level of $W_i(\cdot)$. Since
    Assumption~\ref{ass:dgp}(iii)(a) imposes a fixed
    compact support $\mathcal{B}$ on $\beta_{it}$, the Brownian specifications
    should be read as a convenient shorthand for any compact-state process with this
    same local modulus, such as reflected Brownian motion on $\mathcal{B}$ or a
    diffusion on a compact state space. For such a compact-state process,
    the threshold bound
    $\mathsf{B}_{\mathtt{FS}}\lesssim h^{1/2}(\log(nT))^{1/2}$ is unchanged,
    provided the remaining requirements of Assumption~\ref{ass:dgp}(iii) are maintained.

    The same interpretation applies to case~(iv). There the proof uses
    \begin{align*}
        \max_{i\in[n]}
        \sup_{|u-v|\leq h}
        \left\|
            \int_v^u\bigl(W_i(z)-W_i(v)\bigr)\,\d z
        \right\|
        &\lesssim_\P
        h^{3/2}\bigl(\log(nT)\bigr)^{1/2}.
    \end{align*}
    Thus the displayed integrated-Brownian specification may be replaced by a
    compact-state process with continuously differentiable paths whose
    derivative has the same Brownian local modulus; the rates in case~(iv)
    are unchanged.

\end{enumerate}

\subsection{Proof of Lemma~\ref{lem:first-step-gram}}

Recall $X_s(t)=r_{\mathfrak{p}}(s/H)\otimes(1,f_{t-s}^\top)^\top$ and
$K_{\max}=\sup_{u\in[0,1]}|K(u)|<\infty$.

Define
\begin{align*}
    V_{s,t}
    =\frac{1}{H}X_s(t)X_s(t)^\top
    K\!\left(\frac{s}{H}\right),
\end{align*}
so that $\widehat{A}(t)=\sum_{s=1}^H V_{s,t}$.

Put $r_*=(q+\upsilon)/2>q/2>2$.

\medskip\noindent\textit{Step 1: NED property of $V_{s,t}$.}
Since $X_s(t)=r_{\mathfrak{p}}(s/H)\otimes(1,f_{t-s}^\top)^\top$ is a
function of $f_{t-s}$, Assumption~\ref{ass:dgp}(iv) implies that
$X_s(t)X_s(t)^\top$ is $L_{r_*}$-NED on $\{\xi_t\}$ with geometric
approximation constants. To see this, let
$\mathcal{H}_{t-s,k}=\sigma(\xi_{t-s-k},\ldots,\xi_{t-s+k})$,
$V_{s,t}^{(k)}=\E[V_{s,t}\mid\mathcal{H}_{t-s,k}]$, and
$X_s^{(k)}(t)=\E[X_s(t)\mid\mathcal{H}_{t-s,k}]$. The contraction property
of conditional expectation and the product inequality give
\begin{align*}
    \|V_{s,t}-V_{s,t}^{(k)}\|_{r_*}
    &\leq \frac{2K_{\max}}{H}
    \bigl(\|X_s(t)\|_{q+\upsilon}
    +\|X_s^{(k)}(t)\|_{q+\upsilon}\bigr)
    \|X_s(t)-X_s^{(k)}(t)\|_{q+\upsilon}
    \lesssim H^{-1}\rho_a^k.
\end{align*}

\medskip\noindent\textit{Step 2: Moment bound.}
Uniformly in $s$ and $t$,
$\E[\|V_{s,t}\|^{r_*}]\lesssim H^{-r_*}\E[\|X_s(t)\|^{q+\upsilon}]
\lesssim H^{-r_*}$.

\medskip\noindent\textit{Step 3: Rolling-sum bound.}
Let $Y_u$ be any centered scalar
entry of $(1,f_u^\top)^\top(1,f_u^\top)$. It has a uniformly bounded
$r_*$-th moment and is geometrically $L_{r_*}$-NED. Set
$m=C_0\log T$ and $Y_u^{(m)}=\E[Y_u\mid\mathcal H_{u,m}]$. Because $Y_u$
is centered, so is $Y_u^{(m)}$, and conditional-expectation contraction gives
$\|Y_u^{(m)}\|_{r_*}\leq\|Y_u\|_{r_*}\lesssim1$. Moreover,
$\{Y_u^{(m)}\}$ is strongly mixing: observations separated by more than
$2m$ have mixing coefficient at most $\alpha(j-2m)$. Therefore, at probability
level $v$, its inverse mixing coefficient is at most $2m+C\log(e/v)$, while
Markov's inequality bounds the quantile of $|Y_u^{(m)}|$ by
$Cv^{-1/r_*}$. For sufficiently large $x$, these two bounds give
\begin{align*}
    &\int_{\{0<v<1:\,C\{m+\log(e/v)\}v^{-1/r_*}>x\}}
    Cv^{-1/r_*}\,\d v
    \;\lesssim
    x^{-(r_*-1)}\{m+\log(e+x)\}^{r_*-1}.
\end{align*}
This verifies the tail-integral condition in
\citet[Theorem~6.2]{rio2017}.

For two entries $Y_u$ and $Y_{u+j}$, approximate each one using
$\mathcal H_{u,\lfloor j/3\rfloor}$ and
$\mathcal H_{u+j,\lfloor j/3\rfloor}$. The covariance inequality in
\citet[Theorem~1.1(a)]{rio2017} and the geometric NED bound give, for some
$\rho\in(0,1)$,
\begin{align*}
    |\Cov(Y_u,Y_{u+j})|
    &\lesssim
    \rho_a^{\lfloor j/3\rfloor}
    +\alpha(\lfloor j/3\rfloor)^{1-2/r_*}
    \lesssim \rho^j.
\end{align*}
Thus the sum of the absolute covariances is uniformly bounded. Since
$\|Y_u-Y_u^{(m)}\|_2\lesssim\rho_a^m$, on every interval of length at most
$2H$ the sum of the absolute covariances of $Y_u^{(m)}$ is at most
$CH+CH^2\rho_a^m\lesssim CH$ when $C_0$ is sufficiently large.

The variables $Y_u^{(m)}$ have now been verified to be centered and strongly
mixing, with the moment, tail-integral, and covariance bounds required by
\citet[Theorem~6.2]{rio2017}. Apply that theorem on consecutive blocks of
length $H$. An interval crossing a block boundary splits into at most two
partial sums, and there are $O(T/H)$ blocks. Also, for any fixed $A>0$, $C_0$
can be chosen so that, uniformly for $\lambda\geq1$,
\begin{align*}
    \P\!\left(
        \sum_{u=1}^T|Y_u-Y_u^{(m)}|>\lambda
    \right)
    &\leq
    \lambda^{-1}\sum_{u=1}^T\E|Y_u-Y_u^{(m)}|
    \leq \frac{CT\rho_a^m}{\lambda}
    =o(T^{-A}).
\end{align*}

Choose $C_0$ large enough to absorb this approximation error. Consequently,
for any $a\geq1$ and $1\leq\lambda\leq T$ with $\lambda/a$ sufficiently
large, we have the following bound, where the maximum is over time intervals
$I$:
\begin{equation}\label{eq:first-step-local-maximal}
    \P\!\left(\max_{|I|\leq H}\left|\sum_{u\in I}Y_u\right|>C\lambda\right)
    \lesssim \frac{T}{H}\left(1+\frac{\lambda^2}{aH}\right)^{-a/2}
    +Ta^{r_*-1}\lambda^{-r_*}(\log T)^{r_*-1}.
\end{equation}
Indeed, the tail-integral term on one block is the preceding integral
multiplied by $H/\lambda$ and evaluated at $x=\lambda/a$.

For each matrix entry, the deterministic weights formed from
$r_{\mathfrak p}(s/H)$ and $K(s/H)$ are uniformly bounded and have bounded
total change over $s=1,\ldots,H$. Expanding a weighted sum into partial sums
and successive changes in these weights therefore bounds it by a constant
times the largest unweighted partial sum in
\eqref{eq:first-step-local-maximal}. Take
$a=M\log T$ and
$\lambda=M\{T^{2/q}+\sqrt{H\log T}\}$. The first term in
\eqref{eq:first-step-local-maximal} tends to zero for sufficiently large
$M$. For the second term, the moment gap gives
\begin{align*}
    T\lambda^{-r_*}(\log T)^{2(r_*-1)}
    &\lesssim T^{1-2r_*/q}(\log T)^{2(r_*-1)}
    =T^{-\upsilon/q}(\log T)^{2(r_*-1)}=o(1).
\end{align*}
A finite union over the matrix entries now yields
\begin{align*}
    \max_{H+1\le t\le T}
    \left\|\widehat{A}(t)-\E[\widehat{A}(t)]\right\|
    &\lesssim_\P
    \frac{T^{2/q}}{H}+\sqrt{\frac{\log T}{H}}.
\end{align*}
The covariance bound above also gives
$(T-H)^{-1}\sum_{t=H+1}^T
\|\widehat A(t)-\E[\widehat A(t)]\|^2=O_\P(H^{-1})$ by taking expectations
and applying Markov's inequality. The same rolling-sum argument applies to
the fixed powers of $\|f_t\|$ used below because their available moment is
strictly above the working moment used in the corresponding rate.
\qed

\subsection{Proof of Lemma~\ref{lem:first-step-rate}}

Define
\begin{align*}
    N_i(t)
    =\frac{1}{H}\sum_{s=1}^H
    X_s(t)R_{i(t-s)}K\!\left(\frac{s}{H}\right),
    \qquad\text{and}\qquad
    U_{it}
    =\frac{1}{H}\sum_{s=1}^H
    X_s(t)\varepsilon_{i(t-s)}K\!\left(\frac{s}{H}\right).
\end{align*}
The normal equations give $\widehat{\theta}_{it}=\widehat{A}(t)^{-1}N_i(t)$. Since
\begin{align*}
    N_i(t)
    &=\frac{1}{H}\sum_{s=1}^H
    X_s(t)\big(\alpha_{i(t-s)}+\beta_{i(t-s)}^\top f_{t-s}\big)K\!\left(\frac{s}{H}\right)
    +U_{it},
\end{align*}
the definition of $\breve{\beta}_{it}$ implies the exact identity
\begin{align*}
    \widehat{\beta}_{it}-\breve{\beta}_{it}
    &=S\widehat{A}(t)^{-1}U_{it}.
\end{align*}
\medskip\noindent\textit{Step 1: Bounding $\widehat{A}(t)^{-1}$.}
By Assumptions~\ref{ass:dgp}(iv)
and~\ref{ass:first-step-ls},
$\lambda_{\min}(\E[\widehat{A}(t)])\ge c>0$ uniformly in
$t=H+1,\ldots,T$. Indeed, writing
$\bar f_r=(1,f_r^\top)^\top$, any vector $a$ conformable with $X_s(t)$ can be
written as polynomial coefficients $a_0,\ldots,a_{\mathfrak{p}}\in\mathbb{R}^{d+1}$, so that
$(X_s(t)^\top a)^2=(\bar f_{t-s}^\top a(s/H))^2$, where
$a(u)=\sum_{k=0}^{\mathfrak{p}}a_k u^k$. The uniform lower bound on
$\E[\bar f_r\bar f_r^\top]$ gives a constant multiple of
$H^{-1}\sum_{s=1}^H K(s/H)\|a(s/H)\|^2$, and this discrete polynomial Gram
matrix is bounded away from zero for all large $H$ by the positive definite
integral in Assumption~\ref{ass:first-step-ls}.
By Lemma~\ref{lem:first-step-gram},
$\|\widehat{A}(t)-\E[\widehat{A}(t)]\|\lesssim_\P T^{2/q}/H+\sqrt{\log T/H}=o_\P(1)$
uniformly in $t$. Weyl's inequality then gives
\begin{align*}
    \lambda_{\min}(\widehat{A}(t))
    \ge \lambda_{\min}(\E[\widehat{A}(t)])-\|\widehat{A}(t)-\E[\widehat{A}(t)]\| \ge c-o_\P(1)>0
\end{align*}
uniformly over $t\in[H+1,T]$ with probability approaching one. Hence
\begin{align}\label{eq:first-step-inverse-bound}
    \sup_{H+1\le t\le T}\|\widehat{A}(t)^{-1}\|
    &\lesssim_\P 1.
\end{align}
\medskip\noindent\textit{Step 2: Bounding $U_{it}$.}
Condition on $\mathcal{E}_\infty$. Then the factor history and volatility
scales are fixed, and
$\varepsilon_{i(t-s)}=\varsigma_{i(t-s)}\eta_{i(t-s)}$ with independent
mean-zero innovations. Hence, for fixed $(i,t_0)$, the array
$\{X_s(t_0)\varepsilon_{i(t_0-s)}:s\in[H]\}$ is a martingale difference array
in the index $s$ with respect to the filtration
$\mathcal{H}_{s,it_0}=\mathcal{E}_\infty
\vee\sigma(\eta_{i(t_0-r)}:1\le r\le s)$.
The probability bounds below can therefore be read conditionally on
$\mathcal{E}_\infty$ and then integrated.
Abel's summation formula gives, for any $t_0\in[H+1,T]$,
\begin{align*}
    &\left\|\sum_{s=1}^H K\!\left(\frac{s}{H}\right)
    X_s(t_0)\varepsilon_{i(t_0-s)}\right\|\\
    &\quad\le
    \max_{\ell\in[H]}
    \left\|\sum_{s=1}^\ell X_s(t_0)\varepsilon_{i(t_0-s)}\right\|
    \sum_{s=1}^{H-1}
    \left|K\!\left(\frac{s+1}{H}\right)-K\!\left(\frac{s}{H}\right)\right|
    +\left\|\sum_{s=1}^H X_s(t_0)\varepsilon_{i(t_0-s)}\right\|
    K(1).
\end{align*}
Since $K$ is Lipschitz continuous,
$\sum_{s=1}^{H-1}|K(\frac{s+1}{H})-K(\frac{s}{H})|\le L_K$
for a finite constant $L_K$, and $K(1)\le\|K\|_\infty<\infty$.
Therefore,
\begin{align*}
    &\max_{H+1\le t_0\le T}\max_{i\in[n_{t_0}]}
    \left\|U_{it_0}\right\|
    \lesssim
    \frac{1}{H}
    \max_{H+1\le t_0\le T}\max_{i\in[n_{t_0}]}
    \max_{\ell\in[H]}
    \left\|\sum_{s=1}^\ell X_s(t_0)\varepsilon_{i(t_0-s)}\right\|.
\end{align*}

Fix $t_0$ and $i$, put $\tau=(nT)^{1/q}$, and write
\begin{align*}
    Y_{s,it_0}
    &=X_s(t_0)\varepsilon_{i(t_0-s)},
    &s&\in[H].
\end{align*}
For a fixed constant $M>0$, define the conditionally recentered truncated
increment
\begin{align*}
    Y_{s,it_0}^{(M)}
    &=
    Y_{s,it_0}\IF\!\left(\|Y_{s,it_0}\|\leq M\tau\right)
    -\E\!\left[
        Y_{s,it_0}\IF\!\left(\|Y_{s,it_0}\|\leq M\tau\right)
        \,\middle|\,\mathcal H_{s-1,it_0}
    \right].
\end{align*}
By construction,
\begin{align*}
    \E\!\left[
        Y_{s,it_0}^{(M)}
        \,\middle|\,\mathcal H_{s-1,it_0}
    \right]
    &=0,
    &
    \|Y_{s,it_0}^{(M)}\|
    &\leq 2M\tau.
\end{align*}
Thus the truncated array remains a martingale-difference array with respect to
the original filtration; no conditioning on a future-shock event is involved.

Its predictable quadratic variation satisfies, conditional on
$\mathcal E_\infty$,
\begin{align*}
    \sum_{s=1}^{H}
    \E\!\left[
        \|Y_{s,it_0}^{(M)}\|^2
        \,\middle|\,\mathcal H_{s-1,it_0}
    \right]
    &\lesssim
    \sum_{s=1}^{H}\|X_s(t_0)\|^2,
\end{align*}
because the volatility scales are bounded and $\E[\eta_{it}^2]=1$.
Moreover,
\begin{align*}
    \max_{H+1\leq t\leq T}
    \frac{1}{H}\sum_{s=1}^{H}\|X_s(t)\|^2
    &\lesssim_\P1,
\end{align*}
by applying the same maximal Fuk--Nagaev inequality used in
Lemma~\ref{lem:first-step-gram} to the $L_{q/2}$-NED sequence
$\|\bar f_t\|^2$. Hence, on an $\mathcal E_\infty$-measurable event whose
probability can be made arbitrarily close to one, the coordinatewise maximal
martingale Bernstein inequality \citep{freedman1975tail} gives
\begin{align}
    \P\!\left(
        \max_{\ell\in[H]}
        \left\|
            \sum_{s=1}^{\ell}Y_{s,it_0}^{(M)}
        \right\|>x
        \,\middle|\,\mathcal E_\infty
    \right)
    &\lesssim
    \exp\!\left\{
        -\frac{c x^2}{H+M\tau x}
    \right\}.
    \label{eq:first-step-recentered-truncation}
\end{align}
The dimension of $Y_{s,it_0}$ is fixed, so the coordinatewise inequality and
a finite union bound imply the displayed norm bound.

It remains to transfer this inequality to the original increments. Let
\begin{align*}
    \mathcal A_M
    &=
    \left\{
        \max_{H+1\leq t_0\leq T}
        \max_{i\in[n_{t_0}]}
        \max_{s\in[H]}
        \|Y_{s,it_0}\|
        \leq M\tau
    \right\}.
\end{align*}
Since $r_{\mathfrak p}(\cdot)$ is bounded, the maximum over rolling-window
summands is controlled by the maximum over the $nT$ calendar shock pairs.
Markov's inequality, the finite $q$-th moments of
$(1+\|f_r\|)\varepsilon_{ir}$, and a union bound give
\begin{align}
    \P(\mathcal A_M^c)
    &\lesssim M^{-q}.
    \label{eq:first-step-global-tail-probability}
\end{align}
Because $\E[Y_{s,it_0}\mid\mathcal H_{s-1,it_0}]=0$, on $\mathcal A_M$,
\begin{align*}
    \sum_{s=1}^{\ell}Y_{s,it_0}
    &=
    \sum_{s=1}^{\ell}Y_{s,it_0}^{(M)}
    -\sum_{s=1}^{\ell}
    \E\!\left[
        Y_{s,it_0}
        \IF\!\left(\|Y_{s,it_0}\|>M\tau\right)
        \,\middle|\,\mathcal H_{s-1,it_0}
    \right],
    &\ell&\in[H].
\end{align*}
The recentering term is negligible at the target scale. Indeed, uniformly over
$(i,t_0)$, conditional Markov's inequality and the bounded volatility scales
give
\begin{align*}
    \sum_{s=1}^{H}
    \left\|
    \E\!\left[
        Y_{s,it_0}
        \IF\!\left(\|Y_{s,it_0}\|>M\tau\right)
        \,\middle|\,\mathcal H_{s-1,it_0}
    \right]
    \right\|
    &\leq
    (M\tau)^{1-q}
    \sum_{s=1}^{H}
    \E\!\left[
        \|Y_{s,it_0}\|^q
        \,\middle|\,\mathcal E_\infty
    \right]\\
    &\lesssim_\P
    (M\tau)^{1-q}
    \sum_{r=1}^{T}(1+\|f_r\|)^q
    \lesssim_\P
    M^{1-q}\frac{\tau}{n}
    \lesssim_\P
    M^{1-q}\tau,
\end{align*}
where $\tau^q=nT$ and
$T^{-1}\sum_{r=1}^{T}(1+\|f_r\|)^q=O_\P(1)$ by the uniform $q$-moment
bound in Assumption~\ref{ass:dgp}(iv).

For fixed $M$, taking $x$ to be a sufficiently large multiple of
$\sqrt{H\log(nT)}+\tau\log(nT)$ in
\eqref{eq:first-step-recentered-truncation} and applying a union bound over at
most $nT$ pairs $(i,t_0)$ gives a probability tending to zero. The recentering
term is
$o_\P\!\big(\sqrt{H\log(nT)}+\tau\log(nT)\big)$, while
\eqref{eq:first-step-global-tail-probability} can be made arbitrarily small by
choosing $M$ large. Therefore,
\begin{align*}
    \max_{H+1\leq t_0\leq T}
    \max_{i\in[n_{t_0}]}
    \max_{\ell\in[H]}
    \left\|
        \sum_{s=1}^{\ell}Y_{s,it_0}
    \right\|
    &\lesssim_\P
    \sqrt{H\log(nT)}+(nT)^{1/q}\log(nT).
\end{align*}
Dividing by $H$ in the definition of $U_{it}$ yields
\begin{align*}
    \max_{H+1\le t\le T}\max_{i\in[n_t]}
    \left\|U_{it}\right\|
    &\lesssim_\P
    \sqrt{\frac{\log(nT)}{H}}+\frac{(nT)^{1/q}\log(nT)}{H}.
\end{align*}
\medskip\noindent\textit{Step 3: Combining with the realized bias.}
By the exact identity above, \eqref{eq:first-step-inverse-bound}, and the preceding rate bound,
\begin{align*}
    \max_{H+1\le t\le T}\max_{i\in[n_t]}
    \|\widehat{\beta}_{it}-\breve{\beta}_{it}\|
    &\lesssim_\P
    \sqrt{\frac{\log(nT)}{H}}
    +\frac{(nT)^{1/q}\log(nT)}{H}.
\end{align*}
The triangle inequality and
{Assumption~\ref{ass:first-step-bias}, which gives
$\widehat{\mathsf{B}}_{\mathtt{FS}}\leq\mathsf{B}_{\mathtt{FS}}$ w.p.a.1,} yield
\begin{align*}
    \max_{H+1\le t\le T}\max_{i\in[n_t]}
    \|\widehat{\beta}_{it}-\beta_{it}\|
    &\lesssim_\P
    \sqrt{\frac{\log(nT)}{H}}
    +\frac{(nT)^{1/q}\log(nT)}{H}
    +{\mathsf{B}_{\mathtt{FS}}}.
\end{align*}
\qed

\subsection{Proof of Lemma~\ref{lem:first-step-bias}}

{
We first give an argument common to all four cases. Apply
\eqref{eq:first-step-local-maximal} with the fixed threshold
$\lambda=cH$ and $a=q+2$, both to the rolling Gram matrices and to the
rolling averages of $(1+\|f_t\|)^2$. Because
$T^{2/(q-d-1)}/H=o(1)$, $H$ grows at a positive power of $T$ and hence
$H^{-\upsilon/2}(\log T)^{(q+\upsilon)/2-1}=o(1)$. The two terms in
\eqref{eq:first-step-local-maximal} are therefore bounded by
$CTH^{-q/2}$. It follows that, for a sufficiently large $C$,
\begin{align}
    &\P\!\left(
        \max_{H+1\leq t\leq T}\|\widehat A(t)^{-1}\|>C
    \right)\notag\\
    &\quad+
    \P\!\left(
        \max_{H+1\leq t\leq T}
        \frac{1}{H}\sum_{s=1}^H(1+\|f_{t-s}\|)^2>C
    \right)\notag\\
    &\lesssim
    TH^{-q/2}
    =o(H^{-(d+1)/2}).
    \label{eq:first-step-bias-factor-tail}
\end{align}
Indeed,
\begin{align*}
    H^{(d+1)/2}TH^{-q/2}
    &=
    \left(\frac{T^{2/(q-d-1)}}{H}\right)^{(q-d-1)/2}
    \to0.
\end{align*}
The inverse bound follows from the uniform lower eigenvalue
bound for $\E[\widehat A(t)]$ and the same fixed-threshold inequality.

Now let $a_h>0$ be deterministic. Suppose that, outside an event with
probability $o(H^{-(d+1)/2})$, for every $(i,t)$ there exists a coefficient vector
$\gamma_{it}=(\gamma_{it,0}^\top,\ldots,
\gamma_{it,\mathfrak p}^\top)^\top$ satisfying
$S\gamma_{it}=\beta_{it}$ and, uniformly over $s\in[H]$,
\begin{align}\label{eq:local-polynomial-decomposition}
    \alpha_{i(t-s)}+\beta_{i(t-s)}^\top f_{t-s}
    &=X_s(t)^\top\gamma_{it}+e_{its},&
    |e_{its}|
    &\leq C a_h(1+\|f_{t-s}\|).
\end{align}
Using the definition of $\breve\beta_{it}$ and the normal equations,
\begin{align*}
    \breve{\beta}_{it}-\beta_{it}
    &=
    S\widehat A(t)^{-1}
    \frac{1}{H}\sum_{s=1}^H
    X_s(t)e_{its}K\!\left(\frac{s}{H}\right).
\end{align*}
Consequently, \eqref{eq:first-step-bias-factor-tail} and boundedness of the
kernel imply, for a sufficiently large fixed constant $C$,
\begin{align}
    \P\!\left(
        \widehat{\mathsf{B}}_{\mathtt{FS}}>C a_h
    \right)
    &=o(H^{-(d+1)/2}).
    \label{eq:first-step-bias-global-tail}
\end{align}
Set $\mathsf{B}_{\mathtt{FS}}=C a_h$. If $\sqrt H\,a_h\to0$, then
$\mathsf{B}_{\mathtt{FS}}=o(H^{-1/2})$. Moreover, because
\begin{align*}
    \E\!\left[
        \IF\!\left(\widehat{\mathsf{B}}_{\mathtt{FS}}>
        \mathsf{B}_{\mathtt{FS}}\right)
        \,\middle|\,\mathcal D_{t-1}
    \right],
    \qquad t=H+1,\ldots,T,
\end{align*}
is a nonnegative martingale sampled along the nested design filtration,
Doob's maximal inequality and \eqref{eq:first-step-bias-global-tail} give,
for every $\epsilon>0$,
\begin{align*}
    \P\!\left(
        H^{(d+1)/2}
        \max_{H+1\leq t\leq T}
        \P\!\left(
            \widehat{\mathsf{B}}_{\mathtt{FS}}>
            \mathsf{B}_{\mathtt{FS}}
            \,\middle|\,\mathcal D_{t-1}
        \right)
        >\epsilon
    \right)
    &\leq
    \frac{H^{(d+1)/2}}{\epsilon}
    \P\!\left(
        \widehat{\mathsf{B}}_{\mathtt{FS}}>
        \mathsf{B}_{\mathtt{FS}}
    \right)
    \to0.
\end{align*}
Consequently,
\begin{align*}
    \frac{H^{(d-1)/2}}{\mathsf{B}_{\mathtt{FS}}+H^{-1}}
    \max_{H+1\leq t\leq T}
    \P\!\left(
        \widehat{\mathsf{B}}_{\mathtt{FS}}>
        \mathsf{B}_{\mathtt{FS}}
        \,\middle|\,\mathcal D_{t-1}
    \right)
    &\leq
    H^{(d+1)/2}
    \max_{H+1\leq t\leq T}
    \P\!\left(
        \widehat{\mathsf{B}}_{\mathtt{FS}}>
        \mathsf{B}_{\mathtt{FS}}
        \,\middle|\,\mathcal D_{t-1}
    \right)
    =o_\P(1),
\end{align*}
which verifies Assumption~\ref{ass:first-step-bias}.
Thus it remains only to verify \eqref{eq:local-polynomial-decomposition} with
the stated $a_h$ and tail probability.

\medskip\noindent\textit{Case (i).}
The standard Taylor--H\"older expansion of $\alpha_i(\cdot)$ and each
coordinate of $\beta_i(\cdot)$ around $t/T$ gives a local polynomial in $s/H$
of degree at most $\mathfrak p$, with a deterministic remainder bounded by
$a_h=h^{\min\{s,\mathfrak p+1\}}$ uniformly over $i$, $t$, and $s\in[H]$. The common argument therefore proves
part~(i).

\medskip\noindent\textit{Case (ii).}
For every fixed $M>0$, the Gaussian maximal inequality and the Brownian
modulus of continuity give a fixed constant $C_M$ such that
\begin{align}
    \P\!\left(
        \max_{i\in[n]}
        \sup_{|u-v|\leq h}
        \|W_i(u)-W_i(v)\|
        >
        C_M h^{1/2}\bigl(\log(nT)\bigr)^{1/2}
    \right)
    &\lesssim
    (nT)^{-M}.
    \label{eq:first-step-bias-brownian-tail}
\end{align}
Take $M>(d+1)/2$. Put the current Brownian level in the constant beta coefficient,
so that $S\gamma_{it}=\beta_{it}$. For the intercept component, use the
degree-$\mathfrak p$ local Taylor--H\"older polynomial for
$\tau\mapsto\alpha(\tau,\beta_{it})$ around $t/T$. Its temporal remainder is
$O(h^{\min\{s,\mathfrak p+1\}})$. Lipschitz continuity in the beta argument
adds at most a constant multiple of the Brownian increment in
\eqref{eq:first-step-bias-brownian-tail}. Since
$\min\{s,\mathfrak p+1\}>1/2$,
\eqref{eq:local-polynomial-decomposition} holds outside an
$o(H^{-(d+1)/2})$ event with
$a_h=h^{1/2}\bigl(\log(nT)\bigr)^{1/2}$.
The common argument proves part~(ii).

\medskip\noindent\textit{Case (iii).}
Choose the local polynomial coefficients for the nonrandom beta trend as in
case~(i), and put the current Brownian level in the constant beta coefficient.
The temporal intercept remainder is bounded by $h^s$, while the Lipschitz
contribution from the beta argument is bounded by the beta increment. Its
nonrandom component is of order $h^{\min\{s,1\}}$, which is bounded by the
claimed rate, and its Brownian component satisfies
\eqref{eq:first-step-bias-brownian-tail}. Therefore
\eqref{eq:local-polynomial-decomposition} holds outside an
$o(H^{-(d+1)/2})$ event with
$a_h=\max\{h^s,h^{1/2}\bigl(\log(nT)\bigr)^{1/2}\}$.
The common argument proves part~(iii). The Brownian component is nowhere
differentiable, so increasing $\mathfrak p$ does not improve this component
of the threshold.

\medskip\noindent\textit{Case (iv).}
Write $I_i(u)=\int_0^u W_i(v)\,\d v$. For every fixed $M>0$, the Brownian
maximal and modulus inequalities give a fixed constant $C_M$ such that the
probability that either
\begin{align*}
    \max_{i\in[n]}\sup_{u\in[0,1]}\|W_i(u)\|
    &>
    C_M\bigl(\log(nT)\bigr)^{1/2},\\
    \max_{i\in[n]}\sup_{|u-v|\leq h}
    \left\|
        I_i(u)-I_i(v)-W_i(v)(u-v)
    \right\|
    &>
    C_M h^{3/2}\bigl(\log(nT)\bigr)^{1/2}
\end{align*}
is bounded by a constant multiple of $(nT)^{-M}$. Take $M>(d+1)/2$.

If $\mathfrak p=0$, use the current levels of the beta and conditional
intercept paths as the constant local-polynomial coefficients. Boundedness of
the first derivatives of $\alpha$ gives
\eqref{eq:local-polynomial-decomposition} with
$a_h=h\bigl(\log(nT)\bigr)^{1/2}$.
If $\mathfrak p\geq1$, include the current slope
$\sigma_iW_i(t/T)$ in the degree-one beta coefficient. For
$g_i(u)=\alpha(u,\beta_i(u))$, also include
\begin{align*}
    g_i'(u)
    &=
    \partial_1\alpha(u,\beta_i(u))
    +\nabla_b\alpha(u,\beta_i(u))^\top\sigma_iW_i(u)
\end{align*}
at $u=t/T$ in the degree-one intercept coefficient. The joint
second-derivative bound on $\alpha$, the Brownian bounds above, and
$h\log(nT)=o(1)$ imply
\eqref{eq:local-polynomial-decomposition} with
$a_h=h^{3/2}\bigl(\log(nT)\bigr)^{1/2}$.
Combining the two polynomial orders gives the threshold in part~(iv), and the
common argument completes the proof.
}
\qed

\section{Second-Step Partitioning}\label{sec:classical-partition}

Using the generic partition (portfolios) notation from
Section~\ref{sec:general-linearization} and the concrete design
filtration $\{\mathcal{D}_t\}$ from Section~\ref{sec:classical-filtrations},
we now specialize the estimated partition (portfolio) to the marginal-quantile
scheme used by the classical beta-sorted estimator. Recall that
$\mathcal{B}=[\lb,\ub]$ is a rectangle. The partition is formed by tensor products across coordinates:
\begin{align*}
    \widehat{\Delta}_t
    &=\otimes_{l\in[d]}\widehat{\Delta}_{t,l},\qquad
    \widehat{\Delta}_{t,l}
    =\Big\{[\hat{\tau}_{0t,l},\hat{\tau}_{1t,l}), [\hat{\tau}_{1t,l}, \hat{\tau}_{2t,l}),\ldots,
    [\hat{\tau}_{(J_{t,l}-1)t,l}, \hat{\tau}_{J_{t,l}t,l}]\Big\}.
\end{align*}
Here $\hat{\tau}_{jt,l}$ denotes a possibly random breakpoint and $J_{t,l}$
denotes the number of intervals for coordinate $l\in[d]$, so that
$\bar{J}_t=\prod_{l\in[d]}J_{t,l}$.

For any first-step estimator $\widehat{\beta}_{it}=(\widehat{\beta}_{it,1},\ldots,\widehat{\beta}_{it,d})^\top$, coordinate $l\in[d]$, and time $t=H+1,\ldots,T$, define
\begin{align*}
    \widehat{F}_{\widehat{\beta},t,l}(z)
    =\frac{1}{n_t}\sum_{i=1}^{n_t}\IF(\widehat{\beta}_{it,l}\leq z)
    \qquad\text{and}\qquad
    \bar{F}_{\widehat{\beta},t,l}(z)
    =\frac{1}{n_t}\sum_{i=1}^{n_t}
    \P(\widehat{\beta}_{it,l}\leq z\mid\mathcal{D}_{t-1}).
\end{align*}
Their generalized inverse functions are
\begin{align*}
    \widehat{F}_{\widehat{\beta},t,l}^{-1}(\kappa)
    =\inf\{z\in\mathbb{R}:\widehat{F}_{\widehat{\beta},t,l}(z)\geq \kappa\}
    \qquad\text{and}\qquad
    \bar{F}_{\widehat{\beta},t,l}^{-1}(\kappa)
    =\inf\{z\in\mathbb{R}:\bar{F}_{\widehat{\beta},t,l}(z)\geq \kappa\},
\end{align*}
for $\kappa\in(0,1)$.

Our main inference results focus on partitions formed by tensor products of
marginal empirical quantiles of the components of $\widehat{\beta}_{it}$,
separately on the negative and nonnegative sides of zero. For each $l\in[d]$
and $t=H+1,\ldots,T$, define the split sample sizes
\begin{align*}
    \widehat{N}_{t,l}^{-}
    =\sum_{i=1}^{n_t}\IF(\widehat{\beta}_{it,l}<0)
    \qquad\text{and}\qquad
    \widehat{N}_{t,l}^{+}
    =\sum_{i=1}^{n_t}\IF(\widehat{\beta}_{it,l}\geq0).
\end{align*}
On the event $\min\{\widehat{N}_{t,l}^{-},\widehat{N}_{t,l}^{+}\}>0$, define
the split empirical distribution functions by
\begin{align*}
    \widehat{F}_{\widehat{\beta},t,l,-}(z)
    =\frac{1}{\widehat{N}_{t,l}^{-}}
    \sum_{i=1}^{n_t}\IF(\widehat{\beta}_{it,l}<0)
    \IF(\widehat{\beta}_{it,l}\leq z)
    \qquad\text{and}\qquad
    \widehat{F}_{\widehat{\beta},t,l,+}(z)
    =\frac{1}{\widehat{N}_{t,l}^{+}}
    \sum_{i=1}^{n_t}\IF(\widehat{\beta}_{it,l}\geq0)
    \IF(\widehat{\beta}_{it,l}\leq z),
\end{align*}
with generalized inverse functions
\begin{align*}
    \widehat{F}_{\widehat{\beta},t,l,-}^{-1}(\kappa)
    =\inf\{z<0:\widehat{F}_{\widehat{\beta},t,l,-}(z)\geq \kappa\}
    \qquad\text{and}\qquad
    \widehat{F}_{\widehat{\beta},t,l,+}^{-1}(\kappa)
    =\inf\{z\geq0:\widehat{F}_{\widehat{\beta},t,l,+}(z)\geq \kappa\},
\end{align*}
for $\kappa\in(0,1)$.
If either split sample is empty, define its empirical distribution and
quantiles arbitrarily. The uniform empirical sign-share bound in the proof of
Lemma~\ref{lem:quantile-convergence} shows that this event has probability
approaching zero.

\begin{assumption}[PA][Marginal-quantile-spaced partition]\label{ass:partition}
    For each $t=H+1,\ldots,T$, the partition $\widehat{\Delta}_t=\otimes_{l\in[d]}\widehat{\Delta}_{t,l}$ with
    \begin{align*}
        \widehat{\Delta}_{t,l}=\Big\{&[\hat{\tau}_{0t,l}^-,\hat{\tau}_{1t,l}^-), [\hat{\tau}_{1t,l}^-,\hat{\tau}_{2t,l}^-), \ldots,[\hat{\tau}_{(J_{t,l}^--1)t,l}^-,0),
        [0,\hat{\tau}_{1t,l}^+), [\hat{\tau}_{1t,l}^+, \hat{\tau}_{2t,l}^+), \ldots,
        [\hat{\tau}_{(J_{t,l}^+-1)t,l}^+, \hat{\tau}_{J_{t,l}^+t,l}^+]\Big\}
    \end{align*}
    where {
    $\hat{\tau}_{0t,l}^-=\min_{i:\widehat\beta_{it,l}<0}
    \widehat\beta_{it,l}$ and
    $\hat{\tau}_{J_{t,l}^{+}t,l}^+
    =\max_{i:\widehat\beta_{it,l}\geq0}\widehat\beta_{it,l}$},
    $\hat{\tau}_{jt,l}^-=\widehat{F}_{\widehat{\beta},t,l,-}^{-1}(j/J_{t,l}^-)$ for
    $j\in[J_{t,l}^{-}-1]$,
    $\hat{\tau}_{jt,l}^+=\widehat{F}_{\widehat{\beta},t,l,+}^{-1}(j/J_{t,l}^+)$ for
    $j\in[J_{t,l}^{+}-1]$, and $J_{t,l}^{-}+J_{t,l}^{+}=J_{t,l}$. Assume that
    $J_{t,l}^-, J_{t,l}^+\asymp J$ for all $l\in[d]$ and some
    $J\rightarrow\infty$ as $n\rightarrow\infty$, and $J^d/n=o(1)$.
\end{assumption}

The corresponding reference partition is denoted by
$\Delta_t=\otimes_{l\in[d]}\Delta_{t,l}$ and is nonrandom conditionally on
$\mathcal{D}_{t-1}$. It uses the conditional analogs of the same split
quantiles, and should be read as the conditional reference counterpart of the
estimated quantile partition, rather than as a partition formed from the
distribution of the true $\beta_{it}$. For each coordinate $l\in[d]$,
\begin{align*}
    \Delta_{t,l}
    =\Big\{&(\tau_{0t,l}^-,\tau_{1t,l}^-), [\tau_{1t,l}^-,\tau_{2t,l}^-), \ldots,[\tau_{(J_{t,l}^--1)t,l}^-,0),
    [0,\tau_{1t,l}^+), [\tau_{1t,l}^+, \tau_{2t,l}^+), \ldots,
    [\tau_{(J_{t,l}^+-1)t,l}^+, \tau_{J_{t,l}^+t,l}^+)\Big\},
\end{align*}
where {$\tau_{0t,l}^-=-\infty$ and
$\tau_{J_{t,l}^{+}t,l}^+=+\infty$},
$\tau_{jt,l}^-=\bar{F}_{\widehat{\beta},t,l,-}^{-1}(j/J_{t,l}^-)$ for
$j\in[J_{t,l}^{-}-1]$, and
$\tau_{jt,l}^+=\bar{F}_{\widehat{\beta},t,l,+}^{-1}(j/J_{t,l}^+)$ for
$j\in[J_{t,l}^{+}-1]$, with
\begin{align*}
    \bar\pi_{t,l,-}
    &=\frac{1}{n_t}\sum_{i=1}^{n_t}
    \P(\widehat{\beta}_{it,l}<0\mid\mathcal{D}_{t-1}),
    &
    \bar\pi_{t,l,+}
    &=\frac{1}{n_t}\sum_{i=1}^{n_t}
    \P(\widehat{\beta}_{it,l}\geq0\mid\mathcal{D}_{t-1}),
    \\
    \bar{F}_{\widehat{\beta},t,l,-}(z)
    &=\frac{1}{n_t\bar\pi_{t,l,-}}
    \sum_{i=1}^{n_t}
    \P(\widehat{\beta}_{it,l}\leq z,\widehat{\beta}_{it,l}<0
    \mid\mathcal{D}_{t-1}),
    \qquad z<0,
    \\
    \bar{F}_{\widehat{\beta},t,l,+}(z)
    &=\frac{1}{n_t\bar\pi_{t,l,+}}
    \sum_{i=1}^{n_t}
    \P(0\leq\widehat{\beta}_{it,l}\leq z
    \mid\mathcal{D}_{t-1}),
    \qquad z\geq0.
\end{align*}
Thus $\bar{F}_{\widehat{\beta},t,l,-}$ and
$\bar{F}_{\widehat{\beta},t,l,+}$ are distribution functions on the negative
and nonnegative sides, respectively. Their generalized inverses are defined over
$z<0$ and $z\ge0$, respectively, as for the split empirical distribution
functions above. Accordingly, define marginal split-mixture densities
\begin{align*}
    &\bar{f}_{\widehat{\beta},t,l,-}(z)
    =\frac{1}{n_t\bar\pi_{t,l,-}}\sum_{i=1}^{n_t}
    f_{\widehat{\beta}_{it,l}|\mathcal{D}_{t-1}}(z|\mathcal{D}_{t-1})
    \IF(z<0)\quad\text{and}\\
    &\bar{f}_{\widehat{\beta},t,l,+}(z)
    =\frac{1}{n_t\bar\pi_{t,l,+}}\sum_{i=1}^{n_t}
    f_{\widehat{\beta}_{it,l}|\mathcal{D}_{t-1}}(z|\mathcal{D}_{t-1})
    \IF(z\geq0).
\end{align*}
These ratios are defined as zero if the corresponding pooled sign probability
vanishes. Lemma~\ref{lem:reference-quantile-spacing}(i) implies that
$\bar\pi_{t,l,-}$ and $\bar\pi_{t,l,+}$ are uniformly bounded away from zero
with probability approaching one, so this convention is asymptotically
irrelevant. {Cell membership, counts, and polynomial basis
evaluation all use the raw generated regressor $\widehat\beta_{it}$. The
estimated partition therefore covers the coordinatewise range of the realized
generated regressors, while the reference partition uses the corresponding
complete conditional split distributions. The cells outside $\mathcal{B}^\circ$ may
be unbounded, but they do not intersect the interior evaluation region with
probability approaching one and are not used in any design or inverse-Gram
claim below.} The reference basis $p_t(\cdot)$ is defined from this $\Delta_t$,
and the reference Gram matrix $Q_t$ is the conditional expectation of
$p_t(\widehat\beta_{it})p_t(\widehat\beta_{it})^\top$ as in
Section~\ref{sec:general-linearization}. For the verification with
polynomial degree $\mathfrak{q}\geq0$, the basis normalization is
cell-specific. {For $j\in\mathcal J_t$,} write the paired reference and estimated rectangles as
$P_{jt}=\prod_{l\in[d]}[a_{jtl},b_{jtl})$ and
$\widehat P_{jt}=\prod_{l\in[d]}[\widehat a_{jtl},\widehat b_{jtl})$,
with the endpoint and tie-breaking conventions inherited from the corresponding
partitions. We set
\begin{align*}
    c_{jt,l}
    &=
    \frac{a_{jtl}+b_{jtl}}{2},
    &
    D_{jt}
    &=
    \operatorname{diag}\{b_{jt1}-a_{jt1},\ldots,b_{jtd}-a_{jtd}\},\\
    \widehat c_{jt,l}
    &=
    \frac{\widehat a_{jtl}+\widehat b_{jtl}}{2},
    &
    \widehat D_{jt}
    &=
    \operatorname{diag}\{\widehat b_{jt1}-\widehat a_{jt1},\ldots,
    \widehat b_{jtd}-\widehat a_{jtd}\}.
\end{align*}
Thus, for $b$ in the reference cell $P_{jt}$, the polynomial block is
$\check p_{\mathfrak q}(D_{jt}^{-1}(b-c_{jt}))$, while for $b$ in the
estimated cell $\widehat P_{jt}$, the polynomial block is
$\check p_{\mathfrak q}(\widehat D_{jt}^{-1}(b-\widehat c_{jt}))$.
The block entering the stacked basis is always the block of the cell containing
the argument $b$. When $\mathfrak q=0$, this reduces to the cell indicator.
When $\mathfrak q\geq1$, the monomials are therefore formed from the normalized
coordinates of $b$ within its own rectangle.
For the cell containing the origin, the split construction sets the lower
endpoint equal to zero in every coordinate, both for the reference and estimated
partitions. Hence
$D_{jt}^{-1}(0-c_{jt})=\widehat D_{jt}^{-1}(0-\widehat c_{jt})
=(-1/2,\ldots,-1/2)^\top$ on that cell, and therefore
$\widehat p_t(0)=p_t(0)$ for every $\mathfrak q\geq0$.
This is the specialization of the affine scaling/rotation convention
from Part~\ref{part:general} to the axis-aligned
quantile rectangles used here. On the event where all side lengths are of order
$J^{-1}$, the normalized basis is uniformly bounded and has derivative of order
$J$ in the original beta coordinates. The rest of this section collects the
verification results for this random partitioning scheme and basis function.

Under Assumption~\ref{ass:interior-separation}, $\mathcal J_t$ contains
precisely the estimated cells intersecting $\mathcal B^\circ$. Define the
corresponding interior cell counts
\begin{align*}
    \widehat n_{\min}^{\circ}
    &=
    \min_{H+1\leq t\leq T}\min_{j\in\mathcal J_t}\widehat n_{jt},
    &
    \widehat n_{\max}^{\circ}
    &=
    \max_{H+1\leq t\leq T}\max_{j\in\mathcal J_t}\widehat n_{jt}.
\end{align*}

\subsection{Results}

We present three results that establish the main properties of the quantile-based portfolio construction.

The first result establishes the density regularity of the generated regressor and translates it into density and spacing properties of the split reference quantile portfolios.

\begin{lemma}[Generated-Regressor Density and Reference Quantile Spacing]\label{lem:reference-quantile-spacing}
    Suppose that Assumptions \ref{ass:dgp},
    \ref{ass:first-step-ls}, \ref{ass:partition},
    {\ref{ass:interior-separation}}, and
    \ref{ass:first-step-bias} hold.
    Assume further that $T^{(3+d)/q}/H+\sqrt{\log T/H}=o(1)$.
    Then, with probability approaching one, the following statements hold:
    \begin{enumerate}[label=\textnormal{(\roman*)}]
    \item The conditional density
    $f_{\widehat{\beta}_{it}|\mathcal{D}_{t-1}}(\cdot|\mathcal{D}_{t-1})$
    is uniformly bounded and bounded away from zero on $\mathcal{B}$, uniformly over $i\in[n_t]$ and $t=H+1,\ldots,T$.

    \item The marginal split-mixture densities
    $\bar{f}_{\widehat{\beta},t,l,-}$ and $\bar{f}_{\widehat{\beta},t,l,+}$ are uniformly bounded and bounded away from zero on $[\lb_l,0)$ and $[0,\ub_l]$, respectively, uniformly over $l\in[d]$ and $t=H+1,\ldots,T$.
    \end{enumerate}
    {Consequently, the finite reference cells paired with cells in
    $\mathcal J_t$ satisfy}
    {
    \begin{align*}
        J^{-1}
        \lesssim_\P
        \min_{H+1\leq t\leq T}\min_{j\in\mathcal J_t^\circ}\min_{l\in[d]}
        (b_{jtl}-a_{jtl})
        \leq
        \max_{H+1\leq t\leq T}\max_{j\in\mathcal J_t^\circ}\max_{l\in[d]}
        (b_{jtl}-a_{jtl})
        \lesssim_\P J^{-1},
    \end{align*}
    {where $\mathcal J_t^\circ
    =\{j\in\mathcal J_t:P_{jt}\text{ is bounded}\}$.}
    }
\end{lemma}

The next result compares the empirical breakpoints in $\widehat{\Delta}_t$ with their conditional reference counterparts in $\Delta_t$. The first two bounds control levels of the split quantiles, while the last two control adjacent spacings, which are needed when comparing cell boundaries.

\begin{lemma}[Quantile Convergence]\label{lem:quantile-convergence}
    Suppose that Assumptions \ref{ass:dgp},
    \ref{ass:first-step-ls}, \ref{ass:partition},
    {\ref{ass:interior-separation}}, and
    \ref{ass:first-step-bias} hold.
    Assume further that $T^{(3+d)/q}/H+\sqrt{\log T/H}=o(1)$
    and $\log T/n=o(1)$.
    {Then, uniformly over the empirical cells intersecting
    $\mathcal B^\circ$ and their rank-paired reference cells,}
    {
    \begin{align*}
        &\max_{H+1\leq t\leq T}\max_{j\in\mathcal J_t}\max_{l\in[d]}
        \left(
        |\widehat a_{jtl}-a_{jtl}|
        +|\widehat b_{jtl}-b_{jtl}|
        \right)
        \lesssim_\P \sqrt{\frac{\log T}{n}},\\
        &\max_{H+1\leq t\leq T}\max_{j\in\mathcal J_t}\max_{l\in[d]}
        \left|
        (\widehat b_{jtl}-\widehat a_{jtl})
        -(b_{jtl}-a_{jtl})
        \right|
        \lesssim_\P \sqrt{\frac{\log T}{n}}.
    \end{align*}
    }
    If, in addition, $J^2\log T/n=o(1)$, then the estimated cell widths
    satisfy
    \begin{align*}
        J^{-1}
        \lesssim_\P
        \min_{H+1\leq t\leq T}\min_{j\in\mathcal J_t}\min_{l\in[d]}
        (\widehat b_{jtl}-\widehat a_{jtl})
        \leq
        \max_{H+1\leq t\leq T}\max_{j\in\mathcal J_t}\max_{l\in[d]}
        (\widehat b_{jtl}-\widehat a_{jtl})
        \lesssim_\P J^{-1}.
    \end{align*}
\end{lemma}

With the quantile approximation in hand, we can verify the portfolio geometry, portfolio counts, and Gram behavior used by the subsequent results.

\begin{lemma}[Generated Partition and Basis Moments]\label{lem:partition-basis-moments}
    Suppose that Assumptions \ref{ass:dgp}, \ref{ass:first-step-ls},
    \ref{ass:partition}, {\ref{ass:interior-separation}}, and
    \ref{ass:first-step-bias} hold.
    Assume further that
    \begin{align*}
        \frac{T^{(3+d)/q}}{H}+\sqrt{\frac{\log T}{H}}
        &=o(1),
        &
        \frac{J^d\log(nT)}{n}
        &=o(1).
    \end{align*}
    Then the following results hold.
    \begin{enumerate}
        \item {Generated partition coverage and geometry:
        $\mathcal B^\circ\cup\{\widehat\beta_{it}:i\in[n_t]\}$ is contained in
        $\cup_{j\in[\bar J_t]}\widehat P_{jt}$ uniformly over $t$ w.p.a.1,
        and $\widehat{\mathsf{R}}_{\mathtt{D}}^{\circ}
        \lesssim_\P J^{-1}$.}

        \item Cell counts:
        \begin{align*}
            \frac{n}{J^d}
            &\lesssim_\P
            {\widehat{n}_{\min}^{\circ}}
            \leq
            {\widehat{n}_{\max}^{\circ}}
            \lesssim_\P
            \frac{n}{J^d}.
        \end{align*}

        \item The generated Gram blocks:
        \begin{align*}
            J^{-d}
            \lesssim_\P
            \lambda_{\min}(\widehat{Q}_{jt})
            \leq
            \lambda_{\max}(\widehat{Q}_{jt})
            \lesssim_\P
            J^{-d}
        \end{align*}
        uniformly over $t=H+1,\ldots,T$ and interior cells $j\in\mathcal{J}_t$.

        \item The reference Gram blocks:
        \begin{align*}
            J^{-d}
            \lesssim_\P
            \lambda_{\min}(Q_{jt})
            \leq
            \lambda_{\max}(Q_{jt})
            \lesssim_\P
            J^{-d}
        \end{align*}
        uniformly over $t=H+1,\ldots,T$ and $j\in\mathcal{J}_t$.

        \item The empirical true-exposure reference second moments:
        \begin{align*}
            \lambda_{\max}\!\left(
            \frac{1}{n_t}\sum_{i=1}^{n_t}
            p_{jt}(\beta_{it})p_{jt}(\beta_{it})^\top
            \right)
            &\lesssim_\P J^{-d}
        \end{align*}
        uniformly over $t=H+1,\ldots,T$ and $j\in\mathcal{J}_t$.

        \item The comparison quantities from Lemma \ref{lem:general-linearization}:
        \begin{align*}
            \widehat{\mathsf{R}}_{\mathtt{Q}}
            \lesssim_\P \mathsf{R}_{\mathtt{Q}}
            =
            J^{-d}\sqrt{\frac{J^d\log(nT)}{n}}
            +
            J^{-(d-1)}\sqrt{\frac{\log T}{n}}
            +
            \frac{\log(nT)}{n},
        \end{align*}
        and
        \begin{align*}
            \widehat{\mathsf{R}}_{\mathtt{P}}
            \lesssim_\P \mathsf{R}_{\mathtt{P}}
            &=
            J^{-d}
            \min\left\{
            J\left(\mathsf{R}_{\mathtt{FS}}
            +\sqrt{\frac{\log T}{n}}\right),1\right\}
            +\sqrt{
            \frac{
            J^{-d}\min\Big\{
            J\left(\mathsf{R}_{\mathtt{FS}}
            +\sqrt{\frac{\log T}{n}}\right),1\Big\}
            \log(nT)}{n}
            }\\
            &\qquad
            +\frac{\log(nT)}{n}
            +\IF(\mathfrak{q}\geq1)
            J^{-d}
            \min\left\{
            J^2\left(\mathsf{R}_{\mathtt{FS}}
            +\sqrt{\frac{\log T}{n}}\right)^2,1\right\}
            .
        \end{align*}
    \end{enumerate}
\end{lemma}

\subsection{Discussion}

\begin{enumerate}[label=(\roman*)]
    \item The rate $\mathsf{R}_{\mathtt{Q}}$ controls the inverse-Gram perturbation. It compares the empirical generated Gram matrix with the conditional reference Gram matrix of $\widehat{\beta}_{it}$ evaluated on the reference partition. Its terms correspond to ordinary cross-sectional sampling variation within cells, movement of the empirical quantile endpoints, and empirical discretization. The condition $\mathsf{R}_{\mathtt{Q}}=o(J^{-d})$ is implied by
    \begin{align*}
        \frac{J^d\log(nT)}{n}=o(1)
        \qquad\text{and}\qquad
        \frac{J^2\log T}{n}=o(1).
    \end{align*}
    If these conditions hold, then the construction gives the measurability and coverage components of the partition-design condition \ref{cond:partition-design}, while Lemma \ref{lem:partition-basis-moments} verifies the diameter, design, and Gram-convergence components in conditions \ref{cond:partition-design} and \ref{cond:gram-convergence} of Lemma \ref{lem:general-linearization}. The stochastic linearization for $\mathfrak{q}=0$ can therefore be obtained through the comparison below.

    \item The rate $\mathsf{R}_{\mathtt{P}}$ controls the basis-switching variance term $\mathscr{R}_2(\beta)$. It compares $\widehat p_t(\widehat\beta_{it})$ with $p_t(\beta_{it})$. The switch set is therefore a boundary strip between paired estimated and reference cells, with width determined by both the first-step displacement $\widehat\beta_{it}-\beta_{it}$ and the endpoint quantile error. The condition $\mathsf{R}_{\mathtt{P}}=o(J^{-d})$ is implied by the same two displays above and by $J\mathsf{R}_{\mathtt{FS}}=o(1)$.

    \item For $\mathfrak{q}\geq1$, $\mathsf{R}_{\mathtt{P}}$ also contains a nonswitch polynomial-coordinate term. On observations that remain in paired cells, the indicators agree but the normalized coordinates differ because the estimated and reference cell endpoints are not identical and because $\widehat\beta_{it}$ differs from $\beta_{it}$. Bounded derivatives of the normalized basis convert the combined displacement into the final term in $\mathsf{R}_{\mathtt{P}}$.

\end{enumerate}

\subsection{Proof of Lemma~\ref{lem:reference-quantile-spacing}}

Because $3+d>2$, the condition
$T^{(3+d)/q}/H+\sqrt{\log T/H}=o(1)$ implies the rate condition in
Lemma~\ref{lem:first-step-gram}. Hence, on an event with probability
approaching one, $\widehat A(t)$ is invertible and
\begin{align*}
    1
    &\lesssim
    \lambda_{\min}\left(\frac{1}{H}\sum_{l=1}^{H}c_{itl}c_{itl}^\top\right)
    \leq
    \lambda_{\max}\left(\frac{1}{H}\sum_{l=1}^{H}c_{itl}c_{itl}^\top\right)
    \lesssim 1,
\end{align*}
uniformly over $i\in[n_t]$ and $t=H+1,\ldots,T$, where
\begin{align*}
    c_{itl}
    =
    S\widehat{A}(t)^{-1}X_l(t)K\!\Big(\frac{l}{H}\Big)\varsigma_{i(t-l)}.
\end{align*}
It remains to verify the higher-order coefficient-moment bound used below.
The uniform inverse bound for $\widehat A(t)$ and the boundedness of the kernel
and $\varsigma_{i(t-l)}$ give
\begin{align*}
    \max_{i\in[n_t]}
    \frac{1}{H}\sum_{l=1}^H\|c_{itl}\|^{3+d}
    \lesssim_\P
    \frac{1}{H}\sum_{l=1}^H\|X_l(t)\|^{3+d}
    \lesssim
    1+\frac{1}{H}\sum_{l=1}^H\|f_{t-l}\|^{3+d}.
\end{align*}
The process $\{\|f_t\|^{3+d}\}$ has a finite moment of order $q/(3+d)>2$
and inherits the geometric near-epoch dependence in
Assumption~\ref{ass:dgp}(iv). Applying the same maximal Fuk--Nagaev argument
as in Lemma~\ref{lem:first-step-gram}, now with moment exponent $q/(3+d)$,
yields
\begin{align*}
    \max_{H+1\leq t\leq T}
    \left|
    \frac{1}{H}\sum_{l=1}^H
    \left(
    \|f_{t-l}\|^{3+d}
    -\E[\|f_{t-l}\|^{3+d}]
    \right)
    \right|
    \lesssim_\P
    \frac{T^{(3+d)/q}}{H}
    +\sqrt{\frac{\log T}{H}}
    =o_\P(1).
\end{align*}
Since $\sup_t\E[\|f_t\|^{3+d}]<\infty$, intersecting these events gives an
event $\mathcal A$ with probability approaching one on which
\begin{align*}
    \max_{H+1\leq t\leq T}\max_{i\in[n_t]}
    \frac{1}{H}\sum_{l=1}^H\|c_{itl}\|^{3+d}
    &\lesssim 1
\end{align*}
and the preceding inverse and eigenvalue bounds all hold.
On this event, let
\begin{align*}
    \mathcal{H}_{it}^{\circ}
    =
    \sigma\!\left(
    (\alpha_{i(t-s)},\beta_{i(t-s)}^\top,\varsigma_{i(t-s)}):s\in[H]
    \right)
    \vee\mathcal{D}_{t-1}.
\end{align*}
Conditional on $\mathcal{H}_{it}^{\circ}$, the only remaining randomness in the first-step variance component comes from the i.i.d. innovations $\{\eta_{i(t-s)}:s\in[H]\}$. Let $f_{\widehat{\beta}_{it}|\mathcal{H}_{it}^{\circ}}(\cdot|\mathcal{H}_{it}^{\circ})$ be the corresponding conditional density. Then
\begin{align*}
    f_{\widehat{\beta}_{it}|\mathcal{D}_{t-1}}(z|\mathcal{D}_{t-1})
    =
    \E\!\left[
    f_{\widehat{\beta}_{it}|\mathcal{H}_{it}^{\circ}}(z|\mathcal{H}_{it}^{\circ})
    \big|\mathcal{D}_{t-1}\right]
    =
    H^{d/2}\E\!\left[
    f_{G_{it}}\big(\sqrt{H}(z-\breve{\beta}_{it})\big)
    \big|\mathcal{D}_{t-1}\right],
\end{align*}
where $G_{it}=H^{-1/2}\sum_{l=1}^H c_{itl}\eta_{i(t-l)}$ and $\breve{\beta}_{it}$ is defined in \eqref{eq:first-step-decomposition}.
Given $\mathcal{H}_{it}^{\circ}$, the coefficients $c_{itl}$ and the signal
$\breve{\beta}_{it}$ are fixed, and the independence of the innovations in
Assumption~\ref{ass:dgp}(ii) gives the displayed
convolution representation.
We record the needed properties of $f_{G_{it}}$ directly. Let $\varphi_\eta$ denote the characteristic function of $\eta_{it}$. Conditional on $\mathcal{H}_{it}^{\circ}$,
\begin{align*}
    \varphi_{G_{it}}(u)
    =
    \prod_{l=1}^{H}
    \varphi_\eta\left(\frac{c_{itl}^\top u}{\sqrt H}\right).
\end{align*}
The condition $g_\eta^{(1)}\in L^1(\mathbb{R})$ implies polynomial decay of
$\varphi_\eta$ away from the origin, while the moment condition on $\eta_{it}$
gives the usual expansion
$\varphi_\eta(v)=1-v^2/2+O(|v|^{2+\delta})$ for some $\delta>0$ as
$v\to0$. Together with the eigenvalue bounds on
$H^{-1}\sum_l c_{itl}c_{itl}^\top$ and the moment average bound on the
$c_{itl}$'s, the standard three-region argument for products of
characteristic functions applies: near the origin the product has Gaussian
decay, on intermediate shells a fixed fraction of the projected arguments is
bounded away from zero, and in the tails the decay from
$g_\eta^{(1)}\in L^1(\mathbb{R})$ is applied to enough projected factors. Hence
\begin{align*}
    \int_{\mathbb{R}^d}(1+\|u\|)
    |\varphi_{G_{it}}(u)|\d{u}
    \lesssim 1.
\end{align*}
Fourier inversion therefore gives a continuously differentiable density
\begin{align*}
    f_{G_{it}}(x)
    =
    (2\pi)^{-d}\int_{\mathbb{R}^d}
    \exp(-\mathrm{i} u^\top x)\varphi_{G_{it}}(u)\d{u}
\end{align*}
with $\sup_x\{f_{G_{it}}(x)+\|\nabla f_{G_{it}}(x)\|\}\lesssim 1$ uniformly.

The same Fourier calculation also gives a uniform lower bound near the origin. Indeed, for fixed $r$ and $\|u\|\le r$, the expansion of $\varphi_\eta$ gives
\begin{align*}
    \varphi_{G_{it}}(u)
    =
    \exp\left\{
    -\frac{1}{2H}\sum_{l=1}^{H}(c_{itl}^\top u)^2
    \right\}+o(1),
\end{align*}
uniformly on $\|u\|\le r$. The matrix inside the quadratic form has eigenvalues bounded and bounded away from zero on $\mathcal{A}$, so the real part of the Fourier integral over a sufficiently large fixed ball is close to the corresponding nondegenerate Gaussian integral. The tail outside that ball can be made uniformly small by the preceding integrability bound. Hence $f_{G_{it}}(0)\gtrsim1$ uniformly, and boundedness of $\nabla f_{G_{it}}$ implies that $f_{G_{it}}(x)\gtrsim1$ on a fixed neighborhood of zero.
The same Fourier argument, with integration by parts, also gives a polynomial
tail bound $f_{G_{it}}(x)\lesssim (1+\|x\|)^{-d-1}$ uniformly.
Let
\begin{align*}
    \mathcal A_{it,\mathtt{FS}}
    &=
    \left\{
        \|\breve\beta_{it}-\beta_{it}\|
        \leq \mathsf{B}_{\mathtt{FS}}
    \right\}.
\end{align*}
{
By Assumption~\ref{ass:first-step-bias} and the inclusion
$\mathcal A_{it,\mathtt{FS}}^c
\subseteq\{\widehat{\mathsf{B}}_{\mathtt{FS}}>\mathsf{B}_{\mathtt{FS}}\}$,
\begin{align*}
    \sqrt H\,\mathsf{B}_{\mathtt{FS}}
    &=o(1),&
    H^{d/2}
    \max_{H+1\leq t\leq T}\max_{i\in[n_t]}
    \P\!\left(
        \mathcal A_{it,\mathtt{FS}}^c
        \,\middle|\,\mathcal D_{t-1}
    \right)
    &\lesssim_\P
    \sqrt H\,\mathsf{B}_{\mathtt{FS}}+H^{-1/2}
    =o_\P(1).
\end{align*}
}
We split the conditional convolution into its contributions on
$\mathcal A_{it,\mathtt{FS}}$ and $\mathcal A_{it,\mathtt{FS}}^c$.
Since $\sup_x f_{G_{it}}(x)\lesssim1$, the latter contribution satisfies,
uniformly over $i$, $t$, and $z\in\mathcal B$,
\begin{align}
    H^{d/2}
    \E\!\left[
        f_{G_{it}}\!\left(\sqrt H(z-\breve\beta_{it})\right)
        \IF(\mathcal A_{it,\mathtt{FS}}^c)
        \,\middle|\,\mathcal D_{t-1}
    \right]
    &\lesssim
    H^{d/2}
    \P\!\left(
        \mathcal A_{it,\mathtt{FS}}^c
        \,\middle|\,\mathcal D_{t-1}
    \right)
    =o_\P(1).
    \label{eq:first-step-bias-bad-density}
\end{align}

On $\mathcal A_{it,\mathtt{FS}}$,
$\|\breve\beta_{it}-\beta_{it}\|\leq \mathsf{B}_{\mathtt{FS}}$.
Hence, for every $r>0$,
\begin{align*}
    \P\!\left(
        \|\breve\beta_{it}-z\|\leq r,\,
        \mathcal A_{it,\mathtt{FS}}
        \,\middle|\,\mathcal D_{t-1}
    \right)
    \leq
    \P\!\left(
        \|\beta_{it}-z\|
        \leq r+\mathsf{B}_{\mathtt{FS}}
        \,\middle|\,\mathcal D_{t-1}
    \right)
    \lesssim
    \bigl(r+\mathsf{B}_{\mathtt{FS}}\bigr)^d,
\end{align*}
where the last inequality follows from the uniform upper bound on
$f_{\beta_{it}}(\cdot\mid\mathcal D_{t-1})$.
Decomposing $\mathcal B$ into annuli of width $H^{-1/2}$ around $z$, using
$\sqrt H\,\mathsf{B}_{\mathtt{FS}}=o(1)$ and the polynomial tail of $f_{G_{it}}$,
shows that the contribution on $\mathcal A_{it,\mathtt{FS}}$ is uniformly
bounded. Together with \eqref{eq:first-step-bias-bad-density}, this yields
\begin{align*}
    \sup_{z\in\mathcal B}
    f_{\widehat\beta_{it}\mid\mathcal D_{t-1}}
    (z\mid\mathcal D_{t-1})
    &\lesssim_\P1.
\end{align*}

For the lower bound, fix $r_0>0$ small enough that
$f_{G_{it}}(x)$ is uniformly bounded away from zero on
$\{x:\|x\|\leq2r_0\}$.
Since $\sqrt H\,\mathsf{B}_{\mathtt{FS}}=o(1)$, uniformly in $z\in\mathcal B$,
\begin{align*}
    f_{\widehat\beta_{it}\mid\mathcal D_{t-1}}
    (z\mid\mathcal D_{t-1})
    &\gtrsim
    H^{d/2}
    \P\!\left(
        \|\beta_{it}-z\|\leq r_0H^{-1/2},\,
        \mathcal A_{it,\mathtt{FS}}
        \,\middle|\,\mathcal D_{t-1}
    \right)\\
    &\gtrsim
    H^{d/2}
    \left[
        \P\!\left(
            \|\beta_{it}-z\|\leq r_0H^{-1/2}
            \,\middle|\,\mathcal D_{t-1}
        \right)
        -
        \P\!\left(
            \mathcal A_{it,\mathtt{FS}}^c
            \,\middle|\,\mathcal D_{t-1}
        \right)
    \right].
\end{align*}
Because $\mathcal B$ is a rectangle, its intersection with a ball of radius
$r_0H^{-1/2}$ centered at any $z\in\mathcal B$ contains a cone whose volume
is bounded below by a constant multiple of $H^{-d/2}$. The uniform lower
bound on $f_{\beta_{it}}(\cdot\mid\mathcal D_{t-1})$ and
{Assumption~\ref{ass:first-step-bias}} therefore imply
\begin{align*}
    \inf_{z\in\mathcal B}
    f_{\widehat\beta_{it}\mid\mathcal D_{t-1}}
    (z\mid\mathcal D_{t-1})
    &\gtrsim_\P1.
\end{align*}
This proves the full-vector density statement without requiring a density for
the signal $\breve\beta_{it}$ itself.

The marginal density statement follows by applying the preceding
argument to each coordinate. The eigenvalue bounds for
$H^{-1}\sum_{s=1}^Hc_{its}c_{its}^\top$ bound each coordinate variance above
and away from zero, and the exceptional-event bound above is stronger than
needed for one coordinate.
For the split mixtures,
\begin{align*}
    &\bar{f}_{\widehat{\beta},t,l,-}(z)
    =\frac{1}{n_t\bar\pi_{t,l,-}}\sum_{i=1}^{n_t}
    f_{\widehat{\beta}_{it,l}|\mathcal{D}_{t-1}}(z|\mathcal{D}_{t-1})
    \IF(z<0),\\
    &\bar{f}_{\widehat{\beta},t,l,+}(z)
    =\frac{1}{n_t\bar\pi_{t,l,+}}\sum_{i=1}^{n_t}
    f_{\widehat{\beta}_{it,l}|\mathcal{D}_{t-1}}(z|\mathcal{D}_{t-1})
    \IF(z\geq0).
\end{align*}
Because $0$ is an interior point of $\mathcal{B}$ and the marginal densities are bounded and bounded away from zero on $[\lb_l,\ub_l]$, the denominators are bounded away from zero w.p.a.1. Thus the split densities are bounded and bounded away from zero on $[\lb_l,0)$ and $[0,\ub_l]$, respectively. {The separation of
$\mathcal B^\circ$ from $\partial\mathcal B$ and
$J_{t,l}^-,J_{t,l}^+\asymp J$ imply that every finite reference endpoint
paired with a cell in $\mathcal J_t$ lies in this density-regular region
w.p.a.1. The displayed interior spacing bound therefore follows from the mean
value theorem.} \qed

\subsection{Proof of Lemma~\ref{lem:quantile-convergence}}

For $a\in\{-,+\}$, write $\mathcal{S}_{-}=(-\infty,0)$ and $\mathcal{S}_{+}=[0,\infty)$. Define
\begin{align*}
    \widehat{A}_{tla}(z)
    &=\frac{1}{n_t}\sum_{i=1}^{n_t}
    \IF(\widehat{\beta}_{it,l}\in\mathcal{S}_{a})
    \IF(\widehat{\beta}_{it,l}\leq z),\\
    \bar{A}_{tla}(z)
    &=\frac{1}{n_t}\sum_{i=1}^{n_t}
    \P(\widehat{\beta}_{it,l}\in\mathcal{S}_{a},
    \widehat{\beta}_{it,l}\leq z|\mathcal{D}_{t-1}),
\end{align*}
and let $\widehat{\pi}_{tla}=n_t^{-1}\sum_{i=1}^{n_t}\IF(\widehat{\beta}_{it,l}\in\mathcal{S}_{a})$ and
$\bar{\pi}_{tla}=n_t^{-1}\sum_{i=1}^{n_t}\P(\widehat{\beta}_{it,l}\in\mathcal{S}_{a}|\mathcal{D}_{t-1})$.
Then $\widehat{F}_{\widehat{\beta},t,l,a}=\widehat{A}_{tla}/\widehat{\pi}_{tla}$ and
$\bar{F}_{\widehat{\beta},t,l,a}=\bar{A}_{tla}/\bar{\pi}_{tla}$ on the corresponding side of zero.

Conditional on $\mathcal{D}_{t-1}$, $\widehat{\beta}_{it}$ is independent over $i$. The classes of half-lines intersected with $\mathcal{S}_{-}$ and $\mathcal{S}_{+}$ have finite VC index and envelope one. Therefore the i.n.i.d.\ maximal inequality in \citet[Lemma~SA24]{cattaneo2024uniform}, combined with conditional bounded-difference concentration and followed by a union bound over $t$, coordinates, and signs, gives
\begin{align*}
    &\max_{H+1\leq t\leq T}\max_{l\in[d]}\max_{a\in\{-,+\}}
    \sup_{z\in\mathcal{S}_a}
    |\widehat{A}_{tla}(z)-\bar{A}_{tla}(z)|
    \lesssim_\P \sqrt{\frac{\log T}{n}},\\
    &\max_{H+1\leq t\leq T}\max_{l\in[d]}\max_{a\in\{-,+\}}
    |\widehat{\pi}_{tla}-\bar{\pi}_{tla}|
    \lesssim_\P \sqrt{\frac{\log T}{n}}.
\end{align*}
Lemma \ref{lem:reference-quantile-spacing} and the fact that $0$ lies in the interior of $\mathcal{B}$ imply $\bar{\pi}_{tla}\gtrsim_\P1$ uniformly. Hence $\widehat{\pi}_{tla}\gtrsim_\P1$ uniformly and the ratio map gives
\begin{align*}
    &\max_{H+1\leq t\leq T}\max_{l\in[d]}
    \sup_{z<0}
    |\widehat{F}_{\widehat{\beta},t,l,-}(z)-\bar{F}_{\widehat{\beta},t,l,-}(z)|
    \lesssim_\P \sqrt{\frac{\log T}{n}},\\
    &\max_{H+1\leq t\leq T}\max_{l\in[d]}
    \sup_{z\geq0}
    |\widehat{F}_{\widehat{\beta},t,l,+}(z)-\bar{F}_{\widehat{\beta},t,l,+}(z)|
    \lesssim_\P \sqrt{\frac{\log T}{n}}.
\end{align*}
{
{Since $\log T/n=o(1)$, the uniform split-CDF bounds above,
Assumption~\ref{ass:interior-separation}, and $J\to\infty$ imply that the
first negative and last positive empirical intervals do not intersect
$\mathcal B^\circ$ in any coordinate w.p.a.1. Hence all rank-paired
reference cells indexed by $\mathcal J_t$ have finite endpoints on this event.}
For the probability indices defining cells in $\mathcal J_t$, the associated
reference quantiles lie in the density-regular region identified in
Lemma~\ref{lem:reference-quantile-spacing}. The generalized inverse relation
and the empirical quantile overshoot therefore give the first displayed bound
in the lemma. The side-length bound follows from
\begin{align*}
    \left|
    (\widehat b_{jtl}-\widehat a_{jtl})
    -(b_{jtl}-a_{jtl})
    \right|
    &\leq
    |\widehat b_{jtl}-b_{jtl}|
    +|\widehat a_{jtl}-a_{jtl}|,
\end{align*}
uniformly over $t$, $j\in\mathcal J_t$, and $l\in[d]$.
Under $J^2\log T/n=o(1)$, the right-hand side is
$o_\P(J^{-1})$. Combining this result with the reference width bounds in
Lemma~\ref{lem:reference-quantile-spacing} gives the asserted estimated width
bounds.
}
\qed

\subsection{Proof of Lemma~\ref{lem:partition-basis-moments}}

Let $\widehat{P}_{jt}=[\widehat{\ell}_{jt},\widehat{u}_{jt}]$ denote a generic rectangle in $\widehat{\Delta}_t$ and $P_{jt}=[\ell_{jt},u_{jt}]$ the corresponding rectangle in the reference partition $\Delta_t$. Let $\widehat{\psi}_{jt}(b) =\widehat{p}_{jt}(b)\widehat{p}_{jt}(b)^\top$ and $\psi_{jt}(b) =p_{jt}(b)p_{jt}(b)^\top$.
{The condition $J^d\log(nT)/n=o(1)$ implies $\log T/n=o(1)$,
so the argument in the proof of Lemma~\ref{lem:quantile-convergence} ensures
that all reference cells paired with $j\in\mathcal J_t$ have finite
endpoints w.p.a.1.}
Lemma \ref{lem:reference-quantile-spacing} gives reference side lengths of order $J^{-1}$. {For cells indexed by
$j\in\mathcal J_t$,} the estimated side lengths satisfy the same order. To see this, apply the localized VC/Bernstein inequality to split intervals on each side of zero. Since the split densities are bounded and bounded away from zero, an interval with side length larger than a sufficiently large multiple of $J^{-1}$ has conditional probability larger than a sufficiently large multiple of $J^{-1}$, while an interval with side length smaller than a sufficiently small multiple of $J^{-1}$ has conditional probability smaller than a sufficiently small multiple of $J^{-1}$. The empirical split quantile intervals have empirical mass of order $J^{-1}$ on their corresponding side, up to a discretization error of order $n^{-1}$, and $J^d\log(nT)/n=o(1)$ implies the localized stochastic error is $o_\P(J^{-1})$ uniformly. {Hence the estimated and reference cells indexed by
$\mathcal J_t$ belong w.p.a.1 to a uniformly quasi-uniform class of
rectangles. Moreover, the first and last empirical order statistics lie
outside $\mathcal B^\circ$ coordinatewise w.p.a.1:
Assumption~\ref{ass:interior-separation}
places at least one complete quantile cell between $\mathcal B^\circ$ and
each support boundary, and every sample extremum lies weakly beyond the
corresponding first or last interior empirical quantile. Thus the empirical
partition covers $\mathcal B^\circ$ and, by construction, every realized
$\widehat\beta_{it}$.} {Assumption~\ref{ass:interior-separation} implies that
every cell indexed by $j\in\mathcal{J}_t$ is separated from the outer boundary
of $\mathcal{B}$ by more than the first-step smoothing scale, uniformly over
$t$.} The side-length bounds give
\begin{align*}
    \widehat{\mathsf{R}}_{\mathtt{D}}^{\circ}
    &\lesssim_\P J^{-1}.
\end{align*}
The classes generated by the entries of $\widehat{\psi}_{jt}$ and $\psi_{jt}$ are polynomial transformations of rectangle indicators. Hence they are VC-type with bounded envelopes and conditional variances of order $J^{-d}$.
All empirical-process bounds below are first taken uniformly over deterministic quasi-uniform rectangles equipped with the midpoint and side-length normalization described above. The empirical quantile cells are then substituted into those uniform bounds.

{The same maximal inequality applied only to the cell indicators
indexed by $\mathcal J_t$ gives}
\begin{align*}
    \frac{n}{J^d}
    &\lesssim_\P
    {\widehat{n}_{\min}^{\circ}}
    \leq
    {\widehat{n}_{\max}^{\circ}}
    \lesssim_\P
    \frac{n}{J^d}.
\end{align*}
For a deterministic admissible rectangle $P$ with its midpoint and side-length normalization, write $\psi_P(b)=p_{\mathfrak q}(b,P)p_{\mathfrak q}(b,P)^\top$. Conditional on $\mathcal{D}_{t-1}$, the maximal inequality gives
\begin{align*}
    \max_{H+1\leq t\leq T}
    \sup_{P}
    \left\|
    \frac{1}{n_t}\sum_{i=1}^{n_t}
    \left(
    \psi_P(\widehat\beta_{it})
    -\E[\psi_P(\widehat\beta_{it})|\mathcal{D}_{t-1}]
    \right)
    \right\|
    \lesssim_\P \sqrt{\frac{\log(nT)}{nJ^d}}.
\end{align*}
The supremum is over the deterministic quasi-uniform rectangle class containing the realized estimated cells w.p.a.1, so evaluating the display at $P=\widehat P_{jt}$ controls the empirical part of $\widehat Q_{jt}$. The conditional density of $\widehat{\beta}_{it}$ is bounded and bounded away from zero on interior cells by Lemma \ref{lem:reference-quantile-spacing}. After the affine change of variables to cell-normalized coordinates, the corresponding population generated Gram block has eigenvalues of order $J^{-d}$ because the monomials in $\check p_{\mathfrak{q}}$ are linearly independent on a fixed normalized rectangle. Since $J^d\log(nT)/n=o(1)$, Weyl's inequality gives
\begin{align*}
    J^{-d}
    \lesssim_\P
    \lambda_{\min}(\widehat{Q}_{jt})
    \leq
    \lambda_{\max}(\widehat{Q}_{jt})
    \lesssim_\P
    J^{-d}
\end{align*}
uniformly over $t=H+1,\ldots,T$ and $j\in\mathcal{J}_t$. Multiplying by $n_t/\widehat n_{jt}\asymp_\P J^d$ gives the scaled Gram bound
\begin{align*}
    1
    &\lesssim_\P
    \frac{n_t}{\widehat n_{jt}}\lambda_{\min}(\widehat{Q}_{jt})
    \leq
    \frac{n_t}{\widehat n_{jt}}\lambda_{\max}(\widehat{Q}_{jt})
    \lesssim_\P
    1.
\end{align*}

For the reference Gram blocks, recall that
\begin{align*}
    Q_{jt}
    &=\frac{1}{n_t}\sum_{i=1}^{n_t}
    \E[p_{jt}(\widehat{\beta}_{it})p_{jt}(\widehat{\beta}_{it})^\top|\mathcal{D}_{t-1}].
\end{align*}
After the change of variables $b=c_{jt}+D_{jt}z$, the normalized support for $z$ is a rectangle with nonempty interior. The reference side-length bounds from Lemma \ref{lem:reference-quantile-spacing}, together with the conditional density bounds for $\widehat{\beta}_{it}$ in the same lemma, imply that the Jacobian is of order $J^{-d}$ and that the weight attached to the monomial Gram matrix is bounded and bounded away from zero. Since the monomials up to degree $\mathfrak{q}$ are linearly independent on this rectangle,
\begin{align*}
    J^{-d}
    &\lesssim_\P
    \lambda_{\min}(Q_{jt})
    \leq
    \lambda_{\max}(Q_{jt})
    \lesssim_\P
    J^{-d}
\end{align*}
uniformly over $t=H+1,\ldots,T$ and $j\in\mathcal{J}_t$.

For the empirical true-exposure reference second moments, apply the same VC-type maximal
inequality to the class $\psi_{jt}(\beta_{it})$. Conditional on
$\mathcal{D}_{t-1}$, Assumption~\ref{ass:dgp}(iii)
gives independent unit histories and conditional densities for $\beta_{it}$
that are bounded on $\mathcal{B}$. Since the reference cells have side lengths
of order $J^{-1}$, the population block moments are of order $J^{-d}$. The
empirical deviation is of order $\sqrt{\log(nT)/(nJ^d)}$, which is
$o_\P(J^{-d})$ under $J^d\log(nT)/n=o(1)$. Hence
\begin{align*}
    \lambda_{\max}\!\left(
    \frac{1}{n_t}\sum_{i=1}^{n_t}
    p_{jt}(\beta_{it})p_{jt}(\beta_{it})^\top
    \right)
    &\lesssim_\P J^{-d}
\end{align*}
uniformly over $t=H+1,\ldots,T$ and $j\in\mathcal{J}_t$.

It remains to bound $\widehat{\mathsf{R}}_{\mathtt{Q}}$. Add and subtract the
population counterpart of the estimated rectangle, obtained by evaluating the
deterministic-class map
$P\mapsto n_t^{-1}\sum_i\E[\psi_P(\widehat\beta_{it})|\mathcal{D}_{t-1}]$
at $P=\widehat P_{jt}$. The resulting empirical-process term is already
bounded above by $\sqrt{\log(nT)/(nJ^d)}$.
For the population-comparison term, compare the estimated and reference rectangles while holding the generated regressor fixed. The endpoint quantile bounds in Lemma \ref{lem:quantile-convergence} imply that the symmetric difference between paired rectangles has conditional mass
$J^{-(d-1)}\sqrt{\log T/n}$,
up to the empirical discretization term $\log(nT)/n$. When $\mathfrak{q}\geq1$, the nonswitch change in normalized polynomial coordinates is of the same order: bounded derivatives of $\check p_{\mathfrak{q}}$ give a pointwise perturbation of order $J\sqrt{\log T/n}$ on a cell of mass $J^{-d}$. Combining these bounds gives $\widehat{\mathsf{R}}_{\mathtt{Q}}\lesssim_\P\mathsf{R}_{\mathtt{Q}}$.

For $\widehat{\mathsf{R}}_{\mathtt{P}}$, work on the high-probability event from Lemma~\ref{lem:first-step-rate} and Lemma~\ref{lem:quantile-convergence} on which
$\max_{i,t}\|\widehat\beta_{it}-\beta_{it}\|\lesssim\mathsf{R}_{\mathtt{FS}}$ and all paired endpoints differ by at most a constant multiple of $\sqrt{\log T/n}$. Decompose the full basis discrepancy
into the generated-regressor displacement and the endpoint displacement:
\begin{align*}
    \widehat{p}_{jt}(\widehat{\beta}_{it})-p_{jt}(\beta_{it})
    &=
    \left\{\widehat{p}_{jt}(\widehat{\beta}_{it})
    -p_{jt}(\widehat{\beta}_{it})\right\}
    +
    \left\{p_{jt}(\widehat{\beta}_{it})-p_{jt}(\beta_{it})\right\}.
\end{align*}
Any observation for which the corresponding cell indicators differ in either component belongs to a
neighborhood of $\partial P_{jt}$ (the boundary of $P_{jt}$) with radius proportional to
$\mathsf{R}_{\mathtt{FS}}+\sqrt{\log T/n}$. When
$J(\mathsf{R}_{\mathtt{FS}}+\sqrt{\log T/n})\leq1$, the density bounds give
the boundary-strip probability
$J^{-d}J(\mathsf{R}_{\mathtt{FS}}+\sqrt{\log T/n})$. Otherwise, the squared
basis difference is supported on the union of the paired cells, whose mass is
$O_\P(J^{-d})$. A Bernstein inequality in the first case and the cell-mass
bound in the second give the first three terms in
$\mathsf{R}_{\mathtt{P}}$.

On the complement of the switch set, the paired cell indicators agree. Both basis blocks are then evaluated in corresponding cells, and only the normalized polynomial coordinates can differ. For $\mathfrak{q}\geq1$, bounded derivatives of the normalized polynomial basis, the first-step error bound, and the endpoint quantile bounds in Lemma \ref{lem:quantile-convergence} imply
\begin{align*}
    \max_{H+1\leq t\leq T}\max_{j\in\mathcal{J}_t}\max_{i\in[n_t]}
    \left\|\widehat p_{jt}(\widehat\beta_{it})-p_{jt}(\beta_{it})\right\|
    \IF\!\left(
    \widehat\beta_{it}\in\widehat P_{jt},\
    \beta_{it}\in P_{jt}
    \right)
    \lesssim_\P
    \min\left\{
    J\left(\mathsf{R}_{\mathtt{FS}}+\sqrt{\frac{\log T}{n}}\right),1
    \right\}.
\end{align*}
The squared basis difference on the nonswitch set is therefore bounded by the
corresponding squared displacement times the empirical mass of the paired cell,
which is $\lesssim_\P J^{-d}$ uniformly. This gives the final term in
$\mathsf{R}_{\mathtt{P}}$ when $\mathfrak{q}\geq1$ and proves
$\widehat{\mathsf{R}}_{\mathtt{P}}\lesssim_\P\mathsf{R}_{\mathtt{P}}$.
\qed

\section{Stochastic Linearization}\label{sec:classical-linearization}

This section specializes the general stochastic linearization from Section \ref{sec:general-linearization} to the classical estimator defined in Sections \ref{sec:classical-first-step} and \ref{sec:classical-partition}. The first-step decomposition in \eqref{eq:first-step-decomposition} separates the bias component $\breve{\beta}_{it}-\beta_{it}$ from the first-step variance component, which determines the generated-regressor term $\mathscr{L}(\beta)$.

\subsection{Decomposition}

Using the first-step decomposition in \eqref{eq:first-step-decomposition}, decompose
\begin{align*}
    \mathscr{L}(\beta)
    &=\mathscr{L}_{\mathtt{B}}(\beta)+\mathscr{L}_{\mathtt{V}}(\beta),
\end{align*}
where
\begin{align*}
    \mathscr{L}_{\mathtt{B}}(\beta)
    &=\frac{1}{T-H} \sum_{t=H+1}^T
    \widehat{p}_{t}(\beta)^\top \widehat{Q}_t^{-1}
    \frac{1}{n_t}\sum_{i=1}^{n_t}
    \widehat{p}_t(\widehat{\beta}_{it}) [\nabla\mu_t(\beta_{it})]^\top
    \big(\breve{\beta}_{it}-\beta_{it}\big),\\
    \mathscr{L}_{\mathtt{V}}(\beta)
    &=\frac{1}{T-H} \sum_{t=H+1}^T
    \widehat{p}_{t}(\beta)^\top \widehat{Q}_t^{-1}
    \frac{1}{n_t}\sum_{i=1}^{n_t}
    \widehat{p}_t(\widehat{\beta}_{it}) [\nabla\mu_t(\beta_{it})]^\top
    \big(\widehat{\beta}_{it}-\breve{\beta}_{it}\big).
\end{align*}

$\mathscr{L}_{\mathtt{B}}(\beta)$ is a bias term related to the bias of the first-step estimation, while $\mathscr{L}_{\mathtt{V}}(\beta)$ contains both a bias component (due to the correlation between $\widehat{p}_t(\widehat{\beta}_{it})$ and $\varepsilon_{i(t-s)}$'s) and a variance component.
Let $A(t)=\E[\widehat{A}(t)]$.
The reference leading term for $\mathscr{L}_{\mathtt{V}}(\beta)$ is
\begin{align*}
    \mathscr{A}(\beta)
    =
    \frac{1}{T-H}\sum_{t=1}^{T}
    \sum_{i=1}^{n_t}w_{it}(\beta)\varepsilon_{it}
\end{align*}
with
\begin{align*}
    w_{it}(\beta)
    &=
    \frac{1}{H}\sum_{s=\max\{1,H+1-t\}}^{\min\{H,T-t\}}
    \frac{1}{n_{t+s}}
    \widehat{p}_{t+s}(\beta)^\top Q_{t+s}^{-1}
    p_{t+s}(\beta_{i(t+s)})[\nabla \mu_{t+s}(\beta_{i(t+s)})]^\top
    \\
    &\hspace{2in}\times
    K\!\Big(\frac{s}{H}\Big)\,
    S A(t+s)^{-1}X_s(t+s)\IF(i\leq n_{t+s}).
\end{align*}

Replacing $\beta_{it}$ by $\widehat{\beta}_{it}$ can move the normalized polynomial coordinates even if both points remain in the same cell when $\mathfrak{q}\geq1$.
The proof below isolates the centered part of this smooth-interior movement and controls it by its uniform bound.
We do not keep this component as a separate leading term: it is absorbed into the linear-approximation remainder.
The remaining difference between $\mathscr{L}_{\mathtt{V}}(\beta)$ and its reference approximation is driven by the conditional mean of the same-cell contribution, boundary switching, endpoint movement, and Gram perturbations.

\subsection{Results}

The contribution of the first-step bias $\breve{\beta}_{it}-\beta_{it}$ is a direct consequence of {Assumption~\ref{ass:first-step-bias}}. The preliminary lemma below controls both the conditional mean and conditional variance of the localized block-basis perturbation generated by the first-step variance term. For $\mathfrak{q}=0$, this perturbation is exactly the cell-switching component. For $\mathfrak{q}\geq1$, the same lemma also covers the smooth same-cell movement of the normalized polynomial coordinates.

\begin{lemma}[Generated-Regressor Basis-Perturbation Bias and Variance]\label{lem:basis-perturbation}
    Suppose that Assumptions \ref{ass:dgp}, \ref{ass:first-step-ls},
    \ref{ass:partition}, and \ref{ass:first-step-bias}
    hold and that $s_\mu\geq2$. Assume further that
    \begin{align*}
        \frac{\max\{J^{2/(q-2)}T^{2/q},T^{2(3+d)/q}\}}{H}
        &=o(1).
    \end{align*}
    Let $P=[\ell,u]\subset\mathcal{B}$ satisfy
    \begin{align*}
        \min_{k\in[d]}\min\{|\ell_k-\lb_k|,|\ub_k-u_k|\}
        &\gg \frac{\log(nT)}{\sqrt{H}},
        &
        J^{-1}
        &\lesssim
        \min_{k\in[d]}(u_k-\ell_k)
        \leq
        \max_{k\in[d]}(u_k-\ell_k)
        \lesssim J^{-1}.
    \end{align*}
    Then, the following results hold.
    \begin{enumerate}[label=\textnormal{(\alph*)}]
        \item Conditional mean:
        \begin{align*}
            &\left\|
            \frac{1}{H}\sum_{s=1}^H
            \E\bigg[
            \Big(p_{\mathfrak{q}}(\widehat{\beta}_{it},P)-p_{\mathfrak{q}}(\beta_{it},P)\Big)
            [\nabla\mu_t(\beta_{it})]^\top
            S \widehat{A}(t)^{-1} X_s(t)
            K\!\Big(\frac{s}{H}\Big) \varsigma_{i(t-s)}\eta_{i(t-s)}
            \bigg|\mathcal{D}_{t-1}\bigg]
            \right\|\\
            &\qquad\lesssim_\P
            \frac{1}{HJ^d}+{\frac{\mathsf{B}_{\mathtt{FS}}}{J^d}}.
        \end{align*}

        \item Conditional variance:
        \begin{align*}
            &\left\|\Var\bigg[
            \frac{1}{H}\sum_{s=1}^H
            \Big(p_{\mathfrak{q}}(\widehat{\beta}_{it},P)-p_{\mathfrak{q}}(\beta_{it},P)\Big)
            [\nabla\mu_t(\beta_{it})]^\top
            S \widehat{A}(t)^{-1} X_s(t)
            K\!\Big(\frac{s}{H}\Big) \varsigma_{i(t-s)}\eta_{i(t-s)}
            \bigg|\mathcal{D}_{t-1}\bigg]\right\|\\
            &\qquad\lesssim_\P
            \frac{\log^d(nT)}{HJ^{d-1}\max\{J,\sqrt{H}\}}
            +\Big(\frac{1}{HJ^d}+{\frac{\mathsf{B}_{\mathtt{FS}}}{J^d}}
            \Big)^2
            +\IF(\mathfrak{q}\geq1)\frac{J^{2-d}}{H^2}.
        \end{align*}
    \end{enumerate}
    These results hold uniformly over $i\in[n_t]$,
    $t=H+1,\ldots,T$, and rectangles $P$ satisfying the conditions imposed above.
\end{lemma}

The next lemma refines only the generated-regressor component of Lemma \ref{lem:general-linearization}. The remaining terms in the stochastic decomposition are already controlled by the general result in Part \ref{part:general}. The refinement extracts the leading term $\mathscr{A}(\beta)$ and absorbs the smooth-interior polynomial movement that arises when $\mathfrak{q}\geq1$ into the linear-approximation remainder.

\begin{lemma}[Classical Generated-Regressor Linearization]\label{lem:classical-linearization}
    Suppose that Assumptions \ref{ass:dgp}, \ref{ass:first-step-ls},
    \ref{ass:partition}, {\ref{ass:interior-separation}}, and
    \ref{ass:first-step-bias} hold
    and that $s_\mu\geq2$. Assume further that
    \begin{align}\label{eq:classical-linearization-rate}
        &\frac{J^d\log(nT)}{n}=o(1),\notag\\
        &\frac{J^{d-1}\max\{J,\sqrt{H}\}\log^2(nT)}{\min\{n^{1-2/q}T^{1-4/q}h,n\}}\lesssim 1, \qquad
        \frac{\max\{J^{2/(q-2)}T^{2/q},T^{2(3+d)/q}\}}{H}=o(1).
    \end{align}
    Then, the bias term satisfies
    \begin{align*}
        \sup_{\beta\in\mathcal{B}^\circ}
        |\mathscr{L}_{\mathtt{B}}(\beta)|
        \lesssim_\P
        {\mathsf{B}_{\mathtt{FS}}},
    \end{align*}
    and the variance terms satisfy
    \begin{align*}
        &\sup_{\beta\in\mathcal{B}^\circ}|\mathscr{A}(\beta)|
        \lesssim_\P\sqrt{\frac{J^d\log (nT)}{nH}},
        &|\mathscr{A}(0)|
        \lesssim_\P
        \sqrt{\frac{J^d}{nT}},
    \end{align*}
    and
    \begin{align*}
        \sup_{\beta\in\mathcal{B}^\circ}
        \Big|\mathscr{L}_{\mathtt{V}}(\beta)-\mathscr{A}(\beta)\Big|
        \lesssim_\P
        \mathsf{R}_{\mathtt{LA}}
    \end{align*}
    with
    \begin{align*}
        \mathsf{R}_{\mathtt{LA}}
        &=
        {
        \frac{1}{H}+\mathsf{B}_{\mathtt{FS}}}\\
        &\qquad
        +\sqrt{\frac{J^{2d}\log^{d+1} (nT)}{nHJ^{d-1}\max\{J,\sqrt{H}\}}}
        +\sqrt{\frac{J^{d+1}\sqrt{\log T/n}\log(nT)}{nH}}
        +\min\{J^d\mathsf{R}_{\mathtt{Q}},1\}
        \sqrt{\frac{J^d\log(nT)}{nH}}\\
        &\qquad
        +\IF(\mathfrak{q}\geq1)
        \sqrt{\frac{J^{d+2}\log(nT)}{nH^2}}
        +\IF(\mathfrak{q}\geq1)
        \sqrt{\frac{J^{d+2}\log T\log(nT)}{n^2H}}.
    \end{align*}
\end{lemma}

\subsection{Discussion}

\begin{enumerate}[label=(\roman*)]
    \item The comparison replaces the generated design in $\mathscr{L}_{\mathtt{V}}(\beta)$ by the reference design. For $\mathfrak{q}=0$, the within-cell basis is constant, so only boundary switching matters. For $\mathfrak{q}\geq1$, movement inside a cell also changes the normalized polynomial coordinates. The proof isolates the centered product of this same-cell movement and the first-step shock multiplier, but the lemma absorbs it into $\mathsf{R}_{\mathtt{LA}}$ using its uniform bound. This conservative treatment avoids relying on additional time-averaging for the same-cell component at the origin.
\end{enumerate}

\subsection{Proof of Lemma~\ref{lem:basis-perturbation}\textnormal{(a)}}

The proof is separated into several steps. We first prove the switching-bias bound, which is the case $\mathfrak{q}=0$. The final step extends the same mean calculation to each component of the localized basis $p_{\mathfrak{q}}(\cdot,P)$.

\medskip\noindent\textit{Step 1: Definitions and notation.}
For any generic rectangle of interest $P=[\ell, u]$ satisfying the boundary restrictions in the lemma, we write $\sqrt{H}(P-c)=[\sqrt{H}(\ell-c), \sqrt{H}(u-c)]$ to be the scaled shifted rectangle.
For any vector $v$, $v_{-j}$ denotes the subvector excluding the $j$-th entry; for any rectangle $\mathcal{R}=[\ell, u]$ in $\mathbb{R}^d$, $\mathcal{R}_{-j}=[\ell_{-j}, u_{-j}]$ in $\mathbb{R}^{d-1}$. Also write $\mathcal{B}_{j}=[\lb_j,\ub_j]$ for the $j$-th coordinate projection of $\mathcal{B}$.

Recall that
\begin{align*}
    \breve{\beta}_{it}
    &=\frac{1}{H}\sum_{l=1}^HS \widehat{A}(t)^{-1} X_l(t)
    K\!\Big(\frac{l}{H}\Big)
    \big(\alpha_{i(t-l)}+\beta_{i(t-l)}^\top f_{t-l}\big)
\end{align*}
with {$\max_{i,t}\|\breve{\beta}_{it}-\beta_{it}\|
\leq\mathsf{B}_{\mathtt{FS}}$ w.p.a.1 by
Assumption~\ref{ass:first-step-bias}}, and that $\mathcal{E}_{t-1}$ contains the predictable paths, factor history, and volatility scales used in the conditioning arguments, but excludes the i.i.d. innovations $\eta_{is}$.

{
For each $i$ and $t$, define the local regular event
\begin{align*}
    \mathcal A_{it,\mathtt{FS}}
    &=
    \left\{
        \|\breve\beta_{it}-\beta_{it}\|
        \leq\mathsf{B}_{\mathtt{FS}}
    \right\}.
\end{align*}
{Assumption~\ref{ass:first-step-bias} and
$\mathcal A_{it,\mathtt{FS}}^c
\subseteq\{\widehat{\mathsf{B}}_{\mathtt{FS}}>\mathsf{B}_{\mathtt{FS}}\}$ give}
\begin{align}
    H^{(d-1)/2}
    \max_{H+1\leq t\leq T}\max_{i\in[n_t]}
    \P\!\left(
        \mathcal A_{it,\mathtt{FS}}^c
        \,\middle|\,\mathcal D_{t-1}
    \right)
    &\lesssim_\P
    \mathsf{B}_{\mathtt{FS}}+H^{-1}.
    \label{eq:basis-perturbation-tail-absorption}
\end{align}
Because
$\sqrt H(\mathsf{B}_{\mathtt{FS}}+H^{-1})=o(1)$, this also gives
\begin{align}
    H^{d/2}
    \max_{H+1\leq t\leq T}\max_{i\in[n_t]}
    \P\!\left(
        \mathcal A_{it,\mathtt{FS}}^c
        \,\middle|\,\mathcal D_{t-1}
    \right)
    &\lesssim_\P
    \sqrt H\,\mathsf{B}_{\mathtt{FS}}+H^{-1/2}
    =o_\P(1).
    \label{eq:basis-perturbation-local-tail}
\end{align}
Every $\mathcal D_{t-1}$-conditional calculation below is split over
$\mathcal A_{it,\mathtt{FS}}$ and its complement. On the regular event, the
signal displacement is bounded by the deterministic
$\mathsf{B}_{\mathtt{FS}}$. On the complement, the conditional convolution
representations below, including their leave-one-out and leave-two-out
versions, are bounded by the corresponding uniformly integrable density
kernels multiplied by
$H^{d/2}\P(\mathcal A_{it,\mathtt{FS}}^c\mid\mathcal D_{t-1})=o_\P(1)$.

More explicitly, the uniform conditional density bound for $\beta_{it}$ gives
\begin{align*}
    \P\!\left(
        \mathcal A_{it,\mathtt{FS}}^c,\,
        \beta_{it}\in P
        \,\middle|\,\mathcal D_{t-1}
    \right)
    &\lesssim_\P
    \min\!\left\{
        \P(\mathcal A_{it,\mathtt{FS}}^c\mid\mathcal D_{t-1}),
        J^{-d}
    \right\},
\end{align*}
while conditional convolution on
$\mathcal A_{it,\mathtt{FS}}^c$ gives
\begin{align*}
    \P\!\left(
        \mathcal A_{it,\mathtt{FS}}^c,\,
        \widehat\beta_{it}\in P
        \,\middle|\,\mathcal D_{t-1}
    \right)
    &\lesssim_\P
    \P(\mathcal A_{it,\mathtt{FS}}^c\mid\mathcal D_{t-1})
    \min\!\left\{1,\frac{H^{d/2}}{J^d}\right\}
    =o_\P(J^{-d}).
\end{align*}
These two bounds cover respectively $J<\sqrt H$ and $J\geq\sqrt H$.
For the conditional mean, the leave-one-out Taylor coefficient and conditional
convolution give the sharper exceptional-event bound
\begin{align*}
    &\frac{1}{\sqrt H}
    \P\!\left(
        \mathcal A_{it,\mathtt{FS}}^c
        \,\middle|\,\mathcal D_{t-1}
    \right)
    \min\!\left\{1,\frac{H^{d/2}}{J^d}\right\}\\
    &\qquad\leq
    \frac{1}{J^d}
    \min\!\left\{
        \frac{J^d}{\sqrt H},H^{(d-1)/2}
    \right\}
    \P\!\left(
        \mathcal A_{it,\mathtt{FS}}^c
        \,\middle|\,\mathcal D_{t-1}
    \right)
    \lesssim_\P
    \frac{\mathsf{B}_{\mathtt{FS}}+H^{-1}}{J^d}
    =\frac{\mathsf{B}_{\mathtt{FS}}}{J^d}
    +\frac{1}{HJ^d},
\end{align*}
by \eqref{eq:basis-perturbation-tail-absorption}. For the conditional variance,
the two preceding cell-probability bounds imply that the exceptional-event
contribution is absorbed by
$\{J^{d-1}\max(J,\sqrt H)\}^{-1}$. Thus the drift terms in the calculations
below are evaluated on $\mathcal A_{it,\mathtt{FS}}$ and are controlled by
$\mathsf{B}_{\mathtt{FS}}$, without treating $\widehat{\mathsf{B}}_{\mathtt{FS}}$ as
$\mathcal D_{t-1}$-measurable. To avoid repeating event indicators, the
$\mathcal A_{it,\mathtt{FS}}$ indicator is suppressed in the regular-event
displacement bounds below; the complementary contribution is always controlled
by the bounds just displayed.
}

Write
\begin{align*}
    c_{itl}
    &=S \widehat{A}(t)^{-1} X_l(t)K\!\Big(\frac{l}{H}\Big)\varsigma_{i(t-l)}\in\mathbb{R}^d,\\
    c_{ts}
    &=S \widehat{A}(t)^{-1} X_s(t)K\!\Big(\frac{s}{H}\Big)\in\mathbb{R}^d.
\end{align*}
Define $G_{its}=\frac{1}{\sqrt{H}}\sum_{\substack{l=1\\l\neq s}}^H
    S \widehat{A}(t)^{-1} X_l(t)K\!\Big(\frac{l}{H}\Big)\varsigma_{i(t-l)}\eta_{i(t-l)}$ and $\delta_{its}=\frac{1}{\sqrt{H}}c_{its}\eta_{i(t-s)}$. Let
\begin{align*}
    F_{G_{its}}(z)
    &=\P\bigg(\frac{1}{\sqrt{H}}\sum_{\substack{l=1\\l\neq s}}^H
    S \widehat{A}(t)^{-1} X_l(t)
    K\!\Big(\frac{l}{H}\Big)\varsigma_{i(t-l)}\eta_{i(t-l)}\leq z\Big|\mathcal{E}_{t-1}, \eta_{i(t-s)}
    \bigg)
\end{align*}
be the conditional CDF of $G_{its}$,
with $f_{G_{its}}(z)$ the corresponding conditional density. Its existence follows from the same characteristic-function argument as in Lemma~\ref{lem:reference-quantile-spacing}.
For any rectangle $[\ell, u]$, $F_{G_{its}}([\ell,u])=\P(G_{its}\in [\ell,u]|\mathcal{E}_{t-1},\eta_{i(t-s)})=\P(G_{its}\in [\ell,u]|\mathcal{E}_{t-1})$ by independence of $\eta_{i(t-s)}$.
Use $f_{\beta_{it}}(\cdot|\mathcal{D}_{t-1})$ to denote the conditional density of $\beta_{it}$ given $\mathcal{D}_{t-1}$, which is possibly heterogeneous over $i$ and $t$.

By independence, the (conditional) characteristic function of $G_{its}$ can be written as
\begin{align*}
    \varphi_{G_{its}}(z)=\prod_{\substack{l=1\\l\neq s}}^H\varphi_\eta(c_{itl}^\top z/\sqrt{H}),\quad\text{for}\;z\in\mathbb{R}^d
\end{align*}
where $\varphi_\eta(z)=\E[e^{iz\eta}]$ is the characteristic function of a generic random variable $\eta$ that follows the same distribution as $\eta_{it}$.
By the inversion formula,
\begin{align*}
    f_{G_{its}}(z)=\frac{1}{(2\pi)^d}\int_{\mathbb{R}^d}e^{-iu^\top z}\varphi_{G_{its}}(u)\d{u}, \quad\text{for } z\in\mathbb{R}^d.
\end{align*}
Its derivative is
\begin{align*}
    \nabla f_{G_{its}}(z)=\frac{1}{(2\pi)^d}\int_{\mathbb{R}^d}(-iu)e^{-iu^\top z}\varphi_{G_{its}}(u)\d{u}.
\end{align*}

Before we prove the main results, we need some properties of the density $f_{G_{its}}$. We summarize them in the following claim: with probability approaching one, uniformly over $i\in[n_t]$, $t=H+1,\ldots,T$, $s\in[H]$, and $k=0,1,2,3$,
\begin{align}\label{eq:first-step-density-bounds}
    &\sup_{z\in\mathbb{R}^d}\|\nabla^kf_{G_{its}}(z)\|_\infty\lesssim 1, \qquad
    \|\nabla^k f_{G_{its}}(z)\|_\infty\lesssim \|z\|^{-d-1}\quad\text{for }\|z\|\geq1.
\end{align}
We will prove this claim later in Step 5.

\medskip\noindent\textit{Step 2: Key expansion.}
\begin{align*}
    &\E\bigg[\Big(\IF(\widehat{\beta}_{it}\in P)-\IF(\beta_{it}\in P)\Big)
    (\nabla\mu_t(\beta_{it}))^\top \Big( S \widehat{A}(t)^{-1} X_s(t) \Big)
    K\!\Big(\frac{s}{H}\Big) \varsigma_{i(t-s)}\eta_{i(t-s)}\Big|\mathcal{D}_{t-1}\bigg]\\
    &=\E\bigg[(\nabla\mu_t(\beta_{it}))^\top c_{its}\E\bigg[\IF(\widehat{\beta}_{it}\in P)
     \eta_{i(t-s)}\bigg|\mathcal{E}_{t-1}\bigg]\bigg|\mathcal{D}_{t-1}\bigg]\\
    &=\E\bigg[(\nabla\mu_t(\beta_{it}))^\top c_{its}\E\bigg[\IF\bigg(
     \frac{1}{\sqrt{H}}\sum_{\substack{l=1\\l\neq s}}^Hc_{itl}\eta_{i(t-l)}
     \in \sqrt{H}\Big(P-\breve{\beta}_{it}
    -\frac{1}{H}c_{its}\eta_{i(t-s)}\Big)\bigg)
    \eta_{i(t-s)}\\
    &\hspace{12em}\bigg|\mathcal{E}_{t-1}, \eta_{i(t-s)}\bigg]\bigg|\mathcal{D}_{t-1}\bigg]\\
    &=\E\bigg[(\nabla\mu_t(\beta_{it}))^\top c_{its}\eta_{i(t-s)}
    \Big(F_{G_{its}}\Big(\sqrt{H}(P-\breve{\beta}_{it})\Big)\\
    &\hspace{11em}+F_{G_{its}}\Big(\sqrt{H}(P-\breve{\beta}_{it}
    -\frac{1}{H}c_{its}\eta_{i(t-s)})\Big)
    -F_{G_{its}}\Big(\sqrt{H}(P-\breve{\beta}_{it})\Big)\Big)\bigg|\mathcal{D}_{t-1}\bigg]\\
    &=\E\bigg[(\nabla\mu_t(\beta_{it}))^\top c_{its}\eta_{i(t-s)}
    \Big(F_{G_{its}}\Big(\sqrt{H}(P-\breve{\beta}_{it}
    -\frac{1}{H}c_{its}\eta_{i(t-s)})\Big)
    -F_{G_{its}}\Big(\sqrt{H}(P-\breve{\beta}_{it})\Big)\Big)\bigg|\mathcal{D}_{t-1}\bigg]
\end{align*}
by the independence between $\eta_{i(t-s)}$ and variables in $\mathcal{E}_{t-1}$.

Now, by the density bound \eqref{eq:first-step-density-bounds}, we further have
\begin{align*}
    &\E\bigg[(\nabla\mu_t(\beta_{it}))^\top c_{its}\E\bigg[\IF(\widehat{\beta}_{it}\in P)
     \eta_{i(t-s)}\bigg|\mathcal{E}_{t-1}\bigg]\bigg|\mathcal{D}_{t-1}\bigg]\\
    &=\E\bigg[(\nabla\mu_t(\beta_{it}))^\top c_{its}\eta_{i(t-s)}
     \int_{\sqrt{H}(P-\breve{\beta}_{it})}(f_{G_{its}}(z-\delta_{its})-f_{G_{its}}(z))\d{z}
     \bigg|\mathcal{D}_{t-1}\bigg]\\
    &=\E\bigg[(\nabla\mu_t(\beta_{it}))^\top c_{its}\eta_{i(t-s)}\bigg[
    -\delta_{its}^\top\int_{\sqrt{H}(P-\breve{\beta}_{it})}\nabla f_{G_{its}}(z) \d{z}\\
    &\hspace{7em}+\delta_{its}^\top\int_{\sqrt{H}(P-\breve{\beta}_{it})}\int_0^1
    \Big(\nabla f_{G_{its}}(z) - \nabla f_{G_{its}}(z-v\delta_{its})\Big)\d{v}\d{z}\bigg]\bigg|\mathcal{D}_{t-1}\bigg]\\
    &=-\frac{1}{\sqrt{H}}\E\bigg[(\nabla\mu_t(\beta_{it}))^\top c_{its}c_{its}^\top
    \int_{\sqrt{H}(P-\breve{\beta}_{it})}\nabla f_{G_{its}}(z) dz\bigg|\mathcal{D}_{t-1}\bigg]\\
    &\quad +\frac{1}{H}\E\bigg[(\nabla\mu_t(\beta_{it}))^\top c_{its}c_{its}^\top
    \E\bigg[\eta_{i(t-s)}^3\int_{\sqrt{H}(P-\breve{\beta}_{it})}\int_0^1
    (1-v)\nabla^2 f_{G_{its}}(z-v\delta_{its})c_{its}\d{v}\d{z}\bigg|\mathcal{E}_{t-1}\bigg]\bigg|\mathcal{D}_{t-1}\bigg]\\
    &=-\frac{1}{\sqrt{H}}I_{1,s}+\frac{1}{H}I_{2,s},
\end{align*}
where $I_{1,s}$ and $I_{2,s}$ denote the corresponding terms.

\medskip\noindent\textit{Step 3: Bound $I_{1,s}$.}
\begin{align*}
    &\left|
    \E\bigg[(\nabla\mu_t(\beta_{it}))^\top c_{its}c_{its}^\top
    \int_{\sqrt{H}(P-\breve{\beta}_{it})}\nabla f_{G_{its}}(z)\d{z}
    \bigg|\mathcal{D}_{t-1}\bigg]
    \right|\\
    &\quad =
    \left|
    \E\bigg[(\nabla\mu_t(\beta_{it}))^\top c_{ts}c_{ts}^\top
    \varsigma_{i(t-s)}^2
    \int_{\sqrt{H}(P-\breve{\beta}_{it})}\nabla f_{G_{its}}(z)\d{z}
    \bigg|\mathcal{D}_{t-1}\bigg]
    \right|\\
    &=|c_{ts}^\top M c_{ts}|\\
    &\leq\|c_{ts}\|^2 \|M\|.
\end{align*}
where
\begin{align*}
    M
    &=\E\bigg[\varsigma_{i(t-s)}^2
    \int_{\sqrt{H}(P-\breve{\beta}_{it})}\nabla f_{G_{its}}(z) \d{z} \;
    (\nabla\mu_t(\beta_{it}))^\top \bigg|\mathcal{D}_{t-1}\bigg].
\end{align*}
It is enough to consider each entry of $M$, since $d$ is fixed.

A typical $(j,k)$-th entry of $M$ is
\begin{align*}
    &\E\bigg[\varsigma_{i(t-s)}^2
    \int_{\sqrt{H}(P-\breve{\beta}_{it})}\nabla_j f_{G_{its}}(z)\d{z} \;
    \nabla_k\mu_t(\beta_{it})\bigg|\mathcal{D}_{t-1}\bigg]\\
    &=\E\bigg[\varsigma_{i(t-s)}^2
    \int_{\sqrt{H}(P-\beta_{it})}\nabla_j f_{G_{its}}(z)\d{z} \;
    \nabla_k\mu_t(\beta_{it})\bigg|\mathcal{D}_{t-1}\bigg]\\
    &\quad +\E\bigg[\varsigma_{i(t-s)}^2
    \int_{\sqrt{H}(P-\breve{\beta}_{it})}\nabla_j f_{G_{its}}(z)\d{z} \;
    \nabla_k\mu_t(\beta_{it})
    -\varsigma_{i(t-s)}^2
    \int_{\sqrt{H}(P-\beta_{it})}\nabla_j f_{G_{its}}(z)\d{z} \;
    \nabla_k\mu_t(\beta_{it})\bigg|\mathcal{D}_{t-1}\bigg].
\end{align*}

For the integration by parts in $b_j$, write $f_{G_{its}}(z)\equiv f_{G_{its}}(z_j;z_{-j})$ and define
\begin{align*}
    m(b_j;z_{-j},b_{-j})
    &=\int_{0}^{\sqrt{H}(u_{j}-\ell_{j})}
    f_{G_{its}}(z+\sqrt{H}(\ell-b))\d{z}_j.
\end{align*}
On the one hand,
\begin{align*}
    &\quad\;\E\bigg[\varsigma_{i(t-s)}^2
    \int_{\sqrt{H}(P-\beta_{it})}\nabla_j f_{G_{its}}(z)\d{z} \;
    \nabla_k\mu_t(\beta_{it})\bigg|\mathcal{D}_{t-1}\bigg]\\
    &=\varsigma_{i(t-s)}^2\int_{\mathcal{B}}
    \Big(\int_{0}^{\sqrt{H}(u-\ell)}
    \nabla_j f_{G_{its}}(z+\sqrt{H}(\ell-b))\d{z}\Big)\\
    &\qquad\qquad\qquad\times
    \nabla_k\mu_t(b)
    f_{\beta_{it}}(b|\mathcal{D}_{t-1})\d{b}\\
    &=\varsigma_{i(t-s)}^2
    \int_{\mathcal{B}_{-j}}
    \int_{0}^{\sqrt{H}(u_{-j}-\ell_{-j})}
    \int_{\mathcal{B}_{j}}
    \Big(\int_{0}^{\sqrt{H}(u_{j}-\ell_{j})}\nabla_j f_{G_{its}}(z+\sqrt{H}(\ell-b))
    \;\d{z}_j\Big)\\
    &\qquad\qquad\qquad\times
    \nabla_k\mu_t(b)
    f_{\beta_{it}}(b|\mathcal{D}_{t-1})\d{b}_j\d{z}_{-j}\d{b}_{-j}\\
    &=-\frac{\varsigma_{i(t-s)}^2}{\sqrt{H}}
    \int_{\mathcal{B}_{-j}}\int_{0}^{\sqrt{H}(u_{-j}-\ell_{-j})}
    m(b_j;z_{-j},b_{-j})
    \nabla_k\mu_t(b)\\
    &\qquad\qquad\qquad\times
    f_{\beta_{it}}(b|\mathcal{D}_{t-1})
    \Big|_{\underline{\beta}_j}^{\bar{\beta}_j}
    \d{z}_{-j}\d{b}_{-j}\\
    &\quad +\frac{\varsigma_{i(t-s)}^2}{\sqrt{H}}
    \int_{\mathcal{B}_{-j}}
    \int_{0}^{\sqrt{H}(u_{-j}-\ell_{-j})}
    \int_{\mathcal{B}_{j}}
    \Big(\int_{0}^{\sqrt{H}(u_{j}-\ell_{j})}f_{G_{its}}(z+\sqrt{H}(\ell-b))
    \;\d{z}_j\Big)\\
    &\qquad\qquad\qquad\times
    \nabla_j\Big(\nabla_{k}\mu_t(b)
    f_{\beta_{it}}(b|\mathcal{D}_{t-1})\Big)\d{b}_j\d{z}_{-j}\d{b}_{-j}\\
    &= O\Big(\frac{1}{\sqrt{H}J^d}\Big)\\
    &\quad +\frac{\varsigma_{i(t-s)}^2}{\sqrt{H}}
    \int_{\mathcal{B}_{-j}}
    \int_{0}^{\sqrt{H}(u_{-j}-\ell_{-j})}
    \int_{\mathcal{B}_{j}}
    \Big(\int_{0}^{\sqrt{H}(u_{j}-\ell_{j})}f_{G_{its}}(z+\sqrt{H}(\ell-b))
    \;\d{z}_j\Big)\\
    &\qquad\qquad\qquad\times
    \nabla_j\Big(\nabla_{k}\mu_t(b)
    f_{\beta_{it}}(b|\mathcal{D}_{t-1})\Big)\d{b}_j\d{z}_{-j}\d{b}_{-j}\\
    &= O\Big(\frac{1}{\sqrt{H}J^d}\Big)
    +\frac{\varsigma_{i(t-s)}^2}{\sqrt{H}}
    \int_{\sqrt{H}(P-\mathcal{B})}f_{G_{its}}(z)
    \Big(\int_{\mathcal{B}\cap(P-z/\sqrt{H})}
    \nabla_j\Big(\nabla_{k}\mu_t(b)
    f_{\beta_{it}}(b|\mathcal{D}_{t-1})\Big)\d{b}\Big)\d{z}\\
    &\lesssim_\P \frac{1}{\sqrt{H}J^{d}}
\end{align*}
where the inner integral in the last display is $\lesssim_\P J^{-d}$ uniformly. The above calculation uses the bounded derivatives of $\mu_t$ and the bounded Lipschitz conditional density $f_{\beta_{it}}(\cdot|\mathcal{D}_{t-1})$; the integration-by-parts step is understood coordinatewise using the corresponding bounded weak derivative. We also use that $[\ell, u]$ is at least of order $\log(nT)/\sqrt{H}$ away from the boundary of $\mathcal{B}$.
To see the boundary term, define also
\begin{align*}
    \tilde{m}(b_j,z_{-j})
    &=\int_{0}^{\sqrt{H}(u_{j}-\ell_{j})}
    f_{G_{its}}(z_j+\sqrt{H}(\ell_j-b_j);z_{-j})\d{z}_j.
\end{align*}
\begin{align*}
    &-\frac{1}{\sqrt{H}}
    \int_{\mathcal{B}_{-j}}\int_{0}^{\sqrt{H}(u_{-j}-\ell_{-j})}
    m(b_j;z_{-j},b_{-j})
    \nabla_k\mu_t(b)f_{\beta_{it}}(b|\mathcal{D}_{t-1})
    \Big|_{\underline{\beta}_j}^{\bar{\beta}_j}
    \d{z}_{-j}\d{b}_{-j}\\
    &=\frac{1}{\sqrt{H}}\int_{\sqrt{H}(P_{-j}-\mathcal{B}_{-j})}
    \int_{\mathcal{B}_{-j}\cap(P_{-j}-z_{-j}/\sqrt{H})}\bigg[
    \tilde{m}(\underline{\beta}_j,z_{-j})
    \nabla_k\mu_t(\underline{\beta}_j,b_{-j})f_{\beta_{it}}(\underline{\beta}_j, b_{-j}|\mathcal{D}_{t-1})\\
    &\hspace{10em}-\tilde{m}(\bar{\beta}_j,z_{-j})\nabla_k\mu_t(\bar{\beta}_j, b_{-j})
    f_{\beta_{it}}(\bar{\beta}_j, b_{-j}|\mathcal{D}_{t-1})\bigg]\d{b}_{-j}\d{z}_{-j}\\
    &\lesssim\frac{1}{\sqrt{H}}\,
    \bigg(
    \int_{\sqrt{H}(P_{-j}-\mathcal{B}_{-j})}
    \int_{0}^{\sqrt{H}(u_{j}-\ell_{j})}
    f_{G_{its}}(z_j+\sqrt{H}(\ell_j-\underline{\beta}_j);z_{-j})
    \d{z}_j\d{z}_{-j}\\
    &\qquad\qquad+
    \int_{\sqrt{H}(P_{-j}-\mathcal{B}_{-j})}
    \int_{0}^{\sqrt{H}(u_{j}-\ell_{j})}
    f_{G_{its}}(z_j+\sqrt{H}(\ell_j-\bar{\beta}_j);z_{-j})
    \d{z}_j\d{z}_{-j}
    \bigg)\frac{1}{J^{d-1}}\\
    &\lesssim
    \frac{1}{\sqrt{H}J^{d-1}}
    \P(|G_{its,j}|>\tau_n|\mathcal{E}_{t-1})
    \lesssim_\P\frac{1}{\sqrt{H}J^d},
\end{align*}
where $G_{its,j}$ denotes the $j$-th coordinate of $G_{its}$ and $\tau_n\gtrsim \log(nT)$ denotes the distance between $[\ell,u]$ and the boundary of the support of $\beta_{it}$.
The last inequality follows from Bernstein's inequality and the usual truncation argument: on an $\mathcal{E}_{t-1}$-measurable event that happens w.p.a.1,
each innovation is truncated at $M$ and recentered before applying Bernstein, while Chebyshev's inequality is applied to the centered complementary tail sum.
\begin{align*}
    \P(|G_{its,j}|\geq \tau_n|\mathcal{E}_{t-1})\leq\exp\Big(-\frac{0.5\tau_n^2}{M_0+\tau_nM/3\sqrt{H}}\Big)
    +\frac{1}{\tau_n^2M^{q-2}}\lesssim J^{-1}\qquad \text{for } M\asymp J^{\frac{1}{q-2}}.
\end{align*}
because $J^{1/(q-2)}T^{1/q}/\sqrt{H}=o(1)$ and $\tau_n\gtrsim \log(nT)$.

On the other hand, we will use a partitioning argument to bound the remainder in $I_{1,s}$.
Specifically, if $J\gtrsim H^{1/2}$, we can introduce a multi-index $m\in\mathbb{Z}^d$ and $\hbar\asymp H^{-1/2}$ and define $\mathcal{B}_m=\{b\in\mathcal{B}: b_j\in[\ell_j+m_j\hbar, \ell_j+(m_j+1)\hbar), j\in[d]\}$.
Denote this collection of subrectangles by $\{\mathcal{B}_m: m\in\mathcal{M}\}$ for an index set $\mathcal{M}$.
Since $\mathcal{B}$ is a bounded rectangle, we need at most a constant multiple of $H^{d/2}$ such subrectangles to cover the entire $\mathcal{B}$, i.e., $|\mathcal{M}|\lesssim H^{d/2}$.
Then, we have
\begin{align*}
    &\E\bigg[\varsigma_{i(t-s)}^2
    \int_{\sqrt{H}(P-\breve{\beta}_{it})}\nabla_j f_{G_{its}}(z)\d{z} \;
    \nabla_k\mu_t(\beta_{it})
    -\varsigma_{i(t-s)}^2\int_{\sqrt{H}(P-\beta_{it})}\nabla_j f_{G_{its}}(z)\d{z} \;
    \nabla_k\mu_t(\beta_{it})\bigg|\mathcal{D}_{t-1}\bigg]\\
    &=\varsigma_{i(t-s)}^2\E\bigg[
    \int_{\sqrt{H}(P-\beta_{it})}\Big(\nabla_j f_{G_{its}}(z+\sqrt{H}(\beta_{it}-\breve{\beta}_{it}))-\nabla_j f_{G_{its}}(z)\Big)\d{z} \;
    \nabla_k\mu_t(\beta_{it})\bigg|\mathcal{D}_{t-1}\bigg]\\
    &\lesssim
    \sum_{m\in\mathcal{M}}\E\bigg[\IF(\beta_{it}\in\mathcal{B}_m)
    \Big|\int_{\sqrt{H}(P-\beta_{it})}\Big(\nabla_j f_{G_{its}}(z+\sqrt{H}(\beta_{it}-\breve{\beta}_{it}))-\nabla_j f_{G_{its}}(z)\Big)\d{z} \;\Big|
    \bigg|\mathcal{D}_{t-1}\bigg]\\
    &\lesssim {\frac{\sqrt{H}\mathsf{B}_{\mathtt{FS}}}{J^d}}.
\end{align*}

In the last step we use the following fact: for the first term,
\begin{align*}
    &\sum_{m\in\mathcal{M}}\E\bigg[\IF(\beta_{it}\in\mathcal{B}_m)
    \Big|\int_{\sqrt{H}(P-\beta_{it})}\Big(\nabla_j f_{G_{its}}(z-\sqrt{H}(\breve{\beta}_{it}-\beta_{it}))-\nabla_j f_{G_{its}}(z)\Big)\d{z}\Big| \bigg|\mathcal{D}_{t-1}\bigg]\\
    &=\sum_{m\in\mathcal{M}}\E\bigg[\IF(\beta_{it}\in\mathcal{B}_m)
    \Big|\int_{\sqrt{H}(P-\beta_{it})}[\nabla(\nabla_j f_{G_{its}})(\xi_z)]^\top(\sqrt{H}(\beta_{it}-\breve{\beta}_{it}))\d{z}\Big| \bigg|\mathcal{D}_{t-1}\bigg]\\
    &\lesssim\sum_{m\in\mathcal{M}}(1+\|m\|)^{-d-1}
    (\sqrt{H}/J)^d {\sqrt{H}\mathsf{B}_{\mathtt{FS}}}
    \E\Big[\IF(\beta_{it}\in\mathcal{B}_m)\Big|\mathcal{D}_{t-1}\Big]\lesssim \frac{H^{d/2}}{J^d}\frac{1}{H^{d/2}}{\sqrt{H}\mathsf{B}_{\mathtt{FS}}}
\end{align*}
where $\xi_z$ is some point between $z$ and $z-\sqrt{H}(\breve{\beta}_{it}-\beta_{it})$ and we use the tail bound for $\nabla^2f_{G_{its}}$.

Now, consider the case $J=o(H^{1/2})$.
We slightly redefine the partition: divide the whole support $\mathcal{B}$ into rectangles of length proportional to $H^{-1/2}$, labelling these regions by a multi-index
\begin{align*}
    m
    &=(m_1, m_1', m_2, m_2',\ldots, m_d, m_d')\in\mathbb{Z}^{2d}.
\end{align*}
Define $\mathcal{B}_m$ as the set of $b\in\mathcal{B}$ such that $\sqrt{H}(b_k-\ell_k)\in[m_k,m_k+1]$ and $\sqrt{H}(b_k-u_k)\in[m'_k,m'_k+1]$ for all $k\in[d]$. That is, when a rectangle is associated with a multi-index with some $m_k$ or $m'_k$ large in absolute value, this region is far away from the facets of the hyper-rectangle $[\ell, u]$. Let $\mathcal{M}$ be the collection of possible indices $m$. Again, $|\mathcal{M}|\lesssim H^{d/2}$.
For each $\mathcal{B}_m$, let $\mathcal{J}_m=\{j\in[d]: \exists b\in\mathcal{B}_m\;\text{s.t.}\; b_j\in [\ell_j, u_j]\}$ and $\mathcal{J}_m^c=[d]\setminus\mathcal{J}_m$. For $\beta_{it}\in\mathcal{B}_m$, we have
\begin{align*}
    \int_{\sqrt{H}(P-\beta_{it})}\|\nabla(\nabla_j f_{G_{its}})(\xi_z)\|_\infty\d{z}
    \lesssim \frac{1}{\prod_{j\in\mathcal{J}_m^c}(1+|\min\{m_j,m_j'\}|)^{1+1/d}}\times \Big(\frac{\sqrt{H}}{J}\Big)^{|\mathcal{J}_m^c|}.
\end{align*}
Note that for a fixed $j_0\in[d]$, all $\mathcal{B}_m$'s such that $|\mathcal{J}_m|=j_0$ can be classified into $\lesssim(\sqrt{H}/J)^{j_0}$ groups, each containing $\lesssim\sqrt{H}$ subrectangles.
Repeating the previous argument, we have
\begin{align*}
    &\sum_{m\in\mathcal{M}}\E\bigg[\IF(\beta_{it}\in\mathcal{B}_m)
    \Big|\int_{\sqrt{H}(P-\beta_{it})}[\nabla(\nabla_j f_{G_{its}})(\xi_z)]^\top(\sqrt{H}(\beta_{it}-\breve{\beta}_{it}))\d{z}\Big| \bigg|\mathcal{D}_{t-1}\bigg]\\
    &\lesssim\sum_{j_0=1}^d\Big(\frac{\sqrt{H}}{J}\Big)^{j_0}\times\Big(\frac{\sqrt{H}}{J}\Big)^{d-j_0}{\sqrt{H}\mathsf{B}_{\mathtt{FS}}}\times \frac{1}{H^{d/2}}
    \lesssim{\frac{\sqrt{H}\mathsf{B}_{\mathtt{FS}}}{J^d}}.
\end{align*}

Then, we conclude that
\begin{align*}
    &\frac{1}{H}\sum_{s=1}^HI_{1,s}\lesssim_\P \frac{1}{\sqrt{H}J^d}
    +{\frac{\sqrt{H}\mathsf{B}_{\mathtt{FS}}}{J^d}}.
\end{align*}
\medskip

\medskip\noindent\textit{Step 4: Bound $I_{2,s}$.}
We use a similar argument:
\begin{align*}
    &\E\bigg[(\nabla\mu_t(\beta_{it}))^\top c_{its}c_{its}^\top
    \E\bigg[\eta_{i(t-s)}^3\int_{\sqrt{H}(P-\breve{\beta}_{it})}\int_0^1
    (1-v)\nabla^2 f_{G_{its}}(z-v\delta_{its})c_{its}\d{v}\d{z}\bigg|\mathcal{E}_{t-1}\bigg]\bigg|\mathcal{D}_{t-1}\bigg]\\
    &=c_{ts}^\top
    \E\bigg[\eta_{i(t-s)}^3\varsigma_{i(t-s)}^3\int_{\sqrt{H}(P-\breve{\beta}_{it})}\int_0^1
    (1-v)\nabla^2 f_{G_{its}}(z-v\delta_{its})c_{ts}\d{v}\d{z}(\nabla\mu_t(\beta_{it}))^\top \bigg|\mathcal{D}_{t-1}\bigg]c_{ts}.
\end{align*}
Again, we only need to bound the operator norm of the matrix in the middle. Furthermore,
\begin{align*}
    &\ve\bigg(\E\bigg[\eta_{i(t-s)}^3\varsigma_{i(t-s)}^3\int_{\sqrt{H}(P-\breve{\beta}_{it})}\int_0^1
        (1-v)\nabla^2 f_{G_{its}}(z-v\delta_{its})c_{ts}\d{v}\d{z}(\nabla\mu_t(\beta_{it}))^\top \bigg|\mathcal{D}_{t-1}\bigg]\bigg)\\
        &=\E\bigg[\nabla\mu_t(\beta_{it})\otimes \eta_{i(t-s)}^3\varsigma_{i(t-s)}^3
        \int_{\sqrt{H}(P-\breve{\beta}_{it})}\int_0^1
        (1-v)\nabla^2 f_{G_{its}}(z-v\delta_{its})\d{v}\d{z}\bigg|\mathcal{D}_{t-1}\bigg]c_{ts}.
\end{align*}
We only need to bound each entry of the expected matrix (a $d^2\times d$ matrix).
Consider a generic entry: for some $j$, $j'$, $j''$,
\begin{align*}
    &\E\bigg[\nabla_{j}\mu_t(\beta_{it})\eta_{i(t-s)}^3\varsigma_{i(t-s)}^3
    \int_{\sqrt{H}(P-\breve{\beta}_{it})}\int_0^1
    (1-v)\nabla_{j',j''}^2 f_{G_{its}}(z-v\delta_{its})\d{v}\d{z}\bigg|\mathcal{D}_{t-1}\bigg]\\
    &=\varsigma_{i(t-s)}^3\int_{\mathcal{B}}\nabla_{j}\mu_t(b)
    \int_{\sqrt{H}(P-b)}\E\Big[\eta_{i(t-s)}^3\int_0^1
    (1-v)\nabla_{j',j''}^2 f_{G_{its}}(z-v\delta_{its})\d{v}\Big|\mathcal{E}_{t-1}\Big]\d{z}
    f_{\beta_{it}}(b|\mathcal{D}_{t-1})\d{b}\\
    &+\E\bigg[\nabla_{j}\mu_t(\beta_{it})\bigg(\varsigma_{i(t-s)}^3
    \int_{\sqrt{H}(P-\breve{\beta}_{it})}\eta_{i(t-s)}^3\int_0^1
    (1-v)\nabla_{j',j''}^2 f_{G_{its}}(z-v\delta_{its})\d{v}\d{z}\\
    &\hspace{7em}-\varsigma_{i(t-s)}^3
    \int_{\sqrt{H}(P-\beta_{it})}\eta_{i(t-s)}^3\int_0^1
    (1-v)\nabla_{j',j''}^2 f_{G_{its}}(z-v\delta_{its})\d{v}\d{z}\bigg)
    \bigg|\mathcal{D}_{t-1}\bigg].
\end{align*}

Let $g_\eta$ denote the Lebesgue density of $\eta_{it}$ and let $M_0>0$ be fixed. For the first term,
{\footnotesize
\begin{align*}
    &\quad\int_{\mathcal{B}}\nabla_{j}\mu_t(b)
    \int_{\sqrt{H}(P-b)}\E\Big[\eta_{i(t-s)}^3\int_0^1
    (1-v)\nabla_{j',j''}^2 f_{G_{its}}(z-v\delta_{its})\d{v}\Big|\mathcal{E}_{t-1}\Big]\d{z}\, f_{\beta_{it}}(b|\mathcal{D}_{t-1})\d{b}\\
    &=\;\int_{\mathbb{R}}\eta^3g_\eta(\eta)
    \int_0^1
    (1-v)\int_{\mathcal{B}}\int_{\sqrt{H}(P-b)}\nabla_{j}\mu_t(b)
    \nabla_{j',j''}^2 f_{G_{its}}(z-v\delta_{its})f_{\beta_{it}}(b|\mathcal{D}_{t-1})\d{z}\d{b}\times\\
    &\hspace{12em}\Big(\IF(\|c_{ts}\||\eta|\leq M_0\sqrt{H})+\IF(\|c_{ts}\||\eta|>M_0\sqrt{H})\Big)\d{v}\d{\eta}\\
    &\lesssim_\P J^{-d}.
\end{align*}
}
Note that for $\|c_{ts}\||\eta|\leq M_0\sqrt{H}$, $\|v\delta_{its}\|\lesssim \|c_{ts}\||\eta|/\sqrt{H}\lesssim 1$, and thus in the last line we apply the previous partitioning argument again to obtain
\begin{align*}
    &\Big|\int_{\mathcal{B}}\int_{\sqrt{H}(P-b)}\nabla_{j}\mu_t(b)
    \nabla_{j',j''}^2 f_{G_{its}}(z-v\delta_{its})f_{\beta_{it}}(b|\mathcal{D}_{t-1})\d{z}\d{b}\,
    \IF(\|c_{ts}\||\eta|\leq M_0\sqrt{H})\Big|\\
    &\lesssim\int_{\mathcal{B}}\int_{\sqrt{H}(P-b)}
    |\nabla_{j',j''}^2 f_{G_{its}}(z-v\delta_{its})|\d{b}\d{z}\IF(\|c_{ts}\||\eta|\leq M_0\sqrt{H})\\
    &=\sum_{m\in\mathcal{M}}\int_{\mathcal{B}_m}\int_{\sqrt{H}(P-b)}
    |\nabla_{j',j''}^2 f_{G_{its}}(z-v\delta_{its})|\IF(\|c_{ts}\||\eta|\leq M_0\sqrt{H})\d{b}\d{z}\\
    &\lesssim\; J^{-d}.
\end{align*}

On the other hand,
\begin{align*}
    &\int_{\mathcal{B}}\nabla_{j}\mu_t(b)
    \int_{\sqrt{H}(P-b)}\E\Big[\eta_{i(t-s)}^3\int_0^1
    (1-v)\nabla_{j',j''}^2 f_{G_{its}}(z-v\delta_{its})\d{v}\\
    &\hspace{9em}\times
    \IF(\|c_{ts}\||\eta_{i(t-s)}|> M_0\sqrt{H})\Big|\mathcal{E}_{t-1}\Big]\d{z}\,
    f_{\beta_{it}}(b|\mathcal{D}_{t-1})\d{b}\\
    &\lesssim\;\E[|\eta_{i(t-s)}|^{3+d}]\|c_{ts}\|^dH^{-d/2}
    \times\Big(\frac{\sqrt{H}}{J}\Big)^d\\
    &\lesssim \|c_{ts}\|^dJ^{-d}.
\end{align*}

In addition, repeating the previous argument in Step 3 yields
\begin{align*}
    &\;\E\bigg[\nabla_{j}\mu_t(\beta_{it})\bigg(\varsigma_{i(t-s)}^3
    \int_{\sqrt{H}(P-\breve{\beta}_{it})}\eta_{i(t-s)}^3\int_0^1
    (1-v)\nabla_{j',j''}^2 f_{G_{its}}(z-v\delta_{its})\d{v}\d{z}\\
    &\hspace{5em}-\varsigma_{i(t-s)}^3
    \int_{\sqrt{H}(P-\beta_{it})}\eta_{i(t-s)}^3\int_0^1
    (1-v)\nabla_{j',j''}^2 f_{G_{its}}(z-v\delta_{its})\d{v}\d{z}\bigg)
    \bigg|\mathcal{D}_{t-1}\bigg]\\
    &\lesssim\;
    {\frac{\sqrt{H}\mathsf{B}_{\mathtt{FS}}}{J^d}}.
\end{align*}
Therefore, we conclude that
\begin{align*}
    \frac{1}{H}\sum_{s=1}^HI_{2,s}\lesssim_\P \frac{1}{J^{d}}.
\end{align*}
Also, note that in the above calculation we need $\frac{1}{H}\sum_{s=1}^H\|c_{ts}\|^{3+d}\lesssim_\P 1$.

\medskip\noindent\textit{Step 5: Density bounds for $G_{its}$.}
We prove \eqref{eq:first-step-density-bounds}.

\smallskip\noindent\textit{Step 5.1: Factor moments and projection lower bound.}
The density bounds use two properties, both holding with probability approaching one. First,
\begin{align*}
    \max_{H+1\leq t\leq T}
    \frac{1}{H}\sum_{s=1}^H|f_{t-s}|^m
    &\lesssim 1,
    \qquad m=3+d.
\end{align*}
This is guaranteed by the imposed moment conditions and the maximal Fuk--Nagaev inequality described in Lemma~\ref{lem:first-step-gram}. The process $|f_{t-s}|^m$ is $L_{q/m}$-NED with $q/m>2$, so
\begin{align*}
    \max_{H+1\leq t\leq T}
    \left|
    \frac{1}{H}\sum_{s=1}^H
    \left(|f_{t-s}|^m-\E[|f_{t-s}|^m]\right)
    \right|
    &=o_\P(1)
\end{align*}
by the same argument as in the proof of Lemma \ref{lem:first-step-gram}. Second, for $v\in\mathbb{R}^d$,
$m_v=|\{l: |v^\top c_{itl}|\geq a\|v\|\}|\geq \pi_0(H-1)$ for a constant $a>0$ and a positive sequence $\pi_0$ with $\pi_0H\rightarrow\infty$. To see this,
note that $\frac{1}{H}\sum_{\substack{l=1\\l\neq s}}^H|v^\top c_{itl}|^2/\|v\|^2\geq \lambda_{\min}(\frac{1}{H}\sum_{\substack{l=1\\l\neq s}}^Hc_{itl}c_{itl}^\top)\geq\underline{\lambda}$ for some $\underline{\lambda}$, w.p.a.1.
Let $a=\sqrt{\underline{\lambda}/2}$. We have
\begin{align*}
    \frac{1}{H-1}\sum_{\substack{l=1\\l\neq s}}^H|v^\top c_{itl}|^2/\|v\|^2
    &\leq
    a^2\left(1-\frac{m_v}{H-1}\right)
    +\frac{\bar{c}_f^2m_v}{H-1}
    =
    a^2+\frac{(\bar{c}_f^2-a^2)m_v}{H-1},
\end{align*}
which implies that $\frac{m_v}{H-1}\geq\pi_0$ with $\pi_0=\frac{\underline{\lambda}/2}{\bar{c}_f^2-\underline{\lambda}/2}$. Given the moment condition imposed, $\bar{c}_f\lesssim_\P T^{1/q}$. Therefore, under the rate condition $T^{2/q}/H=o(1)$, we can guarantee $\pi_0H\rightarrow\infty$.

\smallskip\noindent\textit{Step 5.2: Uniform boundedness of density derivatives.}
Use $\nabla^\alpha$ to denote the partial derivative of order $\alpha\in\mathbb{Z}^d_+$.
Note that
\begin{align*}
    \sup_{z\in\mathbb{R}^d}|\nabla^\alpha f_{G_{its}}(z)|=
    \sup_{z\in\mathbb{R}^d}\Big|\frac{1}{(2\pi)^d}\int_{\mathbb{R}^d}(-iu)^\alpha e^{-iu^\top z}\varphi_{G_{its}}(u)\d{u}\Big|
    \lesssim
    \int_{\mathbb{R}^d}\|u\|^{|\alpha|} |\varphi_{G_{its}}(u)|\d{u}.
\end{align*}
The assumption $g_\eta^{(1)}\in L^1(\mathbb{R})$ implies that $\varphi_\eta(u)\rightarrow0$ and $|\varphi_\eta(u)|\lesssim |u|^{-1}$ as $|u|\rightarrow\infty$. Together with the second-order expansion of $\varphi_\eta$ at the origin, there are constants $c_\eta,C_\eta>0$ and $B_\eta<\infty$ such that, for all $x\in\mathbb{R}$,
\begin{align*}
    |\varphi_\eta(x)|
    &\leq \exp\{-c_\eta\min(x^2,1)\},
    \qquad
    |\varphi_\eta(x)|\leq \frac{C_\eta}{|x|}\quad\text{if } |x|\geq B_\eta.
\end{align*}
Let $\bar{c}_f=\max_{itl}\|c_{itl}\|$ and $z_l(u)=c_{itl}^\top u/\sqrt{H}$.
For $\|u\|\leq B\sqrt{H}$ with fixed $B<\infty$, the first bound gives
\begin{align*}
    |\varphi_{G_{its}}(u)|
    &\leq
    \exp\left\{-c_\eta\sum_{\substack{l=1\\l\neq s}}^H\min\big(z_l(u)^2,1\big)\right\}.
\end{align*}
Since $\lambda_{\min}\big(H^{-1}\sum_{\substack{l=1\\l\neq s}}^Hc_{itl}c_{itl}^\top\big)\gtrsim1$ w.p.a.1 and
$\max_l z_l(u)^2\leq \bar{c}_f^2\|u\|^2/H$,
\begin{align*}
    \sum_{\substack{l=1\\l\neq s}}^H\min\big(z_l(u)^2,1\big)
    &\gtrsim
    \frac{\sum_{\substack{l=1\\l\neq s}}^H z_l(u)^2}{1+\max_{\substack{l=1\\l\neq s}}^H z_l(u)^2}
    \gtrsim
    \frac{\|u\|^2}{1+\bar{c}_f^2\|u\|^2/H}
    \gtrsim
    \min\left\{\|u\|^2,\frac{H}{\bar{c}_f^2}\right\}.
\end{align*}
Hence
\begin{align*}
    \int_{\|u\|\leq B\sqrt{H}}\|u\|^{|\alpha|}|\varphi_{G_{its}}(u)|\d{u}
    &\lesssim
    1+H^{(d+|\alpha|)/2}\exp\{-cH/\bar{c}_f^2\}
    \lesssim_\P 1,
\end{align*}
using $\bar{c}_f\lesssim_\P T^{1/q}$ and the imposed rate condition $T^{2(3+d)/q}/H=o(1)$.

It remains to consider $\|u\|>B\sqrt{H}$ for a fixed $B$ large enough. Write $v=u/\sqrt{H}$ and let $\mathcal{I}(u)=\{l: |v^\top c_{itl}|\geq a\|v\|\}$. By Step 5.1, $|\mathcal{I}(u)|\geq \pi_0 H$ for some $a>0$ and a positive sequence $\pi_0$ satisfying $\pi_0H\rightarrow\infty$ w.p.a.1. For $l\in\mathcal{I}(u)$, $|z_l(u)|\geq a\|v\|$. Choosing $B$ so that $aB\geq B_\eta$, the tail bound on $\varphi_\eta$ yields
\begin{align*}
    \int_{\|u\|>B\sqrt{H}}\|u\|^{|\alpha|}|\varphi_{G_{its}}(u)|\d{u}
    &\lesssim
    H^{\frac{d+|\alpha|}{2}}\int_{\|v\|>B}\|v\|^{|\alpha|}
    \Big(\frac{C_\eta}{a\|v\|}\Big)^{\pi_0H}\d{v}
    \lesssim_\P 1,
\end{align*}
for $\pi_0H$ sufficiently large. This proves the boundedness of the density and its derivatives.
\bigskip

\smallskip\noindent\textit{Step 5.3: Polynomial tail bound for density derivatives.}
Fix $\|z\|\geq1$ and choose $r\in[d]$ such that $|z_r|\geq \|z\|/\sqrt{d}$. Let $\gamma=(d+1)e_r$ and define $\Psi(u)=(-iu)^\alpha\varphi_{G_{its}}(u)$. By integration by parts in the $r$-th coordinate,
\begin{align*}
    |z_r|^{d+1}\big|f^{(\alpha)}_{G_{its}}(z)\big|
    &\leq
    \frac{1}{(2\pi)^d}\int_{\mathbb{R}^d}
    \left|\frac{\partial^{d+1}}{\partial u_r^{d+1}}\Psi(u)\right|\d{u}.
\end{align*}
Thus it suffices to show that the last integral is finite uniformly, because then
$|f^{(\alpha)}_{G_{its}}(z)|\lesssim \|z\|^{-d-1}$.

Note that
\begin{align*}
    \left[u^\alpha \varphi_{G_{its}}(u)\right]^{(\gamma)}
    =
    \sum_{\gamma_1\le \gamma}
    \binom{\gamma}{\gamma_1}
    (u^\alpha)^{(\gamma_1)}
    \left(\varphi_{G_{its}}(u)\right)^{(\gamma-\gamma_1)}
\end{align*}
where $\gamma_1\leq \gamma$ means $\gamma_{1,j}\leq \gamma_j$ for every $j\in[d]$ and $\binom{\gamma}{\gamma_1}=\prod_{j\in[d]}\binom{\gamma_j}{\gamma_{1,j}}$.
The derivative of $u^\alpha$ is still polynomial in $u$.
For the derivative of $\varphi_{G_{its}}$, the product rule, the mean-zero expansion of $\varphi_\eta$ around the origin, and the moment condition on $\eta_{it}$ imply the following envelope for every fixed multi-index $\tau$ with $|\tau|\leq d+1+|\alpha|$:
\begin{align*}
    |\varphi_{G_{its}}^{(\tau)}(u)|
    &\lesssim_\P
    (1+\|u\|)^{|\tau|}
    \max_{\substack{A\subset\{1,\dots,H\}\\ |A|\le |\tau|}}
    \prod_{l\notin A}
    \left|
    \varphi_\eta\!\left(\frac{u^\top c_{itl}}{\sqrt{H}}\right)
    \right|.
\end{align*}
The finite set $A$ only removes a fixed number of factors from the product, so the bounds in Step 5.2 continue to hold after replacing $H$ by $H-|A|$. Hence,
for a constant $M_{\alpha d}<\infty$ that depends only on $d$ and $|\alpha|$,
\begin{align*}
    |\Psi^{(\gamma)}(u)|
    &\lesssim_\P
    (1+\|u\|)^{M_{\alpha d}}
    \max_{\substack{A\subset\{1,\dots,H\}\\ |A|\le d+1+|\alpha|}}
    \prod_{l\notin A}
    \left|
    \varphi_\eta\!\left(\frac{u^\top c_{itl}}{\sqrt{H}}\right)
    \right|.
\end{align*}
On $\|u\|\leq B\sqrt H$, the same Gaussian bound as in Step 5.2 applies, with $H$ reduced by at most a fixed constant; on $\|u\|>B\sqrt H$, Step 5.1 gives at least $\pi_0H$ minus a fixed constant projected arguments larger than $a\|u\|/\sqrt H$. Therefore, for some fixed $\bar C<\infty$,
\begin{align*}
    \int_{\mathbb{R}^d}|\Psi^{(\gamma)}(u)|\d{u}
    &\lesssim_\P 1
    +H^{(d+M_{\alpha d})/2}
    \int_{\|v\|>B}(1+\|v\|)^{M_{\alpha d}}
    \Big(\frac{C_\eta}{a\|v\|}\Big)^{\pi_0H-\bar C}\d{v}
    \lesssim_\P 1.
\end{align*}
This proves the polynomial tail bound in \eqref{eq:first-step-density-bounds}.

\medskip\noindent\textit{Step 6: Localized-basis extension.}
Fix a scalar component of $p_{\mathfrak{q}}(\cdot,P)$; the vector statement follows because the number of components is fixed. Conditional on $\mathcal{E}_{t-1}$ and $\eta_{i(t-s)}$, write
\begin{align*}
    \widehat{\beta}_{it}
    =
    \breve{\beta}_{it}
    +H^{-1/2}G_{its}
    +H^{-1}c_{its}\eta_{i(t-s)}.
\end{align*}
The term containing $p_{\mathfrak{q}}(\beta_{it},P)$ has conditional mean zero after multiplication by $\eta_{i(t-s)}$. For the generated-regressor term, define the truncated moment functional
\begin{align*}
    M_P(a;b_0)
    &=
    \int_P
    p_{\mathfrak{q}}(v,P)
    H^{d/2}f_{G_{its}}\!\left(\sqrt H(v-b_0-a)\right)\d{v}.
\end{align*}
Then the conditional expectation is the expectation, over $\beta_{it}$ and $\eta_{i(t-s)}$, of a bounded predictable multiplier times
\begin{align*}
    \eta_{i(t-s)}
    \left[
    M_P\!\left(H^{-1}c_{its}\eta_{i(t-s)};\breve{\beta}_{it}\right)
    -
    M_P(0;\breve{\beta}_{it})
    \right].
\end{align*}
For each fixed component, $M_P(a;b_0)$ is differentiable in $a$ and
\begin{align*}
    \nabla_a M_P(a;b_0)
    &=
    -H^{(d+1)/2}
    \int_P
    p_{\mathfrak{q}}(v,P)
    \nabla f_{G_{its}}\!\left(\sqrt H(v-b_0-a)\right)\d{v}.
\end{align*}
Thus the first-order expansion in $a$ differentiates the density
$f_{G_{its}}$, not the normalized polynomial basis. Equivalently,
\begin{align*}
    M_P\!\left(H^{-1}c_{its}\eta_{i(t-s)};\breve{\beta}_{it}\right)
    -M_P(0;\breve{\beta}_{it})
    =
    H^{-1}\eta_{i(t-s)}c_{its}^\top
    \int_0^1
    \nabla_a M_P\!\left(rH^{-1}c_{its}\eta_{i(t-s)};\breve{\beta}_{it}\right)\d{r}.
\end{align*}
To make explicit why no derivative of the basis appears, let
$a_k(b)=[\nabla\mu_t(b)]_k f_{\beta_{it}}(b|\mathcal{D}_{t-1})$ denote a generic
bounded smooth multiplier. For any coordinate $r$,
\begin{align*}
    &H^{(d+1)/2}
    \int_{\mathcal{B}}\int_P
    p_{\mathfrak{q}}(v,P)
    \partial_r f_{G_{its}}\!\left(\sqrt H(v-b-a)\right)
    a_k(b)\d{v}\d{b}\\
    &\quad =
    H^{d/2}
    \int_P
    p_{\mathfrak{q}}(v,P)
    \int_{\mathcal{B}}
    f_{G_{its}}\!\left(\sqrt H(v-b-a)\right)
    \partial_r a_k(b)\d{b}\d{v}
    +\mathcal{R}_{\partial,P}.
\end{align*}
The boundary term $\mathcal{R}_{\partial,P}$ is negligible uniformly over the
cells in the lemma because $P$ is separated from the boundary of $\mathcal{B}$
by more than $\log(nT)/\sqrt H$ and the same innovation truncation used above
controls the tails of $G_{its}$. The main term is $\lesssim_\P J^{-d}$ because
$p_{\mathfrak{q}}(\cdot,P)$, $\partial_r a_k$, and the conditional convolution
density are uniformly bounded and $\mathrm{vol}(P)\asymp J^{-d}$. Multiplying
by the increment $H^{-1}c_{its}\eta_{i(t-s)}$ gives
$\lesssim_\P (HJ^d)^{-1}$, with no additional derivative factor in $J$. Replacing
$\breve{\beta}_{it}$ by $\beta_{it}$ in the density argument gives the same
deterministic-drift term as before, namely
{$\lesssim_\P\mathsf{B}_{\mathtt{FS}}/J^d$}. Averaging over $s$ gives the displayed
bound for the localized basis. When $\mathfrak{q}=0$, this argument coincides
with the switching-bias calculation above.
\qed

\subsection{Proof of Lemma~\ref{lem:basis-perturbation}\textnormal{(b)}}

The notation is as in part \textnormal{(a)}. We first prove the switching-variance bound, which is the case $\mathfrak{q}=0$, and then add the same-cell polynomial-coordinate movement that arises when $\mathfrak{q}\geq1$.
{
As in \eqref{eq:basis-perturbation-local-tail}, every
$\mathcal D_{t-1}$-conditional moment below is split over
$\mathcal A_{it,\mathtt{FS}}$ and $\mathcal A_{it,\mathtt{FS}}^c$.
The complement is absorbed by the displayed baseline variance terms using
{Assumption~\ref{ass:first-step-bias}}; on
$\mathcal A_{it,\mathtt{FS}}$, the signal displacement is bounded by
$\mathsf{B}_{\mathtt{FS}}$.
}
Define $\tilde{c}_{its}=[\nabla\mu_t(\beta_{it})]^\top c_{its}$. Let $\tilde{P}=\sqrt{H}[\ell-\breve{\beta}_{it}, u-\breve{\beta}_{it}]$ and $\delta_{its}=H^{-1/2}c_{its}\eta_{i(t-s)}$.

\medskip\noindent\textit{Step 1: Switching summand variances.}
Note that
\begin{align*}
    &\Var\left[\Big(\IF(\widehat{\beta}_{it}\in P)-\IF(\beta_{it}\in P)\Big)\sum_{s=1}^{H}\tilde{c}_{its} \eta_{i(t-s)}\Big|\mathcal{D}_{t-1}\right]\\
    &=\sum_{s=1}^H\Var\left[\Big(\IF(\widehat{\beta}_{it}\in P)-\IF(\beta_{it}\in P)\Big)\tilde{c}_{its}\eta_{i(t-s)}\Big|\mathcal{D}_{t-1}\right]\\
    &\quad +\sum_{\substack{s,s'=1\\s\neq s'}}^H\Cov\left[\Big(\IF(\widehat{\beta}_{it}\in P)-\IF(\beta_{it}\in P)\Big)\tilde{c}_{its}\eta_{i(t-s)},\Big(\IF(\widehat{\beta}_{it}\in P)-\IF(\beta_{it}\in P)\Big)\tilde{c}_{its'}\eta_{i(t-s')}\Big|\mathcal{D}_{t-1}\right]
\end{align*}
For the variance of each summand,
first note the crude bound
\begin{align*}
    &\Var\left[\Big(\IF(\widehat{\beta}_{it}\in P)-\IF(\beta_{it}\in P)\Big)\tilde{c}_{its}\eta_{i(t-s)}\Big|\mathcal{D}_{t-1}\right]
    \leq
    \E\left[\tilde{c}_{its}^2\eta_{i(t-s)}^2\Big|\mathcal{D}_{t-1}\right]
    \lesssim {\|c_{ts}\|^2},
\end{align*}
using that the squared indicator difference is bounded by one.
{The last inequality uses the bounded gradient and scale to obtain
$|\tilde c_{its}|\lesssim\|c_{ts}\|$, together with the
$\mathcal D_{t-1}$-measurability of $c_{ts}$ and
$\E[\eta_{i(t-s)}^2|\mathcal D_{t-1}]=1$.}
The sharper bound uses the leave-one-out expansion rather than multiplying dependent conditional expectations.

On the other hand, we also have the following bound
\begin{align*}
    &\Var\left[\Big(\IF(\widehat{\beta}_{it}\in P)-\IF(\beta_{it}\in P)\Big)\tilde{c}_{its}\eta_{i(t-s)}\Big|\mathcal{D}_{t-1}\right]\\
    &\lesssim\|c_{ts}\|^2\E\left[(\IF(\widehat{\beta}_{it}\in P)-\IF(\beta_{it}\in P))^2\eta_{i(t-s)}^2\Big|\mathcal{D}_{t-1}\right]\\
    &=\|c_{ts}\|^2\E\left[\E\left[(\IF(\widehat{\beta}_{it}\in P)-\IF(\beta_{it}\in P))^2\eta_{i(t-s)}^2\Big|\mathcal{E}_{t-1}, \eta_{i(t-s)}\right]
    \Big|\mathcal{D}_{t-1}\right]\\
    &=\|c_{ts}\|^2\E\left[\eta_{i(t-s)}^2\Big(F_{G_{its}}(\tilde{P}-\delta_{its})
    -2\IF(\beta_{it}\in P)F_{G_{its}}(\tilde{P}-\delta_{its})+\IF(\beta_{it}\in P)\Big)\Big|\mathcal{D}_{t-1}\right]\\
    &=\|c_{ts}\|^2\E\bigg[\eta_{i(t-s)}^2\Big[
    \Big(F_{G_{its}}(\tilde{P})\\
    &\quad +\int_{\tilde{P}}\Big(-\nabla f_{G_{its}}(z)^\top \delta_{its}
    +\frac{1}{2}\delta_{its}^\top \nabla^2f_{G_{its}}(\xi_{its})\delta_{its}\Big)\d{z}\Big)\Big(1-2\IF(\beta_{it}\in P)\Big)+\IF(\beta_{it}\in P)\Big]\Big|\mathcal{D}_{t-1}\bigg]\\
    &\lesssim\|c_{ts}\|^2\E\left[\eta_{i(t-s)}^2\Big(F_{G_{its}}(\tilde{P})(1-2\IF(\beta_{it}\in P))+\IF(\beta_{it}\in P)\Big)\Big|\mathcal{D}_{t-1}\right]\\
    &\quad -\frac{\|c_{ts}\|^2}{\sqrt{H}}\E\left[\int_{\tilde{P}}\nabla f_{G_{its}}(z)^\top\d{z}\;c_{its}
    (1-2\IF(\beta_{it}\in P))\eta_{i(t-s)}^3\Big|\mathcal{D}_{t-1}\right]\\
    &\quad +\frac{\|c_{ts}\|^2}{2H}\E\left[c_{its}^\top\int_{\tilde{P}}\nabla^2f_{G_{its}}(\xi_{its})c_{its}\d{z}(1-2\IF(\beta_{it}\in P))\eta_{i(t-s)}^4\Big|\mathcal{D}_{t-1}\right]\\
    &\lesssim\|c_{ts}\|^2\E\left[\Big(F_{G_{its}}(\tilde{P})(1-2\IF(\beta_{it}\in P))+\IF(\beta_{it}\in P)\Big)\Big|\mathcal{D}_{t-1}\right]\\
    &\quad +\Big(\frac{\|c_{ts}\|^3\E[|\eta_{it}|^3]}{\sqrt{H}J^d}+\frac{\|c_{ts}\|^4}{HJ^d}\Big)(1+{\sqrt{H}\mathsf{B}_{\mathtt{FS}}}
    )
\end{align*}
with $\xi_{its}$ some point between $z$ and $z-\delta_{its}$, and the last step follows by the same argument in Step 4 of the proof of part \textnormal{(a)}.

Now, for the leading term, we apply the partitioning argument again.
Specifically, we partition the whole support $\mathcal{B}$ of $\beta_{it}$ into subrectangles of length proportional to $H^{-1/2}$. Recall that we label these subrectangles by a multi-index $m=(m_1, m_1', m_2, m_2',\cdots, m_d, m_d')\in\mathbb{Z}^{2d}$. Define $\mathcal{B}_m$ as the set of $b\in\mathcal{B}$ such that $\sqrt{H}(b_k-\ell_k)\in[m_k,m_k+1]$ and $\sqrt{H}(b_k-u_k)\in[m'_k,m'_k+1]$ for all $k\in[d]$. That is, when a rectangle is associated with a multi-index with some $m_k$ or $m'_k$ large in absolute value, this region is far away from the facets of the hyper-rectangle $[\ell, u]$. Let $\mathcal{M}$ be the collection of possible indices $m$. Note that $|\mathcal{M}|\lesssim H^{d/2}$.
Let
\begin{align*}
    \mathcal{M}_0
    &=\Big\{m\in\mathcal{M}: \forall k, m_km_k'<0,\text{ and } \exists l, |m_l|\leq\tau_n \text{ or } |m_l'|\leq\tau_n\Big\}\\
    &\quad\cup
    \Big\{m\in\mathcal{M}: \exists k, m_km_k'\geq 0;\; |m_l|\leq\tau_n \text{ or } |m_l'|\leq\tau_n \text{ whenever }m_lm_l'\geq 0\Big\}
\end{align*}
for $\tau_n\asymp\log(nT)$. That is, $\cup_{m\in\mathcal{M}_0}\mathcal{B}_m$ is the ``neighborhood'' of every facet of $P$.
Then,
\begin{align*}
    &\E\left[F_{G_{its}}(\tilde{P})(1-2\IF(\beta_{it}\in P))+\IF(\beta_{it}\in P)\Big|\mathcal{D}_{t-1}\right]\\
    &=\sum_{m\in\mathcal{M}}\int_{\mathcal{B}_m}
    \Big(F_{G_{its}}(\sqrt{H}[\ell-b,u-b])(1-2\IF(b\in P))
    +\IF(b\in P)\Big) f_{\beta_{it}}(b|\mathcal{D}_{t-1})\d{b}\;\\
    &\quad +\E\Big[\Big(F_{G_{its}}(\sqrt{H}[\ell-\breve{\beta}_{it},u-\breve{\beta}_{it}])-F_{G_{its}}(\sqrt{H}[\ell-\beta_{it},u-\beta_{it}])\Big)
    (1-2\IF(\beta_{it}\in P))\Big|\mathcal{D}_{t-1}\Big]\\
    &=\sum_{m\in\mathcal{M}_0}\int_{\mathcal{B}_m}
    \Big(F_{G_{its}}(\sqrt{H}[\ell-b,u-b])(1-2\IF(b\in P))
    +\IF(b\in P)\Big) f_{\beta_{it}}(b|\mathcal{D}_{t-1})\d{b}\\
    &\quad +\sum_{m\in\mathcal{M}_0}\E\Big[
    \IF(\beta_{it}\in\mathcal{B}_m)
    \Big(F_{G_{its}}(\sqrt{H}[\ell-\breve{\beta}_{it},u-\breve{\beta}_{it}])
    -F_{G_{its}}(\sqrt{H}[\ell-\beta_{it},u-\beta_{it}])\Big)\\
    &\hspace{13em}\times
    (1-2\IF(\beta_{it}\in P))\Big|\mathcal{D}_{t-1}\Big]\\
    &\quad +\sum_{m\in\mathcal{M}\setminus\mathcal{M}_0}\int_{\mathcal{B}_m}
    \Big(F_{G_{its}}(\sqrt{H}[\ell-b,u-b])(1-2\IF(b\in P))
    +\IF(b\in P)\Big) f_{\beta_{it}}(b|\mathcal{D}_{t-1})\d{b}\\
    &\quad +\sum_{m\in\mathcal{M}\setminus\mathcal{M}_0}\E\Big[
    \IF(\beta_{it}\in\mathcal{B}_m)
    \Big(F_{G_{its}}(\sqrt{H}[\ell-\breve{\beta}_{it},u-\breve{\beta}_{it}])
    -F_{G_{its}}(\sqrt{H}[\ell-\beta_{it},u-\beta_{it}])\Big)\\
    &\hspace{13em}\times
    (1-2\IF(\beta_{it}\in P))\Big|\mathcal{D}_{t-1}\Big]\\
    &\lesssim_\P\,
    \frac{\tau_n^d}{J^{d-1}\max\{J,\sqrt{H}\}}
    +\P(\|G_{its}\|>\tau_n|\mathcal{E}_{t-1}),
\end{align*}
{
Here
\begin{align*}
    \sqrt H\|\breve\beta_{it}-\beta_{it}\|
    &\leq
    \sqrt{H}\mathsf{B}_{\mathtt{FS}}
    =o(1)
    \qquad\text{on }\mathcal A_{it,\mathtt{FS}},
\end{align*}
and the complementary contribution is absorbed by
\eqref{eq:basis-perturbation-local-tail}.
}
The neighborhood of the facets of $P$ has conditional mass bounded by the
smaller of the boundary-strip order $J^{-(d-1)}\tau_n^d/\sqrt H$ and the
whole-cell order $\tau_n^d J^{-d}$. Whenever $b$ lies in $\mathcal{B}_m$ with
$m\not\in\mathcal{M}_0$,
either $F_{G_{its}}(\sqrt{H}[\ell-b,u-b]) \leq\P(\|G_{its}\|_\infty>\tau_n|\mathcal{E}_{t-1})\lesssim H^{-d/2}$ ($b\not\in P$) or
$1-F_{G_{its}}(\sqrt{H}[\ell-b,u-b])\leq \P(\|G_{its}\|_\infty>\tau_n|\mathcal{E}_{t-1})\lesssim H^{-d/2}$ ($b\in P$)
w.p.a.1 (apply the Bernstein inequality as in Step 3 of the proof of part \textnormal{(a)}, but now we set $M=H^{\frac{d}{2(q-2)}}$).
Combining all results together, we conclude that the first term in the variance is $\lesssim_\P\tau_n^d/(J^{d-1}\max\{J,\sqrt{H}\})$.
\bigskip

\medskip\noindent\textit{Step 2: Switching summand covariances.}
Now, consider the covariance between each pair of summands. For $s\neq s'$,
\begin{align*}
    &\Cov\left[\Big(\IF(\widehat{\beta}_{it}\in P)-\IF(\beta_{it}\in P)\Big)
    \tilde{c}_{its}\eta_{i(t-s)},\;
    \Big(\IF(\widehat{\beta}_{it}\in P)-\IF(\beta_{it}\in P)\Big)\tilde{c}_{its'}\eta_{i(t-s')}\Big|\mathcal{D}_{t-1}\right]\\
    &=\E\left[\Big(\IF(\widehat{\beta}_{it}\in P)-\IF(\beta_{it}\in P)\Big)^2\tilde{c}_{its}\tilde{c}_{its'}\eta_{i(t-s)}\eta_{i(t-s')}\Big|\mathcal{D}_{t-1}\right]-\mathfrak{m}_{its}\mathfrak{m}_{its'}\\
    &=\E\bigg[\tilde{c}_{its}\tilde{c}_{its'}\eta_{i(t-s)}\eta_{i(t-s')}
    \Big[\Big(F_{G_{itss'}}(\tilde{P})\\
    &\quad +\int_{\tilde{P}}\Big(-\nabla f_{G_{itss'}}(z)^\top \delta_{itss'}
    +\delta_{itss'}^\top
    \int_0^1(1-v)\nabla^2 f_{G_{itss'}}(z-v\delta_{itss'})\d{v}\delta_{itss'}\Big)\d{z}\Big)\Big(1-2\IF(\beta_{it}\in P)\Big)\\
    &\hspace{5em}+\IF(\beta_{it}\in P)\Big]\Big|\mathcal{D}_{t-1}\bigg]
    -\mathfrak{m}_{its}\mathfrak{m}_{its'}
\end{align*}
where $\mathfrak{m}_{its}=\E[(\IF(\widehat{\beta}_{it}\in P)-\IF(\beta_{it}\in P))
\tilde{c}_{its}\eta_{i(t-s)}|\mathcal{D}_{t-1}]$,
$F_{G_{itss'}}$ and $f_{G_{itss'}}$, analogous to $F_{G_{its}}$ and $f_{G_{its}}$, are the distribution and density of $\frac{1}{\sqrt{H}}\sum_{\substack{l=1\\l\notin\{s,s'\}}}^Hc_{itl}\eta_{i(t-l)}$ conditional on $\sigma(\eta_{i(t-s)}, \eta_{i(t-s')})\vee\mathcal{E}_{t-1}$, and $\delta_{itss'}=\frac{1}{\sqrt{H}}(c_{its}\eta_{i(t-s)}+c_{its'}\eta_{i(t-s')})$.
The density bounds from Step 5 of the proof of part \textnormal{(a)} apply to $f_{G_{itss'}}$ as well, because removing a fixed number of summands preserves the rolling-window Gram lower bound and the characteristic-function product bounds.
When the diagonal and off-diagonal pieces are recombined, the mean-product contribution to the variance of the normalized sum is
$-\big(H^{-1}\sum_{s=1}^H\mathfrak{m}_{its}\big)^2$.
This term is nonpositive. Part \textnormal{(a)} of Lemma~\ref{lem:basis-perturbation} bounds its absolute value by
\begin{align*}
    \left|
    \frac{1}{H}\sum_{s=1}^H\mathfrak{m}_{its}
    \right|^2
    &\lesssim_\P
    {\Big(\frac{1}{HJ^d}+\frac{\mathsf{B}_{\mathtt{FS}}}{J^d}
    \Big)^2},
\end{align*}
which is the squared-mean term retained in the final display.

Since for any $s\neq s'$,
\begin{align*}
    &\E[\eta_{i(t-s)}\eta_{i(t-s')}|\mathcal{E}_{t-1}]=0,\qquad
    \E[\eta_{i(t-s)}^2\eta_{i(t-s')}|\mathcal{E}_{t-1}]
    =
    \E[\eta_{i(t-s)}\eta_{i(t-s')}^2|\mathcal{E}_{t-1}]=0,
\end{align*}
all Taylor terms that are linear in one of the two innovations vanish after multiplication by
$\eta_{i(t-s)}\eta_{i(t-s')}$; the second-order Hessian term remains because it contains products such as $\eta_{i(t-s)}^2\eta_{i(t-s')}^2$. Thus the leading term simplifies to
\begin{align*}
    &\frac{1}{H}\E\bigg[(c_{its}\eta_{i(t-s)}+c_{its'}\eta_{i(t-s')})^\top
    \tilde{c}_{its}\tilde{c}_{its'}\eta_{i(t-s)}\eta_{i(t-s')}\times\\
    &\qquad\int_{\tilde{P}}
    \int_0^1(1-v)\nabla^2 f_{G_{itss'}}(z-v\delta_{itss'})\d{v}\d{z}(1-2\IF(\beta_{it}\in P))(c_{its}\eta_{i(t-s)}+c_{its'}\eta_{i(t-s')})\Big|\mathcal{D}_{t-1}\bigg].
\end{align*}
Consider a generic Hessian element $f_{G_{itss'}}^{(\alpha)}$ with $|\alpha|=2$.
Choose a coordinate $j\in[d]$ such that $\alpha_j\geq1$.
Writing $P=[\ell,u]$ and letting $(x,z_{-j})_j$ denote the vector with $j$-th coordinate $x$ and remaining coordinates $z_{-j}$, the fundamental theorem of calculus gives
\begin{align*}
    \int_{\sqrt{H}(P-\breve{\beta}_{it})}
    f^{(\alpha)}_{G_{itss'}}(z-v\delta_{itss'})\d{z}
    &=
    \int_{\sqrt{H}(P_{-j}-\breve{\beta}_{it,-j})}
    \Big[
    f^{(\alpha-e_j)}_{G_{itss'}}
    \big((\sqrt{H}(u_j-\breve{\beta}_{it,j}),z_{-j})_j-v\delta_{itss'}\big)\\
    &\quad
    -f^{(\alpha-e_j)}_{G_{itss'}}
    \big((\sqrt{H}(\ell_j-\breve{\beta}_{it,j}),z_{-j})_j-v\delta_{itss'}\big)
    \Big]\d{z}_{-j}.
\end{align*}
The bracket contains derivatives of order one, evaluated on two facets of the rectangle.
Keep these facet terms inside the outer conditional expectation.
The density and tail bounds, together with the same innovation truncation,
give
\begin{align*}
    &\E\Bigg[
    \big(1+|\eta_{i(t-s)}|+|\eta_{i(t-s')}|\big)^4\times
    \sup_{v\in[0,1]}
    \left|
    \int_{\sqrt H(P-\breve\beta_{it})}
    f^{(\alpha)}_{G_{itss'}}(z-v\delta_{itss'})\d z
    \right|
    \Biggm|\mathcal D_{t-1}\Bigg]\\
    &\qquad\lesssim_\P
    \begin{cases}
        \tau_n^d/(J^{d-1}\sqrt H),&J\leq\sqrt H,\\
        J^{-d},&J>\sqrt H,
    \end{cases}\\
    &\qquad\lesssim_\P
    \dfrac{\tau_n^d}{J^{d-1}\max\{J,\sqrt H\}}.
\end{align*}
For the first case, integration over $\beta_{it}$ localizes its coordinate
normal to each facet. For the second, Fubini's theorem is applied to the
Hessian integral before separating the two facets. Hence the leading term in
the previous expansion is
$\lesssim_\P\tau_n^d/[HJ^{d-1}\max\{J,\sqrt H\}]$.

Combining all these results yields the desired bound:
\begin{align*}
    &\Var\left[\frac{1}{H}\sum_{s=1}^H\Big(\IF(\widehat{\beta}_{it}\in P)-\IF(\beta_{it}\in P)\Big)\tilde{c}_{its}\eta_{i(t-s)}\Big|\mathcal{D}_{t-1}\right]\\
    &\lesssim_\P\; H^{-2}\Big[H\times \frac{\log^d (nT)}{J^{d-1}\max\{\sqrt{H},J\}}+H^2\times\frac{\log^d (nT)}{HJ^{d-1}\max\{J,\sqrt{H}\}}\Big]
    +{\Big(\frac{1}{HJ^d}+\frac{\mathsf{B}_{\mathtt{FS}}}{J^d}
    \Big)^2}\\
    &\lesssim_\P\; \frac{\log^d (nT)}{HJ^{d-1}\max\{J,\sqrt{H}\}}
    +{\Big(\frac{1}{HJ^d}+\frac{\mathsf{B}_{\mathtt{FS}}}{J^d}
    \Big)^2}.
\end{align*}

The imposed conditions guarantee that, uniformly over $t=H+1,\ldots,T$,
$\frac{1}{H}\sum_{s=1}^H\|c_{ts}\|^4\lesssim 1$ and
$\frac{1}{H^2}\sum_{\substack{s,s'=1\\s\neq s'}}^H\|c_{ts'}\|^2\|c_{ts}\|^2\lesssim 1$ w.p.a.1.

\medskip\noindent\textit{Step 3: Localized-basis extension.}
Fix a scalar component of $p_{\mathfrak{q}}(\cdot,P)$. Write
\begin{align*}
    p_{\mathfrak{q}}(\widehat{\beta}_{it},P)-p_{\mathfrak{q}}(\beta_{it},P)
    =
    \Delta_{itP}^{\mathrm{bd}}
    +
    \Delta_{itP}^{\mathrm{in}},
\end{align*}
where $\Delta_{itP}^{\mathrm{bd}}$ is supported on the event that exactly one of $\widehat{\beta}_{it}$ and $\beta_{it}$ belongs to $P$, and $\Delta_{itP}^{\mathrm{in}}$ is the remaining same-cell coordinate movement. Because $p_{\mathfrak{q}}(\cdot,P)$ is bounded, the boundary part is controlled by the switching calculation above and the squared mean bound from part \textnormal{(a)}. This gives
$\log^d(nT)/\{HJ^{d-1}\max\{J,\sqrt H\}\}
+\{1/(HJ^d)+\mathsf{B}_{\mathtt{FS}}/J^d\}^2$.
When $\mathfrak{q}=0$, the basis is constant on $P$ and $\Delta_{itP}^{\mathrm{in}}=0$. When $\mathfrak{q}\geq1$, the normalized basis has derivative of order $J$, so on the same-cell event
\begin{align*}
    \|\Delta_{itP}^{\mathrm{in}}\|
    &\lesssim
    J\|\widehat{\beta}_{it}-\beta_{it}\|\IF(\beta_{it}\in P).
\end{align*}
This derivative bound is used only for the centered variance. From
\eqref{eq:first-step-decomposition},
\begin{align*}
    \widehat{\beta}_{it}-\breve{\beta}_{it}
    &=
    \frac{1}{H}\sum_{s=1}^H
    S\widehat{A}(t)^{-1}X_s(t)
    K\!\Big(\frac{s}{H}\Big)\varepsilon_{i(t-s)}.
\end{align*}
Conditional Rosenthal bounds for the martingale sum, together with the moment bounds on the local-polynomial weights, give
$\E[\|\widehat{\beta}_{it}-\breve{\beta}_{it}\|^2|\mathcal{E}_{t-1}] \lesssim_\P H^{-1}$ and
$\E[\|\widehat{\beta}_{it}-\breve{\beta}_{it}\|^4|\mathcal{E}_{t-1}] \lesssim_\P H^{-2}$, uniformly over $i$ and $t$.
{On the intersection of the same-cell event with
$\mathcal A_{it,\mathtt{FS}}$,}
\begin{align*}
    \|\Delta_{itP}^{\mathrm{in}}\|
    &\lesssim
    {
    J(\|\widehat{\beta}_{it}-\breve{\beta}_{it}\|+\mathsf{B}_{\mathtt{FS}})
    \IF(\beta_{it}\in P)}
\end{align*}
{and the complementary contribution is absorbed by
\eqref{eq:basis-perturbation-local-tail} as above.}
Using the conditional independence of the innovations in
$\widehat{\beta}_{it}-\breve{\beta}_{it}$ from $\beta_{it}$ given $\mathcal{E}_{t-1}$, and
{the regular-event bound above}, we obtain
\begin{align*}
    \E\!\left[
    \|\Delta_{itP}^{\mathrm{in}}\|^2
    \|\widehat{\beta}_{it}-\breve{\beta}_{it}\|^2
    {\IF(\mathcal A_{it,\mathtt{FS}})}
    \bigm|\mathcal{D}_{t-1}
    \right]
    &\lesssim_\P
    J^2\P(\beta_{it}\in P|\mathcal{D}_{t-1})
    {\left(H^{-2}+\mathsf{B}_{\mathtt{FS}}^2H^{-1}\right)}
    \lesssim_\P
    \frac{J^{2-d}}{H^2}.
\end{align*}
{On $\mathcal A_{it,\mathtt{FS}}^c$, boundedness of the localized
basis, the conditional second-moment bound for
$\widehat\beta_{it}-\breve\beta_{it}$, and the first cell-probability bound
above give
\begin{align*}
    \E\!\left[
        \|\Delta_{itP}^{\mathrm{in}}\|^2
        \|\widehat\beta_{it}-\breve\beta_{it}\|^2
        \IF(\mathcal A_{it,\mathtt{FS}}^c)
        \,\middle|\,\mathcal D_{t-1}
    \right]
    \lesssim_\P
    \frac{1}{H}
    \min\!\left\{
        \P(\mathcal A_{it,\mathtt{FS}}^c\mid\mathcal D_{t-1}),
        J^{-d}
    \right\}
    \lesssim_\P
    \frac{1}{HJ^{d-1}\max\{J,\sqrt H\}}.
\end{align*}}
Therefore, for the same-cell centered component,
\begin{align*}
    &\Var\!\left[
    \Delta_{itP}^{\mathrm{in}}
    [\nabla\mu_t(\beta_{it})]^\top
    \frac{1}{H}\sum_{s=1}^H
    S\widehat{A}(t)^{-1}X_s(t)
    K\!\Big(\frac{s}{H}\Big)\varepsilon_{i(t-s)}
    \biggm|\mathcal{D}_{t-1}
    \right]\\
    &\quad\leq
    \E\!\left[
    \|\Delta_{itP}^{\mathrm{in}}\|^2
    \left\|
    \frac{1}{H}\sum_{s=1}^H
    S\widehat{A}(t)^{-1}X_s(t)
    K\!\Big(\frac{s}{H}\Big)\varepsilon_{i(t-s)}
    \right\|^2
    \biggm|\mathcal{D}_{t-1}
    \right]
    \lesssim_\P
    {
    \frac{1}{HJ^{d-1}\max\{J,\sqrt H\}}}
    +\frac{J^{2-d}}{H^2}.
\end{align*}
Combining the boundary and interior pieces gives the displayed localized-basis variance bound.
\qed

\subsection{Proof of Lemma~\ref{lem:classical-linearization}}

We first record the direct bound for $\mathscr{L}_{\mathtt{B}}(\beta)$, and then focus on the expansion of $\mathscr{L}_{\mathtt{V}}(\beta)$.

By Lemma \ref{lem:partition-basis-moments}, with probability approaching one,
$\|\widehat{Q}_t^{-1}\|\lesssim J^{d}$ and $\|Q_t^{-1}\|\lesssim J^{d}$
uniformly over $t=H+1, \cdots, T$.
Since the gradients are uniformly bounded under Assumption \ref{ass:dgp}, and since the local-polynomial weights have bounded absolute sum under the Gram eigenvalue bound,
\begin{align*}
    |\mathscr{L}_{\mathtt{B}}(\beta)|
    &\lesssim_\P
    \max_{H+1\leq t\leq T}\max_{i\in[n_t]}
    \|\breve{\beta}_{it}-\beta_{it}\|
    {\lesssim_\P\mathsf{B}_{\mathtt{FS}}}.
\end{align*}

\medskip\noindent\textit{Step 1: Bounds for $\mathscr{A}(\beta)$ and $\mathscr{A}(0)$.}
The term $\mathscr{A}(\beta)$ is the re-indexed version of the reference-basis expression with the reference Gram matrix $Q_t$.
It can be written as
\[
\mathscr{A}(\beta)=\frac{1}{T-H}\sum_{t=H+1}^T\widehat p_t(\beta)^\top Q_t^{-1}
\left[\frac{1}{n_t}\sum_{i=1}^{n_t}p_t(\beta_{it})[\nabla\mu_t(\beta_{it})]^\top\frac{1}{H}\sum_{s=1}^HK(s/H)S A(t)^{-1}X_s(t)\varepsilon_{i(t-s)}\right]
\]
The summands in the bracketed double sum are centered and independent conditional on $\mathcal{E}_{t-1}$, are supported on reference cells, and have conditional variance of order $J^{-d}/(nH)$ per time period. The evaluation block and inverse Gram matrix only select and weight elements of the same rectangle-polynomial class; the operator norm contributes at most the factor $J^d$. Since this class is VC-type with bounded envelope, a standard maximal inequality gives
\begin{align*}
    \sup_{\beta\in\mathcal{B}^\circ}|\mathscr{A}(\beta)|
    &\lesssim_\P
    \left(\frac{J^d\log(nT)}{nH}\right)^{1/2}.
\end{align*}

For the special case $\beta=0$, the partition convention and boundary tie-breaking assign $0$ to a unique cell at each date; write its label as $j_0(u)$ at date $u$. For compactness, write $P_{0,u}=P_{j_0(u),u}$, $p_{0,u}=p_{j_0(u),u}$, and $Q_{0,u}=Q_{j_0(u),u}$. By the split-at-zero construction and the midpoint and side-length normalization in Section~\ref{sec:classical-partition}, $\widehat p_{j_0(u),u}(0)=p_{0,u}(0)$ for every $\mathfrak q\geq0$. Let $\mathcal{S}_t=\{s\in[H]:H+1\leq t+s\leq T\}$. Thus,
\begin{align*}
    \mathscr{A}(0)
    &=\frac{1}{T-H} \sum_{t=1}^T \sum_{i=1}^{n_t}
    \frac{1}{H}\sum_{s\in\mathcal{S}_t}
    \frac{1}{n_{t+s}}p_{0,t+s}(0)^\top
    Q_{0,t+s}^{-1}\\
    &\qquad\quad\times
    p_{0,t+s}(\beta_{i(t+s)})
    [\nabla\mu_{t+s}(\beta_{i(t+s)})]^\top
    S A(t+s)^{-1}K\!\Big(\frac{s}{H}\Big) X_s(t+s)
    \IF(i\leq n_{t+s})\varsigma_{it}\eta_{it}.
\end{align*}

The display is obtained by re-indexing the rolling-window shocks: the primitive innovation $\eta_{it}$ enters future first-step estimates at dates $t+s$. After the zero-cell identity above, all coefficients in $\mathscr{A}(0)$ are $\mathcal{E}_\infty$-measurable; conditional on $\mathcal{E}_\infty$, the primitive innovations are independent over $(i,t)$ and have mean zero. For the variance display, write $\Xi_s(u)=S A(u)^{-1}K\!\big(s/H\big)X_s(u)$. Jensen's inequality over the rolling-window average and the bounds
$\|Q_{0,u}^{-1}\|\lesssim_\P J^d$ and
$\|p_{0,u}(0)\|\|p_{0,u}(\beta_{iu})\|
\lesssim \IF(\beta_{iu}\in P_{0,u})$ give
\begin{align*}
    \Var[\mathscr{A}(0)|\mathcal{E}_\infty]
    &\lesssim
    \frac{1}{T^2} \sum_{t=1}^T\sum_{i=1}^{n_t}\bigg(
    \frac{1}{H}\sum_{s\in\mathcal{S}_t}
    \frac{1}{n_{t+s}}p_{0,t+s}(0)^\top
    Q_{0,t+s}^{-1}
    p_{0,t+s}(\beta_{i(t+s)})
    [\nabla\mu_{t+s}(\beta_{i(t+s)})]^\top\\
    &\qquad\qquad\qquad\times
    \Xi_s(t+s)\IF(i\leq n_{t+s})\bigg)^2\\
    &\lesssim\frac{J^{2d}}{T^2}
    \sum_{t=1}^T\sum_{i=1}^{n_t}
    \frac{1}{H}\sum_{s\in\mathcal{S}_t}
    \frac{1}{n_{t+s}^2}
    \IF(\beta_{i(t+s)}\in P_{0,t+s})
    \IF(i\leq n_{t+s}) \|\Xi_s(t+s)\|^2\\
    &\lesssim
    \frac{J^{2d}}{T^2}
    \sum_{u=H+1}^{T}
    \frac{1}{n_u^2}
    \sum_{i=1}^{n_u}
    \IF(\beta_{iu}\in P_{0,u})
    \frac{1}{H}\sum_{s=1}^{H}
    \|\Xi_s(u)\|^2\\
    &\lesssim_\P \frac{J^{2d}}{nT}\frac{1}{J^d}.
\end{align*}
The last line uses the true-exposure reference second-moment bound from
Lemma~\ref{lem:partition-basis-moments} and the local-polynomial Gram bound.
Then, $\mathscr{A}(0)\lesssim_\P \sqrt{J^d/(nT)}$.

\medskip\noindent\textit{Step 2: First-step Gram replacement and inverse-Gram perturbation.}
Let $\mathscr{A}^{\dagger}(\beta)$ denote the same re-indexed expression as
$\mathscr{A}(\beta)$, but with $A(t+s)^{-1}$ replaced by
$\widehat A(t+s)^{-1}$. The mean-square version of
Lemma~\ref{lem:first-step-gram}, the resolvent identity, and the bound for
$\mathscr{A}(\beta)$ in Step 1 give
\begin{align*}
    \sup_{\beta\in\mathcal{B}^\circ}
    |\mathscr{A}^{\dagger}(\beta)-\mathscr{A}(\beta)|
    &\lesssim_\P
    \left(\frac{T^{2/q}}{H}+\frac{1}{\sqrt H}\right)
    \sqrt{\frac{J^d\log(nT)}{nH}}
    =
    o_\P\!\left(\frac{1}{H}\right),
\end{align*}
where the last equality follows from \eqref{eq:classical-linearization-rate} and
$J^d\log(nT)/n=o(1)$. Thus this replacement error is absorbed by the
$H^{-1}$ term in the residual bound.

Next, replace the generated inverse Gram $\widehat{Q}_t^{-1}$ in $\mathscr{L}_{\mathtt{V}}(\beta)$
by the reference inverse Gram $Q_t^{-1}$.
By Lemma \ref{lem:partition-basis-moments},
$\max_{H+1\leq t\leq T}\|\widehat{Q}_t^{-1}-Q_t^{-1}\|
\lesssim_\P J^{2d}\mathsf{R}_{\mathtt{Q}}$.

The cost of this replacement is
$\min\{J^d\mathsf{R}_{\mathtt{Q}},1\}\sqrt{J^d\log(nT)/(nH)}$.
Indeed, if $J^d\mathsf{R}_{\mathtt{Q}}\lesssim1$, the resolvent identity gives an additional relative factor $J^d\mathsf{R}_{\mathtt{Q}}$ compared to the centered reference-basis bound in Step 1; otherwise, the crude bound by the same centered reference-basis scale gives the truncation at one. In the remaining steps we work after this inverse-Gram replacement and add this perturbation cost at the end.

\medskip\noindent\textit{Step 3: Generated-regressor switch term.}
For a fixed admissible partition $\Delta=\{P_j:j\in[\bar J_t]\}$ equipped with its cell normalizations, let $p(\cdot;\Delta)$ denote the corresponding stacked local-polynomial basis.
Consider the following two terms:
\begin{align*}
    \mathscr{T}_{\mathtt{G}}(\beta)
    &=\frac{1}{T-H} \sum_{t=H+1}^T \widehat p_t(\beta)^\top Q_t^{-1}
    \frac{1}{n_t}\sum_{i=1}^{n_t}
    \big[
    p(\widehat{\beta}_{it};\widehat{\Delta}_t)
    -
    p(\beta_{it};\widehat{\Delta}_t)
    \big]
    [\nabla\mu_t(\beta_{it})]^\top
    \big(\widehat{\beta}_{it}-\breve{\beta}_{it}\big),\\
    \mathscr{T}_{\mathtt{P}}(\beta)
    &=\frac{1}{T-H} \sum_{t=H+1}^T \widehat p_t(\beta)^\top Q_t^{-1}
    \frac{1}{n_t}\sum_{i=1}^{n_t}
    \big[
    p(\beta_{it};\widehat{\Delta}_t)
    -
    p(\beta_{it};\Delta_t)
    \big]
    [\nabla\mu_t(\beta_{it})]^\top
    \big(\widehat{\beta}_{it}-\breve{\beta}_{it}\big).
\end{align*}
\medskip

Consider the generated-regressor switch term $\mathscr{T}_{\mathtt{G}}(\beta)$. For any $\beta\in\mathcal{B}^\circ$ (corresponding to an index set $\mathcal{J}_t$ that excludes bins close to the boundary),
let $\mathcal{P}^\circ$ denote the collection of quasi-uniform rectangles $P=[\ell,u]\subset\mathcal{B}$ with $\mathrm{diam}(P)\asymp J^{-1}$ and $\min_{k\in[d]}\min\{|\ell_k-\lb_k|,|\ub_k-u_k|\}\gg\log(nT)/\sqrt{H}$.
\begin{align*}
    |\mathscr{T}_{\mathtt{G}}(\beta)|
    &\lesssim_\P\;
    \max_{H+1\leq t\leq T}\max_{j\in\mathcal{J}_t}
    \Bigg\|
    \frac{J^d}{n}\frac{1}{H}
    \sum_{i=1}^{n_t}\sum_{s=1}^{H}
    \big[
    p_{\mathfrak{q}}(\widehat{\beta}_{it};\widehat{P}_{jt})
    -
    p_{\mathfrak{q}}(\beta_{it};\widehat{P}_{jt})
    \big]\\
    &\qquad\qquad\qquad\times
    [\nabla\mu_t(\beta_{it})]^\top
    S \widehat{A}(t)^{-1} X_s(t)
    K\!\Big(\frac{s}{H}\Big) \varepsilon_{i(t-s)}
    \Bigg\|\\
    &\lesssim_\P\;
    \max_{H+1\leq t\leq T}\sup_{P\in\mathcal{P}^\circ}
    \Bigg\|
    \frac{J^d}{nH}
    \sum_{i=1}^{n_t}\sum_{s=1}^{H}
    \big[
    p(\widehat{\beta}_{it};P)
    -
    p(\beta_{it};P)
    \big]
    [\nabla\mu_t(\beta_{it})]^\top\\
    &\qquad\qquad\qquad\times
    S \widehat{A}(t)^{-1} X_s(t)
    K\!\Big(\frac{s}{H}\Big)\varepsilon_{i(t-s)}
    \Bigg\|\\
    &\lesssim
    \sup_{t,P}
    \Bigg\|
    \frac{J^d}{nH}
    \sum_{i=1}^{n_t}\sum_{s=1}^{H}
    \E[g_{its}|\mathcal{D}_{t-1}]
    \Bigg\|
    +
    \sup_{t,P}
    \Bigg\|
    \frac{J^d}{nH}
    \sum_{i=1}^{n_t}\sum_{s=1}^{H}
    \Big(g_{its}-\E[g_{its}|\mathcal{D}_{t-1}]\Big)
    \Bigg\|
\end{align*}
where
\begin{align*}
    g_{its}
    &=
    \big[
    p_{\mathfrak{q}}(\widehat{\beta}_{it};P)
    -
    p_{\mathfrak{q}}(\beta_{it};P)
    \big]
    [\nabla\mu_t(\beta_{it})]^\top
    S \widehat{A}(t)^{-1} X_s(t)
    K\!\Big(\frac{s}{H}\Big)\varepsilon_{i(t-s)},
\end{align*}
and the shorthand $\sup_{t,P}$ in the preceding display ranges over $H+1\leq t\leq T$ and $P\in\mathcal{P}^\circ$.

By Lemma~\ref{lem:basis-perturbation}\textnormal{(a)},
\begin{align*}
    \max_{H+1\leq t\leq T}\sup_{P\in\mathcal{P}^\circ}
    \Bigg|
    \frac{J^d}{nH}
    \sum_{i=1}^{n_t}\sum_{s=1}^{H}
    \E[g_{its}|\mathcal{D}_{t-1}]
    \Bigg|
    &\lesssim_\P
    {\frac{1}{H}+\mathsf{B}_{\mathtt{FS}}}.
\end{align*}
On the other hand, conditional on $\mathcal{D}_{t-1}$, the second term is a sum of terms independent over $i$ with zero means. By Lemma~\ref{lem:basis-perturbation}\textnormal{(b)},
\begin{align*}
    &\Var\!\left[
    \frac{J^d}{nH}
    \sum_{i=1}^{n_t}\sum_{s=1}^{H}
    g_{its}
    \biggm|\mathcal{D}_{t-1}
    \right]
    \lesssim_\P
    \frac{J^{2d}}{n}
    \left[
    \frac{\log^d(nT)}
    {HJ^{d-1}\max\{J,\sqrt{H}\}}
    +
    \left(
    {
    \frac{1}{HJ^d}
    +\frac{\mathsf{B}_{\mathtt{FS}}}{J^d}}
    \right)^2
    +\IF(\mathfrak{q}\geq1)\frac{J^{2-d}}{H^2}
    \right].
\end{align*}
Define
\begin{align*}
    g_i(t,P)
    &=\frac{1}{H}\sum_{s=1}^Hg_{its},\\
    \mathcal{S}
    &=\{g(\cdot;t,P): P\in\mathcal{P}^\circ, H+1\leq t\leq T\}.
\end{align*}
Here $g(\cdot;t,P)$ is the function that evaluates to $g_i(t,P)$ on the full input history for asset $i$, including the common quantities and all observations entering the rolling window.
The envelope satisfies
\begin{align*}
    \sup_{t,P}\|g_i(t,P)\|
    &\lesssim
    \sup_{H+1\leq t\leq T}
    \left\|\frac{1}{H}\sum_{s=1}^H X_s(t) K\!\Big(\frac{s}{H}\Big) \varepsilon_{i(t-s)}\right\|,\\
    \max_{i\in[n]}\sup_{t,P}|g_i(t,P)|
    &\lesssim_\P
    \frac{\sqrt{\log(nT)}}{\sqrt{H}}
    +\frac{(nT^2)^{1/q}\log (nT)}{H}.
\end{align*}
The class indexed by $P$ is VC-type with a finite exponent because it is generated by rectangular cell indicators and bounded polynomial basis functions on normalized cells.
To transfer this covering bound to the rolling-window summands, fix $t$ and a scalar basis coordinate $r\in[K_{\mathfrak q}]$. Let
\begin{align*}
    b_{P,r}(x,z)
    &=p_{\mathfrak q,r}(x,P)-p_{\mathfrak q,r}(z,P),\\
    M_{it}
    &=[\nabla\mu_t(\beta_{it})]^\top
    \frac{1}{H}\sum_{s=1}^{H}
    S\widehat A(t)^{-1}X_s(t)K(s/H)\varepsilon_{i(t-s)},
\end{align*}
where $p_{\mathfrak q,r}$ is the $r$th coordinate of $p_{\mathfrak q}$. The corresponding coordinate of $g_i(t,P)$ is exactly $M_{it}b_{P,r}(\widehat\beta_{it},\beta_{it})$, and $M_{it}$ does not depend on $P$.
On the full input-history space, denote the functions corresponding to $M_{it}$ and $b_{P,r}(\widehat\beta_{it},\beta_{it})$ by $M$ and $B_{P,r}$. For any probability measure $\mathbb Q$ with $0<\E_{\mathbb Q}[M^2]<\infty$, define the probability measure
\begin{align*}
    \d\mathbb Q_M
    &=\frac{M^2}{\E_{\mathbb Q}[M^2]}\,\d\mathbb Q.
\end{align*}
Then, for any $P,\widetilde P\in\mathcal P^\circ$,
\begin{align*}
    \|M(B_{P,r}-B_{\widetilde P,r})\|_{L_2(\mathbb Q)}
    &=\|M\|_{L_2(\mathbb Q)}
    \|B_{P,r}-B_{\widetilde P,r}\|_{L_2(\mathbb Q_M)}.
\end{align*}
The bounded basis-difference class $\{B_{P,r}:P\in\mathcal P^\circ\}$ has a polynomial covering bound uniformly over probability measures, with an exponent depending only on the fixed $d$ and $\mathfrak q$. The identity transfers this bound to $\{MB_{P,r}:P\in\mathcal P^\circ\}$ with envelope $C|M|$, where $C$ bounds the basis differences. If $\E_{\mathbb Q}[M^2]=0$, the multiplied class is zero $\mathbb Q$-almost surely. Applying the argument to the fixed number of basis coordinates establishes the asserted VC-type bound.

The other index $t$ can only take a finite number ($T-H$) values.
Taking the union over these dates multiplies the covering bound by at most $T-H$. The concentration argument is applied for each fixed $t$ conditional on $\mathcal D_{t-1}$, using the conditional independence across $i$ stated above.
Given these facts, by the maximal inequality for i.n.i.d. empirical processes in \citet[Lemma~SA24]{cattaneo2024uniform},
\begin{align*}
    |\mathscr{T}_{\mathtt{G}}(\beta)|
    &\lesssim_\P{\Big(\frac{1}{H}
    +\mathsf{B}_{\mathtt{FS}}\Big)}
    +\sqrt{\frac{J^{2d}\log^{d+1}(nT)}{nHJ^{d-1}\max\{J,\sqrt{H}\}}}
    +\IF(\mathfrak{q}\geq1)
    \sqrt{\frac{J^{d+2}\log(nT)}{nH^2}},
\end{align*}
under the restriction
\begin{align*}
    \frac{J^{d-1}\max\{J,\sqrt{H}\}\log^2(nT)}
    {\min\{n^{1-2/q}T^{1-4/q}h,n\}}
    &\lesssim 1.
\end{align*}
This restriction ensures that
\begin{align*}
    \frac{\log^{3/2}(nT)}{\sqrt{n}}
    +\frac{(nT^2)^{1/q}\log^2(nT)}{\sqrt{nH}}
    &\lesssim
    \bigl(J^{d-1}\max\{J,\sqrt{H}\}\bigr)^{-1/2}
    \log^{(d+1)/2}(nT).
\end{align*}
\medskip

\medskip\noindent\textit{Step 4: Partition switch term.}
{For $\mathscr{T}_{\mathtt{P}}(\beta)$, by
Lemma~\ref{lem:quantile-convergence}, for all
$\beta\in\mathcal{B}^\circ$,}
\begin{align*}
    |\mathscr{T}_{\mathtt{P}}(\beta)|
    &\lesssim_\P
    {\max_{H+1\leq t\leq T}\max_{j\in\mathcal J_t}}
    \Bigg\|
    \frac{J^d}{n}\frac{1}{H}
    \sum_{i=1}^{n_t}\sum_{s=1}^{H}
    \big[
    p_{\mathfrak{q}}(\beta_{it};\widehat{P}_{jt})
    -
    p_{\mathfrak{q}}(\beta_{it};P_{jt})
    \big]\\
    &\qquad\qquad\qquad\times
    [\nabla\mu_t(\beta_{it})]^\top
    S \widehat{A}(t)^{-1} X_s(t)
    K\!\Big(\frac{s}{H}\Big) \varepsilon_{i(t-s)}
    \Bigg\|\\
    &\lesssim
    {\max_{H+1\leq t\leq T}\max_{j\in\mathcal J_t}}
    \sup_{\tilde{P}\in \mathcal{P}_{jt}}
    \Bigg\|
    \frac{J^d}{n}\frac{1}{H}
    \sum_{i=1}^{n_t}\sum_{s=1}^{H}
    \big[
    p_{\mathfrak{q}}(\beta_{it};\widetilde{P}_{jt})
    -
    p_{\mathfrak{q}}(\beta_{it};P_{jt})
    \big]
    [\nabla\mu_t(\beta_{it})]^\top\\
    &\qquad\qquad\qquad\times
    S \widehat{A}(t)^{-1} X_s(t)
    K\!\Big(\frac{s}{H}\Big)\varepsilon_{i(t-s)}
    \Bigg\|
\end{align*}

\noindent where, by Lemma \ref{lem:quantile-convergence}, we take the supremum over
$\mathcal{P}_{jt}=\{[a,b]: \max\{\|a-\ell\|,\|b-u\|\}\leq\sqrt{\log T/n}\}$ for a generic cell $P_{jt}=[\ell, u]$.
Again, conditional on $\mathcal{E}_{t-1}$, the summands are centered, independent over $i$, and have zero cross terms across different lags by the innovation orthogonality. The symmetric difference between $\widetilde{P}$ and $P_{jt}$ has {$\mathcal{D}_{t-1}$-conditional} probability at most of order $J^{-(d-1)}\sqrt{\log T/n}$, uniformly over $\widetilde{P}\in\mathcal{P}_{jt}$.
{A VC count bound conditional on $\mathcal D_{t-1}$ and a union over dates give}
{
\begin{align*}
    \max_{H+1\leq t\leq T}\max_{j\in\mathcal J_t}
    \sup_{\widetilde P\in\mathcal P_{jt}}
    \sum_{i=1}^{n_t}\IF\big(\beta_{it}\in\widetilde P\mathbin{\triangle}P_{jt}\big)
    &\lesssim_\P
    nJ^{-(d-1)}\sqrt{\frac{\log T}{n}}+\log(nT).
\end{align*}
}
{The same argument and $J^d\log(nT)/n=o(1)$ bound the true-beta count in each reference cell by $O_\P(n/J^d)$, uniformly over dates and cells. Conditional on $\mathcal E_{t-1}$, the realized counts are fixed; the strip and reference-cell count bounds control the boundary and coordinate-normalization contributions, respectively.}
Since $H^{-1}\sum_{s=1}^H\|S\widehat{A}(t)^{-1}X_s(t)K\!\Big(\frac{s}{H}\Big)\|^2\lesssim_\P1$,
the comparison of boundaries and the coordinate-normalization component between two cells gives the following variance bound:
\begin{align*}
    &\frac{J^{2d}}{n^2H^2}\sum_{i=1}^{n_t}\sum_{s=1}^H
    \|\big[
    p_{\mathfrak{q}}(\beta_{it};\widetilde{P}_{jt})
    -
    p_{\mathfrak{q}}(\beta_{it};P_{jt})
    \big]\|^2
    \left\|S \widehat{A}(t)^{-1} X_s(t)
    K\!\Big(\frac{s}{H}\Big)\right\|^2\\
    &\qquad\lesssim_\P
    \frac{J^{d+1}\sqrt{\log T/n}}{nH}
    \mathbin{{+}}{\frac{J^{2d}\log(nT)}{n^2H}}
    +\IF(\mathfrak{q}\geq1)
    \frac{J^{d+2}\log T}{n^2H}.
\end{align*}
In addition, the class of local polynomial basis functions
is of VC-type with a finite exponent{, and the multiplier argument from Step 3 applies because the rolling-score multiplier does not depend on the perturbed cell}. {Using the kernel-weighted rolling-score envelope, datewise truncation, and bounded empirical-process concentration together with} \citet[Lemma~SA24]{cattaneo2024uniform}, {then taking a union over dates and retaining the envelope terms,} we conclude that
\begin{align*}
    |\mathscr{T}_{\mathtt{P}}(\beta)|
    &\lesssim_\P
    \sqrt{\frac{J^{d+1}\sqrt{\log T/n}\log (nT)}{nH}}
    +\IF(\mathfrak{q}\geq1)
    \sqrt{\frac{J^{d+2}\log T\log(nT)}{n^2H}}\\
    &\qquad\mathbin{{+}}
    {\frac{J^d}{n}\left[
    \frac{\log^{3/2}(nT)}{\sqrt H}
    +\frac{(nT^2)^{1/q}\log^2(nT)}{H}\right]},
\end{align*}
and this bound holds uniformly over $\beta\in\mathcal{B}^\circ$.
{The additional boundary-count variance term contributes at most $J^d\log(nT)/(n\sqrt H)$ after concentration and is covered by the displayed envelope terms.}
\medskip

\medskip\noindent\textit{Step 5: Combine the bounds.}
{Multiplying the final inequality in Step 3 by $J^d/\sqrt{nH}$ gives}
{
\begin{align*}
    \frac{J^d}{n}\left[
    \frac{\log^{3/2}(nT)}{\sqrt H}
    +\frac{(nT^2)^{1/q}\log^2(nT)}{H}\right]
    &\lesssim
    \sqrt{\frac{J^{2d}\log^{d+1}(nT)}
    {nHJ^{d-1}\max\{J,\sqrt H\}}}.
\end{align*}
}
{Thus the extra terms retained in Step 4 are absorbed by the existing generated-regressor switch component of $\mathsf R_{\mathtt{LA}}$.}
Combining the first-step Gram replacement and inverse-Gram perturbation from Step 2, the generated-regressor switch bound from Step 3, and the partition-switch bound from Step 4 proves
\begin{align*}
    \sup_{\beta\in\mathcal{B}^\circ}
    |\mathscr{L}_{\mathtt{V}}(\beta)-\mathscr{A}(\beta)|
    &\lesssim_\P
    \mathsf{R}_{\mathtt{LA}}.
\end{align*}

\qed

\section{Convergence Rate}\label{sec:classical-convergence}

We now combine the first-step and partitioning results from Sections~\ref{sec:classical-first-step} and~\ref{sec:classical-partition} with the general results from Part~\ref{part:general}. The first result is the direct specialization of the general beta-sorted portfolio convergence theorem to the classical construction. The second uses the stochastic linearization and is sharper at each fixed evaluation point.

\subsection{Results}

\begin{corollary}[General Beta-Sorted Portfolios]\label{cor:classical-convergence}
    \leavevmode\par\noindent
    Suppose that Assumptions~\ref{ass:dgp}, \ref{ass:first-step-ls},
    \ref{ass:partition}, {\ref{ass:interior-separation}}, and
    \ref{ass:first-step-bias} hold
    and that $s_\mu\geq1$. Assume further that
    \begin{align*}
        &\frac{T^{(3+d)/q}}{H}
        +\sqrt{\frac{\log T}{H}}
        =o(1),\qquad
        \mathsf{R}_{\mathtt{FS}}
        =o(1),\\
        &\frac{J^d\log(nT)}{n}
        =o(1),\qquad
        \frac{(nT)^{2/q}J^d\log(nT)}{n}
        \lesssim1.
    \end{align*}
    {Then, for each fixed $\beta\in\mathcal{B}^\circ$} and any $\ell_n\rightarrow\infty$,
    \begin{align*}
        \big|\widehat{\mu}(\beta)-\bar{\mu}_T(\beta;H)\big|
        &\lesssim_\P
        \ell_n\left[
        \mathsf{R}_{\mathtt{FS}}
        +J^{-\min\{s_\mu,\mathfrak{q}+1\}}
        +\sqrt{\frac{J^d\log(nT)}{n}}
        +\frac{\|\beta\|}{\sqrt{T}}
        \right].
    \end{align*}
\end{corollary}

The next corollary uses the stochastic linearization to recover the time-series averaging in the leading sampling terms. The generated-regressor correction $\mathscr{A}(\beta)$ is the only additional term extracted from $\mathscr{L}_{\mathtt{V}}(\beta)$; the nonconstant-basis movement that appears when $\mathfrak{q}\geq1$ is controlled uniformly and absorbed by $\mathsf{R}_{\mathtt{LA}}$ in Lemma~\ref{lem:classical-linearization}. To state the leading first-step correction compactly, define
\begin{align*}
    \mathsf{R}_{\mathtt{A}}(\beta)
    =
    \begin{cases}
        \sqrt{\dfrac{J^d\log(nT)}{nH}}
        &\text{if } \beta\neq0\\[0.8em]
        \sqrt{\dfrac{J^d}{nT}}
        &\text{if } \beta=0
    \end{cases}.
\end{align*}
This rate collects the pointwise bound for the first-step correction $\mathscr{A}(\beta)$ obtained in Lemma~\ref{lem:classical-linearization}. The Gram-perturbation and basis-switching pieces enter through the explicit $\mathsf{R}_{\mathtt{Q}}$ and $\mathsf{R}_{\mathtt{P}}$ terms displayed below. Under the stronger idiosyncratic-scale localization condition $J\mathsf{R}_{\mathtt{FS}}=o(1)$, these terms are absorbed by $\sqrt{J^d/(nT)}$; fixed nonzero-$\beta$ inference only needs them to be negligible at the factor scale.

\begin{corollary}[Classical Pointwise Convergence Rate]\label{cor:classical-pointwise-convergence}
    Suppose that Assumptions~\ref{ass:dgp}, \ref{ass:first-step-ls},
    \ref{ass:partition}, {\ref{ass:interior-separation}}, and
    \ref{ass:first-step-bias} hold
    and that $s_\mu\geq2$.
    Assume further that
    \begin{align*}
        &\frac{T^{2/q}}{H}
        +\sqrt{\frac{\log T}{H}}=o(1),\qquad
        \mathsf{R}_{\mathtt{FS}}=o(1),\\
        &\frac{J^d\log(nT)}{n}=o(1),\qquad
        \frac{J^2\log T}{n}=o(1),\\
        &\frac{J^{d-1}\max\{J,\sqrt{H}\}\log^2(nT)}
        {\min\{n^{1-2/q}T^{1-4/q}h,n\}}
        \lesssim 1,\qquad
        \frac{\max\{J^{2/(q-2)}T^{2/q},T^{2(3+d)/q}\}}{H}=o(1).
    \end{align*}
    Then, for each fixed $\beta\in\mathcal{B}^\circ$,
    \begin{align*}
        \big|\widehat{\mu}(\beta)-\bar{\mu}_T(\beta;H)\big|
        &\lesssim_\P
        \frac{\|\beta\|}{\sqrt{T}}
        +\sqrt{\frac{J^d}{nT}}
        +\sqrt{\frac{J^{3d}\mathsf{R}_{\mathtt{Q}}^2}{nT}}
        \\
        &\qquad
        +\sqrt{\frac{J^{2d}\mathsf{R}_{\mathtt{P}}}{nT}}
        +\mathsf{R}_{\mathtt{A}}(\beta)
        +\mathsf{R}_{\mathtt{LA}}
        +J^{-\min\{s_\mu,\mathfrak{q}+1\}}
        +\frac{\mathsf{R}_{\mathtt{FS}}}{\sqrt{T}}
        +\mathsf{R}_{\mathtt{FS}}^{2}\\
        &\qquad
        +\IF(\mathfrak{q}=0)\frac{1}{J\sqrt{T}}.
    \end{align*}
\end{corollary}

\subsection{Proof of Corollary~\ref{cor:classical-convergence}}

By the first displayed rate condition, Lemma~\ref{lem:first-step-rate} gives
\begin{align*}
    \widehat{\mathsf{R}}_{\mathtt{FS}}
    &\lesssim_\P\mathsf{R}_{\mathtt{FS}}.
\end{align*}
Because $\mathsf{R}_{\mathtt{FS}}=o(1)$ and $\mathcal{B}^+$ is a fixed neighborhood of $\mathcal{B}$, the first-step estimates lie in $\mathcal{B}^+$ uniformly over $i$ and $t$, w.p.a.1.

{
Next consider the local partition conditions along the cells in
$\mathcal J_t$. The first-step bias condition, the Gram condition for the
rolling-window design, and $J^d\log(nT)/n=o(1)$ imply the assumptions of
Lemmas~\ref{lem:reference-quantile-spacing},
\ref{lem:quantile-convergence}, and
\ref{lem:partition-basis-moments}. These results imply that the estimated
partition covers $\mathcal B^\circ$ and all raw generated regressors, and that
the reference and estimated cells indexed by $\mathcal J_t$ have side lengths
of order $J^{-1}$. Consequently,
}
\begin{align*}
    {\widehat{\mathsf{R}}_{\mathtt{D}}^\circ}
    &\lesssim_\P J^{-1}.
\end{align*}

{
The same lemma gives the required counts only over the relevant interior
cells:
}
\begin{align*}
    \frac{n}{J^d}
    &\lesssim_\P
    {\widehat{n}_{\min}^{\circ}}
    \leq
    {\widehat{n}_{\max}^{\circ}}
    \lesssim_\P
    \frac{n}{J^d}.
\end{align*}
Consequently,
\begin{align*}
    \frac{(nT)^{2/q}\log(nT)}
    {{\widehat{n}_{\min}^{\circ}}}
    &\lesssim_\P
    \frac{(nT)^{2/q}J^d\log(nT)}{n}
    \lesssim 1.
\end{align*}

{
The scaled Gram conclusion in
Lemma~\ref{lem:partition-basis-moments} gives the required eigenvalue bounds
uniformly over $j\in\mathcal J_t$. The pointwise proof of
Theorem~\ref{thm:general-consistency} is therefore applicable cellwise to the
block containing each fixed $\beta\in\mathcal B^\circ$; its argument never
uses an unselected tail block. Applying that local argument and substituting
}
$\widehat{\mathsf{R}}_{\mathtt{FS}}\lesssim_\P\mathsf{R}_{\mathtt{FS}}$,
{$\widehat{\mathsf{R}}_{\mathtt{D}}^\circ\lesssim_\P J^{-1}$},
and the displayed cell-count chain gives the displayed pointwise bound. \qed

\subsection{Proof of Corollary~\ref{cor:classical-pointwise-convergence}}

By the first displayed rate condition, Lemma~\ref{lem:first-step-rate} gives
\begin{align*}
    \widehat{\mathsf{R}}_{\mathtt{FS}}
    &\lesssim_\P\mathsf{R}_{\mathtt{FS}}.
\end{align*}
Since $J^d\log(nT)/n=o(1)$ implies $J\log(nT)/n=o(1)$, the displayed rate conditions imply the conditions of Lemma~\ref{lem:partition-basis-moments}. Thus,
\begin{align*}
    \widehat{\mathsf{R}}_{\mathtt{Q}}
    &\lesssim_\P \mathsf{R}_{\mathtt{Q}},&
    \widehat{\mathsf{R}}_{\mathtt{P}}
    &\lesssim_\P \mathsf{R}_{\mathtt{P}},&
    \widehat{\mathsf{R}}_{\mathtt{D}}^{\circ}
    &\lesssim_\P J^{-1},
\end{align*}
and the generated Gram bounds hold.
The parts of the proof of Lemma~\ref{lem:general-linearization} that keep the generated design fixed give
\begin{align*}
    |\mathscr{G}(\beta)|
    &\lesssim_\P\frac{\|\beta\|}{\sqrt{T}},&
    |\mathscr{B}(\beta)|
    &\lesssim_\P J^{-\min\{s_\mu,\mathfrak{q}+1\}},&
    |\mathscr{R}_3(\beta)|
    &\lesssim_\P \IF(\mathfrak{q}=0)\frac{1}{J\sqrt{T}},\\
    |\mathscr{R}_4(\beta)|
    &\lesssim_\P \frac{\mathsf{R}_{\mathtt{FS}}}{\sqrt{T}},&
    |\mathscr{R}_5(\beta)|
    &\lesssim_\P \mathsf{R}_{\mathtt{FS}}^{2}.
\end{align*}
Lemma~\ref{lem:classical-linearization} gives
\begin{align*}
    |\mathscr{L}_{\mathtt{B}}(\beta)|
    &\lesssim_\P {\mathsf{B}_{\mathtt{FS}}},&
    |\mathscr{A}(\beta)|
    \lesssim_\P \mathsf{R}_{\mathtt{A}}(\beta),
\end{align*}
and the $\mathscr{L}_{\mathtt{B}}(\beta)$ bound is absorbed by $\mathsf{R}_{\mathtt{LA}}$ because $\mathsf{R}_{\mathtt{LA}}$ contains {$\mathsf{B}_{\mathtt{FS}}$}.
The design-comparison argument in Lemma~\ref{lem:general-linearization}, using the generated and reference Gram bounds verified above, yields
\begin{align*}
    |\mathscr{E}(\beta)|
    &\lesssim_\P
    \sqrt{\frac{J^d}{nT}},&
    |\mathscr{R}_1(\beta)|
    &\lesssim_\P
    \sqrt{\frac{J^{3d}\mathsf{R}_{\mathtt{Q}}^2}{nT}},&
    |\mathscr{R}_2(\beta)|
    &\lesssim_\P
    \sqrt{\frac{J^{2d}\mathsf{R}_{\mathtt{P}}}{nT}}.
\end{align*}
The remaining generated-regressor residual is exactly the linear-approximation error controlled in Lemma~\ref{lem:classical-linearization}: $|\mathscr{L}_{\mathtt{V}}(\beta) -\mathscr{A}(\beta)| \lesssim_\P \mathsf{R}_{\mathtt{LA}}$.
Combining these displays with the algebraic decomposition
\begin{align*}
    \widehat{\mu}(\beta)- \bar{\mu}_T(\beta;H)
    &= \mathscr{G}(\beta) + \mathscr{E}(\beta) - \mathscr{L}(\beta) + \mathscr{B}(\beta)\\
    &\qquad + \mathscr{R}_1(\beta)+\mathscr{R}_2(\beta)+\mathscr{R}_3(\beta)+\mathscr{R}_4(\beta) - \mathscr{R}_5(\beta),
\end{align*}
and $\mathscr{L}=\mathscr{L}_{\mathtt{B}}+\mathscr{L}_{\mathtt{V}}$ gives the stated rate. \qed

\section{Pointwise-in-\texorpdfstring{$\beta$}{beta} Distributional Approximation}\label{sec:pointwise-distribution}

We now isolate the numerator for the distributional approximation:
\begin{align}
    \bar{\mathscr{S}}(\beta)
    =
    \mathscr{G}(\beta)
    +\mathscr{E}(\beta)
    -\mathscr{A}(\beta)
    =
    \frac{1}{T-H}\sum_{t=H+1}^{T}
    \beta^{\top}\widetilde{f}_t
    +
    \frac{1}{T-H}\sum_{t=1}^{T}
    \sum_{i=1}^{n_t}\varsigma_{it}\eta_{it}W_{it}(\beta),
    \label{eq:pointwise-leading-sum}
\end{align}
where the direct second-step and first-step primitive-shock weights are
\begin{align*}
    W_{it}(\beta)
    &=
    \frac{1}{n_t}
    \widehat{p}_t(\beta)^{\top}
    Q_t^{-1}
    p_t(\beta_{it})
    \IF(t\geq H+1)
    -w_{it}(\beta).
\end{align*}
Therefore, we obtain
\begin{align*}
    \widehat{\mu}(\beta)-\bar{\mu}_T(\beta;H)
    = \bar{\mathscr{S}}(\beta)+\mathscr{R}(\beta),
\end{align*}
where
\begin{align}
    \mathscr{R}(\beta)
    &=
    \mathscr{B}(\beta)
    +\mathscr{R}_1(\beta)+\mathscr{R}_2(\beta)
    +\mathscr{R}_3(\beta)+\mathscr{R}_4(\beta)
    -\mathscr{R}_5(\beta) \notag\\
    &\qquad -\mathscr{L}_{\mathtt{B}}(\beta)
    -\Big[
    \mathscr{L}_{\mathtt{V}}(\beta)
    -\mathscr{A}(\beta)
    \Big].
    \label{eq:pointwise-clt-remainder}
\end{align}
Thus \eqref{eq:pointwise-clt-remainder} lists all residual terms explicitly: the smoothing bias, the design-comparison and Taylor remainders, the first-step bias contribution, and the residual generated-regressor approximation error from Lemma~\ref{lem:classical-linearization}. The centered nonconstant-basis term that appears when $\mathfrak{q}\geq1$ is included in that last residual and controlled uniformly by $\mathsf{R}_{\mathtt{LA}}$.

For every evaluation point $\beta\in\mathcal B^\circ$, define the baseline
normalization as the sum of the factor variance and the reference
idiosyncratic scale:
\begin{align*}
    \Sigma(\beta)
    &=
    \Sigma_f(\beta)+\Sigma_{\varepsilon}(\beta),\\
    \Sigma_f(\beta)
    &=
    \frac{1}{(T-H)^2}\sum_{t=H+1}^{T}
    \beta^{\top}\Var(f_t\mid\mathcal{G}_{t-1})\beta,
    \\
    \Sigma_{\varepsilon}(\beta)
    &=
    \frac{1}{(T-H)^2}\sum_{t=1}^{T}
    \sum_{i=1}^{n_t}\varsigma_{it}^2\big(W_{it}(\beta)\big)^2,
\end{align*}
where $W_{it}(\beta)$ is the generated-regressor-adjusted reference weight.
At an anchored evaluation point satisfying
$\widehat p_t(\beta)=p_t(\beta)$, the weights $W_{it}(\beta)$ are measurable
with respect to the innovation-free environment, and
$\Sigma_\varepsilon(\beta)$ is the predictable conditional variance of the
idiosyncratic leading term. At a generic nonanchored evaluation point,
$\Sigma_\varepsilon(\beta)$ is used only as a reference quadratic scale. The
proofs below control the corresponding stochastic term through the separate
bounds for $\mathscr E(\beta)$ and $\mathscr A(\beta)$.

\subsection{Results}

The first result isolates the Gaussian approximation for the leading term. The proof treats the origin and nonzero evaluation points separately, but the conclusion is stated as a single Kolmogorov-distance approximation. Away from the origin, the factor variance is nondegenerate under Assumption \ref{ass:dgp}(iv); the only additional variance condition is needed at the origin, where the factor component vanishes.
\begin{lemma}[Gaussian Approximation for the Leading Terms]\label{lem:leading-gaussian-components}
    Suppose that Assumptions \ref{ass:dgp}, \ref{ass:first-step-ls},
    \ref{ass:partition}, {\ref{ass:interior-separation}}, and
    \ref{ass:first-step-bias} hold.
    Assume further that
    \begin{align*}
        &\frac{T^{2/q}}{H}
        +\sqrt{\frac{\log T}{H}}
        =o(1),\\
        &\frac{J^d\log(nT)}{n}=o(1),\\
        &\frac{J^{d-1}\max\{J,\sqrt{H}\}\log^2(nT)}
        {\min\{n^{1-2/q}T^{1-4/q}h,n\}}
        \lesssim 1,\qquad
        \frac{\max\{J^{2/(q-2)}T^{2/q},T^{2(3+d)/q}\}}{H}
        =o(1).
    \end{align*}
    If the fixed evaluation point is $\beta\neq0$, assume additionally that
    \begin{align*}
        \frac{J^d\log(nT)}{nh}
        &=o(1).
    \end{align*}
    If the fixed evaluation point is $\beta=0$, assume additionally that
    \begin{align*}
        \frac{nT}{J^d}\Sigma_{\varepsilon}(0)
        \gtrsim_\P 1.
    \end{align*}
    Then, for each fixed $\beta\in\mathcal{B}^\circ$,
    \begin{align*}
        \sup_{x\in\mathbb{R}}\left|
        \P\left(\frac{\bar{\mathscr{S}}(\beta)}{\sqrt{\Sigma(\beta)}}\le x\right)
        -\Phi(x)\right|
        &\to 0.
    \end{align*}
\end{lemma}

The next theorem combines the Gaussian approximation for the leading term with the pointwise stochastic linearization error. We state the conclusion directly in Kolmogorov distance, so the approximation can be read as the infeasible size control for a pointwise normal test. Let
\begin{align*}
    \mathsf{R}_{\mathtt{CLT}}
    &=J^{-\min\{s_\mu,\mathfrak{q}+1\}}
    +\sqrt{\frac{J^{3d}\mathsf{R}_{\mathtt{Q}}^2}{nT}}
    +\sqrt{\frac{J^{2d}\mathsf{R}_{\mathtt{P}}}{nT}}
    +\mathsf{R}_{\mathtt{LA}}
    +\frac{\mathsf{R}_{\mathtt{FS}}}{\sqrt T}
    +\mathsf{R}_{\mathtt{FS}}^2
    +\IF(\mathfrak{q}=0)\frac{1}{J\sqrt T}.
\end{align*}
This is the non-leading pointwise remainder after removing the leading components $\mathscr{G}(\beta)$, $\mathscr{E}(\beta)$, and $\mathscr{A}(\beta)$.
\begin{theorem}[Distributional Approximation for the Infeasible t-Statistic]\label{thm:pointwise-clt}
    Suppose that the conditions of Lemmas \ref{lem:classical-linearization} and \ref{lem:leading-gaussian-components} hold. For fixed $\beta\in\mathcal{B}^\circ$, suppose further that
    \begin{align*}
        &\mathsf{R}_{\mathtt{FS}}=o(1),\qquad
        \frac{J^2\log T}{n}=o(1),\qquad
        \frac{\mathsf{R}_{\mathtt{CLT}}}
        {\sqrt{\Sigma(\beta)}}
        =o_\P(1).
    \end{align*}
    Then
    \begin{align*}
        &\sup_{x\in\mathbb{R}}\left|
        \P\left(
        \frac{\widehat{\mu}(\beta)-\bar{\mu}_T(\beta;H)}{\sqrt{\Sigma(\beta)}}
        \le x
        \right)
        -\Phi(x)\right|
        \to 0.
    \end{align*}
\end{theorem}

\subsection{Discussion}

\begin{enumerate}[label=(\roman*)]
    \item \emph{Normalization.}
    The theorem imposes the high-level normalization condition
    $\mathsf{R}_{\mathtt{CLT}}/\sqrt{\Sigma(\beta)}=o_\P(1)$ directly.
    For each fixed $\beta\neq0$, Assumption~\ref{ass:dgp}(iv) gives $\Sigma(\beta)\asymp_\P T^{-1}$, so it is enough that
    \begin{align*}
        \sqrt{T}\mathsf{R}_{\mathtt{CLT}}
        &\to0.
    \end{align*}
    At the origin, Lemma~\ref{lem:leading-gaussian-components} imposes
    $(nT/J^d)\Sigma_{\varepsilon}(0)\gtrsim_\P1$, and the leading-weight upper bound used below gives $\Sigma(0)\asymp_\P J^d/(nT)$. Thus it is enough that
    \begin{align*}
        \sqrt{\frac{nT}{J^d}}\mathsf{R}_{\mathtt{CLT}}
        &\to0.
    \end{align*}

    \item \emph{Polynomial order.}
    Higher-order second-step polynomials reduce the smoothing-bias component in the CLT remainder:
    $J^{-\min\{s_\mu,\mathfrak{q}+1\}}$.
    Under the factor nondegeneracy condition in Assumption \ref{ass:dgp}(iv), the relevant bias restrictions are
    \begin{align*}
        &\frac{T}{J^{2\min\{s_\mu,\mathfrak{q}+1\}}}
        \to0
        &&\text{when }\beta\neq0,\\
        &\frac{nT}{J^{d+2\min\{s_\mu,\mathfrak{q}+1\}}}
        \to0
        &&\text{when }\beta=0.
    \end{align*}
    The stronger restriction at the origin appears because the factor component is absent and $\Sigma(0)\asymp_\P J^d/(nT)$. At the origin, the sufficient rate conditions stated here are incompatible with $\mathfrak{q}=0$.
\end{enumerate}

\subsection{Proof of Lemma~\ref{lem:leading-gaussian-components}}

Decompose the leading term into its factor and idiosyncratic components:
\begin{align*}
    \bar{\mathscr{S}}(\beta)
    &=\bar{\mathscr{S}}_f(\beta)+\bar{\mathscr{S}}_{\varepsilon}(\beta),\\
    \bar{\mathscr{S}}_f(\beta)
    &=\frac{1}{T-H}\sum_{t=H+1}^{T}\beta^\top\widetilde{f}_t,\\
    \bar{\mathscr{S}}_{\varepsilon}(\beta)
    &=\frac{1}{T-H}\sum_{t=1}^{T}\sum_{i=1}^{n_t}
    \varsigma_{it}\eta_{it}W_{it}(\beta).
\end{align*}
The displayed rate conditions imply the conditions of Lemma \ref{lem:partition-basis-moments}. Work on the high-probability event from that lemma on which the relevant cell counts, Gram matrices, and local-polynomial basis norms are uniformly controlled. By local support of the partition basis and the quasi-uniform cell-count bounds,
define, only for this proof,
\begin{align*}
    C_{it}(\beta)
    &=
    \frac{1}{n_t}
    \widehat{p}_t(\beta)^{\top}
    Q_t^{-1}
    p_t(\beta_{it})
    \IF(t\geq H+1).
\end{align*}
Then
\begin{align*}
    \sum_{i=1}^{n_t}|C_{it}(\beta)|^m
    &\lesssim_\P \frac{J^{(m-1)d}}{n^{m-1}},
    \qquad m=2,3,
\end{align*}
uniformly in $t$. For the rolling-window weights, Jensen's inequality and the same
local-support argument give, for each $t$,
\begin{align*}
    \sum_{i=1}^{n_t}|w_{it}(\beta)|^m
    &\lesssim_\P \frac{J^{(m-1)d}}{n^{m-1}}(1+\|f_t\|^m),
    \qquad m=2,3.
\end{align*}
It follows from $T^{-1}\sum_{t=1}^T(1+\|f_t\|^m)=O_\P(1)$ that
\begin{align*}
    \frac{1}{T}\sum_{t=1}^T\sum_{i=1}^{n_t}|w_{it}(\beta)|^m
    &\lesssim_\P \frac{J^{(m-1)d}}{n^{m-1}},
    \qquad m=2,3.
\end{align*}
Since $\varsigma_{it}$ is uniformly bounded, it follows that, for each fixed $\beta\in\mathcal{B}^{\circ}$,
\begin{align*}
    \frac{1}{T}\sum_{t=1}^{T}\sum_{i=1}^{n_t}
    |\varsigma_{it}W_{it}(\beta)|^3
    &\lesssim_\P \frac{J^{2d}}{n^2}
    \qquad\text{and}\qquad
    \Sigma_{\varepsilon}(\beta)
    &=\frac{1}{(T-H)^2}\sum_{t=1}^{T}\sum_{i=1}^{n_t}
    \varsigma_{it}^2\big(W_{it}(\beta)\big)^2
    \lesssim_\P \frac{J^d}{nT}.
\end{align*}

At the origin, the split partition fixes the relevant cell containing zero, and the midpoint and side-length normalization gives $\widehat p_t(0)=p_t(0)$. Thus the leading weights at the origin are already expressed in reference zero-cell coordinates and are $\mathcal{E}_\infty$-measurable. Conditional on $\mathcal{E}_\infty$, the shocks $\{\eta_{it}\}$ are independent, mean zero, and have uniformly bounded third moments. Therefore the conditional Berry--Esseen inequality gives
\begin{align*}
    \sup_{x\in\mathbb{R}}\left|
    \P\left(\frac{\bar{\mathscr{S}}_{\varepsilon}(0)}{\sqrt{\Sigma_{\varepsilon}(0)}}\le x\,\Big|\,\mathcal{E}_\infty\right)
    -\Phi(x)\right|
    &\lesssim_\P
    \frac{T J^{2d}/n^2}{\big((T-H)^2\Sigma_{\varepsilon}(0)\big)^{3/2}}.
\end{align*}
At $\beta=0$, the variance nondegeneracy assumption gives $(T-H)^2\Sigma_{\varepsilon}(0)\gtrsim_\P TJ^d/n$, and the displayed bound is $\lesssim_\P\sqrt{J^d/(nT)}=o_\P(1)$. Hence
\begin{align*}
    \sup_{x\in\mathbb{R}}\left|
    \P\left(\frac{\bar{\mathscr{S}}_{\varepsilon}(0)}{\sqrt{\Sigma_{\varepsilon}(0)}}\le x\right)
    -\Phi(x)\right|
    \to 0.
\end{align*}
Since $\bar{\mathscr{S}}_f(0)=0$ and $\Sigma(0)=\Sigma_{\varepsilon}(0)$, the claim follows for $\beta=0$.

Now take $\beta\neq0$. The process
$\{\beta^\top\widetilde f_t\}$ is a martingale-difference sequence with
respect to $\{\mathcal G_t\}$. Define the deterministic reference variance
\begin{align*}
    \Sigma_{f,0}(\beta)
    &:=
    \frac{\beta^\top\Omega_f\beta}{T-H}.
\end{align*}
By \eqref{eq:factor-variance-stabilization}, the sum of the conditional
variances of the normalized martingale differences satisfies
\begin{align*}
    \frac{1}{(T-H)\beta^\top\Omega_f\beta}
    \sum_{t=H+1}^{T}
    \Var(
    \beta^\top\widetilde f_t
    \mid\mathcal G_{t-1}
    )
    &\to_\P1.
\end{align*}
For every $\epsilon>0$, the conditional fourth-moment bound verifies the
Lindeberg condition using the deterministic threshold
$\epsilon\sqrt{(T-H)\beta^\top\Omega_f\beta}$:
\begin{align*}
    &\frac{1}{(T-H)\beta^\top\Omega_f\beta}
    \sum_{t=H+1}^{T}
    \E\left[
    |\beta^\top\widetilde f_t|^2
    \IF\!\left(
    |\beta^\top\widetilde f_t|
    >
    \epsilon\sqrt{(T-H)\beta^\top\Omega_f\beta}
    \right)
    \,\bigg|\,
    \mathcal G_{t-1}
    \right]
    \\
    &\qquad\lesssim_\P
    \frac{
    \sum_{t=H+1}^{T}
    \E[
    |\beta^\top\widetilde f_t|^4
    \mid\mathcal G_{t-1}
    ]
    }
    {(T-H)^2(\beta^\top\Omega_f\beta)^2}
    \lesssim_\P
    (T-H)^{-1}
    =o_\P(1).
\end{align*}
The martingale central limit theorem and Polya's theorem therefore give
\begin{align*}
    \sup_{x\in\mathbb R}
    \left|
    \P\left(
    \frac{\bar{\mathscr S}_f(\beta)}
    {\sqrt{\Sigma_{f,0}(\beta)}}
    \leq x
    \right)
    -\Phi(x)
    \right|
    &\to0.
\end{align*}
Finally,
\begin{align*}
    \frac{\Sigma_f(\beta)}{\Sigma_{f,0}(\beta)}
    &=
    \frac{
    \beta^\top
    \left[
    (T-H)^{-1}\sum_{t=H+1}^{T}
    \Var(f_t\mid\mathcal G_{t-1})
    \right]
    \beta
    }
    {\beta^\top\Omega_f\beta}
    \to_\P1.
\end{align*}
Slutsky's lemma and continuity of the standard normal cdf thus imply
\begin{align*}
    \sup_{x\in\mathbb R}
    \left|
    \P\left(
    \frac{\bar{\mathscr S}_f(\beta)}
    {\sqrt{\Sigma_f(\beta)}}
    \leq x
    \right)
    -\Phi(x)
    \right|
    &\to0.
\end{align*}
For the idiosyncratic component, use the exact decomposition
$\bar{\mathscr S}_{\varepsilon}(\beta)=\mathscr E(\beta)-\mathscr A(\beta)$.
The direct idiosyncratic bound from Lemma~\ref{lem:general-linearization} and
the uniform first-step-correction bound from
Lemma~\ref{lem:classical-linearization} give
\begin{align*}
    |\mathscr E(\beta)|
    &\lesssim_\P
    \sqrt{\frac{J^d}{n(T-H)}},
    &
    |\mathscr A(\beta)|
    &\lesssim_\P
    \sqrt{\frac{J^d\log(nT)}{nH}}.
\end{align*}
Because $\Sigma_f(\beta)\asymp_\P(T-H)^{-1}$, the displayed rate conditions
imply
\begin{align*}
    \frac{\mathscr E(\beta)}{\sqrt{\Sigma_f(\beta)}}
    &=o_\P(1),
    &
    \frac{\mathscr A(\beta)}{\sqrt{\Sigma_f(\beta)}}
    &=o_\P(1),
\end{align*}
and hence
$\bar{\mathscr S}_{\varepsilon}(\beta)/\sqrt{\Sigma_f(\beta)}=o_\P(1)$.
Separately, the reference idiosyncratic scale is negligible relative to the
factor variance because
\begin{align*}
    \frac{\Sigma_{\varepsilon}(\beta)}{\Sigma_f(\beta)}
    &\lesssim_\P \frac{J^d/(nT)}{1/T}
    =\frac{J^d}{n}
    =o(1).
\end{align*}
Thus $\bar{\mathscr{S}}_{\varepsilon}(\beta)/\sqrt{\Sigma(\beta)}=o_\P(1)$ and $\Sigma_f(\beta)/\Sigma(\beta)=1+o_\P(1)$. Hence full joint convergence of the factor and idiosyncratic components is not needed for fixed $\beta\neq0$; Slutsky's argument only requires the factor approximation and the negligibility of the idiosyncratic component. Since the standard normal distribution is continuous, the preceding Kolmogorov approximation is preserved under this $o_\P(1)$ perturbation, and hence
\begin{align*}
    &\sup_{x\in\mathbb{R}}\left|
    \P\left(\frac{\bar{\mathscr{S}}(\beta)}{\sqrt{\Sigma(\beta)}}\le x\right)
    -\Phi(x)\right|
    \to 0.
\end{align*}
This gives the stated conclusion for $\beta\neq0$. \qed

\subsection{Proof of Theorem~\ref{thm:pointwise-clt}}

By the exact decomposition at the beginning of this section,
\begin{align*}
    \frac{\widehat{\mu}(\beta)-\bar{\mu}_T(\beta;H)}{\sqrt{\Sigma(\beta)}}
    &=\frac{\bar{\mathscr{S}}(\beta)}{\sqrt{\Sigma(\beta)}}
    +\frac{\mathscr{R}(\beta)}{\sqrt{\Sigma(\beta)}}.
\end{align*}
The componentwise bounds from Lemma \ref{lem:general-linearization}, the partition
bounds from Lemma \ref{lem:partition-basis-moments}, and the
generated-regressor refinement in Lemma \ref{lem:classical-linearization} imply
\begin{align*}
    |\mathscr{R}(\beta)|
    &\lesssim_\P
    \mathsf{R}_{\mathtt{CLT}}.
\end{align*}
Here the Gram-perturbation and basis-switching pieces are not reabsorbed into
the local-support idiosyncratic scale; they are part of
$\mathsf{R}_{\mathtt{CLT}}$ through the explicit
$\mathsf{R}_{\mathtt{Q}}$ and $\mathsf{R}_{\mathtt{P}}$ terms in its
definition. The normalization condition
$\mathsf{R}_{\mathtt{CLT}}/\sqrt{\Sigma(\beta)}=o_\P(1)$ therefore removes the
full remainder directly. Lemma \ref{lem:leading-gaussian-components} gives the
Kolmogorov approximation for the leading terms, and the standard perturbation
argument for continuous limiting cdfs gives the displayed result. The sufficient
conditions for $\beta\neq0$ and $\beta=0$ follow from
$\Sigma(\beta)\asymp_\P T^{-1}$ and $\Sigma(0)\asymp_\P J^d/(nT)$,
respectively. \qed
\section{Pointwise-in-\texorpdfstring{$\beta$}{beta} Variance Estimation}\label{sec:pointwise-variance}

This section constructs feasible estimators of the components of the baseline
normalization in Section~\ref{sec:pointwise-distribution}. Recall that
$\Sigma(\beta)=\Sigma_f(\beta)+\Sigma_{\varepsilon}(\beta)$.
{The idiosyncratic variance result has two parts. At fixed
evaluation points whose basis coordinates are anchored in the estimated and
reference partitions, the weights $W_{it}(\beta)$ are
$\mathcal E_\infty$-measurable, $\Sigma(\beta)$ is the predictable quadratic
variation of the leading sum in \eqref{eq:pointwise-leading-sum}, and the
factor--idiosyncratic cross term vanishes by
Assumption~\ref{ass:dgp}(ii)(a). At a generic nonanchored point,
$\Sigma_\varepsilon(\beta)$ remains a reference scale. For fixed
$\beta\neq0$, only a local-support upper bound for the jackknife estimator is
needed because the factor variance is of order $T^{-1}$.}

Let $m_t =\E[f_t\mid\mathcal{G}_{t-1}]$,
and let $\widehat{m}_t$ be a $\mathcal{G}_{t-1}$-measurable predictor of $m_t$.
For instance, $\widehat{m}_t$ can be a recursive past sample mean of
$\{f_s:s<t\}$ when
$\E[f_t\mid\mathcal G_{t-1}]=\mu_f$ for every $t$ and the recursive mean is
consistent for $\mu_f$, or a fitted predictive model using only information
in $\mathcal{G}_{t-1}$ when richer factor dynamics are imposed. Define the
estimated factor innovation $\widehat{U}_t=f_t-\widehat{m}_t$. The
factor-variance estimator is
\begin{align*}
    \widehat{\Sigma}_f(\beta)
    &=\frac{1}{(T-H)^2}\sum_{t=H+1}^{T}
    \beta^\top \widehat{U}_t\widehat{U}_t^\top\beta.
\end{align*}

{For a fixed evaluation point, we propose a bin-restricted
delete-$i$ jackknife estimator. At an anchored point it estimates
$\Sigma_{\varepsilon}(\beta)$ at the idiosyncratic scale; at a generic
nonzero point we use only its order bound.}
Let $\widehat{M}_t^{(-i)}(\beta)$ denote the second-step estimator with unit $i$ removed at time $t$, holding the partition $\widehat{\Delta}_t$ fixed. Define the leave-one-out pseudo-value
\begin{align*}
    \widehat{d}_{it}(\beta)
    &=\widehat{M}_t^{(-i)}(\beta)-\widehat{M}_t(\beta),
\end{align*}
with the convention that $\widehat{d}_{it}(\beta)=0$ when unit $i$ is absent from the cross-section at time $t$. The bin-restricted jackknife estimator is
\begin{align*}
    \widehat{\Sigma}_{\varepsilon}(\beta)
    &=\frac{1}{(T-H)^2}\sum_{i=1}^{n}
    \sum_{t=H+1}^{T}\sum_{s=H+1}^{T}
    \widehat{d}_{it}(\beta)\widehat{d}_{is}(\beta)\IF(|t-s|\leq H),
\end{align*}
where $[n]$ is the common index set from the general setup and absent units have
zero pseudo-values.
The Sherman--Morrison formula for deleting the $i$-th observation from the period-$t$ least-squares normal equations gives
\begin{align*}
    \widehat{d}_{it}(\beta)
    &=-\widehat{C}_{it}(\beta)\widehat{\varepsilon}_{it},
    \qquad
    \widehat{C}_{it}(\beta)
    =
    \frac{1}{n_t}
    \frac{
    \widehat{p}_t(\beta)^\top
    \widehat{Q}_t^{-1}
    \widehat{p}_t(\widehat{\beta}_{it})
    }
    {1-
    n_t^{-1}
    \widehat{p}_t(\widehat{\beta}_{it})^\top
    \widehat{Q}_t^{-1}
    \widehat{p}_t(\widehat{\beta}_{it})},
    \qquad
    \widehat{\varepsilon}_{it}
    =
    R_{it}-\widehat{M}_t(\widehat{\beta}_{it}).
\end{align*}
When $\mathfrak{q}=0$, this reduces to the more familiar bin-count expression
\begin{align*}
    \widehat{\Sigma}_{\varepsilon}(\beta)
    &=\frac{1}{(T-H)^{2}}\sum_{i=1}^{n}\sum_{t=H+1}^{T}\sum_{s=H+1}^{T}
    \frac{\IF_{it}^{(\widehat{\beta})}\IF_{is}^{(\widehat{\beta})}}
    {(\widehat{n}_{j_{0}(t)t}-1)(\widehat{n}_{j_{0}(s)s}-1)}
    \widehat{\varepsilon}_{it}\widehat{\varepsilon}_{is}\IF(|t-s|\leq H),
\end{align*}
where $\IF_{it}^{(\widehat{\beta})}=\IF(\widehat{\beta}_{it}\in\widehat{P}_{j_{0}(t)t})$ and $j_0(t)$ denotes the cell containing the evaluation point $\beta$ at time $t$.

Combining the two components, the feasible variance estimator is
\begin{align*}
    \widehat{\Sigma}(\beta)
    &=\widehat{\Sigma}_f(\beta)+\widehat{\Sigma}_{\varepsilon}(\beta).
\end{align*}
Recall that for $\beta=0$, $\Sigma_f(0)=0$ and $\widehat{\Sigma}(0)=\widehat{\Sigma}_{\varepsilon}(0)$
suffices; for each fixed $\beta\neq 0$,
$\Sigma_f(\beta)\asymp T^{-1}$.

\subsection{Results}

The first result controls the factor component of the variance estimator and isolates the cost of replacing the predictable factor mean by a fitted predictor.

\begin{lemma}[Factor Variance Estimation]\label{lem:factor-variance}
    Suppose that Assumption~\ref{ass:dgp}(iv) holds {and $H\leq cT$ for some fixed $c\in(0,1)$}. If
    $\widehat{m}_t$ is $\mathcal{G}_{t-1}$-measurable and
    \begin{align*}
        &\frac{1}{T-H}\sum_{t=H+1}^{T}\|m_t-\widehat{m}_t\|^2
        =o_\P(T),
    \end{align*}
    then, for each fixed $\beta\in\mathcal{B}$,
    \begin{align*}
        \widehat{\Sigma}_f(\beta)-\Sigma_f(\beta)-\mathscr{B}_{\mathtt{F}}(\beta)
        &=
        \begin{cases}
            o_\P(\|\beta\|^2T^{-1}),&\beta\neq0,\\
            0,&\beta=0,
        \end{cases}
    \end{align*}
    where
    \begin{align*}
        \mathscr{B}_{\mathtt{F}}(\beta)
        =
        \frac{1}{(T-H)^2}\sum_{t=H+1}^{T}
        \beta^\top(m_t-\widehat{m}_t)(m_t-\widehat{m}_t)^\top\beta
        \geq0
    \end{align*}
    is the squared factor-prediction bias.
\end{lemma}

{The next result first gives, at every fixed evaluation point, the
upper bound needed to make the entire idiosyncratic component negligible
relative to the factor variance. This bound does not require the selected
block to be anchored. The split-at-zero construction is anchored at
$\beta=0$, and the generalized construction in
Section~\ref{sec:factor-cancelling} is anchored at each prespecified contrast
point. At these points the result further gives sharp additive consistency at
the idiosyncratic scale.}

\begin{lemma}[Bin-Restricted Jackknife]\label{lem:bin-restricted-jackknife}
    {Fix $\beta\in\mathcal B^\circ$.} Suppose Assumptions~\ref{ass:dgp},
    \ref{ass:first-step-ls}, \ref{ass:partition},
    {\ref{ass:interior-separation}}, and
    \ref{ass:first-step-bias} hold, and that $s_\mu\geq2$.
    Assume further that
    \begin{align*}
        &\frac{T^{2/q}}{H}
        +\sqrt{\frac{\log T}{H}}
        =o(1),\\
        &\frac{J^2\log T}{n}=o(1),\\
        &\frac{J^{d-1}\max\{J,\sqrt{H}\}\log^2(nT)}{\min\{n^{1-2/q}T^{1-4/q}h,n\}}\lesssim 1,\qquad
        \frac{\max\{J^{2/(q-2)}T^{2/q},T^{2(3+d)/q}\}}{H}=o(1).
    \end{align*}
    Assume further that
    \begin{align}
        &H\log(nT)\big(J^{-2\min\{s_\mu,\mathfrak{q}+1\}}+\mathsf{R}_{\mathtt{FS}}^4+{\mathsf{B}_{\mathtt{FS}}^2}\big)
        +\frac{\log(nT)}{H}
        +\frac{J^d\log^{d+2}(nT)}{n}
        =o(1).
        \label{eq:jackknife-conditions}
    \end{align}
    {Then the following conclusions hold.
    \begin{enumerate}[label=(\roman*)]
        \item {For every fixed $\beta\in\mathcal{B}^\circ$,}
        \begin{align*}
            \left|\widehat{\Sigma}_{\varepsilon}(\beta)\right|
            &=O_\P\!\left(\frac{J^d}{nT}\right).
        \end{align*}

        \item If, in addition, $J\mathsf R_{\mathtt{FS}}=o(1)$ and
        \begin{align}\label{eq:jackknife-anchored-basis}
            \widehat p_t(\beta)&=p_t(\beta),
            \qquad t=H+1,\ldots,T,
        \end{align}
        then
        \begin{align*}
            \widehat{\Sigma}_{\varepsilon}(\beta)
            &=\Sigma_{\varepsilon}(\beta)
            +o_\P\!\left(\frac{J^d}{nT}\right).
        \end{align*}
    \end{enumerate}}
\end{lemma}

Combining the two pieces gives the feasible variance estimator used in the pointwise Gaussian approximation.

\begin{theorem}[Feasible Variance Estimation]\label{thm:variance-estimation}
    Fix $\beta\in\mathcal B^\circ$. Suppose Assumptions~\ref{ass:dgp},
    \ref{ass:first-step-ls}, \ref{ass:partition},
    {\ref{ass:interior-separation}}, and
    \ref{ass:first-step-bias} hold, and that
    $s_\mu\geq2$. Assume that $\widehat{m}_t$ is $\mathcal{G}_{t-1}$-measurable,
    \begin{align*}
        &\frac{1}{T-H}\sum_{t=H+1}^{T}\|m_t-\widehat{m}_t\|^2
        =o_\P(T),
    \end{align*}
    and that the following rate restrictions hold:
    \begin{align*}
        &\frac{T^{2/q}}{H}
        +\sqrt{\frac{\log T}{H}}
        =o(1),\\
        &\frac{J^2\log T}{n}=o(1),\\
        &\frac{J^{d-1}\max\{J,\sqrt{H}\}\log^2(nT)}{\min\{n^{1-2/q}T^{1-4/q}h,n\}}\lesssim 1,\qquad
        \frac{\max\{J^{2/(q-2)}T^{2/q},T^{2(3+d)/q}\}}{H}=o(1),\\
        &H\log(nT)\big(J^{-2\min\{s_\mu,\mathfrak{q}+1\}}+\mathsf{R}_{\mathtt{FS}}^4+{\mathsf{B}_{\mathtt{FS}}^2}\big)
        +\frac{\log(nT)}{H}
        +\frac{J^d\log^{d+2}(nT)}{n}
        =o(1).
    \end{align*}
    Then
    \begin{align*}
        \widehat{\Sigma}(\beta)-\Sigma(\beta)-\mathscr{B}_{\mathtt{F}}(\beta)
        &= o_\P(T^{-1}).
    \end{align*}
    For $\beta=0$, if, in addition,
    equation \eqref{eq:jackknife-anchored-basis} holds,
    $J\mathsf R_{\mathtt{FS}}=o(1)$, and $(nT/J^d)\Sigma_{\varepsilon}(0)\gtrsim_\P1$, then
    \begin{align*}
        \widehat{\Sigma}(0)-\Sigma(0)
        &=o_\P\Big(\frac{J^d}{nT}\Big).
    \end{align*}
\end{theorem}

\subsection{Fama-MacBeth Variance Estimator}

The Fama-MacBeth variance estimator is widely used because it treats the time-series dispersion of the per-period second-step estimates as the sampling variance. Its target generally differs from the baseline normalization in Section~\ref{sec:pointwise-distribution}, which equals the predictable variance of the leading sum at anchored evaluation points. Recall that
\begin{align*}
    \mathscr{B}(\beta)
    &=\frac{1}{T-H}\sum_{t=H+1}^{T}\mathscr{B}_t(\beta),
\end{align*}
where $\mathscr{B}_t(\beta)$ is the period-$t$ smoothing-bias component from Section~\ref{sec:classical-linearization}. The classical Fama-MacBeth estimator and its oracle bias-corrected variant are
\begin{align*}
    \widehat{\Sigma}_{\mathtt{FM}}(\beta)
    &=\frac{1}{(T-H)^2}\sum_{t=H+1}^{T}
    \bigl(\widehat{M}_t(\beta)-\widehat{\mu}(\beta)\bigr)^2,\\
    \widetilde{\Sigma}_{\mathtt{FM}}(\beta)
    &=\frac{1}{(T-H)^2}\sum_{t=H+1}^{T}
    \Bigl[
    \bigl(\widehat{M}_t(\beta)-\mathscr{B}_t(\beta)\bigr)
    -\bigl(\widehat{\mu}(\beta)-\mathscr{B}(\beta)\bigr)
    \Bigr]^2.
\end{align*}
The second estimator is an oracle benchmark because $\mathscr{B}_t(\beta)$ depends on the unknown function $\mu_t(\cdot)$.

For the comparison at the origin, define the direct oracle idiosyncratic term
\begin{align*}
    \mathscr{E}_{\mathtt{o},t}(\beta)
    &=
    \sum_{i=1}^{n_t}
    \varsigma_{it}\eta_{it}
    \left(
    \frac{1}{n_t}\widehat{p}_t(\beta)^\top Q_t^{-1}p_t(\beta_{it})
    \right),
\end{align*}
and its pre-shock conditional variance
\begin{align*}
    \Sigma_{\varepsilon,\mathtt{o}}(\beta)
    &=
    \frac{1}{(T-H)^2}
    \sum_{t=H+1}^{T}
    \E\!\left[
    \mathscr{E}_{\mathtt{o},t}(\beta)^2
    \mid
    \mathcal{C}_t
    \right]
    =
    \frac{1}{(T-H)^2}\sum_{t=H+1}^{T}\sum_{i=1}^{n_t}
    \varsigma_{it}^2
    \left(
    \frac{1}{n_t}\widehat{p}_t(\beta)^\top Q_t^{-1}p_t(\beta_{it})
    \right)^2.
\end{align*}
Here $\mathcal C_t$, defined in Section~\ref{sec:classical-conditioning-discussion}, fixes the current innovation-free scales and the past generated design while excluding the current primitive innovations.
This is the variance of the direct cross-sectional idiosyncratic component. It differs from $\Sigma_{\varepsilon}(\beta)$ in Section~\ref{sec:pointwise-distribution}, whose weight $W_{it}(\beta)$ also subtracts the first-step generated-regressor correction.

The following theorem makes the comparison precise by characterizing the probability limit of the Fama-MacBeth estimator.
Define
\begin{align*}
    \mathsf{R}_{\mathtt{FM}}
    &=\mathsf{R}_{\mathtt{LA}}+\frac{\log(nT)}{H}
    +\sqrt{\frac{J^{2d}\mathsf{R}_{\mathtt{P}}}{n}}
    +\frac{\IF(\mathfrak{q}=0)}{J}
    +\mathsf{R}_{\mathtt{FS}}^2.
\end{align*}

\begin{theorem}[Fama-MacBeth Variance Estimation]\label{thm:fama-macbeth-variance}\leavevmode\par\noindent
    Suppose Assumptions~\ref{ass:dgp},
    \ref{ass:first-step-ls}, \ref{ass:partition},
    \ref{ass:interior-separation}, and
    \ref{ass:first-step-bias} hold, and that $s_\mu\geq2$.
    Assume further that
    \begin{align*}
        &\frac{T^{2/q}}{H}
        +\sqrt{\frac{\log T}{H}}=o(1),\qquad
        \mathsf R_{\mathtt{FS}}=o(1),\\
        &\frac{J^2\log T}{n}=o(1),\\
        &\frac{J^{d-1}\max\{J,\sqrt{H}\}\log^2(nT)}{\min\{n^{1-2/q}T^{1-4/q}h,n\}}\lesssim 1,\qquad
        \frac{\max\{J^{2/(q-2)}T^{2/q},T^{2(3+d)/q}\}}{H}=o(1).
    \end{align*}
    The anchored-basis condition
    is required for the origin statement below and holds automatically under
    Assumption~\ref{ass:partition}.
    If, in addition,
    \begin{align*}
        J^{-\min\{s_\mu,\mathfrak{q}+1\}}
        +\mathsf{R}_{\mathtt{FM}}
        &=o(1),
    \end{align*}
    then, for each fixed $\beta\neq0$ in $\mathcal{B}^\circ$,
    \begin{align*}
        \widehat{\Sigma}_{\mathtt{FM}}(\beta)
        &=\Sigma_f(\beta)+\Sigma_\mu(\beta)+o_\P(T^{-1}),
    \end{align*}
    where
    \begin{align*}
        \Sigma_\mu(\beta)
        &=\frac{1}{(T-H)^2}\sum_{t=H+1}^{T}
        \bigl(\mu_t(\beta)-\bar{\mu}_T(\beta;H)\bigr)^2.
    \end{align*}
    For $\beta=0$, if, in addition,
    $J\mathsf R_{\mathtt{FS}}=o(1)$,
    \begin{align*}
        \left|\widetilde{\Sigma}_{\mathtt{FM}}(0)
        -\Sigma_{\varepsilon,\mathtt{o}}(0)-\Sigma_\mu(0)\right|
        &\lesssim_\P
        \frac{\mathsf{R}_{\mathtt{FM}}^2}{T}
        +\left(\sqrt{\frac{J^d}{nT}}
        +\frac{\mathsf{R}_{\mathtt{FM}}}{\sqrt{T}}\right)
        \sqrt{\Sigma_\mu(0)}
        +o_\P\Big(\frac{J^d}{nT}\Big).
    \end{align*}
\end{theorem}

\subsection{Discussion}

\begin{enumerate}[label=(\roman*)]
    \item \emph{Truncated jackknife.}
    The estimator $\widehat{\Sigma}_{\varepsilon}(\beta)$ uses the delete-$i$ jackknife perturbations only for time pairs with $|t-s|\leq H$. After the rolling-window first-step correction is isolated, pairs farther apart do not share primitive idiosyncratic shocks and are conditionally mean zero. The truncation therefore keeps the serial dependence induced by the first-step window while reducing the number of nonzero pairs from order $T^2$ to order $TH$.

    \item \emph{Scale.}
    {At every fixed evaluation point, the selected-block argument
    gives
    $|\widehat\Sigma_\varepsilon(\beta)|=O_\P(J^d/(nT))$. For fixed
    $\beta\neq0$, this is sufficient: $J^d/n=o(1)$ makes both the estimator
    and the reference idiosyncratic scale negligible relative to the
    first-order factor variance. At an anchored evaluation point,
    Lemma~\ref{lem:bin-restricted-jackknife} further gives additive
    consistency for the idiosyncratic variance at the scale $J^d/(nT)$. At
    $\beta=0$, this is the relevant first-order variance scale, so relative
    consistency additionally requires the nondegeneracy condition
    $(nT/J^d)\Sigma_{\varepsilon}(0)\gtrsim_\P1$.}

    \item \emph{Factor prediction.}
    The factor variance estimator separates sampling variation from prediction bias. Lemma~\ref{lem:factor-variance} controls the centered estimation error, while $\mathscr{B}_{\mathtt{F}}(\beta)$ records the bias from using $\widehat m_t$ in place of $m_t$. Thus relative consistency of $\widehat{\Sigma}(\beta)$ for fixed $\beta\neq0$ requires this prediction bias to be negligible relative to $\Sigma(\beta)$; otherwise the estimator is conservative.

    \item \emph{Approximation bias in the jackknife.}
    The jackknife condition \eqref{eq:jackknife-conditions} uses the actual second-step approximation order $J^{-\min\{s_\mu,\mathfrak{q}+1\}}$. {Up to the displayed $\log(nT)$ factor, it gives the familiar $HJ^{-2}$ restriction for $\mathfrak{q}=0$, and $HJ^{-4}$ for a degree-one second step when $s_\mu\ge2$.} This variance-estimation restriction is distinct from the same-cell nonconstant-basis term absorbed by $\mathsf{R}_{\mathtt{LA}}$ in the stochastic linearization.

    \item \emph{Primitive conditions.}
    {The displayed conditions $J^2\log T/n=o(1)$ and
    $T^{2/q}/H+\sqrt{\log T/H}=o(1)$ verify the common partition and
    first-step moment requirements. The additional condition
    $J\mathsf R_{\mathtt{FS}}=o(1)$ in the anchored case verifies the
    comparison with a single reference block, including
    $\Delta_{\mathtt{JK}}=o(1)$ and
    $\mathsf{R}_{\mathtt{JKMD}}=o(1)$. The selected-block upper bound does not
    make that comparison and does not require this stronger localization.}

    \item \emph{Fama-MacBeth.}
    Theorem~\ref{thm:fama-macbeth-variance} shows that the Fama-MacBeth variance has a different target. For fixed $\beta\neq0$, it includes the time-series variation $\Sigma_\mu(\beta)$ of the conditional mean path. At $\beta=0$, even the oracle bias-corrected version targets $\Sigma_{\varepsilon,\mathtt{o}}(0)+\Sigma_\mu(0)$, where $\Sigma_{\varepsilon,\mathtt{o}}(0)$ is based on the direct second-step idiosyncratic weights rather than the generated-regressor-adjusted weights in $\Sigma_{\varepsilon}(0)$.
    The origin comparison uses
    \eqref{eq:fm-periodwise-residual-squares} and conditions on current
    innovation-free information and earlier primitive shocks. It does not
    require the sharp jackknife rate conditions. The extra condition
    $J^{-\min\{s_\mu,\mathfrak{q}+1\}}+\mathsf{R}_{\mathtt{FM}}=o(1)$ is only
    needed for the fixed nonzero-$\beta$ probability limit.
\end{enumerate}

\subsection{Proof of Lemma~\ref{lem:factor-variance}}

Let $\widetilde{\Sigma}_{f}(\beta)
 =
\frac{1}{(T-H)^2}
\sum_{t=H+1}^{T}
\beta^{\top}\,\widetilde{f}_t\,\widetilde{f}_t^{\top}\,\beta$,
and recall $\widetilde{f}_t = f_t - \E[f_t\mid\mathcal{G}_{t-1}]$.
Decompose $\widehat{\Sigma}_{f}(\beta)=\widetilde{\Sigma}_{f}(\beta)
+\mathscr{B}_{\mathtt{F}}(\beta)+\mathscr{V}_{\mathtt{F}}(\beta)$, where
\begin{align*}
    \mathscr{V}_{\mathtt{F}}(\beta)
    &=
    \frac{2}{(T-H)^2}\sum_{t=H+1}^{T}
    \beta^{\top}\widetilde{f}_t\,
    (m_t-\widehat{m}_t)^\top\beta.
\end{align*}
Conditional on ${\mathcal{F}}_{t-1}$, the summands in $\mathscr{V}_{\mathtt{F}}(\beta)$ form a martingale-difference sequence, and their predictable quadratic variation is bounded by
\begin{align*}
    {\frac{4\|\beta\|^4}{(T-H)^4}\sum_{t=H+1}^{T}
    \|m_t-\widehat{m}_t\|^2
    \big\|\Var(f_t\mid\mathcal{F}_{t-1})\big\|}
    &\mathrel{{\lesssim_\P}}
    \frac{\|\beta\|^4}{(T-H)^4}\sum_{t=H+1}^{T}
    \|m_t-\widehat{m}_t\|^2\\
    &=
    \|\beta\|^4o_\P(T^{-2}).
\end{align*}
{Assumption~\ref{ass:dgp}(iv) and conditional Cauchy--Schwarz give
$\max_{1\leq t\leq T}\|\Var(f_t\mid\mathcal{F}_{t-1})\|\lesssim_\P1$.}
{The Lenglart--Rebolledo inequality, applied after localization, gives}
$\mathscr{V}_{\mathtt{F}}(\beta)=\|\beta\|^2o_\P(T^{-1})$, while $\mathscr{V}_{\mathtt{F}}(0)=0$ by construction.
It remains to compare $\widetilde{\Sigma}_{f}(\beta)$ with $\Sigma_f(\beta)$.
Write $\xi_t=(\beta^\top\widetilde{f}_t)^2
-\E[(\beta^\top\widetilde{f}_t)^2\mid\mathcal{G}_{t-1}]$, so that
$\{\xi_t,\mathcal{G}_t\}$ is a martingale difference
sequence and
\begin{align*}
    \widetilde{\Sigma}_{f}(\beta)-\Sigma_f(\beta)
    = \frac{1}{(T-H)^2}\sum_{t=H+1}^T \xi_t.
\end{align*}
By {the unconditional higher-moment bound in} Assumption~\ref{ass:dgp}(iv),
$\sup_t{\E[\|\widetilde{f}_t\|^4]}\lesssim 1$,
so $\E[\xi_t^2]\leq\E[(\beta^\top\widetilde{f}_t)^4]
\lesssim\|\beta\|^4$. The {variance of $\widetilde{\Sigma}_f(\beta)-\Sigma_f(\beta)$}
satisfies
\begin{align*}
    \frac{1}{(T-H)^4}\sum_{t=H+1}^T\E[\xi_t^2]
    \lesssim \frac{\|\beta\|^4}{(T-H)^3}.
\end{align*}
For $\beta\neq 0$, applying Chebyshev's inequality gives
\begin{align*}
    \Big|\widetilde{\Sigma}_{f}(\beta)-\Sigma_f(\beta)\Big|
    \lesssim_\P \frac{\|\beta\|^2}{T^{3/2}}.
\end{align*}
For $\beta=0$, both $\widetilde{\Sigma}_{f}(0)$
and $\Sigma_f(0)$ vanish identically, so the result
is immediate.
\qed

{
\subsection{Preliminary Results for Proof of Lemma~\ref{lem:bin-restricted-jackknife}}

We next record the unit-level residual bounds used in the jackknife proof.
They are stated separately because the aggregate linearization in
Lemmas~\ref{lem:general-linearization} and~\ref{lem:classical-linearization}
does not by itself control the banded products of fitted residuals.  Put
$T_0=T-H$ and, for $B=C$ or $B=\widehat C$, write
\begin{align*}
    \mathscr H_B(U,V)
    &=
    \frac{1}{T_0^2}\sum_{i=1}^{n}
    \sum_{t=H+1}^{T}\sum_{s=H+1}^{T}
    B_{it}(\beta)B_{is}(\beta)U_{it}V_{is}
    \IF(|t-s|\leq H),
\end{align*}
where coefficients and arrays are set to zero for invalid unit--date pairs.
The map $\mathscr H_B$ is only a convenient signed bilinear form; the flat
band matrix $[\IF(|t-s|\leq H)]_{t,s}$ is not positive semidefinite, so
$\mathscr H_B$ is not used as an inner product.

For any fixed $\beta$, define
\begin{align*}
    v_{it}
    &=
    \frac{1}{H}\sum_{l=1}^{H}
    K\!\left(\frac{l}{H}\right)
    S A(t)^{-1}X_l(t)\varepsilon_{i,t-l},
    \\
    Y_{it}
    &=
    \varepsilon_{it}
    -\nabla\mu_t(\beta_{it})^\top v_{it}.
\end{align*}
When $\beta$ is anchored, also define
\begin{align*}
    C_{it}(\beta)
    &=
    \frac{1}{n_t}p_t(\beta)^\top Q_t^{-1}p_t(\beta_{it}).
\end{align*}
For $l\in[H]$, define, whenever $t+l\in\{H+1,\ldots,T\}$ and
$i\in[n_{t+l}]$,
\begin{align*}
    a_{itl}(\beta)
    &=
    \nabla\mu_{t+l}(\beta_{i,t+l})^\top
    K\!\left(\frac{l}{H}\right)
    S A(t+l)^{-1}X_l(t+l),
\end{align*}
and set it equal to zero otherwise.
On the high-probability event from
Lemma~\ref{lem:partition-basis-moments}, local support gives, for every fixed
$\beta$,
\begin{align}
    \max_{t\in[T]}\max_{i\in[n]}
    |\widehat C_{it}(\beta)|
    &\lesssim_\P \frac{J^d}{n},\notag\\
    \max_{t\in[T]}\sum_{i=1}^{n}
    |\widehat C_{it}(\beta)|^m
    &\lesssim_\P \frac{J^{(m-1)d}}{n^{m-1}},
    \qquad m=2,3,
    \label{eq:jackknife-weight-bounds}
\end{align}
because $\widehat p_t(\beta)$ selects only one well-conditioned interior
block and its normalized polynomial coordinates are bounded. The same bounds
and scalar Cauchy--Schwarz give, uniformly over $t,s\in[T]$,
\begin{align}
    \sum_{i=1}^{n}\widehat C_{it}(\beta)^2
    \widehat C_{is}(\beta)^2
    &\lesssim_\P \frac{J^{3d}}{n^3},
    &
    \sum_{i=1}^{n}|\widehat C_{it}(\beta)
    \widehat C_{is}(\beta)|
    &\lesssim_\P \frac{J^d}{n}.
    \label{eq:jackknife-cross-weight-bounds}
\end{align}
Under \eqref{eq:jackknife-anchored-basis}, all these bounds also hold with
$\widehat C_{it}(\beta)$ replaced by $C_{it}(\beta)$.
For a period-$t$ vector $z=(z_1,\ldots,z_{n_t})^\top$ and a row satisfying
$\widehat C_{it}(\beta)\neq0$, let
\begin{align*}
    (\widehat\Pi_tz)_i
    &=
    \widehat p_t(\widehat\beta_{it})^\top\widehat Q_t^{-1}
    \frac{1}{n_t}\sum_{k=1}^{n_t}
    \widehat p_t(\widehat\beta_{kt})z_k,
\end{align*}
and set $(\widehat\Pi_tz)_i=0$ on the unused rows.  Thus only the interior
block selected by the jackknife weight is inverted below.  Recall
$M_t(b)=\mu_t(b)+b^\top\widetilde f_t$ and, finally, write
\begin{align*}
    \delta_{it}
    &=\widehat\beta_{it}-\beta_{it},
    \\
    \mathscr T_{it}^{(\mu)}
    &=
    \mu_t(\widehat\beta_{it})-\mu_t(\beta_{it})
    -\nabla\mu_t(\beta_{it})^\top\delta_{it}.
\end{align*}
On every row with $\widehat C_{it}(\beta)\neq0$, the following is an exact
definition of the residual remainder; set
$\widetilde{\mathscr R}_{it}=0$ on the unused rows:
\begin{align}
    \widehat\varepsilon_{it}
    &=Y_{it}+\widetilde{\mathscr R}_{it},
    \qquad \widehat C_{it}(\beta)\neq0,
    \label{eq:jackknife-residual-expansion}\\
    \widetilde{\mathscr R}_{it}
    &=
    \left[(I-\widehat\Pi_t)
    M_t(\widehat\beta_{\cdot t})\right]_i
    -(\widehat\Pi_t\varepsilon_{\cdot t})_i
    -\nabla\mu_t(\beta_{it})^\top(\delta_{it}-v_{it})
    \notag\\
    &\qquad
    -\widetilde f_t^\top\delta_{it}
    -\mathscr T_{it}^{(\mu)}
    +\left\{
    \widehat\Pi_t\left[
    \nabla\mu_t(\beta_{\cdot t})^\top\delta_{\cdot t}
    +\widetilde f_t^\top\delta_{\cdot t}
    +\mathscr T_{\cdot t}^{(\mu)}
    \right]
    \right\}_i.
    \notag
\end{align}
Here $M_t(\widehat\beta_{\cdot t})$ and the expressions inside
$\widehat\Pi_t$ denote the corresponding period-$t$ vectors.
{Number the six terms in the second equality of
\eqref{eq:jackknife-residual-expansion} in their displayed order. We use the
following exact split. The block
$\widetilde{\mathscr R}_{it}^{(\mathrm{md})}$ contains term~(2); the part of
term~(3) obtained by replacing $\delta_{it}-v_{it}$ with the summation in
\eqref{eq:jackknife-first-step-residual}; and, after writing
$\delta_{it}=v_{it}+(\delta_{it}-v_{it})$, the two parts of term~(4)
containing $v_{it}$ and that same summation.
The block $\widetilde{\mathscr R}_{it}^{(\mathrm{pr})}$ contains all the
remaining pieces. Thus $\widetilde{\mathscr R}_{it}
=\widetilde{\mathscr R}_{it}^{(\mathrm{md})}
+\widetilde{\mathscr R}_{it}^{(\mathrm{pr})}$.}
{The current-shock pieces of the first block are handled by
ordering dates. Its remaining piece and the second block are handled by
direct squared-sum bounds.}

Define
\begin{align*}
    \Delta_{\mathtt{JK}}
    &=
    \frac{J^d}{n}
    +J^{2d}\mathsf R_{\mathtt Q}^{2}
    +J^d\mathsf R_{\mathtt P},
    \\
    \mathsf R_{\mathtt{JKMD}}
    &=
    \frac{J^d}{n}
    +J^{2d}\mathsf R_{\mathtt Q}^{2}
    +J^d\mathsf R_{\mathtt P}
    +\frac{1}{H}+\frac{1}{\sqrt T}
    +\left(
    \frac{T^{2/q}}{H}+\sqrt{\frac{\log T}{H}}
    \right)^2,
    \\
    \mathsf R_{\mathtt{JKPR}}
    &=
    H\left(
    \mathsf R_{\mathtt{FS}}^4
    +J^{-2\min\{s_\mu,\mathfrak q+1\}}
    +\mathsf B_{\mathtt{FS}}^2
    \right)
    +\frac{J^d\log^{d+1}(nT)}{n}
    +\frac{1}{H}.
\end{align*}

\begin{lemma}[Anchored Residual Arrays]\label{lem:jackknife-residual-arrays}
    Suppose that the assumptions and rate conditions of
    Lemma~\ref{lem:bin-restricted-jackknife} hold and that
    $J\mathsf R_{\mathtt{FS}}=o(1)$. For every fixed
    evaluation point satisfying~\eqref{eq:jackknife-anchored-basis},
    \begin{align}
        &\sum_{t=H+1}^{T}\sum_{i=1}^{n_t}
        \big(\widehat C_{it}(\beta)-C_{it}(\beta)\big)^2
        \left(
        \varepsilon_{it}^2
        +\frac{1}{H}\sum_{l=1}^{H}
        a_{i,t-l,l}(\beta)^2\varepsilon_{i,t-l}^2
        \right)
        \lesssim_\P
        T\frac{J^d}{n}\Delta_{\mathtt{JK}},
        \label{eq:jackknife-weighted-hat-removal}
    \end{align}
    and
    \begin{align}
        {\left|
        \mathscr H_{\widehat C}(Y,Y)-\mathscr H_C(Y,Y)
        \right|}
        &{\lesssim_\P
        \frac{J^d}{nT}\Delta_{\mathtt{JK}}^{1/4}},
        \label{eq:jackknife-leading-replacement}\\
        \left|\mathscr H_{\widehat C}
        (Y,\widetilde{\mathscr R}^{(\mathrm{md})})\right|
        &\lesssim_\P
        \frac{J^d}{nT}\mathsf R_{\mathtt{JKMD}}^{1/2},
        \notag\\
        \left|\mathscr H_{\widehat C}
        (\widetilde{\mathscr R}^{(\mathrm{md})},
        \widetilde{\mathscr R}^{(\mathrm{md})})\right|
        &\lesssim_\P
        \frac{J^d}{nT}\mathsf R_{\mathtt{JKMD}},
        \notag\\
        \left|\mathscr H_{\widehat C}
        (Y,\widetilde{\mathscr R}^{(\mathrm{pr})})\right|
        &\lesssim_\P
        \frac{J^d}{nT}
        \sqrt{\mathsf R_{\mathtt{JKPR}}\log(nT)},
        \notag\\
        \left|\mathscr H_{\widehat C}
        (\widetilde{\mathscr R}^{(\mathrm{pr})},
        \widetilde{\mathscr R}^{(\mathrm{pr})})\right|
        &\lesssim_\P
        \frac{J^d}{nT}\mathsf R_{\mathtt{JKPR}},
        \notag\\
        \left|\mathscr H_{\widehat C}
        (\widetilde{\mathscr R}^{(\mathrm{md})},
        \widetilde{\mathscr R}^{(\mathrm{pr})})\right|
        &\lesssim_\P
        \frac{J^d}{nT}
        \sqrt{\mathsf R_{\mathtt{JKMD}}
        \mathsf R_{\mathtt{JKPR}}}.
        \label{eq:jackknife-array-bounds}
    \end{align}
\end{lemma}

\begin{proof}
\textit{Step 1: exact residual expansion.}
On every row with $\widehat C_{it}(\beta)\neq0$,
\begin{align*}
    \widehat\varepsilon_{it}
    &=
    \left[(I-\widehat\Pi_t)
    \{M_t(\beta_{\cdot t})+\varepsilon_{\cdot t}\}\right]_i.
\end{align*}
The unused rows never enter $\mathscr H_{\widehat C}$.  Write
\begin{align*}
    M_t(\beta_{it})
    &=
    M_t(\widehat\beta_{it})
    -\nabla\mu_t(\beta_{it})^\top\delta_{it}
    -\widetilde f_t^\top\delta_{it}
    -\mathscr T_{it}^{(\mu)}.
\end{align*}
Substitution and the definition of $Y_{it}$ give
\eqref{eq:jackknife-residual-expansion}.  Moreover, the exact first-step
decomposition gives
\begin{align}
    \delta_{it}-v_{it}
    &=
    \breve\beta_{it}-\beta_{it}
    +\frac{1}{H}\sum_{l=1}^{H}
    K\!\left(\frac{l}{H}\right)
    S\{\widehat A(t)^{-1}-A(t)^{-1}\}
    X_l(t)\varepsilon_{i,t-l}.
    \label{eq:jackknife-first-step-residual}
\end{align}
{Thus the first-step bias and the error from replacing
$A(t)^{-1}$ by $\widehat A(t)^{-1}$ enter the remainder separately.} If
$\mathfrak q\geq1$, polynomial reproduction makes
the factor part of
$[(I-\widehat\Pi_t)M_t(\widehat\beta_{\cdot t})]_i$ identically zero.  If
$\mathfrak q=0$, that part is the within-cell deviation of
    $\widehat\beta_{it}^\top\widetilde f_t$ and is bounded by
    $J^{-1}\|\widetilde f_t\|$.  {Together with
    \eqref{eq:jackknife-first-step-residual}, this verifies that the two
    blocks defined after~\eqref{eq:jackknife-residual-expansion} contain every
    part of the remainder exactly once.}

\medskip\noindent\textit{Step 2: shock-weighted generated-design bounds.}
Put
\begin{align*}
    \omega_{it}
    &=
    \varepsilon_{it}^2
    +\frac{1}{H}\sum_{l=1}^{H}
    a_{i,t-l,l}(\beta)^2\varepsilon_{i,t-l}^2.
\end{align*}
We first make the leave-out construction explicit. For a set
$\mathcal I$ containing one or two distinct primitive-shock indices, define
\begin{align}
    \widehat\beta_{kr}^{[-\mathcal I]}
    &=
    \widehat\beta_{kr}
    -
    \sum_{(i,u)\in\mathcal I}
    \frac{\IF(k=i,\ 1\leq r-u\leq H)}{H}
    K\!\left(\frac{r-u}{H}\right)
    S\widehat A(r)^{-1}X_{r-u}(r)\varepsilon_{iu}.
    \label{eq:jackknife-leaveout-beta}
\end{align}
Every partition, basis, Gram block, leverage, and coefficient carrying the
superscript $[-\mathcal I]$ is rebuilt from the entire trajectory in
\eqref{eq:jackknife-leaveout-beta}. Thus a removed shock is deleted from
all future first-step estimates whose rolling windows contain it.
Conditional on $\mathcal E_\infty$ and the remaining innovations, the rebuilt
objects do not depend on the removed shock or pair.

Let $j_t$ be the anchored block containing $\beta$. Apply the deterministic
quasi-uniform rectangle bounds used in
Lemma~\ref{lem:partition-basis-moments} to the leave-out trajectories before
substituting the empirical cells. The conditional convolution and
boundary-strip calculations in
Lemmas~\ref{lem:reference-quantile-spacing} and
\ref{lem:basis-perturbation} then give
\begin{align}
    &\sum_{t=H+1}^{T}\sum_{i=1}^{n_t}\omega_{it}
    \left\{
    \|\widehat p_{j_t t}(\widehat\beta_{it})\|^2
    +\|p_{j_t t}(\beta_{it})\|^2
    \right\}
    \lesssim_\P \frac{Tn}{J^d},
    \notag\\
    &\sum_{t=H+1}^{T}\sum_{i=1}^{n_t}\omega_{it}
    \|\widehat p_{j_t t}(\widehat\beta_{it})
    -p_{j_t t}(\beta_{it})\|^2
    \lesssim_\P Tn\mathsf R_{\mathtt P}.
    \label{eq:jackknife-weighted-basis}
\end{align}
To pass from a leave-out empirical partition to the original one, restore the
removed observation at every affected date. Each marginal empirical
quantile moves by at most one rank. {At each affected date, only a
fixed number of observations can cross a boundary of the selected block.
The conditional moment bounds used above show that their total weighted
contribution over all dates is $O_\P(T\log(nT))$. After division by $Tn$,
this is of order $\log(nT)/n$, which is already included in
$\mathsf R_{\mathtt Q}$ and $\mathsf R_{\mathtt P}$.}
For $\mathfrak q\geq1$, the nonswitch coordinate displacement is handled by
the order-$J$ derivative of the normalized polynomial basis. The weights
containing $a_{i,t-l,l}(\beta)^2$ are controlled by the factor fourth-moment
and NED time-average bounds. This proves
\eqref{eq:jackknife-weighted-basis} without replacing the boundary
probability by a fractional power.

For completeness, set
\begin{align*}
    \widetilde C_{it}(\beta)
    &=
    \frac{1}{n_t}p_t(\beta)^\top\widehat Q_t^{-1}
    \widehat p_t(\widehat\beta_{it}),
    &
    h_{it}
    &=
    \frac{1}{n_t}
    \widehat p_t(\widehat\beta_{it})^\top\widehat Q_t^{-1}
    \widehat p_t(\widehat\beta_{it}).
\end{align*}
On the selected block, the exact decomposition
\begin{align*}
    \widehat C_{it}(\beta)-C_{it}(\beta)
    ={}&
    \frac{\widetilde C_{it}(\beta)h_{it}}{1-h_{it}}
    \\
    &+
    \frac{1}{n_t}p_t(\beta)^\top
    (\widehat Q_t^{-1}-Q_t^{-1})
    \widehat p_t(\widehat\beta_{it})
    \\
    &+
    \frac{1}{n_t}p_t(\beta)^\top Q_t^{-1}
    \{\widehat p_t(\widehat\beta_{it})-p_t(\beta_{it})\}
\end{align*}
separates leverage, inverse-Gram, and basis perturbations. The leverage
bound, the resolvent identity, and
\eqref{eq:jackknife-weighted-basis} yield
\begin{align*}
    \sum_{t=H+1}^{T}\sum_{i=1}^{n_t}\omega_{it}
    \big(\widehat C_{it}(\beta)-C_{it}(\beta)\big)^2
    &\lesssim_\P
    T\frac{J^d}{n}
    \left\{
    \frac{J^d}{n}
    +J^{2d}\mathsf R_{\mathtt Q}^2
    +J^d\mathsf R_{\mathtt P}
    \right\}.
\end{align*}
This is~\eqref{eq:jackknife-weighted-hat-removal}.

{To bound the terms in $\mathscr H_C(Y,Y)$ that contain
$\varepsilon_{iu}\varepsilon_{iv}$ with $u\neq v$ in the proof of
Lemma~\ref{lem:bin-restricted-jackknife}, define}
{
\begin{align*}
    \Gamma_{iuv}(C)
    &=
    \sum_{r=H+1}^{T}\sum_{s=H+1}^{T}
    C_{ir}(\beta)C_{is}(\beta)
    \left\{
    \IF(u=r)-\frac{\IF(1\leq r-u\leq H)}{H}
    a_{iu,r-u}(\beta)
    \right\}
    \\
    &\qquad\qquad\times
    \left\{
    \IF(v=s)-\frac{\IF(1\leq s-v\leq H)}{H}
    a_{iv,s-v}(\beta)
    \right\}
    \IF(|r-s|\leq H).
\end{align*}
Then
\begin{align*}
    T_0^2\mathscr H_C(Y,Y)
    &=
    \sum_{i=1}^{n}\sum_{u=1}^{T}\sum_{v=1}^{T}
    \Gamma_{iuv}(C)\varepsilon_{iu}\varepsilon_{iv}.
\end{align*}
}
{
We now prove~\eqref{eq:jackknife-leading-replacement}. It bounds the difference
between $\mathscr H_{\widehat C}(Y,Y)$ and $\mathscr H_C(Y,Y)$. Write
$\ell_{it}=\nabla\mu_t(\beta_{it})^\top v_{it}$, so
$Y_{it}=\varepsilon_{it}-\ell_{it}$.

\medskip\noindent\textit{Squared-sum bounds for the rolling term.}
Use the exact leverage, inverse-Gram, and basis decomposition displayed
immediately before~\eqref{eq:jackknife-weighted-hat-removal}, now with unit
weights. It gives
\begin{align*}
    \sum_{t=H+1}^{T}\sum_{i=1}^{n_t}
    \big(\widehat C_{it}(\beta)-C_{it}(\beta)\big)^2
    &\lesssim_\P
    T\frac{J^d}{n}\Delta_{\mathtt{JK}},
    &
    \max_{i,t}\big|\widehat C_{it}(\beta)-C_{it}(\beta)\big|
    &\lesssim_\P\frac{J^d}{n}.
\end{align*}
It follows that
\begin{align*}
    \sum_{t=H+1}^{T}\sum_{i=1}^{n_t}
    \big(\widehat C_{it}(\beta)-C_{it}(\beta)\big)^4
    &\lesssim_\P
    T\frac{J^{3d}}{n^3}\Delta_{\mathtt{JK}}.
\end{align*}
We next control $\ell_{it}$ for observations that receive nonzero weight.
Anchoring makes the reference and estimated bases select the same cell index
at $\beta$. On the high-probability events in
Lemmas~\ref{lem:first-step-rate}, \ref{lem:quantile-convergence},
and~\ref{lem:partition-basis-moments}, the cell-endpoint error and the maximum
first-step error are both $o(J^{-1})$. Hence, if $C_{it}(\beta)$ or
$\widehat C_{it}(\beta)$ is nonzero, $\beta_{it}$ lies in the reference cell
containing $\beta$ or in one of a fixed number of neighboring reference cells.
The union of these cells is $\mathcal E_\infty$-measurable and contains
$O_\P(n/J^d)$ units at each date.

We can therefore sum over this union without conditioning on the random
selection induced by $\widehat C_{it}(\beta)$. The bounded gradient, the
conditional fourth moment of $\varepsilon_{it}$, and independence of the
primitive shocks give
\begin{align*}
    \E[\ell_{it}^4\mid\mathcal E_\infty]
    &\lesssim_\P
    \frac{1}{H^4}
    \left[
    \left\{\sum_{l=1}^{H}(1+\|f_{t-l}\|^2)\right\}^2
    +\sum_{l=1}^{H}(1+\|f_{t-l}\|^4)
    \right].
\end{align*}
To sum this bound over the union and over time, use
$(\sum_l x_l)^2\leq H\sum_l x_l^2$ and the time-average fourth-moment bound
for the factors. Conditional Markov's inequality then gives
\begin{align*}
    \sum_{t=H+1}^{T}\sum_{i=1}^{n_t}
    \ell_{it}^4
    \IF\!\left(
    |C_{it}(\beta)|+|\widehat C_{it}(\beta)|>0
    \right)
    &\lesssim_\P
    \frac{Tn}{J^dH^2}.
\end{align*}
The weights vanish outside the same union. Their pointwise bounds imply
\begin{align*}
    \sum_{t=H+1}^{T}\sum_{i=1}^{n_t}
    \left(C_{it}(\beta)^4+\widehat C_{it}(\beta)^4\right)
    &\lesssim_\P T\frac{J^{3d}}{n^3}.
\end{align*}
Scalar Cauchy--Schwarz now gives the first bound below. For the second, use
the earlier fourth-power bound for $\widehat C_{it}(\beta)-C_{it}(\beta)$ in
the same way:
\begin{align}
    \sum_{t=H+1}^{T}\sum_{i=1}^{n_t}
    \left(C_{it}(\beta)^2+\widehat C_{it}(\beta)^2\right)\ell_{it}^2
    &\lesssim_\P
    T\frac{J^d}{nH},
    \notag\\
    \sum_{t=H+1}^{T}\sum_{i=1}^{n_t}
    \big(\widehat C_{it}(\beta)-C_{it}(\beta)\big)^2\ell_{it}^2
    &\lesssim_\P
    T\frac{J^d}{nH}\Delta_{\mathtt{JK}}^{1/2}.
    \label{eq:jackknife-rolling-squares}
\end{align}

\medskip\noindent\textit{Current-shock windows.}
At date $t$, the weights $C_{it}(\beta)$ and
$\widehat C_{it}(\beta)$ depend only on the innovation-free environment and
innovations dated before $t$. Thus their products with
$\varepsilon_{it}$ are martingale differences when the dates are ordered.
The same is true after replacing either weight by their difference.
Successive conditioning, \eqref{eq:jackknife-weight-bounds}, and the
unweighted comparison above give
\begin{align}
    \sum_{i=1}^{n}\sum_{s=H+1}^{T}
    \left(
    \sum_{\substack{H+1\leq t\leq T\\|t-s|\leq H}}
    B_{it}(\beta)\varepsilon_{it}
    \right)^2
    &\lesssim_\P TH\frac{J^d}{n},
    \qquad B=C,\widehat C,
    \notag\\
    \sum_{i=1}^{n}\sum_{s=H+1}^{T}
    \left(
    \sum_{\substack{H+1\leq t\leq T\\|t-s|\leq H}}
    \big(\widehat C_{it}(\beta)-C_{it}(\beta)\big)
    \varepsilon_{it}
    \right)^2
    &\lesssim_\P
    TH\frac{J^d}{n}\Delta_{\mathtt{JK}}.
    \label{eq:jackknife-current-window-squares}
\end{align}
For unequal dates, condition at the later date; the cross term then has
conditional mean zero. Each diagonal term is counted only $O(H)$ times.

In the remaining displays, the common argument $\beta$ is suppressed.
We first compare the current-shock terms. For $B=C,\widehat C$, order the
off-diagonal sum by its later date. Its summed conditional variance is at
most the squared pointwise weight bound times the corresponding one-sided
sum of squares. The same successive-conditioning calculation as
\eqref{eq:jackknife-current-window-squares} bounds this sum of squares at the
same order. Hence
\begin{align*}
    \left|
    \mathscr H_B(\varepsilon,\varepsilon)
    -\frac{1}{T_0^2}\sum_{i,t}B_{it}^2\varepsilon_{it}^2
    \right|
    &\lesssim_\P
    \sqrt{\frac{HJ^{3d}}{n^3T^3}}.
\end{align*}
For the diagonal part, use
$\widehat C_{it}^2-C_{it}^2
=2C_{it}(\widehat C_{it}-C_{it})
+(\widehat C_{it}-C_{it})^2$.
Equation~\eqref{eq:jackknife-weighted-hat-removal} and scalar
Cauchy--Schwarz give
\begin{align}
    \left|
    \mathscr H_{\widehat C}(\varepsilon,\varepsilon)
    -\mathscr H_C(\varepsilon,\varepsilon)
    \right|
    &\lesssim_\P
    \frac{J^d}{nT}
    \left\{
    \Delta_{\mathtt{JK}}^{1/2}
    +\Delta_{\mathtt{JK}}
    +\sqrt{\frac{HJ^d}{nT}}
    \right\}.
    \label{eq:jackknife-current-comparison}
\end{align}

For the current--rolling terms, use
\begin{align*}
    \widehat C_{it}\widehat C_{is}-C_{it}C_{is}
    &=
    (\widehat C_{it}-C_{it})\widehat C_{is}
    +C_{it}(\widehat C_{is}-C_{is}).
\end{align*}
Aggregate each current-shock window before applying scalar
Cauchy--Schwarz. Equations~\eqref{eq:jackknife-rolling-squares}
and~\eqref{eq:jackknife-current-window-squares} give
\begin{align}
    &\left|
    \mathscr H_{\widehat C}(\varepsilon,\ell)
    -\mathscr H_C(\varepsilon,\ell)
    \right|
    \notag\\
    &\quad\leq
    \frac{1}{T_0^2}
    \left\{
    \sum_{i,s}\widehat C_{is}^2\ell_{is}^2
    \right\}^{1/2}
    \left\{
    \sum_{i,s}
    \left(
    \sum_{|t-s|\leq H}
    (\widehat C_{it}-C_{it})\varepsilon_{it}
    \right)^2
    \right\}^{1/2}
    \notag\\
    &\qquad+
    \frac{1}{T_0^2}
    \left\{
    \sum_{i,s}(\widehat C_{is}-C_{is})^2\ell_{is}^2
    \right\}^{1/2}
    \left\{
    \sum_{i,s}
    \left(
    \sum_{|t-s|\leq H}C_{it}\varepsilon_{it}
    \right)^2
    \right\}^{1/2}
    \notag\\
    &\quad\lesssim_\P
    \frac{J^d}{nT}
    \left(
    \Delta_{\mathtt{JK}}^{1/2}
    +\Delta_{\mathtt{JK}}^{1/4}
    \right).
    \label{eq:jackknife-current-rolling-comparison}
\end{align}

Finally, use the order-$H$ absolute row-sum bound rather than treating the
signed flat band as positive semidefinite. This gives
\begin{align}
    &\left|
    \mathscr H_{\widehat C}(\ell,\ell)
    -\mathscr H_C(\ell,\ell)
    \right|
    \notag\\
    &\quad\lesssim
    \frac{H}{T_0^2}
    \left\{
    \sum_{i,t}(\widehat C_{it}-C_{it})^2\ell_{it}^2
    \right\}^{1/2}
    \left[
    \left\{\sum_{i,t}\widehat C_{it}^2\ell_{it}^2\right\}^{1/2}
    +\left\{\sum_{i,t}C_{it}^2\ell_{it}^2\right\}^{1/2}
    \right]
    \lesssim_\P
    \frac{J^d}{nT}\Delta_{\mathtt{JK}}^{1/4}.
    \label{eq:jackknife-rolling-comparison}
\end{align}

Expand $Y=\varepsilon-\ell$ and combine
\eqref{eq:jackknife-current-comparison}--
\eqref{eq:jackknife-rolling-comparison}. Here
$J^d/n\leq\Delta_{\mathtt{JK}}$ and $H\leq T$. Moreover,
$\Delta_{\mathtt{JK}}=o(1)$. Hence all the displayed terms are bounded by
$\Delta_{\mathtt{JK}}^{1/4}$. This proves
\eqref{eq:jackknife-leading-replacement}.
}

\medskip\noindent{\textit{Step 3: current-shock and
inverse-matrix terms.}}
{
Here and in Step~4, the term numbers refer to the six terms in the second
equality of~\eqref{eq:jackknife-residual-expansion}. This step controls
$\widetilde{\mathscr R}^{(\mathrm{md})}$: term~(2), the part of term~(3)
obtained from the summation in~\eqref{eq:jackknife-first-step-residual}, and
the two parts of term~(4) containing $v_{it}$ and that same summation.
For this proof only, write $\widetilde{\mathscr R}^{(\mathrm{md})}
=\widetilde{\mathscr R}^{(\mathrm{md,cur})}
+\widetilde{\mathscr R}^{(\mathrm{md,lag})}$.
$\widetilde{\mathscr R}^{(\mathrm{md,cur})}$ contains term~(2) and the two
parts of term~(4). Each part of term~(4) contains the date-$t$ factor
innovation multiplying a coefficient fixed before that innovation.
$\widetilde{\mathscr R}^{(\mathrm{md,lag})}$ is the stated part of term~(3).

We first give the calculation for term~(2). Conditional on $\mathcal C_t$,
$\widehat C_{it}(\beta)$ and $\widehat\Pi_t$ are fixed, whereas the current
idiosyncratic innovations are independent and centered. The selected
least-squares projection has fixed dimension and does not increase the
squared norm. Together with~\eqref{eq:jackknife-weight-bounds}, this gives
\begin{align}
    \sum_{i=1}^{n_t}
    \E\!\left[
    \widehat C_{it}(\beta)^2
    \{(\widehat\Pi_t\varepsilon_{\cdot t})_i\}^2
    \,\middle|\,\mathcal C_t
    \right]
    &\lesssim_\P
    \left(\frac{J^d}{n}\right)^2.
    \label{eq:jackknife-projected-current-square}
\end{align}
The diagonal part of its banded quadratic form is therefore
$O_\P(J^{2d}/(n^2T))$. For unequal dates, order each pair by its later date.
The earlier projected error and both weights are then fixed, while the later
projected error has conditional mean zero. After division by $T_0^4$, the
sum of conditional variances is
$O_\P(\frac{H}{T^3}(\frac{J^d}{n})^4)$.
Thus the off-diagonal part is also
$O_\P(J^{2d}/(n^2T))$, since $H\leq T$.

For the two parts of term~(4), the factor innovation is common across units.
Order unequal dates as above and condition on the past at the later date.
The conditional mean of the later factor innovation is zero. The second
bound in~\eqref{eq:jackknife-cross-weight-bounds} controls the sum over units
of the weighted coefficient multiplying that innovation. The conditional
second moment of $v_{it}$ is of order $H^{-1}$. Therefore, the time-diagonal
part of the banded product for $-\widetilde f_t^\top v_{it}$ is bounded by
$J^d/(nTH)$. For unequal dates, first sum over the lags shared by the two
rolling windows and then take the conditional variance. The factor
fourth-moment bound yields
$O_\P((J^d/n)T^{-3/2})$ for the off-diagonal part. Relative to the common
factor $J^d/(nT)$, these are the $H^{-1}$ and $T^{-1/2}$ contributions in
$\mathsf R_{\mathtt{JKMD}}$. For the coefficient containing
$\widehat A(t)^{-1}-A(t)^{-1}$, use
$\widehat A(t)^{-1}-A(t)^{-1}
=\widehat A(t)^{-1}\{A(t)-\widehat A(t)\}A(t)^{-1}$ and
Lemma~\ref{lem:first-step-gram}. This gives the additional factor
$\{T^{2/q}/H+\sqrt{\log T/H}\}^2$. Combining these calculations with
\eqref{eq:jackknife-projected-current-square} gives
\begin{align}
    \sum_{i,s}
    \left(
    \sum_{|t-s|\leq H}\widehat C_{it}(\beta)
    \widetilde{\mathscr R}_{it}^{(\mathrm{md,cur})}
    \right)^2
    &\lesssim_\P
    TH\frac{J^d}{n}\mathsf R_{\mathtt{JKMD}},
    \notag\\
    \left|\mathscr H_{\widehat C}
    (\widetilde{\mathscr R}^{(\mathrm{md,cur})},
     \widetilde{\mathscr R}^{(\mathrm{md,cur})})\right|
    &\lesssim_\P
    \frac{J^d}{nT}\mathsf R_{\mathtt{JKMD}},
    \notag\\
    \left|\mathscr H_{\widehat C}
    (\varepsilon,\widetilde{\mathscr R}^{(\mathrm{md,cur})})\right|
    &\lesssim_\P
    \frac{J^d}{nT}\mathsf R_{\mathtt{JKMD}}^{1/2}.
    \label{eq:jackknife-current-shock-bounds}
\end{align}

Consider the part of term~(3) in
\eqref{eq:jackknife-residual-expansion} obtained by substituting the
summation in~\eqref{eq:jackknife-first-step-residual}. This is precisely
$\widetilde{\mathscr R}_{it}^{(\mathrm{md,lag})}$. On the enlarged-support
event used in Step~2, at most $O_\P(n/J^d)$ units enter at each date and
$\max_i\widehat C_{it}(\beta)^2\lesssim_\P(J^d/n)^2$. The inverse-matrix
identity and Lemma~\ref{lem:first-step-gram} bound the squared matrix
difference by $\{T^{2/q}/H+\sqrt{\log T/H}\}^2$. Conditional on
$\mathcal E_\infty$, the rolling sum has second moment of order $H^{-1}$.
Conditional Markov's inequality therefore yields
\begin{align}
    \sum_{t=H+1}^{T}\sum_{i=1}^{n_t}
    \widehat C_{it}(\beta)^2
    \{\widetilde{\mathscr R}_{it}^{(\mathrm{md,lag})}\}^2
    &\lesssim_\P
    \left(\frac{J^d}{n}\right)^2
    T\frac{n}{J^d}\frac{1}{H}
    \left(\frac{T^{2/q}}{H}
    +\sqrt{\frac{\log T}{H}}\right)^2
    \notag\\
    &=
    T\frac{J^d}{nH}
    \left(\frac{T^{2/q}}{H}
    +\sqrt{\frac{\log T}{H}}\right)^2
    \lesssim_\P
    T\frac{J^d}{nH}\mathsf R_{\mathtt{JKMD}}.
    \label{eq:jackknife-inverse-matrix-squares}
\end{align}
This calculation sums conditional second moments over dates and therefore
does not require different rolling windows to be independent.

Cauchy--Schwarz within each lag window and
\eqref{eq:jackknife-inverse-matrix-squares} give the same squared-window bound
as the first line of~\eqref{eq:jackknife-current-shock-bounds}. Hence
\begin{align}
    \sum_{i,s}
    \left(
    \sum_{|t-s|\leq H}\widehat C_{it}(\beta)
    \widetilde{\mathscr R}_{it}^{(\mathrm{md})}
    \right)^2
    &\lesssim_\P
    TH\frac{J^d}{n}\mathsf R_{\mathtt{JKMD}}.
    \label{eq:jackknife-first-block-window-squares}
\end{align}
The band has at most $2H+1$ entries in each row. Thus
\eqref{eq:jackknife-inverse-matrix-squares} bounds the lag--lag product.
Scalar Cauchy--Schwarz, using the current-part window bound and the lagged
squared sum, bounds the current--lag product. Together with
\eqref{eq:jackknife-current-shock-bounds}, this gives the second line
below. For $\varepsilon$ paired with the current array, use the third line of
\eqref{eq:jackknife-current-shock-bounds}. For $\varepsilon$ paired with
the lagged array, use~\eqref{eq:jackknife-current-window-squares} and
Cauchy--Schwarz. For $\ell$, use~\eqref{eq:jackknife-rolling-squares} with the
corresponding current or lagged squared-sum bound. Therefore
\begin{align}
    \left|\mathscr H_{\widehat C}
    (Y,\widetilde{\mathscr R}^{(\mathrm{md})})\right|
    &\lesssim_\P
    \frac{J^d}{nT}\mathsf R_{\mathtt{JKMD}}^{1/2},
    \notag\\
    \left|\mathscr H_{\widehat C}
    (\widetilde{\mathscr R}^{(\mathrm{md})},
     \widetilde{\mathscr R}^{(\mathrm{md})})\right|
    &\lesssim_\P
    \frac{J^d}{nT}\mathsf R_{\mathtt{JKMD}}.
    \label{eq:jackknife-first-block-products}
\end{align}
}

\medskip\noindent{\textit{Step 4: terms controlled by squared sums.}}
{
This step controls $\widetilde{\mathscr R}^{(\mathrm{pr})}$: term~(1), the
first-step-bias parts of terms~(3) and~(4), term~(5), and every projected
piece in term~(6).
We first bound these terms before forming their banded products. Polynomial
reproduction and local approximation bound term~(1) by
$J^{-2\min\{s_\mu,\mathfrak q+1\}}$ after squaring. The first-step-bias
parts of terms~(3) and~(4) contribute $\mathsf B_{\mathtt{FS}}^2$, and the
second-order Taylor remainder in term~(5) contributes
$\mathsf R_{\mathtt{FS}}^4$.

For the projected rolling terms in~(6), we bound the conditional mean induced
by sorting on the estimated betas. First consider
$z_{it}=\nabla\mu_t(\beta_{it})^\top
(\widehat\beta_{it}-\breve\beta_{it})$ for $i\in[n_t]$ and
$t=H+1,\ldots,T$.
For a deterministic admissible rectangle $P\in\mathcal P^\circ$, with
$\mathcal P^\circ$ as in the proof of
Lemma~\ref{lem:classical-linearization}, define the vector
\begin{align*}
    Z_t(P)&=\frac{J^d}{n_t}\sum_{i=1}^{n_t}
    p_{\mathfrak q}(\widehat\beta_{it};P)z_{it}.
\end{align*}
For the reference basis, conditioning first on the innovation-free
environment makes the rolling term centered. Applying
Lemma~\ref{lem:basis-perturbation}(a) then gives
\begin{align*}
    \max_{H+1\leq t\leq T}\sup_{P\in\mathcal P^\circ}
    \|\E[Z_t(P)\mid\mathcal D_{t-1}]\|
    &\lesssim_\P H^{-1}+\mathsf B_{\mathtt{FS}}.
\end{align*}
This mean need not have order $H^{-1}$ alone.

We next bound the centered part before substituting the empirical cell.
Conditional on $\mathcal D_{t-1}$, the unit histories and primitive
innovations entering $Z_t(P)$ are independent over $i$. The reference-basis
term has conditional second moment of order $J^{-d}/H$.
Since $J\mathsf R_{\mathtt{FS}}=o(1)$ implies $J/\sqrt H=o(1)$,
Lemma~\ref{lem:basis-perturbation}(b), together with
$\mathsf B_{\mathtt{FS}}=o(H^{-1/2})$, bounds the conditional second
moment of each generated-basis summand by $\log^d(nT)/(HJ^d)$.
The rectangle and normalized-polynomial class is VC-type with fixed
exponent. The conditional $L_2$ maximal bound, obtained by combining
the empirical-process and Hoffmann--J\o rgensen inequalities in
\citet[Lemmas~SA24 and~SA41]{cattaneo2024uniform}, gives
\begin{align*}
    \E\!\left[\sup_{P\in\mathcal P^\circ}
    \|Z_t(P)-\E[Z_t(P)\mid\mathcal D_{t-1}]\|^2
    \mid\mathcal D_{t-1}\right]
    \lesssim_\P
    \frac{J^d\log^{d+1}(nT)}{nH}
    +\frac{J^{2d}\log^2(nT)}{n^2}
    \E\!\left[\max_{i\in[n_t]}|z_{it}|^2
    \mid\mathcal D_{t-1}\right].
\end{align*}
Localize the $\mathcal D_{t-1}$-measurable constants in these conditional
bounds and the first-step Gram bound. The empirical-cell and second-step
inverse-Gram events are used only in the pathwise projection bound below.
Independent-innovation truncation gives
\begin{align*}
    \E\!\left[\max_{i\in[n_t]}|z_{it}|^2
    \mid\mathcal D_{t-1}\right]
    &\lesssim
    \frac{\log(nT)}{H}
    +\frac{n^{2/q}\log^2(nT)}{H^2}
    \left\{\sum_{l=1}^H(1+\|f_{t-l}\|)^q\right\}^{2/q}.
\end{align*}
To verify this bound, truncate each coordinate's primitive summands at
$H^{-1}\{n\sum_{l=1}^H(1+\|f_{t-l}\|)^q\}^{1/q}$ and center the
truncated and discarded parts separately. Bernstein's inequality bounds
the squared maximum of the truncated sums by the displayed right-hand side.
Independence over lags and the $q$th innovation moment bound the squared
$L_2$ norm of the maximum discarded sum by the squared truncation level.
The time average of the last factor is $O_\P(H^{2/q})$ by the factor
moment bound and concavity. Thus the time average of the envelope term is
bounded by
\begin{align*}
    \frac{J^{2d}}{n^2}
    \left\{\frac{\log^3(nT)}{H}
    +\frac{(nH)^{2/q}\log^4(nT)}{H^2}\right\}
    &\lesssim \frac{J^d\log^{d+1}(nT)}{nH}.
\end{align*}
After division by the right-hand side, the two ratios are bounded by
constants times $(J/\sqrt H)\log^{-d}(nT)$ and
$(J/\sqrt H)(H/T^2)^{2/q}\log^{1-d}(nT)$, respectively. This follows from the
common ratio restriction, using each of its two denominator branches.
Taking expectations after localization and applying Markov's inequality
to the nonnegative sum over dates now give
\begin{align*}
    \sum_{t=H+1}^T\sup_{P\in\mathcal P^\circ}
    \|Z_t(P)-\E[Z_t(P)\mid\mathcal D_{t-1}]\|^2
    &\lesssim_\P \frac{TJ^d\log^{d+1}(nT)}{nH}.
\end{align*}
This is a bound for a sum of squares; it does not require independence
between rolling windows.

The estimated selected cell belongs to $\mathcal P^\circ$ w.p.a.1.
Bounded basis coordinates, the selected inverse-Gram bound, and
\eqref{eq:jackknife-weight-bounds} imply, pathwise on the same events,
\begin{align*}
    \sum_{i=1}^{n_t}\widehat C_{it}(\beta)^2
    |(\widehat\Pi_t z_{\cdot t})_i|^2
    &\lesssim \frac{J^d}{n}
    \sup_{P\in\mathcal P^\circ}\|Z_t(P)\|^2.
\end{align*}
Separate $Z_t(P)$ into its conditional mean and centered part in this
inequality. Squaring the mean bound
retains $\mathsf B_{\mathtt{FS}}^2$ in addition to $H^{-2}$.

To replace $\widehat\beta_{it}-\breve\beta_{it}$ by $v_{it}$, use the
inverse-Gram remainder in~\eqref{eq:jackknife-first-step-residual}.
Projection contraction bounds the weighted squared norm by a constant
times $(J^d/n)^2$ times the sum of squared unprojected terms over the
innovation-free enlarged selected support used in Step~2. This support
contains $O_\P(n/J^d)$ units at each date, giving
\begin{align*}
    &\sum_{t=H+1}^{T}\sum_{i=1}^{n_t}\widehat C_{it}(\beta)^2
    \left|\left\{\widehat\Pi_t
    [\nabla\mu_t(\beta_{\cdot t})^\top
    (\widehat\beta_{\cdot t}-\breve\beta_{\cdot t}-v_{\cdot t})]
    \right\}_i\right|^2
    \lesssim_\P \frac{TJ^d}{nH^2}.
\end{align*}
For the last bound, conditional on $\mathcal E_\infty$ the rolling
innovation sum has covariance bounded by a constant times $H^{-1}$
on the uniform Gram event. The resolvent identity and the time-averaged
squared Gram bound at the end of the proof of
Lemma~\ref{lem:first-step-gram} give
\begin{align*}
    \frac{1}{T_0}\sum_{t=H+1}^{T}
    \|\widehat A(t)^{-1}-A(t)^{-1}\|^2
    &\lesssim_\P H^{-1}.
\end{align*}
Conditional Markov's inequality and the support count then establish the
displayed projected inverse-Gram bound.
Combining the mean, centered, and Gram-replacement bounds proves
\begin{align*}
    &\sum_{t=H+1}^{T}\sum_{i=1}^{n_t}
    \widehat C_{it}(\beta)^2
    \left|\left\{\widehat\Pi_t
    [\nabla\mu_t(\beta_{\cdot t})^\top v_{\cdot t}]
    \right\}_i\right|^2
    \lesssim_\P
    T\frac{J^d}{n}
    \left\{
    \frac{J^d\log^{d+1}(nT)}{nH}+\frac{1}{H^2}
    +\mathsf B_{\mathtt{FS}}^2
    \right\}.
\end{align*}
For each coordinate of $\widehat\beta_{it}-\breve\beta_{it}$, the same argument applies:
the basis-perturbation mean and variance calculations use a constant
coordinate multiplier in place of $\nabla\mu_t(\beta_{it})$.
The resulting projected arrays and weights are
$\mathcal F_{t-1}$-measurable. Conditioning at date $t$ and using the
conditional factor moment bound therefore gives the same displayed order
for $\{\widehat\Pi_t[\widetilde f_t^\top v_{\cdot t}]\}_i$.
The projected inverse-Gram terms, with or without the current factor,
are covered by the inverse-Gram bound above. Finally, bounded absolute row
sums of the selected projection give contributions
$TJ^d\mathsf B_{\mathtt{FS}}^2/n$ and $TJ^d\mathsf R_{\mathtt{FS}}^4/n$
for the projected first-step-bias and Taylor terms, respectively; use
the same conditional factor argument for the factor-times-bias term.
The square of their sum is at most a fixed constant times the sum of their
squares. Hence
\begin{align}
    \sum_{t=H+1}^{T}\sum_{i=1}^{n_t}
    \widehat C_{it}(\beta)^2
    \{\widetilde{\mathscr R}_{it}^{(\mathrm{pr})}\}^2
    &\lesssim_\P
    T\frac{J^d}{n}
    \left\{
    J^{-2\min\{s_\mu,\mathfrak q+1\}}
    +\mathsf B_{\mathtt{FS}}^2
    +\mathsf R_{\mathtt{FS}}^4
    +\frac{J^d\log^{d+1}(nT)}{nH}
    +\frac{1}{H^2}
    \right\}
    \notag\\
    &=T\frac{J^d}{nH}\mathsf R_{\mathtt{JKPR}}.
    \label{eq:jackknife-second-block-squares}
\end{align}
Since $2|xy|\leq x^2+y^2$ and each date has at most $2H+1$ partners in the
band, this gives
\begin{align}
    \left|\mathscr H_{\widehat C}
    (\widetilde{\mathscr R}^{(\mathrm{pr})},
     \widetilde{\mathscr R}^{(\mathrm{pr})})\right|
    &\lesssim_\P
    \frac{J^d}{nT}\mathsf R_{\mathtt{JKPR}}.
    \label{eq:jackknife-second-block-products}
\end{align}

Move the factor indexed by $s$ outside the inner sum and apply scalar
Cauchy--Schwarz.

Equation~\eqref{eq:jackknife-second-block-squares} controls
the first factor below. Equation~\eqref{eq:jackknife-first-block-window-squares}
controls the second:
\begin{align}
    &\left|\mathscr H_{\widehat C}
    (\widetilde{\mathscr R}^{(\mathrm{md})},
     \widetilde{\mathscr R}^{(\mathrm{pr})})\right|
    \notag\\
    &\quad\leq
    \frac{1}{T_0^2}
    \left\{\sum_{i,s}\widehat C_{is}(\beta)^2
    \{\widetilde{\mathscr R}_{is}^{(\mathrm{pr})}\}^2\right\}^{1/2}
    \left\{\sum_{i,s}
    \left(\sum_{|t-s|\leq H}\widehat C_{it}(\beta)
    \widetilde{\mathscr R}_{it}^{(\mathrm{md})}\right)^2
    \right\}^{1/2}
    \lesssim_\P
    \frac{J^d}{nT}
    \sqrt{\mathsf R_{\mathtt{JKMD}}
    \mathsf R_{\mathtt{JKPR}}}.
    \label{eq:jackknife-mixed-direct}
\end{align}
Equation~\eqref{eq:jackknife-current-window-squares} controls the lag windows
of $\varepsilon$. Cauchy--Schwarz within each window and
\eqref{eq:jackknife-rolling-squares} give the same bound for $\ell$. Thus
\begin{align*}
    \sum_{i,s}
    \left(\sum_{|t-s|\leq H}\widehat C_{it}(\beta)Y_{it}\right)^2
    &\lesssim_\P TH\frac{J^d}{n}.
\end{align*}
Applying the same scalar Cauchy--Schwarz bound with $Y$ in place of
$\widetilde{\mathscr R}^{(\mathrm{md})}$ first gives the right-hand side
below without the factor $\sqrt{\log(nT)}$. We retain the weaker displayed
bound to match the sufficient condition in~\eqref{eq:jackknife-conditions}:
\begin{align}
    \left|\mathscr H_{\widehat C}
    (Y,\widetilde{\mathscr R}^{(\mathrm{pr})})\right|
    &\lesssim_\P
    \frac{J^d}{nT}
    \sqrt{\mathsf R_{\mathtt{JKPR}}\log(nT)}.
    \label{eq:jackknife-leading-second-block-products}
\end{align}
Equations~\eqref{eq:jackknife-first-block-products},
\eqref{eq:jackknife-second-block-products},
\eqref{eq:jackknife-mixed-direct}, and
\eqref{eq:jackknife-leading-second-block-products} prove
\eqref{eq:jackknife-array-bounds}.
}
\end{proof}

\begin{lemma}[Selected-Block Residual Bound]
\label{lem:jackknife-selected-block}
    Suppose that the assumptions and common rate conditions stated before
    parts~\textnormal{(i)}--\textnormal{(ii)} of
    Lemma~\ref{lem:bin-restricted-jackknife} hold. For every fixed
    $\beta\in\mathcal B^\circ$, without imposing
    \eqref{eq:jackknife-anchored-basis},
    \begin{align*}
        \left|\mathscr H_{\widehat C}
        (\widehat\varepsilon,\widehat\varepsilon)\right|
        &\lesssim_\P
        \frac{J^d}{nT}.
    \end{align*}
\end{lemma}

\begin{proof}
On the rows used by $\widehat C_{it}(\beta)$, the exact identity in Step~1 of
the preceding proof can be written as
\begin{align*}
    \widehat\varepsilon_{it}
    &=e_{it}+r_{it},
    \qquad
    e_{it}
    =\left[(I-\widehat\Pi_t)\varepsilon_{\cdot t}\right]_i,
    \qquad
    r_{it}
    =\left[(I-\widehat\Pi_t)M_t(\beta_{\cdot t})\right]_i,
\end{align*}
and all three quantities are set to zero on unused rows. Conditional on $\mathcal C_t$, the
selected empirical block, $\widehat C_{it}(\beta)$, and
$\widehat\Pi_t$ are fixed, whereas the current innovations entering $e_{it}$
are cross-sectionally independent and centered.

{We first record two weighted squared-sum bounds.} Projection contraction, the
uniform cell-count and inverse-Gram bounds, and conditional second moments give
\begin{align}
    \sum_{t=H+1}^{T}\sum_{i=1}^{n_t}
    \widehat C_{it}(\beta)^2e_{it}^2
    &\lesssim_\P
    T\frac{J^d}{n}.
    \label{eq:jackknife-selected-current-squares}
\end{align}
For $r_{it}$, polynomial reproduction removes the factor part of
$M_t(\widehat\beta_{it})$ when $\mathfrak q\geq1$, while local polynomial
approximation controls its conditional-mean part. When $\mathfrak q=0$, the
entire within-cell variation is of order $J^{-1}$. {Apply the
first-step decomposition \eqref{eq:jackknife-first-step-residual} inside
$r_{it}$. The local polynomial residual contributes
$J^{-2\min\{s_\mu,\mathfrak q+1\}}$ after squaring. The linear first-step
term has selected-cell mean square of order $H^{-1}$, its bias contributes
$\mathsf B_{\mathtt{FS}}^2$, and the Taylor remainder contributes
$\mathsf R_{\mathtt{FS}}^4$.
For the projected rolling terms, the coarser $H^{-1}$ order suffices here.
On the selected block at each date $t=H+1,\ldots,T$, projection contraction and
$\max_{i\in[n_t]}\widehat C_{it}(\beta)^2\lesssim_\P(J^d/n)^2$ bound the
weighted squared norm of a projection by a constant times $(J^d/n)^2$
times the unprojected block's squared norm. The selected-cell rolling
moment bound just used therefore also bounds the projected rolling
contribution by $O_\P(TJ^d/(nH))$. Bounded absolute projection row sums
handle the bias and Taylor pieces, and conditioning at date $t$ handles
current factor multipliers. This argument does not use the sharp anchored
projected bound or impose $J\mathsf R_{\mathtt{FS}}=o(1)$.
These bounds hold uniformly over the
deterministic quasi-uniform rectangle class, so they may be evaluated at the
realized selected block. Hence}
\begin{align}
    \sum_{t=H+1}^{T}\sum_{i=1}^{n_t}
    \widehat C_{it}(\beta)^2r_{it}^2
    &\lesssim_\P
    T\frac{J^d}{n}
    \left\{
    \frac{1}{H}
    +J^{-2\min\{s_\mu,\mathfrak q+1\}}
    +\mathsf R_{\mathtt{FS}}^4
    +\mathsf B_{\mathtt{FS}}^2
    +\frac{J^d\log^{d+1}(nT)}{nH}
    +\frac{1}{H^2}
    \right\}
    \notag\\
    &\lesssim_\P
    T\frac{J^d}{nH}.
    \label{eq:jackknife-selected-signal-squares}
\end{align}
The first term in braces covers both the unprojected and projected
rolling first-step terms. The remaining nonnegative terms are bounded by
$\mathsf R_{\mathtt{JKPR}}/H$, so the last inequality follows from
\eqref{eq:jackknife-conditions}. This calculation does not compare the cell
containing $\beta$ with a reference cell.
The underlying weighted moment bounds hold
uniformly over the deterministic quasi-uniform rectangle class before the
realized empirical block is selected.

The current-innovation array
$\{\widehat C_{it}(\beta)e_{it}\}$ is a chronological martingale-difference
array. Martingale orthogonality and
\eqref{eq:jackknife-selected-current-squares} therefore give
\begin{align}
    \sum_{s=H+1}^{T}\sum_{i=1}^{n}
    \left(
    \sum_{\substack{H+1\leq t\leq T\\|t-s|\leq H}}
    \widehat C_{it}(\beta)e_{it}
    \right)^2
    &\lesssim_\P
    TH\frac{J^d}{n}.
    \label{eq:jackknife-selected-window-squares}
\end{align}
In the displays below, $\sum_{s,i}$ uses these same ranges, and invalid
unit--date pairs continue to contribute zero.
This aggregation is important: bounding the individual time pairs first would
lose a factor $H$. By scalar Cauchy--Schwarz after the lag aggregation,
\eqref{eq:jackknife-selected-signal-squares} and
\eqref{eq:jackknife-selected-window-squares} imply
\begin{align*}
    \left|\mathscr H_{\widehat C}(e,r)\right|
    &\leq
    \frac{1}{T_0^2}
    \left\{
    \sum_{s,i}\widehat C_{is}(\beta)^2r_{is}^2
    \right\}^{1/2}
    \left\{
    \sum_{s,i}
    \left(
    \sum_{|t-s|\leq H}
    \widehat C_{it}(\beta)e_{it}
    \right)^2
    \right\}^{1/2}
    \lesssim_\P
    \frac{J^d}{nT}.
\end{align*}
The absolute band bound and
\eqref{eq:jackknife-selected-signal-squares} likewise give
\begin{align*}
    \left|\mathscr H_{\widehat C}(r,r)\right|
    &\lesssim
    \frac{H}{T_0^2}
    \sum_{t,i}\widehat C_{it}(\beta)^2r_{it}^2
    \lesssim_\P
    \frac{J^d}{nT}.
\end{align*}

Finally, split $\mathscr H_{\widehat C}(e,e)$ into its time-diagonal and
off-diagonal parts. {Conditional second moments and
\eqref{eq:jackknife-selected-current-squares} show that the diagonal is
$O_\P(J^d/(nT))$. For the off-diagonal part, martingale orthogonality, the
leverage bound, and~\eqref{eq:jackknife-cross-weight-bounds} give}
\begin{align*}
    \left|\textnormal{off-diagonal part of }
    \mathscr H_{\widehat C}(e,e)\right|
    &=O_\P\!\left(
    \sqrt{\frac{HJ^{3d}}{n^3T^3}}
    \right)
    =o_\P\!\left(\frac{J^d}{nT}\right),
\end{align*}
where the last equality uses $H\leq T$ and $J^d/n=o(1)$. Expanding
$\mathscr H_{\widehat C}(e+r,e+r)$ and combining the preceding bounds proves
the result. The proof controls only magnitude and therefore does not require
the selected block to be innovation-free.
\end{proof}
}

\subsection{Proof of Lemma~\ref{lem:bin-restricted-jackknife}}

{
\medskip\noindent\textit{Step 1: common reduction.}
By the Sherman--Morrison identity above and the convention for invalid
unit--date pairs, $\widehat\Sigma_{\varepsilon}(\beta)
=\mathscr H_{\widehat C}(\widehat\varepsilon,\widehat\varepsilon)$.
Expanding \eqref{eq:jackknife-residual-expansion} in the symmetric bilinear
form gives
\begin{align*}
    \widehat\Sigma_{\varepsilon}(\beta)
    ={}&
    \mathscr H_{\widehat C}(Y,Y)
    +2\mathscr H_{\widehat C}
    (Y,\widetilde{\mathscr R}^{(\mathrm{md})})
    +2\mathscr H_{\widehat C}
    (Y,\widetilde{\mathscr R}^{(\mathrm{pr})})
    \\
    &+
    \mathscr H_{\widehat C}
    (\widetilde{\mathscr R}^{(\mathrm{md})},
     \widetilde{\mathscr R}^{(\mathrm{md})})
    +2\mathscr H_{\widehat C}
    (\widetilde{\mathscr R}^{(\mathrm{md})},
     \widetilde{\mathscr R}^{(\mathrm{pr})})
    +\mathscr H_{\widehat C}
    (\widetilde{\mathscr R}^{(\mathrm{pr})},
     \widetilde{\mathscr R}^{(\mathrm{pr})}).
\end{align*}

\medskip\noindent\textit{Step 2: rough bound for every fixed evaluation point.}
For any fixed $\beta\in\mathcal B^\circ$, do not impose either
$J\mathsf R_{\mathtt{FS}}=o(1)$ or
\eqref{eq:jackknife-anchored-basis}. Apply
Lemma~\ref{lem:jackknife-selected-block} directly to the common reduction in
Step~1. It gives
\begin{align*}
    \left|\widehat\Sigma_{\varepsilon}(\beta)\right|
    &\lesssim_\P
    \frac{J^d}{nT}.
\end{align*}
{The proof of
Lemma~\ref{lem:leading-gaussian-components} establishes
$\sum_{t,i}\varsigma_{it}^2W_{it}(\beta)^2
\lesssim_\P TJ^d/n$. By the definition of
$\Sigma_{\varepsilon}(\beta)$ following
\eqref{eq:pointwise-leading-sum}, this gives
$\Sigma_{\varepsilon}(\beta)\lesssim_\P J^d/(nT)$.}
The triangle inequality therefore yields
\begin{align*}
    \widehat\Sigma_{\varepsilon}(\beta)
    -\Sigma_{\varepsilon}(\beta)
    &=O_\P\!\left(\frac{J^d}{nT}\right).
\end{align*}
Finally, $J^d/n=o(1)$ under \eqref{eq:jackknife-conditions}, so the last
display is $o_\P(T^{-1})$. This argument controls only the magnitude of the
realized selected block and therefore requires neither anchoring nor the
stronger localization $J\mathsf R_{\mathtt{FS}}=o(1)$.

\medskip\noindent\textit{Step 3: anchored evaluation points.}
Suppose \eqref{eq:jackknife-anchored-basis} holds and
$J\mathsf R_{\mathtt{FS}}=o(1)$. Re-indexing the primitive shocks in
$Y_{it}$ gives the reference weight
\begin{align}
    W_{it}(\beta)
    &=
    C_{it}(\beta)
    -\frac{1}{H}\sum_{l=1}^{H}
    C_{i,t+l}(\beta)a_{itl}(\beta).
    \label{eq:jackknife-reference-weight}
\end{align}
This is exactly the weight $W_{it}(\beta)$ defined in
Section~\ref{sec:pointwise-distribution}, and it is
$\mathcal E_\infty$-measurable.
Consequently, Lemma~\ref{lem:jackknife-residual-arrays} yields
\begin{align}
    \widehat\Sigma_{\varepsilon}(\beta)
    &=
    \mathscr H_C(Y,Y)
    +\mathscr R_{\mathtt{JK}}^\star(\beta),
    \notag\\
    \left|\mathscr R_{\mathtt{JK}}^\star(\beta)\right|
    &\lesssim_\P
    \frac{J^d}{nT}\rho_{\mathtt{JK}},
    \label{eq:jackknife-remainder-bound}\\
    {\rho_{\mathtt{JK}}}
    &{=
    \Delta_{\mathtt{JK}}^{1/4}}
    +\mathsf R_{\mathtt{JKMD}}^{1/2}
    +\mathsf R_{\mathtt{JKMD}}
    +\mathsf R_{\mathtt{JKPR}}
    +\sqrt{\mathsf R_{\mathtt{JKPR}}\log(nT)}.
    \notag
\end{align}
{The mixed term in
\eqref{eq:jackknife-array-bounds} is absorbed here by
$\sqrt{ab}\leq(a+b)/2$.}
Lemma~\ref{lem:partition-basis-moments},
$J^2\log T/n=o(1)$, $J\mathsf R_{\mathtt{FS}}=o(1)$, and
$T^{2/q}/H+\sqrt{\log T/H}=o(1)$ imply
$\Delta_{\mathtt{JK}}=o(1)$ and
$\mathsf R_{\mathtt{JKMD}}=o(1)$, while
\eqref{eq:jackknife-conditions} implies
$\mathsf R_{\mathtt{JKPR}}\log(nT)=o(1)$. Hence
$\rho_{\mathtt{JK}}=o(1)$.

It remains to identify the leading part of $\mathscr H_C(Y,Y)$.
After re-indexing its primitive shocks, write
\begin{align*}
    \mathscr H_C(Y,Y)
    &=
    \mathscr D_{\mathtt{JK}}^\circ(\beta)
    +\mathscr O_{\mathtt{JK}}^\circ(\beta),
    \\
    \mathscr D_{\mathtt{JK}}^\circ(\beta)
    &=
    \frac{1}{(T-H)^2}
    \sum_{i=1}^{n}\sum_{t=1}^{T}
    W_{it}(\beta)^2\varepsilon_{it}^2,
    \\
    \mathscr O_{\mathtt{JK}}^\circ(\beta)
    &:=
    \mathscr H_C(Y,Y)-\mathscr D_{\mathtt{JK}}^\circ(\beta).
\end{align*}
The lag restriction is automatically satisfied between any two appearances
of the same primitive shock, so its coefficient is precisely
$W_{it}(\beta)^2$. Every term in
$\mathscr O_{\mathtt{JK}}^\circ(\beta)$ is instead a product of two distinct
primitive shocks.

Conditional on $\mathcal E_\infty$, these shocks are independent and
centered, and therefore
\begin{align*}
    \E[
    \mathscr D_{\mathtt{JK}}^\circ(\beta)
    \mid\mathcal E_\infty]
    &=
    \Sigma_{\varepsilon}(\beta),
    &
    \E[
    \mathscr O_{\mathtt{JK}}^\circ(\beta)
    \mid\mathcal E_\infty]
    &=0.
\end{align*}
The local-support bounds in
\eqref{eq:jackknife-weight-bounds}, Jensen's inequality over the rolling
window, and $T^{-1}\sum_t(1+\|f_t\|^4)=O_\P(1)$ give
\begin{align*}
    \sum_{i=1}^{n}\sum_{t=1}^{T}W_{it}(\beta)^2
    &\lesssim_\P T\frac{J^d}{n},
    &
    \sum_{i=1}^{n}\sum_{t=1}^{T}W_{it}(\beta)^4
    &\lesssim_\P T\frac{J^{3d}}{n^3}.
\end{align*}
The conditional fourth-moment bound thus gives
\begin{align}
    \mathscr D_{\mathtt{JK}}^\circ(\beta)
    -\Sigma_{\varepsilon}(\beta)
    &=
    O_\P\!\left(
    \sqrt{\frac{J^{3d}}{n^3T^3}}
    \right).
    \label{eq:jackknife-diagonal-rate}
\end{align}
For the off-diagonal term, \eqref{eq:jackknife-cross-weight-bounds},
Jensen's inequality, and
$T^{-1}\sum_t(1+\|f_t\|^4)=O_\P(1)$ {give the following bound on
the sum of the squared coefficients:}
\begin{align*}
    \sum_{i=1}^{n}\sum_{u\neq v}\Gamma_{iuv}(C)^2
    &\lesssim_\P
    TH\frac{J^{3d}}{n^3}.
\end{align*}
Conditional orthogonality of distinct-shock products therefore gives
\begin{align}
    \mathscr O_{\mathtt{JK}}^\circ(\beta)
    &=
    O_\P\!\left(
    \sqrt{\frac{HJ^{3d}}{n^3T^3}}
    \right).
    \label{eq:jackknife-offdiagonal-rate}
\end{align}
Combining the preceding displays preserves the explicit leading expansion
\begin{align}
    \widehat\Sigma_{\varepsilon}(\beta)
    -\Sigma_{\varepsilon}(\beta)
    ={}&
    \frac{1}{(T-H)^2}
    \sum_{i=1}^{n}\sum_{t=1}^{T}
    W_{it}(\beta)^2
    \bigl(\varepsilon_{it}^2-\varsigma_{it}^2\bigr)
    \notag\\
    &+
    \mathscr O_{\mathtt{JK}}^\circ(\beta)
    +\mathscr R_{\mathtt{JK}}^\star(\beta).
    \label{eq:jackknife-leading-expansion}
\end{align}
Since $H\leq T$ and $J^d/n=o(1)$, the two leading fluctuations in
\eqref{eq:jackknife-diagonal-rate}--\eqref{eq:jackknife-offdiagonal-rate}
are $o_\P(J^d/(nT))$; the remainder has the same conclusion by
\eqref{eq:jackknife-remainder-bound}. Thus, for every fixed anchored
evaluation point---both $\beta=0$ and an anchored $\beta\neq0$---
\begin{align*}
    \widehat\Sigma_{\varepsilon}(\beta)
    &=
    \Sigma_{\varepsilon}(\beta)
    +o_\P\!\left(\frac{J^d}{nT}\right).
\end{align*}

\qed
}

\subsection{Proof of Theorem~\ref{thm:variance-estimation}}

{For $\beta\neq0$, part~(i) of
Lemma~\ref{lem:bin-restricted-jackknife} and
$\Sigma_\varepsilon(\beta)\lesssim_\P J^d/(nT)$ give
\begin{align*}
    \widehat{\Sigma}_{\varepsilon}(\beta)-\Sigma_{\varepsilon}(\beta)
    &=
    O_\P\Big(\frac{J^d}{nT}\Big)
    =o_\P(T^{-1}).
\end{align*}
Lemma~\ref{lem:factor-variance} then gives
\begin{align*}
    \widehat{\Sigma}(\beta)-\Sigma(\beta)-\mathscr{B}_{\mathtt{F}}(\beta)
    &=o_\P(T^{-1}).
\end{align*}
For $\beta=0$, $\Sigma_f(0)=\widehat{\Sigma}_f(0)=\mathscr{B}_{\mathtt{F}}(0)=0$. The additive claim follows directly from part~(ii) of Lemma~\ref{lem:bin-restricted-jackknife}. If $(nT/J^d)\Sigma_{\varepsilon}(0)\gtrsim_\P1$, then $\Sigma(0)=\Sigma_{\varepsilon}(0)\gtrsim_\P J^d/(nT)$, so the additive error is also $o_\P(\Sigma(0))$.}
\qed

\subsection{Proof of Theorem~\ref{thm:fama-macbeth-variance}}

Fix an evaluation point $\beta$ and
let $u_t$ be the periodwise estimation error entering the Fama--MacBeth
dispersion, with $\bar u=(T-H)^{-1}\sum_{t=H+1}^{T}u_t$.
For any variance target $V_u$, adding and subtracting $\mu_t(\beta)$ and
$\bar{\mu}_T(\beta;H)$ gives
\begin{align*}
    \left|
    \frac{1}{(T-H)^2}\sum_{t=H+1}^{T}
    \Big[
    u_t-\bar u+\mu_t(\beta)-\bar{\mu}_T(\beta;H)
    \Big]^2
    -V_u-\Sigma_\mu(\beta)
    \right|
    \leq
    \mathscr{V}_{\mathtt{FM},1}
    +\mathscr{V}_{\mathtt{FM},2}
    +\mathscr{V}_{\mathtt{FM},3},
\end{align*}
where
\begin{align*}
    \mathscr{V}_{\mathtt{FM},1}
    &=
    \left|
    \frac{1}{(T-H)^2}\sum_{t=H+1}^{T}u_t^2
    -V_u
    \right|,
    \\
    \mathscr{V}_{\mathtt{FM},2}
    &=
    \frac{\bar u^2}{T-H},
    \\
    \mathscr{V}_{\mathtt{FM},3}
    &=
    \left|
    \frac{2}{(T-H)^2}\sum_{t=H+1}^{T}
    u_t\big(\mu_t(\beta)-\bar{\mu}_T(\beta;H)\big)
    \right|.
\end{align*}

\medskip\noindent\textit{Step 1: The case $\beta\neq0$.}
Take
$u_t=\widehat{M}_t(\beta)-\mu_t(\beta)$ and
$V_u=\Sigma_f(\beta)+\Sigma_{\varepsilon}(\beta)$. The left-hand side in the
common display is
$|\widehat{\Sigma}_{\mathtt{FM}}(\beta)-\Sigma_f(\beta)
-\Sigma_{\varepsilon}(\beta)-\Sigma_\mu(\beta)|$.

For $\mathscr{V}_{\mathtt{FM},1}$, the per-period linearization in
Lemma~\ref{lem:classical-linearization} gives
\begin{align*}
    \widehat{M}_t(\beta)-\mu_t(\beta)
    &=
    \beta^\top\widetilde{f}_t
    +\mathscr{I}_{\mathtt{FM},t}(\beta)
    +\mathscr{R}_{\mathtt{FM},t}(\beta),
\end{align*}
where
$\mathscr{I}_{\mathtt{FM},t}(\beta)$ collects the idiosyncratic leading terms and $\mathscr{R}_{\mathtt{FM},t}(\beta)$ collects the remaining linearization errors. The same weight calculations as in the proof of Lemma~\ref{lem:bin-restricted-jackknife} give
\begin{align*}
    \frac{1}{(T-H)^2}\sum_{t=H+1}^T \mathscr{I}_{\mathtt{FM},t}(\beta)^2
    &\lesssim_\P
    \frac{J^d}{nT}
    =
    o_\P(T^{-1}),
\end{align*}
and the periodwise residual bounds apply. Since
\begin{align*}
    J^{-\min\{s_\mu,\mathfrak{q}+1\}}
    +\mathsf{R}_{\mathtt{FM}}
    &=o(1),
\end{align*}
they give
\begin{align*}
    \frac{1}{(T-H)^2}\sum_{t=H+1}^T
    \mathscr{R}_{\mathtt{FM},t}(\beta)^2
    &=o_\P(T^{-1}).
\end{align*}
Hence all terms involving $\mathscr{I}_{\mathtt{FM},t}(\beta)$ or $\mathscr{R}_{\mathtt{FM},t}(\beta)$ are $o_\P(T^{-1})$ by Cauchy--Schwarz. For the factor square,
\begin{align*}
    &\frac{1}{(T-H)^2}\sum_{t=H+1}^T
    \left[
    \big(\beta^\top\widetilde{f}_t\big)^2
    -\Var(\beta^\top f_t\mid\mathcal{G}_{t-1})
    \right]
    =
    o_\P(T^{-1}),
\end{align*}
because the summands are martingale differences with conditional second moments bounded by Assumption~\ref{ass:dgp}(iv). Since $\Sigma_{\varepsilon}(\beta)\lesssim_\P J^d/(nT)=o(T^{-1})$, we obtain
\begin{align*}
    \mathscr{V}_{\mathtt{FM},1}
    =
    o_\P\!\Big(\frac{1}{T}\Big),
\end{align*}
for $\beta\neq 0$.

Next, the aggregate expansion gives
$\widehat{\mu}(\beta)-\bar{\mu}_T(\beta;H)=o_\P(1)$:
the factor term is $O_\P(T^{-1/2})$, the idiosyncratic term is
$O_\P(\sqrt{J^d/(nT)})=o_\P(T^{-1/2})$, and the residual terms are
$o_\P(1)$ under the displayed rate condition. Therefore
\begin{align*}
    \mathscr{V}_{\mathtt{FM},2}
    =
    o_\P\!\Big(\frac{1}{T}\Big).
\end{align*}

Finally, for $\mathscr{V}_{\mathtt{FM},3}$, write
\begin{align*}
    \mathscr{V}_{\mathtt{FM},3}
    &=
    \Big|
    \frac{2}{(T-H)^2}\sum_{t=H+1}^{T}
    \big(\widehat{M}_t(\beta)-\mu_t(\beta)\big)\,\mu_t(\beta)
    \;-\;
    \frac{2\,\bar{\mu}_T(\beta;H)}{T-H}
    \big(\widehat{\mu}(\beta)-\bar{\mu}_T(\beta;H)\big)
    \Big|.
\end{align*}
For the first term, by the linearization in
Lemma~\ref{lem:classical-linearization}, the leading term
in $\widehat{M}_t(\beta)-\mu_t(\beta)$ is a martingale
difference $\beta^\top\widetilde{f}_t$, and thus the corresponding component in the first term of $\mathscr{V}_{\mathtt{FM},3}$ is
\begin{align*}
    \frac{2}{(T-H)^2}\sum_{t=H+1}^T
    \beta^\top\widetilde{f}_t\mu_t(\beta).
\end{align*}
Since $\mu_t(\beta)$ is uniformly bounded,
the predictable quadratic variation of this term is bounded by
\begin{align*}
    \frac{\|\beta\|^2}{(T-H)^4}\sum_{t=H+1}^T
    \mu_t(\beta)^2\E[\|\widetilde{f}_t\|^2|\mathcal{G}_{t-1}]
    \;\lesssim\;
    \frac{1}{T^3},
\end{align*}
so that by the Lenglart--Rebolledo inequality,
\begin{align*}
    \frac{2}{(T-H)^2}\sum_{t=H+1}^T\beta^\top\widetilde{f}_t\mu_t(\beta)
    &\lesssim_\P T^{-3/2}.
\end{align*}
The idiosyncratic leading term and the remaining linearization errors contribute $o_\P(T^{-1})$ by the same predictable-variation and Cauchy--Schwarz bounds used for $\mathscr{V}_{\mathtt{FM},1}$, plus the fact that the deterministic periodwise residual scale is $o(1)$.

For the second term in $\mathscr{V}_{\mathtt{FM},3}(\beta)$, since $\bar{\mu}_T(\beta;H)\lesssim_\P1$ and
$\widehat{\mu}(\beta)-\bar{\mu}_T(\beta;H)=o_\P(1)$ by the preceding aggregate expansion,
\begin{align*}
    \frac{\bar{\mu}_T(\beta;H)}{T-H}
    \big(\widehat{\mu}(\beta)-\bar{\mu}_T(\beta;H)\big)
    &=o_\P(T^{-1}).
\end{align*}
Thus,
\begin{align*}
    \mathscr{V}_{\mathtt{FM},3}
    &=o_\P\!\Big(\frac{1}{T}\Big).
\end{align*}
Combining the bounds for $\mathscr{V}_{\mathtt{FM},1}$, $\mathscr{V}_{\mathtt{FM},2}$, and
$\mathscr{V}_{\mathtt{FM},3}$, and using $\Sigma_{\varepsilon}(\beta)\lesssim_\P J^d/(nT)=o(T^{-1})$, yields the desired result for $\beta\neq 0$.

\medskip\noindent\textit{Step 2: The case $\beta=0$.}
Consider the bias-corrected estimator
$\widetilde{\Sigma}_{\mathtt{FM}}(0)$.
The periodwise version of Lemmas~\ref{lem:general-linearization} and~\ref{lem:classical-linearization}, with the smoothing-bias component $\mathscr{B}_t(0)$ removed exactly, gives
\begin{align}
    \widehat{M}_t(0)-\mathscr{B}_t(0)-\mu_t(0)
    &=
    \mathscr{E}_{\mathtt{o},t}(0)+\mathscr{R}_{\mathtt{FM},t},
    \label{eq:fm-origin-periodwise-expansion}
\end{align}
The remainder $\mathscr{R}_{\mathtt{FM},t}$ is detailed below.
Equation~\eqref{eq:fm-periodwise-residual-squares} will give the time-averaged
squared bound used in the proof.
The rate $\mathsf{R}_{\mathtt{FM}}$ collects the periodwise versions of the residual terms in the stochastic linearization: the first-step bias and smooth-interior generated-regressor terms in $\mathsf{R}_{\mathtt{LA}}$, the first-step time-series approximation term $\log(nT)/H$, and, when $\mathfrak{q}=0$, the curvature and endpoint-comparison terms.

In the common decomposition, take
\begin{align*}
    u_t
    &=
    \widehat{M}_t(0)-\mathscr{B}_t(0)-\mu_t(0),
    &
    V_u
    &=
    \Sigma_{\varepsilon,\mathtt{o}}(0).
\end{align*}
Then $\bar u=\widehat{\mu}(0)-\mathscr{B}(0)-\bar{\mu}_T(0;H)$, and the
left-hand side is
$|\widetilde{\Sigma}_{\mathtt{FM}}(0)
-\Sigma_{\varepsilon,\mathtt{o}}(0)-\Sigma_\mu(0)|$.

For $\mathscr{V}_{\mathtt{FM},1}$, insert \eqref{eq:fm-origin-periodwise-expansion}. Expanding the square gives
\begin{align*}
    &\frac{1}{(T-H)^2}\sum_{t=H+1}^{T}
    \big(\widehat{M}_t(0)-\mathscr{B}_t(0)-\mu_t(0)\big)^2\\
    &=
    \frac{1}{(T-H)^2}\sum_{t=H+1}^{T}\mathscr{E}_{\mathtt{o},t}(0)^2
    +\frac{2}{(T-H)^2}\sum_{t=H+1}^{T} \mathscr{E}_{\mathtt{o},t}(0)\mathscr{R}_{\mathtt{FM},t}
    +\frac{1}{(T-H)^2}\sum_{t=H+1}^{T}\mathscr{R}_{\mathtt{FM},t}^2 .
\end{align*}
At the origin, $\widehat p_t(0)=p_t(0)$, so the periodwise leading weights below use the same reference zero-cell coordinates as Lemma~\ref{lem:classical-linearization}. The periodwise weights and volatility scales are $\mathcal{E}_\infty$-measurable; the primitive innovations in $\mathscr{E}_{\mathtt{o},t}(0)$ are conditionally independent over $i$ with mean zero and unit variance. Hence
\begin{align*}
    \E\!\left[
    \mathscr{E}_{\mathtt{o},t}(0)^2
    \mid
    \mathcal{E}_\infty
    \right]
    &=
    \sum_{i=1}^{n_t}
    \varsigma_{it}^2
    \left(
    \frac{1}{n_t}p_t(0)^\top Q_t^{-1}p_t(\beta_{it})
    \right)^2.
\end{align*}
Therefore,
\begin{align*}
    \frac{1}{(T-H)^2}\sum_{t=H+1}^{T}
    \mathscr{E}_{\mathtt{o},t}(0)^2
    &=
    \Sigma_{\varepsilon,\mathtt{o}}(0)
    +o_\P\!\left(\frac{J^d}{nT}\right),
\end{align*}
where the deviation follows from conditional independence given $\mathcal{E}_\infty$ and the same fourth-moment calculation used for the jackknife variance terms.
Also, $\Sigma_{\varepsilon,\mathtt{o}}(0)\lesssim_\P J^d/(nT)$. We next control the cross product without applying Cauchy--Schwarz to the full remainder. Write the periodwise summands of the linearization remainders as follows:
\begin{align*}
    \mathscr{L}_{\mathtt{B},t}(0)
    &=
    \widehat{p}_{t}(0)^\top \widehat{Q}_t^{-1}
    \frac{1}{n_t}\sum_{i=1}^{n_t}
    \widehat{p}_t(\widehat{\beta}_{it}) [\nabla\mu_t(\beta_{it})]^\top
    \big(\breve{\beta}_{it}-\beta_{it}\big),\\
    \mathscr{L}_{\mathtt{V},t}(0)
    &=
    \widehat{p}_{t}(0)^\top \widehat{Q}_t^{-1}
    \frac{1}{n_t}\sum_{i=1}^{n_t}
    \widehat{p}_t(\widehat{\beta}_{it}) [\nabla\mu_t(\beta_{it})]^\top
    \big(\widehat{\beta}_{it}-\breve{\beta}_{it}\big),\\
    \mathscr{A}_{t}(0)
    &=
    p_t(0)^\top Q_t^{-1}
    \frac{1}{n_t}\sum_{i=1}^{n_t}
    p_t(\beta_{it})[\nabla\mu_t(\beta_{it})]^\top
    \frac{1}{H}\sum_{s=1}^H
    S A(t)^{-1}X_s(t)K\!\Big(\frac{s}{H}\Big)\varepsilon_{i(t-s)} .
\end{align*}
The five periodwise remainders from Lemma~\ref{lem:general-linearization} are
\begin{align*}
    \mathscr{R}_{1,t}(0)
    &=
    \widehat{p}_{t}(0)^\top
    \left(\widehat{Q}_t^{-1} - Q_{t}^{-1} \right)
    \frac{1}{n_t} \sum_{i=1}^{n_t} \widehat{p}_t(\widehat{\beta}_{it}) \varepsilon_{it},\\
    \mathscr{R}_{2,t}(0)
    &=
    \widehat{p}_{t}(0)^\top Q_{t}^{-1}
    \frac{1}{n_t} \sum_{i=1}^{n_t}
    \big(\widehat{p}_t(\widehat{\beta}_{it})-p_t(\beta_{it})\big) \varepsilon_{it},\\
    \mathscr{R}_{3,t}(0)
    &=
    \widehat{p}_{t}(0)^\top \widehat{Q}_{t}^{-1}
    \left(\frac{1}{n_t}\sum_{i=1}^{n_t}
    \widehat{p}_t(\widehat{\beta}_{it})\widehat{\beta}_{it}^\top\right)
    \widetilde{f}_t,\\
    \mathscr{R}_{4,t}(0)
    &=
    \widehat{p}_{t}(0)^\top \widehat{Q}_{t}^{-1}
    \left(\frac{1}{n_t}\sum_{i=1}^{n_t}
    \widehat{p}_t(\widehat{\beta}_{it})(\beta_{it}-\widehat{\beta}_{it})^\top\right)
    \widetilde{f}_t,\\
    \mathscr{R}_{5,t}(0)
    &=
    \frac{1}{2}
    \widehat{p}_{t}(0)^\top\widehat{Q}_t^{-1}
    \frac{1}{n_t}\sum_{i=1}^{n_t}\widehat{p}_t(\widehat{\beta}_{it})
    \big(\widehat{\beta}_{it} - \beta_{it} \big)^\top
    \nabla^2\mu_t(\tilde{\beta}_{it})
    \big(\widehat{\beta}_{it} - \beta_{it} \big).
\end{align*}
Thus the remainder in \eqref{eq:fm-origin-periodwise-expansion} is
\begin{align*}
    \mathscr{R}_{\mathtt{FM},t}
    &=
    -\mathscr{A}_{t}(0)
    -\mathscr{L}_{\mathtt{B},t}(0)
    -\big(\mathscr{L}_{\mathtt{V},t}(0)-\mathscr{A}_{t}(0)\big)
    +\sum_{j=1}^{4}\mathscr{R}_{j,t}(0)-\mathscr{R}_{5,t}(0).
\end{align*}
The periodwise bounds in Lemmas~\ref{lem:general-linearization} and~\ref{lem:classical-linearization}, with the sharper origin calculation for $\mathscr{A}_{t}(0)$, imply
\begin{align}
    \frac{1}{T-H}\sum_{t=H+1}^{T}
    \Big[
    \mathscr{A}_{t}(0)^2
    +\mathscr{L}_{\mathtt{B},t}(0)^2
    +\big(\mathscr{L}_{\mathtt{V},t}(0)-\mathscr{A}_{t}(0)\big)^2
    +\sum_{j=1}^{5}\mathscr{R}_{j,t}(0)^2
    \Big]
    &\lesssim_\P
    \mathsf{R}_{\mathtt{FM}}^2+o_\P\!\left(\frac{J^d}{n}\right).
    \label{eq:fm-periodwise-residual-squares}
\end{align}

We spell out the argument for $\mathscr{L}_{\mathtt{B},t}(0)$.
{This term is $\mathcal C_t$-measurable, while
$\E[\mathscr{E}_{\mathtt{o},t}(0)\mid\mathcal C_t]=0$.} Hence the
predictable quadratic variation of the cross product satisfies
\begin{align*}
    &\frac{1}{(T-H)^4}\sum_{t=H+1}^T
    \E\!\left[
    \mathscr{E}_{\mathtt{o},t}(0)^2
    \mathscr{L}_{\mathtt{B},t}(0)^2
    \mid \mathcal C_t
    \right]
    \lesssim_\P
    \frac{1}{(T-H)^4}\frac{J^d}{n}
    \sum_{t=H+1}^T \mathscr{L}_{\mathtt{B},t}(0)^2.
\end{align*}
Therefore,
\begin{align*}
    \left|
    \frac{1}{(T-H)^2}\sum_{t=H+1}^T
    \mathscr{E}_{\mathtt{o},t}(0)\mathscr{L}_{\mathtt{B},t}(0)
    \right|
    &\lesssim_\P
    \frac{1}{T^{3/2}}\sqrt{\frac{J^d}{n}}
    \left(
    \frac{1}{T-H}\sum_{t=H+1}^T
    \mathscr{L}_{\mathtt{B},t}(0)^2
    \right)^{1/2}\\
    &\lesssim_\P
    \frac{J^d}{nT^2}
    +
    \frac{1}{T(T-H)}\sum_{t=H+1}^T
    \mathscr{L}_{\mathtt{B},t}(0)^2\\
    &=
    o_\P\!\left(\frac{J^d}{nT}\right)
    +O_\P\!\left(
    \frac{1}{T(T-H)}\sum_{t=H+1}^T
    \mathscr{L}_{\mathtt{B},t}(0)^2
    \right),
\end{align*}
where the last line uses Young's inequality. {To bound the
components of
$\sum_{t=H+1}^T\mathscr{E}_{\mathtt{o},t}(0)\mathscr{R}_{\mathtt{FM},t}$
that contain
$\mathscr{A}_{t}(0)$,
$\mathscr{L}_{\mathtt{V},t}(0)-\mathscr{A}_{t}(0)$,
$\mathscr{R}_{3,t}(0)$, $\mathscr{R}_{4,t}(0)$, and
$\mathscr{R}_{5,t}(0)$, apply the preceding predictable-variation argument
used for $\mathscr{L}_{\mathtt{B},t}(0)$ after conditioning on
$\mathcal E_\infty$ and the primitive shocks dated before $t$. Use
Cauchy--Schwarz for any part without a conditionally mean-zero shock.}
Summing over these displayed components and using
\eqref{eq:fm-periodwise-residual-squares} gives a contribution bounded by
$\mathsf{R}_{\mathtt{FM}}^2/T+o_\P(J^d/(nT))$.

It remains to handle the two design-comparison terms. For $\mathscr{R}_{1,t}(0)$, Cauchy--Schwarz after the inverse-Gram decomposition and the Gram-convergence bound from Lemma~\ref{lem:partition-basis-moments} give $o_\P(J^d/(nT))$. For $\mathscr{R}_{2,t}(0)$, open the product with $\mathscr{E}_{\mathtt{o},t}(0)$ into diagonal and off-diagonal pieces. The predictable diagonal piece is
\begin{align*}
    \frac{2}{(T-H)^2}\sum_{t=H+1}^{T}\sum_{i=1}^{n_t}
    \varsigma_{it}^2
    &\left(
    \frac{1}{n_t}p_t(0)^\top Q_t^{-1}p_t(\beta_{it})
    \right)
    \left(
    \frac{1}{n_t}p_t(0)^\top Q_t^{-1}
    \big(\widehat{p}_t(\widehat{\beta}_{it})-p_t(\beta_{it})\big)
    \right),
\end{align*}
and is bounded by the switch-probability and cell-count rates entering $\mathsf{R}_{\mathtt{FM}}^2/T$. The centered diagonal remainder and the off-diagonal piece are conditionally mean-zero sums and are $o_\P(J^d/(nT))$ by the same predictable-variation calculation as above. Therefore,
\begin{align*}
    \left|
    \frac{2}{(T-H)^2}\sum_{t=H+1}^{T}
    \mathscr{E}_{\mathtt{o},t}(0)\mathscr{R}_{\mathtt{FM},t}
    \right|
    &\lesssim_\P
    \frac{\mathsf{R}_{\mathtt{FM}}^2}{T}
    +o_\P\!\left(\frac{J^d}{nT}\right),
\end{align*}
and
\begin{align*}
    \frac{1}{(T-H)^2}\sum_{t=H+1}^{T}\mathscr{R}_{\mathtt{FM},t}^2
    &\lesssim_\P
    \frac{\mathsf{R}_{\mathtt{FM}}^2}{T}
    +o_\P\!\left(\frac{J^d}{nT}\right).
\end{align*}
Hence, we conclude that
\begin{align*}
    \mathscr{V}_{\mathtt{FM},1}
    &\lesssim_\P
    \frac{\mathsf{R}_{\mathtt{FM}}^2}{T}
    +o_\P\!\left(\frac{J^d}{nT}\right).
\end{align*}

For $\mathscr{V}_{\mathtt{FM},2}$, the aggregate deviation satisfies
$\widehat{\mu}(0)-\mathscr{B}(0)-\bar{\mu}_T(0;H)
\lesssim_\P \sqrt{J^d/(nT)}+\mathsf{R}_{\mathtt{FM}}$
by \eqref{eq:fm-origin-periodwise-expansion}, so
\begin{align*}
    \mathscr{V}_{\mathtt{FM},2}
    &\lesssim_\P
    \frac{J^d}{nT^2}+\frac{\mathsf{R}_{\mathtt{FM}}^2}{T}
    =
    o_\P\!\left(\frac{J^d}{nT}\right)
    +\frac{\mathsf{R}_{\mathtt{FM}}^2}{T}.
\end{align*}

For $\mathscr{V}_{\mathtt{FM},3}$, by Cauchy--Schwarz,
\begin{align*}
    \mathscr{V}_{\mathtt{FM},3}^2
    &\lesssim
    \left(
    \frac{1}{(T-H)^2}\sum_{t=H+1}^{T}
    \big(\widehat{M}_t(0) - \mathscr{B}_t(0)- \mu_t(0)\big)^2
    \right)
    \left(
    \frac{1}{(T-H)^2}\sum_{t=H+1}^{T}
    \big(\mu_t(0) - \bar{\mu}_T(0;H)\big)^2
    \right),
\end{align*}
where the first factor is $\lesssim_\P J^d/(nT)+\mathsf{R}_{\mathtt{FM}}^2/T$ by \eqref{eq:fm-origin-periodwise-expansion}, and the second equals $\Sigma_\mu(0)$. Thus,
\begin{align*}
    \mathscr{V}_{\mathtt{FM},3}
    &\lesssim_\P
    \left(
    \sqrt{\frac{J^d}{nT}}
    +\frac{\mathsf{R}_{\mathtt{FM}}}{\sqrt{T}}
    \right)
    \sqrt{\Sigma_\mu(0)}.
\end{align*}

Combining the bounds for $\mathscr{V}_{\mathtt{FM},1}$,
$\mathscr{V}_{\mathtt{FM},2}$, and $\mathscr{V}_{\mathtt{FM},3}$ gives the
asserted bound for
$\widetilde{\Sigma}_{\mathtt{FM}}(0)-\Sigma_{\varepsilon,\mathtt{o}}(0)-\Sigma_\mu(0)$.
\qed

\section{Uniform-in-\texorpdfstring{$\beta$}{beta} Inference}\label{sec:uniform-beta-inference}

The pointwise-in-$\beta$ Gaussian approximation in
Theorem~\ref{thm:pointwise-clt} and the feasible variance estimators
above give the usual confidence intervals at each fixed nonsingular evaluation
point. This section converts those pointwise intervals into a family of
confidence intervals whose
coverage is controlled uniformly over $\beta\in\mathcal{B}^\circ$, where
$\mathcal{B}^\circ$ is the interior evaluation region used in the
stochastic-linearization results. The result is not a simultaneous coverage
statement for the entire function; it is a uniform statement about the coverage
probability of the interval indexed by the evaluation point.

The only nonstandard feature is the transition in the variance as $\beta$
approaches the origin. Away from zero, the factor variance is first order and
$\Sigma(\beta)\asymp_\P \|\beta\|^2/T$. At the origin, the factor component
vanishes and the idiosyncratic variance is at the local-support scale
$J^d/(nT)$. We therefore use the usual feasible standard error outside a
shrinking neighborhood of the origin and a conservative variance floor inside
that neighborhood. The floor argument deliberately avoids using the pointwise
origin Gaussian approximation, so no lower bound on
$\Sigma_{\varepsilon}(0)$ is needed in this section.
Throughout this section, take $\mathcal{B}^\circ$ to contain a fixed
neighborhood of $0$.

\subsection{Result}

We impose the stochastic-linearization and factor-CLT conditions used above,
but the variance requirements are weaker than full pointwise jackknife
consistency. For the factor component, we use the factor-variance estimator
condition in Lemma~\ref{lem:factor-variance}. For the idiosyncratic
component, the conservative floor only needs the local-support upper bound
{verified in the proof of
Theorem~\ref{thm:uniform-coverage}.} Thus this section does not
import the primitive jackknife localization condition
$J\mathsf{R}_{\mathtt{FS}}=o(1)$. We also do not impose the origin-specific
lower-bound condition
$(nT/J^d)\Sigma_{\varepsilon}(0)\gtrsim_\P1$ from
Lemma~\ref{lem:leading-gaussian-components}; the conservative floor below
replaces that requirement. Set
\begin{align*}
    \mathsf{R}_{\mathtt{UI}}
    = \sqrt{T-H}\,\mathsf{R}_{\mathtt{CLT}}.
\end{align*}
The rate $\mathsf{R}_{\mathtt{UI}}$ is the threshold scale obtained by comparing
the pointwise linearization remainder with the factor standard deviation. By
definition, it includes the same-cell nonconstant-basis component in
$\mathsf{R}_{\mathtt{LA}}$ when $\mathfrak{q}\geq1$.
The first-step correction $\mathscr{A}(\beta)$ is part of the adjusted
idiosyncratic leading term, but the uniform bound for this
term is the stronger scale from Lemma~\ref{lem:classical-linearization}. Define
\begin{align*}
    \mathsf{R}_{\mathtt{A}}^\circ
    &=
    \sqrt{T-H}
    \left(
    \frac{J^d\log(nT)}{nH}
    \right)^{1/2}.
\end{align*}
Thus
$\mathsf{R}_{\mathtt{A}}^\circ/\sqrt{T-H}$ is the uniform scale for
$\mathscr{A}(\beta)$ over $\mathcal{B}^\circ$.
The factor-variance estimation error is handled differently: it is a quadratic
form in $\beta$ and is therefore controlled relative to $\Sigma_f(\beta)$ by
the factor-prediction condition in Lemma~\ref{lem:factor-variance}.

Let $\delta>0$ be a possibly data-dependent factor-variance threshold.
The estimated factor variance $\widehat\Sigma_f(\beta)$ is invariant to
nonsingular rescalings of the beta coordinates and factor normalization. The
associated normalized factor-risk scale is
$\sqrt{(T-H)\widehat\Sigma_f(\beta)}$. The condition below compares all error
scales that can bind with the corresponding boundary scale
$\sqrt{(T-H)\delta}$.
We require $\delta$ to satisfy
\begin{align}\label{eq:uniform-delta-conditions}
    (T-H)\delta=o_\P(1),
    \qquad
    \frac{
    \mathsf{R}_{\mathtt{UI}}
    +\mathsf{R}_{\mathtt{A}}^\circ}
    {\sqrt{(T-H)\delta}}
    =o_\P(1)
\end{align}

Let $z_{1-\alpha/2}$ denote the $1-\alpha/2$ quantile of the standard normal
distribution.
The uniform-in-$\beta$ $100(1-\alpha)\%$ confidence interval is
\begin{align*}
    [\widehat{L}_T(\beta),\widehat{U}_T(\beta)]
    &=
    \widehat{\mu}(\beta)
    \pm z_{1-\alpha/2}\sqrt{\widehat{\Sigma}_{\mathtt{UI}}(\beta)},
\end{align*}
where
\begin{align*}
    \widehat{\Sigma}_{\mathtt{UI}}(\beta)
    &=
    \begin{cases}
        \widehat{\Sigma}(\beta)&
        \quad\text{if }\widehat\Sigma_f(\beta)>\delta\\
        \max\{\widehat{\Sigma}(\beta),\widehat{\Sigma}_{\delta}\}&
        \quad\text{if }\widehat\Sigma_f(\beta)\leq\delta
    \end{cases},
\end{align*}
and, when the factor-variance level set is nonempty,
\begin{align*}
    \widehat{\Sigma}_{\delta}
    =
    \sup_{\substack{b\in\mathcal{B}^\circ\\
    \widehat\Sigma_f(b)=\delta}}
    \max\{\widehat{\Sigma}(b),\delta\}
    =
    \delta+
    \sup_{\substack{b\in\mathcal{B}^\circ\\
    \widehat\Sigma_f(b)=\delta}}
    \max\{\widehat{\Sigma}_{\varepsilon}(b),0\}.
\end{align*}

The floor $\widehat{\Sigma}_{\delta}$ adds the threshold factor variance
$\delta$ to the largest nonnegative feasible idiosyncratic component on the
boundary of the shrinking factor-variance neighborhood.
On the negligible complement event where the level set is empty, set
$\widehat{\Sigma}_{\delta}=\delta$; this convention has no asymptotic effect.
The level set is nonempty w.p.a.1 because $\widehat\Sigma_f(0)=0$,
$\widehat\Sigma_f(\cdot)$ is continuous, and $\mathcal{B}^\circ$ contains a fixed
neighborhood of zero. The uniform quadratic-form bound used below, factor
nondegeneracy, and $\mathscr{B}_{\mathtt{F}}(b)\geq0$ give
$\widehat\Sigma_f(b)\gtrsim_\P(T-H)^{-1}$ uniformly on any fixed small sphere
around zero. Since $(T-H)\delta=o_\P(1)$, continuity then gives a point inside
the sphere at which $\widehat\Sigma_f(b)=\delta$.

For a family of random variables $\{X_n(\beta):\beta\in\mathcal{I}_n\}$ and a
positive deterministic or data-dependent rate $r_n$, write
$X_n(\beta)=o_{\P,\mathcal{I}_n}(r_n)$ if, for every $\epsilon>0$,
\begin{align*}
    \sup_{\beta\in\mathcal{I}_n}
    \P\!\left(|X_n(\beta)|>\epsilon r_n\right)
    &\to0,
\end{align*}
and $X_n(\beta)\lesssim_{\P,\mathcal{I}_n}r_n$ if
\begin{align*}
    \lim_{M\to\infty}\limsup_{n,T\to\infty}
    \sup_{\beta\in\mathcal{I}_n}
    \P\!\left(|X_n(\beta)|>M r_n\right)
    &=0.
\end{align*}
These are pointwise-uniform probability statements; they are the right notion
for the infimum of pointwise coverage probabilities below and do not assert a
realized supremum bound unless one is displayed explicitly.

\begin{theorem}[Uniform Coverage of Confidence Intervals]\label{thm:uniform-coverage}
    Assume that the conditions of Corollary
    \ref{cor:classical-pointwise-convergence} hold, {that
    \eqref{eq:jackknife-conditions} holds,} and that the
    factor-variance estimator satisfies the conditions of
    Lemma~\ref{lem:factor-variance}.
    If
    $\delta$ satisfies~\eqref{eq:uniform-delta-conditions}, then
    \begin{align*}
        \liminf_{n,T\to\infty}
        \inf_{\beta\in\mathcal{B}^\circ}
        \P\Bigl(
        \bar{\mu}_T(\beta;H)
        \in[\widehat{L}_T(\beta),\widehat{U}_T(\beta)]
        \Bigr)
        &\geq 1-\alpha.
    \end{align*}
\end{theorem}

The theorem does not require a pointwise Gaussian approximation at $\beta=0$, and the origin-specific lower-bound condition $(nT/J^d)\Sigma_{\varepsilon}(0)\gtrsim_\P1$ is not imposed.
Inside the factor-variance neighborhood
$\widehat\Sigma_f(\beta)\leq\delta$, the interval is calibrated using
the boundary variance floor $\widehat\Sigma_\delta$. The condition
\eqref{eq:uniform-delta-conditions}, together with local support of the portfolio
weights, ensures that the idiosyncratic leading term, the uniform first-step
correction, and the full linearization remainder are negligible relative to
that boundary standard-error scale.

\subsection{Proof of Theorem~\ref{thm:uniform-coverage}}

We first record the uniform consequences of the linearization and variance
bounds. The pointwise Gaussian approximation at the origin is not used.

The stochastic linearization in Lemmas~\ref{lem:general-linearization}
and~\ref{lem:classical-linearization}, together with Section~\ref{sec:pointwise-distribution}, gives
\begin{align*}
    \widehat{\mu}(\beta)-\bar{\mu}_T(\beta;H)
    &=
    \mathscr{G}(\beta)+\bar{\mathscr{S}}_{\varepsilon}(\beta)
    +\mathscr{R}(\beta),
    \qquad
    \bar{\mathscr{S}}_{\varepsilon}(\beta)
    =
    \mathscr{E}(\beta)-\mathscr{A}(\beta).
\end{align*}
Thus the first-step correction $\mathscr{A}(\beta)$ is not part of the
non-leading remainder, but it must be controlled at its own uniform scale. The
rate $\mathsf{R}_{\mathtt{CLT}}$ controls only
$\mathscr{R}(\beta)$, while Lemma~\ref{lem:classical-linearization} controls
$\mathscr{A}(\beta)$:
\begin{align*}
    |\mathscr{R}(\beta)|
    &\lesssim_{\P,\mathcal{B}^\circ}
    \mathsf{R}_{\mathtt{CLT}},
    \\
    |\mathscr{A}(\beta)|
    &\lesssim_{\P,\mathcal{B}^\circ}
    \frac{\mathsf{R}_{\mathtt{A}}^\circ}{\sqrt{T-H}}.
\end{align*}
{The proofs of Lemmas~\ref{lem:leading-gaussian-components}
and~\ref{lem:jackknife-selected-block} take their local-support and
selected-block bounds uniformly over the deterministic quasi-uniform rectangle
class before the block containing $\beta$ is selected. Their constants
therefore do not depend on $\beta$, so
\begin{align*}
    \Sigma_{\varepsilon}(\beta)
    +|\widehat{\Sigma}_{\varepsilon}(\beta)|
    &\lesssim_{\P,\mathcal{B}^\circ}
    \frac{J^d}{n(T-H)}.
\end{align*}}
This is dominated by
$(\mathsf{R}_{\mathtt{A}}^\circ)^2/(T-H)$ under the maintained rolling-window
scaling. Hence \eqref{eq:uniform-delta-conditions} and the definitions of
$\mathsf{R}_{\mathtt{UI}}$ and $\mathsf{R}_{\mathtt{A}}^\circ$ imply
\begin{align*}
    \mathscr{R}(\beta)
    &=
    o_{\P,\mathcal{B}^\circ}(\sqrt{\delta}),
    \\
    \mathscr{A}(\beta)
    &=
    o_{\P,\mathcal{B}^\circ}(\sqrt{\delta}),
    \\
    \Sigma_{\varepsilon}(\beta)
    +|\widehat{\Sigma}_{\varepsilon}(\beta)|
    &=
    o_{\P,\mathcal{B}^\circ}(\delta).
\end{align*}
The bound for $\mathscr A(\beta)$ is the uniform first-step-correction bound
from Lemma~\ref{lem:classical-linearization}. For the direct idiosyncratic term
$\mathscr E(\beta)$, condition period by period on the field
$\mathcal C_t$ from Section~\ref{sec:classical-conditioning-discussion}. Its
coefficients are fixed under this conditioning, and its predictable variance
is the direct variance $\Sigma_{\varepsilon,\mathtt{o}}(\beta)$ defined in the
Fama--MacBeth variance subsection above, not the adjusted reference scale
$\Sigma_\varepsilon(\beta)$. The same local-support calculation used for
$\mathscr E(\beta)$ in Lemma~\ref{lem:general-linearization} gives
\begin{align*}
    \Sigma_{\varepsilon,\mathtt{o}}(\beta)
    &\lesssim_{\P,\mathcal B^\circ}
    \frac{J^d}{n(T-H)}
    =
    o_{\P,\mathcal B^\circ}(\delta).
\end{align*}
Because $\delta$ may be data dependent, choose a deterministic sequence
$a_n\to\infty$ such that
$\P\big(\delta<a_nJ^d/[n(T-H)]\big)\to0$. The preceding conditional-variance
bound and conditional Chebyshev's inequality at the deterministic threshold
then imply that for every $\epsilon>0$,
\begin{align*}
    \sup_{\beta\in\mathcal B^\circ}
    \P\left(
    |\mathscr E(\beta)|>\epsilon\sqrt\delta
    \right)
    &\leq
    \P\left(\delta<\frac{a_nJ^d}{n(T-H)}\right)
    +\sup_{\beta\in\mathcal B^\circ}
    \P\left(
    |\mathscr E(\beta)|>\epsilon
    \sqrt{\frac{a_nJ^d}{n(T-H)}}
    \right)
    \to0,
\end{align*}
or equivalently $\mathscr E(\beta)=o_{\P,\mathcal B^\circ}(\sqrt\delta)$.
Here the supremum is taken over the pointwise probabilities after applying
Chebyshev's inequality; no realized supremum of
$\Sigma_{\varepsilon,\mathtt{o}}(\beta)$ or
$\Sigma_\varepsilon(\beta)$ is invoked. Consequently,
$\bar{\mathscr{S}}_{\varepsilon}(\beta)
=\mathscr{E}(\beta)-\mathscr{A}(\beta)
=o_{\P,\mathcal{B}^\circ}(\sqrt{\delta})$.

Next, record the uniform quadratic-form version of
Lemma~\ref{lem:factor-variance}. Write
\begin{align*}
    \widehat{\Sigma}_f(\beta)-\Sigma_f(\beta)
    -\mathscr{B}_{\mathtt{F}}(\beta)
    =
    \beta^\top\Delta_{\mathtt{F}}\beta,
\end{align*}
where
\begin{align*}
    \Delta_{\mathtt{F}}
    =
    \frac{1}{(T-H)^2}\sum_{t=H+1}^{T}
    \left\{
    \widetilde f_t\widetilde f_t^\top
    -\E[\widetilde f_t\widetilde f_t^\top\mid\mathcal{G}_{t-1}]
    \right\}
    +
    \frac{2}{(T-H)^2}\sum_{t=H+1}^{T}
    \widetilde f_t(m_t-\widehat m_t)^\top .
\end{align*}
Because $d$ is fixed and $\mathcal{B}^\circ$ is bounded, the entrywise
martingale bounds in the proof of Lemma~\ref{lem:factor-variance}
and the factor-prediction condition in Lemma~\ref{lem:factor-variance}
give
\begin{align*}
    \|\Delta_{\mathtt{F}}\|
    &=
    o_\P\!\left((T-H)^{-1}\right).
\end{align*}
Then, by the factor nondegeneracy condition in Assumption \ref{ass:dgp}(iv), we have
\begin{align*}
    \omega_n
    &:=
    \sup_{\substack{\beta\in\mathcal{B}^\circ\\ \beta\neq0}}
    \frac{\left|
    \widehat{\Sigma}_f(\beta)-\Sigma_f(\beta)
    -\mathscr{B}_{\mathtt{F}}(\beta)
    \right|}
    {\Sigma_f(\beta)}
    =
    o_\P(1),
\end{align*}
while the numerator is zero at $\beta=0$.
Since $\mathscr{B}_{\mathtt{F}}(\beta)\geq0$, this implies
\begin{align*}
    \left(
    \Sigma_f(\beta)-\widehat{\Sigma}_f(\beta)
    \right)_+
    &\leq
    \omega_n\Sigma_f(\beta),
\end{align*}
where $(a)_+:=\max\{a, 0\}$ for any $a\in\mathbb{R}$.
Combining this display with
$|\widehat{\Sigma}_{\varepsilon}(\beta)|=o_{\P,\mathcal{B}^\circ}(\delta)$ and
$\widehat{\Sigma}(\beta)=\widehat{\Sigma}_f(\beta)
+\widehat{\Sigma}_{\varepsilon}(\beta)$ gives the pointwise-uniform lower
bound
\begin{align}\label{eq:uniform-floor-outer-bound}
    \left(
    1-
    \frac{\widehat{\Sigma}_{\mathtt{UI}}(\beta)}
    {\max\{\Sigma_f(\beta),\delta\}}
    \right)_+
    &=
    o_{\P,\mathcal{B}^\circ}(1).
\end{align}
To see this, consider the two regions used in the construction. If
$\widehat\Sigma_f(\beta)>\delta$, then
$\widehat{\Sigma}_{\mathtt{UI}}(\beta)=\widehat{\Sigma}(\beta)$ and the
centered factor-variance bound gives
\begin{align*}
    \widehat{\Sigma}_{\mathtt{UI}}(\beta)
    &\geq
    (1-\omega_n)\Sigma_f(\beta)
    -o_{\P,\mathcal{B}^\circ}(\delta),
    \qquad
    \widehat{\Sigma}_{\mathtt{UI}}(\beta)
    \geq
    \delta-o_{\P,\mathcal{B}^\circ}(\delta).
\end{align*}
Thus \eqref{eq:uniform-floor-outer-bound} holds on the outer region.
If
$\widehat\Sigma_f(\beta)\leq\delta$, since
$\widehat\Sigma_f(\beta)\geq(1-\omega_n)\Sigma_f(\beta)$,
$\Sigma_f(\beta)\leq(1+o_\P(1))\delta$ uniformly on this region. Also,
$\widehat{\Sigma}_{\mathtt{UI}}(\beta)\geq\widehat{\Sigma}_{\delta}\geq\delta$,
where the last inequality follows from the definition of the variance floor.
This gives the same ratio lower bound on the inner region. In particular,
\begin{align}\label{eq:uniform-floor-inner-bound}
    \left(
    1-
    \frac{\widehat{\Sigma}_{\mathtt{UI}}(\beta)}{\delta}
    \right)_+
    &=
    o_{\P,\mathcal{B}^\circ}(1).
\end{align}

Next, write
\begin{align*}
    \mathscr{G}(\beta)
    &=
    \beta^\top
    \frac{1}{T-H}\sum_{t=H+1}^{T}\widetilde{f}_t .
\end{align*}
Let
\begin{align*}
    S_T
    &:=
    \sum_{t=H+1}^{T}\widetilde f_t,
    &
    V_T
    &:=
    \sum_{t=H+1}^{T}
    \Var(\widetilde f_t\mid\mathcal G_{t-1}).
\end{align*}
Then
\begin{align*}
    \Sigma_f(\beta)
    &=
    \frac{\beta^\top V_T\beta}{(T-H)^2},
    &
    \left\|
    \frac{V_T}{T-H}-\Omega_f
    \right\|
    &=
    o_\P(1)
\end{align*}
by \eqref{eq:factor-variance-stabilization}.

For $u\in\mathbb S^{d-1}$, define the deterministically normalized process
\begin{align*}
    Z_T^0(u)
    &:=
    \frac{u^\top S_T}
    {\sqrt{(T-H)u^\top\Omega_fu}}.
\end{align*}
For each fixed $u$, the sum of the conditional variances of its martingale
increments converges in probability to one:
\begin{align*}
    \frac{
    u^\top V_Tu
    }
    {(T-H)u^\top\Omega_fu}
    &\to_\P1.
\end{align*}
The conditional Lyapunov bound is uniform over directions because
$\lambda_{\min}(\Omega_f)>0$:
\begin{align*}
    \sup_{u\in\mathbb S^{d-1}}
    \frac{
    \sum_{t=H+1}^{T}
    \E[
    |u^\top\widetilde f_t|^4
    \mid\mathcal G_{t-1}
    ]
    }
    {(T-H)^2(u^\top\Omega_fu)^2}
    &\lesssim_\P
    (T-H)^{-1}
    =o_\P(1).
\end{align*}
The scalar martingale CLT therefore gives
\begin{align*}
    \sup_{x\in\mathbb R}
    \left|
    \P\left(
    Z_T^0(u)\leq x
    \right)
    -\Phi(x)
    \right|
    &\to0
\end{align*}
for each fixed $u$.

We now make this approximation uniform over directions. Fix $\epsilon>0$ and
let $\mathcal N_\epsilon$ be a finite $\epsilon$-net of
$\mathbb S^{d-1}$. The preceding display holds uniformly over the fixed finite
set $\mathcal N_\epsilon$. Positive definiteness of $\Omega_f$ and
martingale orthogonality give
\begin{align*}
    \sup_{\|u-v\|\leq\epsilon}
    |Z_T^0(u)-Z_T^0(v)|
    &\lesssim
    \epsilon\frac{\|S_T\|}{\sqrt{T-H}}
    =
    O_\P(\epsilon),
    &
    \frac{\|S_T\|}{\sqrt{T-H}}
    &=O_\P(1).
\end{align*}
The finite-net argument and normal anticoncentration thus imply
\begin{align}
    \sup_{u\in\mathbb S^{d-1}}
    \sup_{x\in\mathbb R}
    \left|
    \P\left(
    Z_T^0(u)\leq x
    \right)
    -\Phi(x)
    \right|
    &\to0.
    \label{eq:uniform-factor-reference-normality}
\end{align}

Finally, stabilization gives
\begin{align*}
    \sup_{u\in\mathbb S^{d-1}}
    \left|
    \frac{u^\top V_Tu}
    {(T-H)u^\top\Omega_fu}
    -1
    \right|
    &\leq
    \frac{
    \|V_T/(T-H)-\Omega_f\|
    }
    {\lambda_{\min}(\Omega_f)}
    =
    o_\P(1).
\end{align*}
Since
\(\sup_{u\in\mathbb S^{d-1}}|Z_T^0(u)|
\lesssim\|S_T\|/\sqrt{T-H}=O_\P(1)\),
\begin{align*}
    \sup_{u\in\mathbb S^{d-1}}
    \left|
    \frac{u^\top S_T}{\sqrt{u^\top V_Tu}}
    -
    Z_T^0(u)
    \right|
    &=
    o_\P(1).
\end{align*}
Combining this uniform replacement with
\eqref{eq:uniform-factor-reference-normality} yields
\begin{align*}
    \sup_{u\in\mathbb S^{d-1}}
    \sup_{x\in\mathbb R}
    \left|
    \P\left(
    \frac{u^\top S_T}{\sqrt{u^\top V_Tu}}
    \leq x
    \right)
    -\Phi(x)
    \right|
    &\to0.
\end{align*}
Thus the factor Gaussian approximation is uniform after writing
$\beta=\|\beta\|u$; at $\beta=0$, $\mathscr{G}(0)=0$. Since
$\max\{\Sigma_f(\beta),\delta\}\geq\Sigma_f(\beta)$, the event with the
conservative normalizer contains the event with the true factor normalizer for
every realization of $\delta$. Therefore, for any fixed $c\geq0$,
\begin{align*}
    \inf_{\beta\in\mathcal{B}^\circ}
    \P\!\left(
    |\mathscr{G}(\beta)|
    \leq
    c\sqrt{\max\{\Sigma_f(\beta),\delta\}}
    \right)
    &\geq
    2\Phi(c)-1+o(1).
\end{align*}
This statement does not condition on the data-dependent inner or outer region;
the conservative scale $\max\{\Sigma_f(\beta),\delta\}$ dominates the true
factor variance for every $\beta$ and every realization.

Finally, \eqref{eq:uniform-floor-inner-bound} and
$\bar{\mathscr{S}}_{\varepsilon}(\beta)+\mathscr{R}(\beta)
=o_{\P,\mathcal{B}^\circ}(\sqrt{\delta})$ give
\begin{align*}
    \frac{
    |\bar{\mathscr{S}}_{\varepsilon}(\beta)+\mathscr{R}(\beta)|}
    {\sqrt{\widehat{\Sigma}_{\mathtt{UI}}(\beta)}}
    &=
    o_{\P,\mathcal{B}^\circ}(1).
\end{align*}
To assemble the coverage event, fix $\eta>0$ with
$\eta<z_{1-\alpha/2}$. The probability of the following event tends to one,
uniformly over $\beta\in\mathcal{B}^\circ$ in the pointwise sense:
\begin{align*}
    |\bar{\mathscr{S}}_{\varepsilon}(\beta)+\mathscr{R}(\beta)|
    &\leq
    \eta\sqrt{\widehat{\Sigma}_{\mathtt{UI}}(\beta)}
    \quad\text{and}\quad
    \widehat{\Sigma}_{\mathtt{UI}}(\beta)
    \geq
    (1-\eta)\max\{\Sigma_f(\beta),\delta\}.
\end{align*}
On this event,
\begin{align*}
    \left\{
    |\mathscr{G}(\beta)|
    \leq
    (z_{1-\alpha/2}-\eta)\sqrt{1-\eta}\,
    \sqrt{\max\{\Sigma_f(\beta),\delta\}}
    \right\}
    &\subseteq
    \left\{
    |\widehat{\mu}(\beta)-\bar{\mu}_T(\beta;H)|
    \leq
    z_{1-\alpha/2}\sqrt{\widehat{\Sigma}_{\mathtt{UI}}(\beta)}
    \right\}.
\end{align*}
The factor bound above with
$c=(z_{1-\alpha/2}-\eta)\sqrt{1-\eta}$ gives
\begin{align*}
    \inf_{\beta\in\mathcal{B}^\circ}
    \P\!\Bigl(
    |\widehat{\mu}(\beta)-\bar{\mu}_T(\beta;H)|
    \leq
    z_{1-\alpha/2}\sqrt{\widehat{\Sigma}_{\mathtt{UI}}(\beta)}
    \Bigr)
    &\geq
    2\Phi\!\left((z_{1-\alpha/2}-\eta)\sqrt{1-\eta}\right)-1+o(1).
\end{align*}
Letting $\eta\downarrow0$ yields the claimed lower bound $1-\alpha$.
The boundary $\widehat\Sigma_f(\beta)=\delta$ is covered by the inner formula
in the definition of $\widehat{\Sigma}_{\mathtt{UI}}(\beta)$, so no separate
boundary argument is needed. The displayed lower bound is the claimed uniform
coverage.
\qed

\section{Factor-Cancelling Linear Contrasts}\label{sec:factor-cancelling}

This section develops inference for finite linear contrasts that eliminate the
factor component in the pointwise stochastic expansion. Let $K_{\Psi}$ be fixed,
let $\mathsf{b}_1,\ldots,\mathsf{b}_{K_{\Psi}}\in\mathcal{B}^\circ$, and let
$a_1,\ldots,a_{K_{\Psi}}$ be fixed real weights. The contrast of interest is
\begin{align*}
    \widehat{\Psi}
    &=
    \sum_{k=1}^{K_{\Psi}}a_k\widehat{\mu}(\mathsf{b}_k),
    \qquad
    \bar{\Psi}_T
    =
    \sum_{k=1}^{K_{\Psi}}a_k\bar{\mu}_T(\mathsf{b}_k;H).
\end{align*}
The key restriction is the factor-cancelling condition
\begin{align}\label{eq:factor-cancelling-contrast}
    \sum_{k=1}^{K_{\Psi}}a_k\mathsf{b}_k
    &=0.
\end{align}
For example, the butterfly spread corresponds to $K_{\Psi}=3$,
$(a_1,a_2,a_3)=(1,-2,1)$, and
$\mathsf{b}_2=(\mathsf{b}_1+\mathsf{b}_3)/2$.

\subsection{Setup and Linearization}

For any scalar term $A(\cdot)$, write its contrast as
\begin{align*}
    A_{\Psi}
    &=
    \sum_{k=1}^{K_{\Psi}}a_kA(\mathsf{b}_k).
\end{align*}
Applying this operator to the pointwise linearization in
Section~\ref{sec:pointwise-distribution} gives
\begin{align}\label{eq:factor-cancelling-expansion}
    \widehat{\Psi}-\bar{\Psi}_T
    &=
    \bar{\mathscr{S}}_{\varepsilon\Psi}
    +\mathscr{R}_{\Psi},
\end{align}
because
\begin{align*}
    \bar{\mathscr{S}}_{f\Psi}
    &=
    \frac{1}{T-H}\sum_{t=H+1}^{T}
    \left(\sum_{k=1}^{K_{\Psi}}a_k\mathsf{b}_k\right)^\top
    \widetilde{f}_t
    =0.
\end{align*}
The idiosyncratic leading term can be written as
\begin{align*}
    \bar{\mathscr{S}}_{\varepsilon\Psi}
    &=
    \frac{1}{T-H}\sum_{t=1}^{T}\sum_{i=1}^{n_t}
    \varsigma_{it}\eta_{it}W_{it,\Psi},
    \qquad
    W_{it,\Psi}
    =
    \sum_{k=1}^{K_{\Psi}}a_kW_{it}(\mathsf{b}_k).
\end{align*}
Since $K_{\Psi}$ is fixed, the leading-weight bounds used in
Lemma~\ref{lem:leading-gaussian-components} imply
\begin{align*}
    \left|\bar{\mathscr{S}}_{\varepsilon\Psi}\right|
    &\lesssim_\P
    \sqrt{\frac{J^d}{nT}}.
\end{align*}
Because $K_{\Psi}$ and the contrast weights are fixed, define the
deterministic rate
\begin{align*}
    \mathsf{R}_{\mathtt{\Psi}}
    &:={}
    \left(\sum_{k=1}^{K_{\Psi}}|a_k|\right)
    \mathsf{R}_{\mathtt{CLT}}.
\end{align*}
The triangle inequality and the common pointwise remainder bound give
\begin{align*}
    |\mathscr{R}_{\Psi}|
    &\lesssim_\P
    \mathsf{R}_{\mathtt{\Psi}}.
\end{align*}
The condition needed
for idiosyncratic-rate inference is
\begin{align}\label{eq:factor-cancelling-rate}
    \sqrt{\frac{nT}{J^d}}\mathsf{R}_{\mathtt{\Psi}}
    &\to 0.
\end{align}
In particular, the second-step smoothing-bias component requires
\begin{align*}
    \frac{nT}{J^{d+2\min\{s_\mu,\mathfrak{q}+1\}}}
    &\to 0,
\end{align*}
the same smoothing-bias restriction as in the pointwise $\beta=0$ theory.
Thus, $\mathfrak{q}=0$ cannot satisfy the sufficient rate conditions for factor-cancelling contrast inference.

\subsection{Evaluation-Point-Anchored Partitioning}

For factor-cancelling contrasts, the second-step partition is constructed so
that the evaluation points themselves are fixed marginal breakpoints. This is
the analogue of the split-at-zero construction in
Section~\ref{sec:classical-partition}: the cell containing each evaluation point
has a nonrandom index, because each coordinate of the evaluation point is
assigned to the subsample immediately to its right.
For notational simplicity, we state the construction under the convention that
the evaluation points have distinct coordinates in every dimension. If several
evaluation points share a coordinate in some dimension, the same construction
uses the sorted unique coordinates in that dimension; the number of fixed
breakpoints, and hence the number of marginal subsamples, is then smaller than
$K_{\Psi}$ and $K_{\Psi}+1$, respectively.

We describe the construction first. For each coordinate $l\in[d]$, order the
fixed evaluation-point coordinates as
\begin{align*}
    \lb_l
    <\mathsf{b}_{\pi_l(1),l}
    <\cdots
    <\mathsf{b}_{\pi_l(K_{\Psi}),l}
    <\ub_l,
\end{align*}
where $\pi_l$ is a permutation of $[K_{\Psi}]$. Set
$\gamma_{0,l}=\lb_l$, $\gamma_{r,l}=\mathsf{b}_{\pi_l(r),l}$ for
$r=1,\ldots,K_{\Psi}$, and $\gamma_{K_{\Psi}+1,l}=\ub_l$. These fixed
breakpoints split coordinate $l$ into the $K_{\Psi}+1$ subsamples
\begin{align*}
    \mathcal{I}_{r,l}
    &=
    [\gamma_{r,l},\gamma_{r+1,l}),
    \qquad r=0,\ldots,K_{\Psi}-1,\qquad
    \mathcal{I}_{K_{\Psi},l}
    =
    [\gamma_{K_{\Psi},l},\gamma_{K_{\Psi}+1,l}],
\end{align*}
with ties at a fixed breakpoint assigned to the interval on the right. Thus, if
$\mathsf{b}_{k,l}=\gamma_{r,l}$, then $\mathsf{b}_{k,l}\in\mathcal{I}_{r,l}$.

Within each subsample, form marginal empirical quantile-spaced intervals of the
generated regressor. Specifically, for $t=H+1,\ldots,T$, $l\in[d]$, and
$r=0,\ldots,K_{\Psi}$, let
\begin{align*}
    \widehat{N}_{t,l,r}
    &=
    \sum_{i=1}^{n_t}\IF(\widehat{\beta}_{it,l}\in\mathcal{I}_{r,l}),
    \\
    \widehat{F}_{\widehat{\beta},t,l,r}(z)
    &=
    \frac{1}{\widehat{N}_{t,l,r}}
    \sum_{i=1}^{n_t}
    \IF(\widehat{\beta}_{it,l}\in\mathcal{I}_{r,l})
    \IF(\widehat{\beta}_{it,l}\leq z),
    \qquad z\in\mathcal{I}_{r,l},
\end{align*}
on the event $\widehat{N}_{t,l,r}>0$, with generalized inverse on
$\mathcal{I}_{r,l}$. Given integers $J_{t,l,r}$, define the empirical
breakpoints by
\begin{align*}
    \hat{\tau}_{0t,l,r}
    &=
    \gamma_{r,l},
    \qquad
    \hat{\tau}_{J_{t,l,r}t,l,r}
    =
    \gamma_{r+1,l},
    \qquad
    \hat{\tau}_{jt,l,r}
    =
    \widehat{F}_{\widehat{\beta},t,l,r}^{-1}
    \!\left(\frac{j}{J_{t,l,r}}\right),
    \qquad
    j=1,\ldots,J_{t,l,r}-1.
\end{align*}
The marginal partition $\widehat{\Delta}_{t,l}$ is the ordered collection of
intervals generated by these breakpoints over $r=0,\ldots,K_{\Psi}$, and the
full partition is the tensor product
$\widehat{\Delta}_t=\otimes_{l\in[d]}\widehat{\Delta}_{t,l}$.

\begin{assumption}[PA-G][Evaluation-point-anchored partition]\label{ass:partition-generalized}\leavevmode
	$\mathsf{b}_1,\ldots,\mathsf{b}_{K_{\Psi}}$ are fixed evaluation points with
	distinct coordinates in each dimension and lie strictly inside
	$\mathcal{B}^\circ$. For every $t=H+1,\ldots,T$, the estimated partition
	$\widehat\Delta_t$ is formed by the
	evaluation-point-based marginal-quantile construction described above, with
	$J_{t,l,r}\asymp J$ uniformly over $t,l,r$, $J\rightarrow\infty$, and
	$J^d/n=o(1)$.
\end{assumption}

In this scenario, the reference partition is the conditional counterpart of the same construction.
For each $t,l,r$, define
\begin{align*}
    \bar{\pi}_{t,l,r}
    &=
    \frac{1}{n_t}\sum_{i=1}^{n_t}
    \P(\widehat{\beta}_{it,l}\in\mathcal{I}_{r,l}
    \mid\mathcal{D}_{t-1}),
    \\
    \varrho_{it,l,r}
    &=
    \frac{
    \P(\widehat{\beta}_{it,l}\in\mathcal{I}_{r,l}
    \mid\mathcal{D}_{t-1})
    }{\bar{\pi}_{t,l,r}},
    \\
    \bar{F}_{\widehat{\beta},t,l,r}(z)
    &=
    \frac{1}{n_t}\sum_{i=1}^{n_t}
    \varrho_{it,l,r}
    \P(\widehat{\beta}_{it,l}\leq z
    \mid
    \widehat{\beta}_{it,l}\in\mathcal{I}_{r,l},
    \mathcal{D}_{t-1}),
    \qquad z\in\mathcal{I}_{r,l}.
\end{align*}
Let 
\begin{align*}
    \tau_{0t,l,r}=\gamma_{r,l},\qquad
    \tau_{J_{t,l,r}t,l,r}=\gamma_{r+1,l}, \qquad
    \tau_{jt,l,r}
    &=
    \bar{F}_{\widehat{\beta},t,l,r}^{-1}
    \!\left(\frac{j}{J_{t,l,r}}\right),
    \qquad
    j=1,\ldots,J_{t,l,r}-1.
\end{align*}
The reference marginal partition $\Delta_{t,l}$ and tensor-product partition
$\Delta_t$ are formed from these conditional breakpoints. Cell normalizations,
{raw-regressor membership conventions}, and reference-basis definitions are the same as
in Section~\ref{sec:classical-partition}.

Under Assumption~\ref{ass:partition-generalized}, each evaluation point
$\mathsf{b}_k$ belongs to a cell with a deterministic tensor-product label. On
that cell, every coordinate of $\mathsf{b}_k$ is a fixed lower endpoint of both
the estimated and reference cell. Therefore the midpoint and side-length
normalization gives
\begin{align*}
	\widehat p_t(\mathsf{b}_k)
	&=
	p_t(\mathsf{b}_k),
	\qquad
	k=1,\ldots,K_{\Psi},
\end{align*}
for every polynomial degree $\mathfrak q\geq0$, exactly as the split-at-zero
construction gives $\widehat p_t(0)=p_t(0)$ in
Section~\ref{sec:classical-partition}.

\subsection{Result}

The relevant variance is the variance of the factor-cancelling idiosyncratic
leading term,
\begin{align*}
    \Sigma_{\varepsilon\Psi}
    &=
    \frac{1}{(T-H)^2}\sum_{t=1}^{T}\sum_{i=1}^{n_t}
    \varsigma_{it}^2W_{it,\Psi}^2.
\end{align*}
The bin-restricted jackknife from Section~\ref{sec:pointwise-variance}
extends by applying the same contrast to the delete-$i$ pseudo-values:
\begin{align*}
    \widehat{\Sigma}_{\varepsilon\Psi}
    &=
    \frac{1}{(T-H)^2}
    \sum_{i=1}^n\sum_{t=H+1}^T\sum_{s=H+1}^T
    \Delta_{it\Psi}\Delta_{is\Psi}\IF(|t-s|\leq H),
\end{align*}
where
\begin{align*}
    \Delta_{it\Psi}
    &=
    \sum_{k=1}^{K_{\Psi}}a_k\Delta_{it}(\mathsf{b}_k),
    \qquad
    \Delta_{it}(\beta)
    =
    \widehat{M}_t^{(-i)}(\beta)-\widehat{M}_t(\beta).
\end{align*}

\begin{theorem}[Factor-Cancelling Contrast Inference]\label{thm:factor-cancelling-inference}\leavevmode\par\noindent
    Let $K_{\Psi}$ be fixed, $\mathsf{b}_1,\ldots,\mathsf{b}_{K_{\Psi}}\in\mathcal{B}^\circ$ be fixed evaluation points, and $a_1,\ldots,a_{K_{\Psi}}$ be fixed contrast weights satisfying \eqref{eq:factor-cancelling-contrast}.
    Suppose Assumptions~\ref{ass:dgp}, \ref{ass:first-step-ls},
    \ref{ass:partition-generalized}, {\ref{ass:interior-separation}}, and
    {\ref{ass:first-step-bias}} hold, and that $s_\mu\geq2$.
    Assume that the rate conditions \eqref{eq:classical-linearization-rate}, \eqref{eq:jackknife-conditions}, and \eqref{eq:factor-cancelling-rate} hold, that
    \begin{align*}
        \frac{T^{2/q}}{H}
        +\sqrt{\frac{\log T}{H}}
        &=o(1),
    \end{align*}
    and assume further that $J\mathsf{R}_{\mathtt{FS}}=o(1)$ and $J^2\log T/n=o(1)$. If
    \begin{align*}
        \frac{nT}{J^d}\Sigma_{\varepsilon\Psi}
        &\gtrsim_\P
        1,
    \end{align*}
    then
    \begin{align*}
        \sup_{x\in\mathbb{R}}\left|
        \P\left(
        \frac{\widehat{\Psi}-\bar{\Psi}_T}{\sqrt{\Sigma_{\varepsilon\Psi}}}
        \leq x
        \right)
        -\Phi(x)\right|
        &\to 0,
    \end{align*}
    and
    \begin{align*}
        \widehat{\Sigma}_{\varepsilon\Psi}
        &=
        \Sigma_{\varepsilon\Psi}
        +o_\P\!\left(\frac{J^d}{nT}\right).
    \end{align*}
\end{theorem}

\subsection{Discussion}

\begin{enumerate}[label=(\roman*)]
    \item \emph{Confidence interval.}
    The theorem gives both the Gaussian approximation and relative consistency
    of $\widehat{\Sigma}_{\varepsilon\Psi}$ under the nondegeneracy condition.
    Therefore Slutsky's lemma implies that
    \begin{align*}
        \widehat{\Psi}
        &\pm
        z_{1-\alpha/2}\sqrt{\widehat{\Sigma}_{\varepsilon\Psi}}
    \end{align*}
    has asymptotic coverage $1-\alpha$.

    \item \emph{Butterfly spreads.}
    The butterfly contrast is the special case
    $\widehat{\Psi}=\widehat{\mu}(\mathsf{b}_1)-2\widehat{\mu}(\mathsf{b}_2)+\widehat{\mu}(\mathsf{b}_3)$ with $\mathsf{b}_2=(\mathsf{b}_1+\mathsf{b}_3)/2$. The factor component vanishes because
    $\mathsf{b}_1-2\mathsf{b}_2+\mathsf{b}_3=0$, so no factor variance estimator is needed.

    \item \emph{Pairwise differences.} The pairwise difference $\widehat{D}=\widehat{\mu}(\mathsf{b}_2)-\widehat{\mu}(\mathsf{b}_1)$ does not cancel the factor component:
    \begin{align*}
        \bar{\mathscr{S}}_{fD}
        &=
        \frac{1}{T-H}\sum_{t=H+1}^{T}
        (\mathsf{b}_2-\mathsf{b}_1)^\top\widetilde{f}_t.
    \end{align*}
    Its factor variance contribution is
    \begin{align*}
        \frac{1}{(T-H)^2}
        (\mathsf{b}_2-\mathsf{b}_1)^\top
        \sum_{t=H+1}^{T}
        \Var(\widetilde{f}_t\mid\mathcal{G}_{t-1})
        (\mathsf{b}_2-\mathsf{b}_1).
    \end{align*}
    A two-point difference with weights $(-1,1)$ cancels the factor term only in the degenerate case $\mathsf{b}_2=\mathsf{b}_1$, so a nontrivial two-point difference is outside Theorem~\ref{thm:factor-cancelling-inference}.

    \item \emph{Other difference-in-differences.}
    A quadruple contrast with weights $(1,-1,-1,1)$ also removes the factor
    component whenever $\mathsf{b}_1+\mathsf{b}_4=\mathsf{b}_2+\mathsf{b}_3$. It targets the difference-in-differences restriction
    \begin{align*}
        \bar{\mu}_T(\mathsf{b}_1;H)-\bar{\mu}_T(\mathsf{b}_2;H)
        &=
        \bar{\mu}_T(\mathsf{b}_3;H)-\bar{\mu}_T(\mathsf{b}_4;H).
    \end{align*}
    The theorem applies by taking $K_{\Psi}=4$ and the corresponding weights.
\end{enumerate}

\subsection{Proof of Theorem~\ref{thm:factor-cancelling-inference}}

By \eqref{eq:factor-cancelling-contrast}, the factor component in the pointwise
linearization cancels exactly, giving \eqref{eq:factor-cancelling-expansion}. Under
\eqref{eq:factor-cancelling-rate}, the residual satisfies
\begin{align*}
    |\mathscr{R}_{\Psi}|
    &=
    o_\P\!\left(\sqrt{\frac{J^d}{nT}}\right).
\end{align*}
It remains to verify the Gaussian approximation and the jackknife variance
consistency for the combined idiosyncratic weights.

The partition, Gram, and basis-switching bounds from
Section~\ref{sec:classical-partition} carry over under
Assumption~\ref{ass:partition-generalized}: the number of fixed subsamples is
finite, and the interior fixed breakpoints and density bounds give each
subsample positive conditional mass, so the marginal quantile-spacing and
cell-count arguments apply with constants depending only on $K_{\Psi}$.

Under Assumption~\ref{ass:partition-generalized}, the anchored construction
gives $\widehat p_t(\mathsf b_k)=p_t(\mathsf b_k)$ for every contrast point.
Thus the combined reference weights $W_{it,\Psi}$ are
$\mathcal E_\infty$-measurable by the same argument as in
Lemma~\ref{lem:bin-restricted-jackknife}. Conditional on
$\mathcal{E}_\infty$,
$\bar{\mathscr{S}}_{\varepsilon\Psi}$ is a sum of independent mean-zero
primitive shocks with weights $W_{it,\Psi}/(T-H)$. Since the number of contrast
points is fixed, the single-point local-support bounds, Jensen's inequality over
the rolling windows, and $T^{-1}\sum_{t=1}^T(1+\|f_t\|^3)=O_\P(1)$ imply
\begin{align*}
    \sum_{t=1}^{T}\sum_{i=1}^{n_t} W_{it,\Psi}^2
    &\lesssim_\P \frac{TJ^d}{n},\qquad
    \sum_{t=1}^{T}\sum_{i=1}^{n_t} |W_{it,\Psi}|^3
    &\lesssim_\P
    \frac{TJ^{2d}}{n^2}.
\end{align*}
The conditional Berry--Esseen bound therefore gives
\begin{align*}
    \sup_{x\in\mathbb{R}}\left|
    \P\left(
    \frac{\bar{\mathscr{S}}_{\varepsilon\Psi}}
    {\sqrt{\Sigma_{\varepsilon\Psi}}}\le x
    \,\Big|\,\mathcal{E}_\infty
    \right)-\Phi(x)\right|
    &\lesssim_\P
    \frac{T J^{2d}/n^2}
    {\big((T-H)^2\Sigma_{\varepsilon\Psi}\big)^{3/2}}
    =
    o_\P(1),
\end{align*}
where the last equality uses
$(nT/J^d)\Sigma_{\varepsilon\Psi}\gtrsim_\P1$ and $J^d/n=o(1)$. Removing the
conditioning gives the infeasible Gaussian approximation for
$\bar{\mathscr{S}}_{\varepsilon\Psi}$, and the residual bound above transfers it
to $\widehat{\Psi}-\bar{\Psi}_T$.

For the variance estimator, apply the proof of Lemma~\ref{lem:bin-restricted-jackknife}
to the combined delete-$i$ perturbation $\Delta_{it\Psi}$. Each diagonal,
off-diagonal, and remainder term in that proof is quadratic in the relevant
weights. Replacing $W_{it}(\beta)$ by the fixed finite sum
$W_{it,\Psi}=\sum_{k=1}^{K_{\Psi}}a_kW_{it}(\mathsf{b}_k)$ only changes the constants
in the second- and fourth-power sums and the squared-coefficient bound in that proof.
The same argument therefore yields
\begin{align*}
    \widehat{\Sigma}_{\varepsilon\Psi}
    &=
    \Sigma_{\varepsilon\Psi}
    +o_\P\!\left(\frac{J^d}{nT}\right).
\end{align*}
This proves the theorem. \qed

\section{Fixed-Grid Test of Linearity}
\label{sec:fixed-grid-linearity}

\subsection{Testing Problem}
We test whether the conditional mean $\E[R_{it}|\mathcal{F}_{t-1}]$ is linear in $\beta_{it}$ at every date. 
Specifically, the null hypothesis is
\begin{align}
    \mathsf{H}_{0}^{\mathtt{LIN}}:
    \qquad
    \mu_t(b)&=a_t+c_t^\top b,
    \qquad b\in\mathcal{B},\quad t=1,\ldots,T,
    \label{eq:forward-linearity-null}
\end{align}
for some $a_t$ and $c_t$ that may vary over time and are
$\mathcal G_{t-1}$-measurable. Under the null, realized returns satisfy
\begin{align*}
    R_{it}
    &=a_t+\varpi_t^\top\beta_{it}+\varepsilon_{it},
    \qquad
    \varpi_t=c_t+\widetilde f_t.
\end{align*}
Thus, the cross-sectional return slope is $\varpi_t$, not $c_t$: the linear
fit must absorb the realized factor innovation.

The test uses two adjacent first-step windows. The partition at date $t$ is
formed from $\widehat\beta_{it}$, whose window ends at $t-1$. The test outcome
is dated $t+H$, and $\widehat\beta_{i(t+H)}$ uses dates
$t,\ldots,t+H-1$. The two windows do not overlap. The linear fit uses a
generic instrument available before the future window; it need not be
$\widehat\beta_{it}$.

For $t=H+1,\ldots,T-H$, define the future-window conditioning field
\begin{align*}
    \mathcal K_t
    &=
    \mathcal E_\infty\vee
    \sigma\!\left(\eta_{jr}:j\in[n],\ r\leq t-1\right).
\end{align*}

To accommodate an unbalanced panel, for each base date
$t=H+1,\ldots,T-H$, let $\mathcal I_t^{\mathtt{LIN}}\subseteq[n]$ collect the
units observed throughout both rolling windows and at the outcome date
$t+H$, and write
$n_t^{\mathtt{LIN}}=|\mathcal I_t^{\mathtt{LIN}}|$. Thus both
$\widehat\beta_{it}$ and
$\widehat\beta_{i(t+H)}$, as well as $R_{i(t+H)}$, are available for every
$i\in\mathcal I_t^{\mathtt{LIN}}$. 
All date-$t$ cross-sectional objects in this section, including the partition and Gram
matrix, are constructed on $\mathcal I_t^{\mathtt{LIN}}$. We require
$\mathcal I_t^{\mathtt{LIN}}$ to be measurable with respect to the conditioning
field $\mathcal K_t$, so sample availability does not select on
the future primitive innovations, and assume that, for some fixed $c>0$,
\begin{align*}
    \min_{H+1\leq t\leq T-H}n_t^{\mathtt{LIN}}
    &\geq cn
\end{align*}
with probability approaching one. The DGP, partition, and Gram conditions
invoked below are transferred to these matched cross-sections through the
following random-missingness condition. For this purpose,
$\{(\alpha_{ir},\beta_{ir},\varsigma_{ir},\eta_{ir}):
i\in[n],\,r\in[T]\}$ is
understood as a latent full-data array satisfying
Assumption~\ref{ass:dgp} before selection. The auxiliary variables entering
$\mathcal E_\infty$ are also understood to include the innovation-free
availability process.

\begin{assumption}[LIN-M][Persistent random matched-sample availability]
\label{ass:linearity-missingness}\leavevmode
    Let
    $D_{it}^{\mathtt{LIN}}=\IF(i\in\mathcal I_t^{\mathtt{LIN}})$ and let
    $\mathcal Z_t^{\mathtt{LIN}}$ be the sigma-field generated by the
    full-data variables
    \begin{align*}
        \{f_r:t-H\leq r\leq t+H\}
        \quad\text{and}\quad
        \{(\alpha_{ir},\beta_{ir},\varsigma_{ir},\eta_{ir}):
        i\in[n],\ t-H\leq r\leq t+H\}.
    \end{align*}
    Uniformly over $t=H+1,\ldots,T-H$:
    \begin{enumerate}[label=(\roman*)]
        \item The availability vector
        $(D_{1t}^{\mathtt{LIN}},\ldots,D_{nt}^{\mathtt{LIN}})$ is
        $\mathcal K_t$-measurable. Given $\mathcal D_{t-1}$, this vector has
        mutually independent coordinates and is independent of
        $\mathcal Z_t^{\mathtt{LIN}}$.
        \item For some fixed $\pi_0>0$,
            $\min_{i\in[n]}
            \P(D_{it}^{\mathtt{LIN}}=1\mid\mathcal D_{t-1})
            \geq\pi_0$ 
        almost surely.
    \end{enumerate}
\end{assumption}

Define the matched design field
\begin{align*}
    \mathcal D_{t-1}^{\mathtt{LIN}}
    &=
    \mathcal D_{t-1}\vee
    \sigma(D_{1t}^{\mathtt{LIN}},\ldots,D_{nt}^{\mathtt{LIN}}).
\end{align*}
Conditional on this field, the retained old-window predictable histories
and old first-step estimates remain independent over $i$ and retain their
original conditional laws. Future-window primitive shocks remain independent
conditional on $\mathcal K_t$. Although
$\mathcal D_{t-1}^{\mathtt{LIN}}$ need not be contained in
$\mathcal F_{t-1}$, it is contained in $\mathcal K_t$. The matched versions
of Lemmas~\ref{lem:reference-quantile-spacing},
\ref{lem:quantile-convergence}, and
\ref{lem:partition-basis-moments} therefore follow by the same conditional
arguments, with $n_t$ replaced by $n_t^{\mathtt{LIN}}$ and with the
reference partitions conditioning on
$\mathcal D_{t-1}^{\mathtt{LIN}}$. Whenever a sum is written over the common
index set $[n]$, the contribution of unit $i$ is set to zero if
$i\notin\mathcal I_t^{\mathtt{LIN}}$.

\subsection{IV Fit and Test Statistic}

For $t=H+1,\ldots,T-H$, define the future regressor
\begin{align*}
    \widehat x_{i(t+H)}
    &=(1,\widehat\beta_{i(t+H)}^\top)^\top.
\end{align*}
Let $\zeta_{it}\in\mathbb R^{d+1}$ be a generic instrument, whose first coordinate is one to accommodate the intercept. Set
\begin{align}
    \widehat{\mathsf M}_t
    =
    \frac{1}{n_t^{\mathtt{LIN}}}
    \sum_{i\in\mathcal I_t^{\mathtt{LIN}}}
    \zeta_{it}\widehat x_{i(t+H)}^\top,
    \quad
    \mathsf M_t^\circ
    =
    \E[\widehat{\mathsf M}_t\mid\mathcal K_t].
    \label{eq:forward-instrument-matrices}
\end{align}

We impose the following high-level conditions on the instrument.
\begin{assumption}[LIN-IV][High-level instrument]\label{ass:linearity-instrument}\leavevmode
    The following conditions hold uniformly over
    $t=H+1,\ldots,T-H$:
    \begin{enumerate}[label=(\roman*)]
        \item \textbf{Predetermination.}
        $\zeta_{it}$ is $\mathcal F_{t-1}$-measurable for every
        $i\in\mathcal I_t^{\mathtt{LIN}}$, and hence is
        $\mathcal K_t$-measurable. 

        \item \textbf{Moment and leverage regularity.}
        \begin{align*}
            \max_{H+1\leq t\leq T-H}
            \frac{1}{n_t^{\mathtt{LIN}}}
            \sum_{i\in\mathcal I_t^{\mathtt{LIN}}}\|\zeta_{it}\|^4
            &\lesssim_\P1,
            &
            \max_{H+1\leq t\leq T-H}
            \max_{i\in\mathcal I_t^{\mathtt{LIN}}}\|\zeta_{it}\|
            &\lesssim_\P1.
        \end{align*}

        \item \textbf{Relevance and concentration.}
        \begin{align*}
            \max_{H+1\leq t\leq T-H}
            \left\{\|\mathsf M_t^\circ\|+\|(\mathsf M_t^\circ)^{-1}\|\right\}
            &\lesssim_\P1,
            &
            \max_{H+1\leq t\leq T-H}
            \|\widehat{\mathsf M}_t-\mathsf M_t^\circ\|
            &\lesssim_\P
            \sqrt{\frac{\log(nT)}{n}}.
        \end{align*}
    \end{enumerate}
\end{assumption}

Predetermination is the exclusion condition. 
It ensures that the instrument is fixed before the innovations entering
$\widehat\beta_{i(t+H)}$ and $R_{i(t+H)}$ are observed.
Moment and leverage regularity control both aggregate moments and
cell-specific instrument leverage in the covariance proof. The first bound in part~(iii) is
the relevance condition
for the intercept and realized cross-sectional slope; the second is a
high-level uniform cross-sectional concentration condition on the sample
instrument--regressor matrix.

One natural choice is
$\zeta_{it}=(1,\widehat\beta_{it}^\top)^\top$.
Lemma~\ref{lem:forward-old-beta-instrument} below verifies
Assumption~\ref{ass:linearity-instrument} for this choice under an oracle
old--future beta relevance condition and the primitive first-step
conditions. Other predetermined characteristics or generated instruments
may be used whenever Assumption~\ref{ass:linearity-instrument} holds.

Given this generic instrument $\zeta_{it}$, define
\begin{align}
    \widehat\theta_{t+H}^{\mathtt{IV}}
    &=
    \widehat{\mathsf M}_t^{-1}
    \frac{1}{n_t^{\mathtt{LIN}}}
    \sum_{i\in\mathcal I_t^{\mathtt{LIN}}}
    \zeta_{it}R_{i(t+H)}.
    \label{eq:forward-iv-fit}
\end{align}
Write
$\widehat\theta_{t+H}^{\mathtt{IV}}
=(\widehat a_{t+H},\widehat\varpi_{t+H}^\top)^\top$ and
\begin{align*}
    \widehat e_{i(t+H)}
    &=
    R_{i(t+H)}
    -\widehat x_{i(t+H)}^\top
    \widehat\theta_{t+H}^{\mathtt{IV}},
    \qquad i\in\mathcal I_t^{\mathtt{LIN}},\quad
    t=H+1,\ldots,T-H.
\end{align*}
Set $\widehat e_{i(t+H)}=0$ when
$i\notin\mathcal I_t^{\mathtt{LIN}}$.

For any evaluation point $b\in\mathcal B^\circ$, define the projection weight
\begin{align*}
    \widehat w_{it}(b)
    &=
    \frac{1}{n_t^{\mathtt{LIN}}}\widehat p_t(b)^\top
    \widehat Q_t^{-1}\widehat p_t(\widehat\beta_{it}).
\end{align*}
Set $\widehat w_{it}(b)=0$ when
$i\notin\mathcal I_t^{\mathtt{LIN}}$.
When $\mathfrak q=0$, this is the usual equal weight in the cell containing
$b$. The test statistic is
\begin{align}
    \widehat\nu(b)
    &=
    \frac{1}{T-2H}
    \sum_{t=H+1}^{T-H}\sum_{i\in\mathcal I_t^{\mathtt{LIN}}}
    \widehat w_{it}(b)\widehat e_{i(t+H)}.
    \label{eq:forward-linearity-statistic}
\end{align}
More generally, for distinct evaluation points
$\mathsf{b}_1,\ldots,\mathsf{b}_{K_{\mathtt{LIN}}}\in\mathcal B^\circ$ 
with $K_{\mathtt{LIN}}$ fixed, define
\begin{align*}
    \widehat{\boldsymbol\nu}_{\mathtt{LIN}}
    &=
    \big(
    \widehat\nu(\mathsf{b}_1),\ldots,
    \widehat\nu(\mathsf{b}_{K_{\mathtt{LIN}}})
    \big)^\top.
\end{align*}
Let $\Sigma_{\mathtt{LIN}}$ denote the predictable covariance matrix of the
fixed-grid leading term under the null. The variance-stabilization condition
below makes it asymptotically equivalent to an
$\mathcal E_\infty$-measurable reference covariance
$\overline\Sigma_{\mathtt{LIN}}$. Let
$\widehat\Sigma_{\mathtt{LIN}}$ denote the feasible estimator of the
predictable covariance; these objects are defined explicitly below. The Wald
test statistic is
\begin{align}
    \mathsf W_{\mathtt{LIN}}
    &:=
    \widehat{\boldsymbol\nu}_{\mathtt{LIN}}^\top
    \widehat\Sigma_{\mathtt{LIN}}^{-1}
    \widehat{\boldsymbol\nu}_{\mathtt{LIN}}.
    \label{eq:forward-linearity-wald}
\end{align}
Its null validity is established in
Theorem~\ref{thm:forward-linearity}.

\subsection{Decomposition}

From~\eqref{eq:first-step-decomposition}, write the centered first-step error
as
\begin{align*}
    v_{it}
    &=\widehat\beta_{it}-\breve\beta_{it}
    =\frac{1}{H}\sum_{s=1}^H w_s(t)\varepsilon_{i(t-s)}.
\end{align*}
Under the null, let $\theta_t=(a_t,\varpi_t^\top)^\top$. The future residual
has the exact decomposition
\begin{align}
    \widehat e_{i(t+H)}
    &=
    \varepsilon_{i(t+H)}
    -\varpi_{t+H}^\top v_{i(t+H)}
    -\varpi_{t+H}^\top
    (\breve\beta_{i(t+H)}-\beta_{i(t+H)})
    \notag\\
    &\quad
    -\widehat x_{i(t+H)}^\top
    (\widehat\theta_{t+H}^{\mathtt{IV}}-\theta_{t+H}).
    \label{eq:forward-residual-decomposition}
\end{align}
Substitution into~\eqref{eq:forward-linearity-statistic} gives
\begin{align}
    \widehat\nu(b)
    &=\mathscr S_{\mathtt{LIN}}(b)
    +\mathscr B_{\mathtt{LIN}}(b)
    +\mathscr R_{\mathtt{LIN}}(b),
    \label{eq:forward-linearity-decomposition}
\end{align}
where
\begin{align*}
    \mathscr S_{\mathtt{LIN}}(b)
    &=
    \frac{1}{T-2H}
    \sum_{t=H+1}^{T-H}\sum_{i\in\mathcal I_t^{\mathtt{LIN}}}
    \widehat w_{it}(b)
    \left[
    \varepsilon_{i(t+H)}
    -\varpi_{t+H}^\top v_{i(t+H)}
    \right],
    \\
    \mathscr B_{\mathtt{LIN}}(b)
    &=
    -\frac{1}{T-2H}
    \sum_{t=H+1}^{T-H}\sum_{i\in\mathcal I_t^{\mathtt{LIN}}}
    \widehat w_{it}(b)\varpi_{t+H}^\top
    (\breve\beta_{i(t+H)}-\beta_{i(t+H)}),
    \\
    \mathscr R_{\mathtt{LIN}}(b)
    &=
    -\frac{1}{T-2H}
    \sum_{t=H+1}^{T-H}
    \left(
    \sum_{i\in\mathcal I_t^{\mathtt{LIN}}}\widehat w_{it}(b)
    \widehat x_{i(t+H)}^\top
    \right)
    (\widehat\theta_{t+H}^{\mathtt{IV}}-\theta_{t+H}).
\end{align*}
The first term is centered and determines the limiting distribution. The
second is the smooth first-step bias. The third is the error from the IV fit.

Note that the forward split removes the leave-in bias from the first term.
By Assumption~\ref{ass:linearity-missingness}(i), the matched set is also
fixed conditional on $\mathcal K_t$. Thus $\widehat w_{it}(b)$ and
$\zeta_{it}$ are fixed, while
$v_{i(t+H)}$ uses innovations dated $t,\ldots,t+H-1$ and
$\varepsilon_{i(t+H)}$ is dated $t+H$. Hence
\begin{align}
    \E\!\left[
    \widehat w_{it}(b)v_{i(t+H)}
    \mid\mathcal K_t
    \right]
    &=0,
    \\
    \E\!\left[
    \widehat w_{it}(b)\varepsilon_{i(t+H)}
    \mid\mathcal K_t
    \right]
    &=0.
    \label{eq:forward-no-leave-in}
\end{align}
By Assumption~\ref{ass:linearity-instrument}(i), the same argument gives
\begin{align}
    \E\!\left[
    \zeta_{it}v_{i(t+H)}^\top
    \mid\mathcal K_t
    \right]=0,
    \qquad
    \E\!\left[
    \zeta_{it}\varepsilon_{i(t+H)}
    \mid\mathcal K_t
    \right]
    =0.
    \label{eq:forward-instrument-exogeneity}
\end{align}
Therefore the generic IV score is centered at the true realized slope, up to
the smooth first-step bias.

For the limiting distribution, set $\widehat w_{it}(b)=0$ when
$t\notin\{H+1,\ldots,T-H\}$ or
$i\notin\mathcal I_t^{\mathtt{LIN}}$. Re-indexing the shocks in
$\mathscr S_{\mathtt{LIN}}(b)$ gives
\begin{align}
    \mathscr S_{\mathtt{LIN}}(b)
    &=
    \frac{1}{T-2H}
    \sum_{r=1}^{T}\sum_{i=1}^{n}
    W_{ir}(b)\varepsilon_{ir},
    \label{eq:forward-leading-sum}
\end{align}
where
\begin{align}
    W_{ir}(b)
    &=
    \widehat w_{i,r-H}(b)
    -\frac{1}{H}\sum_{s=1}^H
    \widehat w_{i,r+s-H}(b)
    \varpi_{r+s}^\top w_s(r+s).
    \label{eq:forward-primitive-weight}
\end{align}
Invalid terms are set to zero. Every coefficient on $\varepsilon_{ir}$ uses
a partition whose first-step window ends before $r$. Hence $W_{ir}(b)$ is
predictable. For $b,b'\in\mathcal B^\circ$, define
\begin{align}
    \Sigma_{\mathtt{LIN}}(b,b')
    &=
    \frac{1}{(T-2H)^2}
    \sum_{r=1}^{T}\sum_{i=1}^{n}
    \varsigma_{ir}^2W_{ir}(b)W_{ir}(b').
    \label{eq:forward-linearity-variance}
\end{align}
Let
\begin{align*}
    \Sigma_{\mathtt{LIN}}
    &=
    \big(
    \Sigma_{\mathtt{LIN}}(\mathsf{b}_k,\mathsf{b}_l)
    \big)_{k,l\in[K_{\mathtt{LIN}}]},
\end{align*}
which is the terminal predictable covariance of the fixed-grid leading term
under the null.

\subsection{Results}

The first lemma gives the infeasible fixed-grid Gaussian limit for a
generic instrument satisfying Assumption~\ref{ass:linearity-instrument}.

\begin{lemma}[Forward-Split Asymptotic Normality]
\label{lem:forward-linearity-normality}
    Suppose Assumptions~\ref{ass:dgp}, \ref{ass:first-step-ls},
    \ref{ass:partition}, \ref{ass:linearity-missingness},
    \ref{ass:linearity-instrument},
    {\ref{ass:interior-separation}}, and
    \ref{ass:first-step-bias} hold.
    Assume further that:
    \begin{enumerate}[label=(\roman*)]
        \item $T^{(3+d)/q}/H+\sqrt{\log T/H}=o(1)$.

        \item $J^d\log(nT)/n=o(1)$.

        \item The following rate restrictions hold:
        \begin{align}
            \frac{T-2H}{nH}
            &=o(1),
            \label{eq:forward-linearity-rough-multiplier-rate}
            \\
            \sqrt{\frac{n(T-2H)}{J^d}}
            \left(
            {\mathsf{B}_{\mathtt{FS}}}
            +\frac{\log(nT)}{n}
            \right)
            &=o(1).
            \label{eq:forward-linearity-remainder-rate}
        \end{align}

        \item \emph{Variance stabilization.} There exists an
        $\mathcal E_\infty$-measurable positive-definite matrix
        $\overline\Sigma_{\mathtt{LIN}}$ such that
        \begin{align}
            \lambda_{\min}\!\left(
            \frac{n(T-2H)}{J^d}
            \overline\Sigma_{\mathtt{LIN}}
            \right)
            &\gtrsim_\P1,
            &
            \lambda_{\max}\!\left(
            \frac{n(T-2H)}{J^d}
            \overline\Sigma_{\mathtt{LIN}}
            \right)
            &\lesssim_\P1,
            \label{eq:forward-linearity-nondegeneracy}
            \\
            \left\|
            \overline\Sigma_{\mathtt{LIN}}^{-1/2}
            \Sigma_{\mathtt{LIN}}
            \overline\Sigma_{\mathtt{LIN}}^{-1/2}
            -I_{K_{\mathtt{LIN}}}
            \right\|
            &=o_\P(1).
            \label{eq:forward-linearity-variance-stabilization}
        \end{align}
    \end{enumerate}
    Then under the null hypothesis \eqref{eq:forward-linearity-null},
    \begin{align*}
        \Sigma_{\mathtt{LIN}}^{-1/2}
        \widehat{\boldsymbol\nu}_{\mathtt{LIN}}
        &\rightsquigarrow
        \mathcal N(0,I_{K_{\mathtt{LIN}}}).
    \end{align*}
\end{lemma}

For feasible inference, define
\begin{align}
    \widehat\Sigma_{\mathtt{LIN}}(b,b')
    &=
    \frac{1}{(T-2H)^2}
    \sum_{i=1}^{n}
    \sum_{t=H+1}^{T-H}\sum_{s=H+1}^{T-H}
    \widehat w_{it}(b)\widehat w_{is}(b')
    \widehat e_{i(t+H)}\widehat e_{i(s+H)}
    \IF(|t-s|\leq H),
    \label{eq:forward-linearity-hac}
\end{align}
and set
\begin{align*}
    \widehat\Sigma_{\mathtt{LIN}}
    &=
    \big(
    \widehat\Sigma_{\mathtt{LIN}}(\mathsf{b}_k,\mathsf{b}_l)
    \big)_{k,l\in[K_{\mathtt{LIN}}]}.
\end{align*}
The outer sum in~\eqref{eq:forward-linearity-hac} uses the common index set
$[n]$; the zero-weight and zero-residual conventions remove a unit whenever
it is unavailable at either base date.
The truncation at $H$ is exact for the
leading term because two forward residuals share a primitive idiosyncratic
shock only when their base dates are at most $H$ periods apart.

The next lemma shows that this feasible variance estimator is consistent.
\begin{lemma}[Feasible Variance Consistency]
\label{lem:forward-linearity-variance}
    Suppose Assumptions~\ref{ass:dgp}, \ref{ass:first-step-ls},
    \ref{ass:partition}, \ref{ass:linearity-missingness},
    \ref{ass:linearity-instrument},
    {\ref{ass:interior-separation}}, \ref{ass:first-step-bias}, and conditions
    \textnormal{(i)}--\textnormal{(ii)} of Lemma~\ref{lem:forward-linearity-normality} hold. 
    Also, assume
    \begin{align}
        \frac{H\log(nT)}{n}
        &=o(1).
        \label{eq:forward-linearity-variance-rates}
    \end{align}
    Then under the null hypothesis \eqref{eq:forward-linearity-null},
    \begin{align}
        \|\widehat\Sigma_{\mathtt{LIN}}
        -\Sigma_{\mathtt{LIN}}\|
        &=o_\P\!\left(\frac{J^d}{n(T-2H)}\right).
        \label{eq:forward-linearity-variance-consistency}
    \end{align}
\end{lemma}

The next theorem establishes the validity of the Wald test under the null.
\begin{theorem}[Validity of the Forward-Split Wald Test]
\label{thm:forward-linearity}
    Suppose the conditions of
    Lemma~\ref{lem:forward-linearity-normality} hold and
    \eqref{eq:forward-linearity-variance-rates} holds. Then
    $\widehat\Sigma_{\mathtt{LIN}}$ is nonsingular with probability
    approaching one and the statistic in
    \eqref{eq:forward-linearity-wald} satisfies
    $\mathsf W_{\mathtt{LIN}}
    \rightsquigarrow\chi^2_{K_{\mathtt{LIN}}}$ under the null hypothesis
    \eqref{eq:forward-linearity-null}.
    Consequently, the test that rejects when
    $\mathsf W_{\mathtt{LIN}}>
    \chi^2_{K_{\mathtt{LIN}},1-\alpha}$
    has asymptotic size $\alpha$.
\end{theorem}

Finally, the following lemma verifies Assumption~\ref{ass:linearity-instrument} for the natural choice 
of instrument $\zeta_{it}=(1,\widehat\beta_{it}^\top)^\top$ under the primitive
first-step conditions and an oracle relevance condition.
\begin{lemma}[Old-Beta Instrument Verification]
\label{lem:forward-old-beta-instrument}
    Let
    \begin{align*}
        x_{it}
        =(1,\beta_{it}^\top)^\top,
        \qquad
        \mathsf M_{t,\beta}^0
        =
        \frac{1}{n_t^{\mathtt{LIN}}}
        \sum_{i\in\mathcal I_t^{\mathtt{LIN}}}
        x_{it}x_{i(t+H)}^\top.
    \end{align*}
    Suppose Assumptions~\ref{ass:dgp}, \ref{ass:first-step-ls},
    {\ref{ass:first-step-bias}}, and
    \ref{ass:linearity-missingness} hold,
    \begin{align*}
        \frac{T^{2/q}}{H}
        +\sqrt{\frac{\log T}{H}}
        =o(1),
        \qquad
        \mathsf R_{\mathtt{FS}}
        +\frac{\log(nT)}{n}
        =o(1),
    \end{align*}
    and the oracle old--future beta cross-moment is uniformly relevant:
    \begin{align}
        \max_{H+1\leq t\leq T-H}
        \left\{
        \|\mathsf M_{t,\beta}^0\|
        +\|(\mathsf M_{t,\beta}^0)^{-1}\|
        \right\}
        &\lesssim_\P1.
        \label{eq:forward-old-beta-relevance}
    \end{align}
    Then the choice
    \begin{align*}
        \zeta_{it}
        &=(1,\widehat\beta_{it}^\top)^\top
    \end{align*}
    satisfies Assumption~\ref{ass:linearity-instrument}.
\end{lemma}

Condition~\eqref{eq:forward-old-beta-relevance} isolates the only substantive extra requirement for the old-beta choice: the predetermined beta must retain enough cross-sectional information about the future beta. Predetermination alone does not imply this cross-time relevance.

\begin{remark}[Linear specification for excess returns]
\label{rem:forward-linearity-excess-returns}
When $R_{it}$ denotes an excess return, the absence of arbitrage imposes the stronger zero-intercept restriction
\begin{align*}
    \mu_t(b)
    &=
    c_t^\top b,
    \qquad
    b\in\mathcal B,\quad t=1,\ldots,T.
\end{align*}
The corresponding test is obtained by omitting the intercept from the period-by-period cross-sectional IV fit. Specifically, set
$\widehat x_{i(t+H)}=\widehat\beta_{i(t+H)}$ and use a $d$-dimensional predetermined instrument $\zeta_{it}$ without a constant, with
$\zeta_{it}=\widehat\beta_{it}$ as the natural choice. Define
\begin{align*}
    \widehat{\mathsf M}_t
    &=
    \frac{1}{n_t^{\mathtt{LIN}}}
    \sum_{i\in\mathcal I_t^{\mathtt{LIN}}}
    \zeta_{it}\widehat\beta_{i(t+H)}^\top,
    \\
    \widehat\varpi_{t+H}^{\mathtt{IV}}
    &=
    \widehat{\mathsf M}_t^{-1}
    \frac{1}{n_t^{\mathtt{LIN}}}
    \sum_{i\in\mathcal I_t^{\mathtt{LIN}}}
    \zeta_{it}R_{i(t+H)},
\end{align*}
and construct the residuals as
\begin{align*}
    \widehat e_{i(t+H)}
    &=
    R_{i(t+H)}
    -
    \widehat\beta_{i(t+H)}^\top
    \widehat\varpi_{t+H}^{\mathtt{IV}}.
\end{align*}
The fixed-grid moments, feasible covariance estimator, and Wald statistic are then constructed as above using these residuals. Under the corresponding $d$-dimensional version of Assumption~\ref{ass:linearity-instrument}, including nonsingularity of the $d\times d$ instrument--regressor matrix, and the remaining conditions of Theorem~\ref{thm:forward-linearity}, the same argument yields
\[
    \mathsf W_{\mathtt{LIN}}
    \rightsquigarrow
    \chi^2_{K_{\mathtt{LIN}}}.
\]
\end{remark}

\subsection{Discussion}

\begin{enumerate}[label=(\roman*)]
    \item \emph{Realized versus conditional slope.}
    The fitted slope estimates $\varpi_{t+H}=c_{t+H}+\widetilde f_{t+H}$.
    Estimating only $c_{t+H}$ would leave the realized factor component in the
    residual and would not give an idiosyncratic-scale test.

    \item \emph{Why IV is necessary.}
    The forward split makes $v_{i(t+H)}$ independent of the old basis, but this
    fact alone does not make OLS valid. 
    Specifically, for the OLS estimator of $\theta_{t+H}$ from regressing 
    $R_{i(t+H)}$ on $\widehat{x}_{i(t+H)}$, 
    its population score at the true coefficient contains a component
    \begin{align*}
        \left\|\E\!\left[
        v_{i(t+H)}v_{i(t+H)}^\top
        \mid\mathcal E_\infty
        \right]\varpi_{t+H}\right\|
        &=O_\P(H^{-1}).
    \end{align*}
    Thus OLS would add an $O_\P(H^{-1})$ attenuation bias term to
    $\mathscr R_{\mathtt{LIN}}(b)$ even though $v_{i(t+H)}$ is independent of the
    old partition.    
    Any instrument satisfying
    Assumption~\ref{ass:linearity-instrument} removes this term. The old beta
    estimate is one example because its first-step window does not overlap the
    future window.

    \item \emph{A primitive verification of old--future relevance.}
    Condition~\eqref{eq:forward-old-beta-relevance} can be verified under
    genuine sample-path smoothness of the beta process and conditionally
    random missingness. Suppose, for example, that there are possibly random
    paths $B_{iT}:[0,1]\to\mathcal B$ such that
    $\beta_{it}=B_{iT}(t/T)$ and, for some $s_\beta>0$, the paths are
    H\"older-$s_\beta$ on $[0,1]$ uniformly over $i\in[n]$, with their
    uniform bounds and H\"older constants bounded in probability.
    Then, since $H/T=o(1)$,
    \begin{align*}
        \delta_{\beta,H}
        &:=
        \max_{H+1\leq t\leq T-H}
        \max_{i\in\mathcal I_t^{\mathtt{LIN}}}
        \|\beta_{i(t+H)}-\beta_{it}\|
        \lesssim_\P
        \left(\frac{H}{T}\right)^{s_\beta\wedge1}
        =o_\P(1).
    \end{align*}
    Define the old-beta Gram matrix on the matched cross-section by
    \begin{align*}
        Q_{t,\beta}^{0}
        &=
        \frac{1}{n_t^{\mathtt{LIN}}}
        \sum_{i\in\mathcal I_t^{\mathtt{LIN}}}
        x_{it}x_{it}^\top.
    \end{align*}
    Compactness of $\mathcal B$ gives
    \begin{align*}
        \max_{H+1\leq t\leq T-H}
        \|\mathsf M_{t,\beta}^0-Q_{t,\beta}^{0}\|
        &\lesssim_\P \delta_{\beta,H}
        =o_\P(1).
    \end{align*}
    Under Assumption~\ref{ass:linearity-missingness}, conditioning on the
    realized matched set preserves the original conditional beta-density
    bounds and cross-sectional independence. The density lower bound on the
    full-dimensional rectangle $\mathcal B$,
    $n_t^{\mathtt{LIN}}\asymp n$, and uniform cross-sectional concentration
    under $\log(nT)/n=o(1)$ therefore imply
    \begin{align*}
        \min_{H+1\leq t\leq T-H}
        \lambda_{\min}(Q_{t,\beta}^{0})
        &\gtrsim_\P1,
        &
        \max_{H+1\leq t\leq T-H}
        \|Q_{t,\beta}^{0}\|
        &\lesssim_\P1.
    \end{align*}
    Singular-value perturbation therefore yields
    \begin{align*}
        \max_{H+1\leq t\leq T-H}
        \left\{
        \|\mathsf M_{t,\beta}^0\|
        +\|(\mathsf M_{t,\beta}^0)^{-1}\|
        \right\}
        &\lesssim_\P1,
    \end{align*}
    which is~\eqref{eq:forward-old-beta-relevance}. We retain the high-level
    relevance condition in Lemma~\ref{lem:forward-old-beta-instrument}
    because it also accommodates other persistence and missingness mechanisms.

    \item \emph{Generated partition.}
    The proof conditions on the old generated partition. It does not compare
    that partition with a true-beta partition. The basis-switching restriction
    $J\mathsf R_{\mathtt{FS}}=o(1)$ from the contemporaneous theory is therefore
    not needed for null size.

    \item \emph{Rate conditions.}
    Consider the proposed MSE-optimal tuning choices in
    Part~\ref*{applied-part:tuning}. For $(\mathfrak p,\mathfrak q)=(0,0)$, they give
    \begin{align*}
        h&\asymp T^{-1/3},
        &
        H&\asymp T^{2/3},
        &
        J&\asymp n^{1/(d+2)}.
    \end{align*}
    The bias component of~\eqref{eq:forward-linearity-remainder-rate} then
    requires
    \begin{align}
        J\sqrt T\,{\mathsf{B}_{\mathtt{FS}}}&=o(1),
        \qquad\text{or equivalently}\qquad
        {\mathsf{B}_{\mathtt{FS}}}
        =o\!\left(n^{-1/(d+2)}T^{-1/2}\right).
        \label{eq:forward-linearity-mse-bias-rate}
    \end{align}
    The implications depend on the order of the first-step local polynomial.

    \smallskip\noindent\emph{Local-constant first step.}
    With a nonzero leading bias, {if
    $\mathsf{B}_{\mathtt{FS}}\asymp h\asymp T^{-1/3}$, then}
    $J\sqrt T\,{\mathsf{B}_{\mathtt{FS}}}\asymp JT^{1/6}$ does not vanish, so
    the proposed MSE-optimal bandwidth does not satisfy
    \eqref{eq:forward-linearity-mse-bias-rate}.

    \smallskip\noindent\emph{Local-linear first step.}
    If $s\geq2$ {and
    $\mathsf{B}_{\mathtt{FS}}\lesssim h^2\asymp T^{-2/3}$}, write
    $T=n^\eta$, suppress logarithmic factors, and maintain the high-level
    instrument and covariance nondegeneracy conditions. The bias restriction
    gives the lower bound $\eta>6/(d+2)$, while the other Gaussian
    approximation rates give the upper bound
    $\eta<2(d+1)/(d+2)$. Thus
    \begin{align*}
        \frac{6}{d+2}<\eta<\frac{2(d+1)}{d+2}.
    \end{align*}
    For the feasible Wald statistic,
    \eqref{eq:forward-linearity-variance-rates} further requires
    $\eta<3/2$. The resulting window is
    \begin{align}
        \frac{6}{d+2}<\eta<
        \min\left\{\frac{2(d+1)}{d+2},\frac32\right\}.
        \label{eq:forward-linearity-mse-window}
    \end{align}
    This window is nonempty when $d>2$. It contains a subrange with
    $\eta<1$, and hence permits $T\ll n$, only when $d>4$.

    \smallskip\noindent\emph{Local-quadratic first step.}
    If $s\geq3$ {and
    $\mathsf{B}_{\mathtt{FS}}\lesssim h^3\asymp T^{-1}$}, the bias restriction
    now gives $\eta>2/(d+2)$, so the feasible window becomes
    \begin{align*}
        \frac{2}{d+2}<\eta<
        \min\left\{\frac{2(d+1)}{d+2},\frac32\right\}.
    \end{align*}
    This window is nonempty for every $d\geq1$. In particular, when $d=1$,
    it contains the subrange $2/3<\eta<1$, which permits $T\ll n$.
    Thus, under the proposed MSE-optimal bandwidth, a higher-order first step is
    needed to accommodate $T\ll n$ in the lower-dimensional cases.

    \item \emph{Variance stabilization.}
    Condition~\eqref{eq:forward-linearity-variance-stabilization} concerns the
    full rolling covariance functional, not pointwise convergence of the
    generated weights. One verification separates dates at which each grid
    point is away from reference-cell boundaries from boundary-near dates:
    the former have a stable cell label, while the latter are negligible
    under a time-averaged margin condition or are covered by local equivalence
    of the adjacent-cell rolling covariance kernels. Alternatively, the
    evaluation-point-anchored construction in
    Section~\ref{sec:factor-cancelling}, particularly
    Assumption~\ref{ass:partition-generalized}, fixes each grid point's cell
    label and basis coordinate; conditional quantile and Gram concentration
    can then control the remaining empirical-to-reference covariance
    difference. We retain the high-level condition here to accommodate either
    route.

    \item \emph{Power against fixed persistent alternatives.}
    Let $\boldsymbol\nu_{0,T,\mathtt{LIN}}$ denote the time-averaged
    population forward moment, after the population cross-time IV linear fit,
    evaluated on the fixed grid. If
    $\boldsymbol\nu_{0,T,\mathtt{LIN}}\to_\P
    \boldsymbol\nu_{0,\mathtt{LIN}}\neq0$ and
    $\widehat{\boldsymbol\nu}_{\mathtt{LIN}}
    -\boldsymbol\nu_{0,T,\mathtt{LIN}}=o_\P(1)$, while
    $\widehat\Sigma_{\mathtt{LIN}}$ is positive definite with probability
    approaching one and
    $\lambda_{\max}(\widehat\Sigma_{\mathtt{LIN}})=o_\P(1)$, then
    $\mathsf W_{\mathtt{LIN}}\to_\P\infty$ and the test is consistent.
    The nonzero-moment condition requires persistence and rules out
    cancellation over dates and grid points; without it, a nonlinear fixed
    alternative need not be detected.

    \item \emph{Scope of the test.}
    Taking $K_{\mathtt{LIN}}=1$ gives a pointwise specification test; fixed
    $K_{\mathtt{LIN}}$ gives the Wald test above. A diverging-grid omnibus test
    requires a separate high-dimensional Gaussian approximation. The test
    uses an exactly identified $(d+1)$-vector of instruments.
    An overidentified specification can be accommodated after replacing
    \eqref{eq:forward-iv-fit} by the corresponding GMM fit and imposing the
    analogous Jacobian and weight-matrix conditions.

        \item \emph{A fixed-period OLS alternative.}
    At one prespecified date $t>H$, one could instead compare the
    degree-one beta-sorted fit with the same-date full-cross-sectional OLS fit.
    Degree-one reproduction removes the within-cell discrepancy under the
    linear null. The nonparametric component has standard deviation
    $\sqrt{J^d/n}$, whereas the OLS sampling error is of smaller order $n^{-1/2}$. 
    Unlike the forward-split test, this
    approach does not remove the generated-regressor and OLS attenuation
    distortions; it instead requires the sufficient centering condition
    ${\mathsf{B}_{\mathtt{FS}}}+H^{-1}=o\!\left(\sqrt{J^d/n}\right)$.
    Use the paper's proposed local-constant point-estimation MSE tuning,
    \begin{align*}
        h&\asymp T^{-1/3},
        &
        H&\asymp T^{2/3},
        &
        J&\asymp n^{1/(d+2)},
    \end{align*}
    but construct both steps with degree-one fits. Then
    {if $\mathsf{B}_{\mathtt{FS}}\lesssim T^{-2/3}$,} the main restriction
    reduces, up to logarithmic factors from uniform remainder bounds, to
    $T\gg n^{3/[2(d+2)]}$.
    Thus $T=n^\eta$ with
    $3/[2(d+2)]<\eta<1$ is admissible for every $d\geq1$, permitting
    $T\ll n$. 
\end{enumerate}

\subsection{Proof of Lemma~\ref{lem:forward-linearity-normality}}

\medskip\noindent\textit{Step 1: Remainders.}
Put
\begin{align*}
    u_{it}^\circ
    &=
    \varepsilon_{i(t+H)}
    -\varpi_{t+H}^\top v_{i(t+H)},
    &
    g_t
    &=
    \frac{1}{n_t^{\mathtt{LIN}}}
    \sum_{i\in\mathcal I_t^{\mathtt{LIN}}}
    \zeta_{it}u_{it}^\circ,
    \\
    A_t(b)
    &=
    \sum_{i\in\mathcal I_t^{\mathtt{LIN}}}
    \widehat w_{it}(b)\widehat x_{i(t+H)}^\top,
    &
    A_t^\circ(b)
    &=
    \E[A_t(b)\mid\mathcal K_t].
\end{align*}
For notational convenience, set $u_{it}^\circ=0$ when
$i\notin\mathcal I_t^{\mathtt{LIN}}$.
The IV normal equations give
\begin{align}
    \widehat\theta_{t+H}^{\mathtt{IV}}-\theta_{t+H}
    &=
    \widehat{\mathsf M}_t^{-1}g_t+
    \mathscr B_{t+H}^{\mathtt{IV}},
    \label{eq:forward-iv-error-expansion}
    \\
    \mathscr B_{t+H}^{\mathtt{IV}}
    &:=
    -\widehat{\mathsf M}_t^{-1}
    \frac{1}{n_t^{\mathtt{LIN}}}
    \sum_{i\in\mathcal I_t^{\mathtt{LIN}}}\zeta_{it}
    \varpi_{t+H}^\top
    (\breve\beta_{i(t+H)}-\beta_{i(t+H)}).
    \label{eq:forward-iv-first-step-bias}
\end{align}
Thus $\mathscr B_{t+H}^{\mathtt{IV}}$ is the component of the IV estimation
error induced by the smooth first-step bias. In particular,
\begin{align*}
    \mathscr R_{\mathtt{LIN}}(b)
    &=
    -\frac{1}{T-2H}\sum_{t=H+1}^{T-H}
    A_t(b)\widehat{\mathsf M}_t^{-1}g_t
    -\frac{1}{T-2H}\sum_{t=H+1}^{T-H}
    A_t(b)\mathscr B_{t+H}^{\mathtt{IV}}.
\end{align*}
The partition, Gram, first-step moment, and instrument moment bounds imply
\begin{align*}
    \max_{\substack{H+1\leq t\leq T-H\\
                    k\in[K_{\mathtt{LIN}}]}}
    \left\{
    \|A_t(\mathsf{b}_k)\|+\|A_t^\circ(\mathsf{b}_k)\|
    \right\}
    &\lesssim_\P1.
\end{align*}
Hence the bias terms in
\eqref{eq:forward-linearity-decomposition} and
\eqref{eq:forward-iv-error-expansion} satisfy
\begin{align}
    \max_{k\in[K_{\mathtt{LIN}}]}
    \left\{
    |\mathscr B_{\mathtt{LIN}}(\mathsf{b}_k)|
    +\left|
    \frac{1}{T-2H}\sum_{t=H+1}^{T-H}
    A_t(\mathsf{b}_k)\mathscr B_{t+H}^{\mathtt{IV}}
    \right|
    \right\}
    &\lesssim_\P{\mathsf{B}_{\mathtt{FS}}}.
    \label{eq:forward-iv-bias-bound}
\end{align}

Conditional on $\mathcal K_t$, $g_t$ is centered by
\eqref{eq:forward-instrument-exogeneity}. Cross-sectional independence,
which is preserved after selection by
Assumption~\ref{ass:linearity-missingness},
Assumption~\ref{ass:linearity-instrument}(ii), the first-step moment bounds,
and the weight bounds give
\begin{align}
    &\E[\|g_t\|^2\mid\mathcal K_t]
    \lesssim_\P
    \frac{1+\|\varpi_{t+H}\|^2/H}{n_t^{\mathtt{LIN}}}
    \lesssim_\P
    \frac{1+\|\varpi_{t+H}\|^2/H}{n},
    \label{eq:forward-iv-conditional-moments}
    \\
    &\max_{\substack{H+1\leq t\leq T-H\\
                     k\in[K_{\mathtt{LIN}}]}}
    \E[\|A_t(\mathsf{b}_k)-A_t^\circ(\mathsf{b}_k)\|^2
       \mid\mathcal K_t]
    \lesssim_\P\frac{J^d}{nH}.
    \notag
\end{align}
Boundedness of $\mu_t$ on the fixed rectangle $\mathcal B$ bounds $c_t$
under the null. Assumption~\ref{ass:dgp}(iv) and Markov's inequality therefore
give
$(T-2H)^{-1}\sum_{t=H+1}^{T-H}
(1+\|\varpi_{t+H}\|^4)\lesssim_\P1$.
The multiplier
$A_t^\circ(\mathsf{b}_k)(\mathsf M_t^\circ)^{-1}\zeta_{it}$ attached to each
summand of $g_t$ is fixed before the future-window shocks begin. Re-indexing
the direct outcome shocks therefore gives a martingale sum. The rolling-window
component is also centered; it has periodwise variance
$O_\P((nH)^{-1})$ and overlaps for at most $H$ adjacent base dates. Thus, for
fixed $K_{\mathtt{LIN}}$,
\begin{align}
    \max_{k\in[K_{\mathtt{LIN}}]}
    \left|
    \frac{1}{T-2H}\sum_{t=H+1}^{T-H}
    A_t^\circ(\mathsf{b}_k)(\mathsf M_t^\circ)^{-1}g_t
    \right|
    &\lesssim_\P
    \frac{1}{\sqrt{n(T-2H)}}.
    \label{eq:forward-iv-predictable-score}
\end{align}

It remains to replace the conditional multipliers by their sample
counterparts. Assumption~\ref{ass:linearity-instrument}(iii) and a matrix
perturbation bound imply that $\widehat{\mathsf M}_t$ and $\widehat{\mathsf M}_t^{-1}$ are
uniformly bounded in probability and
\begin{align*}
    &r_{\mathtt M}
    :=
    \max_{H+1\leq t\leq T-H}
    \|\widehat{\mathsf M}_t-\mathsf M_t^\circ\|
    \lesssim_\P\sqrt{\frac{\log(nT)}{n}},
    \\
    &\max_{H+1\leq t\leq T-H}
    \|\widehat{\mathsf M}_t^{-1}-(\mathsf M_t^\circ)^{-1}\|
    \lesssim_\P r_{\mathtt M}.
\end{align*}
Use the exact decomposition
\begin{align*}
    A_t\widehat{\mathsf M}_t^{-1}-A_t^\circ(\mathsf M_t^\circ)^{-1}
    &=
    (A_t-A_t^\circ)(\mathsf M_t^\circ)^{-1}
    +A_t\{\widehat{\mathsf M}_t^{-1}-(\mathsf M_t^\circ)^{-1}\}.
\end{align*}
For the first term on the right-hand side,
\eqref{eq:forward-iv-conditional-moments}, the uniform bound on
$(\mathsf M_t^\circ)^{-1}$, the time-averaged factor-moment bound, and Markov's
inequality give, uniformly over the fixed grid,
\begin{align*}
    \frac{1}{T-2H}\sum_{t=H+1}^{T-H}
    \|A_t(\mathsf{b}_k)-A_t^\circ(\mathsf{b}_k)\|^2
    &\lesssim_\P\frac{J^d}{nH},
    &
    \frac{1}{T-2H}\sum_{t=H+1}^{T-H}\|g_t\|^2
    &\lesssim_\P\frac{1}{n}.
\end{align*}
Cauchy--Schwarz over the base dates therefore yields
\begin{align}
    \max_{k\in[K_{\mathtt{LIN}}]}
    \left|
    \frac{1}{T-2H}\sum_{t=H+1}^{T-H}
    \{A_t(\mathsf{b}_k)-A_t^\circ(\mathsf{b}_k)\}
    (\mathsf M_t^\circ)^{-1}g_t
    \right|
    &\lesssim_\P
    \frac{1}{n}\sqrt{\frac{J^d}{H}}.
    \label{eq:forward-iv-rough-multiplier}
\end{align}
Relative to the local standard-deviation scale, this bound satisfies
\begin{align*}
    \sqrt{\frac{n(T-2H)}{J^d}}\,
    \frac{1}{n}\sqrt{\frac{J^d}{H}}
    &=
    \sqrt{\frac{T-2H}{nH}}
    =o(1)
\end{align*}
by~\eqref{eq:forward-linearity-rough-multiplier-rate}.

For the inverse-matrix term,
\eqref{eq:forward-iv-conditional-moments}, the time-averaged factor-moment
bound, and Markov's inequality give
\begin{align*}
    \frac{1}{T-2H}\sum_{t=H+1}^{T-H}\|g_t\|
    &\leq
    \left\{
    \frac{1}{T-2H}\sum_{t=H+1}^{T-H}\|g_t\|^2
    \right\}^{1/2}
    \lesssim_\P \frac{1}{\sqrt n}.
\end{align*}
Together with the uniform bound on $A_t$, this gives
\begin{align*}
    \max_{k\in[K_{\mathtt{LIN}}]}
    \left|
    \frac{1}{T-2H}\sum_{t=H+1}^{T-H}
    A_t(\mathsf{b}_k)
    \{\widehat{\mathsf M}_t^{-1}-(\mathsf M_t^\circ)^{-1}\}g_t
    \right|
    &\lesssim_\P
    \frac{r_{\mathtt M}}{\sqrt n}
    \lesssim_\P
    \frac{\sqrt{\log(nT)}}{n}.
\end{align*}
The last bound is absorbed by $O_\P\{\log(nT)/n\}$. Combining the two
parts of the exact multiplier decomposition gives
\begin{align}
    \max_{k\in[K_{\mathtt{LIN}}]}
    \left|
    \frac{1}{T-2H}\sum_{t=H+1}^{T-H}
    \left\{
    A_t(\mathsf{b}_k)\widehat{\mathsf M}_t^{-1}
    -A_t^\circ(\mathsf{b}_k)(\mathsf M_t^\circ)^{-1}
    \right\}g_t
    \right|
    &\lesssim_\P
    \frac{1}{n}\sqrt{\frac{J^d}{H}}
    +\frac{\log(nT)}{n}.
    \label{eq:forward-iv-random-multiplier}
\end{align}
Combining~\eqref{eq:forward-iv-bias-bound}--
\eqref{eq:forward-iv-random-multiplier} gives
\begin{align}
    \max_{k\in[K_{\mathtt{LIN}}]}
    \left\{
    |\mathscr R_{\mathtt{LIN}}(\mathsf{b}_k)|
    +|\mathscr B_{\mathtt{LIN}}(\mathsf{b}_k)|
    \right\}
    &\lesssim_\P
    {\mathsf{B}_{\mathtt{FS}}}
    +\frac{1}{\sqrt{n(T-2H)}}
    +\frac{1}{n}\sqrt{\frac{J^d}{H}}
    +\frac{\log(nT)}{n}.
    \label{eq:forward-fit-remainder-bound}
\end{align}
The restriction $J^{-d/2}=o(1)$ controls the normalized
$\{n(T-2H)\}^{-1/2}$ term.
Equations~\eqref{eq:forward-linearity-rough-multiplier-rate} and
\eqref{eq:forward-linearity-remainder-rate} control the remaining terms.
Thus both non-leading terms in~\eqref{eq:forward-linearity-decomposition},
uniformly over the fixed grid, are
$o_\P(\sqrt{J^d/(n(T-2H))})$.

\medskip\noindent\textit{Step 2: Leading term.}
The matched version of
Lemma~\ref{lem:partition-basis-moments} described after
Assumption~\ref{ass:linearity-missingness}, Jensen's inequality over the $H$
rolling lags, the first-step Gram bound, and
$cn\leq n_t^{\mathtt{LIN}}\leq n$ imply, under the absent-unit convention and
uniformly over
$k\in[K_{\mathtt{LIN}}]$,
\begin{align}
    \sum_{r=1}^{T}\sum_{i=1}^{n}
    W_{ir}(\mathsf{b}_k)^2
    &\lesssim_\P
    \frac{(T-2H)J^d}{n},
    &
    \sum_{r=1}^{T}\sum_{i=1}^{n}
    |W_{ir}(\mathsf{b}_k)|^3
    &\lesssim_\P
    \frac{(T-2H)J^{2d}}{n^2}.
    \label{eq:forward-weight-moments}
\end{align}
Consequently,
$\lambda_{\max}(\Sigma_{\mathtt{LIN}})
\lesssim_\P J^d/\{n(T-2H)\}$. Condition~\textnormal{(iv)} implies that all
eigenvalues of both $\overline\Sigma_{\mathtt{LIN}}$ and
$\Sigma_{\mathtt{LIN}}$ have this order.

Let
$\boldsymbol W_{ir}
=\big(W_{ir}(\mathsf b_1),\ldots,
W_{ir}(\mathsf b_{K_{\mathtt{LIN}}})\big)^\top$.
For fixed $a\in\mathbb R^{K_{\mathtt{LIN}}}$, put
\begin{align*}
    \widetilde W_{ir}(a)
    &:=
    a^\top\overline\Sigma_{\mathtt{LIN}}^{-1/2}
    \boldsymbol W_{ir}.
\end{align*}
The reference covariance is $\mathcal E_\infty$-measurable. Hence, after
ordering the primitive innovations first by date and then by unit,
$\widetilde W_{ir}(a)$ is measurable before $\eta_{ir}$ is observed. 
The sum of the conditional variances of the corresponding 
normalized martingale differences 
(terminal predictable quadratic variation of their sum) is
\begin{align*}
    \frac{1}{(T-2H)^2}
    \sum_{r=1}^{T}\sum_{i=1}^{n}
    \varsigma_{ir}^2\widetilde W_{ir}(a)^2
    &=
    a^\top\overline\Sigma_{\mathtt{LIN}}^{-1/2}
    \Sigma_{\mathtt{LIN}}
    \overline\Sigma_{\mathtt{LIN}}^{-1/2}a
    \to_\P\|a\|^2
\end{align*}
by~\eqref{eq:forward-linearity-variance-stabilization}.
Let
\begin{align*}
    \mathcal J_{ir-}^{\mathtt{LIN}}
    &:=
    \mathcal E_\infty\vee
    \sigma(\eta_{js}:s<r,\text{ or }s=r\text{ and }j<i).
\end{align*}
Since $K_{\mathtt{LIN}}$ is fixed, the third-moment bound in
\eqref{eq:forward-weight-moments} and
\eqref{eq:forward-linearity-nondegeneracy} also give the conditional
Lyapunov bound
\begin{align*}
    \frac{1}{(T-2H)^3}
    \sum_{r=1}^{T}\sum_{i=1}^{n}
    \E\!\left[
    |\widetilde W_{ir}(a)\varepsilon_{ir}|^3
    \mid\mathcal J_{ir-}^{\mathtt{LIN}}
    \right]
    &\lesssim_\P
    \sqrt{\frac{J^d}{n(T-2H)}}
    =o_\P(1).
\end{align*}
The martingale central limit theorem and the Cram\'er--Wold device therefore
give
\begin{align*}
    \overline\Sigma_{\mathtt{LIN}}^{-1/2}
    \boldsymbol{\mathscr S}_{\mathtt{LIN}}
    &\rightsquigarrow
    \mathcal N(0,I_{K_{\mathtt{LIN}}}),
\end{align*}
where $\boldsymbol{\mathscr S}_{\mathtt{LIN}}$ stacks the leading terms over
the fixed grid. Condition~\eqref{eq:forward-linearity-nondegeneracy} and the
Step~1 remainder bound then imply
\begin{align}
    \overline\Sigma_{\mathtt{LIN}}^{-1/2}
    \widehat{\boldsymbol\nu}_{\mathtt{LIN}}
    &\rightsquigarrow
    \mathcal N(0,I_{K_{\mathtt{LIN}}}).
    \label{eq:forward-linearity-reference-normality}
\end{align}
Finally, \eqref{eq:forward-linearity-nondegeneracy} and
\eqref{eq:forward-linearity-variance-stabilization}, together with a
matrix square-root perturbation bound, give
\begin{align*}
    \left\|
    \Sigma_{\mathtt{LIN}}^{-1/2}
    \overline\Sigma_{\mathtt{LIN}}^{1/2}
    -I_{K_{\mathtt{LIN}}}
    \right\|
    &=o_\P(1).
\end{align*}
Therefore
\begin{align*}
    \Sigma_{\mathtt{LIN}}^{-1/2}
    \widehat{\boldsymbol\nu}_{\mathtt{LIN}}
    &=
    \Sigma_{\mathtt{LIN}}^{-1/2}
    \overline\Sigma_{\mathtt{LIN}}^{1/2}
    \left(
    \overline\Sigma_{\mathtt{LIN}}^{-1/2}
    \widehat{\boldsymbol\nu}_{\mathtt{LIN}}
    \right)
    \rightsquigarrow
    \mathcal N(0,I_{K_{\mathtt{LIN}}})
\end{align*}
by~\eqref{eq:forward-linearity-reference-normality} and Slutsky's theorem.
\qed

\subsection{Proof of Lemma~\ref{lem:forward-linearity-variance}}

For this proof, put
\begin{align*}
    \mathcal T_H
    &=
    \{H+1,\ldots,T-H\},
    &
    L_T
    &=
    |\mathcal T_H|=T-2H,
    &
    \ell_{nT}
    &=
    \frac{\log(nT)}{n},
    \\
    e_{nT}^2
    &=
    {\mathsf{B}_{\mathtt{FS}}^2}+\ell_{nT},
    &
    \rho_{nT}^2
    &=
    \frac{1}{H}+\ell_{nT},
    &
    u_{it}^\circ
    &=
    \varepsilon_{i(t+H)}
    -\varpi_{t+H}^\top v_{i(t+H)},
    \\
    g_t
    &=
    \frac{1}{n_t^{\mathtt{LIN}}}
    \sum_{i\in\mathcal I_t^{\mathtt{LIN}}}
    \zeta_{it}u_{it}^\circ.
\end{align*}
Unless stated otherwise, sums over $t$ and $s$ below are over
$\mathcal T_H$, and shifted dates outside $\mathcal T_H$ are omitted.
The generated weight $\widehat w_{it}$ and the fitted residual-error
components introduced below are set to zero outside the corresponding
matched set. The latent block $u_{it}^\circ$ and its primitive shocks remain
defined for every $i\in[n]$; zero weights remove unavailable units from all
sample sums.

The assumptions used in this proof imply
\begin{align}
    \frac{H}{L_T}
    &=o(1),
    &
    \frac{J^d}{n}
    &=o(1),
    &
    He_{nT}^2
    &=o_\P(1),
    &
    H\rho_{nT}e_{nT}
    &=o_\P(1),
    \notag\\
    \frac{H}{n}
    &=o(1),
    &
    \sqrt{\frac{HJ^d}{nL_T}}
    &=o(1),
    &
    H\sqrt{\frac{J^d}{n^2L_T}}
    &=o(1).
    \label{eq:forward-variance-proof-rates}
\end{align}
For example,
$H^2\rho_{nT}^2e_{nT}^2
={(1+H\ell_{nT})(H\mathsf{B}_{\mathtt{FS}}^2+H\ell_{nT})}=o_\P(1)$.
The matched-sample local-support and Gram bounds also give, uniformly over
the fixed grid,
\begin{align}
    \max_{t,k}\sum_i\widehat w_{it}(\mathsf b_k)^2
    &\lesssim_\P\frac{J^d}{n},
    &
    \max_{i,t,k}|\widehat w_{it}(\mathsf b_k)|
    &\lesssim_\P\frac{J^d}{n},
    \notag\\
    \max_{t,s,k,\ell}
    \sum_i
    \widehat w_{it}(\mathsf b_k)^2
    \widehat w_{is}(\mathsf b_\ell)^2
    &\lesssim_\P\frac{J^{3d}}{n^3}.
    \label{eq:forward-variance-proof-weight-bounds}
\end{align}

We use two primitive-shock calculations repeatedly. First, define
\begin{align*}
    \kappa_{tr}
    &:=
    \IF(r=t+H)
    -\frac{1}{H}\IF(t\leq r\leq t+H-1)
    \varpi_{t+H}^{\top}w_{t+H-r}(t+H).
\end{align*}
The uniform first-step Gram bound gives
\begin{align*}
    \max_{s\in\mathcal T_H}
    \frac{1}{H}\sum_{a=1}^H\|w_a(s+H)\|^2
    &\lesssim_\P1.
\end{align*}
Therefore
$u_{it}^\circ=\sum_{r=1}^T\kappa_{tr}\varepsilon_{ir}$ and
\begin{align}
    \sum_{r=1}^T\kappa_{tr}^2
    &\lesssim_\P
    1+\frac{\|\varpi_{t+H}\|^2}{H}.
    \label{eq:forward-residual-primitive-energy}
\end{align}
More generally, if $Z_{is}$ is $\mathcal K_s$-measurable, then
\begin{align}
    \sum_s Z_{is}u_{is}^\circ
    &=
    \sum_{r=1}^T\Gamma_{ir}(Z)\varepsilon_{ir},
    \notag\\
    \Gamma_{ir}(Z)
    &:=
    Z_{i,r-H}
    -\frac{1}{H}\sum_{a=1}^H
    Z_{i,r+a-H}\varpi_{r+a}^{\top}w_a(r+a),
    \notag\\
    \sum_{i,r}|\Gamma_{ir}(Z)|^2
    &\lesssim_\P
    \sum_{i,s}
    (1+\|\varpi_{s+H}\|^2)|Z_{is}|^2,
    \label{eq:forward-linear-shock-sum-reindexing}
\end{align}
where invalid subscripts are zero. Each $\Gamma_{ir}(Z)$ is measurable
before $\varepsilon_{ir}$ is observed.

Second, index primitive shocks by $a=(i,r)$ and $b=(j,r')$, write
$\varepsilon_a=\varepsilon_{ir}$ and
$\varepsilon_b=\varepsilon_{jr'}$, and order these compound indices
lexicographically by date and unit. Let
\begin{align*}
    \mathcal J_{ir-}^{\mathtt{LIN}}
    &:=
    \mathcal E_\infty\vee
    \sigma(\eta_{js}:s<r,\text{ or }s=r\text{ and }j<i).
\end{align*}
Suppose that $a<b$, $A_{ab}$ is measurable before $\varepsilon_a$, and, for a
deterministic sequence $r_n$,
$\sum_{a<b}|A_{ab}|^2=O_\P(r_n^2)$. Then bounded conditional fourth moments and
successive conditioning give
\begin{align}
    \sum_{a<b}A_{ab}\varepsilon_a\varepsilon_b
    &=O_\P(r_n).
    \label{eq:forward-quadratic-shock-sum-bound}
\end{align}
For a quadratic sum over $a\ne b$, we first combine the coefficients on
$(a,b)$ and $(b,a)$, which multiply the same shock product. We also use the
analogous bound for a predictable linear shock sum:
$\sum_{i,r}\Gamma_{ir}\varepsilon_{ir}=O_\P(r_n)$ whenever the coefficients
are predictable and $\sum_{i,r}|\Gamma_{ir}|^2=O_\P(r_n^2)$.

\medskip\noindent\textit{Step 1: Oracle covariance.}
For
$k,\ell\in[K_{\mathtt{LIN}}]$, let
$\widetilde\Sigma_{k\ell}$ denote
\eqref{eq:forward-linearity-hac} with
$(b,b')=(\mathsf{b}_k,\mathsf{b}_\ell)$ and
$\widehat e_{i(t+H)}$ replaced by $u_{it}^\circ$. Because
$u_{it}^\circ$ is a linear combination of primitive shocks dated in
$[t,t+H]$, the truncation at $H$ is exact. For example, if $t+H<s$, then
$\widehat w_{it}(\mathsf b_k)u_{it}^\circ
\widehat w_{is}(\mathsf b_\ell)$ is $\mathcal K_s$-measurable, whereas
$\E[u_{is}^\circ\mid\mathcal K_s]=0$; the reverse ordering is symmetric.
Re-indexing the retained products by primitive shock gives
\begin{align}
    \widetilde\Sigma_{k\ell}
    &=\mathscr D_{k\ell}+\mathscr O_{k\ell},
    \label{eq:forward-oracle-variance-decomposition}\\
    \mathscr D_{k\ell}
    &=\frac{1}{L_T^2}
    \sum_{r=1}^{T}\sum_{i=1}^{n}
    W_{ir}(\mathsf{b}_k)W_{ir}(\mathsf{b}_\ell)\varepsilon_{ir}^2,
    \notag\\
    \mathscr O_{k\ell}
    &=\frac{1}{L_T^2}
    \sum_{i=1}^{n}\sum_{r\ne r'}
    M_{i,rr'}^{k\ell}\varepsilon_{ir}\varepsilon_{ir'}.
    \notag
\end{align}
Here $M_{i,rr'}^{k\ell}$ collects the coefficients on products of distinct
primitive shocks. The diagonal coefficient is exactly the product of the
primitive weights in~\eqref{eq:forward-primitive-weight}: every pair of
residual terms containing the same primitive shock is retained because their
base dates differ by at most $H$.

\smallskip\noindent\emph{Diagonal component.}
The first two bounds in
\eqref{eq:forward-variance-proof-weight-bounds},
$|x+y|^4\leq8(|x|^4+|y|^4)$, Jensen's inequality over the rolling average,
and $L_T^{-1}\sum_t(1+\|\varpi_{t+H}\|^4)\lesssim_\P1$ give
\begin{align}
    \sum_{r=1}^{T}\sum_{i=1}^{n}
    |W_{ir}(\mathsf{b}_k)|^4
    &\lesssim_\P
    \frac{L_T J^{3d}}{n^3}.
    \label{eq:forward-primitive-fourth-moment}
\end{align}
Since
$W_{ir}(\mathsf b_k)W_{ir}(\mathsf b_\ell)$ is
$\mathcal J_{ir-}^{\mathtt{LIN}}$-measurable,
$W_{ir}(\mathsf b_k)W_{ir}(\mathsf b_\ell)
(\varepsilon_{ir}^2-\varsigma_{ir}^2)$ is a martingale difference and
\begin{align*}
    \mathscr D_{k\ell}
    -\Sigma_{\mathtt{LIN}}(\mathsf b_k,\mathsf b_\ell)
    &=
    \frac{1}{L_T^2}
    \sum_{r=1}^{T}\sum_{i=1}^{n}
    W_{ir}(\mathsf b_k)W_{ir}(\mathsf b_\ell)
    \left(\varepsilon_{ir}^2-\varsigma_{ir}^2\right).
\end{align*}
Its conditional variance is bounded by
\eqref{eq:forward-primitive-fourth-moment} and the conditional fourth-moment
bound for $\varepsilon_{ir}$. Therefore
\begin{align}
    \mathscr D_{k\ell}
    -\Sigma_{\mathtt{LIN}}(\mathsf{b}_k,\mathsf{b}_\ell)
    &=O_\P\!\left(
    \sqrt{\frac{J^{3d}}{n^3L_T^3}}
    \right)
    =o_\P\!\left(\frac{J^d}{nL_T}\right).
    \label{eq:forward-oracle-diagonal-rate}
\end{align}

\smallskip\noindent\emph{Off-diagonal component.}
For $t<s\leq t+H$, define the pre-$s$ and $s$-onward components by
\begin{align*}
    u_{it}^{<s}
    &:=
    -\frac{1}{H}\sum_{r=t}^{s-1}
    \varpi_{t+H}^{\top}w_{t+H-r}(t+H)\varepsilon_{ir},
    &
    u_{it}^{\geq s}
    &:=
    u_{it}^{\circ}-u_{it}^{<s}.
\end{align*}
Set $u_{it}^{<s}=0$ and $u_{it}^{\geq s}=u_{it}^{\circ}$ when $s\leq t$;
the reverse ordering is handled by interchanging $t$ and $s$.
Put $\Lambda_{t,h}=1+\|\varpi_{t+h+H}\|^2$. The first-step Gram and
factor-moment bounds, $2|xy|\leq x^2+y^2$, and
$L_T^{-1}\sum_t(1+\|\varpi_{t+H}\|^4)\lesssim_\P1$ give
\begin{align}
    \max_{k\in[K_{\mathtt{LIN}}]}
    \sum_{h=1}^H\frac{1}{L_T}\sum_t\sum_i
    \Lambda_{t,h}
    \widehat w_{it}(\mathsf b_k)^2
    \bigl(u_{it}^{<t+h}\bigr)^2
    &\lesssim_\P\frac{J^d}{n},
    \notag\\
    \sum_{h=1}^H\frac{1}{L_T}\sum_t
    \frac{\Lambda_{t,h}}{n}\sum_i
    \bigl(u_{it}^{<t+h}\bigr)^2
    &\lesssim_\P1,
    \label{eq:forward-pre-s-energy}
\end{align}
where boundary terms are omitted.

For the products involving the $s$-onward component $u_{it}^{\geq s}$, all
primitive shocks are dated
at or after the later base date, so each coefficient is measurable before
the earlier of the two shocks. Let
$\mathscr O_{k\ell}^{\mathrm F}$ denote their off-diagonal contribution and
define $\{M_{i,rr'}^{k\ell,\mathrm F}\}$ as the collected coefficients in
\begin{align*}
    \mathscr O_{k\ell}^{\mathrm F}
    &:=
    \frac{1}{L_T^2}
    \sum_{i=1}^n\sum_{r\ne r'}
    M_{i,rr'}^{k\ell,\mathrm F}\varepsilon_{ir}\varepsilon_{ir'}.
\end{align*}
The coefficients include the matched-set indicators and hence vanish when
the relevant unit is unavailable. Equations
\eqref{eq:forward-variance-proof-weight-bounds} and
\eqref{eq:forward-residual-primitive-energy}, together with Jensen's
inequality over the rolling averages, give
\begin{align}
    \sum_{r\ne r'}\sum_{i=1}^{n}
    |M_{i,rr'}^{k\ell,\mathrm F}|^2
    &\lesssim_\P
    \frac{L_T HJ^{3d}}{n^3}.
    \label{eq:forward-off-diagonal-coefficients}
\end{align}
Combining the coefficients on $(r,r')$ and $(r',r)$, which multiply the
same product $\varepsilon_{ir}\varepsilon_{ir'}$,
\eqref{eq:forward-quadratic-shock-sum-bound} bounds this predictable
quadratic shock sum by
$O_\P\{\sqrt{L_T HJ^{3d}/n^3}/L_T^2\}$.

For the pre-$s$ component when $s$ is the later base date, put
\begin{align*}
    P_{is}^{k\ell}
    &:=
    \widehat w_{is}(\mathsf b_\ell)
    \sum_{h=1}^H
    \widehat w_{i,s-h}(\mathsf b_k)
    u_{i,s-h}^{<s}.
\end{align*}
This coefficient is $\mathcal K_s$-measurable. The maximum-weight bound in
\eqref{eq:forward-variance-proof-weight-bounds}, Cauchy--Schwarz over $h$,
and \eqref{eq:forward-pre-s-energy} imply
\begin{align*}
    \max_{k,\ell}
    \sum_i\sum_s
    (1+\|\varpi_{s+H}\|^2)|P_{is}^{k\ell}|^2
    &\lesssim_\P\frac{L_T HJ^{3d}}{n^3}.
\end{align*}
Equation~\eqref{eq:forward-linear-shock-sum-reindexing}, applied with
$Z_{is}=P_{is}^{k\ell}$, therefore rewrites
$L_T^{-2}\sum_{i,s}P_{is}^{k\ell}u_{is}^\circ$ as a predictable linear
shock sum whose coefficient-square sum is
$O_\P(L_T HJ^{3d}/n^3)$. The same argument applies after interchanging the base
dates. Combining the two orderings of the base dates with the preceding
off-diagonal contribution yields
\begin{align}
    \mathscr O_{k\ell}
    &=O_\P\!\left(
    \sqrt{\frac{HJ^{3d}}{n^3L_T^3}}
    \right)
    =o_\P\!\left(\frac{J^d}{nL_T}\right).
    \label{eq:forward-oracle-off-diagonal-rate}
\end{align}

\medskip\noindent\textit{Step 2: Fitted-residual error.}
We next replace $u_{it}^\circ$ by the fitted residual. From
\eqref{eq:forward-iv-error-expansion}, the IV matrix bounds, and
\eqref{eq:forward-iv-conditional-moments},
\begin{align}
    \frac{1}{L_T}\sum_{t=H+1}^{T-H}
    \|\widehat\theta_{t+H}^{\mathtt{IV}}-\theta_{t+H}\|^2
    &\lesssim_\P
    e_{nT}^2.
    \label{eq:forward-iv-mean-square}
\end{align}
Moreover,
\begin{align*}
    \widehat e_{i(t+H)}-u_{it}^\circ
    &=-\varpi_{t+H}^\top
    (\breve\beta_{i(t+H)}-\beta_{i(t+H)})
    -\widehat x_{i(t+H)}^\top
    (\widehat\theta_{t+H}^{\mathtt{IV}}-\theta_{t+H}).
\end{align*}
For $i\in\mathcal I_t^{\mathtt{LIN}}$, let the left-hand side be
$\delta_{it}$, and set $\delta_{it}=0$ otherwise. The first bound in
\eqref{eq:forward-variance-proof-weight-bounds}, the time-averaged factor
moments, and
\eqref{eq:forward-iv-mean-square} imply
\begin{align}
    \max_{k\in[K_{\mathtt{LIN}}]}
    \frac{1}{L_T}\sum_{t=H+1}^{T-H}
    \sum_{i\in\mathcal I_t^{\mathtt{LIN}}}
    \widehat w_{it}(\mathsf{b}_k)^2\delta_{it}^2
    &\lesssim_\P
    \frac{J^d}{n}
    e_{nT}^2.
    \label{eq:forward-residual-weighted-mean-square}
\end{align}
To make the residual replacement explicit, for arrays $a=\{a_{it}\}$ and
$b=\{b_{it}\}$ define the banded bilinear form
\begin{align*}
    \mathcal H_{k\ell}(a,b)
    &:=
    \frac{1}{L_T^2}
    \sum_{i=1}^n
    \sum_{t=H+1}^{T-H}\sum_{s=H+1}^{T-H}
    \widehat w_{it}(\mathsf b_k)
    \widehat w_{is}(\mathsf b_\ell)
    a_{it}b_{is}\IF(|t-s|\leq H),
\end{align*}
using the zero-weight convention for unavailable units. Since
$\|a\|_k^2:=
L_T^{-1}\sum_{t=H+1}^{T-H}\sum_{i=1}^n
\widehat w_{it}(\mathsf b_k)^2a_{it}^2$, banded Cauchy--Schwarz gives
\begin{align}
    |\mathcal H_{k\ell}(a,b)|
    &\leq
    \frac{2H+1}{L_T}\|a\|_k\|b\|_\ell.
    \label{eq:forward-residual-banded-cauchy}
\end{align}
Since
$\widehat e_{i(t+H)}=u_{it}^\circ+\delta_{it}$,
\begin{align}
    \widehat\Sigma_{\mathtt{LIN}}(\mathsf b_k,\mathsf b_\ell)
    -\widetilde\Sigma_{k\ell}
    &=
    \mathcal H_{k\ell}(u^\circ,\delta)
    +\mathcal H_{k\ell}(\delta,u^\circ)
    +\mathcal H_{k\ell}(\delta,\delta).
    \label{eq:forward-residual-bilinear-decomposition}
\end{align}
The inequality $2|xy|\leq x^2+y^2$, the fact that each base date has at
most $2H+1$ retained partners, and
\eqref{eq:forward-residual-weighted-mean-square} give
\begin{align}
    \max_{k,\ell}
    |\mathcal H_{k\ell}(\delta,\delta)|
    &\lesssim_\P
    \frac{J^d}{nL_T}
    He_{nT}^2.
    \label{eq:forward-residual-delta-delta}
\end{align}

To decompose the two remaining cross terms, put
\begin{align*}
    d_{it}
    &=
    \breve\beta_{i(t+H)}-\beta_{i(t+H)},
    &
    \breve x_{i(t+H)}
    &=(1,\breve\beta_{i(t+H)}^\top)^\top,
    &
    C_t^\circ
    &=(\mathsf M_t^\circ)^{-1},
    \\
    h_t
    &:=
    \frac{1}{n_t^{\mathtt{LIN}}}
    \sum_{j\in\mathcal I_t^{\mathtt{LIN}}}
    \zeta_{jt}\varpi_{t+H}^\top d_{jt},
    &
    \mathscr B_{t+H}^{\circ,\mathtt{IV}}
    &:=
    -C_t^\circ h_t,
    \\
    b_{it}^\circ
    &:=
    -\varpi_{t+H}^\top d_{it}
    -\breve x_{i(t+H)}^\top
    \mathscr B_{t+H}^{\circ,\mathtt{IV}},
    &
    q_{it}^\circ
    &:=
    -\breve x_{i(t+H)}^\top C_t^\circ g_t.
\end{align*}
For $i\in\mathcal I_t^{\mathtt{LIN}}$, also define
\begin{align*}
    r_{it}
    &:=
    -\left\{
    \widehat x_{i(t+H)}^\top\widehat{\mathsf M}_t^{-1}
    -\breve x_{i(t+H)}^\top C_t^\circ
    \right\}(g_t-h_t).
\end{align*}
Set $b_{it}^\circ=q_{it}^\circ=r_{it}=0$ outside the matched set. Direct
substitution gives the exact identity
$\delta_{it}=b_{it}^\circ+q_{it}^\circ+r_{it}$.
Consequently, the two cross terms in
\eqref{eq:forward-residual-bilinear-decomposition} satisfy
\begin{align}
    &\mathcal H_{k\ell}(u^\circ,\delta)
    +\mathcal H_{k\ell}(\delta,u^\circ)
    \notag\\
    &\quad=
    \mathcal H_{k\ell}(u^\circ,b^\circ)
    +\mathcal H_{k\ell}(b^\circ,u^\circ)
    +\mathcal H_{k\ell}(u^\circ,q^\circ)
    +\mathcal H_{k\ell}(q^\circ,u^\circ)
    \notag\\
    &\qquad
    +\mathcal H_{k\ell}(u^\circ,r)
    +\mathcal H_{k\ell}(r,u^\circ).
    \label{eq:forward-residual-cross-decomposition}
\end{align}

\medskip\noindent\textit{Step 3: Cross terms.}
\smallskip\noindent\emph{Multiplier remainder.}
Assumption~\ref{ass:linearity-instrument}(iii) gives
\begin{align*}
    \max_{H+1\leq t\leq T-H}
    \|\widehat{\mathsf M}_t^{-1}-(\mathsf M_t^\circ)^{-1}\|
    &\lesssim_\P
    \sqrt{\frac{\log(nT)}{n}}.
\end{align*}
For $i\in\mathcal I_t^{\mathtt{LIN}}$,
\begin{align*}
    \widehat x_{i(t+H)}^\top\widehat{\mathsf M}_t^{-1}
    -\breve x_{i(t+H)}^\top C_t^\circ
    &=
    (0,v_{i(t+H)}^\top)\widehat{\mathsf M}_t^{-1}
    +\breve x_{i(t+H)}^\top
    (\widehat{\mathsf M}_t^{-1}-C_t^\circ).
\end{align*}
The conditional score bound
\eqref{eq:forward-iv-conditional-moments},
{$\max_{i,t}\|d_{it}\|\leq\mathsf{B}_{\mathtt{FS}}$ w.p.a.1}, and the time-averaged
factor-moment bound imply
\begin{align*}
    \frac{1}{L_T}\sum_t\|g_t-h_t\|^2
    &\lesssim_\P e_{nT}^2,
    &
    \frac{1}{L_T}\sum_t
    \E[\|g_t-h_t\|^4\mid\mathcal K_t]
    &\lesssim_\P e_{nT}^4.
\end{align*}
Together with
$\E[\|v_{i(t+H)}\|^2\mid\mathcal K_t]\lesssim_\P H^{-1}$,
the corresponding conditional fourth-moment bounds, and the
maximum-instrument bound in
Assumption~\ref{ass:linearity-instrument}(ii), the multiplier identity above
gives
\begin{align}
    \max_{k\in[K_{\mathtt{LIN}}]}
    \frac{1}{L_T}\sum_{t=H+1}^{T-H}
    \sum_{i\in\mathcal I_t^{\mathtt{LIN}}}
    \widehat w_{it}(\mathsf b_k)^2r_{it}^2
    &\lesssim_\P
    \frac{J^d}{n}\rho_{nT}^2e_{nT}^2.
    \label{eq:forward-residual-random-multiplier-mse}
\end{align}
The corresponding weighted mean square for $u^\circ$ is
$O_\P(J^d/n)$. Hence
\eqref{eq:forward-residual-banded-cauchy} gives
\begin{align}
    \max_{k,\ell}
    \left\{
    |\mathcal H_{k\ell}(u^\circ,r)|
    +|\mathcal H_{k\ell}(r,u^\circ)|
    \right\}
    &\lesssim_\P
    \frac{J^d}{nL_T}
    H\rho_{nT}e_{nT}
    =o_\P\!\left(\frac{J^d}{nL_T}\right),
    \label{eq:forward-residual-r-cross}
\end{align}
by~\eqref{eq:forward-variance-proof-rates}.

\smallskip\noindent\emph{First-step bias.}
Use the pre-$s$/$s$-onward split introduced in Step~1.
By the definition of $\kappa_{tr}$ above,
\begin{align*}
    u_{it}^{\geq s}
    &=
    \sum_{r=1}^{T}
    \kappa_{tr}\varepsilon_{ir}
    \IF(r\geq\max(t,s)).
\end{align*}
Define the $s$-onward and pre-$s$ pieces explicitly by
\begin{align*}
    \mathscr F_{k\ell}^{u,b}
    &:=
    \frac{1}{L_T^2}
    \sum_{i=1}^{n}\sum_t\sum_s
    \widehat w_{it}(\mathsf b_k)
    \widehat w_{is}(\mathsf b_\ell)
    u_{it}^{\geq s}b_{is}^{\circ}
    \IF(|t-s|\leq H),
    \\
    \mathscr P_{k\ell}^{u,b}
    &:=
    \frac{1}{L_T^2}
    \sum_{i=1}^{n}\sum_t\sum_s
    \widehat w_{it}(\mathsf b_k)
    \widehat w_{is}(\mathsf b_\ell)
    u_{it}^{<s}b_{is}^{\circ}
    \IF(|t-s|\leq H),
    \\
    \mathcal H_{k\ell}(u^\circ,b^\circ)
    &=
    \mathscr F_{k\ell}^{u,b}
    +\mathscr P_{k\ell}^{u,b}.
\end{align*}
The two weights and $b_{is}^\circ$ use only innovations dated before their
base dates, and $\kappa_{tr}$ is $\mathcal E_\infty$-measurable. Define
\begin{align*}
    F_{ir}^{k\ell}
    &:=
    \sum_t\sum_s
    \widehat w_{it}(\mathsf b_k)
    \widehat w_{is}(\mathsf b_\ell)
    b_{is}^{\circ}\kappa_{tr}
    \IF(r\geq\max(t,s))
    \IF(|t-s|\leq H).
\end{align*}
Then $F_{ir}^{k\ell}$ is measurable before $\varepsilon_{ir}$, and
\begin{align*}
    \mathscr F_{k\ell}^{u,b}
    &=
    \frac{1}{L_T^2}
    \sum_{i=1}^{n}\sum_{r=1}^{T}
    F_{ir}^{k\ell}\varepsilon_{ir}.
\end{align*}
Equation~\eqref{eq:forward-variance-proof-weight-bounds}, the bound
{$\max_{i,t}\|d_{it}\|\leq\mathsf{B}_{\mathtt{FS}}$ w.p.a.1}, and the instrument
moment, inverse-matrix, and time-averaged factor-moment bounds give
\begin{align*}
    \max_{k,\ell}
    \sum_{i=1}^{n}\sum_{r=1}^{T}
    |F_{ir}^{k\ell}|^2
    &\lesssim_\P
    \frac{L_T H^2J^{3d}}{n^3}{\mathsf{B}_{\mathtt{FS}}^2}.
\end{align*}
The predictable linear-shock-sum bound therefore yields
\begin{align*}
    \max_{k,\ell}
    |\mathscr F_{k\ell}^{u,b}|
    &\lesssim_\P
    \sqrt{\frac{H^2J^{3d}}{n^3L_T^3}}\,
    {\mathsf{B}_{\mathtt{FS}}}.
\end{align*}

For the pre-$(t+h)$ component, we use a direct bound rather than martingale
cancellation. For
$h=1,\ldots,H$, let
\begin{align*}
    E_{k,h}^{u}
    &:=
    \frac{1}{L_T}\sum_t\sum_i
    \widehat w_{it}(\mathsf b_k)^2
    (u_{it}^{<t+h})^2,
    \\
    E_{\ell,h}^{b}
    &:=
    \frac{1}{L_T}\sum_t\sum_i
    \widehat w_{i,t+h}(\mathsf b_\ell)^2
    (b_{i,t+h}^{\circ})^2,
\end{align*}
where boundary terms are omitted. Equation~\eqref{eq:forward-pre-s-energy},
the maximum-instrument bound, $n_t^{\mathtt{LIN}}\asymp n$, and
$\sup_t\|C_t^\circ\|\lesssim_\P1$ give
\begin{align*}
    \sum_{h=1}^H E_{k,h}^{u}
    &\lesssim_\P\frac{J^d}{n},
    &
    \sum_{h=1}^H E_{\ell,h}^{b}
    &\lesssim_\P
    \frac{HJ^d}{n}{\mathsf{B}_{\mathtt{FS}}^2}.
\end{align*}
Cauchy--Schwarz first over $(i,t)$ and then over $h$ therefore bounds the
pre-$s$ piece explicitly as
\begin{align*}
    \max_{k,\ell}
    |\mathscr P_{k\ell}^{u,b}|
    &\leq
    \frac{1}{L_T}\sum_{h=1}^H
    \sqrt{E_{k,h}^{u}E_{\ell,h}^{b}}
    \lesssim_\P
    \frac{J^d}{nL_T}\sqrt H\,{\mathsf{B}_{\mathtt{FS}}}.
\end{align*}
By interchanging $t$ and $s$,
\begin{align*}
    \mathcal H_{k\ell}(b^\circ,u^\circ)
    &=
    \mathcal H_{\ell k}(u^\circ,b^\circ)
    =
    \mathscr F_{\ell k}^{u,b}
    +\mathscr P_{\ell k}^{u,b}.
\end{align*}
Consequently,
\begin{align}
    \max_{k,\ell}
    \left\{
    |\mathcal H_{k\ell}(u^\circ,b^\circ)|
    +|\mathcal H_{k\ell}(b^\circ,u^\circ)|
    \right\}
    &\lesssim_\P
    \frac{J^d}{nL_T}
    \left\{
    \sqrt H\,{\mathsf{B}_{\mathtt{FS}}}
    +H{\mathsf{B}_{\mathtt{FS}}}
       \sqrt{\frac{J^d}{nL_T}}
    \right\}
    =o_\P\!\left(\frac{J^d}{nL_T}\right).
    \label{eq:forward-residual-bias-cross}
\end{align}
The last equality follows from
{$\sqrt H\,\mathsf{B}_{\mathtt{FS}}=o(1)$} and
\eqref{eq:forward-variance-proof-rates}.

\smallskip\noindent\emph{IV score.}
The maximum-instrument bound in
Assumption~\ref{ass:linearity-instrument}(ii) and
\eqref{eq:forward-variance-proof-weight-bounds} imply
\begin{align}
    \max_{t,k}
    \sum_{i\in\mathcal I_t^{\mathtt{LIN}}}
    \widehat w_{it}(\mathsf b_k)^2\|\zeta_{it}\|^2
    &\lesssim_\P\frac{J^d}{n}.
    \label{eq:forward-instrument-local-leverage}
\end{align}
Substituting
$g_s=(n_s^{\mathtt{LIN}})^{-1}
\sum_{j\in\mathcal I_s^{\mathtt{LIN}}}\zeta_{js}u_{js}^\circ$ into
$q_{is}^\circ$, for $i\in\mathcal I_s^{\mathtt{LIN}}$, gives
\begin{align*}
    q_{is}^{\circ,\mathrm{same}}
    &:=
    -\frac{
    \breve x_{i(s+H)}^\top C_s^\circ\zeta_{is}
    }{n_s^{\mathtt{LIN}}}
    u_{is}^\circ,
    \\
    q_{is}^{\circ,\mathrm{cross}}
    &:=
    -\frac{1}{n_s^{\mathtt{LIN}}}
    \sum_{\substack{j\in\mathcal I_s^{\mathtt{LIN}}\\j\ne i}}
    \breve x_{i(s+H)}^\top C_s^\circ\zeta_{js}u_{js}^\circ,
    \qquad
    q_{is}^\circ
    =
    q_{is}^{\circ,\mathrm{same}}
    +q_{is}^{\circ,\mathrm{cross}}.
\end{align*}
Set both components to zero for
$i\notin\mathcal I_s^{\mathtt{LIN}}$. Thus the displayed decomposition is
exact for every $i\in[n]$.
Accordingly, define the same-unit and cross-unit pieces of the two score
cross-products by
\begin{align*}
    \mathscr Q_{k\ell}^{\mathrm{same}}
    &:=
    \mathcal H_{k\ell}(u^\circ,q^{\circ,\mathrm{same}})
    +\mathcal H_{k\ell}(q^{\circ,\mathrm{same}},u^\circ),
    \\
    \mathscr Q_{k\ell}^{\mathrm{cross}}
    &:=
    \mathcal H_{k\ell}(u^\circ,q^{\circ,\mathrm{cross}})
    +\mathcal H_{k\ell}(q^{\circ,\mathrm{cross}},u^\circ),
    \\
    \mathcal H_{k\ell}(u^\circ,q^\circ)
    +\mathcal H_{k\ell}(q^\circ,u^\circ)
    &=
    \mathscr Q_{k\ell}^{\mathrm{same}}
    +\mathscr Q_{k\ell}^{\mathrm{cross}}.
\end{align*}
The same-unit piece need not be centered. The multiplier
$\breve x_{i(s+H)}^\top C_s^\circ\zeta_{is}$ is uniformly bounded in
probability and $(n_s^{\mathtt{LIN}})^{-1}\lesssim n^{-1}$.
Applying $2|xy|\leq x^2+y^2$ to the banded sum, then using
\eqref{eq:forward-variance-proof-weight-bounds} and the weighted
second-moment bound for $u^\circ$ used above, gives
\begin{align}
    \max_{k,\ell}
    |\mathscr Q_{k\ell}^{\mathrm{same}}|
    &\lesssim_\P
    \frac{J^d}{nL_T}\frac{H}{n}.
    \label{eq:forward-residual-same-unit-score}
\end{align}
For $j\ne i$, Assumption~\ref{ass:linearity-missingness} preserves
conditional independence of future primitive shocks across units after
selection. For each pair $t,s$, split the residual associated with the
earlier base date into its component before the later base date and its
component from the later base date onward. In the latter component, all
weights, instruments, and IV multipliers are fixed before both primitive
shocks. Let
$\mathscr Q_{k\ell}^{\mathrm{cross},\mathrm F}$ denote this contribution and
define $\{N_{ij,rr'}^{k\ell,\mathrm F}\}$ as the collected coefficients in
\begin{align*}
    \mathscr Q_{k\ell}^{\mathrm{cross},\mathrm F}
    &:=
    \frac{1}{L_T^2}
    \sum_{\substack{i,j\in[n]\\i\ne j}}
    \sum_{r=1}^T\sum_{r'=1}^T
    N_{ij,rr'}^{k\ell,\mathrm F}\varepsilon_{ir}\varepsilon_{jr'}.
\end{align*}
These coefficients include all matched-set indicators and are zero if a
required unit is unavailable. Assumption~\ref{ass:linearity-instrument}(ii),
\eqref{eq:forward-variance-proof-weight-bounds},
\eqref{eq:forward-residual-primitive-energy}, and Jensen's inequality give
\begin{align}
    \sum_{\substack{i,j\in[n]\\i\ne j}}
    \sum_{r=1}^{T}\sum_{r'=1}^{T}
    |N_{ij,rr'}^{k\ell,\mathrm F}|^2
    &\lesssim_\P
    \frac{L_T H^2J^{3d}}{n^4}.
    \label{eq:forward-residual-cross-unit-score-coefficients}
\end{align}
After ordering the compound shock indices, combine the coefficients on each
pair and its reversal, which multiply the same shock product. The predictable
quadratic-shock-sum bound in
\eqref{eq:forward-quadratic-shock-sum-bound} then gives
\begin{align*}
    \max_{k,\ell}
    |\mathscr Q_{k\ell}^{\mathrm{cross},\mathrm F}|
    &\lesssim_\P
    \sqrt{\frac{H^2J^{3d}}{n^4L_T^3}}.
\end{align*}

Consider first
$\mathcal H_{k\ell}(u^\circ,q^{\circ,\mathrm{cross}})$ in the case $t<s$,
so the IV-score term indexed by $s$ has the later base date. Its pre-$s$
contribution is linear in $u_{js}^{\circ}$. For
$j\in\mathcal I_s^{\mathtt{LIN}}$, its $\mathcal K_s$-measurable
coefficient is
\begin{align*}
    P_{js}^{k\ell,t<s}
    &:=
    -\frac{1}{n_s^{\mathtt{LIN}}}
    \sum_{h=1}^H
    \sum_{\substack{i\in\mathcal I_s^{\mathtt{LIN}}\\i\ne j}}
    \widehat w_{i,s-h}(\mathsf b_k)
    \widehat w_{is}(\mathsf b_\ell)
    \bigl\{\breve x_{i(s+H)}^\top C_s^\circ\zeta_{js}\bigr\}
    u_{i,s-h}^{<s}.
\end{align*}
As stipulated above, shifted base dates outside $\mathcal T_H$ are omitted.
Set this coefficient to zero for
$j\notin\mathcal I_s^{\mathtt{LIN}}$. Since
$n_s^{\mathtt{LIN}}\asymp n$, Cauchy--Schwarz,
\eqref{eq:forward-pre-s-energy},
\eqref{eq:forward-variance-proof-weight-bounds}, and the uniform bounds on
$\zeta_{js}$, $\breve x_{i(s+H)}$, and $C_s^\circ$ give
\begin{align*}
    \max_{k,\ell}
    \sum_j\sum_s
    (1+\|\varpi_{s+H}\|^2)
    |P_{js}^{k\ell,t<s}|^2
    &\lesssim_\P\frac{L_T HJ^{3d}}{n^3}.
\end{align*}
Equation~\eqref{eq:forward-linear-shock-sum-reindexing} therefore gives
$O_\P\!\left(\sqrt{HJ^{3d}/(n^3L_T^3)}\right)$ for the corresponding
predictable linear shock sum.

Next consider the same bilinear form in the case $s<t$, so the IV-score term
indexed by $s$ has the earlier base date. For
$s,t\in\mathcal T_H$ with $s<t$ and
$i\in\mathcal I_s^{\mathtt{LIN}}$, define
\begin{align*}
    G_{i,s,t}^{<t}
    &:=
    \frac{1}{n_s^{\mathtt{LIN}}}
    \sum_{\substack{j\in\mathcal I_s^{\mathtt{LIN}}\\j\ne i}}
    \breve x_{i(s+H)}^\top C_s^\circ\zeta_{js}u_{js}^{<t}.
\end{align*}
Set $G_{i,s,t}^{<t}=0$ outside this domain. For
$t\in\mathcal T_H$ and $i\in\mathcal I_t^{\mathtt{LIN}}$, define
\begin{align*}
    P_{it}^{k\ell,s<t}
    &:=
    -\widehat w_{it}(\mathsf b_k)
    \sum_{h=1}^H
    \widehat w_{i,t-h}(\mathsf b_\ell)
    G_{i,t-h,t}^{<t}.
\end{align*}
Set $P_{it}^{k\ell,s<t}=0$ outside this domain. Then
$P_{it}^{k\ell,s<t}$ is $\mathcal K_t$-measurable and is the
coefficient multiplying the residual $u_{it}^\circ$ at the later base date
$t$.
Assumption~\ref{ass:linearity-instrument}(ii) gives
$|G_{i,s,t}^{<t}|^2\lesssim_\P
n^{-1}\sum_{j\in\mathcal I_s^{\mathtt{LIN}}}(u_{js}^{<t})^2$.
Cauchy--Schwarz over $h$,
\eqref{eq:forward-variance-proof-weight-bounds}, and the second bound in
\eqref{eq:forward-pre-s-energy} therefore give
\begin{align*}
    \max_{k,\ell}\sum_i\sum_t
    (1+\|\varpi_{t+H}\|^2)
    |P_{it}^{k\ell,s<t}|^2
    &\lesssim_\P\frac{L_T HJ^{3d}}{n^3}.
\end{align*}
Equation~\eqref{eq:forward-linear-shock-sum-reindexing} gives the same rate
in the case $s<t$. The identity
$\mathcal H_{k\ell}(q^{\circ,\mathrm{cross}},u^\circ)
=\mathcal H_{\ell k}(u^\circ,q^{\circ,\mathrm{cross}})$ covers the other
cross term. Consequently,
\begin{align}
    \max_{k,\ell}
    |\mathscr Q_{k\ell}^{\mathrm{cross}}|
    &\lesssim_\P
    \sqrt{\frac{H^2J^{3d}}{n^4L_T^3}}
    +\sqrt{\frac{HJ^{3d}}{n^3L_T^3}}
    =o_\P\!\left(\frac{J^d}{nL_T}\right).
    \label{eq:forward-residual-cross-unit-score}
\end{align}
Relative to $J^d/(nL_T)$, the same-unit, cross-unit $s$-onward, and
cross-unit pre-$s$ bounds are
\begin{align*}
    \frac{H}{n}, \qquad
    H\sqrt{\frac{J^d}{n^2L_T}}, \qquad\text{and}\qquad
    \sqrt{\frac{HJ^d}{nL_T}}.
\end{align*}
All three are $o(1)$ by~\eqref{eq:forward-variance-proof-rates}.

\smallskip\noindent\emph{Conclusion.}
The quadratic and random-multiplier terms are controlled by
\eqref{eq:forward-residual-delta-delta} and
\eqref{eq:forward-residual-r-cross}. The bias and score terms are controlled
by \eqref{eq:forward-residual-bias-cross},
\eqref{eq:forward-residual-same-unit-score}, and
\eqref{eq:forward-residual-cross-unit-score}. Substitution into
\eqref{eq:forward-residual-bilinear-decomposition} and
\eqref{eq:forward-residual-cross-decomposition} gives, uniformly over the
fixed grid,
\begin{align}
    \max_{k,\ell\in[K_{\mathtt{LIN}}]}
    \left|
    \widehat\Sigma_{\mathtt{LIN}}(\mathsf{b}_k,\mathsf{b}_\ell)
    -\widetilde\Sigma_{k\ell}
    \right|
    &\lesssim_\P
    \frac{J^d}{nL_T}
    He_{nT}^2
    +o_\P\!\left(\frac{J^d}{nL_T}\right)
    \notag\\
    &=
    o_\P\!\left(\frac{J^d}{nL_T}\right).
    \label{eq:forward-residual-replacement-rate}
\end{align}
Step~1, specifically
\eqref{eq:forward-oracle-diagonal-rate} and
\eqref{eq:forward-oracle-off-diagonal-rate}, also gives
\begin{align*}
    \max_{k,\ell\in[K_{\mathtt{LIN}}]}
    \left|
    \widetilde\Sigma_{k\ell}
    -\Sigma_{\mathtt{LIN}}(\mathsf b_k,\mathsf b_\ell)
    \right|
    &=
    o_\P\!\left(\frac{J^d}{nL_T}\right).
\end{align*}
The triangle inequality and fixed $K_{\mathtt{LIN}}$ therefore give
\begin{align*}
    \left\|
    \widehat\Sigma_{\mathtt{LIN}}-\Sigma_{\mathtt{LIN}}
    \right\|
    &\leq
    K_{\mathtt{LIN}}
    \max_{k,\ell\in[K_{\mathtt{LIN}}]}
    \left|
    \widehat\Sigma_{\mathtt{LIN}}(\mathsf b_k,\mathsf b_\ell)
    -\Sigma_{\mathtt{LIN}}(\mathsf b_k,\mathsf b_\ell)
    \right|
    \\
    &=
    o_\P\!\left(\frac{J^d}{nL_T}\right),
\end{align*}
which is~\eqref{eq:forward-linearity-variance-consistency}.
\qed

\subsection{Proof of Theorem~\ref{thm:forward-linearity}}

By Lemma~\ref{lem:forward-linearity-variance},
\eqref{eq:forward-linearity-nondegeneracy}, and
\eqref{eq:forward-linearity-variance-stabilization},
$\lambda_{\min}(\Sigma_{\mathtt{LIN}})
\gtrsim_\P J^d/\{n(T-2H)\}$ and hence
\begin{align*}
    \left\|
    \Sigma_{\mathtt{LIN}}^{-1/2}
    \widehat\Sigma_{\mathtt{LIN}}
    \Sigma_{\mathtt{LIN}}^{-1/2}
    -I_{K_{\mathtt{LIN}}}
    \right\|
    &=o_\P(1).
\end{align*}
It follows that $\widehat\Sigma_{\mathtt{LIN}}$ is nonsingular with
probability approaching one. Put
$Z_{\mathtt{LIN}}
=\Sigma_{\mathtt{LIN}}^{-1/2}
\widehat{\boldsymbol\nu}_{\mathtt{LIN}}$ and
$A_{\mathtt{LIN}}
=\Sigma_{\mathtt{LIN}}^{-1/2}
\widehat\Sigma_{\mathtt{LIN}}
\Sigma_{\mathtt{LIN}}^{-1/2}$.
Then
\begin{align*}
    \widehat{\boldsymbol\nu}_{\mathtt{LIN}}^\top
    \widehat\Sigma_{\mathtt{LIN}}^{-1}
    \widehat{\boldsymbol\nu}_{\mathtt{LIN}}
    &=
    Z_{\mathtt{LIN}}^\top A_{\mathtt{LIN}}^{-1}Z_{\mathtt{LIN}},
    \\
    Z_{\mathtt{LIN}}^\top A_{\mathtt{LIN}}^{-1}Z_{\mathtt{LIN}}
    &\rightsquigarrow
    \chi^2_{K_{\mathtt{LIN}}}.
\end{align*}
The convergence follows from
Lemma~\ref{lem:forward-linearity-normality}, Slutsky's theorem, and the
continuous mapping theorem.
Continuity of the chi-square distribution at its
$1-\alpha$ quantile gives the asserted asymptotic size.
\qed

\subsection{Proof of Lemma~\ref{lem:forward-old-beta-instrument}}

Write
\begin{align*}
    \widehat x_{it}^{\,\mathtt{old}}
    =(1,\widehat\beta_{it}^\top)^\top,
    \qquad
    \breve x_{it}
    =(1,\breve\beta_{it}^\top)^\top.
\end{align*}
Because the rolling window defining $\widehat\beta_{it}$ ends at $t-1$,
$\widehat x_{it}^{\,\mathtt{old}}$ is $\mathcal F_{t-1}$-measurable and hence
$\mathcal K_t$-measurable. This proves
Assumption~\ref{ass:linearity-instrument}(i).

Lemma~\ref{lem:first-step-rate}, the boundedness of $\beta_{it}$ in
Assumption~\ref{ass:dgp}(iii), and $\mathsf R_{\mathtt{FS}}=o(1)$ give
\begin{align*}
    \max_{H+1\leq t\leq T-H}
    \max_{i\in\mathcal I_t^{\mathtt{LIN}}}
    \|\widehat\beta_{it}\|
    &\lesssim_\P1.
\end{align*}
Consequently,
\begin{align*}
    \max_{H+1\leq t\leq T-H}
    \frac{1}{n_t^{\mathtt{LIN}}}
    \sum_{i\in\mathcal I_t^{\mathtt{LIN}}}
    \|\widehat x_{it}^{\,\mathtt{old}}\|^4
    &\lesssim_\P1.
\end{align*}
The last two displays verify
Assumption~\ref{ass:linearity-instrument}(ii).

The future-window decomposition and conditional mean-zero property imply
\begin{align*}
    \mathsf M_t^\circ
    &=
    \frac{1}{n_t^{\mathtt{LIN}}}
    \sum_{i\in\mathcal I_t^{\mathtt{LIN}}}
    \widehat x_{it}^{\,\mathtt{old}}\breve x_{i(t+H)}^\top.
\end{align*}
The first-step rate and boundedness of the true betas therefore give
\begin{align*}
    \max_{H+1\leq t\leq T-H}
    \|\mathsf M_t^\circ-\mathsf M_{t,\beta}^0\|
    &\lesssim_\P
    \mathsf R_{\mathtt{FS}}
    =o_\P(1).
\end{align*}
Equation~\eqref{eq:forward-old-beta-relevance} and a matrix perturbation bound
then yield
\begin{align*}
    \max_{H+1\leq t\leq T-H}
    \left\{\|\mathsf M_t^\circ\|+\|(\mathsf M_t^\circ)^{-1}\|\right\}
    &\lesssim_\P1.
\end{align*}

Finally, with $(0,v_{i(t+H)}^\top)$ interpreted as a row vector,
\begin{align*}
    \widehat{\mathsf M}_t-\mathsf M_t^\circ
    &=
    \frac{1}{n_t^{\mathtt{LIN}}}
    \sum_{i\in\mathcal I_t^{\mathtt{LIN}}}
    \widehat x_{it}^{\,\mathtt{old}}
    (0,v_{i(t+H)}^\top).
\end{align*}
Conditional on $\mathcal K_t$, the summands are centered and independent
across units; Assumption~\ref{ass:linearity-missingness} preserves this
property after restriction to the matched set. The first-step Gram bound and
the conditional moment bounds in
Assumption~\ref{ass:dgp}(ii) give
\begin{align*}
    \E[\|v_{i(t+H)}\|^2\mid\mathcal K_t]
    &\lesssim_\P H^{-1}.
\end{align*}
Applying the conditional Fuk--Nagaev inequality and a union bound over the
base dates, using $T^{2/q}/H=o(1)$ and
$cn\leq n_t^{\mathtt{LIN}}\leq n$, yields
\begin{align*}
    \max_{H+1\leq t\leq T-H}
    \|\widehat{\mathsf M}_t-\mathsf M_t^\circ\|
    &\lesssim_\P
    \sqrt{\frac{\log(nT)}{nH}}
    +\frac{(nT)^{1/q}\log(nT)}{nH}
    \lesssim_\P
    \sqrt{\frac{\log(nT)}{n}}.
\end{align*}
This proves the concentration part of
Assumption~\ref{ass:linearity-instrument}(iii) and completes the verification.
\qed

\bibliography{CCFW_2026_AER--bib}
\bibliographystyle{jfe}

%% file: arxiv-crossrefs.tex
\makeatletter
\newcommand{\arxiveqref}[1]{\textup{\tagform@{\ref*{#1}}}}
\makeatother

%% file: tables/predictive-is.tex
(log) Dividend Price Ratio & 0.09 & 0.31 & 0.71 & 1.45 \\
(log) Dividend Yield & 0.11 & 0.32 & 0.74 & 1.48 \\
(log) Earnings Price Ratio & 0.03 & 0.04 & 0.06 & 0.25 \\
(log) Dividend Payout Ratio & 0.02 & 0.20 & 0.56 & 0.74 \\
Stock Variance & 1.01 & 0.11 & 0.15 & 0.65 \\
Book-to-Market Ratio & 0.00 & 0.01 & 0.07 & 0.12 \\
Net Equity Expansion & 0.15 & 0.22 & 0.48 & 1.25 \\
Treasury Bill Yield & 0.40 & 0.79 & 1.09 & 1.57 \\
Long Term Treasury Yield & 0.13 & 0.17 & 0.14 & 0.00 \\
Long Term Treasury Return & 1.16 & 0.71 & 1.88 & 1.34 \\
Term Spread & 0.52 & 1.38 & 2.50 & 6.99 \\
Default Yield Spread & 0.26 & 0.82 & 2.31 & 2.88 \\
Default Return Spread & 0.30 & 0.24 & 0.29 & 0.02 \\
(lagged) Inflation & 0.01 & 0.46 & 1.84 & 2.29 \\[1 ex]
CGP & 1.26 & 3.04 & 3.65 & 4.06 \\[1 ex]
BCF & 2.89 & 7.87 & 10.82 & 12.68 \\

%% file: tables/CCFW_2026_AER--credit-main-contrasts.tex
\begin{tabular}{lrrr}
\hline\hline
Variance estimator & SE & 90\% CI & $p$-value \\
\hline
JN & 0.304 & $[-1.049,-0.050]$ & 0.070 \\
Modified FM & 0.285 & $[-1.019,-0.080]$ & 0.054 \\
$\widehat\Sigma_f$ only & 0.234 & $[-0.934,-0.165]$ & 0.019 \\
\hline\hline
\end{tabular}

%% file: tables/predictive-oos.tex
(log) Dividend Price Ratio & -1.79 & -5.51 & -12.00 & -27.91 \\
(log) Dividend Yield & -1.94 & -5.35 & -12.12 & -26.96 \\
(log) Earnings Price Ratio & -0.90 & -3.70 & -8.55 & -14.87 \\
(log) Dividend Payout Ratio & -1.50 & -4.51 & -6.51 & -5.08 \\
Stock Variance & {\bf 0.79} & -4.15 & -1.54 & -0.35 \\
Book-to-Market Ratio & -0.60 & -2.01 & -4.68 & -11.79 \\
Net Equity Expansion & -2.35 & -7.20 & -16.86 & -24.49 \\
Treasury Bill Yield & {\bf 0.02} & -0.08 & -0.38 & -0.55 \\
Long Term Treasury Yield & -0.03 & -0.77 & -2.51 & -8.44 \\
Long Term Treasury Return & -0.14 & -1.10 & -0.84 & -0.98 \\
Term Spread & -0.98 & -2.12 & -1.68 & {\bf 4.26} \\
Default Yield Spread & -2.43 & -5.89 & -6.84 & -5.46 \\
Default Return Spread & -0.57 & -1.09 & -0.43 & -0.95 \\
(lagged) Inflation & -0.61 & -0.08 & {\bf 2.56} & {\bf 2.58} \\[1 ex]
CGP & {\bf 1.67} & {\bf 3.98} & {\bf 4.41} & {\bf 3.88} \\[1 ex]
BCF & {\bf 3.00} & {\bf 9.60} & {\bf 13.90} & {\bf 11.88} \\

%% file: tables/CCFW_2026_AER--llm-capital-share-contrasts.tex
\begin{tabular}{lrrr}
\hline\hline
Variance estimator & SE & 90\% CI & $p$-value \\
\hline
JN & 1.030 & $[-1.162,2.227]$ & 0.605 \\
Modified FM & 0.634 & $[-0.511,1.576]$ & 0.401 \\
$\widehat\Sigma_f$ only & 0.874 & $[-0.904,1.969]$ & 0.542 \\
\hline\hline
\end{tabular}